\documentclass[twoside]{article}
\usepackage[dvips]{graphicx,color} 
\usepackage{amsmath,amssymb}   
\newcommand{\bce}{\begin{center}}
\newcommand{\ece}{\end{center}}
\newcommand{\bea}{\begin{eqnarray}}
\newcommand{\eea}{\end{eqnarray}}
\newcommand{\be}{\begin{equation}}
\newcommand{\ee}{\end{equation}}
\newcommand{\bd}{\begin{displaymath}}
\newcommand{\ed}{\end{displaymath}}
\newcommand{\bit}{\begin{itemize}}
\newcommand{\eit}{\end{itemize}}
\newcommand {\ben}{\begin{enumerate}}
\newcommand{\een}{\end{enumerate}}
\newcommand{\bdes}{\begin{description}}
\newcommand{\edes}{\end{description}}

\newcommand{\E}{\> = \>}
\newcommand{\EA}{&=&}
\newcommand{\EQ}{\> \equiv \>}
\newcommand{\no}{\nonumber}
\newcommand{\non}{\nonumber\\}
\newcommand{\To}{\> \longrightarrow \>}
\newcommand{\Def}{\> := \>}
\newcommand{\deF}{\> =: \>}
\newcommand{\Tint}{\int_{-T}^{+T} dt}
\newcommand{\sgn}{{\rm sgn}}
\newcommand{\bfx}[1]{\mbox{\boldmath $#1$}}
\newcommand{\fb}{{\bf b}}             
\newcommand{\fk}{{\bf k}}             
\newcommand{\fK}{{\bf K}}             
\newcommand{\fp}{{\bf p}}             
\newcommand{\fq}{{\bf q}}             
\newcommand{\fr}{{\bf r}}             
\newcommand{\fv}{{\bf v}}             
\newcommand{\fw}{{\bf w}}             
\newcommand{\fx}{{\bf x}}             
\newcommand{\fy}{{\bf y}}             
\newcommand{\fz}{{\bf z}}             
\newcommand{\fxi}{\mbox{\boldmath $\xi$ }}       
\newcommand{\fdet}{{\cal D}{\rm et}}  
\newcommand{\A}{\mathbb A}            
\newcommand{\B}{\mathbb B}            
\newcommand{\He}{\mathbb H}           
\newcommand{\C}{\mathbb C}            
\newcommand{\dslash}{\partial \hspace{-5pt}/}
\newcommand{\diffslash}{\partial \hspace{-6pt}/}
\newcommand{\kslash}{k \hspace{-5pt}/}
\newcommand{\pslash}{p \hspace{-5pt}/}
\newcommand{\Aslash}{A \hspace{-6pt}/}
\newcommand{\ddslash}{D \hspace{-7pt}/}
\newcommand{\dddslash}{D \hspace{-5pt}/}
\newcommand{\vslash}{v\hspace{-5pt}/}
\newcommand{\vvslash}{v\hspace{-4pt}/}
\newcommand{\la}{\left\langle \,}
\newcommand{\ra}{\, \right\rangle}
\newcommand{\lrp}{\left ( \, }    
\newcommand{\rrp}{\, \right ) }   
\newcommand{\lsp}{\left [ \, }    
\newcommand{\rsp}{\, \right ] }   
\newcommand{\lcp}{\left \{ \, }   
\newcommand{\rcp}{\, \right \} }  
\def\lvl{\, \left | \, }     
\def\rvl{\, \right | \, }    
\newcommand{\rot}{\textcolor{red}}
\newcommand{\blau}{\textcolor{blue}}
\newcommand{\purpur}{\textcolor{magenta}}

\newcommand{\cy}{\textcolor{cyan}}
\newcommand{\latop}[2]{{#1\atop#2}}
\renewcommand{\baselinestretch}{1.2}

\newcounter{abb}
\newcounter{ueb}                  
\newcounter{tief}                 

\begin{document}

\definecolor{mygreen}{rgb}{0.4,0.6,0}
\newcommand{\meingruen}{\textcolor{mygreen}}
\definecolor{gruenblau}{rgb}{0,0.4,0.6}

\thispagestyle{empty}
\setcounter{page}{1}
\setcounter{section}{-1}
\setcounter{footnote}{0}
\setcounter{abb}{-1}
\setlength{\fboxrule}{0.7mm}


\vspace{0.3cm}
\bce
\fcolorbox{red}{yellow}{\parbox{13cm}
{
\bea
&& \non
&& \hspace{-1.2cm}\mbox{\LARGE  \bf \qquad Path Integrals $\>\>$  in $\>\>$  
Quantum Physics \qquad} \non
&& \no
\eea
}}

\vspace{0.8cm}
Lectures given at ETH Zurich 

\vspace{0.5cm}
R.~Rosenfelder

\vspace{0.3cm}
Paul Scherrer Institute, CH-5232 Villigen PSI, Switzerland

\end{center}

\vspace{1.5cm}

\begin{abstract}                                       
\noindent                                              
These lectures are intended for graduate students who want to acquire 
a working knowledge of path integral methods
in a wide variety of fields in physics. In general the presentation is elementary and path integrals are developed
in the usual heuristic, non-mathematical way for application in
many diverse problems in quantum physics.
Three main parts deal with path integrals in non-relativistic 
quantum mechanics, many-body physics and field theory and contain standard 
examples (quadratic Lagrangians, tunneling, description of bosons and fermions etc.) as well as specialized topics (scattering, dissipative systems, spin \& color in the path integral, lattice methods etc.).
In each part simple Fortran programs which can be run on a PC, illustrate the numerical evaluation of 
(Euclidean) path integrals by Monte-Carlo or variational methods.
Also included are the set of problems which accompanied the lectures and their solutions.                

\end{abstract}

\newpage  

\pagestyle{myheadings}
\markboth{\textcolor{green}{Content}}{\textcolor{green}{R. Rosenfelder : 
Path Integrals in Quantum Physics}}

\section{\textcolor{red}{Contents}}
\label{sec1: Inhalt}
\renewcommand{\theequation}{0.\arabic{equation}}
\vspace{0.3cm}

First, an overview over the planned topics.
The subsections marked by $ \> \> ^{\star}$ are optional and may be left out if there is 
no time available whereas the chapters printed in \blau{blue} deal with basic concepts.
Problems from the optional chapters or referring to "\blau{Details}" are marked by a $ \> \> ^{\star}$ as well.

\vspace{0.4cm}
\bdes
\item[\rot{1.}]\rot{\bf Path Integrals in Non-Relativistic Quantum Mechanics of a Single Particle}
\ben
\item \blau{Action Principle and Sum over all Paths} \hspace{0.2cm}......................................................................\hspace{0.2cm}6
\item \blau{Lagrange, Hamilton and other Path-Integral Formulations} \hspace{0.2cm}........................................\hspace{0.2cm}10
\item \blau{Quadratic Lagrangians}  \hspace{0.2cm}................................................................................................\hspace{0.2cm}22
\item Perturbation Theory $^{\star}$  \hspace{0.2cm}................................................................................................\hspace{0.2cm}26
\item \blau{Semiclassical Expansions}\hspace{0.2cm}..............................................................................................\hspace{0.2cm}29
\item Potential Scattering and Eikonal Approximation $^{\star}$ \hspace{0.2cm}.....................................................\hspace{0.2cm}33
\item \blau{Green Functions as Path Integrals} \hspace{0.2cm}..............................................................................\hspace{0.2cm}44
\item Symmetries and Conservation Laws $^{\star}$\hspace{0.2cm}..........................................................................\hspace{0.2cm}53
\item Numerical Treatment of Path Integrals $^{\star}$\hspace{0.2cm}.....................................................................\hspace{0.2cm}56
\item Tunneling and Instanton Solutions $^{\star}$\hspace{0.2cm}............................................................................\hspace{0.2cm}63	
\een
\item[\rot{2.}] \rot{\bf Path Integrals in
Statistical Mechanics  and Many-Body Physics}
\ben
\item \blau{Partition Function}\hspace{0.2cm}.........................................................................................................\hspace{0.1cm}72
\item \blau{The Polaron}\hspace{0.2cm}.................................................................................................................\hspace{0.2cm}74
\item Dissipative Quantum Systems $^{\star}$\hspace{0.2cm}...................................................................................\hspace{0.2cm}79
\item \blau{Particle Number Representation and Path Integrals over Coherent States}\hspace{0.2cm}.................\hspace{0.25cm}88
\item \blau{Description of Fermions: Grassmann Variables}\hspace{0.2cm}...........................................................\hspace{0.2cm}95
\item Perturbation Theory and Diagrams $^{\star}$\hspace{0.2cm}..........................................................................\hspace{0.2cm}98
\item Auxiliary Fields and Hartree Approximation $^{\star}$\hspace{0.2cm}..........................................................\hspace{0.2cm}105
\item Asymptotic Expansion of a Class of Path Integrals $^{\star}$\hspace{0.2cm}................................................\hspace{0.2cm}109  
\een
\item[\rot{3.}] \rot{\bf Path Integrals in Field Theory}
\ben
\item \blau{Generating Functionals and Perturbation Theory }\hspace{0.2cm}...................................................\hspace{0.2cm}120   
\item Effective Action $^{\star}$ \hspace{0.2cm}....................................................................................................\hspace{0.2cm}135            
\item \blau{Quantization of Gauge Theories}\hspace{0.2cm}...............................................................................\hspace{0.2cm}139
\item Worldlines and Spin in the  Path Integral $^{\star}$\hspace{0.2cm}..............................................................\hspace{0.2cm}149
\item Anomalies $^{\star}$\hspace{0.2cm}..............................................................................................................\hspace{0.2cm}157
\item Lattice Field Theories $^{\star}$\hspace{0.2cm}............................................................................................\hspace{0.1cm}162
\een

\edes

\noindent
Additional Literature\hspace{0.2cm}................................................................................................................\hspace{0.2cm}175\\
Original Publications \hspace{0.2cm}...............................................................................................................\hspace{0.2cm}176\\
Problems\hspace{0.2cm}...................................................................................................................................\hspace{0.2cm}180\\
Solutions\hspace{0.2cm}...................................................................................................................................\hspace{0.2cm}195

\newpage
\noindent
{\bf \large Literature:}\\
\noindent
There is a large number of text books which are dealing with path integrals either
as a tool for specific problems or as basis for an unified treatment.
In the following a small selection is presented in which my favorite books (whom I mostly follow \footnote{"\textsf{(My father used to say: 'If you steal from one book, you are condemned as a plagiarist, but if you steal from ten books, you are considered a scholar, and if you steal from thirty or fourty books, a distinguished scholar.')}" {\bf \{Oz\}}, p.129.})
are marked by the symbol
$\meingruen{\clubsuit}$.
\vspace{0.5cm}

\noindent
\rot{\bf Section 1 :} 

\bit
\item {\meingruen{\bf R. P. Feynman and A. R. Hibbs}}: {\it Quantum Mechanics and
Path Integrals}, McGraw-Hill (1965).
\item {\meingruen{\bf L. S. Schulman}}: {\it Techniques and Applications of Path
Integration}, John Wiley (1981).  $ \> \meingruen{\clubsuit}$ 
\item {\meingruen{\bf J. Glimm and A. Jaffe}}: {\it Quantum Physics: A Functional Point
of View}, Springer (1987).
\item {\meingruen{\bf G. Roepstorff}}: {\it Path Integral Approach to Quantum Physics},
Springer (1994).
\item {\meingruen{\bf G. W. Johnson and M. L. Lapidus}}: {\it The Feynman Integral and 
Feynman's Operational Calculus}, Oxford University Press (2000).
\item{\meingruen{\bf J. Zinn-Justin}}: {\it Path Integrals in Quantum Mechanics}, Oxford University
Press (2006).

\eit
 
\vspace{0.2cm}

\noindent
\rot{\bf Section 2 :}

\bit
\item {\meingruen{\bf R. P. Feynman}}: {\it Statistical Mechanics}, Benjamin (1976).
\item {\meingruen{\bf V. N. Popov}}: {\it Functional Integrals in Quantum Field Theory and
Statistical Physics}, Reidel (1983).
\item {\meingruen{\bf J. W. Negele and H. Orland}}: {\it Quantum Many-Particle 
Systems}, Addison-Wesley (1987).  $\> \meingruen{\clubsuit}$
\item {\meingruen{\bf J. Zinn-Justin}}: {\it Quantum Field Theory and Critical Phenomena}, 4th ed.,
Oxford U. Press (2002).
\item {\meingruen{\bf H. Kleinert}}: {\it Path Integrals in Quantum Mechanics, Statistics,
Polymer Physics, and Financial Markets}, 3rd ed., World Scientific (2004).
\eit

\vspace{ 0.2cm}

\noindent
\rot{\bf Section 3 :}

\bit
\item {\meingruen{\bf C. Itzykson and J.-B. Zuber}}: {\it Quantum Field Theory}, McGraw-Hill
(1980), ch. 9. 
\item {\meingruen{\bf T.-P. Cheng and L.-F. Li}}: {\it Gauge Theory of Elementary Particle
Physics}, Clarendon (1988).
\item {\meingruen{\bf R. J. Rivers}}: {\it Path Integral Methods in Quantum Field 
Theory}, Cambridge University Press (1990).
\item {\meingruen{\bf M. E. Peskin and D. V. Schroeder}}: {\it An Introduction to 
Quantum Field Theory}, Addison-Wesley \\(1995), ch. 9. $\> \meingruen{\clubsuit}$ 
\item{\meingruen{\bf V. Parameswaran Nair}}: {\it Quantum Field Theory. A Modern Perspective},
Springer (2005). $ \> \meingruen{\clubsuit}$
\item{\meingruen{\bf H. J. Rothe}}: {\it Lattice Gauge Theories. An Introduction}, 3rd ed., 
World Scientific Lecture Notes in Physics, Vol. 74,
World Scientific (2005).
\item{\meingruen{\bf A. Das}}: {\it Field Theory. A Path Integral Approach}, 2nd ed.,
World Scientific Lecture Notes in Physics, Vol. 75,
World Scientific (2006). $\meingruen{\clubsuit}$
\eit

\vspace{0.5cm}
\noindent
{\bf \large Prerequisites: }

\noindent
Quantum Mechanics: A two-semester course, a little bit of Statistical Mechanics, 
basic concepts of Field Theory. 
\vspace{0.5cm}

\noindent
\purpur{\bf \large Practice Lessons:} 

\noindent
Participation in the Practice Sessions is strongly recommended -- according to the motto

\vspace{0.6cm}

\refstepcounter{abb}
\begin{figure}[hbtp]
\bce
\includegraphics[angle=0,scale=0.45]{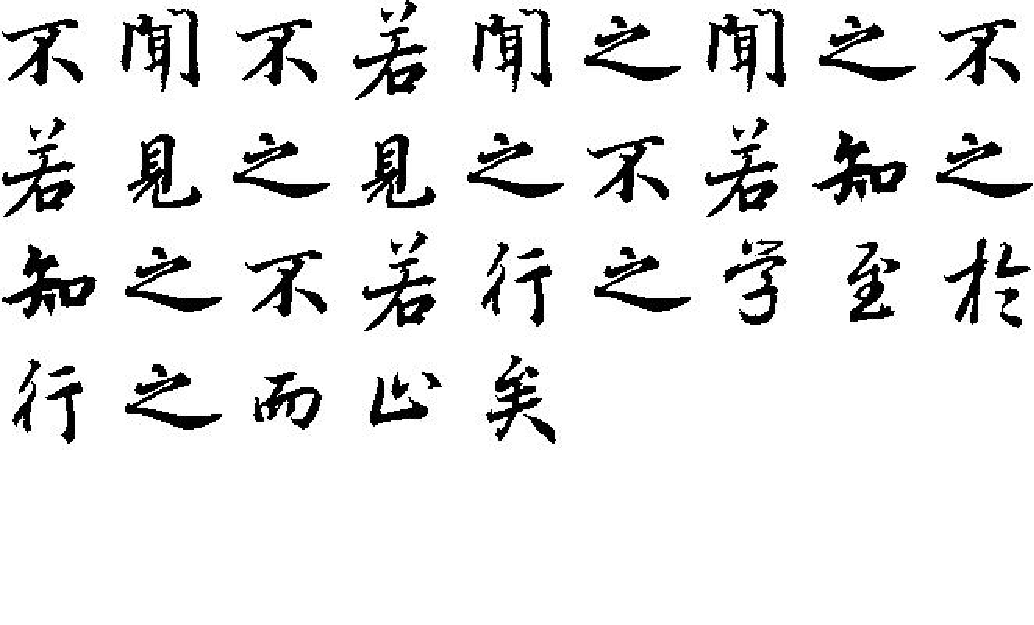}
\label{Xunzi Chin}
\ece
\end{figure}
\vspace{-2.5cm}

\bce
\bea
&&  \mbox{\textsf{ ``I hear -- and I forget,}} \non
&& \mbox{\textsf{  I see -- and I remember,}} 
 \non
&&\mbox{\textsf{   I do -- and I understand !''}} \no
\label{Xunzi Englisch}
\eea

\vspace{0.2cm}

\quad(Chinese proverb \footnote{Attributed to Xun Kuang 
(c. 310 –- c. 235 BC) known as Xunzi ("Master Xun"), a Chinese Realist Confucian philosopher 
(see https://en.wikipedia.org/wiki/Xun$_-$Kuang).})

\ece

\vspace{0.6cm}

\noindent
{\bf \large Remarks:} 

\noindent
In order to find some orientation in the jungle
of (over 1300 listed) equations, the most important equations are framed
according to the following scheme
\vspace{0.6cm}

\bce
\fcolorbox{red}{white}{\parbox{4cm}
{
\mbox{\bf \qquad fundamental}  \no
}}
\vspace{-0.5cm}

\hspace{9cm} \qquad (\mbox{ 9 times})
\vspace{1cm}

\fcolorbox{blue}{white}{\parbox{4cm}
{
\mbox{\bf \qquad very important}  \no

}}
\vspace{-0.5cm}

\hspace{9cm} \qquad (\mbox{ 24 times})
\vspace{1cm}

\fcolorbox{black}{white}{\parbox{4cm}
{
\mbox{\bf \qquad \quad important}   \no

}}
\vspace{-0.5cm}

\hspace{9.5cm} \qquad (\mbox{ 101 times}) \quad ,     

\ece
\vspace{0.3cm}
where, of course, my rating may be debatable $\ldots $ \\
\vspace{0.1cm} 

\noindent
The ``\blau{\bf Details}'' in small print give more in-depth information or detailed derivations and 
can be omitted in a first study.
Additional literature is referenced in the text in curly brackets, e.g. {\bf \{Weiss\}} and is listed
at the end in aphabetical order. Original publications are cited by consecutive numbers, e.g. [11], which
are collected at the end of the text just before the \purpur{\bf Problems}. \\
\noindent
The following {\bf abbreviations} are frequently used in the text:\\
\hspace*{0.3cm} {\it w.r.t.} : with respect to  \qquad {\it i.e.} : (Latin: {\it id est}) that is \qquad {\it viz.} : (Latin: {\it videlicet}) namely \\
\hspace*{0.3cm} {\it l.h.s.} : left-hand side \hspace{0.9cm} {\it r.h.s.} : right-hand side 
\hspace*{1.2cm} {\it e.g.} : (Latin: {\it exemplum gratum}) for example \\
\hspace*{0.3cm} {\it cf.} : (Latin: {\it confer}) compare \hspace{1.3cm} {\it q.e.d.} : (Latin: {\it quod erat demonstrandum}) which was to be proved.\\
\vspace{0.3cm}

\noindent
Despite the word "Integral" appearing in the title no knowledge
of particular integrals is needed -- the only integral which appears again and again
in various (multi-dimensional) generalizations is
the \blau{\bf Gaussian integral}
\bce

\fcolorbox{red}{white}{\parbox{8cm}
{
\bea
G(a) \EA \int_{-\infty}^{+\infty} dx \> e^{- a x^2} \E 
\sqrt{\frac{\pi}{a}}\> , \quad {\rm Re} \> \> a > 0 \> . \no
\eea
}}
\ece
\vspace{-2cm}

\bea
\label{Gauss 1}
\eea
\vspace{0.2cm}

\vspace{0.4cm}

\renewcommand{\baselinestretch}{0.9}
\begin{subequations}
\scriptsize
\noindent
{\bf Proof:} Evaluate 
\be
G^2(a) \E  \int_{-\infty}^{+\infty} \int_{-\infty}^{+\infty}
dx \> dy \> \exp \lsp- a (x^2 + y^2) \rsp
\ee
in polar coordinates $x = r \cos \phi , y = r \sin \phi, \> dx dy = r dr d\phi \> $. 
Then one obtains
\be
G^2(a) \E \int_0^{2 \pi} d\phi \> \int_0^{\infty} dr \, r \, \exp(-a r^2)
\E 2 \pi \int_0^{\infty} dr \left ( - \frac{1}{2 a}  \frac{d}{dr} \right)
\> \exp(-a r^2) 
\E \frac{\pi}{a}\> .
\ee
The constraint for the complex parameter $ \> a \ $ is needed for convergence of the
integral or for vanishing of the integrated term at infinity, respectively.

\end{subequations}
\renewcommand{\baselinestretch}{1.2}
\normalsize
\vspace{1.4cm}

\renewcommand{\baselinestretch}{0.9}
\small
\noindent
{\bf Some Personal Words at the End of Section 0:}
\vspace{0.5cm}

\noindent
These notes have their origin in material handed over to students who attended my lectures. These I have 
given first at the universities of Mainz and Hannover in Germany and later for many years 
at the ETH Zurich in Switzerland.
\vspace{0.2cm}

\noindent
Since a long time I have been fascinated by the elegance and versatility of this particular
description of the quantum world and I hope that also these notes can convey
both aspects -- beauty and utility -- to some of the readers.
In addition to the "canonical" topics (which one has to deal with in a lecture on Path Integrals)
I have added several chapters on themes in which I was particularly interested or to which I have added 
own contributions. In this way a rather personal collection of path integral 
methods in quantum physics \footnote{I have left out applications in financial industry for
obvious reasons} has been created which may be attractive for some readers but obviously cannot
replace a detailed text book.
\vspace{0.2cm}

\noindent
Due to practical reasons (and insufficient knowledge) I have not tried to derive Feynman's path integral
with strict mathematical rigor but I have chosen the usual heuristic and descriptive time-splitting approach. 
In order to show that one can use these functional integrals also for numerical work 
I have included small (and unsophisticated) FORTRAN programs in which Monte-Carlo methods 
play an important role.
\vspace{0.2cm}

\noindent
I am indebted to all students asking "silly questions" which I could not answer immediately but which
gave me (and hopefully them) a better understanding of the topic; of course, all remaining errors
and deficiencies of these notes remain in my responsibility.
Nadia Fettes and Mirko Birbaumer detected errors in the original lattice program which I hadn't seen.
I would like to thank Qiang Li who unearthed the original chinese proverb for me
when he was in the Theory Group of PSI. Valeri Markushin came to my rescue when
my proven drawing program suddenly became obsolete ... With Julien Carron I had a very agreeable 
collaboration on complex Gaussian integrals and other curiosities. Thanks also to Matthias
who gave me an important hint about the proper use of the German language ...
\vspace{0.3cm} 

\noindent
The present notes are (a rough)  English version of the German text arXiv:1209.1315 v3  which 
includes many corrections and additions compared to the previous versions. In addition, (my) Solutions to the Problems are also attached.

\noindent
This is a good opportunity to thank Manfred Stingl for his constant support and encouragement and 
to appreciate Ingrid's skills for finding a particular, proper citation.            
\vspace{0.5cm}

\hspace{2cm} Villigen PSI, July 2017              \hspace{4cm} Roland Rosenfelder
\vspace{1.4cm}

\renewcommand{\baselinestretch}{1.2}
\normalsize

\newpage

\pagestyle{myheadings}
\markboth{\textcolor{green}{Section 1 : Quantum Mechanics}}{\textcolor{green}{R. Rosenfelder : 
Path Integrals in Quantum Physics}}



\section{\textcolor{red}{Path Integrals in Non-Relativistic Quantum Mechanics of a Single Particle}}

\setcounter{equation}{0}

\renewcommand{\thesubsection}{\textcolor{blue}{1.\arabic{subsection}}}
\renewcommand{\theequation}{1.\arabic{equation}}

\subsection{\textcolor{blue}{Action Principle and Sum over all Paths}}
\label{sec1: Summe ueber Pfade}


There are essentially three formulations \footnote{The article \cite{Styer}
counts nine!} of quantum mechanics:

\bit
\item Matrix mechanics (Heisenberg 1925)
\item Wave mechanics (Schr\"odinger 1926)
\item Path integrals (\blau {\bf Feynman 1942/1948)}
\eit

\noindent 
The first two versions whose equivalence was shown very soon by Schr\"odinger,
Dirac et al., are presented in every quantum mechanics course, or textbook; the last one
is the subject of this lecture. It was developed by
\blau{\bf Feynman} \cite{Feyn1}
and is based on a work of \blau{\bf Dirac} \cite{Dirac} (both reprinted in \cite{Schwing}).
For a long time this formulation was considerd as too difficult
and useless for practical purposes, so that it was not included in a textbook.
However, conceptually the Feynman path integral is easier to grasp
and requires far less radical departure from the ideas of
classical physics (to which we are all used) than the  usual
quantum mechanics with their operators and state vectors in a Hilbert space.

As for the practical side of the various formulations, it is
undisputed that for many problems Schr\"odinger's wave mechanics is
the fastest and easiest way to obtain result: For example, when calculating
the stationary  states of a particle in a given potential. But
in many-particle problems the Schr\"odinger equation is
of much less importance and Heisenberg's matrix mechanics of Heisenberg plays a
much larger role (e.g. when diagonalization the Hamiltonian matrix in a
certain subspace). Similarly, the method of the path integrals
has become an indispensable tool in field theory: The Feynman rules
for non-abelian gauge theories were first derived in this way and the
attempts to treat numerically the field theory of the strong interaction are based
directly on the (Euclidean) path integral representation of
Quantum Chromodynamics.

Moreover, what this method distinguishes from the other formulations, is its applicability in
many areas of physics. This ``unification'' of diverse fields
not only helps in understanding, but also gives suggestions to
apply new methods that have been successful in other areas.

One caveat: The Feynman path integral is not defined mathematically in a rigorous manner
except for special cases
(Euclidean formulation, i.e. for imaginary times $ \To $
Brownian motion $\To$ Wiener integral). There are numerous attempts and formulations 
to provide a mathematically sound basis (see, e.g. 
\meingruen{\bf Johnson \& Lapidus} in the list of textbooks or \cite{GiZa}).
This will not worry us in the following
-- in most cases mathematical rigor is far behind physical intuition.

\vspace {0.2cm}
\noindent 
Feynman developed his formulation of quantum mechanics in close analogy
to classical mechanics. First, therefore, a brief outline of the essential
aspects of classical mechanics and quantum mechanics.

\vspace {0.5cm}
\noindent
{\bf Classical Mechanics} \\

\noindent 
A system (for simplicity: A point particle in one spatial dimension)
with coordinate(s) $ q (t) $ is described by the \textcolor{blue}{\bf Lagrangian}
\be
L (q(t), \dot q(t), t)
\ee
Its dynamic development, that is, its trajectory (or ``path'') from
an initial point $ q(t_a) = q_a$  to an end point $ q(t_b) = q_b $ 
proceeds in such a way that among all possible paths the one is selected
that makes the \textcolor{blue} {\bf action}
\bce
\vspace {0.2cm}

\fcolorbox{blue}{white}{\parbox{6cm}{
\bea
S \EA \int_{t_a}^{t_b} dt \> L(q(t), \dot q(t), t) \no
\eea
}}
\ece
\vspace{-1.5cm}

\bea
\label{Lagrange Wirkung}
\eea
\vspace{0.2cm}

\noindent 
minimal (or more precisely: extremal):
\be
\boxed{
\qquad \delta S \E 0 \qquad
}
\ee
(Hamilton's principle or principle of ``least'' effect). From that 
the equations of motion (Euler-Lagrange equations)  follow
\be
\frac{d}{dt} \frac{\partial L}{\partial \dot q} \> - \>
\frac{\partial L}{\partial q} \ E 0 \>.
\ee
An alternative formulation is the one which uses the \textcolor{blue}{\bf Hamiltonian}
of the system: We define the conjugated momentum
\be
p \E \frac{\partial L}{\partial \dot q}
\ee
and define the Hamiltonian as
\be
H (p (t), q (t), t) \E \left [ \> p \cdot \dot q \> - \> L (q, \dot q, t) \>
\right]_{\dot q = \dot q (p)} \>.
\ee
Hamilton's equations
\bea
\dot q (t) \EA \frac{\partial H}{\partial p} \non
\dot p (t) \EA - \frac{\partial H}{\partial q}
\eea
can also be obtained from an extremum of the action
\be
\boxed{
\qquad \delta S \E \delta \int_{t_a}^{t_b} dt \> \left [\> p \cdot \dot q \> - \>
H (p (t), q (t), t) \> \right] \E 0 \qquad
}
\label {Hamilton Wirkung}
\ee
by independent variation of $ q $ and $ p $.

\vspace {2.5cm}
\noindent 
{\bf Quantum Mechanics} \\

\begin{figure}[hbtp]
\bce
\includegraphics[angle=0,scale=0.6]{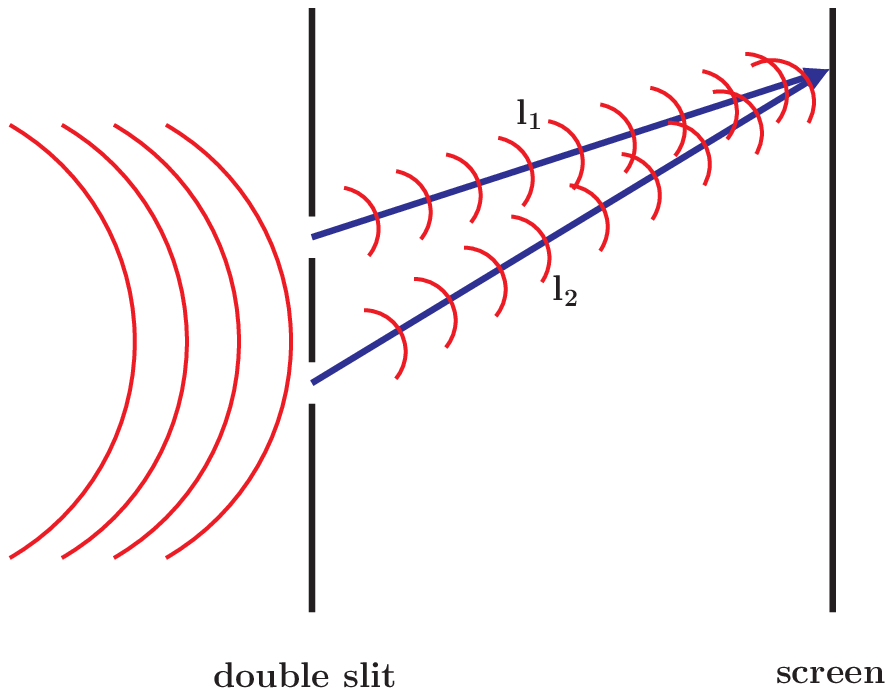}
\label{abb:1.1.1}
\vspace*{-9.5cm}

{\bf Fig. \arabic{abb}} : Diffraction of waves through a double slit.
\ece
\end{figure}

\noindent
In classical mechanics Nature seems to
``compare'' different paths  and then to select the path of extremal action. This is
also the case in optics, as long as the wavelength of light is much
smaller than the typical dimensions of the system. However, if both
sizes are comparable, we observe the typical interference behavior
of waves, e.g.  diffraction by a double slit as depicted in Fig. \ref{abb:1.1.1}.
\refstepcounter{abb}

\vspace {0.2cm}

\noindent 
The different ways now play a crucial role, since the
scattered wave is composed of two parts
\be
\Phi \propto e^{i k l_1} + e^{i k l_2}
\ee
($ k $: wavenumber). The intensity at the detector (screen) is proportional
to $ |\Phi |^2 $ and shows the typical diffraction minima and maxima.

\vspace {0.5cm}
\noindent 
It is now a crucial experimental fact that even
matter particles (like electrons or neutrons) show diffraction effects
when passing through slits or scatter from crystals. This happens when their
de Broglie wavelength is comparable to the
dimensions of the object to be imaged (for electrons, this was 
demonstrated for the first time by Davisson and Germer 1927).

\noindent 
The quantum mechanical description of this effect postulates a
\textcolor{blue}{\bf probability amplitude} which is  a superposition of the two
possible amplitudes (for the passage through the respective slit):
\be
\Phi \E \Phi_1 \> + \> \Phi_2 \>,
\ee
and the probability of finding the electron at the detector is,
\be
 W \> \sim \> | \Phi |^2 \>.
\ee
In other words:
\vspace {0.1cm}

\color[rgb]{0.2,0,0.8}
\noindent 
{\it One has to sum {\bf coherently} over the various (unobserved) alternative
ways an event can happen.}
\color{black}

\vspace {0.2cm}
Let us imagine that we now drill more and more holes in the screen -- until
it no longer exists --  and that we put more and more screens in the
space between source and detector which we also treat in the same way,
then we have
\be
\Phi(a, b) \E \sum _{{\rm all \> path \> from} \> a \> {\rm to} \> b} \>
\Phi_i \>.
\ee

\refstepcounter{abb}
\begin{figure}[hbtp]
\bce
\includegraphics[angle=90,scale=0.45]{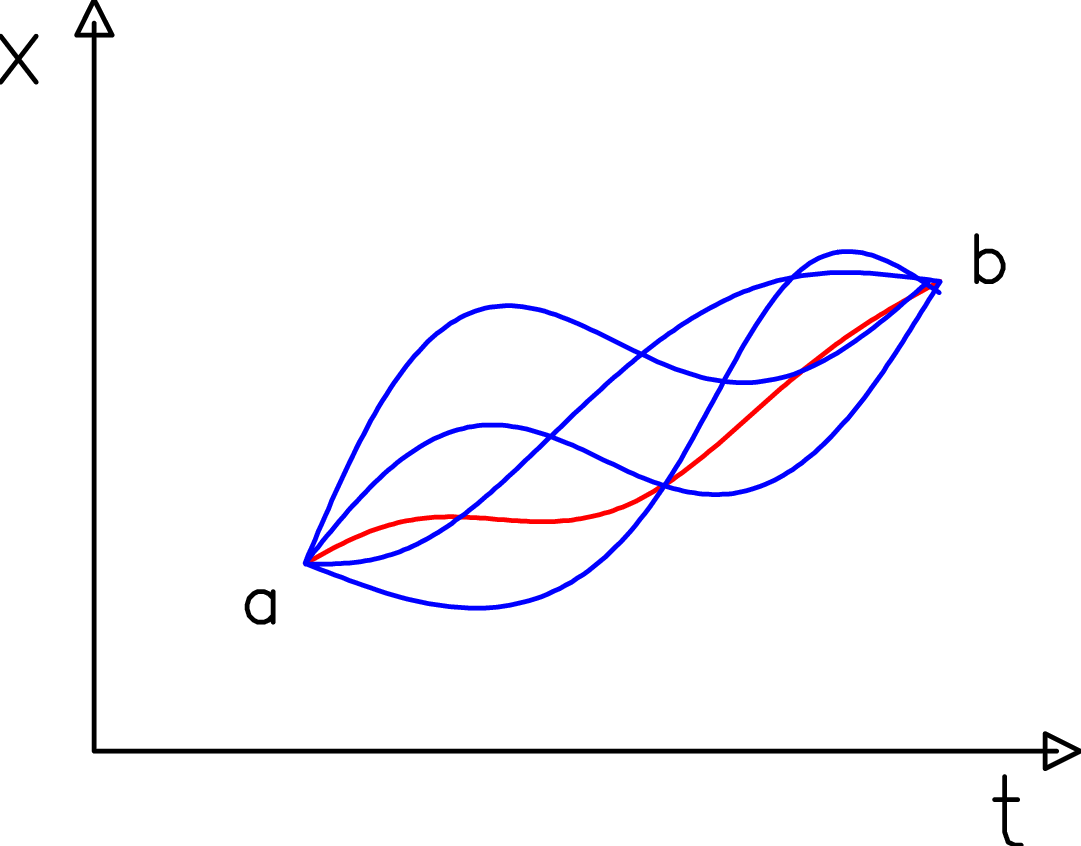}
\label{abb:1.1.2}
\ece
{\bf Fig. \arabic{abb}} : Quantum mechanical propagation of a particle from 
$a$ to $b$. The classical path is indicated \\
\hspace*{1.4cm}  by a red line.
\end{figure}
\vspace{0.2cm}

\noindent 
Of course, it remains to specify more precisely how  this sum over
``all paths from $ a $ to $ b $ '' has to be performed and with which weight
each path contributes. The answer to the first question leads to
\textcolor{blue}{\bf functional integrals}, which we will treat 
in the next chapter while the problem of the weighting is solved in a 
heuristic way: Since Planck's constant $ \hbar $
has the dimension of an action and determines the quantum effects it is not surprising 
that the rule is:

\vspace {0.2cm}
\noindent
\blau{\it All paths contribute with the same absolute magnitude but with different
phase: The phase of every path $ x (t) $ is its classical action
$ S [x (t)] \> $ \footnote{The notation means that
$ S $ is a \textcolor{blue}{\bf functional} of the  path $ x (t) $, i.e. a number 
associated with each trajectory.} divided by $ \hbar $.}
\be
\boxed{
\qquad \Phi(a, b) \E \sum_{{\rm all \> path \> of} \> a \> {\rm to} \> b}
\> {\rm const.} \> \> e^{i S [x (t)] / \hbar} \>. \qquad
}
\ee

\noindent 
With this prescription the \textcolor{blue}{\bf classical limit} (formally:
$ \hbar \to $ 0) can be performed immediately: If all actions $ S \gg \hbar $
this leads to immense oscillations of the exponential function, that is, to each
path with a positive contribution, there is an adjacent path which 
gives a negative contribution  and therefore deletes it.
There remains only the special path for which this doesn't happen, {\it viz.}
whose phase is stationary 
\be
\delta S_{\rm classical} \E 0
\ee
and only one trajectory, the classical trajectory, contributes to the sum over
all paths.
\vspace {0.4cm}

\renewcommand{\baselinestretch}{0.9}
\scriptsize
\refstepcounter{tief}
\noindent
\blau{\bf Detail \arabic{tief}:} {\bf Orders of Magnitude}\\

\begin{subequations}
\noindent 
It is instructive to estimate the magnitude of the action for two different systems by 
simple dimensional analysis: First, we take a 
mechanical watch, the moving parts of it having  an approximate size 
$ d \sim 10^{-4} $ m , mass $ m \sim 10^{-4} $ kg and typical time $ t  \sim 1 $ s.
Then, the characteristic action  of this system is
\bea
S_1 \> \sim \> m \, d^2 \, t^{- 1} \> \sim \> 10^{- 12} \> {\rm J}
\> \sim \> 10^{22} \, \hbar \>. 
\eea
However, if we
consider a microprocessor, the centerpiece of every computer,
then we know that its integrated circuits typically are
$ d \sim 0.2 \mu$m $ = 2 \times 10^{-7} $ m thick and that it operates with electrons
($ m \sim 10^{-30} $ kg). With a clock frequency of 1 GHz, corresponding to an
angular frequency of $ \omega \sim 6 $ GHz, the typical action then is
\bea
S_2 \> \sim \> m \, d^2 \, \omega \> \sim \> 2.4 \times 10^{-34} \> {\rm J} \>
 \sim \> 2 \, \hbar \>. 
\eea
Thus, although both systems possess comparable (outer)
dimensions, the traditional watchmaker doesn't have to know 
anything about 
\vspace*{-0.1cm}

\noindent
quantum theory while the developers of microprocessors 
definitely need it for their work.
\end{subequations}
\renewcommand{\baselinestretch}{1.2}
\normalsize
\vspace {0.5cm}

Of course, the quantum mechanical description cannot be derived 
in this way from classical mechanics, the classical limit is
only a necessary condition of the larger theory. Therefore, in
the next chapter we will proceed in the opposite way and derive the path integral
description from the usual quantum mechanical one. This will give us directly
both the weighting of each path as well as the prescription how to do their summation.
\vspace {0.5cm}


\subsection{\textcolor{blue}{Lagrange, Hamilton and Other Formulations}}
\label{sec1: Lagr,Ham}
In the following all quantum mechanical operators are denoted by
an ``hat ''. We also assume that the Hamiltonian
\be
\hat H \E \hat T \> + \> \hat V, \> \> \> \> \hat V \> = V(\hat x)
\ee
is time-independent and consider matrix elements
of the \textcolor{blue}{\bf time-evolution operator} 
\footnote{Frequently the notation $ \> \la x_b, t_b \,  | \, x_a, t_a \ra \> $ 
is used for this matrix element.}
\bce
\vspace{0.2cm}

\fcolorbox{blue}{white}{\parbox{12cm}{
\bea
U(x_b, t_b ; x_a, t_a) \EA \la x_b \, \left | \> \hat U(t_b,t_a) \> \right | \, x_a \ra 
\E  \la x_b \, \left | \> e^{  -i (t_b - t_a) \hat H /\hbar }\> \right | 
\, x_a \ra \>. \no
\eea
}}
\ece
\vspace{-1.5cm}

\bea
\label{U operator}
\eea
\vspace{0.2cm}

\noindent
As we will see, the time-evolution operator contains all information
about the system that we need.

\vspace{1cm}
\noindent
{\bf Feynman's Path Integral}\\

\noindent
The path integral representation of
$ \, U \, $ is based on the decomposition of the time interval $ t_b - t_a $ into $ N $
single, infinitesimal small intervals
\be
\epsilon = \frac{t_b - t_a}{N},
\ee
so that
\be
\exp \left [ -i(t_b - t_a) \hat H /\hbar \right ] \> \E \lim_{N \to \infty}
\prod_{k=1}^N \> 
e^{-i \epsilon \hat H/\hbar} \>.
\label{Trotter}
\ee
This is also called Trotter's product formula.
Now it is a fact that
\be
\exp \left [ -i \epsilon ( \hat T  + \hat V) /\hbar \right ]
\> = 
\exp \left [-i \epsilon \hat T / \hbar \right ] \> \cdot \>
\exp \left [ -i \epsilon \hat V / \hbar \right ] \> + {\cal O}(\epsilon^2) 
\> ,
\label{kurz Zeit 1}
\ee
i.e. for infinitesimal times the non-commutativity of the  the quantum mechanical operators 
can (\blau {\bf almost}) be neglected.
If we use this expression
in \eqref{Trotter} and insert the identity operator 
\be
1 \E \int_{-\infty}^{+\infty} dx_j \> \left | x_j \ra  \la x_j \right| \hspace{1.5cm}
j = 1,2 \ldots (N-1)
\ee
between each factor, it is only a small step to obtain the 
path integral: Since for
a local potential \\ 
$ V(\hat x) | \, x > = V(x) | \, x > $ is valid, we obtain
\bea
U(x_b, t_b; x_a, t_a) \EA \lim_{N \to \infty}
\int_{-\infty}^{+\infty} dx_1 \> dx_2 \> ... \> dx_{N-1} \>   
\la x_b \, \left | \, e^{-i\epsilon \hat T/\hbar} \, \right |\, x_{N-1} \ra \> 
e^{-i \epsilon V(x_{N-1})/\hbar} \non
&& \hspace{4cm}\cdot \>
\> \la  x_{N-1} \, \left | \, e^{-i\epsilon \hat T/\hbar} \, \right | \, x_{N-2} \ra \> 
e^{-i \epsilon V(x_{N-2})/\hbar} \non
&& \hspace{4cm} \vdots \non
&&  \hspace{4cm} \cdot \> 
\la x_1 \, \left | \, e^{-i\epsilon\hat T/\hbar} \, \right | \, x_a \ra  \> 
e^{-i \epsilon V(x_a)/\hbar}.
\eea
We now need the matrix elements of $ \exp (-i \epsilon \, \hat T / \hbar) $
between position eigenstates
\be
\la x_{j+1} \, \left | \, e^{-i\epsilon \hat T/\hbar}\, \right  | \, x_j \ra \E 
\frac{1}{2 \pi \hbar}
\int_{-\infty}^{+\infty} dp_j \> \exp \left (-i 
\frac{\epsilon}{\hbar} \frac{p_j^2}{2m}
\right ) \> e^{i p_j \cdot (x_{j+1} - x_j)/\hbar} \> .
\label{Impulsintegral}
\ee
Here we have used that $\hat T = \hat p^2/(2 m)$ is diagonal in momentum space.
The integral in \eqref{Impulsintegral} is an extended
Gaussian integral which can be derived from the basic integral (\ref{Gauss 1}) 
by completion of the square:
\bce
\vspace{0.2cm}

\fcolorbox{red}{white}{\parbox{11cm}{
\bea
\int_{-\infty}^{+\infty} dx \> \exp \left ( - a x^2 + b x \right)
\EA \sqrt{ \frac{\pi}{a}} \> \exp \left ( \frac{b^2}{4a} \right ) \quad , \> \> 
 a \ne 0 \> , \> \> {\rm Re} \> a \ge 0 
\quad . \no
\eea
}}
\ece
\vspace{-2cm}

\bea
\label{Gauss 2}
\eea
\vspace{0.5cm}

\renewcommand{\baselinestretch}{0.9}
\scriptsize
\refstepcounter{tief}
\noindent
\blau{\bf Detail \arabic{tief}:} {\bf Complex Gaussian Integrals}\\

\noindent
\begin{subequations}
To be more precise the integral in Eq. (\ref{Impulsintegral}) is an (extended)
\textcolor{blue}{\bf Fresnel integral} which arises from the basic integral \eqref{Gauss 1}
through the limit $ \> {\rm Re} \, a \to 0 \> $: 
\bea
\int_{-\infty}^{+\infty} dx \> e^{i a x^2} \EA \lim_{\delta \to 0^+} \, 
\int_{-\infty}^{+\infty} dx \> e^{- ( \delta - i a) \,  x^2} \E 
\lim_{\delta \to 0^+} \, 
\sqrt{\frac{\pi}{\delta - i a}} \qquad (a \> \> {\rm reeal}) \non
\EA \sqrt{\frac{\pi}{|a|}}
\, \exp \left [ \frac{i}{2} \, \lim_{\delta \to 0^+} \arctan \left ( \frac{a}{\delta} \right ) 
\right ]
\E \sqrt{\frac{\pi}{|a|}} \, \exp \left [ i \, \frac{\pi}{4} \, \sgn (a) \right ] 
\non
\EA  \sqrt{\frac{\pi}{|a|}} \, \left [ \, \frac{1}{\sqrt{2}} +  i \,    
\frac{\sgn(a)}{\sqrt{2}} \, \right ] \> .
\label{Fresnel}
\eea
Note that the real parameters $ \, a \,  $ can have any sign.
Here and below we understand the root of a complex variable $ z $ as
the \textcolor {blue} {\bf principal value}, i.e. the branch of the two-valued square-root function which has 
a positive real part  ({\ bf \{Handbook \}}, eq. 3.7.26)
\be
\sqrt{z} \Def \sqrt{ | z |} \, \exp \left ( \, \frac{1}{2} \, i \, {\rm arg} \, z \,
\right ) \> \> , \> \> - \pi \, < {\rm arg} \, z  \, \le \, \pi \> .
\ee
This value one already has to take in Eq. \eqref{Gauss 1}, as in the real case the above definition coincides 
with the positive root which one has take for a positive integrand.
Therefore the Gaussian integral with complex parameters is to be understood 
as an \textcolor {blue}{\bf analytic continuation} of the result in the neighborhood
of the real axis.
\vspace{0.3cm}

\noindent
Another way to calculate the Fresnel integral with infinite limits 
is to regard it as the limit of finite Fresnel integrals:
\be
\int_{-\infty}^{+\infty} dx \> e^{i a x^2} \E \lim_{R \to \infty} \, \int_{-R}^{+R} dx \, 
\left [ \, \cos (a x^2) + i \sin (a x^2) \, \right ] \> .  
\ee 
The transformation $ x = \sqrt{\pi/(2 |a|)} \, t $ gives
\bea
\int_{-\infty}^{+\infty} dx \> e^{i a x^2} \EA 2 \sqrt{\frac{\pi}{2 |a|}} \lim_{T \to \infty} 
\, \left [ \, \int_0^{T} dt \,
\cos \left ( \frac{\pi}{2} \, t^2 \right ) + i \, \sgn(a)  \, \int_0^{T} dt \, \sin \left ( 
\frac{\pi}{2} \, t^2 \right ) \, \right ] \non
\EA 2 \sqrt{\frac{\pi}{2 |a|}} \lim_{T \to \infty} \, \Bigl [ \, C \left ( T \right ) - i 
\, \sgn(a) \, S  \left ( T \right )\, \Bigr ]
\eea
where $T = R \sqrt{ 2 |a|/\pi}$ and  $C(T), \> S(T) $ denote the standard Fresnel integrals 
as they are defined in {\bf \{Handbook\}}, eq. (7.3.1) and (7.3.2).
Since these tend to the asymptotic limit
$1/2$ for  $T \to \infty$ ({\it ibid.}, eq. (7.3.20)) we obtain
\be
\int_{-\infty}^{+\infty} dx \> e^{i a x^2} \E \sqrt{\frac{\pi}{2 |a|}} \> 
\Bigl [ \, 1 + i \, \sgn(a) \, \Bigr ] \> ,     
\ee
in agreement with  Eq. \eqref{Fresnel}. The extended Fresnel integral ($a, b\> $ reell)
\be 
\int_{-\infty}^{+\infty} dx \> e^{i a x^2 + i b x} \E \int_{-\infty}^{+\infty} dx \> \exp \lsp
i a \lrp x + \frac{b}{2a} \rrp^2 - i \frac{b^2}{4 a} \rsp
\ee
follows from simple completion of the square
\be 
\boxed{
\qquad \int_{-\infty}^{+\infty} dx \> \exp \lrp i \, x a x + i \, b x \rrp 
\E \sqrt{\frac{\pi}{-i \, a}} \, 
\exp \lrp - \frac{i}{4} \> b \, a^{-1} \, b \rrp \> . \qquad 
}
\label{erw Fresnel}
\ee
Eq. \eqref{erw Fresnel} has been written in  mnemonic form \footnote{see https://en.wikipedia.org/wiki/Mnemonic}
 which makes the generalization to the multi- and infinite-dimensional case more obvious.

\end{subequations}
\renewcommand{\baselinestretch}{1.2}
\normalsize
\vspace{0.5cm}

\noindent
With this result we get
\be
\la x_{j+1} \left | \> e^{-i\epsilon \hat T/\hbar} \> \right  | x_j \ra \E 
\sqrt{\frac{m}{2 \pi i \epsilon \hbar}} \> \exp \left [ \frac{i m}{2 \hbar}
\frac{ (x_{j+1} - x_j)^2}{\epsilon} \right ]
\ee
and if, for notational simplification, we set 
\be 
x_0 = x_a \> , \> \> x_N = x_b
\ee
we have
\bea
U(x_b, t_b; x_a, t_a) \EA \lim_{N \to \infty} \> 
\left (\frac{m}{2 \pi i \epsilon \hbar} \right )^{N/2}
\int_{-\infty}^{+\infty} dx_1 \> dx_2 \> ... \> dx_{N-1} \non
&& \cdot \> \prod_{j=0}^{N-1} \> 
\exp \left [ \> \frac{i m}{2 \epsilon \hbar}
(x_{j+1} - x_j)^2 \right ] \> 
\exp \left [ \frac{-i \epsilon}{\hbar} V(x_j) \right ] \> .
\eea
The product of exponential terms is the exponent of their sum and we obtain
\bce
\vspace{0.2cm}

\fcolorbox{red}{white}{\parbox{12cm}
{
\bea
\hspace{-1cm} U(x_b, t_b; x_a, t_a) \EA \lim_{N \to \infty} \>
\left (\frac{m}{2 \pi i \epsilon \hbar} \right )^{N/2}
\int_{-\infty}^{+\infty} dx_1 \> dx_2 \> \ldots \> dx_{N-1} \non
&& \cdot \exp \left \{ \frac{i \epsilon}{\hbar} \sum_{j=0}^{N-1} \left [
\frac{m}{2} \left ( \frac{x_{j+1} - x_j}{\epsilon} \right )^2 - 
V(x_j) \right ] \> \right \}  \non
&\EQ & \int_{x(t_a)=x_a}^{x(t_b)=x_b} {\cal D}x(t) \> e^{i S[x(t)]/\hbar} \> \> . \no
\eea
}}
\ece
\vspace{-3cm}

\bea
\label{Lagrange diskret}
\eea
\vspace{-1cm}

\bea
\label{Lagrange PI}
\eea
\vspace{0.2cm}

\noindent
The last line is a {\it symbolic} shorthand for the
limit
\be
N \to \infty \>,  \> \> \epsilon \to 0 \>,
\hspace {0.5cm} {\rm but} \> \> \> \epsilon N \E t_b - t_a \hspace {0.5cm}
{\rm fixed}
\ee
of the $ (N-1) $ - dimensional integral.
It is indeed the action which appears in the argument of the exponential function
in this limit because the discrete sums approximate the Riemann integral
\bea
\epsilon \sum_{j=0}^{N-1} \left [
\frac{m}{2} \left ( \frac{(x_{j+1} - x_j}{\epsilon} \right )^2 -
V(x_j) \right ]  &\longrightarrow& 
\int_{t_a}^{t_b} dt \> \left [ \, \frac{m}{2} \dot x(t)^2 - V(x(t)) \, \right ] 
\non
\EA \int_{t_a}^{t_b} dt \> L(x(t),\dot x(t)) 
\E S[x(t)] \>.
\eea
Note that the short-time approximation
\be
\exp \left [ -i \epsilon ( \hat T  + \hat V) /\hbar \right ]
\> = \exp \left [ -i \epsilon \hat V /\hbar \right ] \> \cdot
\exp \left [-i \epsilon \hat T / \hbar \right ] \> + {\cal O}(\epsilon^2)
\label{kurz Zeit 2}
\ee
leads to an action in which $ \> V(x_{j+1}) \> $ appears instead of
$ \> V(x_j) \> $ .
In a similar way the more symmetric form
\be
\exp \left [ -i \epsilon ( \hat T  + \hat V) /\hbar \right ]
\> = \exp \left [ -i \epsilon \hat V / (2\hbar) \right ] \> \cdot \>
\exp \left [-i \epsilon \hat T / \hbar \right ] \> \cdot \>
\exp \left [ -i \epsilon \hat V / (2\hbar) \right ] \> + {\cal O}(\epsilon^3)
\> ,
\label{kurz Zeit sym}
\ee
would lead to a slightly different discrete action.
In the simple case under discussion (one particle in a scalar potential)
this does not lead to any consequence: The differences are of order 
$\epsilon$ and vanish in the continuum limit $ \epsilon \to 0 $.
The situation is different if one considers velocity-dependent interactions
because the velocity is represented by $ (x_{j+1} - x_j)/\epsilon $.
More on that under the heading ``{\bf Ordering Problem}''.
\vspace{0.3cm}

\refstepcounter{abb}
\begin{figure}[hbtp]
\bce
\includegraphics[angle=90,scale=0.5]{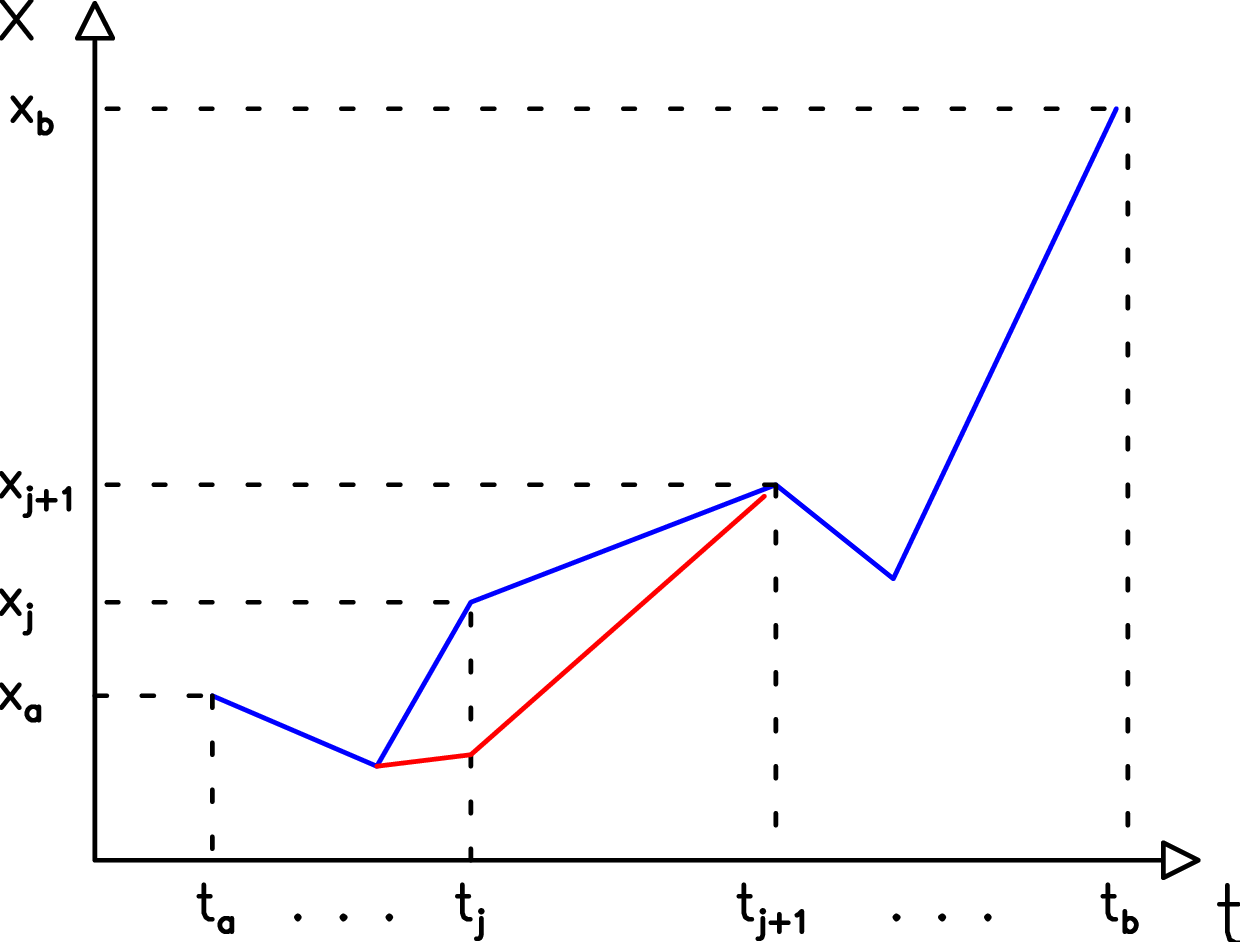}
\label{abb:1.2.1}
\ece
{\bf Fig. \arabic{abb}} : Discretization and summation over the
quantum mechanical paths. The red path \\
\hspace*{1.3cm} arises from a different value of $ x_j $.
\vspace{0.4cm}
\end{figure}

\noindent
The expression (\ref{Lagrange diskret}) gives an exact prescription how to sum over 
``all paths from $ a $ to $ b $ '' : One discretizes the time
and approximates each possible path from
$ a $ to $ b $ by a \textcolor {blue} {\bf polygon path}. Integration over an
intermediate point $ x_j $ corresponds to the summation over all paths from
$ x_{j-1} $ to $ x_{j + 1} $, because with fixed $ x_ {j-1}, x_{j + 1} $ a new path
is generated by a different value of $ x_j$.
There is no integration over the endpoints because they are fixed by the boundary conditions.

An extraordinary aspect of the path integrals is that they deal with
\textcolor {blue}{\bf ordinary numbers} and not with non-commuting
operators! This is an enormous advantage compared to the usual formulation 
of quantum mechanics. The price one has to pay for that is hidden in 
the complicated definition of the path integral: It is an
infinite-dimensional integral, a so-called
\textcolor {blue}{\bf functional integral} \footnote {Other
notations for it are $ \int \, [dx] $ or $ \int Dx $.}. 

\vspace{0.5cm}

It can be shown that the usual Schr\"odinger theory follows from
Eq. (\ref {Lagrange diskret}). This proceeds in the following manner:

\ben
\item The wave function at an infinitesimal later time
$ T + \epsilon $ is obtained from the wave function at the time $ t $ by
applying the time-evolution operator
\bea
\psi(x,t+\epsilon) \EA \left < x \, \left | \, \hat U (t + \epsilon, t ) \, \right | \, 
\psi \right > \E \int dy \> U ( x, \epsilon; y, 0 ) \, \psi(y,t) \non
\EA  \int_{-\infty}^{+\infty} dy \> \left ( \frac{m}{2 \pi i \epsilon \hbar} 
\right )^{1/2} 
\, \exp \left \{ \,  \frac{i \epsilon}{\hbar} \left [ \frac{m}{2} 
\frac{(x-y)^2}{\epsilon^2} - V(y) \right ] \, \right \} \, \psi(y,t)\> ,
\label{Schroed 1}
\eea
where in the last line the discretized form (\ref{Lagrange diskret}) 
for the time-evolution operator has been employed with $N = 1$ .

\item Now we substitute $ \xi = y - x $ in the integral 
(\ref{Schroed 1}) and obtain
\be
\psi(x,t+\epsilon) \E \left ( \frac{m}{2 \pi i \epsilon \hbar} 
\right )^{1/2} \, \int_{-\infty}^{+\infty}  d\xi \> \exp \left ( 
\frac{i m \xi^2}{2 \epsilon \hbar} 
\right ) \, \exp \left [ - \frac{i \epsilon}{\hbar} V(x+\xi) \right ] \, 
\psi(x+\xi,t) \> .
\label{psi etwas spaeter}
\ee
Due to the Gaussian factor in the integral, which restricts the values of
$ \, \xi \, $ to
$\, {\cal O} \left ( \epsilon^{1/2} \right )\, $, we can expand in powers of
$\xi$ and $\epsilon$: 
\bea 
\psi(x,t+\epsilon) \! \EA \!\left ( \frac{m}{2 \pi i \epsilon \hbar} 
\right )^{1/2} \int\limits_{-\infty}^{+\infty}  d\xi \, \exp \left ( 
\frac{i m \xi^2}{2 \epsilon \hbar} 
\right ) \left [ 1 - \frac{i \epsilon}{\hbar} V(x) - 
\frac{i \epsilon}{\hbar} \xi \frac{\partial}{\partial x} V(x) + 
{\cal O} \left ( \epsilon^2, \epsilon \xi^2 \right ) 
\right ] \non
&& \hspace{0.8cm} \cdot \left [ \, \psi(x,t) + \xi 
\frac{\partial}{\partial x} 
\psi(x,t) + \frac{1}{2} \xi^2  \frac{\partial^2}{\partial x^2} \psi(x,t) + 
{\cal O} \left ( \xi^3 \right ) 
\right ]  \> .
\eea

\item We now perform the Gaussian integrals: The ones over odd powers vanish whereas the 
integrals over even powers of $ \xi $ lead to 
\be
\int_{-\infty}^{+\infty}  d\xi \> \exp \left ( \frac{i m \xi^2}{2 \epsilon 
\hbar} \right ) \E \left ( \frac{2 \pi i \epsilon \hbar}{m} \right )^{1/2} 
\> , \quad \quad
\int_{-\infty}^{+\infty}  d\xi \> \xi^2 \, \exp \left ( \frac{i m \xi^2}{2 
\epsilon \hbar} \right ) \E \frac{i \epsilon \hbar}{m} \, \left ( 
\frac{2 \pi i \epsilon \hbar}{m} \right )^{1/2}
\ee
(the last integral is easily obtained by differentiation of the Gaussian
integral (\ref{Gauss 1}) with respect to the parameter $a$).
This gives
\be
\psi(x,t+\epsilon) \E \psi(x,t)  - \frac{i \epsilon}{\hbar} V(x) \, \psi(x,t)
 + \frac{i \epsilon \hbar}{2m} \frac{\partial^2}{\partial x^2} \psi(x,t) +  
{\cal O} \left ( \epsilon^2 \right ) \> .
\ee
The neglected terms come from higher powers of
$ \epsilon V(x) $, higher terms in the expansion of $V(x+\xi)$ and from
the Gaussian integral over the fourth power of $\xi$. 
It is now seen that \blau{\bf all} short-time approximations
(\ref{kurz Zeit 1}, \ref{kurz Zeit 2}, \ref{kurz Zeit sym}) only differ
by higher-order terms and thus lead to the same result in the limit $ \epsilon \to 0 $.

\item If we now multiply by $ i \hbar/\epsilon $ we obtain
\be
i \hbar \, \frac{\psi(x,t+\epsilon) - \psi(x,t)}{\epsilon} \E 
- \frac{\hbar^2}{2m} \frac{\partial^2}{\partial x^2} \psi(x,t) + V(x) 
\psi(x,t) 
+ {\cal O} \left ( \epsilon \right ) \> .
 \ee
In the limit $\epsilon \to 0 $ this gives exactly \blau{\bf Schr\"odinger's equation}
\be
\boxed{
\qquad i \hbar \, \dot \psi(x,t) \E  - \frac{\hbar^2}{2m} \, 
\frac{\partial^2}{\partial x^2} \psi(x,t) + V(x) \,  \psi(x,t) \qquad 
}
\label{Schroed 2}
\ee
for a particle in the potential $ V(x) $.
As mentioned before one must be more cautious when dealing with velocity-dependent
interactions: For example,
the interaction of 
a charged particle in a magnetic field contains a term $ {\bf A} (\fx) \cdot \dot
\fx $ (where $ {\bf A} (\fx) $ is the corresponding vector potential) and it is now
essential which discrete form of the path integral
is used in the above derivation: \\
$ \>  {\bf A}(\fx) \cdot (\fx-{\bf y})/\epsilon \> , \quad $ or \\
$ \> [ {\bf A}(\fx) +  {\bf A}({\bf y})] \cdot (\bf{x}-\bf{y})/(2\epsilon) \> ,  
\quad $ or \\ $ \>  {\bf A}((\fx+{\bf y})/2) \cdot (\bf{x}-{\bf y}) /\epsilon \quad $ ? 
\quad
(see \purpur{\bf Problem \ref{Magnetfeld Eich}}).

\een 
\vspace{0.6cm}

\noindent 
Some other properties of the time-evolution operator can be obtained from 
the path integral form \eqref{Lagrange diskret}. For example,
the law for consecutive events can be derived in the following way:
Let $ t_c $ be an intermediate time with $ t_a < t_c < t_b $ .
If we divide the intervals
$ t_c - t_a = L \epsilon \> , t_b - t_c = M \epsilon $ 
we get
\bea
&& U(x_b, t_b; x_a, t_a)  \E \non
&&\lim_{L \to \infty} 
\left (\frac{m}{2 \pi i \epsilon \hbar} \right )^{L/2}
\int\limits_{-\infty}^{+\infty} dx_1 \ldots dx_{L-1} \,
\int\limits_{-\infty}^{+\infty} dx_L \, 
\exp \left \{ \frac{i \epsilon}{\hbar} \sum_{j=0}^{L-1} \left [
\frac{m}{2} \left ( \frac{x_{j+1} - x_j}{\epsilon} \right )^2 - V(x_j) 
\right ] \right \}  \non
&& \cdot \lim_{M \to \infty} 
\left (\frac{m}{2 \pi i \epsilon \hbar} \right )^{M/2}
\int\limits_{-\infty}^{+\infty} dx_{L+1} 
\ldots dx_{L+M-1} \, \exp \left \{ \frac{i \epsilon}{\hbar} 
\sum_{k=L}^{L+M-1} 
\left [ \frac{m}{2} \left ( \frac{x_{k+1} - x_k}{\epsilon} \right )^2 - 
V(x_k) \right ] \> \right \} \non
&& \hspace{2.5cm} \EQ  \int_{-\infty}^{+\infty} dx_c \> 
U(x_b, t_b; x_c, t_c ) \> U(x_c, t_c; x_a, t_a ) \> ,
\label{Kompositionsgesetz}
\eea
where we have set $ x_c \EQ x_L $. This is the
\textcolor{blue}{\bf composition law} of the time-evolution operator
which, of course, is quite obvious 
in the operator form \eqref{U  operator}. A special case is the 
unitarity of the time evolution operator
\be
 \hat U(t',t)  \, \hat U^{\dagger} (t',t) \E  \hat U(t',t)  \, \hat U(t,t') 
\E \hat 1 \> .
\ee
The above example illustrates that
path-integral methods can be more complicated and intransparent than operator 
methods when it comes to proving such properties. However,
``\blau{\it nobody is perfect}''!  

\noindent
Furthermore, one can show that the path-integral representation
\eqref{Lagrange diskret}  remains valid also for time-dependent potentials 
$ V( x , t ) $
if at each time step the corresponding value
of the potential is taken
(\purpur{\bf Problem \ref{U fuer zeitabh. Pot}}~) .
This is again a
simplification compared to the operator form of quantum mechanics : As is well known,
there one has to order the operators chronologically to get 
a formal solution of the Schr\"odinger equation
\be
i \hbar \, \frac{\partial}{\partial t '} \hat U ( t ' , t ) \E \hat H ( t ' ) \,
\hat U ( t ' , t )
\ee
for a time-dependent Hamiltonian .


\vspace{2cm}

\noindent
{\bf Phase Space Path Integral}\\

\noindent
We obtain another form of the path integral if the $ p$-integration
of the Gaussian integral is not performed:
\vspace{0.5cm}

\fcolorbox{red}{white}{\parbox{15cm}
{
\bea
\hspace{-1cm} U(x_b, t_b; x_a, t_a) \EA \lim_{N \to \infty} \>
\int_{-\infty}^{+\infty} dx_1 \> dx_2 \> ... \> dx_{N-1} \> 
\int_{-\infty}^{+\infty} \frac{dp_1}{2 \pi \hbar} \> \frac{dp_2}{2 \pi \hbar} 
\> \ldots \> \frac{dp_N}{2 \pi \hbar} \non
&& \cdot \exp \left \{ \frac{i \epsilon}{\hbar} \sum_{j=1}^N \left [
p_j \cdot \frac{x_j - x_{j-1}}{\epsilon} - 
\frac{p_j^2}{2m} - V(x_{j-1})\right ]
\> \right \} \non 
&\EQ & \int_{x(t_a)=x_a}^{x(t_b)=x_b} \frac{ {\cal D}'x(t) \> {\cal D}p(t)}
{2 \pi \hbar} \> \exp \left \{ \frac{i}{\hbar} 
\int_{t_a}^{t_b} dt \> \left [ \, p(t) \cdot \dot x(t) - H(p(t),x(t)) \, \right ] \> 
\right \} \> . \no
\eea
}}
\vspace{-3cm}

\bea
\label{Hamilton diskret}
\eea
\vspace{-0.8cm}

\bea
\label{Hamilton PI}
\eea
\vspace{1.2cm}


\noindent
Some comments about the ``phase space''  or Hamiltonian  formulation:
\ben
\item The argument of the exponential is again the classical action
divided by the action quantum, but this time formulated
in coordinates and canonical momenta (see. Eq. (\ref{Hamilton Wirkung}).
\item The ``integration measure'' $ dx_j dp_j / (2 \pi \hbar) $ is  different from
the Lagrangian formulation; for distinction  we therefore use 
the notation $ {\cal D} 'x (t) $. This is descriptive and 
easily remembered as dividing the 
phase space into cells of the size $ h = 2 \pi \hbar $ (as in statistical mechnaics).
\item There is an \textcolor{blue}{\bf additional integration} over $ dp_N $ and
\textcolor{blue}{\bf no} boundary conditions for the momenta.
\item The interpretation is more difficult: Is this a sum over trajectories
in phase space? Classically, however, a single \textcolor{blue}{\bf point} in
phase space determines the trajectory.
\item While in the Lagrangian formulation the paths are continous 
and only the velocities are discontinuous (see Fig. \ref{abb:1.2.1}),
in the phase-space path integral the paths are discontinuous as well. Therefore 
caution is needed when performing canonical transformations and other manipulations
(an instructive example is given in \meingruen{\bf Schulman}, p. 306-309).

\item Klauder \cite{Klau} has shown that by introducing a regularizing factor        
\be
\lim_{\nu \to \infty} \, \int {\cal D}x \, {\cal D}p \> \> \exp \lcp 
\frac{i}{\hbar}  S[x,p] \rcp 
\> \cdot \> \exp \lsp - \frac{1}{2 \nu \hbar} \int_{t_a}^{t_b} dt \, \lrp m^2 \dot x^2 + 
\frac{\dot p^2}{m^2} \rrp \rsp
\ee
the phase space path integral can be defined unambigously.
\een

\vspace{1cm}

\noindent
{\bf An Example: The Free Particle}\\

\noindent
We now want to calculate explicitly the path integral for the propagator of the free
particle using the Hamiltonian formulation.
We write the discretized action in Eq. (\ref{Hamilton diskret}) as
\be
S[x,p] \> = \sum_{j=1}^{N-1} \left ( p_j - p_{j+1} \right ) \cdot x_j
\> + p_N \cdot x_N - p_1 \cdot x_0 - \epsilon \sum_{j=1}^N \frac{p_j^2}{2 m} 
\> ,
\ee
and first perform the  $x_j$ integrations. This gives
$(N-1) \> \> \> \delta$-functions which allow to perform
the 
$p_j$ integrations ($ j = 1,2 ... N-1$): 
$p_1 = p_2 = ... p_{N-1} = p_N$. There remains
\be
U_0(x_b, t_b; x_a, t_a) = \lim_{N \to \infty} \int_{-\infty}^{+\infty} 
\frac{dp_N}{2 \pi \hbar} \> \exp \left [ \frac{i}{\hbar} \left ( p_N \cdot
(x_b - x_a) - \epsilon N \frac{p_N^2}{2 m} \right ) \> \right ] \> .
\ee 
Since $ \> \epsilon N = t_b - t_a \> $, the limit can be performed trivially
and we obtain 
\bce
\vspace{0.2cm}

\fcolorbox{black}{white}{\parbox{12cm}
{
\bea
U_0(x_b, t_b; x_a, t_a) \EA \int_{-\infty}^{+\infty}\frac{dp}{2 \pi \hbar} 
\exp \left [ \frac{i}{\hbar} \left ( p \cdot
(x_b - x_a) - \frac{p^2}{2 m} T \right ) \> \right ] \non 
\EA \sqrt{ \frac{m}{2 \pi i \hbar T}} \> 
\exp \left [ \frac{i m}{2 \hbar T} \left (x_b - x_a \right )^2 \right ]  \no
\eea
}}
\ece
\vspace{-3.2cm}

\bea
\label{freier Prop 1}
\eea
\vspace{-0.8cm}

\bea
\label{freier Prop 2}
\eea
\vspace{1cm}


where the abbreviation $ T = t_b - t_a $ has been used.
\vspace{0.5cm}

\noindent
{\bf Discussion}:

\ben
\item Comparing Eq. (\ref{freier Prop 1})
with the \textcolor{blue}{\bf spectral representation} \footnote{Of course, here the 
spectrum is completely continous and the summation over discrete states $\psi_n$ has to be replaced
by an integral over the continous parameter $p$.}
of the time-evolution operator
\be
U(x_b, t_b; x_a, t_a) \E  \la x_b \lvl \> \sum_n  \lvl \psi_n \ra \la \psi_n \rvl 
e^{-i \hat H T/\hbar} \rvl x_a \ra \E \sum_n \psi_n (x_b) \> 
e^{-i E_n T/\hbar} \> \psi_n^{\star} (x_a)
\label{spektral}
\ee
allows to read off the normalized eigenfunctions
\be
\psi_p (x) \> = \frac{1}{\sqrt{2 \pi \hbar}} \> e^{i p \cdot x/\hbar}
\ee
and eigenenergies
\be
E_p = \frac{p^2}{2 m} \> .
\ee
This is, of course, the expected result.

\item For a free particle the classical action is
 $ \, S_{\rm kl}/\hbar = m (x_b - x_a)^2/(2 \hbar T) \, $ 
(\purpur{\bf Problem \ref{klass Wirk} a)} ) which agrees with the phase of
Eq. (\ref{freier Prop 2}). 
We will
see that this is no coincidence, but is valid for all actions which are
quadratic  in coordinates and momenta.
\een

\vspace{1cm}
\noindent
{\bf The Ordering Problem}\\

\noindent
If a classical Hamiltonian is given, in which products of
coordinates and momenta occur, the question arises how to quantize such forms:

\vspace{0.2cm}
\blau{\it 
%
%
``Aucune r\`egle bas\'ee sur la correspondance avec la M\'echanique Classique ne peut r\'esoudre de telles 
ambigu\"{\i}t\'es, puisque ces derni\`eres proviennent de la {\bf non-commutativit\'es} des op\'erateurs, 
elle-m\^eme li\'ee au caract\`ere fini et non nul de $\hbar$ ... Dans tout les cas d'inter\^et pratique, 
il faut se conformer aux prescriptions suivantes: ... la fonction de Hamilton \'etant mise sous ... la forme
$\> \sum_i p_i f_i(q_1 ... q_n)\> $ ... , on y remplace le dernier term de la somme par l'expression 
$<\!\!<${\bf symetris\'ee}$>\!\!>$
$ \> \frac{1}{2} \sum_i \left [ \hat p_i \hat f_i + \hat f_i \hat p_i \right ]$ ...}
\vspace{0.2cm}

\hspace{3cm} ( {\bf\{Messiah\}}, p. 59 )
\footnote{
%
%
"No rule  based on the correspondence with Classical Mechanics can resolve such
ambiguities, since the latter arise from the {\bf non-commutability}  of operators,
which in turn is tied to the finite and non-zero character of $ \hbar $ ...
In all cases of practical interest, one must conform to the following prescriptions:  
... Having put the Hamiltonian function into the form $ \sum_i p_i f_i(q_1 ... q_n) $
one replaces... (this) term... by the {\bf symmetrized} expression
$\frac{1}{2} \sum_i \left [ \hat p_i \hat f_i + \hat f_i \hat p_i \right ]$ ''
 {\bf\{Messiah 1\}}, p. 70) }
\vspace{0.5cm}

How do these ambiguities show up in the path integral? Obviously
not in the symbolic notation as used in 
Eq. (\ref{Lagrange PI}) and (\ref{Hamilton PI})! 
Actually, they appear in the innocent-looking problem
how the Hamilton function is treated in the discretized form of the
path integral, e.g.
\bdes
\item[a)]{outer averaging: $\frac{1}{2} \left [ H(p_j,x_j) + 
H(p_j,x_{j-1}) \right ]$}
\item[b)]{inner averaging: $H \left (p_j,
\frac{x_j + x_{j-1}}{2} \right ) $}
\item[c)]{right hand: $H(p_j,x_j) $}
\item[d)]{left hand: $H(p_j,x_{j-1}) \> .$}
\edes

\noindent
If  there are no products of $ x $ and $ p $ in the Hamiltonian, then
- as discussed before  -- the various procedures give identical results and
one  can chose  the simplest formulation. In the general case
one can show that for the treatment of products of the form
\be
A \Def p^m x^n
\label{pm xn}
\ee
in the Hamilton function, there exists the following correspondence 
between the above path integral prescriptions and the operator ordering:
\bdes
\item[a)] $ \frac{1}{2} \left [ \> \hat p^m \hat x^n \> + 
\> \hat x^n \hat p^m \> \right ] $
\item[b)] $ \frac{1}{2^n} \sum_{k=0}^n \> {n \choose k} \> \hat x^k \> 
\hat p^m \> \hat x^{n-k} $
\item[c)] $\hat x^n \> \hat p^m $
\item[d)] $\hat p^m \> \hat x^n $  .
\edes
Since $\hat x, \hat p$ are hermitean operators, only the symmetric forms a) and b)
also give hermitean operators.
Rule~b) is called \blau{\bf Weyl's quantization rule} or 
``\blau{\bf mid-point rule}'' and is obtained automatically if
Eq. (\ref{pm xn}) is considered as  \textcolor{blue}{\bf Wigner transform}
\be
\boxed{
\qquad A \EQ  A_W(x,p) \E \int_{-\infty}^{+\infty} dy \> 
\left < \> x - \frac{y}{2} \>
\left | \> \hat A \> \right | \>  x + \frac{y}{2} \> \right >
\> e^{i p \cdot y/ \hbar} \qquad
}
\label{def Wigner}
\ee
of the corresponding quantum mechanical operator $ \hat A $. 
The reverse transform is
(\purpur{\bf Problem \ref{Wigner Trans}})
\be
\la x \, \left | \> \hat A \> \right | \, x' \ra \E \int_{-\infty}^{+\infty} 
\frac{d p}{2 \pi \hbar} \> 
A_W \left ( \frac{x+x'}{2}, p \right ) \, e^{i p \cdot (x-x')/\hbar} \> .
\label{Wigner reverse}
\ee
If we apply this to each factor in Trotter's product formula, we obtain
\bea
\la x_{j+1} \, \left | \, e^{-i \epsilon \hat H/\hbar} \, \right | \, x_j \ra
\EA \int_{-\infty}^{+\infty} 
\frac{d p_j}{2 \pi \hbar} \> \underbrace{
\Biggl ( \, e^{-i \epsilon \hat H/\hbar} \, \Biggr )_W
 \left ( \frac{x_{j+1}+x_j}{2}, p_j \right ) }_{= 1 - 
i \epsilon H_W((x_{j+1}+x_j)/2, p_j)/\hbar + \ldots}
\, e^{i p_j \cdot 
\left ( x_{j+1}-x_j \right )  /\hbar}  \non 
& \simeq &  \int\limits_{-\infty}^{+\infty} 
\frac{d p_j}{2 \pi \hbar}  \exp \left [ - \frac{i \epsilon}{\hbar} \, H_W  
\left ( \frac{x_{j+1}+x_j}{2}, p_j \right ) \,\right ] 
\, e^{i p_j \cdot 
\left ( x_{j+1}-x_j \right )  /\hbar} ,  
\eea
which leads to a generalization of Eq. (\ref{Hamilton diskret}) for
arbitrary Hamilton functions and which contains the mid-point rule.

\noindent
The ordering prescription becomes important for curvi-linear coordinates
or for quantization in curved spaces. A less exotic application 
is the case of a charged particle in a magnetic field
because the minimal substitution for the electromagnetic field leads
to the velocity-dependent Lorentz force
(\purpur{\bf Problem \ref{Magnetfeld Eich}}).


\vspace{0.8cm}
\noindent
{\bf Fourier Path Integral}\\
 
\noindent
In the following we will determine the propagator for the harmonic oscillator, 
i.e. for the potential
\be
V(x) = \frac{1}{2} m \omega^2 \> x^2
\ee
by a method that illustrates how to integrate functionally over paths
which are parametrized by {\bf Fourier series} and not by polygon paths.
At the same time we want to use the symbolic notation for the path integrals more frequently,
but will return to the discretized form when required or in case of doubt.
First, in the path integral 
\be
U(x_b, t_b; x_a, t_a) \E \int_{x(t_a)=x_a}^{x(t_b)=x_b} {\cal D}x(t) \> 
\exp \left [ \> \frac{i}{\hbar} \int_{t_a}^{t_b} dt \left ( \frac{m}{2} 
\dot x^2 - \frac{m}{2} \omega^2 x^2 \right ) \> \right ]
\ee
we introduce a new integration variable
\be
y(t) = x(t) - x_{\rm cl}(t)
\ee
where $ x_{\rm cl}(t) $ is the  \textcolor{blue}{\bf classical path}. 
It fulfills the equation of motion
$ \, \ddot x_{\rm cl} + \omega^2 x_{\rm cl} = 0 \, $ with the boundary conditions
$ \, x_{\rm cl}(t_a) = x_a \, $ and $ \, x_{\rm cl}(t_b) = x_b \, $.
The Jacobi determinant of the transformation is $ \, 1 \, $, since in any
$ x_i $-integral only a shift takes place.

\vspace{0.8cm}

\renewcommand{\baselinestretch}{0.9}
\scriptsize
\refstepcounter{tief}
\noindent
\blau{\bf Detail \arabic{tief}:} {\bf Jacobi Determinant and Functional Derivative}\\

\noindent
\begin{subequations}
If in a  $n$-dimensionalen integral 
$ \, I = \int_V dx_1 \ldots dx_n \, f \left ( x_1, \ldots x_n \right ) \, $
a substitution of variables is performed
$ \, x_i =  x_i \left (\xi_1, \ldots \xi_n \right ) \, $,  $ i = 1 \ldots n $
then
\be
I \E \int_{V'} d\xi_1 \ldots d\xi_n \> \left | {\cal J} \right | \> 
f \left (\xi_1, \ldots 
\xi_n \right ) \> \> , \hspace{1cm}  {\cal J} \E {\rm det}_n \, 
\frac{\partial x_i}{\partial \xi_j}
\ee
where ${\cal J}$ denotes \textcolor{blue}{\bf Jacobi determinant} (or Jacobian) and 
$V'$ is the transformed integration region expressed in the new variables.

\vspace{0.1cm}
\noindent
The \textcolor{blue}{\bf functional derivative} is the generalization of the 
partial derivative of a function of several variables to the case of
infinitely many variables.
It can be defined by
\be
\int d\sigma \> \frac{\delta S[x]}{\delta x(\sigma)} \, \eta(\sigma) 
\E \lim_{\epsilon \to 0} \, 
\frac{ S[x + \epsilon \eta] - S[x]}{\epsilon} 
\ee
where $ \eta(\sigma) $ is an arbitrary test function. In the physical literature
one can also find the expression
\be 
\frac{\delta S[x]}{\delta x(\sigma)} \Def \lim_{\epsilon \to 0} \, 
\frac{ S[x + \epsilon \delta(\sigma)] - S[x]}{\epsilon} 
\ee
which resembles the definition of the ordinary derivative of a function.
From these definitions it immediately follows that
\be
\frac{\delta x(t)}{\delta x(\sigma)} \E \delta(t-\sigma) 
\label{func x x delta}
\ee
(cf. $ \, \partial x_i/\partial x_j = \delta_{i j} \, $ in the finite-dimensional case). 
Keeping this property in mind all usual rules of differentiaion like 
product rule or chain rule
are valid (see
\purpur{\bf Problem \ref{functionalableit}}).

\end{subequations}
\renewcommand{\baselinestretch}{1.2}
\normalsize

\vspace{0.5cm}
\noindent
In the new integration variables the action is
\bea
S[x(t)] \E S[x_{\rm cl}(t) + y(t)] \EA 
 S[x_{\rm cl}(t)] \> + \> \int_{-\infty}^{+\infty} d\sigma \> 
\frac{ \delta S}{\delta x_{\rm cl}(\sigma)} \> y(\sigma) \non
&& +  
\frac{1}{2} \int_{-\infty}^{+\infty} d\sigma \,  \int_{-\infty}^{+\infty}
d\sigma' \> \frac{ \delta^2 S}{\delta x_{\rm cl}(\sigma) 
\delta x_{\rm cl}(\sigma')} \> y(\sigma) \> y(\sigma') \> .
\label{functional Taylor}
\eea
In our case the (functional) Taylor expansion stops after the second term since the action is quadratic.
The first term in Eq. (\ref{functional Taylor}) is the classical action of the
harmonic oscillator, for which one finds
\be
\boxed{
\qquad S_{\rm cl}^{\rm h.O.} \E \frac{m}{2} \frac{\omega}{\sin \omega T} \> \Bigl [ \> 
(x_a^2 + x_b^2) \cos \omega T - 2 x_a \cdot x_b \> \Bigr ] \qquad
}
\label{S kl fuer HO}
\ee
(\purpur{\bf Problem \ref{klass Wirk} b)} ). 
Here again $ T = t_b - t_a $ is the time difference. 
The second term in Eq. (\ref{functional Taylor}) vanishes
since we have expanded around the classical path which is determined by
$ \delta S_{\rm cl} = 0$.
One easily finds 
(\purpur{\bf Problem \ref{functionalableit}~b)})
\be
\frac{1}{2} \int_{-\infty}^{+\infty} \int_{-\infty}^{+\infty} d\sigma \>
d\sigma' \frac{ \delta^2 S}{\delta x_{\rm cl}(\sigma)
\delta x_{\rm cl}(\sigma')} \> y(\sigma) \> y(\sigma') \>  = \> 
\int_{t_a}^{t_b} dt \> \left [ \> \frac{m}{2} \dot y^2(t) - \frac{m}{2} 
\omega^2 y^2(t) \> \right ] \> .
\ee
After these steps the calculation of the harmonic oscillator propagator
\be
\boxed{
\qquad U (x_b, t_b; x_a, t_a) \E F(t_b,t_a) \> e^{i S_{\rm cl}/\hbar} \qquad
}
\label{HO propagator}
\ee
is reduced to the calculation of the simpler path integral
\be
F(t_b,t_a) \E \int_{y(t_a)=0}^{y(t_b)=0} {\cal D}y(t) \exp \left [ \> 
\frac{i}{\hbar} \, \int_{t_a}^{t_b} dt \, \left ( 
\> \frac{m}{2} \dot y^2 - \frac{m}{2} \omega^2
y^2 \right ) \> \right ] \> .
\ee
Of course, due to time translation invariance the prefactor
can only depend on the difference $T = t_b - t_a $.

\vspace{0.5cm}
Since all paths $y(t)$ start at the point $ 0 $ at $t = t_a$ and end at  the point $0$ at $t = t_b$ 
they also can be written as a Fourier sine series:
\be
y(t) = \sum_{k=1}^{N-1} b_k \>  \sin \left ( \> \frac{k \pi (t-t_a)}
{T} \> \right ) \> ,
\ee

\vspace{0.8cm}

\renewcommand{\baselinestretch}{0.9}
\scriptsize
\refstepcounter{tief}
\noindent
\blau{\bf Detail \arabic{tief}:} {\bf Mathematical Background}\\

\noindent
\begin{subequations}
Note that this is {\bf not} the usual (well-known) Fourier series of a function 
 $f(t)$ which reads
\be
f(t) \E \frac{a_0}{2} + \sum_{k=1}^{\infty} a_k \, \cos \left ( \frac{2 k \pi t}{T} 
\right) + \sum_{k=1}^{\infty} b_k \, \sin \left ( \frac{2 k \pi t}{T} \right ) \E 
\sum_{k=-\infty}^{+\infty} c_k \, \exp \left (  \frac{2 k \pi i t}{T} \right) 
\ee 
and which represents  the (periodic) function in the interval $ 0 < t < T $. Due to 
this perodicity different function values at the endpoints are not possible.
However, since the fluctuations around the classical path vanish at the endpoints and 
therefore the function $ y(t) $ indeed is periodic, one also could use the {\it ansatz}
\be
y(t) \E \sum_{k=1}^{\infty} a_k \, \left [ \, \cos \left ( \frac{2 k \pi t}{T} 
\right ) - 1 
\, \right ]  + \sum_{k=1}^{\infty} b_k \, \sin \left ( \frac{2 k \pi t}{T} 
\right) \> .
\ee 
Less known \footnote{See, e.g., {\bf\{Bronstein-Semendjajew\}}, S. 475}
are the Fourier series with double period
\be 
f(t) \E \frac{a_0}{2} + \sum_{k=1}^{\infty} a_k \, \cos \left ( \frac{k \pi t}{T} 
\right) \> , 
\ee 
which represents a function in the full interval
$ 0 \le t \le T $ and
\be
f(t) \E \sum_{k=1}^{\infty} b_k \, \sin \left ( \frac{k \pi t}{T} 
\right) \> ,
\ee 
which is valid for an arbitrary function in the interval $ 0 < t < T $ , i.e. except the
endpoints. Since the fluctuations vanish at the endpoints, the last form is
most convenient for our purposes as this behaviour is already built in.

\end{subequations}
\renewcommand{\baselinestretch}{1.2}
\normalsize

\vspace{0.8cm}

\noindent
Instead of integrating over the intermediate points $ \> y_j, \> \> j = 1 ... (N-1) \> $
in the discretized path integral
we may integrate over 
the Fourier coefficients $b_k ,  \> \> k = 1 ... (N-1) $. Indeed, this is a 
linear transformation whose Jacobian is constant and independent
of $ \omega, m $ and $ \hbar $. In this calculation, however,
we keep the action as {\bf full} time integral and do not use
its approximation as Riemann sum as before -- the difference is of higher order and
disappears in the continuum limit $ \, N \to \infty, \epsilon \to 0 $.
By using the orthogonality relations of
the trigonometric functions in the interval $ [0, T] $ one finds for the kinetic energy
\be
\int_{t_a}^{t_b} dt \> \frac{m}{2} \dot y^2(t) \> = \frac{m}{4} T \> 
\sum_k \left ( \frac{k \pi}{T} \right )^2  b_k^2 \> ,
\ee
and for the potential energy
\be
\int_{t_a}^{t_b} dt \> \frac{m}{2} \omega^2 y^2(t) \> = \frac{m}{4} 
\omega^2 T \> \sum_k \> b_k^2 \> .
\ee
Therefore
\be
F(T) \E {\rm const.} \> \int_{-\infty}^{+\infty} db_1 \> db_2 \> ... \> 
db_{N-1} \> \exp \left \{ \>  \frac{i m T}{4 \hbar} \sum_{k=1}^{N-1}
\left [ \left ( \frac{k \pi}{T} \right )^2 - \omega^2  \right ] 
\> b_k^2 \> \right \}
\> .
\ee
This simply is the $(N-1)$-fold  product of a Gaussian integral so that
we obtain
\be
F(T) \E {\rm const'.} \> \left [ \>  \prod_{k=1}^{N-1} \left ( 1 \> - \> 
\frac{\omega^2 T^2}{k^2 \pi^2} \right ) \> \right ]^{-1/2} 
\stackrel{N \to \infty}{\longrightarrow} \> {\rm const'.} \> \left ( 
\frac{\omega T}{\sin \omega T} \right )^{1/2}.
\label{HO Vor prod}
\ee
In the last step we have used a product representation of $ \sin x/x $ which is due
to Euler {\bf \{Euler\}}. The constant is determined by requiring that 
for $\omega = 0$ the result (\ref{freier Prop 2}) for the free particle should follow. 
Therefore we have
\be
\boxed{
\qquad F^{\rm h.o.}(T) \E \left (\frac{m \omega }{2 \pi i \hbar \sin \omega T} 
\right )^{1/2} \qquad 
}\> .
\label{HO Vorfaktor}
\ee
However, this only holds for $ T < \pi/\omega $: In this case all factors in the product are 
positive. For $ \pi/\omega < T < 2 \pi/\omega $, $ F(T) $ acquires  an additional factor 
$ (-1)^{-1/2} = - i $, as the $k = 1$ term of the produkt has become negative~\footnote{Another possibility 
to prove that is by using the composition law of
the time-evolution operator, see \purpur{\bf Problem \ref {Maslov}}.}
; for $ 2\pi/\omega < T < 3 \pi/\omega $ a factor $(-i)^2$ etc.
In general, Eq. (\ref{HO Vorfaktor}) is multiplied by the \textcolor{blue}{\bf Maslov correction}
\be
\boxed{
\qquad e^{-i n \frac{\pi}{2} } \qquad
} 
\ee
where  $n$ counts how many times the trajectory has passed through a {\bf focal point} where the prefactor diverges 
\cite{Horv}. 
At these points the quantum mechanical probability amplitude becomes singular 
which in real systems
is prevented by anharmonic terms, so that the intensity only becomes maximal.
The spatial accumulation of such points is called a \textcolor {blue}{\bf caustic},
a phenomenon freqently encountered in optics (e.g.  when a water glass is illuminated by sunlight 
(see  https://en.wikipedia.org/wiki/Caustic\_(optics)).
The Maslov correction is a result of the so-called
index theorem and we will encounter it again when studying semi-classical approximations.

\vspace{0.4cm}

From the result in Eqs. (\ref{HO propagator}, \ref{S kl fuer HO})
and (\ref{HO Vorfaktor}) we can again extract the energy eigenvalues and eigenfunctions
of the (one-dimensional) system:
Using
\begin{subequations}
\be 
\cos \omega T \E \frac{1}{2} \, e^{i \omega T} \left ( 1 + 
e^{-2 i \omega T} \right )\>, \qquad 
i \sin \omega T \E \frac{1}{2} \, e^{i \omega T} 
\left ( 1 - e^{-2 i \omega T} \right ) \no
\ee
one obtains
\bea
U^{\rm h.o.} \! &=&\! \sqrt{ \frac{m \omega}{\pi \hbar}} e^{-i \omega T/2}
\left ( 1 - e^{-2 i \omega T} \right )^{-1/2} \hspace{-0.2cm}
\exp \left \{ - \frac{m \omega}{2 \hbar} \left [
(x_a^2+x_b^2) \frac{1 +e^{-2 i \omega T}}{1 - e^{-2 i \omega T}}  - 
4 x_a \! \cdot \! x_b \frac{e^{-i \omega T}}{1 - e^{-2 i \omega T}} \right ] 
\right \} \non
\! \EA\!  \sqrt{ \frac{m \omega}{\pi \hbar}} \> e^{-i \omega T/2} \> 
\left ( 1 + \frac{1}{2} e^{-2 i \omega T} + ... \right ) 
\> \exp \left [ \> - \frac{m \omega}{2 \hbar}(x_a^2+x_b^2) \left(
1 + 2 e^{-2 i \omega T} \right ) \right ] \non
&& \hspace{6cm} \cdot \> \exp \left [ \frac{2 m \omega}{\hbar} x_a \cdot 
x_b e^{-i \omega T} ( 1 + ... ) \right ] \no
\eea
and the comparison with the spectral representation (\ref{spektral}) yields
\bea
\psi_0(x) \EA \left ( \frac{m \omega}{\pi \hbar} \right )^{1/4}
\> \exp \left ( - \frac{m \omega}{2 \hbar} x^2 \right ) \> \> , \hspace{2cm}
E_0 = \frac{1}{2} \hbar \omega 
\label{psi0}\\
\psi_1(x) \EA \left ( \frac{m \omega}{\pi \hbar} \right )^{1/4}
\> \sqrt{ \frac{2 m \omega}{\hbar} }  \, x \, 
\> \exp \left ( - \frac{m \omega}{2 \hbar} x^2 \right ) \> \> , \hspace{0.4cm}
E_1 = \frac{3}{2} \hbar \omega \> .
\label{psi1}
\eea
\end{subequations}
Since we are expanding in powers of $ e^{- i \omega T} $ and an additional factor 
$ e^{- i \omega T/2} $ is always coming from the prefactor, it is obvious that the general
rule for the energies is  
\be
E_n \E \left ( n + \frac{1}{2} \right ) \hbar \omega \> \> , \hspace{0.3cm}
n = 0, 1, ...
\ee
Note that \textcolor{blue}{\bf zero-point energy} 
$ \> \> \frac{1}{2} \hbar \omega \> \> $ arises from the prefactor, i.e. from the
quadratic fluctuations around the classical path.
\newpage

\renewcommand{\baselinestretch}{0.9}
\scriptsize
\refstepcounter{tief}
\noindent
\blau{\bf Detail \arabic{tief}:} {\bf All Eigenfunctions}\\

\noindent
\begin{subequations}
The general form of the eigenfunctions can be derived from Mehler's formula({\bf \{Bateman Proj. 2\}}, 
ch. 10.13, eq. (22)) for the Hermite polynomials
\be
\sum_{n=0}^{\infty} \, \frac{(z/2)^n}{n!} \, H_n(\xi) \, H_n(\eta) \E \lrp 1 - z^2 \rrp^{-1/2} \, 
\exp \lsp \frac{2 \xi \eta z - (\xi^2 + \eta^2) z^2}{1 - z^2} \rsp \> 
\ee
If one sets
$ \> \xi = x_a/b \> , \eta = x_b/b \> , (\> b = \hbar/(m \omega) \> $ is the oscillator length), 
$\> z = \exp(-i \omega T)$ and multiplies both sides of Mehler's formula by
$ \> \exp [ - (\xi^2 + \eta^2)/2 ] \> $ then one finds by comparing with the spectral representation
\eqref{spektral} 
\be
\psi_n(x) \E \lrp \sqrt{\pi b} \>  2^n \, n!  \rrp^{-1/2} \, H_n \lrp \frac{x}{b} \rrp
\, \exp \lrp - \frac{x^2}{2 b^2} \rrp \> ,
\label{ho wfn}
\ee
which is in agreement with {\bf \{Messiah 1\}}, eq. (B.70).

\end{subequations}
\renewcommand{\baselinestretch}{1.2}
\normalsize
                
\vspace{0.5cm}

\subsection{\textcolor{blue}{Quadratic Lagrangians}}
\label{sec1: quadrat Lagr}

We now consider the general \footnote{Except for requiring  a constant mass. General time-dependent 
quadratic Hamiltonian system are treated, for example, in
\cite{YLUGP}.}
quadratic Lagrangian
\be
\boxed{
\qquad L \E \frac{1}{2} m \> \dot x^2 \> + \> b(t) \> x \cdot \dot x \> - \> 
\frac{1}{2} c(t) \> x^2 \> + \> d(t) \> \dot x  \> - \> e(t) \> x \> .
}
\label{L quadrat 1}
\ee
Without restriction of generality one may omit the terms with the
coefficient functions $ b(t) $ and $ d(t) $ since they may be added 
to the terms $ c(t) $ and $ e(t) $ , respectively, after an integration by parts in the action.
Thus we only need to consider
\be
L \E \frac{1}{2} m \> \dot x^2 \> - \> \frac{1}{2} c(t) \> x^2 \> - 
\> e(t) \> x \> .
\label{L quadrat 2}
\ee
Special cases are the free particle ($ c = e = 0 $),
the harmonic oscillator ($ e = 0, \> c = m \omega^2 $) or the  
forced harmonic oscillator ($ c = m \omega^2 , \> e \neq 0$). 
As in the treatment of the harmonic oscillator in the last chapter we introduce
the deviation from the classical path
\be
y(t) \E x(t) \> - \> x_{\rm cl}(t)
\ee
as integration variable. $ x_{\rm cl}(t) $ is solution of the classical
equation of motion
\be
m \ddot x_{\rm cl} + c(t) x_{\rm cl} + e(t) \E 0
\ee
with boundary conditions
\be
x_{\rm cl}(t_a) \> = x_a \hspace{2cm} x_{\rm cl}(t_b) \> = x_b \> .
\ee
Functional expansion of the action $ \> S [x(t)] = S [x_{\rm cl}(t) + y(t)] $
gives again
\be
U(x_b, t_b; x_a, t_b) \E e^{i S_{\rm cl}/\hbar} 
\int_{y(t_a)=0}^{y(t_b)=0} {\cal D}y(t) \> \exp \left \{ \> \frac{i}{\hbar}
\int_{t_a}^{t_b} dt \> \left [ \frac{m}{2} \dot y^2 - \frac{1}{2} c(t) y^2
\right ] \> \right \} \> = e^{i S_{\rm cl}/\hbar} \> F(t_b,t_a) \> .
\label{U quadrat 1}
\ee
The prefactor explicitly reads
\be
F(t_b,t_a) = \lim_{N \to \infty} {\cal N}_{\epsilon}^{N+1} \int dy_1 \> dy_2 \> 
\ldots \> dy_N 
\> \exp \left \{ \frac{i}{\hbar} \sum_{j=1}^{N+1} \left [
\frac{m}{2 \epsilon} \left  (y_j - y_{j-1} \right )^2 - \frac{1}{2}
\epsilon \> c_j \> y_j^2 \right ] \> \right \} \> ,
\label{Vorfaktor quadr 1}
\ee
where (to simplify the notation) we have replaced $ N $ by $ N + 1 $ in the discretized path integral.
In addition, 
\be
{\cal N}_{\epsilon} = \left ( \, \frac{m}{2 \pi i \epsilon \hbar} \, \right )^{1/2}
\ee
is the normalization factor of the Lagrange path integral
and $ c_j = c(t_j) = c(t_a + j \epsilon) $. 
Eq. (\ref{Vorfaktor quadr 1}) is again a Gaussian integral of the type
\be
F(t_b,t_a) = \lim_{N \to \infty} {\cal N}^{N+1}_{\epsilon} \int dy_1 \> dy_2 \> 
\ldots \> dy_N
\> e^{ i {\bf y}^{t} \mathbb{B}_N \> {\bf y} },
\ee
where $ \, {\bf y}^{t} \, $ denotes the row vector $ \, (y_1 \ldots y_N)\,  $,
$\, {\bf y}\, $ the corresponding column vector and
\be
\mathbb{B}_N \E \frac{m}{2  \epsilon \hbar } \> \left (
\begin{array}{rrrrrr}
 2 & -1 &  0 & \ldots & 0 & 0 \\
-1 &  2 & -1 & \ldots & 0 & 0 \\
 0 & -1 &  2 & \ldots & 0 & 0\\
   &    &    & \ddots &   &  \\
 0 & 0  &  0 & \ldots & -1& 2 
\end{array}
\right ) \> - \> \frac{ \epsilon}{2 \hbar} \> \left (
\begin{array}{rrrrr}
 c_1 & 0  &  0 & \ldots & 0 \\
 0   &c_2 &  0 & \ldots & 0 \\
 0   & 0  &c_3 & \ldots & 0 \\
     &    &    & \ddots &   \\
 0 & 0  &  0   & \ldots & c_N
\end{array}
\right ) \>.
\ee
In contrast to previous applications this integral does not factorize into
individual integrals of the type of Eq. (\ref{Gauss 1}).
This only happens if we have diagonalized the real, symmetric matrix
$ \mathbb{B}_N $ 
by an orthogonal transformation. 
The product of eigenvalues in the denominator of the square-root expression
of Eq. (0.1) then yields the determinant of the matrix.

In general, the \blau{\bf Gaussian integral for a real, symmetric matrix}
$ \A $ therefore is
\bce
\vspace{0.2cm}

\fcolorbox{blue}{white}{\parbox{9cm}
{
\
\bea
G_N(\A) \Def \int_{-\infty}^{+\infty} dx_1 \> dx_2 \> \ldots \> dx_N \> e^{- \fx^t \> 
 {\mathbb A} \> \fx}
\EA \frac{ \> \> \pi^{N/2} }{\sqrt{\det {\A} }} \> . \no
\eea
}}
\ece
\vspace{-2cm}

\bea
\label{Gauss 3}
\eea
\vspace{0.4cm}

\noindent
In addition, the matrix $  A $ has to be 
\textcolor{blue}{\bf positive definite} \footnote{i.e., all eigenvalues are $ > 0 $. 
More about positive definite matrices, see, e.g., {\bf \{Horn-Johnson\}}, 
ch. 7.} 
in order that the integral converges.
In the present Fresnel-like case where 
$  \A = 0^+ I - i \B_N \> $ the sign of the eigenvalues 
of $ \B_N $ is again irrelevant.
The prefactor  then is
\be
F(t_b,t_a) = \lim_{N \to \infty} \> \frac{ \pi^{N/2} {\cal N}_{\epsilon}^{N+1}}
{\sqrt{\det \lrp - i \B_N \rrp }} \E 
\sqrt{\frac{m}{2 \pi i \hbar} \frac{1}{f(t_b,t_a)}}
\label{prefactor}
\ee
with
\be
f(t_b,t_a) \E \lim_{N \to \infty} \epsilon \left ( 
\frac{2  \hbar \epsilon}{m} \right )^N \> \det \B_N \> .
\label{Vorfaktor quadr 2}
\ee

\vspace{0.2cm}
\noindent
There remains the problem to evaluate the determinant of the matrix
$\B_N$  and to perform the limit
$ \, N \to \infty \> , \epsilon \to 0 \, $~.
\blau{Gel'fand and Yaglom} \cite{GelYag}
have shown that this is possible by solving a differential equation:
We have to calculate the determinant
\be
p_N  =  \left ( \frac{2  \hbar \epsilon}{m} \right )^N \hspace{-0.3cm}
\det \B_N  =  \det \left \{ \left (
\begin{array}{rrrrr}
 2 & -1 &  \ldots & 0 & 0\\
-1 &  2 &  \ldots & 0 & 0\\
 0 & -1 &  \ldots & 0 & 0\\
   &    &  \ddots &   &  \\
 0 & 0  &  \ldots & -1& 2
\end{array}
\right )  -  \frac{ \epsilon^2}{m} \> \left (
\begin{array}{rrrrr}
 c_1 & 0  &  0 & \ldots & 0 \\
 0   &c_2 &  0 & \ldots & 0 \\
 0   & 0  &c_3 & \ldots & 0 \\
     &    &    & \ddots &   \\
 0 & 0  &  0   & \ldots & c_N
\end{array}
\right )\right \}
\ee
This is done by expanding $ p_{j+1} $ with respect to the
$ (j+1) ^{\rm th}$ column. The remaining determinant is then expanded with respect to 
$j ^{\rm th} $ row. In this way one obtains the recursion relation
\be
p_{j+1} \E \left ( 2 - \frac{ \epsilon^2}{m} c_{j+1} \right ) \> 
p_j \> - \> p_{j-1} \> , \hspace{0.5cm} j = 2,3 \ldots N-1 \> ,
\label{Rekurs}
\ee
which is also valid for $ j = 1 $ if the convention
$ p_0 = 1 $ is adopted. If we write 
Eq. (\ref{Rekurs}) as 
\be
\frac{ p_{j+1} - 2 p_j + p_{j-1} }{\epsilon^2} \E - 
\frac{c_{j+1}}{m} \> p_j \> ,
\ee
then we see that in the limit $ \epsilon \to 0 $ the desired function
$ \> \epsilon p_j \To f(t,t_a) \> $ is given by the solution of the 
differential equation
\be
\boxed{
\qquad \frac{\partial^2 f_{\rm GY}(t,t_a)}{\partial t^2} \E -
\frac{c(t)}{m} \> f_{\rm GY}(t,t_a) \qquad \> .
}
\label{GY Diffgl.}
\ee
The initial conditions are
\bea
f_{\rm GY}(t_a,t_a) \EA \epsilon p_0 \E \epsilon \To0 
\non
\frac{\partial f_{\rm GY}(t,t_a)}{\partial t} \Bigr |_{t=t_a} \EA \epsilon 
\frac{p_1 - p_0}{\epsilon} \E 2 - \frac{\epsilon^2}{m} c_1 - 1
\To 1 \> .
\label{GY Randbed.}
\eea

\vspace{2cm}

\noindent
{\bf Examples :}
\ben
\item As a test we consider the harmonic oscillator, i.e. 
$ \, c = m \omega^2 \, $. Gel'fand-Yaglom's differential equation 
(\ref{GY Diffgl.}) then simply is a free wave equation
with the general solution
$ \> f_{\rm GY}(t,t_a) = C \sin \left ( \omega (t-t_a) + \gamma \right )$. 
The initial conditions (\ref{GY Randbed.}) give 
$ \> \gamma = 0, \> \omega C = 1 \> $ for the integration constants. 
Using that we indeed obtain the correct result
\be
f^{ \> h.o.}_{\rm GY} (t_b,t_a) = \frac{1}{\omega} \> \sin (\omega T) \> .
\ee

\item The forced harmonic oscillator
($ \> c = m \omega^2, \> e(t) \> $
arbitrary) may be treated in the same way. Its classical action is
(\purpur{\bf Problem \ref{functionalableit} c)} )
\vspace{0.2cm}

\fcolorbox{black}{white}{\parbox{14cm}
{
\bea
S_{\rm cl} \E \frac{m \omega}{2 \sin \omega T} \> \Biggl \{ && 
\hspace{-0.5cm}
(x_b^2+x_a^2) \cos \omega T - 2 x_a \cdot x_b - \frac{2 x_b}{m \omega}
\> \int_{t_a}^{t_b} dt \> e(t) \sin \omega (t-t_a) \non
&& \hspace{4.5cm} - \frac{2 x_a}{m \omega} \> \int_{t_a}^{t_b} dt \> 
e(t) \sin \omega (t_b-t) \non
&& \hspace{-0.5cm} 
- \frac{2}{m^2 \omega^2} \> \int_{t_a}^{t_b} dt \> \int_{t_a}^t dt'
\> e(t) \> e(t') \sin \omega (t_b-t) \sin \omega(t'-t_a) \> \Biggr \} \> . \no
\eea
}}
\vspace{-2cm}

\bea
\label{erzwung HO}
\eea
\vspace{0.2cm}


\noindent
The function $ \> f_{\rm GY}(t_b,t_a) \> $ and thus the prefactor remain 
the same as for the free
harmonic oscillator since  $ \> e(t) \> $ does not appear in the
Gel'fand-Yaglom equation. In this way the problem has been solved completely 
by "quadrature", i.e. by evaluating ordinary integrals!
\een

\vspace{0.5cm}
\noindent
A further simplification results from the fact that there is no need to solve the
Gel'fand-Yaglom equation if one already knows 
the classical action. That is because one can show
(\purpur{\bf Problem \ref{Jacobi Eq.}}) that
\be
\frac{1}{f_{\rm GY}(t_b,t_a)} \E -  \frac{1}{m} \> \frac{\partial^2 S_{\rm cl}}
{\partial x_a \partial x_b} 
\label{Vor durch S cl}
\ee
holds -- meaning  that the time-evolution operator for a quadratic Lagrangian 
is completely determined 
by the classical action! If the classical trajectory goes through a 
\textcolor{blue}{\bf focal point} in which $ \> f_{\rm GY}(t_b,t_a) \> $ vanishes
an additional phase is generated -- as for the simple harmonic oscillator.
Therefore we have the result
\be
\boxed{
\qquad U(x_b, t_b; x_a, t_a) \E \sqrt{ \frac{1}{2 \pi i \hbar} \Biggl |
\frac{\partial^2 S_{\rm cl}}{\partial x_a \partial x_b}  \Biggr | }
\> \exp \left \{ i \left [ S_{\rm cl}(x_b,x_a)/\hbar - n \frac{\pi}{2} 
\right ] \> \right \} \quad ,
} 
\label{U quadrat 2}
\ee
where $ n $ is the number of focal points (including their multiplicity).
\vspace{0.5cm}


\noindent
It is instructive to evaluate the prefactor in the symbolic
notation for the path integral:
\be
F(t_b,t_a) \E \int_{y(t_a)=0}^{y(t_b)=0} {\cal D}y \> \exp \left [ \, 
\frac{i}{2 \hbar} 
\int dt \, dt' \> y(t) \frac{\delta^2 S}{\delta x(t) \delta x(t')} 
\Biggr |_{\rm cl} y(t') \, \right ] \E 
\frac{{\rm const}}{\sqrt{\fdet \, \delta^2 S_{\rm cl}}} \> .
\ee
Here $ \> \fdet \, \delta^2 S \> $ is the  \textcolor{blue}{\bf functional determinant}
of the operator
\be
\delta^2 S_{\rm cl} \EQ \frac{\delta^2 S}{\delta x(t) \delta x(t')} 
\Biggr |_{\rm cl} \E \left [ \, 
- \frac{m}{2} \frac{\partial^2}{\partial t^2} -  \frac{1}{2} \, c(t) \, \right ] \, 
\delta(t-t') \> ,  
\ee
i.e. the product of its eigenvalues. Therefore $ n $ can be understood as
\blau{\bf number of negative eigenvalues} of the operator
$ \> \delta^2 S_{\rm cl} \> $ (Morse's index theorem).

\vspace{0.8cm}

\renewcommand{\baselinestretch}{0.9}
\scriptsize
\refstepcounter{tief}
\noindent
\blau{\bf Detail \arabic{tief}:} {\bf Evaluation of Functional Determinants}\\   
\vspace{0.1cm}

\noindent
\begin{subequations}
One has to calculate the eigenvalues of 
$ \, \delta^2 S_{\rm cl}  \, $ in the space of functions which fulfill the 
boundary conditions for the path integral, i.e. in the present case functions which vanish 
at $ t_a $ as well at $ t_b $.
As an example we take the simple case
$ c(t) = m \omega^2 $ for which the eigenvalue equation reads
\be
-\frac{m}{2} \, \ddot f(t) - \frac{m}{2} \, \omega^2 \, f(t) \E \lambda \, f(t)
\ee
The eigenfunction, which vanishes at $ t_a $, obviously is
\be
f(t) \E C \, \cdot \, \sin \lsp \lrp 2 \frac{\lambda}{m} + \omega^2 \rrp^{1/2}  \,  
\biggl ( t-t_a \biggr ) \rsp \> .
\ee
The boundary condition at $ t = t_b$ gives as eigenvalues
\bea
\lrp 2 \frac{\lambda}{m} + \omega^2 \rrp^{1/2} T \EA k \pi \> ,  \qquad k = 1,2, \ldots \non
\Longrightarrow \quad \lambda_k(\omega)  \EA \frac{m}{2} \, \frac{k^2 \pi^2}{T^2} \, 
\lsp 1 - \lrp \frac{\omega T}{k \pi} 
\rrp^2  \rsp  \E \lambda_k(\omega=0) \cdot \lsp 1 - \lrp \frac{\omega T}{k \pi} \rrp^2  \rsp \> ,
\eea
which immediately leads to the product representation
(\ref{HO Vor prod}) for the prefactor of the harmonic oscillator.
In the general case this infinite product of eigenvalues has to be calculated 
without Euler's help ...
While this method to evaluate functional determinants is quite general, it obviously
is much more difficult than the Gel'fand-Yaglom method which, however, is only applicable 
for one-dimensional systems. To calculate functional determinants
for radial operators see \cite{DuKi}.

\end{subequations}
\renewcommand{\baselinestretch}{1.2}
\normalsize
\vspace{0.8cm}

\noindent
It is easy to generalize Eq. (\ref{U quadrat 2}) to any space dimension $ d $:   
For each dimension the quadratic
fluctuations give a prefactor in the form of a root expression and the
second derivative of the classical action w.r.t. the endpoints 
is replaced by the \textcolor{blue}{\bf van-Vleck-Pauli-Morette determinant} 
\be
\frac{\partial^2 S_{\rm cl}}{\partial x_a \partial x_b} \To
 \det \frac{\partial^2 S_{\rm cl}}{\partial {\bf x_a} \partial {\bf x_b}} 
\ee
Note that this is an ordinary $ d \times d $-determinant
and not a functional determinant (which we write as $\fdet$ ). 
Therefore the result is
\be
\boxed{
\qquad U(x_b, t_b; x_a, t_a) \E \left (\frac{1}{2 \pi i \hbar} \right )^{d/2}
\sqrt{ \Biggl |
\det \frac{\partial^2 S_{\rm cl}}{\partial {\bf x_a} \partial {\bf x_b}}  
\Biggr | \> }
\> \exp \left \{ i \left [ S_{\rm cl}({\bf x_b,x_a})/\hbar \> - 
\> n \, d \,  \frac{\pi}{2} \right ] \> \right \} \quad .
} 
\label{U quadrat 3}
\ee
\vspace{0.5cm}


\subsection{\textcolor{blue}{Perturbation Theory}}
\label{sec1: stoerung}
\begin{figure}[htbp]
\vspace*{-5cm}
\bce
\includegraphics[angle=0,scale=0.95]{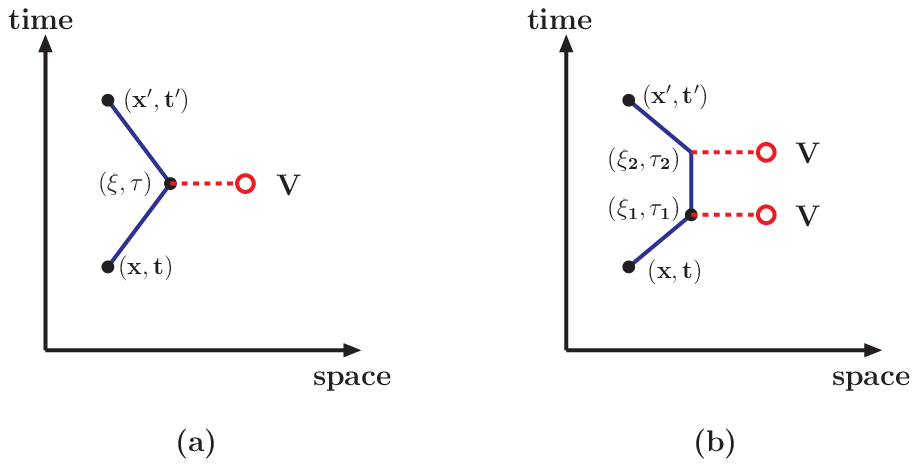}
\label{abb:Stoerung}
\ece
\vspace*{-16cm}
{\bf Fig. \arabic{abb}} : (a) Feynman diagram in $1^{\rm st}$ order perturbation theory
for the time-evolution operator in Eq. \eqref{U1}.\\
\hspace*{1.8cm} The solid blue lines represent the free propagator 
between two space-time points, the dashed\\
\hspace*{1.8cm} red line the interaction of the potential $ V $. 
Intermediate  points are to be integrated over. \\
\hspace*{1.4cm}(b) Feynman diagram for the $2^{\rm nd} $ order, see Eq. (\ref{U2}).
\end{figure}

We now consider the (for simplicity again: one-dimensional) motion of a particle 
in a general potential $\> V(x) \> $, i.e. Lagrange functions of the form
\be
L \E\frac{m}{2} \dot x^2 \> - \> V(x) \> .
\ee
In general the path integral cannot be evaluated anymore \footnote{For some cases, 
such as the Coulomb potential, special methods have been found
to obtain the (already known) solutions from the path integral formalism, see, e.g. \cite{DuKl}.
However, these methods do not help
in the general case.}. A widely used method for weak interactions
(potentials) is perturbation theory, which we may obtain
from the phase space formula (\ref{Hamilton PI}) by expansion in powers of the potential:
\bea
U\left ( x',t';x,t \right ) \EA \int_{x(t)=x}^{x(t')=x'}  
\frac{{\cal D}'x \, {\cal D}p}{2 \pi \hbar} \> \exp \left [ \, \frac{i}{\hbar} 
\int_t^{t'} d\sigma \left ( p \, \dot x - \frac{p^2}{2m} \right ) \right ]\, 
\sum_{j=0}^{\infty} 
\left ( - \frac{i}{\hbar} \right )^j \frac{1}{j !} \left ( \int_t^{t'} d\tau \, 
V(x(\tau)) \right )^j  \non
&=:& \sum_{j=0}^{\infty} \left ( - \frac{i}{\hbar} \right )^j \, U_j 
\left ( x',t';x,t \right ) \> .
\label{def Uj}
\eea
The first term is the free propagator which we already have determined 
in Eq. (\ref{freier Prop 2}):
\be
U_0\left ( x',t';x,t \right ) \E \int_{-\infty}^{+\infty} \frac{dp}{2 \pi \hbar}
\> \exp \left \{ \, \frac{i}{\hbar} 
\left [p \, (x'-x) - \frac{p^2}{2m} (t'-t) \right ] \right \} \> .
\ee
Let us now consider the next term of the perturbation expansion
\be
U_1\left ( x',t';x,t \right ) \E \int_{x(t)=x}^{x(t')=x'} 
\frac{ {\cal D}'x \,{\cal D}p}{2 \pi \hbar} \> \exp \left [ \, \frac{i}{\hbar} 
\int_t^{t'} d\sigma \left ( p \, \dot x - \frac{p^2}{2m} \right ) \right ] \, 
\int_t^{t'} d\tau \, V(x(\tau)) \> .  
\ee
In the discretized form the last factor is
\be
\int_t^{t'} d\tau \> V(x(\tau)) \E  \epsilon \sum_{j=1}^N V\left ( x_j \right ) \> , 
\hspace{0.5cm} {\rm with} \> \> \> \epsilon \E \frac{t'-t}{N}
\ee
and thus
\be
U_1\left ( x',t';x,t \right ) = \lim_{N \to \infty}  \epsilon \sum_{j=1}^N 
\prod_{k=1}^{N-1} \left ( \int\limits_{-\infty}^{+\infty} \frac{dx_k dp_k}{2 \pi \hbar} 
\right )  
\int\limits_{-\infty}^{+\infty} \frac{dp_N }{2 \pi \hbar} V\left ( x_j \right ) \, 
\exp \left \{  \frac{i}{\hbar} \sum_{l=1}^N 
\left [ p_l  ( x_l - x_{l-1})  - \epsilon \frac{p_l^2}{2m} \right ] \right \} .
\ee
Similarly as in the evaluation of the free propagator we write the exponent as
\be
\frac{i}{\hbar} \left [ \, \sum_{l=1}^{N-1} \left (p_l - p_{l+1} \right ) x_l 
+ p_N \,x' - p_1 \,x - \epsilon \sum_{l=1}^N \frac{p_l^2}{2m} \, \right ] 
\ee
and perform all $x_k$-integrations (except $x_j$ ). This gives
$(N-2)$ momentum $\delta$-functions
\bea
U_1\left ( x',t';x,t \right ) \EA \lim_{N \to \infty} \, \epsilon \sum_{j=1}^N 
\int_{-\infty}^{+\infty} dx_j \, V\left ( x_j \right ) 
\prod_{k=1}^N \left ( \int_{-\infty}^{+\infty} \frac{dp_k}{2 \pi \hbar} 
\right ) \, (2 \pi \hbar)^{N-2} \non
&& \cdot \, \delta \left ( p_1 - p_2 \right ) \, 
 \delta \left ( p_2 - p_3 \right ) \ldots   \delta \left ( p_{j-1} - p_j \right ) \, 
\delta \left ( p_{j+1} - p_{j+2} \right ) \ldots  
\delta \left ( p_{N-1} - p_N \right ) \non
&& \cdot \, \exp \left \{ \, \frac{i}{\hbar} \left [ (p_j - p_{j+1} ) x_j 
+ p_N x' - p_1 x \right ] - \epsilon \sum_{l=1}^N \frac{p_l^2}{2m} \, \right \} \> .
\eea
These enforce $ p_1 = p_2 = \ldots p_{j-1} = p_j$ and
$ p_{j+1} = p_{j+2} = \ldots p_{N-1} = p_N $, so that we obtain
\bea
U_1\left ( x',t';x,t \right ) \EA \lim_{N \to \infty} \, \epsilon \sum_{j=1}^N 
\int_{-\infty}^{+\infty} dx_j \, V\left ( x_j \right ) 
\int_{-\infty}^{+\infty} \frac{dp_j dp_N}{(2 \pi \hbar)^2} \, 
\exp \left \{ \, \frac{i}{\hbar} \left [ (p_j - p_N ) x_j 
+ p_N x' - p_j x \right ]\, \right \} \non
&& \hspace{5cm}  \cdot \, \exp \left [ - \frac{i}{\hbar} \epsilon 
\left (   \frac{p_j^2}{2m}j  + \frac{p_N^2}{2m}(N-j) \right ) \, \right ] \> .
\eea
If we now write $ \xi = x_j \> , \> p = p_j \> $ and $p' = p_N$ , the result is
\bea
U_1\left ( x',t';x,t \right ) \EA \lim_{N \to \infty} \, \epsilon \sum_{j=1}^N 
\int_{-\infty}^{+\infty} d\xi \, V\left ( \xi \right ) \, 
\underbrace{\int_{-\infty}^{+\infty} \frac{dp'}{2 \pi \hbar} \> 
\exp \left \{ \, \frac{i}{\hbar} \left [ (x'-\xi) p' - \frac{p'^2}{2m} \, 
\epsilon (N-j)
\right ] \, \right \} }_{=U_0(x',t';\xi, t+ j \epsilon)} \non
&& \hspace{4cm} \cdot \, \underbrace{\int_{-\infty}^{+\infty} \frac{dp}{2 \pi \hbar} 
\> 
\exp \left \{ \, \frac{i}{\hbar} \left [ (\xi-x) p -   \frac{p^2}{2m} \, j \epsilon
\right ] \, \right \} }_{=U_0(\xi, t+ j \epsilon;x,t)} \> .
\eea
It is now possible to perform the limit which gives a Riemannian integral
over the intermediate time $\tau \> , (\tau_j = t + j \epsilon) $
\be
U_1\left ( x',t';x,t \right ) \E \int_t^{t'} d\tau \, 
\int_{-\infty}^{+\infty} d\xi \>  U_0(x',t';\xi,\tau) \, V(\xi) \, U_0(\xi, \tau;x,t) 
\> .
\label{U1}
\ee
Read from right to left this contribution to the  perturbative expansion can be 
depicted as a 
\textcolor{blue}{\bf Feynman diagram} as shown in Fig. \ref{abb:Stoerung}a .

\refstepcounter{abb}
\vspace{0.5cm}

\noindent
Similarly one proceeds in higher orders
\be
U_j(x',t';x,t) \E \frac{1}{j!} \int_{x(t)=x}^{x(t')=x'} 
\frac{{\cal D}'x \, {\cal D}p}{2 \pi \hbar} \> \exp \left [ \, \frac{i}{\hbar} 
\int_t^{t'} d\sigma \, \left ( p \, \dot x - \frac{p^2}{2m} \right ) \, \right ]
\, \left ( \int_t^{t'} d\tau \, V ( x(\tau) ) \right )^j \> .
\ee
By induction one can prove that
\be
\left ( \int_t^{t'} d\tau \, V ( x(\tau) ) \right )^j \E j! \, \int_t^{t'} 
d\tau_1 \, 
\int_t^{\tau_1}  d\tau_2 \ldots \int_t^{\tau_{j-1}}  d\tau_j \> 
V \left ( x(\tau_1) \right ) \, V \left ( x(\tau_2) \right )\ldots 
V \left ( x(\tau_j) \right )
\label{Intpotenz}
\ee
which -- with a similar calculation -- leads to
\be
U_j\left ( x',t';x,t \right ) \E \int_t^{t'} d\tau \, 
\int_{-\infty}^{+\infty} d\xi \, U_0(x',t';\xi,\tau) \, V(\xi) \, 
U_{j-1}(\xi, \tau;x,t) \> \> , \> \> j = 1, 2, \ldots
\label{Uj}
\ee
The graphical representation for $ j = 2 $ 
\bea
U_2\left ( x',t';x,t \right ) \EA \int_t^{t'} d\tau_2 \int_{-\infty}^{+\infty} d\xi_2 \, 
U_0\left ( x',t';\xi_2,\tau_2 \right ) \, V(\xi_2) \non
&& \hspace{0.5cm} \cdot \, \int_t^{\tau_2} d\tau_1 
\int_{-\infty}^{+\infty} d\xi_1 \, U_0\left (\xi_2,\tau_2;\xi_1,\tau_1 \right ) \, 
V(\xi_1) \,  U_0\left (\xi_1,\tau_1;x,t \right )
\label{U2}
\eea
is shown in Fig. \ref{abb:Stoerung}b .
Insertion of the recursion relation (\ref{Uj}) into Eq. (\ref{def Uj}) gives an
integral equation for the time-evolution operator
\be
U\left ( x',t';x,t \right ) \E U_0 \left ( x',t';x,t \right ) - \frac{i}{\hbar}
\int_t^{t'} d\tau \, 
\int_{-\infty}^{+\infty} d\xi \, U_0(x',t';\xi,\tau) \, V(\xi) \, 
U(\xi, \tau;x,t) \> .
\label{U Intgl}
\ee 
This is the usual equation for the 
time-evolution operator in the  \textcolor{blue}{\bf interaction picture}
as one can see if one writes Eq. (\ref{U Intgl}) in operator form
\be 
\hat U(t',t) \E \hat U_0(t',t) - \frac{i}{\hbar}
\int_t^{t'} d\tau \,  \hat U_0(t',\tau) \, \hat V \, \hat U(\tau,t) \> .
\label{Uhat Intgl}
\ee
Defining
\be
\hat U(t',t) \deF \hat U_0(t',t) \, \hat U_I(t',t) \> ,
\ee
one obtains
\be
\hat U_0(t',t) \, \hat U_I(t',t) \E  \hat U_0(t',t) - \frac{i}{\hbar}
\int_t^{t'} d\tau \hspace{-0.5cm} \underbrace{\hat U_0(t',\tau)}_
{= \hat U_0(t',t) \hat U_0(t,\tau)
= \hat U_0(t',t) \hat U_0^{\dagger}(\tau,t)}
\hspace{-0.5cm} \hat V \, \hat U(\tau,t) \> .
\ee
Here the composition law for the free propagator $ \hat U_0$ has been used.
Multiplying from the left with
$ \hat U_0^{-1}(t',t) $ one obtains
\be
\hat U_I(t',t) \E 1 - \frac{i}{\hbar}
\int_t^{t'} d\tau \,  \underbrace{\hat U_0^{\dagger} (\tau,t) \, \hat V \,  
\hat U_0(\tau,t)}_{\deF \hat V_I(t',t)} \, \hat U_I(\tau,t) \> ,
\ee
which is identical with the equation of motion for the 
time-evolution operators in the interaction picture
\footnote{See, e.g., {\bf \{Messiah 2\}}, p. 723 .}.

\vspace{0.3cm}

From the derivation of Eq. (\ref {U Intgl}) it is clear that this equation is also valid for
time-dependent potentials. However, if the potential is \textcolor {blue}{\bf time-independent}
then the integral term on the right side of Eq. (\ref{Uhat Intgl}) 
is a convolution integral, since the evolution operators then only depend on the time 
difference. According to well-known  theorems for the Fourier transformation
we then immediately obtain
for the time-independent Green function
\be 
\hat G(E) \E \int_{-\infty}^{+\infty} dt \> \hat G(t) \, e^{i E t/\hbar} 
\> \> , \> \> \> {\rm with} \hspace{0.4cm} \hat G(t) \Def   - \frac{i}{\hbar} 
\Theta (t) \, \hat U(t,0)
\label{def G}
\ee
the \textcolor{blue}{\bf Lippmann-Schwinger equation}
\be
\hat G(E) \E \hat G_0(E) + \hat G_0(E) \, \hat V \, \hat G(E) \> ,
\ee
which simply is the operator identity
\be
\frac{1}{E-\hat H + i \, 0^+} \E \frac{1}{E-\hat H_0 + i \, 0^+} + \frac{1}{E-\hat H_0 + i \, 0^+} 
\, \hat V \, \frac{1}{E-\hat H + i \, 0^+} \> \> .
\ee
The prescription that the singularity in these operators has to be 
regularized by adding $ + \,i \, \times $ an infinitesimal positive quantity in the 
denominator
follows from the definition (\ref{def G}) in which the energy $ E $ must have a small, 
positive
imaginary part in order that the time integral converges
\footnote{Obviously only propagation with increasing time ($t \ge 0$ ) is possible, i.e.
we deal here with the \textcolor{blue}{\bf advanced} 
Green function which frequently is denoted by $ \hat G^{(+)}(E)$.}.

\vspace{0.3cm}
Finally it should be mentioned that, of course, one can derive other perturbative expensions;
e.g. one may write  $ \, V = V_0 + (V-V_0) \, $ to obtain
\bea
U \left (x',t';x,t \right) \EA \sum_{j=0}^{\infty} \frac{1}{j!} \, 
\left ( - \frac{i}{\hbar} \right )^j \, 
\int_{x(t)=x}^{x(t')=x'} \frac{{\cal D}'x \, {\cal D}p}{2 \pi \hbar} \, 
\left \{ \int_t^{t'} d\tau 
\left [ V(x(\tau)) - V_0(x(\tau)) \right ] \, \right \}^j \non
&&  \hspace{2.5cm} \cdot \,
\exp \left \{ \, \frac{i}{\hbar} \int_t^{t'} d\sigma \, 
\left [ p(\sigma) \, \dot x(\sigma) 
- \frac{p^2(\sigma)}{2m} - V_0(x(\sigma)) \right ] \, \right \} \> .
\eea
This only makes sense if the time-evolution operator for the
potential $V_0$ can be evaluated.


\vspace{0.5cm}

\subsection{\textcolor{blue}{Semi-classical Expansions}}
\label{sec1: halbklass}

It is possible to evaluate 
the path integral for a particle in an arbitrary  potential
in an approximate way 
for those situations where the classical description gives the 
main contribution. Path integrals are the natural
starting point for such semi-classical approximations because the notion of a trajectory 
and the classical limit are "built in".

The obvious method is to take into account only the
\textcolor {blue} {\bf quadratic fluctuations} around the classical path. 
This corresponds to the generalization of the well-known
\textcolor {blue} {\bf saddle point approximation} or \blau{stationary phase method} 
from ordinary integrals to functional integrals.
\vspace{0.8cm}

\renewcommand{\baselinestretch}{0.9}
\scriptsize
\refstepcounter{tief}
\noindent
\blau{\bf Detail \arabic{tief}:} {\bf Asymptotic Expansion of Integrals}\\

\noindent
\begin{subequations}
A (real) integral of the form
\be
I \E \int_a^b dt \> \exp \left [ \, - f(t)/\epsilon \, \right ] 
\ee
can be evaluated in the limit $\epsilon \to 0$ by \blau{\bf Laplace's method}:  Find the
\textcolor{blue}{\bf minimum} of the function $f(t)$ at the point
$t_0 \in [a,b] $ and expand the function in a  Taylor series around this minimum up 
to $2^{\rm nd}$ order. 
This gives
\be
I \> \simeq \>  \exp \left [ \, - f(t_0)/\epsilon \, \right ] \> 
\int_a^b dt \>  \exp \left [ - \frac{1}{2\epsilon} f''(t_0) \, (t-t_0)^2 
\right ] \> \> . 
\ee
Since  $ f''(t_0) > 0 $ we can substitute
$ \, s = \sqrt{f''(t_0)/(2 \epsilon)} (t-t_0) \, $ as new integration variable. For
$\epsilon \to 0$ the limits of the $s$-Integrals become $ -\infty, +\infty $, 
if $t_0$ is between $a, b$ and $-\infty, 0$, or $0,+\infty$ if
$t_0$ happens to coincide with one of the integration limits . After performing 
the $s$-integration we therefore have
\be
I \> \simeq \>  \exp \left [ \, - f(t_0)/\epsilon \, \right ] \> 
\sqrt{ \frac{2 \pi \epsilon}{f''(t_0)}} \cdot A  \> ,
\label{Int a la Laplace} 
\ee
in which the factor $\, A \, $ takes the value 
$ \, A = 1 $ if the minimum is within the integration interval 
and $ \, A = 1/2 $ if the minimum occurs at the boundary.
If one writes in Eq. (\ref{Int a la Laplace})
$ \, \epsilon^{1/2} = \exp( \frac{1}{2} \ln \epsilon) \, $ then
one sees that the quadratic expansion around the minimum is subdominant
compared to the leading term $ \, (- f (t_0) / \epsilon) $ :
$ \, \ln \epsilon $ diverges weaker  than $ 1 / \epsilon $.
Eq. (\ref {Int a la Laplace}) can be seen as the beginning of a
systematic (asymptotic) expansion of the integral $ I $ which can be obtained
by higher terms in the Taylor expansion of
$ f(t) $ around the minimum. The standard example for such an expansion is Euler's
integral representation of the Gamma function
\be
\Gamma(x+1) \E \int_0^{\infty} dt \> t^x \, e^{-t} \> 
\ee
in which $f(t) = t - x \ln t $. The minimum of this function occurs at
$ t_0 = x $ and we have $ f''(t_0) = 1/x $ so that Stirling's formula is obtained
for large positive $x$ :
\be
x! \EQ \Gamma(x + 1) \> \stackrel{x \to \infty}{\longrightarrow} 
\> \sqrt{2 \pi} \, x^{x + 1/2} \, e^{-x} \> . 
\label{Stirling}
\ee
\vspace{0.1cm}

\noindent
A generalization of this method to (complex) integrals of the form
\be
I \E \int_{C} dz \> g(z) \, \exp \left [ \, f (z) / \epsilon \, \right] \>,
\ee
has been developed by P. Debye. Here
$ \, f(z), g(z) $ are analytic functions in a region of the complex plane which contains
the path of integration $ C $. The crucial idea is to deform $ \, C $ in such a way 
that on a part $ \, C_0 $ of $ \, C $ the following conditions
are fulfilled:
(1) Along $ \, C_0 \> \>  \> {\rm Im} \, f(z) $ is constant, (2) $\,  C_0 $ goes through a
saddle point $ \, z_0 $ where $ \, df(z)/dz = 0, $  and (3) at $ \, z =  z_0 $
$ \, {\rm Re} \, f (z) $ goes through a relative maximum, that is $ \, C_0 $ is the path of
\blau{\bf steepest descent}. If one writes
$ \, f(z) = f(z_0) - \tau^2 $ , expands the product
$ \, g(z(\tau)) \, dz (\tau) / d\tau = \sum_m c_m \tau^m $ in a power series
and integrates term by term
then one can derive the asymptotic expansion \footnote{An excellent and particularly clear account
can be found in {\bf \{Dennery-Krzywicki\}}, ch. I.31.}
\be
I \> \simeq \> \exp \left [ \, f(z_0)/\epsilon \, \right  ] \,  
\sqrt{\epsilon} \, \sum_{m=0}^{\infty}  \, c_{2 m} \, \Gamma(m+1/2) \, 
\epsilon^m \> .
\ee
From the Taylor expansion of $f(z), g(z) $ around $ z = z_0$ it is easy to find the
lowest coefficient as
\be
c_0 \E \frac{g(z_0)}{\sqrt{-f''(z_0)/2}} \> .
\ee

\end{subequations}
\renewcommand{\baselinestretch}{1.2}
\normalsize
\vspace{0.6cm}


\noindent
The classical path obeys the equation of motion
\be
m \ddot x_{\rm cl}(t) \> + V'\left ( x_{\rm cl}(t) \right ) \E 0\> ,
\hspace{1cm} x_{\rm cl}(t_a) = x_a \> , \> x_{\rm cl}(t_b) = x_b \> .
\label{klass Beweg-Eq.}
\ee
With $ \> y(t) = x(t) - x_{\rm cl}(t) \> $ we obtain
\bea
U(x_b, t_b; x_a, t_b) \EA e^{i S_{\rm cl}/\hbar}
\int_{y(t_a)=0}^{y(t_b)=0} {\cal D}y(t) \> \exp \left \{ \> \frac{i}{\hbar}
\left ( S [x_{\rm cl}+y] - S [x_{\rm cl}] \right ) \> \right \}\non
&& \hspace{-1cm} \simeq e^{i S_{\rm cl}/\hbar}
\int_{y(t_a)=0}^{y(t_b)=0} \hspace{-0.2cm} {\cal D}y(t) 
\exp \left \{ \> \frac{i}{2 \hbar}
\int d\sigma d\sigma' \> 
\frac{ \delta^2 S}{\delta x_{\rm cl}(\sigma)
\delta x_{\rm cl}(\sigma')} \> y(\sigma) \> y(\sigma') \> \right \} ,
\label{U halbkl 1}
\eea
if we stop the functional Taylor expansion after the quadratic term. 
A simple calculation ( \purpur{\bf problem 4 b)} gives
\be
\frac{1}{2} \int d\sigma d\sigma' \> \frac{ \delta^2 S}
{\delta x_{\rm cl}(\sigma) \delta x_{\rm cl}(\sigma')} \> y(\sigma) \> 
y(\sigma') \E \int_{t_a}^{t_b} dt \>\left [ \frac{m}{2} \dot y^2(t)
\> - \> \frac{1}{2} V''\left ( x_{\rm cl}(t) \right ) y^2(t) \right ] \> ,
\ee
i.e. the action of a harmonic oscillator with a time-dependent oscillator 
constant
\be
\omega^2(t) \E \frac{1}{m} \> V''\left ( x_{\rm cl}(t) \right ) \> .
\ee
Employing the results from the previous chapter we immediately can write 
down the result
\be
U(x_b, t_b; x_a, t_a) \E \sqrt{ \frac{m}{2 \pi i \hbar} \Bigl |
f(t_b,t_a) \Bigr |^{-1} }
\> \exp \left \{ i \left [ S_{\rm cl}(x_b, t_b; x_a,t_a)/\hbar - 
n \frac{\pi}{2} \right ] \> \right \} \> ,
\label{U halbkl 2}
\ee
in which  $ \> f(t_b,t_a) $ fulfills the Gel'fand-Yaglom equation 
\footnote{In the calculus of variations this equation is also
known as \textcolor{blue}{\bf Jacobi equation}. It determines that displacement
of the path that makes
$ \delta^2 S $ minimal. If this minimum is positive, we know
that for all displacements the classical path is a minimum, not only
an extremal value of the action. It can be shown (e.g. \meingruen {\bf Schulman}, p. 81
- 83), that for sufficiently small $ T = t_b - t_a $ the classical path
indeed is a minimum but that the second functional derivative of the action changes sign 
($ \delta^2 S <0 $)  as the particle passes through a
focal point. This means that the classical path in general is only an
extremum of the action.}
\be
m \frac{\partial^2 f(t,t_a)}{\partial t^2}  \> + \> V'' \left ( 
x_{\rm cl}(t) \right ) \> f(t,t_a) \E 0 \> , \hspace{0.3cm} 
{\rm with} \> \> \>  
f(t_a,t_a) \E 0 \> , \> \> \frac{\partial f(t,t_a)}{\partial t} 
\Biggr |_{t=t_a} \E 1
\ee
Of course, it is also possible to express the  prefactor of 
the propagator by the second derivative of the classical action with respect to
the boundary points, i.e. by the van-Vleck determinant
(see Eq. (\ref{Vor durch S cl})).

\noindent 
If there
are several stationary  points in the functional integral (sufficiently far apart)
their contributions must be summed up:
\be
U(x_b, t_b; x_a, t_a) \> \simeq \> \sum_k \> 
\left (\frac{1}{2 \pi i \hbar} \Biggl |
\frac{\partial^2 S_k}{\partial x_a \partial x_b}  \Biggr |
\right )^{1/2}
\> \exp \left \{ i \left [ S_k/\hbar \> - \> n_k \frac{\pi}{2}
\right ] \> \right \} \> .
\label{U halbkl 3}
\ee

\begin{figure}[hbtp]
\bce
\vspace*{-8cm}
\includegraphics[angle=0,scale=0.8]{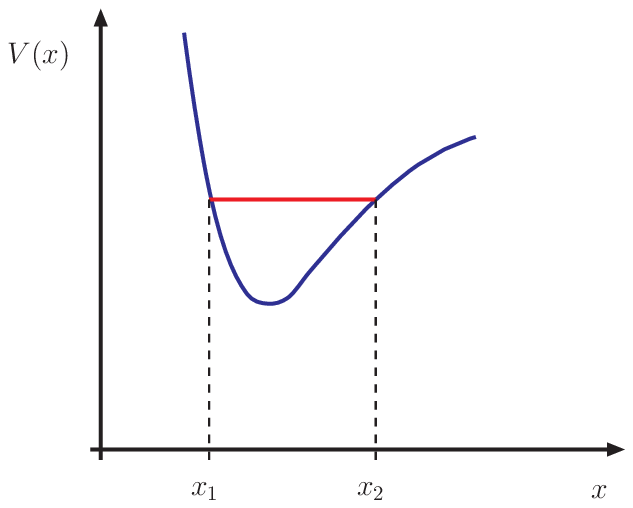}
\label{abb:1.4.1}
\ece
\vspace{-7cm}                  
\bce
{\bf Fig. \arabic{abb}} : A potential with two turning points.
\ece
\end{figure}

\vspace{0.5cm}
\noindent
We now want to determine the energy eigenvalues of possible bound states of the
system using the semi-classical result (\ref {U halbkl 3}).
In principle, this would be possible again by comparing
the semi-classical result with the spectral representation of the
time-evolution operator. However, it is generally not possible to solve
the classical equation of motion (\ref {klass Beweg-Eq.}) in closed
form and thereby to determine the classical action.
Instead, we look at the time-independent Green function
\be
G(x_b,x_a;E) \E \int_0^{\infty} d(t_b - t_a) \> \left ( - \frac{i}{\hbar}
\right ) \> U(x_b t_b; x_a t_a) \> e^{i E (t_b-t_a)/\hbar} \>  
= \> \sum_n \> \frac{\psi_n(x_b) \> 
\psi_n^{\star}(x_a)}{E - E_n + i \, 0^+} \> ,
\ee
the poles of which give us the energy eigenvalues and the residues
give us the eigenfunctions. If we are not interested in the latter ones, the simplest quantity 
to study is
\be
{\rm tr} \> G(E) \E \int_{-\infty}^{+\infty} dx \> G(x,x;E) \>
= \> \sum_n \> \frac{1}{E - E_n + i \, 0^+} \> \> .
\label{Sp G(E) 1}
\ee
The  Fourier transform will be performed again by using 
the stationary phase method. The stationary points of the time integral
obey
\be
\frac{\partial}{\partial T} \> \Bigl [ \> S_{\rm cl}(xT,x0) + E \, T\> 
\Bigr ] \E 0 \> ,
\label{Stat Bed.}
\ee
which is a well-known relation for the energy of a classical trajectory.
In the same way we can evaluate the integral over $ x $ (i.e. the trace)
by this method. Since the derivative of the action with respect to the endpoints 
gives us the momentum of the particle at these points
(\purpur{\bf Problem \ref{klass Anfang/End}}) we find the relation
\be
p(T) \> - \> p(0) \E 0 \>.
\ee
In other words: We search for classical trajectories with energy
$ \> E $ , which are periodic both in the coordinates
$ x $ as well in the momenta $ p $ .
Eq. (\ref{Sp G(E) 1})  then becomes
\be 
{\rm tr} \> G(E) \> \simeq \> {\rm const.} \sum_k A_k  \> e^{i W(T_k)/\hbar}
\label{Sp G(E) 2}
\ee
in which 
\be 
W(T) \E E T \> + S_{\rm cl}(T) \> = E T + \int_0^T dt \> \left (
p \cdot \dot x - H \right ) \E \int_0^T dt \> p \cdot \dot x \> .
\label{Def W}
\ee
The factor $ \> A_k \> $ contains all pre-factors arising from the semi-classical 
approximation and the quadratic fluctuations around the stationary points.
The result
(\ref{Sp G(E) 2}) is particularly simple for a potential with a single minimum
as sketched in Fig.  \ref{abb:1.4.1}.

\refstepcounter{abb}

\vspace{0.3 cm}
\noindent
The particle then executes a periodic classical motion between the  
\blau{turning points} $x_1$ and $x_2$; as derived from energy
conservation $ \> m \dot x^2/2 + V(x) = E \> $, one obtains for the period
\be
T(E) \E 2 \int_{x_1}^{x_2} dx \> \sqrt{\frac{m}{2 [E - V(x)] }} \> .
\ee
Any multiple value of $ \> T(E) \> $ also fulfills the
stationary condition (\ref{Stat Bed.}) so that $ \> T_k (E) = k T(E) \> $ holds. 
The exponent in Eq. (\ref{Sp G(E) 2}) 
can then be written as a multiple value of
$ \> W[T(E)]/\hbar \> $ . 
If the prefactor  $ A_k $ is ignored
then one obtains
\be 
{\rm tr} G(E) \E {\rm const.} \>  \sum_{k=1}^{\infty}
 \left ( e^{i W[T(E)]/\hbar} 
\right )^k
\E {\rm const.} \frac{\exp \left (i W[T(E)]/\hbar \right )}
{ 1 - \exp \left (i W[T(E)]/\hbar \right )} \> .
\label{Sp G(E) 3}
\ee
This expression has poles at energies which fulfill
the quantization condition
\be
W[T(E)] = \> \oint p \, dx = 2 \pi n \> \hbar \> .
\ee
It can be shown (\meingruen{\bf Schulman}, ch. 18) that  prefactors
which have been neglected up to now give a phase $ \pi/2 $ each.
This means that the minus sign in the denominator of Eq.
(\ref{Sp G(E) 3}) is converted into a plus sign and that, finally, 
we have obtained the old \textcolor{blue}{\bf Bohr-Sommerfeld quantization rule} 

\be
\boxed{
\qquad \oint p \> dx \E 
2 \int_{x_1}^{x_2} dx  \> \sqrt{ 2 m \left [ E - V(x) \right ] } \E 
\pi \left ( 2 n + 1 \right ) \> \hbar \qquad \> .
}
\ee

\newpage

\subsection{\textcolor{blue}{Potential Scattering  and Eikonal Approximation}}
\label{sec1: Potstreu}

We now want to consider scattering in a local potential
\be
\hat V \E V(\hat \fr) 
\ee
which leads to a continous spectrum of the Hamiltonian
if it falls off (sufficiently fast) for large values of
$ \, \fr $~. Let the
initial momentum of the particle be $ \> {\bf k}_i \> \> \>  
( \hbar = 1) \> $ and the final one  $ \> {\bf k}_f \> $.
Time-dependent scattering is formulated in the 
\textcolor{blue}{\bf interaction picture} in which the free propagation
of the particle has been removed.
The $S$ matrix then simply is the matrix element of
the time-evolution operator in the interaction picture
$ \> \> \hat U_I(t_b,t_a) \> = \exp(i\hat H_0 t_b)
\hat U(t_b,t_a) \exp(-i\hat H_0 t_a) \> \> $
between the scattering states and evaluated at asymptotic times:
\vspace{0.2cm}

\fcolorbox{black}{white}{\parbox{14cm}
{
\bea         
S_{i \to f} \EA \lim_{T \to \infty} \> \left < {\bf k}_f \left | \, 
\hat U_I (T,-T) \, \right | {\bf k}_i \right > \E\lim_{T \to \infty} \> 
e^{i (E_i + E_f) T} \, \left < {\bf k}_f \left | \,
\hat U(T,-T) \, \right | {\bf k}_i \right >   \non
&=:& 
(2 \pi)^3 \delta^{(3)} \left ( {\bf k}_i - {\bf k}_f \right )  - 2 \pi i \,  
\delta\left ( E_i - E_f \right ) \, T_{i \to f} \> . \no
\eea
}}
\vspace{-2.5cm}

\bea
\label{S-Matrix}
\eea
\vspace{-1cm}

\bea
\label{T aus S}
\eea
\vspace{0.2cm}


\noindent
The second line defines the $T$ matrix in which the energy-conserving
$\delta$-function \footnote{Since the scattering potential is not
translationally invariant there is only energy and not momentum conservation:\\
\qquad $  E_i = {\bf k}_i^2/(2m) = \> E_f = {\bf k}_f^2/(2m) \> $. 
The scattering states are normalized as  $ \> \left < {\bf k}_f | {\bf k}_i \right >
= (2 \pi)^3 \delta^{(3)}( {\bf k}_f - {\bf k}_i ) \> $.} has been
removed. It is customary to define a scattering amplitude
\be 
f(\Omega) \E- \frac{m}{2 \pi} \> T_{i \to f}
\label{Streuamp}
\ee
so that the differential cross section is given by 
\be
\frac{d \sigma}{d \Omega} \E\left | f(\Omega) \right |^2 \> .
\ee
Here $ \> \Omega  \> $ is the solid angle of the outgoing particle.
For a spherically symmetric potential the scattering amplitude is only a function
of the polar angle $ \theta $.
\vspace{0.2cm}

We now want to find a path-integral representation of the $T$ matrix
\footnote{This derivation follows Ref. \cite{Ros1}.}.
We start from the formulation (\ref{Lagrange PI})
for the time-evolution operator
$U \left (\fx_b,t_b; \fx_a,t_a \right )$
and use a trick to switch to an \blau{\bf integration over velocities}: 
We multiply the path integral with the following "One"
\vspace*{-0.2cm}

\be
1 = \prod_{k=1}^N \, \int \, d^3 v_k \> \delta \left ( \frac{\fx_k -
\fx_{k-1}}{\epsilon} - \fv_k \right ) \E\epsilon^{3N} 
\prod_{k=1}^N \, \int \, d^3 v_k \> \delta \left ( \fx_k - \fx_{k-1} 
- \epsilon \fv_k \right ) \> .
\ee
\noindent
Now we can perform the $\fx_k$-integrations ($ k = 1, \ldots N-1 $) 
This gives 
$\fx_j = \fx_0 + \epsilon \sum_{i=1}^j \, \fv_j $, or in the continous 
notation the trajectory
$ \fx_B(t) \E \fx_a + \int_{t_a}^t dt' \> \fv(t') $, 
in which the boundary condition at $ t = t_a $ was used. 
The boundary condition at $ t = t_b $ gives 
$x_b = x_a + \int_{t_a}^{t_b} dt' \fv(t')$ , 
which -- after addition of the resulting expressions
for $ \fx_B(t) $  and dividing by two  -- leads to the symmetric  form
\be
\fx_B(t) \E \frac{x_a + x_b}{2} + \frac{1}{2} \left [ \, \int_{t_a}^t dt' \> \fv(t') 
-  \int_t^{t_b} dt' \> \fv(t') \, \right ] \deF \frac{x_a + x_b}{2} +  \fx_v(t) \> .
\label{Bahn}
\ee 

However, when integrating over $\fx_k $ there remains 
one $\delta$-function so that we obtain
\vspace{0.3cm}

\fcolorbox{black}{white}{\parbox{14cm}
{
\bea
\vspace*{-0.3cm}
U(\fx_b, t_b; \fx_a, t_a) \EA \lim_{N \to \infty}
\left ( \frac{\epsilon m}{2 \pi i} \right )^{\frac{3 N}{2}} \!
 \int d^3v_1 \ldots d^3v_N \> \delta^{(3)} \left (
\fx_b - \fx_a - \epsilon \sum_{j=1}^N \fv_j \right )\non
&& \cdot \exp \left \{\>  i \epsilon \sum_{j=1}^N \Bigl [ \, \frac{m}{2}
\fv_j^2 - V \left (\fx_j = \fx_a + \epsilon \sum_{i=1}^j \fv_i
\right ) \, \Bigr ] \> \right \} \non
&& \hspace*{-2.5cm} \EQ 
\int {\cal D}^3 v(t)\,                
\delta^{(3)} \left ( \, \fx_b - \fx_a -\int\limits_{t_a}^{t_b} dt \> 
\fv(t)\, \right ) \> 
\exp \left \{ \, i \int\limits_{t_a}^{t_b} dt \> \Bigl [  \frac{m}{2} 
\fv^2(t) - V( \fx_B(t) )
\, \Bigr ] \> \right \}  . \no 
\eea
}}
\vspace{-3.2cm}

\bea
\label{Geschwind PI}
\eea
\vspace{-0.6cm}

\bea
\label{velo PI}
\eea
\vspace{0.5cm}

\noindent
Here the "measure" ${\cal D} \fv$ is defined in such a way that the Gaussian
integral yields unity
\be  
\int {\cal D}^3 v(t) \> \exp \left \{ \, i \int_{t_a}^{t_b} dt \>  
\frac{m}{2} \fv^2(t) \,\right \}  \E 1 \> , 
\label{Normierung velo}
\ee
as may be seen from the discretized form. Of course, for a general potential
this path-integral representation of the time-evolution operator
is analytically as unsolvable as the original form. However,
because of the physical interpretation of the auxiliary variable $ \fv $ as
velocity  Eq. (\ref{velo PI}) is a good starting point for approximations,
especially in the high-energy case.
Note that the functional integral over $ \fv $ has no boundary conditions
since they are all contained in the remaining 
$ \delta $-function and the path $ \fx_B (t) $ . 

\noindent
We now write Eq. (\ref{S-Matrix}) as
\be
S_{i \to f} \E \lim_{T \to \infty} \> 
e^{i (E_i + E_f) T} \, \int d^3x \, d^3y \> e^{- i {\bf k}_f \cdot \fx }
\, U(\fx,T; \fy,-T) \, e^{ i {\bf k}_i \cdot \fy }
\ee
and insert the representation (\ref{Geschwind PI}). If we use the coordinates
$ \> \fr = (\fx + \fy)/2 \> , \> \> 
{\bf s} = \fx - \fy \> $, then we obtain
\bea
S_{i \to f} \EA \lim_{T \to \infty} \exp \left [ \, i (E_i + E_f) T \, 
\right ] \> \int d^3 r \> e^{- i \fq \cdot \fr} \> \int {\cal D}^3   
\fv \> \exp \left \{ \, 
i \int_{-T}^{+T} dt \left [ \, \frac{m}{2} \fv^2(t) -   
{\bf K} \cdot \fv(t)
 \, \right ]  \> \right \} \non
&& \hspace{7cm} \cdot \exp \left \{ \> - i \int_{-T}^{+T} dt \> V \left ( \, 
\fr + \fx_v(t) \, \right ) \> \right \} \> ,
\label{S PI1}
\eea
since the relative coordinate $ \> {\bf s} \> $ is fixed by the $\delta$-function
in Eq. (\ref{Geschwind PI}). Here we have introduced the momentum transfer
and the average momentum
\be
\fq \E{\bf k}_f - {\bf k}_i \> , \hspace{0.3cm} {\bf K} \E
\frac{1}{2} \left ( {\bf k}_i + {\bf k}_f \right ) \> .
\ee
In addition, the sign function $ \> {\rm sgn}(x) = x/|x| \> $ has been used 
to write the argument of the potential compactly as
\be
\fx_v(t) \Def \frac{1}{2} \int_{-T}^{+T} dt' \> {\rm sgn}(t-t') \,  \fv(t') \> .
\ee
($\fx_v(t)$ ist the deviation of the trajectory from the mean position, see
 Eq. (\ref{Bahn})). The shift 
\be
\fv(t) \To  \fv(t) + \frac{{\bf K}}{m} 
\label{v verschieb}
\ee
eliminates the linear term in the exponent of the functional integral.
Since  $ \quad \int_{-T}^{+T} dt' \, {\rm sgn}(t-t') = 2 t \quad $ 
and $E_i + E_f - {\bf K}^2/m = \fq^2/(4 m) $  the result is
\bea
S_{i \to f} \EA \lim_{T \to \infty} \exp \left ( \, i \frac{\fq^2}{4 m} T
\, \right ) \> \int d^3 r \> 
e^{- i \fq \cdot \fr} \>  \> \int {\cal D}^3 v \> 
\exp \left [ \, 
i \int_{-T}^{+T} dt \, \frac{m}{2} {\bf v}^2(t)    
 \, \right ]   \non
&& \hspace{4.5cm} \cdot \exp \left \{ \> - i \int_{-T}^{+T} dt \>   
V \left ( \, \fr + \frac{{\bf K}}{m} t + \fx_v(t)
\right ) \right \} \> .
\label{S PI2}
\eea
If the interaction is weak one can expand in powers of the potential.
In zeroth order one has
\be
S^{(0)}\E \lim_{T \to \infty} \exp \left ( \, i \frac{\fq^2}{4 m} T\, \right ) \> 
\int d^3 r \> 
e^{- i \fq \cdot \fr} \E (2 \pi)^3 \delta^{(3)} \left ( {\bf k}_i - {\bf k}_f \right ) \> 
\ee
and thus we may write
\bea
\left ( S - 1 \right )_{i \to f} \EA \lim_{T \to \infty} \exp \left ( \, 
i \frac{\fq^2}{4 m} T \, \right ) \> \int d^3 r \> 
e^{- i \fq \cdot \fr} \>  \> \int {\cal D}^3 v \> 
\exp \left [ \, 
i \int_{-T}^{+T} dt \, \frac{m}{2} {\bf v}^2(t)    
 \, \right ]   \non
&& \hspace{3cm} \cdot \left \{ \, \exp \left [ \> - i \int_{-T}^{+T} dt \>
V \left ( \, \fr + \frac{{\bf K}}{m} t + \fx_v(t) \right ) \right ] \, - 1 \right \} \> .
\label{S PI3}
\eea
The path-integral representation (\ref{S PI3}) has the disadvantage that a phase 
$ \> \fq^2 T/(4m) \>  $ appears which in the limit
$ T \to \infty $ seems to diverge. Of course, in each order of perturbation theory
this phase cancels so that the limit   $ T \to \infty $ actually exists
but one would like to have a formulation where this property is explicitly built in.
This can be achieved in the following manner: First one uses the fact
that any power of $ \fq^2 $ in the $ \fr $-integral 
can be generated by applying the Laplace operator $ \, - \Delta \, $
on  $\exp (i \fq \cdot \fr) $. In other
words:
\be
\exp \left ( \, i \frac{\fq^2}{4 m} T \, \right )
\, e^{- i \fq \cdot \fr } 
\E \exp \left ( \, - i \frac{\Delta}{4 m} T \, \right ) \, 
e^{- i \fq \cdot \fr} \> .
\ee
An integration by parts (which doesn't produce any boundary terms if the
potential falls off rapidly enough) leads to the exponential function with
the Laplace operator acting to  the right, i.e. on the potential term.

Finally, we use a trick which goes under the name
\blau{undoing the square} \footnote{In many-body physics this is called the
``Hubbard-Stratonovich''  transformation, see
 {\bf chapter} $\>$ {\bf \ref{sec2: Hilfsfelder}}. } and which linearizes the square
(of an operator) in the exponent. In the simplest case of an ordinary integral 
this simply is the identity
\be
e^{-ia x^2} \E \sqrt{\frac{i}{4 \pi a}} \, \int_{-\infty}^{+\infty} dy \> 
\exp \left ( - \frac{i}{4 a} y^2 \right ) \, e^{- x \, y } \hspace{1cm}    
(a \> {\rm real} ) \> \> ,
\label{aufheb quad}
\ee
which can be proved by completing the square.
This can be extended to path integrals and one may write
\be
\exp \left ( -\frac{i}{4 m} T \, \Delta \right ) \E \int {\cal D}^3 w(t) \> 
\exp \left [ - i \int_{-T}^{+T} dt  \, \frac{m}{2} \, 
\fw^2 (t) - \int_{-T}^{+T} dt \, f(t) \, 
\fw(t) \cdot \nabla \> \right ] \> ,
\label{undoing1}
\ee
where the "measure" has again been chosen such that the Gaussian integral
is normalized to one.

The arbitrary function  $ f(t) $ only has to fulfill $ \quad \Tint \, f^2(t) \E T/2 \quad  $ 
and we chose it as  $ \> f(t) = {\rm sgn}(t)/2 \>  $. Our representation now reads
 \be
\exp \left ( -\frac{i}{4 m} T \, \Delta \right ) \E \int {\cal D}^3 w(t) \> 
\exp \left [ - i \Tint  \, \frac{m}{2} \, 
\fw^2 (t) - \fx_w(0)\cdot \nabla \> \right ] \> .
\label{undoing2}
\ee
Note that the sign of the quadratic 
$\fw$-term in Eq. (\ref{undoing2}) necessarily is negative -- therefore we may 
call  the 3-dimensional auxiliary variable $\fw(t)$ an "{\bf anti-velocity}" \footnote{Formulations 
of potential scattering  \textcolor{blue}{\bf without} anti-velocity can be found in Refs.
\cite{CaRo}, \cite{Rosen}.}.

The advantage of linearizing the exponent is that we now have a differential
operator which simply shifts the argument of the potential function by
$ - \fx_w(0) $ (that is nothing else than Taylor's theorem
$ \> \exp(a \, \frac{d}{dx}) f(x) = f(x+a) \> $ ). The result of these manipulations
is therefore the expression
\bea
\left ( S - 1 \right )_{i \to f} \EA i \lim_{T \to \infty}  \int d^3 r \,
e^{- i \fq \cdot \fr} 
\int {\cal D}^3 v \, {\cal D}^3 w
\, \exp \left \{  i \frac{m}{2} \int_{-\infty}^{+\infty}  dt \, 
\left [  \fv^2(t)-\fw^2 (t)  \right ] \right \} \non
&& \hspace{1.5cm} \cdot  
\Biggl \{ \, \exp \left [ i \Tint \, V \left ( \fr + \frac{\fK}{m} t + \fx_v(t) - 
\fx_w(0) \right ) \, \right ] - 1 \, \Biggr \} \> .
\label{S PI4}
\eea
That is free of the "dangerous" phase which allows us to perform formally
the limit $ T \to \infty $. The price to pay is an additional  
(functional) integration over the "anti-velocity".
\vspace{0.2cm}

However, we have not yet displayed the energy-conserving $ \delta $-function
to obtain the $ T $ matrix  according to Eq. (\ref{T aus S}).
One can check that this is achieved when Eq. (\ref{S PI3}) is expanded 
in powers of the interaction: In each term of this perturbative expansion
we find such an energy-conserving $ \delta $-function.

\noindent 
How to achieve this without expanding in powers of the interaction?
For that purpose one may use a
trick which \blau{Faddeev and Popov} have introduced in field theory to
quantize non-abelian gauge theories (see {\bf chapter} \ref{sec3: Eichtheo}).
We first note that in the limit $ T  \to \infty $ the action
in the path integral (\ref{S PI2}) is invariant under the transformation
\be
t \E \bar t + \tau \> , \hspace{1cm} \fr \E \bar \fr - \frac{{\bf K}}{m} \tau \> , 
\hspace{1cm} 
\fv(t) \E \bar \fv(\bar t)   
\label{Zeit trans}
\ee
since
\be
\int_{-T}^{+T} dt \> V \left ( \, \fr
+ \frac{{\bf K}}{m} t + \fx_v (t) \right) \E
\int_{-T-\tau}^{T-\tau} d \bar t \> V \Biggl ( \, 
\bar \fr + \frac{{\bf K}}{m} \bar t  
+\frac{1}{2} \int\limits_{-T-\tau}^{T-\tau} d \bar t' 
\, \bar {\bf v}(\bar t') \, {\rm sgn}(\bar t- \bar t') 
\Biggr ) \> .
\ee
For finite $\tau$ and $ T \to \infty $ the change of the integration limits 
is irrelevant and the action remains invariant under the transformation
(\ref{Zeit trans})
 \footnote{Actually this is a delicate interchange of limits whose justification has to be checked.}.

However, this means that the
component of the vector $ \fr $ which is parallel to $ {\bf K} $
is not fixed leading to a singularity when we integrate over this component,
This singularity is just
the energy-conserving $ \delta $-function we are looking for. We can it
extract, if we first fix the longitudinal component and then
integrate over all possible values -- that is, we multiply the
path integral (\ref {S PI3}), for example with the following "one"
\be
1 \E \frac{|{\bf K}|}{m} \int_{-\infty}^{+\infty} d\tau \> 
\delta \left ( \> \hat
{\bf K} \cdot \left [ \fr + \frac{{\bf K}}{m} \tau \right ] \> \right ) 
\label{FP}
\ee
Now we perform the transformation (\ref{Zeit trans}) in the path integral 
and obtain
\bea
\left ( S - 1 \right )_{i \to f} \EA  \frac{|{\bf K}|}{m} 
\lim_{T \to \infty} \, \int_{-\infty}^{+\infty} d\tau\,  \int d^3 r \> 
\exp \left (- i \fq \cdot \fr + i \fq \cdot \frac{{\bf K}}{m}   
\tau \right ) \, \delta \left ( \hat {\bf K} \cdot \fr \right ) \non
&& \cdot \int {\cal D}^3 v \, {\cal D}^3 w \> \exp \left \{  \, 
i \Tint \, \frac{m}{2} \left [ \fv^2(t) - \fw^2(t) \right ] \, \right \} \non   
&& \cdot  \left \{ \, \exp \left [ \, - i \Tint \, V 
\left ( \, \fr + \frac{{\bf K}}{m} t + \fx_v(t) - \fx_w(0) \, \right ) \, \right ] \, 
 - 1 \, \right \} \, ,
\label{S PI5}
\eea
where we have replaced the variables with a bar by the original ones in order to
simplify our notation. The only dependence on $\tau$ which remains in the integrand
is now in the factor
$ \exp (  i \tau \fq \cdot {\bf K}/m ) $, so that integration over it exactly generates  
the energy-conserving $\delta$-function:
\be
2 \pi \, \delta \left ( \, \frac{\fq \cdot \fK}{m} \, \right ) \E 2 \pi \, \delta \left ( \, 
\frac{\fk_f^2}{2 m} - \frac{\fk_i^2}{2 m} \, \right ) \> .
\ee
In addition, after the transformation the longitudinal component of $ \fr $ in the path 
integral is set to zero. If we keep in mind that
$ q_{\parallel} = 0 $ then we obtain the following expression for the $ T $ matrix
\vspace{2cm}

\fcolorbox{black}{white}{\parbox{14.7cm}
{
\bea
T_{i \to f} \EA i \frac{K}{m} \,  
\int d^2 b \, e^{- i \fq \cdot \fb } \!
\int {\cal D}^3 v \, {\cal D}^3 w \, \exp \left \{ \, 
i \int\limits_{-\infty}^{+\infty} dt \,  \frac{m}{2} \left [ \fv^2(t) - \fw^2(t)\, 
\right ] \, \right \} \, 
 \Biggl \{ \,  e^{ i \chi( \fb, K; \fv, \fw)} \, - \, 1 \, \Biggr \} \> \> .\no
\eea
}}
\vspace{-2cm}

\bea
\label{T PI}
\eea
\vspace{0.4cm}


\noindent
Here a phase
\be
\chi(\fb, K; \fv, \fw) \E -  \int_{-\infty}^{+\infty}
dt \> V \left ( \fb + \frac{{\bf K}}{m} t + \fx_v (t)- \fx_w(0) \right )  
\label{chi(v,w)}
\ee
has been defined which depends on the velocities  $ \fv(t), \fw(t) $
and the transverse component $\fb$ of the vector $\fr$.
$ E = E_i = E_f = k^2/(2m) $ is the scattering energy. In addition,
the magnitudes of momentum transfer and average momentum are given by
\be
q \EQ | \fq| \E2 k \sin \frac{\theta}{2} \> ,
\hspace{0.4cm} K \EQ  | {\bf K}| \E k \cos \frac{\theta}{2} \> .
\ee
\vspace{0.1cm}

One can check that Eq. (\ref {T PI}) reproduces the exact Born series  in all orders,
i.e. all manipulations have been correct. More interesting than a 
derivation of perturbation theory, however, is that this result can serve as a starting point
for high-energy approximations. In this case, we expect that
the particle moves preferentially along a
\textcolor {blue}{\bf straight line path} with the (constant) velocity $ {\bf} K /m $
and that the
functional integral over $ \fv $ and $ \fw $ only describes the 
fluctuations around this path. That this is indeed the case
can be seen when 
\be
t \E \frac{m}{K} z \> , \hspace{0.3cm} \fv(t)
\E \frac{\sqrt{K}}{m} \, \bar \fv(z) \> , \hspace{0.3cm}  
\fw(t) \E \frac{\sqrt{K}}{m} \, \bar \fw(z) 
\ee
is substituted in
the path integral (\ref{T PI}): It basically keeps its form in the new variables
 \footnote{Except that the kinetic energies of velocity and anti-velocity now read
  $ (\bar \fv^2(t) - \bar \fw^2(t))/2 $ , i.e. do not contain anymore the mass of the particle.}
but the phase becomes
\be
\chi \left (\fb, K;\bar \fv ,\bar \fw \right ) \E
 -  \frac{m}{K}\int\limits_{-\infty}^{+\infty} dz \> 
V \left ( \, \fb +  \hat \fK z +  \frac{1}{\sqrt{K}}  \left [ \fx_{\bar v}(z) -        
\fx_{\bar w}(0)
\right ] \, \right ) \> .
\ee
This shows that for a fixed momentum transfer a \blau{\bf systematic}
expansion in inverse powers of $ K = k \cos (\theta / 2) $ is obtained
by expanding the phase
$ \chi $ simultaneously in powers of $ \fv (t) $ and $ \fw (t) $
and integrating functionally term by term.
One can expect
that this will be valid for large values of the 
energy, small scattering angles $ \theta $ and weakly varying 
potentials.

This can be seen as follows: Obviously the correction from the next term
of the Taylor expansion of the potential should be small compared to the leading term,
i.e. $ | \nabla V \cdot \fx_{\bar v}/\sqrt {K} | \ll V $.
Assuming now that the velocity fluctuations are only relevant within the range 
$ R $ of the potential, then we find that $ \bar v = {\cal O} (1/\sqrt {R}) $
and $ x_{\bar v} = {\cal O} (\sqrt {R}) $ and thus
\be
K R \> \gg \> \left ( \, R \, \frac{\nabla V}{V} \, \right )^2 \> \simeq \>  
\left ( \, \frac{R}{a}\, \right )^2 \> ,
\label{eik Krit}
\ee
where $ a $ is the distance over which the potential changes appreciably. 
So for the application of the eikonal approximation it is 
not necessary that the velocity $ \, k/m $ of the scattered particle should be very large; 
in atomic physics, for example, this is almost never the case as the 
masses involved are large and the energies small.
On the other hand, for large scattering angles 
$ K = k \cos (\theta / 2) $ gets smaller and smaller and one sees that 
the criterion (\ref{eik Krit}) cannot be fulfilled anymore. 
\vspace{0.3cm}

The leading approximation is simply obtained by setting
$ \fv = \fw = 0 $ in the argument of the potential:
\be
T_{i \to f} \> \simeq \>  i \frac{K}{m} \> \int d^2 b \> 
e^{- i \fq \cdot \fb} \> \Biggl \{  \>  
\exp \left [ \, - i \frac{m}{K} \int\limits_{-\infty}^{+\infty} dz \,
V \left ( \fb + \hat {\bf K} z \right ) \, \right ]
- 1 \> \Biggr \} \> ,
\label{AI 0}
\ee
as the functional integrals are trivially one by normalization.
In forward direction the difference between $ K = k \cos(\theta/2) $ 
and $k$ is irrelevant so that one obtains the usual
\textcolor{blue}{\bf eikonal approximation} \footnote{The variant (\ref{AI 0}) is due to
Ref. \cite{AI}.}
\be
T_{i \to f}^{\rm eik} \E i \frac{k}{m}
\int d^2 b \> e^{- i \fq \cdot \fb} \left ( \> e^{i \chi_0(\fb)} 
- 1 \> \right )  \> .
\label{T eik0}
\ee
Here 
\be
\chi_0(\fb) \E
 - \int_{-\infty}^{+\infty} dt \> V \left ( \fb + \hat {\bf K} 
\frac{k}{m} t 
\right ) \E - \frac{m}{k} \int_{-\infty}^{+\infty} dz \> V \left ( 
\fb, z \right )
\label{chi0}
\ee
is the phase which the particle has "accumulated" along its straight line path  
 \footnote{The index indicates the inverse power of $K$ while $m/K$ may have an arbitrary value.}.
Note that the first Born approximation for the $T$ matrix follows from
Eq. (\ref{T eik0}) if -- for weak potentials -- $ \exp(i \chi_0) $ is expanded till first order:
\be
T_{i \to f}^{\rm eik} \To  i \frac{k}{m} \int d^2 b \> e^{- i \fq \cdot \fb} \, \left [
- i \frac{m}{k} \int_{-\infty}^{+\infty} dz \> V \left ( 
\fb, z \right ) \, \right ] \E \int d^3r \> e^{- i \fq \cdot \fr} \, V(\fr) \E
\tilde V(\fq) \EQ T_{i \to f}^{\rm 1^{st} Born} \> .
\label{Born 1}
\ee
However, unlike the first Born approximation the eikonal approximation 
yields a {\bf complex}
scattering amplitude as required by  \textcolor {blue}{\bf unitarity}
(the \textcolor{blue} {\bf ``optical theorem''}).

\vspace{0.8cm}

\renewcommand{\baselinestretch}{0.9}
\scriptsize
\refstepcounter{tief}
\noindent
\blau{\bf Detail \arabic{tief}:} {\bf Unitarity}\\

\noindent
\begin{subequations}

Unitarity is one of the` `sacred '' principles of quantum physics; roughly speaking, 
it requires that in every process no more gets out than what entered in the beginning.
More formally (but more accurately) it means that the $ S $ matrix should fulfill
\be
\hat S^{\dagger} \, \hat S \E \hat S \, \hat S^{\dagger} \E  1 \> .
\label{Unitar}
\ee
If the relation (\ref{T aus S}) is written representation free as
\be
\hat S \E 1 - 2 \pi i \, \delta \left ( E_i - \hat H_0 \right ) \, \hat T 
\ee
then the  unitarity (\ref{Unitar}) of the  $S$ matrix requires 
the $ T $ matrix to obey
\be
\hat T - \hat T^{\dagger} \E \frac{2 \pi}{i} \, \hat T^{\dagger} \,  \delta \left ( E_i - 
\hat H_0 \right ) \, \hat T \> .
\label{Unitar fuer T}
\ee
Taking matrix elements for the special case $ i = f $, i.e. for forward scattering
one obtains
\be
{\rm Im} \, T_{i \to i} \E - \frac{m k}{8 \pi^2} \, \int d\Omega_p \> \left | \, 
T_{i \to p} \, \right |^2_{p = k} \> ,
\ee
because the $\delta$-function in Eq. (\ref{Unitar fuer T}) fixes the magnitude 
of the internal momentum $\fp$ 
to be the magnitude of the exterior momentum
$k = |\fk_i| = |\fk_f| $. Usually one discusses the unitarity relation (the optical theorem)
for the scattering amplitude (\ref{Streuamp})
\be
{\rm Im} \, f \left ( \theta = 0 \right ) \E \frac{k}{4 \pi} \, \int d\Omega \, 
| f(\theta) |^2 
\E \frac{k}{4 \pi} \, \sigma_{\rm tot} \quad .
\label{opt Theor}
\ee
Since the first Born approximation (\ref{Born 1}) gives a real scattering amplitude
it is clear that the optical theorem is violated in this frequently employed
approximation.
However, the scattering amplitude of the eikonal approximation has an imaginary part
in forward direction
\be
{\rm Im} f^{\rm eik} ( \theta = 0 ) \E \frac{k}{2 \pi} \, \int d^2b \> \left [ \,  1 - 
\cos \chi_0(\fb) \, \right ] \> ,
\ee
so that the l.h.s. of the optical theorem is non-zero for this approximation.
The total cross section on the r.h.s. is
\be
\sigma_{\rm tot}^{\rm eik} \E \frac{k^2}{4 \pi^2} \, \int d\Omega \, \int d^2 \, d^2b' \> 
e^{i \fq \cdot (\fb - \fb')} \, \left ( e^{i \chi_0(\fb)} - 1 \right ) \> . 
\left ( e^{i \chi_0(\fb')} - 1 \right ) \> .
\ee
Replacing the angle integration by an integration over the momentum transfer
$ q = 2 k \sin (\theta/2) $ 
\be
d\Omega \E 2 \pi \, \sin \theta \, d\theta \E \frac{2 \pi}{k^2} \, q dq \E \frac{1}{k^2} 
\, d^2q
\ee
one obtains
\be
\sigma_{\rm tot}^{\rm eik} \E \frac{1}{4 \pi^2} \,  \int d^2b \, d^2b' \, \int_{q < 2 k} 
d^2q \> \, 
e^{i \fq \cdot (\fb - \fb')} \, \left ( e^{i \chi_0(\fb)} - 1 \right ) \, 
\left ( e^{i \chi_0(\fb')} - 1 \right ) \> .
\ee
If we split up the 2-dimensional $q$-integral into one part in which the momentum transfer
is unrestricted and a "defect" which corrects this assumption, the first part gives a 
$\delta$-function 
\be
\int_{q < 2 k} d^2q \> e^{i \fq \cdot ( \fb - \fb')}  \E (2 \pi)^2 \, \delta ( \fb - 
\fb' ) - 
\int_{q > 2 k} d^2q \> e^{i \fq \cdot ( \fb - \fb')} \>,
\ee
which exactly leads to the fulfillment of the optical theorem:
\be
\sigma_{\rm tot}^{\rm eik} \E   \int d^2b \> \underbrace{\left | \, e^{i \chi_0(\fb)} - 
1 \, \right |^2}_{= 2 \, (1 - \cos \chi_0)} \> + \> \Delta^{\rm eik} \> . 
\ee
The defect of the total eikonal cross section obviously is negative
\be
\Delta^{\rm eik} \E -  \frac{1}{4 \pi^2} \, \int_{q > 2 k} d^2q \, \left | \, \int d^2b \, 
e^{ - i \fq \cdot \fb} \, \left ( e^{i \chi_0(\fb)} - 1 \right ) \, \right |^2
\ee
and small if the cross section has its maximum in forward direction and falls off rapidly
at larger scattering angles (momentum transfers) which usually is the case at high energy.
Let us estimate roughly the defect by using the first Born approximation, i.e. we assume that 
the phase $\chi_0(\fb)$ is small. Then we obtain
\be
\Delta^{\rm eik} \simeq  \frac{1}{4 \pi^2} \, \frac{m^2}{k^2} \, \int_{q > 2 k} d^2q \> 
\left [ \,  \tilde V(\fq) \,\right ]^2
\label{defekt}
\ee
and we have to evaluate the Fourier transform of a (for simplicity assumed spherically 
symmetric) potential
\be
\tilde V(q) \E \frac{4 \pi}{q} \, \int_0^{\infty} dr \> r \, \sin (q r) \, V(r)
\ee
for large values of $ q = |\fq| $ . This is done most easily by subsequent 
integrations by parts ($ \sin (q r) = -  [\cos(qr)]'/q \> ; \> 
\cos (qr) = [\sin(qr)]'/q$) and gives
\be
\tilde V(q) \E \frac{4 \pi}{q} \, \lcp  \, \frac{[r V(r)](0)}{q} - 
\frac{[r V(r)]''(0)}{q^3} + \ldots \, \rcp \> .
\ee
Depending on the behaviour of the potential at the origin, the power of the fall-off 
is different; for example, the Fourier transform of a Yukawa potential
$ \> V(r) = V_0 \, \exp(-\mu r)/r \> $ falls off asymptotically like $1/q^2$. 
More general, the behaviour of the potential at the origin determines the asymptotic
fall-off:
\be
V(r) \> \stackrel{r \to 0}{\longrightarrow} \quad V_0 \, r^{\alpha} \qquad 
\> \Longrightarrow \quad \tilde V(q) \> \stackrel{q \to \infty}{\longrightarrow} \quad
\frac{\rm const.}{q^{\alpha+3}} \> .
\ee 
From Eq. (\ref{defekt}) we then conclude that for these classes of potentials the defect
\be
\Delta^{\rm eik} \> \sim \>  \frac{1}{k^{2 \alpha+6}} \> .
\ee
vanishes rapidly at high energies: The eikonal approximation is then (nearly) unitary!

\end{subequations}
\renewcommand{\baselinestretch}{1.2}
\normalsize
\vspace{0.5cm}

\noindent
If one considers scattering in a spherically symmetric potential, one can perform the
angular integration 
in Eq. (\ref{T eik0}) \footnote{See, e.g., {\bf \{Handbook\}}, eq. 9.1.21~.} 
and one obtains for the scattering amplitude
\be
f^{\rm eik}(\theta) \E\frac{k}{i} \, \int_0^{\infty} db \> b \, J_0(q b)
\> \left ( e^{i \chi_0(b)} - 1 \right ) \> .
\label{f eik}
\ee
Here $ J_0(x) $ is a Bessel function of order zero  and the eikonal phase
explicitly reads
\be
\chi_0(b) \E- \frac{m}{k} \int_{-\infty}^{+\infty} dz \> V \left ( 
\sqrt{b^2 + z^2} \right ) \> .
\label{Eik Phase}
\ee
Comparing with the well-known partial wave expansion of the scattering 
amplitude
\be
f(\theta) \E\frac{1}{2 i k} \sum_{l=0}^{\infty} \> ( 2 l + 1 ) 
\, P_l(\cos \theta) \, \left ( e^{2 i \delta_l(k)} - 1 \right ) \> ,
\ee
reveals a close connection: For large scattering energies so many partial
waves contribute that summation over  $l$ can be replaced by an integral over the
\textcolor{blue}{\bf impact parameter} $ \> b = (l + 1/2)/k \> $.
If, in addition an asymptotic expansion 
\footnote{{\bf \{Handbook\}}, eq. 22.15.1 .} of the Legendre polynomials
$ \, P_l(\cos \theta) \, $ is used for large values of  $l$, then one finds complete
agreement with the form of the eikonal result (\ref{f eik}) and the simple relation 
$ \, 2 \delta_l(k) \simeq  \chi_0(b) \, $ between scattering phase and eikonal phase.
\vspace{0.2cm}

It is (relatively) easy to calculate
the first correction to the eikonal result: We expand  
the phase (\ref{chi(v,w)}) up to first order
\bea
\chi(\fb,\fK;\fv,\fw) \EA - \int_{-\infty}^{+\infty} dt \, V \left ( \fb + 
\frac{{\bf K}}{m} t \right ) \non
&&-  \int\limits_{-\infty}^{+\infty} dt \, \nabla V  \left ( \fb + 
\frac{{\bf K}}{m} t \right )
\, \frac{1}{2} \int_{-\infty}^{+\infty} ds \, \biggl [  \fv(s) \, 
\sgn(t-s) - \fw(s) \, \sgn(-s)  \biggr ]
\eea
and perform the Gaussian integrals over $\fv$ and 
$\fw$. The corrected result
\be
T_{i \to f} \> \simeq \>  i \frac{K}{m} \> \int d^2 b \> 
e^{- i \fq \cdot \fb} \> \Biggl \{  \>  e^{ i\chi_0(\fb) + i \chi_1(\fb)}  - 1 \> 
\Biggr \} 
\label{T eik1}
\ee
now contains an additional phase
\be
\chi_1(\fb) \E - \frac{1}{8 m } \, 
\lim_{T \to \infty} \int_{-T}^{+T} ds 
\int_{-T}^{+T} dt_1 dt_2 \>   \nabla V_1 \cdot \nabla V_2                
\, \biggl [ \, \sgn(t_1-s) \, \sgn(t_2-s) - \sgn^2(-s) \, \biggr ] \> . 
\ee
Here the second term in the square bracket is due to the integration over
the anti-velocity and  $V_{1/2}$ is a shortform for  $ V( \fb + {\bf K} t_{1/2}/m )$ .
The  $s$-integration can be performed with the help of
\be
\int_{-T}^{+T} ds \>  \sgn(t-s) \, \sgn(s-t') \E 2 \Bigl [ \, |t-t'| - T \, \Bigr ]  
\label{sign int}
\ee
and one obtains
\be
\chi_1(\fb) \E \frac{1}{4 m } \, 
\lim_{T \to \infty} \int_{-T}^{+T}  dt_1 dt_2 \>           
\nabla V_1 \cdot \nabla V_2 \> |t_1 - t_2| \> .
\label{eik 1}
\ee
As expected the $T$-dependent terms have been canceled completely by the contribution 
from the anti-velocity.
For a spherically symmetric potential Eq. (\ref{eik 1}) can be simplified by using 
the relations
$ \> \partial V(r)/\partial z = z V'(r)/r \> , \>  \partial V(r)/\partial b 
= b V'(r)/r \> $ and by integrations by parts. One obtains
\be
\chi_1 (b) \E -\frac{1}{2 K} \left (\frac{m}{K} \right )^2\, \left [ \, 1 + b 
\frac{\partial}{\partial b} \, \right ] \,  \int_{-\infty}^{+\infty} dz \> 
V^2(r) \> , \hspace{0.3cm}
r \EQ \sqrt{b^2 + z^2} \> ,
\label{eik 1 symm}
\ee
which shows explicitly that this phase is $ {\cal O} (1/K)$, 
if one allows arbitrary values for the velocity $K/m$.
The result is identical (except for the appearance of $K = k \cos(\theta/2) $ 
instead of $k$ which, however, is irrelevant in this order) with
the next term in a systematic eikonal expansion where also higher orders 
have been calculated by quantum mechanical methods \cite{Wal}, \cite{Sar}.

\vspace{1cm}

\noindent
{\bf Example: Scattering in a Coulomb Potential}
\vspace{0.3cm}

\noindent
The Coulomb potential
\be
V_C(r) \E\frac{Z \alpha}{r}
\ee
causes well-known difficulties in scattering theory due to its slow decrease  
at infinity. From the solution of  Schr\"odinger's equation,
for example, it is known that the scattered wave in the asymptotic region
not only exhibits the characteristic spherical wave
but an additional logarithmic $ r $-dependency
\footnote{See, e.g., {\bf \{Messiah 1\}}, p. 430.}:
\be
\psi_{\rm scattered} \> \stackrel{r \to \infty}{\longrightarrow} \> 
\frac{1}{r} \exp \left [ \> i \left (k r - \gamma \ln 2 k r \right ) \> 
\right ]\, \cdot \, f_C(\theta) \> ,
\label{Streuwelle Coul}
\ee
where $ \gamma = Z \alpha m/k $ is the Sommerfeld parameter and
\be
f_C(\theta) \E- \frac{\gamma}{2 k \sin^2 (\theta/2)} \, 
\exp \left [ \> - i \gamma \ln \left (\sin^2 \frac{\theta}{2} \right )
+ 2 i \sigma_0 \> \right ] \> , \> \sigma_0 \E{\rm arg} \>  
\Gamma (1+i \gamma)
\label{F Coul}
\ee
the Coulomb scattering amplitude. In eikonal approximation the long range 
of the potential
leads to a divergent eikonal phase $\chi_0$ (see Eq. 
(\ref{Eik Phase})). In order to be able to treat this potential anyway we introduce 
a cut-off 
at large radius $ R $
\be
V_C(r) \To V_C^{(R)} (r) \E\frac{Z \alpha}{r} \, 
\Theta(R - r) \> .
\ee
Indeed, in nature this screening is always realized, e.g. in an atom
where the bare charge of the nucleus is neutralized by the electrons which predominantly 
are far away at atomic distances.
Having modified (regularized) the pure Coulomb potential 
we can now calculate  the eikonal phase (\ref{Eik Phase}) and find
\be
\chi_C^{(R)}(b) = - 2 \gamma \Theta(R - b) \, \int_b^R \! 
\frac{dr}{\sqrt{r^2 - b^2}}
 =  - 2 \gamma  \Theta(R - b) \ln \, \frac{R + \sqrt{R^2 - b^2}}{b} \>  
\stackrel{R \gg b}{\longrightarrow}  \> - 2 \gamma  \ln \frac{2 R}{b} \> .
\label{chi Coulomb}
\ee
\vspace{0.2cm}

\renewcommand{\baselinestretch}{0.9}
\footnotesize
\noindent
An alternative regularization is to consider the pure Coulomb potential
as limit $ \> \mu \to 0 \> $ of a Yukawa potential $ \> V_Y(r) = Z \alpha \,  \, \exp(-\mu r)/r \> $. 
With {\bf \{Handbook\}}, eq. 9.6.23 its
eikonal phase is  $ \> \chi_Y(b) = 2 \gamma \, K_0(\mu b) \> $, where $ K_0 $  denotes the 
modified Bessel function of second type. For small $ \> \mu b \> $ one may use the expansion
{\bf \{Handbook\}}, eq. 9.6.13 to find the same logarithmic impact-parameter dependence as in Eq. 
\eqref{chi Coulomb} but with a cut-off radius $ \> R_Y = \exp(\gamma_E)/\mu \> $
where $ \> \gamma_E = 0.57721566...$ is Euler's constant.

\renewcommand{\baselinestretch}{1.2}
\normalsize
\vspace{0.2cm}

\noindent
Since
\be
e^{ i \chi_C^{(R)}(b)}  \> \simeq \>  \left (\frac{k b}{2 k R} 
\right)^{2 i \gamma} \> , \hspace{0.3cm} b \ll R
\label{Profil Coul}
\ee
is then a power in $b$ , the impact-parameter integral in Eq. (\ref{f eik}) can be 
performed analytically \footnote{See, e.g., {\bf \{Gradshteyn-Ryzhik\}}, eq. 6.561.14 .} 
and one finds
\be
f_C^{(R) \> {\rm eik}} (\theta) \E\frac{1}{2 i k} \, 
\frac{ (2 k R)^{- 2 i \gamma}}{(\sin \theta/2)^{2 + 2 i \gamma}}
\, \frac{\Gamma(1 + i \gamma)}{\Gamma(-i \gamma)} \E e^{ - 2 i \gamma 
\ln 2 kR} \, f_C(\theta) \> .
\label{f Coul}
\ee
Except for the additional phase from the shielding 
the eikonal approximation thus reproduces the \textcolor{blue}{exact result for the
Coulomb scattering amplitude}. An additional phase $ \, - \gamma \ln 2 kR \, $
is also obtained from Eq. (\ref{Streuwelle Coul}) when the potential is truncated
at $ r = R $ 
and the spherical wave is then  allowed to spread out freely into the  asymptotic region 
\footnote {A closer examination of the truncated Coulomb potential
can be found in Ref. \cite {GGSW}.}.


\vspace{0.8cm}
\renewcommand{\baselinestretch}{0.9}
\scriptsize
\refstepcounter{tief}
\noindent
\blau{\bf Detail \arabic{tief}:} {\bf Correct Treatment of the Coulomb Potential}\\

\noindent
\begin{subequations}
In the derivation of Eq. (\ref{f Coul}) there was some "cheating" to obtain 
the correct result: First, we have assumed $ b \ll R $, 
then used the formula
\be
\int_0^{\infty} dx \> x^{\mu} \, J_0 (a x) \E 2^{\mu} \, a^{-\mu-1} \,
\frac{\Gamma \left ( (1 +\mu)/2 \right )}{\Gamma \left ( (1 -\mu)/2 \right )}
\ee
with $ \mu = 1 + 2 i \gamma  $ which only holds for 
$ -1 < {\rm Re} \, \mu < 1/2 $ and finally have heedlessly omitted the 
``1'' in the integrand following the exponential function ...
A correct derivation would keep the cut-off radius $ R $ finite
during integration and let it tend to infinity only
\textcolor{blue}{\bf after} performing the (now convergent) integral. 
That is more complicated but can be done as the indefinite integral
\be
\int_0^1 dx \> x^{\mu} \, J_0 (a x) \E a^{-\mu-1} \, \Biggl [ \, \left ( \mu - 
1 \right ) a \, J_0(a) \, S_{\mu - 1,-1}(a) 
+ a  \, J_1(a) \, S_{\mu,0}(a) \, + 2^{\mu} \, 
\frac{\Gamma \left ( (1 +\mu)/2 \right )}{\Gamma \left ( (1 -\mu)/2 \right )}
\, \Biggr ]
\label{Int endlich}
\ee
can be expressed analytically \footnote{{\bf \{Gradshteyn-Ryzhik\}},
eq. 6.56.13; caution misprint!} by Lommel functions $S_{\mu,\nu}(z)$
and converges for $ {\rm Re} \, \mu > - 1 $ .

\noindent
If we set  $ x = b/R $ the scattering amplitude for the screened
Coulomb potential in eikonal approximation therefore is given by
\be
f_C^{(R) \, {\rm eik}} \E - i k R^2 \, \int_0^1 dx \> x \, 
\left ( \, \frac{x}{1+\sqrt{1-x^2}} \, 
\right )^{2 i \gamma} \, J_0 \left (q R x \right ) \> - \> \left ( \, \gamma = 0 
\, \right ) 
\ee
where the previously missing "1" is generated by subtracting the same expression
with $\gamma = 0$ . The expansion
\be
\left ( \, \frac{x}{1+\sqrt{1-x^2}} \, 
\right )^{2 i \gamma} \E \left ( \frac{x}{2} \right )^{2 i \gamma} \,
\sum_{n=0}^{\infty} a_n \, x^{2 n}
\label{Entwick}
\ee
now allows term-by-term integration over  $x$ using formula
(\ref{Int endlich}) and we obtain
\bea
f_C^{(R)  \, {\rm eik}}  \EA  - i k R^2  \, 2^{-2 i \gamma} \,  
\sum_{n=0}^{\infty} \, \frac{a_n}{(q R)^{1+ 2n+2i \gamma}}  \, \Biggl [ 
\, (2n +2i \gamma) \, J_0(qR) \, S_{2n+2 i \gamma,-1} (q R) \non
&& \hspace{1.5cm} + \, J_1(qR) \, S_{1+2n + 2 i \gamma,0}(qR) 
\, + \, 2^{1+2n+2i\gamma} \frac{1}{qR} 
\frac{\Gamma(n+1+i\gamma)}{\Gamma(-n-i\gamma)} \, \Biggr ]
\> - \> \left ( \, \gamma = 0 \, \right ) \> .
\label{f Coul R}
\eea
Now we may use the asymptotic behaviour of the Lommel and Bessel
functions \footnote{{\bf \{Gradshteyn-Ryzhik\}}, eq. 8.576, and eq. 8.451, respectively.}
\be
S_{\mu,\nu}(z) \stackrel{z \to \infty}{\longrightarrow}  z^{\mu-1} 
\left [ 1 + {\cal O} \left ( \frac{1}{z} \right ) \right ] \> ,
\hspace{1cm} 
J_{\nu}(z) \stackrel{z \to \infty}{\longrightarrow} 
\sqrt{\frac{2}{\pi z}} \, \cos \left ( z - \frac{\pi}{2} \nu - \frac{\pi}{4}
\right ) 
\ee
to let the cut-off radius $R$ (more precisely: $q R$, therefore not for
$\theta = 0$ ) go to infinity. In the last term of
Eq. (\ref{f Coul R}) only the term with $n = 0 $ contributes while
the second term is dominant and the first subdominant. This gives
\be
f_C^{(R) \, {\rm eik}} \to - i k R^2  \, 2^{-2 i \gamma} 
\sum_{n=0}^{\infty} a_n 
\Biggl [ \frac{J_1(qR)}{qR} + \frac{\delta_{n,0}}{2} \left ( \frac{qR}{2} 
\right )^{-2-2i\gamma} \frac{\Gamma(1+i\gamma)}{\Gamma(-i\gamma)} 
+ {\cal O} \left ( \frac{\cos(qR - \pi/4)}{(qR)^{5/2}} \right )  \Biggr ] 
- \left (\gamma = 0  \right ) \, . 
\ee
From the expansion (\ref{Entwick}) we read off 
for $x \to 0 \, : \> a_0 = 1$
and for $x = 1 \, : \> \sum_n a_n = 2^{2 i \gamma} $. 
With this we obtain
\be
f_C^{(R) \, {\rm eik}} \to - i k R^2 \, \left [ \, \frac{J_1(qR)}{qR} + 
\frac{2}{(qR)^{2+2i\gamma}} \, \frac{\Gamma(1+i\gamma)}{\Gamma(-i\gamma)} \, 
\right ] \> - \> \left ( \, \gamma = 0 \, \right ) \> , 
\ee
and we see that subtracting the term with  $ \gamma = 0 $, i.e. the ``1'' 
in the impact-parameter integral precisely cancels the (unwanted) term
 $J_1(qR)/(qR)$. Thus we indeed find the expression
(\ref{f Coul}) for the Coulomb scattering amplitude.

\end{subequations}
\renewcommand{\baselinestretch}{1.2}
\normalsize

\vspace{1cm}

\noindent
By an analytic continuation of the scattering amplitude in the wavenumber 
$k$ we also can obtain the energies of bound states:
It is clear that as a function of $ k $,  Eq. (\ref{f Coul}) has simple poles 
exactly where the Gamma function in the numerator diverges.
This happens at
\be 
1 + i \gamma_n \E  1 + i \frac{Z \alpha m}{k_n} = - (n - 1) \> , \> \> \> 
n = 1, 2, \ldots  \quad k_n \E - i \frac{Z \alpha m}{n}
\ee
and, indeed, gives the well-known binding energies of hydrogenlike atoms 
\footnote{Note that bound states only exist for $ Z < 0 $ as Im $ k_n $ has to be positive 
for a wavefunction $ \, \sim \exp(i k_n r) $ vanishing at infinity.}
\be
E_n \E \frac{k_n^2}{2m} \E -  \frac{(Z \alpha)^2}{2 n^2} \, m \> .
\ee
Of course, it is a coincidence that the (lowest-order) eikonal approximation
gives the exact result for the Coulomb potential -- the general estimates for 
the region of validity of this approximation remain unchanged. Note that the first 
correction (\ref{eik 1 symm}) indeed vanishes for this potential. 
\vspace{0.5cm}

\begin{figure}[hbtp]

\bce
\includegraphics[angle=90,scale=0.55]{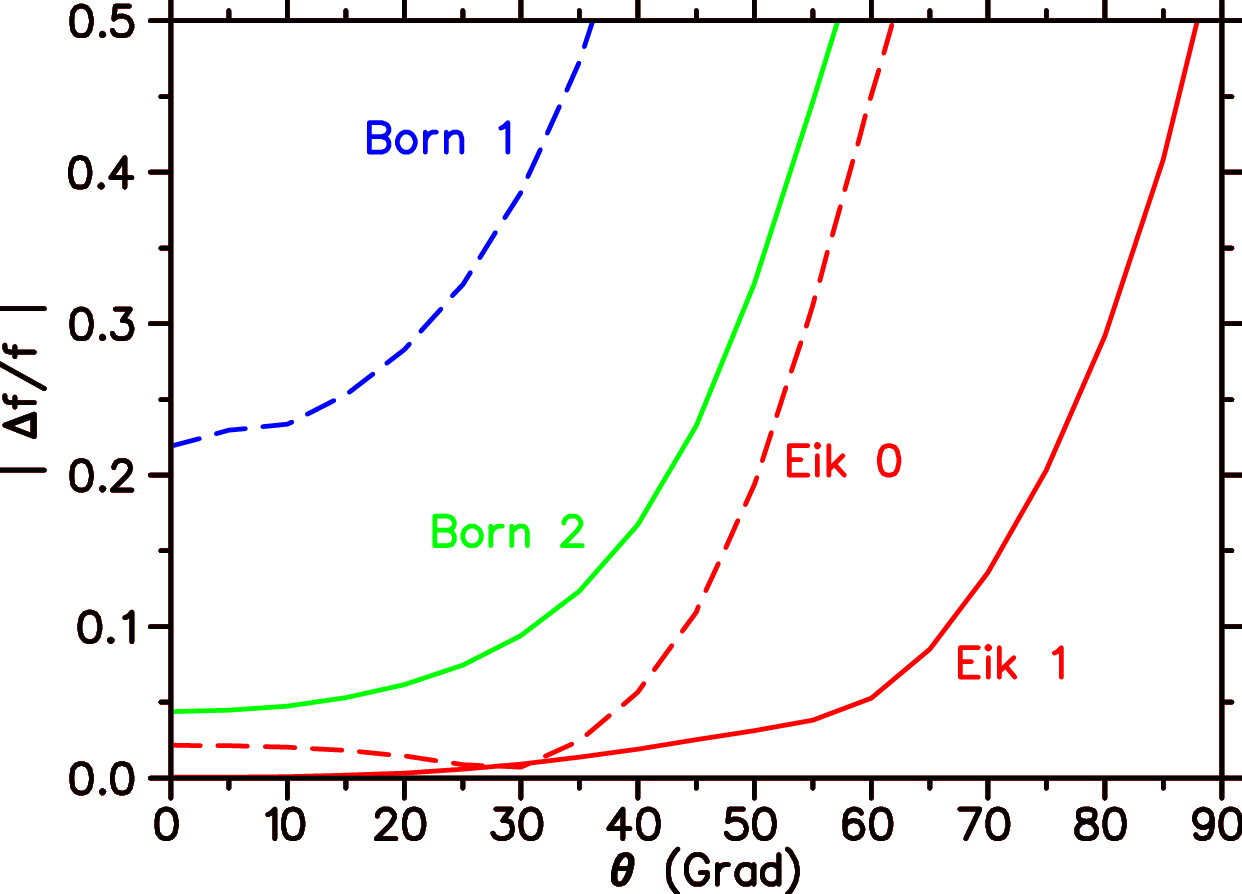}
\ece
\vspace{0.5cm}

{\bf Fig. \arabic{abb}} : Relative deviation \eqref{rel df} of several approximations
for the scattering amplitude from the exact  
\\ \hspace*{1.3cm} partial-wave amplitude for scattering from the Gaussian potential
\eqref{Gauss Pot}. The 0$^{\rm th}$-order eikonal \\ \hspace*{1.3cm}   approximation \eqref{T eik0}
is denoted by ``Eik 0'', the 1$^{\rm st}$-order eikonal approximation \eqref{T eik1} 
by ``Eik 1''. \\ \hspace*{1.3cm} Also depicted 
are the 1$^{\rm st}$ Born  (``Born 1'') and the 2$^{\rm nd}$ Born approximation (``Born 2'').
\label{abb: querschnitt}
\vspace*{0.4cm}
\end{figure}

\noindent              
As a quantitative test of the various eikonal approximations
Fig. \ref{abb: querschnitt} shows the relative deviation of the corresponding
scattering amplitude
\be
\left | \, \frac{\Delta f}{f} \, \right | \Def \left | \, \frac{f_{\rm approx} - f_{\rm exact}}
{f_{\rm exact}} \, \right |
\label{rel df}
\ee
from the exact result. In this case the potential is an attractive Gaussian potential
\be
V(r) \E V_0 \, e^{-r^2/R^2} \> , \qquad \mbox{with} \quad 2 m \, V_0 \, R^2 \E - 4 \> .
\label{Gauss Pot}
\ee
at a ("high") energy so that $ \, k R = 4 \, $. The relative difference \eqref{rel df}
is sensitive for the correct phase of the scattering amplitude whereas the deviation
of the cross sections may be misleading.
The comparison with the numerically calculated exact partial-wave amplitude 
\footnote{The phase-shift method for central potentials is well explained in
{\bf \{Messiah 1\}}, ch. 10.}
shows that the eikonal approximation is excellent in forward direction but that the 
deviations increase systematically with increasing scattering angle.
\noindent
Although  it is known that large-angle scattering for a Gaussian potential 
is particularly difficult to describe, it can be seen that the eikonal approximation
is far superior when compared with the 1$^{\rm st}$ and 2$^{\rm nd}$ Born approximation.
\vspace{1cm}

\refstepcounter{abb}

\vspace{0.8cm}

\noindent
The good description of potential scattering
by the eikonal approximation is used in the so-called {\bf Glauber theory} 
to describe high-energy scattering of strongly interacting particles from
composite targets (say, nuclei) in terms of the elementary scattering of the projectile
from the constituents (in the example: proton and neutron). This turns out to 
be quite successful although there is actually no good reason
to expect that these processes may be adequately described in terms of potential sacttering.
\vspace{1.5cm}

\renewcommand{\baselinestretch}{0.9}
\scriptsize
\refstepcounter{tief}
\noindent
\blau{\bf Detail \arabic{tief}:} {\bf Basics of Glauber Theory}\\
\vspace{0.2cm}

\noindent
\begin{subequations}
The elimination of the potential is possible because the relation
\eqref{T eik0} is a 
2-dimensional Fourier transform which can be inverted
\be
\Gamma(\fb) \Def e^{i \chi_0(\fb)} - 1 \E \frac{m}{ik} \, \int \frac{d^2 q}{(2 \pi)^2} \> 
e^{i \fq \cdot \fb} \, T(\fq) \E \frac{i}{2 \pi k} \, \int d^2 q \> e^{i \fq \cdot \fb} \,
 f(\fq) \> .
\ee
Therefore the "profile function" $ \, \Gamma(\fb) \> $ may be obtained directly
from (a parametrization of) the scattering amplitude.
In the case of nucleon-nucleon scattering at a fixed, high energy this is usually done 
by writing
\be
f(q) \E \frac{i k}{4 \pi} \, \sigma_{\rm tot} \, \lrp 1 - i \rho \rrp \, e^{- \beta^2 q^2/2} \> ,
\ee
which has the following parameters: The measured total cross section
$ \sigma_{\rm tot} $ (obviously the optical theorem \eqref{opt Theor} is built in), the
ratio $ \> \rho \> $ of real to imaginary part of the scattering amplitude in forward direction
(determinable from interference with the Coulomb scattering amplitude)
and the parameter $\beta $ which describes the fall-off of the differential cross section
with momentum transfer. All these parameters are empirical and measurable quantities
-- and the potential description of the interaction has disappeared! 
This assumption only comes back into play if we assume that for scattering from a 
composite object (say, an atomic nucleus) the interaction of the projectile with the
$A$ constituents is given by
\be
V(\fr) \E \sum_{i=1}^A \, V \left ( \fr - \fx_i \right ) \> ,
\ee
as is customary in non-relativistic many-body physics
(see Eq. \eqref{H 1. Quant}). As the eikonal phase $ \> \chi_0(\fb) \> $ in Eq. \eqref{chi0} 
is {\bf linear} in the potential, the profile function for the composite target reads
\be
\Gamma\left ( \fx_1 \ldots \fx_A; \fb \right ) \E  \exp \lsp i \sum_{i=1}^A \chi_0(\fb-\fx_i )\rsp - 1 
\E \prod_{i=1}^A \, \Bigl [ 1 + \Gamma \left ( \fb - \fx_i \right ) \Bigr ] - 1 
\ee
and the total scattering amplitude at fixed scattering centers $ \fx_i $ becomes
\bea
f_{i \to f}\left ( \fx_1 \ldots \fx_A \right )  \EA \frac{k}{2 \pi i} \, \int d^2b \> 
e^{-i \fq \cdot \fb} \>
\lcp \prod_{i=1}^A \, \Bigl [ 1 + \Gamma \left ( \fb - \fx_i \right ) \Bigr ] - 1 \rcp \\
\EA \frac{k}{2 \pi i} \, \int d^2b \> e^{-i \fq \cdot \fb} \> \lcp \sum_{i=1}^A  
\Gamma \left ( \fb - \fx_i \right) + \sum_{i > j }^A \Gamma \left ( \fb - \fx_i \right) \,   
\Gamma \left ( \fb - \fx_j  \right) + \ldots \rcp \> .
\label{f 1-A}
\eea
In the second line the product has been expanded in powers of the individual profile functions
-- this gives a (finite) series of {\bf multiple scatterings} from the individual constituents.
If, in addition, one assumes that at high energies the target does not change during the ultra-short
scattering process ("frozen nucleus approximation"), then the total scattering amplitude 
simply is the matrix element of Eq. \eqref{f 1-A} taken between the wave function
$ \Psi_I $ of the initial target state and the one $ \Psi_F $ of the final target state
\be
f_{i,I \to f,F} \E \la  \Psi_F \, \left | \, f_{i \to f} \left ( \fx_1 \ldots \fx_A \right) \, \right | \, 
 \Psi_I \ra \> .
\ee
For mehr details see, e.g.,  {\bf \{Eisenberg-Koltun\}} or {\bf \{Scheck\}}.

\end{subequations}
\renewcommand{\baselinestretch}{1.2}
\normalsize

\vspace{0.8cm}

\subsection{\textcolor{blue}{Green Functions as Path Integrals}}
\label{sec1: greensche Funk}

\vspace{0.3cm}
Up to now we have discussed the path-integral representation of the
time-evolution operators
$ \hat U $ in different formulations. Now we want to consider
{\bf Green functions}
\be
\boxed{
\qquad G_{AB}(t_1,t_2) \Def \left <0 \left | \> {\cal T} \left [ \> 
\hat A_H(t_1) \, \hat B_H(t_2) \> \right ] \> \right | 0 \right > \qquad
}
\label{def G_AB}
\ee
where  $ | \, 0 > \, $ is the  exact ground state of the system,
\be
\hat A_H(t) \E e^{i \hat H t/ \hbar} \> \hat A_H(0) \>  
e^{- i \hat H t/ \hbar}
\ee
a quantum mechanical operator in the Heisenberg picture 
(as is $ \hat B_H(t) $) and
\be
{\cal T} \left [ \> \hat A_H(t_1) \, \hat B_H(t_2) \> \right ] = 
\> \Theta(t_1-t_2) \, \hat A_H(t_1) \, \hat B_H(t_2) \> + \> \Theta(t_2-t_1) \, 
\hat B_H(t_2) \, \hat A_H(t_1)
\ee
the time-ordered product of these two operators.
Such objects play a central role in many-body physics and in field theory, as
it is straight-forward to evaluate them in perturbation theory and all important 
quantities (ground-state energy, expectation values of operators,
$ S $ matrix elements) can be calculated from them.
\vspace{0.2cm}

If we insert into the definition (\ref{def G_AB}) a complete set of position eigenstates
\be 
1 \E \int dq \> |\, q> <q \,| \E \int dq \> e^{ i \hat H t/ \hbar}
\> |\, q> <q \,| \> e^{ -i \hat H t/ \hbar} \> ,
\ee
then we obtain
\bea
G_{AB}(t_1,t_2) \EA \int dq \, dq' \> \left < 0 \left | \> 
e^{ i \hat H t'/ \hbar} 
\> \right | q' \right > \, \left < q' \left | \> e^{ -i \hat H t'/ \hbar}
{\cal T} \left [ \> \hat A_H(t_1) \, \hat B_H(t_2) \> \right ] \> 
e^{ i \hat H t/ \hbar} \right | q \right > \non
&& \hspace{4.2cm} \cdot \left < q \left | \> e^{ -i \hat H t/ \hbar} \>
\right | \, 0 \right >  \> .
\label{G_AB 1}
\eea
The individual factors in this expression may be rewritten in the following way: 
The first one
\be
\left < 0 \left | \> e^{ i \hat H t'/ \hbar} \>
\right | \, q' \right >  \E e^{ i E_0 t'/ \hbar} <0 \, |\, q '> \> \E  
\> e^{ i E_0 t'/ \hbar} \> \psi_0^{\star}(q')
\ee
can be expressed by the wave function of the ground state 
and the ground-state energy -- similarly, also the last one.

Therefore we now concentrate on a path-integral representation for the
middle factor and first assume that $ \> \hat A = \hat B = \hat x \> $ 
and that $ \> t_1 > t_2 \> $. After inserting two complete sets of position
eigenstates we then obtain
\bea
&& \left < q' \left | \> e^{ -i \hat H t'/ \hbar}
\> \hat x_H(t_1) \, \hat x_H(t_2)  \>
e^{ i \hat H t/ \hbar} \right | q \right > \> =\>  \left < q' \left | \> 
e^{ -i \hat H (t'-t_1)/ \hbar} \> \hat x \> e^{ -i \hat H (t_1-t_2)/ \hbar} 
\> \hat x \> e^{ -i \hat H (t_2-t)} \> \right | q \right > \non
&& =   \int dq_1 dq_2 \> q_1 \, q_2 \> \left < q' \left | \,
e^{ -i \hat H (t'-t_1)/ \hbar} \, \right | q_1 \right > \> 
\left < q_1 \left |
\, e^{ -i \hat H (t_1-t_2)/ \hbar} \, \right | q_2 \right > \> 
\left < q_2 \left |
\, e^{ -i \hat H (t_2-t)/ \hbar} \, \right | q \right > \> .
\eea
Now we may use the path-integral representation of the individual
time-evolution operators and obtain in the Hamilton formulation
\bea
&& \left < q' \left | \> e^{ -i \hat H t'/ \hbar}
\> \hat x_H(t_1) \, \hat x_H(t_2)  \>
e^{ i \hat H t/ \hbar} \right | q \right > \non   
&& \hspace{1cm} = \> \int dq_1 dq_2 \> q_1 q_2 \>
\int_{x(t_1)=q_1}^{x(t')=q'} \frac{{\cal D}'x {\cal D}p}{2 \pi \hbar}
\> \exp \left \{ \> \frac{i}{\hbar} \int_{t_1}^{t'} d\tau \> 
[ \, p \dot x - H(x,p) \, ] \> \right \}
\non
&& \hspace{3.7cm} \cdot \int_{x(t_2)=q_2}^{x(t_1)=q_1} 
\frac{{\cal D}'x {\cal D}p}{2 \pi \hbar}
\> \exp \left \{ \> \frac{i}{\hbar} \int_{t_2}^{t_1} d\tau \> 
[ \, p \dot x - H(x,p) \, ] \> \right \} \non
&& \hspace{3.7cm} \cdot
\int_{x(t)=q}^{x(t_2)=q_2} \frac{{\cal D}'x {\cal D}p}{2 \pi \hbar}
\> \exp \left \{ \> \frac{i}{\hbar} \int_{t}^{t_2} d\tau \>  
[ \, p \dot x - H(x,p) \, ] \> \right \} \> .
\eea

Similar as in the proof of the composition law (\ref{Kompositionsgesetz}) 
we can concentrate the exponents into one exponent and all integrations
can be combined into one functional integral if one integrates additionally
over the endpoints  $ q_1, q_2 $. For $ \> t_1 > t_2 \> $ the result therefore is
\bea
 && \left < q' \left | \> e^{ -i \hat H t'/ \hbar}
\> {\cal T} \left [ \> \hat x_H(t_1) \, \hat x_H(t_2)  \> \right ] \> 
e^{ i \hat H t/ \hbar} \right | q \right > \non
&& \hspace{2cm} = \> \int_{x(t)=q}^{x(t')=q'} 
\frac{{\cal D}'x {\cal D}p}{2 \pi \hbar} \> x(t_1) x(t_2) \> 
\> \exp \left \{ \> \frac{i}{\hbar} \int_{t}^{t'} d\tau \>
[ \, p \dot x - H(x,p) \, ] \> \right \} \> .
\label{Zeitordnung x x}
\eea

For $ \> t_2> t_1 \> $, a similar calculation gives the same
expression, i.e. the time-ordering operator is {\bf automatically} built into the
path-integral representation! This is not quite so surprising as it first seems,
since the time ordering is necessary because the Heisenberg operator do not commute
at different times -- the path integral, however, deals with ordinary numbers.
Eq. (\ref{Zeitordnung x x}) is easily generalized to any operator
$ \> \hat A \E A \left (\hat x, \hat p \right) \>,
\hspace {0.2cm} \hat B \E B \left (\hat x, \hat p \right) \> $
and then reads
\bea
 && \left < q' \left | \> e^{ -i \hat H t'/ \hbar}
\> {\cal T} \left [ \> \hat A_H(t_1) \, \hat B_H(t_2)  \> \right ] \> 
e^{ i \hat H t/ \hbar} \right | q \right >  \> \non
&&  = \> \int_{x(t)=q}^{x(t')=q'} 
\frac{{\cal D}'x {\cal D}p}{2 \pi \hbar} \> A \left ( p(t_1),x(t_1) \right )
\, B \left ( p(t_2),x(t_2) \right )
\> \exp \left \{ \> \frac{i}{\hbar} S \left [ x(t),p(t) \right ] \> \right \}
\> .
\label{Zeitordnung A B }
\eea

Let us return to the formula (\ref{G_AB 1}) for the Green function
in which we now can insert the result (\ref{Zeitordnung A B }).
However, the appearance of the ground-state wave function
$ \> \psi_0(q) \> $ is unsatisfactory since the exact ground state is unknown
in general.
We would like to have an expression available in which the ground state is 
generated at the same time.
This is possible by ``filtering'' out the lowest lying state of the system.
Indeed, in an expression of the form
\be
\left < q' \left |\>  e^{-i \hat H t'/\hbar} \hat {\cal O} 
e^{i \hat H t/\hbar}
\> \right | q \right > \E \sum_{n,m} \> \psi_m(q') \,
e^{-i E_m t'/\hbar} \left < m \left | \> \hat {\cal O} \> \right | n 
\right > \> 
e^{i E_n t/\hbar} \, \psi_n^{\star}(q)
\ee
only the ground state contribution survives if one formally lets approach
\be
t \to i \infty, \hspace{1cm} t' \to -i\infty 
\label{Randbedingungen}
\ee
as is easily seen:
\bea
\left < q' \left |\>  e^{-i \hat H t'/\hbar} \hat {\cal O} 
e^{i \hat H t/\hbar} 
\> \right | q \right > &
\stackrel{\latop{t \to i \infty}{t' \to -i\infty}}
{\longrightarrow} &
 \psi_0(q') \psi_0^{\star}(q) \> e^{ - E_0 ( |t| + |t'| )/\hbar} 
\left < 0 \left | \> \hat {\cal O} \> \right | 0 \right > \non
\EA \lim_{\latop{t \to i \infty}{t' \to -i\infty}} 
\left < q' \left | \> 
e^{- i \hat H (t'-t)/\hbar } \> \right | q \right > \> 
\left < 0 \left | \> \hat {\cal O} \> \right | 0 \right >  \> .
\label{Filter}
\eea
This is due to the fact that
for $ \> n > 0 \>,   \> E_n > E_0 \> $ ,
so that the contributions of the excited states decay more rapidly than
those from the ground state.
If we use Eq. (\ref{Filter}) for the Green function of the operators
$ \> \hat A, \hat B \> $ we obtain
\be
G_{AB}(t_1,t_2) \E  
\lim_{\latop{t \to i \infty}{t' \to -i\infty}} \frac{
\left < q' \left | \> e^{- i \hat H t'/\hbar } {\cal T} \left [ \> 
\hat A_H(t_1) \, \hat B_H(t_2) \> \right ]  \, e^{ i \hat H t/\hbar} \> 
\right | q \right > }{\left < q' \left | \>
e^{- i \hat H (t'-t)/\hbar } \> \right | q \right > }
\ee
or as path integral
\bce
\vspace{0.2cm}

\fcolorbox{blue}{white}{\parbox{12cm}
{
\bea
G_{AB}(t_1,t_2) \EA
\lim_{\latop{t \to i \infty}{t' \to -i\infty}} \frac{
\int {\cal D}'x {\cal D}p \> A(x(t_1),p(t_1)) \, B(x(t_2),p(t_2))
\> e^{i S[x,p]/\hbar} }
{\int {\cal D}'x {\cal D}p \> e^{i S[x,p]/\hbar} } \> \> \> . \hspace*{3cm}
\label{Green als Pfad}
\eea
}}
\ece

\vspace{0.5cm}


\noindent
Note that in the expression
(\ref{Green als Pfad}) all normalization factors cancel!
The unphysical limits (\ref{Randbedingungen})
are quite natural for {\bf Euclidean} Green functions
where the time $ t $ is replaced by $ - i \tau$ everywhere. 
Instead of oscillations one has damping everywhere and 
the path integral ist well-defined mathematically.
Therefore one frequently investigates Euclidean Green functions 
and transforms back -- if possible -- to real time by an
analytic continuation. 
It is clear that in order to project out the ground state by
the boundary conditions \eqref{Randbedingungen} one doesn't necessarily have to go to
infinity along the imaginary axis  but that it is sufficient
to do that along a suitable ray in the complex time plane 
(see, e.g.  ch. 7 in Ref. \cite{MacK}).
\vspace{0.2cm} 

The operators $ \hat A, \hat B $ can be expressed in terms of the fundamental
operators $ \, \hat x, \> \hat p \, $, say by a Taylor expansion. However, it is not necessary
that in the matrix element the products of operators
$ \, \hat x, \> \hat p \, $ all are taken at the same time: One can define
\textcolor{blue}{\bf n-point functions} 
\be
G \left (t_1 \ldots  t_j,t_{j+1} \ldots t_n \right ) \Def 
\left < 0 \left | \> {\cal T} \left [ \> 
\hat x_H(t_1) \ldots \hat x_H(t_j) \> \hat p_H(t_{j+1}) \ldots \hat p_H(t_n) 
\> \right ] \> \right | 0 \right >
\label{def n-Punkt} 
\ee
and write as path integral (in the following we do not indicate 
the limits anymore explicitly):
\be
G \left (t_1 \ldots t_j, t_{j+1} \ldots  t_n \right ) \E  \frac{
\int {\cal D}'x {\cal D}p \> x(t_1) \ldots x(t_j) \> p(t_{j+1}) \ldots p(t_n)
\> e^{i S[x,p]/\hbar} }
{\int {\cal D}'x {\cal D}p \> e^{i S[x,p]/\hbar} } \> \> \> .
\label{n-Punkt als Pfad}
\ee
The full set of $n$-point functions can be obtained from a 
\textcolor{blue}{\bf generating functional} 
\be
\boxed{
\qquad Z[J,K] \E \int {\cal D}'x {\cal D}p \> \exp \left \{ \> \frac{i}{\hbar} 
S[x,p] + \frac{i}{\hbar} \int dt \, \left [ \, J(t) \, x(t) + K(t) \, p(t) \, 
\right ] \> \right \} \qquad
}
\label{erzeug Funk}
\ee
by functional differentiation with respect to  artificially introduced sources
which are set to zero thereafter
 \footnote{ \blau{\textsf{''Der Mohr hat seine Arbeit [Schuldigkeit] getan, der Mohr kann 
gehn'' -- "The moor has done his obligation, the moor can go"}}, in {\bf \{Schiller\}},  3. Aufzug, 4. Auftritt .}
\be
\boxed{
\qquad G \left (t_1 \ldots  t_j, t_{j+1} \ldots t_n \right ) \E (-i \hbar)^n 
\> \frac{\delta^j}{\delta J(t_1)
\ldots \delta J(t_j)} \> \>  \frac{\delta^{n-j}}{\delta K(t_{j+1})
\ldots \delta K(t_n)} \> \> \frac{Z[J,K]}{Z[0,0]} \> \Biggr |_{J=K=0} \> . \quad
\label{n-Punkt aus erzeug funk}
}
\ee
\vspace{0.4cm}

\renewcommand{\baselinestretch}{0.9}
\scriptsize
\refstepcounter{tief}
\noindent
\blau{\bf Detail \arabic{tief}:} {\bf Generating Functions}
\vspace{0.4cm}

\begin{subequations}
\noindent
The generating functional is a generalization of the notion of a generating function  
which exists for a multitude of orthogonal polynomials
( {\bf \{Handbook\}}, Table 22.9).
For example,
\be 
g(x,t)\E \exp \lrp 2 x t - t^2 \rrp \E \sum_{n=0}^{\infty} H_n(x)\, \frac{t^n}{n!} 
\ee 
is the generating function for the Hermite polynomials $H_n(x)$. Obviously we have
\be 
H_n(x) \E \frac{\partial^n}{\partial t^n} \,   g(x,t) \Bigr |_{t=0} 
\ee
and it easily is found that $ H_0 = 1 , \> H_1 = 2 x , \>  H_2 = 4x^2 - 2 , \> H_3 = 8 x^3 - 12 x , \> \ldots $.
Also quite known is the generating function for the Legendre polynomials                
\be
\frac{1}{\sqrt{1 - 2 x t + t^2}} \E \sum_{n=0}^{\infty} \, P_n(x) \, t^n  \> , \quad |t|, |x| < 1 \> ,
\ee
which is important for the multipole expansion of the
Coulomb potential or the phase-shift method in scattering.
 
\end{subequations}
\renewcommand{\baselinestretch}{1.2}
\normalsize
\vspace{0.5cm}

In general the generating functional cannot be evaluated exactly.
However, if one can split up the Hamiltonian
\be
\hat H \E \hat H_0 \> + V(\hat x,\hat p)
\ee
in such a way that the generating  functional $ \> Z_0[J,K] \> $ is known
then one can give a closed expression for the full generating
functional: First one also splits up the exponent in 
the path integral as 
\be
\int dt \> \left [ \> p \, \dot x - H_0(x,p) + J(t) \, x(t) + K(t) \, p(t) \> 
\right ] - \int dt \> V(x(t),p(t))
\ee
and then one uses the fact that $ \> x(t) \> $ can be generated
by functional differentiation w.r.t. the source  $ \> J(t) \> $, $ \> p(t) \> $ 
by differentiation w.r.t. $ \> K(t) \> $ . With this procedure the exponent of the
``perturbation'' $ V $  can be taken out from the path integral and we obtain the representation
\bea
Z[J,K]  \! \! \EA \! \! \int {\cal D}'x {\cal D}p \, 
\exp \left [  - \frac{i}{\hbar}
\int dt \> V(x(t),p(t))  \right ] \> \exp \left \{  \frac{i}{\hbar} 
S_0 + \frac{i}{\hbar} \int dt \, \left [ J(t) x(t) + K(t) p(t) \right ]
\right \} \non
\EA \exp \left [ \, - \frac{i}{\hbar}
\int dt \> V\left ( \frac{\hbar}{i} \frac{\delta}{\delta J(t)} ,
\frac{\hbar}{i} \frac{\delta}{\delta K(t)}\right )\, \right ] \> \> Z_0[J,K]
\> .
\label{volles erzeug funk}
\eea
Of course, this is only a formal solution and in most cases one has to expand the
exponential function in powers of the perturbation in order to perform the
functional differentiations. This produces the perturbation series for the generating functional and 
therefore for alle $n$-point functions.

\vspace{1cm}
\noindent
{\bf Examples: }\\

\noindent
The Hamiltonian for the anharmonic oscillator in one dimension
($ \> m = \hbar = 1 \> $) 
\be 
\hat H \E \underbrace{\frac{1}{2} \hat p^2 + 
\frac{\omega^2}{2} \hat x^2}_{= H_0}
 + \lambda \hat x^4 
\label{anharmon}
\ee
is a prototype for many similar problems. Indeed 
Eq. (\ref{anharmon}) can be seen as a
$\phi^4$ field theory in (0+1) (space + time)-dimensions. Also the Hamiltonian
of a non-relativistic many-body system (in ``2$^{\rm nd}$ quantization'') has a similar 
structure.

\noindent 
The free generating functional can be determined easily by completing the square:
First, one writes
\be
p \, \dot x \> - \> \frac{1}{2} p^2 \> + \> K \,p \E - \frac{1}{2}
\left [ \> p - ( \dot x + K ) \> \right ]^2 \> + \frac{1}{2} ( \dot x + K )^2
\ee
and performs the functional momentum integration which simply gives a constant.
The remaining integral is a Gaussian one again:
\be 
Z_0[J,K] \E {\rm const.} \> \int {\cal D}x \> e^{i S_0[x,J,K]}
\ee
with
\bea
S_0[x,J,K] \EA \int dt \> \left [ \> \frac{1}{2} ( \dot x + K )^2 - 
\frac{\omega^2}{2} x^2 + J \, x \> \right ] \non
\EA \int dt \> \left [ \> \frac{1}{2} x \Bigl ( \underbrace{-\frac{\partial^2}
{\partial t^2} - \omega^2}_{:={\cal O}} \Bigr) \, x \, + \, (J - \dot K) \, x
+ \frac{1}{2} K^2 \> \right ] \non
&\EQ & \frac{1}{2} \left ( x, \, {\cal O} \, x \right ) \, + \, 
\left (J - \dot K, x \right ) +  \frac{1}{2} \left ( K, K \right ) \> .
\eea
Here an integration by parts has been performed in the second line
and we have assumed that no boundary terms appear at
$ \> t_a = i \infty, \, t_b = - i\infty \> $;
The third line is a shorthand, e.g. for
\be
\left ( x, \, {\cal O} \, x \right ) \Def \int dt \, dt' \> x(t) \, {\cal O}(t,t') \,
x(t') \hspace{0.5cm} {\rm with} \hspace{0.3cm}
{\cal O}(t,t') \E \left ( -\frac{\partial^2}
{\partial t^2} - \omega^2 \right ) \, \delta(t-t') \> .
\ee
Completing the square another time allows to perform the Gaussian
$x$-integral and gives the result
\be
Z_0[J,K] \E {\rm const.} \> \exp \left \{ \> \frac{i}{2} (K,K)
- \frac{i}{2} \left ( J - \dot K, {\cal O}^{-1}, J - \dot K \right ) \> 
\right \} \> .
\ee
We now have to evaluate the inverse operator $ \> {\cal O}^{-1}(t,t') \> $. 
This can be done by Fourier transformation
\be
{\cal O}^{-1}(t,t') \>  = \> \int_{-\infty}^{+\infty} \frac{dE}{2 \pi} \> 
\tilde {\cal O}^{-1}(E) \> e^{- iE(t-t')}
\ee
and one obtains an algebraic equation with the solution
$ \> {\cal O}^{-1}(E) \E 1/(E^2 - \omega^2) \> $ .\\
How to treat the pole at $ \> E = \omega \> $ ? The easiest way is
to require a damping of the functional integral in the factor
\be
e^{i S_0} \E \exp \left [ \> - \frac{i}{2} \left ( x, 
\frac{\partial^2}{\partial t^2} + \omega^2, x \right ) + \ldots \> \right ]
\ee
which leads to Feynman's prescription
\be
\boxed{
\omega^2 \To\omega^2  - i \, 0^+ 
}
\label{feyn regel}
\ee
($ i \, 0^+ $ is a shorthand for a small, positive imaginary part
which is set to zero at the end of the calculation).
%
%
\vspace{1cm}

\renewcommand{\baselinestretch}{0.9}
\scriptsize
\refstepcounter{tief}
\noindent
\blau{\bf Detail \arabic{tief}:} {\bf Derivation of Feynman's Rule}\\
\label{tief}

\noindent
\begin{subequations}
In some sense, this comes out of the blue and we would better look more carefully into the previous result
for the forced harmonic oscillator. First, we notice that the prefactor
\eqref{HO Vorfaktor} does not matter, because the Green functions are ratios of path integrals.
Then we consider the result \eqref{erzwung HO}
for the classical action  with a linear perturbation $ \> e (t) = - J (t) \> $ and the boundary conditions
$ \> x_a = x_b = 0 \> $. These conditions we have already used to perform an integration 
by parts without 
troublesome boundary terms (for the current discussion this is just pure convenience --
as we shall see below, one could use arbitrary boundary conditions). We now have
\be
S_{\rm cl} \E \frac{m}{2 \sin \omega (t_b - t_a)} \, \lrp - \frac{2}{m^2 \omega^2} \rrp 
\int_{t_a}^{t_b} dt \> J(t) \sin \omega (t_b-t)\, \int_{t_a}^t dt' \> J(t') \sin \omega (t'-t_a) 
\> .
\label{S rand}
\ee 
Assuming that the perturbation $ \> J(t) \> $ only acts during the time interval
$ \> |t| < \tau \quad (\tau \to + \infty)\> $ we now make an analytic continuation of initial 
and final times into the complex time plane
\be
t_a \E -\tau + i \kappa \, \tau \> , \qquad t_b \E \tau - i \kappa \, \tau
\qquad \Longrightarrow \> e^{ - i \hat H (t_b - t_a)} \E 
e^{-2 \hat H \kappa \tau} \> e^{-2 i \hat H \tau}
\ee
with $ \> \kappa > 0 \> $. 
The integration path in the complex $t$-plane thus looks as shown in Fig.
\ref{Intweg im Komplexen}. 

\refstepcounter{abb}
\begin{figure}[hbtp]
\bce
\vspace*{-2.8cm}
\includegraphics[angle=0,scale=0.45]{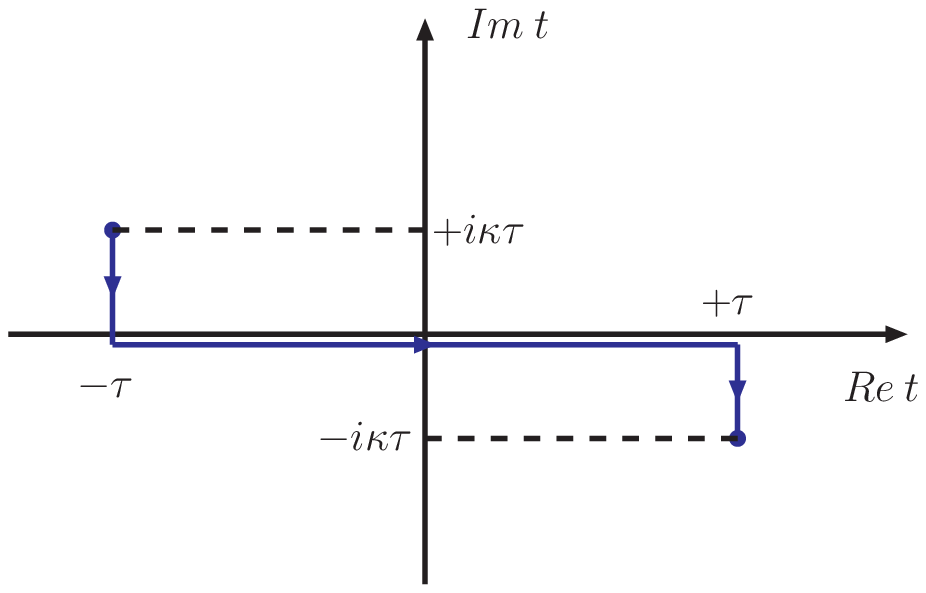}
\label{Intweg im Komplexen}
\ece
\vspace{-6.5cm}
\bce
{\scriptsize\bf Fig. \arabic{abb}}: {\scriptsize Deformation of the integration path in the complex
$t$-plane for projecting out the ground state.}
\ece
\end{figure}
\vspace{0.3cm}
\noindent
However, the trigonometric functions will now increase expenentially
\be
\sin \omega(t_b -t_a) \stackrel{\tau \to \infty}{\longrightarrow}  \frac{1}{2 i} \, 
e^{2 \omega \kappa \tau } \, e^{2 i \omega \tau } \> , \quad
\sin \omega (t_b - t) \stackrel{\tau \to \infty}{\longrightarrow}   \frac{1}{2 i} \, 
e^{\omega \kappa \tau} \, e^{i \omega (\tau - t)} \> , \quad
\sin \omega (t' -t_a) \stackrel{\tau \to \infty}{\longrightarrow}  \frac{1}{2 i} \, 
e^{\omega \kappa \tau} \, e^{i \omega (t'+\tau)}\> ,
\ee
and the classical action takes the form
\be
S_{\rm cl} \stackrel{\tau \to \infty}{\longrightarrow}  \frac{i}{2 m \omega} \, 
\int_{-\infty}^{+\infty} dt \> J(t) \, e^{-i \omega t} \, \int_{-\infty}^t dt' \> J(t')
\, e^{i \omega t'} \E  \frac{i}{4 m \omega} \, 
\int_{-\infty}^{+\infty} dt dt' \> J(t) \, J(t') \, 
e^{-i \omega |t-t'|} \> .
\ee
The generating functional for the Green function of the harmonic oscillator therefore is 
(for $ m = 1 $ )
\be
Z_0[J] \E {\rm const.} \, e^{i S_{\rm cl}} \E  {\rm const.} \,\exp \lsp - \frac{1}{4 \omega} \, 
\int_{-\infty}^{+\infty} dt dt' \> J(t) \, J(t') \, e^{-i \omega |t-t'|} \rsp \> .
\ee
But this is exactly the same result as if we perform the Gaussian integral for the generating functional
(disregarding all boundary conditions) and apply Feynman's rule for treating the pole while doing the complex integration:
\be
Z_0[J] \E  {\rm const.} \,\exp \lsp - \frac{i}{2} \lrp J, {\cal O}^{-1} J \rrp \rsp 
\E  {\rm const.} \,\exp \lcp - \frac{i}{2} \int_{-\infty}^{+\infty} dt dt' \> J(t) \, J(t') \, 
\int_{-\infty}^{+\infty} \frac{dE}{2 \pi} \, \exp \lsp - i \omega (t - t') \rsp \, 
\frac{1}{E^2 - \omega^2 + i 0^+} \rcp \> .
\ee
The derivation given above also shows that the boundary terms $ x_a, x_b $ are irrelevant: 
The terms linear in $ J(t) $ 
appearing in Eq. \eqref{erzwung HO} do not contribute anything in the limit $ \> \tau \to \infty \> $ 
since they contain only {\it one} complex
time $ t_{a/b} $ and the first, $ J(t)$-independent term indeed makes a (constant) contribution
(because it is multiplied by a factor $ \> \cos \omega (t_b - t_a) \> $) but not for 
the Green functions  after dividing by $ \> Z_0[0] \> $.

\end{subequations}
\renewcommand{\baselinestretch}{1.2}
\normalsize
\vspace{0.6cm}

\noindent
{\bf a) Ground-state Energy for the Anharmonic Oscillator}
\vspace{0.2cm}

\noindent
We now concentrate our interest on the special 2-point function
\bea
G_{xx}(E) &:=& \int_{-\infty}^{+\infty} dt \> e^{i E t} \> \left < 0 \left | \> 
{\cal T} \left [ \> \hat x_H(t) \, \hat x_H(0) \> \right ] \> \right | 0 
\right > \non
\EA \int_{-\infty}^{+\infty} dt \> e^{i E t} \> (-) 
\frac{\delta^2}{\delta J(t)
\delta J(0)} \> \frac{Z[J,0]}{Z[0,0]} \> \Biggr |_{J=0} \> ,
\eea
which is also called ``single-particle Green function''.
This quantity is particularly important because the ground-state energy of the system
can be derived from it in the most simple way (see, e.g. {\bf \{Fetter-Walecka\}}, p. 66 - 68 .):
\be
E_0 \E \frac{1}{4} \lim_{\epsilon \to 0}
\int_{-\infty}^{+\infty} \frac{dE}{2 \pi} \> e^{- i \epsilon E} \> 
\left ( \omega^2 + 3 E^2 \right ) \> G_{xx}(E) \> ,
\label{E0 aus Green}
\ee
\vspace{0.4cm}

\renewcommand{\baselinestretch}{0.9}
\scriptsize
\refstepcounter{tief}
\noindent
\blau{\bf Detail \arabic{tief}:} {\bf Ground-State Energy from the Single-Particle Green Function}\\

\noindent
\begin{subequations}
This relation is based on the equations of motion $ p_H(t) = \dot x_H(t)$ and
$ \dot p_H(t) = \ddot x_H(t) = - \omega^2 x_H(t) - 4 \lambda x_H^3(t)$. From the first one, 
one obtains
\be
\frac{1}{2} < 0 | p^2 + \omega^2 x^2 | 0 > \E \frac{1}{2} \lim_{t_1,t_2 \to 0}
 \left ( \, \frac{\partial^2}{\partial t_1 \partial t_2} + \omega^2 \, \right ) \, 
< 0 | x_H(t_1) x_H(t_2) | 0 > \E
 \frac{1}{2} \lim_{\epsilon \to 0} \, \int_{-\infty}^{+\infty} \frac{dE}{2 \pi}
\, e^{-i E \epsilon} \left ( E^2 + \omega^2 \right ) \, G_{xx}(E) \> ,
\ee
from the second one after multiplication with $x \EQ x_H(0) $
\be
< 0 | \lambda x^4 | 0 > \E - \frac{1}{4} \lim_{t \to 0} 
\left ( \, \frac{\partial^2}{\partial t^2} + \omega^2 \, \right ) \, 
< 0 | x_H(t) x_H(0) | 0 > 
\E \frac{1}{4} \lim_{\epsilon \to 0} \, \int_{-\infty}^{+\infty} 
\frac{dE}{2 \pi} \, e^{-i E \epsilon} \left ( E^2 - \omega^2 \right ) \, G_{xx}(E) 
\> . 
\ee
Adding both leads to Eq. (\ref{E0 aus Green}).

\end{subequations} 
\renewcommand{\baselinestretch}{1.2}
\normalsize
\vspace{0.4cm}

\noindent
A simple calculation gives
\be
G_{xx}^{(0)}(E) \E i \int_{-\infty}^{+\infty} dt \> e^{i E t} \> 
{\cal O}^{-1}(t,0) \E i \tilde {\cal O}^{-1}(E) \E \frac{i}{E^2 - 
\omega^2 + i \, 0^+}
\ee
and therefore the expected result is  obtained from Eq. (\ref{E0 aus Green})
\be
E_0^{(0)} \E \frac{i}{8 \pi} \lim_{\epsilon \to 0} \> 
\int_{-\infty}^{+\infty} dE \> e^{- i \epsilon E} \> 
\frac{3 E^2 + \omega^2}{E^2 -\omega^2 + i \, 0^+} \E \frac{1}{2} \omega \>. 
\ee
Here the theorem of residues has been used after closing the 
integration contour in the lower complex $E$-plane.
In first-order perturbation theory we have to expand the exponential function 
$ \> \exp \left (- i \lambda \int_{-\infty}^{+\infty} dt \> \delta^4/\delta 
J(t)^4 \right ) \> $ 
(see Eq. (\ref{volles erzeug funk})) in the numerator and the denominator of Eq.
(\ref{n-Punkt aus erzeug funk}), differentiate functionally and perform the 
Fourier transformation. The result of this -- admittingly a bit lengthy -- calculation is
\bea
G_{xx}^{(1)}(E) \EA G_{xx}^{(0)}(E) \> - 12 i \lambda \left [ G_{xx}^{(0)}(E) \right ]^2
\> \int_{-\infty}^{+\infty} \frac{dE'}{2 \pi}  \> G_{xx}^{(0)}(E')
\non
\EA \frac{i}{E^2 - \omega^2 - i \, 0^+} \> + \> \frac{6 i \lambda}{\omega}
\frac{1}{(E^2 - \omega^2 - i \, 0^+)^2} \> .
\label{G1}
\eea   
With that we find for the ground state energy        
\be
E_0 \E \frac{1}{2} \omega + \frac{3 i \lambda}{2 \omega}
\> \underbrace{ \int_{-\infty}^{+\infty} \frac{dE}{2 \pi} \> e^{-i \epsilon E}
\> \frac{\omega^2 + 3 E^2}{(E^2 - \omega^2 - i \, 0^+)^2}}_{= - 
\frac{i}{2 \omega}} \E \frac{1}{2} \omega + \frac{3 \lambda}{4 \omega^2}
\> .
\label{Stoerung}
\ee
Of course, this is exactly the result which we would obtain in ordinary
perturbation theory:
\be
\Delta E_0 \E \left < 0 \left | \> \lambda \hat x^4 \> \right | 0 \right >
\E \lambda \left < 0 \left | \> \left [ \frac{1}{\sqrt{2 \omega}} ( 
\hat a  + \hat a^{\dagger} ) \, \right ]^4 \> \right | 0 \right > \E 
\frac{3 \lambda}{4 \omega^2} \> .
\ee

\vspace{0.6cm}
\noindent
{\bf b) Structure Function of the Harmonic Oscillator}
\vspace{0.1cm}

\noindent
An important, experimentally measurable quantity is the so-called \blau {\bf structure function}
of a bound sytem (e.g. a proton, a nucleus or a quantum liquid)
which can be determined by inelastic scattering of electrons or neutrons and gives us
-- as the name implies -- information on the structure of the object under investigation.
In this (inclusive) process the projectile transfers momentum
$ q $ and energy $ \nu $ to the target which gets excited but whose 
final states are not observed: Only the final energy $ E_f $ and the scattering angle 
$ \Omega_f $ of the projectile are measured.
Therefore, one has to sum {\bf incoherently} over all (energetically accessible) 
final states $ \lvl n \ra $: 
\bea
\frac{d^2 \sigma}{d \Omega_f dE_f} \EA \lrp\frac{d\sigma}{d\Omega_f}\right )_0 \cdot S(q,\nu) \\
S(q, \nu) \EA \sum_n \, \delta(\nu - (E_n - E_0)) \, 
\lvl \la 0 \,  \left | \, \hat {\cal O}\, \right | \, n \ra \rvl^2 \> .
\label{def Struk}
\eea
Here the excitation operator essentially is the Fourier transform of the
density operator
\be
\hat {\mathcal{O}} \E e^{-i q \cdot \hat x}
\ee
since the elementary interaction is already taken care of by the factor
$ \> \lrp d\sigma/d\Omega_f \rrp_0 \> $ 
(which, e.g., is the Rutherford or Mott cross section in the case of electromagnetic
interactions) \footnote{Eq. \eqref{def Struk} only holds under some simplifying assumptions:
The interaction must be weak and scalar or transferred by a single impulse. Velocity-dependent
(magnetic) parts and the recoil of the target are neglected, see, e.g.  Ref. \cite{Ros1a}.}.

The {\bf inelastic} structure function (for which the elastic contribution $ n = 0 $ is excluded)
may be determined by the Green function \footnote{See, e.g., {\bf \{Fetter-Walecka\}}, eq. 17.13 . 
Note that in this work the so-called  {\bf polarisations propagator} $ \Pi = - i G $ is used
which requires a corresponding modification of Eq. \eqref{struk aus G OO+}.}
\be
G_{\mathcal{O} \mathcal{O}^{\dagger}}(T) \Def \la 0 \lvl \hat{\mathcal{O}}_H(T) \, 
\hat {\mathcal{O}}_H^{\dagger}(0)\rvl 0 \ra 
\label{def G OO+}
\ee
where
\be
\hat{\mathcal{O}}_H(T) \E e^{i \hat H T} \, e^{- i q \cdot x} \, e^{-i \hat H T}
\ee
is the excitation operator in the Heisenberg picture. Due to the well-known
relation
\be
\frac{1}{x - x_0 \pm i 0^+} \E {\cal P} \, \frac{1}{x - x_0} \mp i \pi \, \delta \lrp  x - x_0 \rrp
\label{Zerleg Green}
\ee
(${\cal P} $ denotes the principal value) we indeed have for $ \nu > 0 $ 
\be
S^{\> {\rm inelast.}}(q,\nu) \E \frac{1}{\pi} \, {\rm Re} \, \int_{-\infty}^{+\infty} dT \> e^{i \nu T} \, 
G_{\mathcal{O} \mathcal{O}^{\dagger}}(T) \>  .
\label{struk aus G OO+}
\ee
\vspace{0.3cm}

\renewcommand{\baselinestretch}{0.9}
\scriptsize
\noindent
{\bf Proof:}\\   
\vspace{0.1cm}

\noindent
\begin{subequations}
Completeness of the final states ( $ \sum_n \lvl n \ra \la n \rvl = 1 $ ) 
allows to write the Fourier transform of the Green function
in  Eq. \eqref{def G OO+} as
\bea 
\int_{-\infty}^{+\infty} dT \> e^{i \nu T} \, 
G_{\mathcal{O} \mathcal{O}^{\dagger}}(T) \EA
\sum_{n=0} \lcp \int_0^{\infty} dT \> e^{i T( \nu - (E_n-E_0) + i 0^+)} \, \lvl \la 0 \lvl 
{\mathcal{O}} \rvl n \ra \rvl^2 + 
\int_{-\infty}^0 dT \> e^{i T( \nu + (E_n-E_0)-i 0^+)} \, \lvl \la n \lvl {\mathcal{O}} \rvl 0 \ra 
\rvl^2 \rcp \non 
\EA
i \sum_{n=0} \lcp \frac{\lvl \la 0 \lvl \hat {\mathcal{O}} \rvl n \ra \rvl^2}{ \nu - (E_n-E_0) + i 0^+} \,  
+ \frac{\lvl \la n \lvl \hat {\mathcal{O}} \rvl 0 \ra \rvl^2}{ \nu + (E_n-E_0)-i 0^+} \,  \rcp \> .
\label{FT von G OO+}
\eea
Applying the relation \eqref{Zerleg Green} then gives
\be
{\rm Re} \, \int_{-\infty}^{+\infty} dT \> e^{i \nu T} \, 
G_{\mathcal{O} \mathcal{O}^{\dagger}}(T) \E \pi \, 
\sum_{n=0} \lcp \lvl \la 0 \lvl \hat {\mathcal{O}} \rvl n \ra \rvl^2 \, \delta \lrp \nu - (E_n-E_0) \rrp   
+ \lvl \la n \lvl \hat {\mathcal{O}} \rvl 0 \ra \rvl^2 \, \delta \lrp \nu + (E_n-E_0) \rrp   \rcp  \> ,
\ee
which leads to Eq. \eqref{struk aus G OO+} for positive excitation energies. For $ \nu > 0^- $
the elastic line is also included
\be
\frac{1}{\pi} \, {\rm Re} \, \int_{-\infty}^{+\infty} dT \> e^{i \nu T} \, 
G_{\mathcal{O} \mathcal{O}^{\dagger}}(T) \E
S^{\> {\rm inelast.}}(q,\nu) + 2 \, F_{00}^2(q) \, \delta ( \nu ) \> , 
\ee
where
\be
F_{00}(q) \E \la 0 \lvl \hat {\mathcal{O}} \rvl 0 \ra 
\label{def form}
\ee
is the elastic {\bf form factor}.

\end{subequations}

\renewcommand{\baselinestretch}{1.2}
\normalsize

\vspace{0.8cm}

\noindent
This Green function is easily evaluated for the harmonic oscillator:
\bea
G^{\> {\rm h.O.}}_{{\cal O} {\cal O}^{\dagger}}(T) \EA \int {\cal D} x \> 
\exp \lcp i \int_{-\infty}^{+\infty} dt  \> \lsp \frac{m}{2} \dot x^2(t) - \frac{m}{2} \omega^2 \, x^2(t) 
\rsp - i q \cdot x(T) + i q \cdot x(0) \rcp \non
\EA \int {\cal D} x \> 
\exp \lcp i  \int_{-\infty}^{+\infty} dt \> \lsp \frac{m}{2} x(t) \lrp -\partial^2_t - \omega^2 \, + i 0^+ 
\rrp x(t)  - b(t)  \cdot  x(t) \rsp \rcp \> ,
\label{Green hO}
\eea
where we have defined
\be
b(t) \Def q \lsp \delta(t-T) - \delta(t) \rsp \> .
\label{def b(t)}
\ee
Inevitably for problems involving the harmonic oscillator. the functional integral is a Gaussian
one
giving the result (the normalization is such that  $ G = 1 $ for $ q = 0 $ )
\bea
G^{\> {\rm h.O.}}_{{\cal O} {\cal O}^{\dagger}}(T) \EA \exp \lsp - \frac{i}{2m} \, 
\int_{-\infty}^{+\infty} dt  \, \int_{-\infty}^{+\infty} dt' \, b(t) \lrp t \lvl \frac{1}{-\partial^2_t 
- \omega^2 \, + i 0^+ }\rvl t' \rrp 
\, b(t') \rsp \non
\EA \exp \lsp - i \frac{q^2}{2m} \, \int_{-\infty}^{+\infty} \frac{dE}{\pi} \> 
\frac{1}{E^2 - \omega^2 \, + i 0^+} \, \lrp 1 - \cos(ET) \rrp
\rsp \> .
\label{G HO}
\eea
The $E$-integral can be easily calculated by the theorem of residues and one obtains
\be
G^{\rm h.O.}_{{\cal O} {\cal O}^{\dagger}}(T) \E \exp \lsp - \frac{q^2}{2m \omega} \,
\lrp 1 - e^{-i \omega |T|} \rrp \rsp \> .
\ee
Inserted into Eq. \eqref{struk aus G OO+} this gives
\be 
S^{\> {\rm inelast. h.O.}} (q,\nu) \E  \, \exp \lrp -\frac{q^2}{2 m \omega} \rrp  \, \sum_{n=1}^{\infty}
\lrp \frac{q^2}{2 m \omega} \rrp^n \frac{1}{n!} \> \delta \lrp \nu - n \omega \rrp \> .
\label{ho Struk}
\ee
The square of the elastic form factor is given by the $T$-independent part in Eq. 
\eqref{G HO} and reads
\be
\lrp F_{00}^{\> {\rm h.O. }}(q) \rrp^2 \E \exp \lsp - i \frac{q^2}{2m} \, \int_{-\infty}^{+\infty} 
\frac{dE}{\pi} \> \frac{1}{E^2 - \omega^2 \, + i 0^+} \rsp \E 
\exp \lrp -\frac{q^2}{2 m \omega} \rrp \> .
\label{elast form HO} 
\ee
This supplies exactly the missing  $ (n = 0)$-contribution in the sum of Eq. \eqref{ho Struk}
and agrees with the result which one obtains if one calculates the matrix element
 $ \la 0 \lvl \mathcal{O} \rvl n \ra $ in Eq. \eqref{def Struk} 
with the explicit wave functions \eqref{ho wfn} and then incoherently sums over all
excited states. It goes without saying that such a calculation is much more involved 
than the derivation 
presented here.
However, the biggest advantage of Green function methods is that they can easily be
extended to many-body and field-theoretical problems.
\vspace{0.2cm}

\noindent
From Eq. \eqref {ho Struk} one sees that
the structure function of a harmonically bound particle consists -- as expected~--
of a collection of $ \delta $- functions because the entire spectrum is discrete. 
In general, however, the 
interactions in the target are such that also a continuous spectrum exists; 
then the structure function contains -- besides the excitation of individual discrete levels
-- at large momentum transfer a broad, continous "quasi-elastic peak" that 
corresponds to the knock-out of the bound particle into the continuum.
Although the model of the harmonic oscillator is unrealistic
(even for confined quarks) Eq. \eqref {ho Struk} also contains this infomation 
in terms of the weight which a single $ \delta $-functions carries:
If $ q^2 \gg 2m \omega $, we can estimate the excitation number of maximal strength by using
Stirling's formula \eqref{Stirling} and we obtain
$ \> n_{max} \simeq q^2/(2m \omega) \> $, i.e. the strongest excitation
occurs at $ \nu_{max} \simeq q^2/(2m) \> $, which exactly is the energy of a particle initially
at rest after it has absorbed a momentum  $ q $ .


\vspace{0.5cm}

\subsection{\textcolor{blue}{Symmetries and Conservation Laws}}
\label{sec1: Symm}

From the usual operator formulation of quantum mechanics we know 
that the time-dependence of an Heisenberg operator is given by
\be
\frac{d \hat A_H(t)}{dt} \E \frac{d}{dt} \> \left ( \, e^{i \hat H t /\hbar} \, 
\hat A \, e^{-i \hat H t/\hbar } \, \right ) \E \frac{i}{\hbar} \, 
\left [ \, \hat H , \hat A_H(t) \, \right ] \E \frac{i}{\hbar} \,  
e^{i \hat H t /\hbar} \, \left [ \, \hat H , \hat A \, \right ] \,  
e^{-i \hat H t/\hbar } 
\label{Heis Beweggl}
\ee
if the operator $ \hat A $ does not depend explicitly on time.
By that it follows in particular that operators which commute with the
Hamiltonian of the system are \textcolor{blue}{\bf conserved}
\be
\frac{d \hat A_H(t)}{dt} \E 0 \hspace{0.5cm} , \quad {\rm if} \hspace{0.5cm} 
\left [ \, \hat H , \hat A \, \right ]  \E 0 \> ,
\ee
i.e. they don't change during the evolution of the system in time.
Since in most cases of interest an exact solution of the equation of motion
(\ref{Heis Beweggl}) is not possible such constants of motion
are of highest importance in quantum physics.

This is also the case in classical physics where \blau{\bf Noether's theorem} 
links the existence of a conserved quantity to the invariance of the 
action under a (continous) symmetry transformation. In the usual proof
\footnote{See, e.g. {\bf \{Landau-Lifschitz 1\}}, ch. 2. This chapter 
mainly follows \meingruen{\bf Peskin \& Schroeder}, ch. 2.2  and 9.6 .} 
one uses the Euler-Lagrange equations for the system: Let
\be
x(t) \To x'(t) = x(t) + \alpha \, \Delta x(t) 
\ee 
be an infinitesimal transformation of the classical path with a constant parameter
$\alpha$, which leaves the action invariant (up to boundary terms which do not change
the equation of motion). This means that the Lagrange function  changes at most by a 
total time derivative:
\be
L \To L + \alpha \, \frac{d}{dt} \Lambda(t) \> .    
\label{L+totder}
\ee
Here $\Lambda(t)$ is a quantity which can be calculated from the
symmetry transformation and the Lagrange function under discussion.
However, the change of $L(x, \dot x) $ can be also calculated from the
change of the path
\bea
\alpha \, \Delta L \EA \frac{\partial L}{\partial x} \, \alpha \, \Delta x
+ \frac{\partial L}{\partial \dot x} \,  \frac{d}{dt} \left ( \alpha \, \Delta x 
\right ) \non
&=& \alpha \, \frac{d}{dt} \left ( \, \frac{\partial L}{\partial \dot x} \Delta x 
\, \right )
\, + \alpha \, \left [ \, \frac{\partial L}{\partial x} - \frac{d}{dt} \left ( 
 \frac{\partial L}{\partial \dot x} \right ) \, \right ] \Delta x  \> .
\eea
Due to the equation of motion the term in the square bracket vanishes.
If we equate the two changes of the  Lagrange function then we obtain
\be
\frac{d}{dt} j(t) \E 0 \hspace{0.5cm} {\rm with} \> \> \> \> \> j(t) \Def   
 \frac{\partial L}{\partial \dot x} \Delta x - \Lambda(t)  \> , 
\label{Noether-Strom}
\ee
i.e.  $ j(t) $ is a conserved "current" (the normalization is arbitrary).
\vspace{0.8cm}

\renewcommand{\baselinestretch}{1.0}
\small
\noindent
{\bf Example : Time-Translation Invariance}
\vspace{0.2cm}

\noindent
Consider a particle in a time-independent potential with the Lagrange function
 $ L = m \dot x^2/2 - V(x) $. It is obvious that its motion does not depend 
on the moment when we start counting the time, i.e.
 we can shift $t \to t + \alpha$ . Infinitesimally this induces 
 the transformation
\be
x(t) \To x(t) + \alpha \, \dot x(t) \> ,
\ee 
under which the Lagrange function changes as
\bea
\alpha \, \Delta L \E m \dot x \, \alpha \ddot x - V'(x) \, \alpha \dot x \E 
\alpha \, \frac{d}{dt} \left ( \, \frac{m}{2} \dot x^2 - V(x) \, \right ) \> .
\eea
From Eq. (\ref{Noether-Strom}) we therefore find that
\be 
j \E m \dot x \, \dot x - \left (  \, \frac{m}{2} \dot x^2 - V(x) \, \right ) 
\E \frac{m}{2} \dot x^2 +  V(x) \E {\rm const}_t\> , 
\ee
i.e. the {\bf energy} of the system is conserved due to the homogenity in time.

\renewcommand{\baselinestretch}{1.2}
\normalsize
\vspace{0.4cm}

How does that come out in the path integral formalism which doesn't know anything about
operators or classical equations of motion?
To answer this question we will investigate the behavior of the path integral (in the Lagrange form) under 
the general \textcolor{blue}{\bf time-dependent} transformation
\be
x(t) \To x'(t) \E x(t) + \alpha(t) \, \Delta x(t)   \> .  
\label {Trans}
\ee
We will assume again that the transformation with a constant $ \alpha $ is 
a symmetry transformation, i.e. 
as in Eq. (\ref{L+totder}) 
it changes the Lagrangian 
only by a total derivative. With a
time-dependent parameter, however, we now obtain
\bea   
L(x,\dot x)  \To  L \lrp x +\alpha \Delta x,\dot x + \alpha \Delta \dot x) + \dot \alpha \,
\Delta x \rrp \EA  L \lrp x + \alpha \Delta x, \dot x + \alpha \Delta \dot x \rrp
+ \dot \alpha \frac{\partial L}{\partial \dot x} \Delta x \non
\EA \alpha \frac{d}{dt} \Lambda(t) + \dot \alpha \, 
\frac{\partial L}{\partial \dot x} \Delta x \> ,
\eea
because we have to consider the change in velocity induced by $\alpha(t)$.
Of course, the transformation (\ref{Trans}) is not allowed to change the fixed
boundary conditions $\> x(t_a) = x_a , \> x(t_b) = x_b \> $ 
(we recall that  there is {\bf no} integration
over the boundary points in the path integral) so that we have to require that
\be
\alpha(t_a) \E  \alpha(t_b) \E 0 \> .
\label{Rand fuer Trans}
\ee
As in the classical case we will restrict ourselves to infinitesimal
transformations, i.e. in the following  we will only consider terms up to order
 $\alpha$. Since the path integral does not change we have
\be
\int_{x(t_a) = x_a}^{x(t_b) = x_b} {\cal D} x \, e^{i S[x]/\hbar} \E   
\int_{x'(t_a) = x_a}^{x'(t_b) = x_b} {\cal D} x' \> | {\cal J} | \, 
e^{i S[x' + \alpha \Delta \xi]/\hbar} \> ,
\label{Pfad trans 1}
\ee
where 
\be
{\cal J} \E \fdet_{t,t'} \left ( \frac{\delta x(t)}
{\delta x'(t')} \right )
\ee
is the Jacobi determinant of the transformation. Here we will assume that 
$  {\cal J} = 1 $ , i.e. that the "measure" $ {\cal D} x $ of the path integral
is also invariant under the symmetry transformation.
If we expand the r.h.s. of  Eq. (\ref{Pfad trans 1}) up to order
$\alpha$, we therefore obtain
\be
\int_{x'(t_a) = x_a}^{x'(t_b) = x_b} {\cal D} x' \> \frac{i}{\hbar} \,
\Delta S[x'] \, e^{i S[x']/\hbar} \E 
\int_{x't_a) = x_a}^{x'(t_b) = x_b} {\cal D} x'\> \frac{i}{\hbar} \, 
\int_{t_a}^{t_b} dt \, \left \{ \alpha(t) \frac{d}{dt} \Lambda(t) + \dot \alpha(t)  \, 
\frac{\partial L}{\partial \dot x'} \Delta x' \, \right \} \, 
e^{i S[x']/\hbar} \E 0 \> .
\label{Pfad trans 2}
\ee
Now we perform an integration by parts in the second term: Due to  Eq. (\ref{Rand fuer Trans})
the boundary terms at $t = t_a$ and $t = t_b$ vanish and we get
\be 
\int_{t_a}^{t_b} dt \, \alpha(t) \,
\int_{x(t_a) = x_a}^{x(t_b) = x_b} {\cal D} x \> \frac{i}{\hbar} \, \frac{d}{dt} \, 
\left \{ \,  \Lambda(t) - \frac{\partial L}{\partial \dot x} \Delta x\, 
\right \} \, e^{i S[x]/\hbar} \E 0 \> ,
\label{Pfad trans 3}
\ee
where we again have written $x(t)$ as integration variable in the  functional integral.
As $\alpha(t) $ is arbitrary the integrand in the curly bracket has to vanish at 
any time $ t $, i.e. 
\be
\boxed{
\qquad \frac{d}{dt} \, \left < \, j(t) \, \right > \Def {\cal N} \, 
\int_{x(t_a) = x_a}^{x(t_b) = x_b} {\cal D} x \> \frac{d}{dt} \, 
\left \{ \,  \frac{\partial L}{\partial \dot x} \Delta x  - \Lambda(t) \, \right \} \, 
e^{i S[x]/\hbar} \E 0 \quad
}
\hspace{1cm} \mbox{for all} \> \> \> t 
\label{Pfad Noether}
\ee
must hold. Here  ${\cal N}^{-1} = \int {\cal D}x \, \exp(i S[x]/\hbar) $   
is the normalization which ensures that  $ \>  < 1 > \, = 1$ . 
\vspace{1.2cm}

\noindent
{\bf Example: Space-Translation Invariance}
\renewcommand{\baselinestretch}{1.1}
\small
\vspace{0.2cm}

\noindent
As simplest example we will consider a free particle when its coordinate is shifted
 by a constant: $ x(t) \to x(t) + \alpha $. Obviously, the Jacobi determinant here
 is $ {\cal J} = 1 $ and
 the Lagrange function  $ m \dot x^2/2 $ remains unchanged under this transformation.
Therefore we have $\Lambda(t) = 0 $ and the path-integral version of Noether's theorem 
(\ref{Pfad Noether}) says, that the momentum
\be
P \Def \left < \, m \dot x(t)\, \right > \E {\rm const}_t 
\ee 
is conserved.

\noindent
This we may check by calculating the path integral exactly.
The easiest way is to introduce an external force
 $ e(t) $, to differentiate functionally w.r.t. this source
and then to switch it off. For that purpose we can use 
the explicit results 
obtained in {\bf chapter} {\bf \ref{sec1: quadrat Lagr}}: Since the 
prefactor in Eq. (\ref{U quadrat 1}) does not depend on $e(t)$, we have
to calculate 
\be
P^{\> \rm free} \E m \, e^{-i S_{\rm cl}/\hbar} \, \frac{\partial}{\partial t} \, 
\left ( - \frac{\hbar}{i} \right )  
 \, \frac{\delta}{\delta e(t)} \,  e^{i S_{\rm cl}/\hbar}
\Biggr |_{e=0} \E - m \frac{\partial}{\partial t} \, 
\frac{\delta S_{\rm cl}}{\delta e(t)} \Biggr |_{e=0} \> .
\label{p frei} 
\ee
The classical action of a particle under the influence of a time-dependent
force $e(t)$ is obtained from Eq. (\ref{erzwung HO}) in the limit $\omega \to 0$ :
\be
S_{\rm cl} \E \frac{m}{2 T} \, \left [ \, \left (x_b - x_a \right )^2 - 
\frac{2x_b}{m} \, \int_{t_a}^{t_b} ds \, e(s) \, \left ( s - t_a \right )
- \frac{2x_a}{m} \, 
\int_{t_a}^{t_b} ds \, e(s) \, \left ( t_b - s \right ) + {\cal O} \left ( 
e^2 \right ) \, \right ] \> . 
\ee
Differentation in Eq. (\ref{p frei}) then gives
\be
P^{\> \rm free} \E m \, \frac{x_b - x_a}{T} \> , \> \> \> T \EQ  t_b - t_a \> , 
\ee
which, indeed, is independent of the arbitrary time
 $t \in [t_a,t_b] $ and illustrates the classical result
``momentum = mass $\times$ (constant) velocity between $a$ and $b$''. 
\vspace{0.1cm}

A less trivial example is given by a system of two particles (although this rather belongs 
into \rot{\bf section 2})  interacting via a potential which only depends on their 
mutual distance:
\be
L \E \frac{m_1}{2} \dot x_1^2 + \frac{m_2}{2} \dot x_2^2 - V \left (x_1 - x_2 
\right ) \> . 
\label{zwei Teilchen}
\ee
This system is again translation invariant as the previusly considered free particle  and therefore 
Noether's theorem (\ref{Pfad Noether}) (with obvious 
modifications  
for several particles) tells us  that the total momentum
\be
P^{\> \rm total} \Def \left < \, m_1 \, \dot x_1 + m_2 \, \dot x_2 \, \right > \E 
{\rm const}_t  
\label{P gesamt}
\ee
is conserved.
However, the quantum dynamics of the system described by Eq. 
(\ref{zwei Teilchen}) is far from being so simple as in the previous example.
Fortunately, the path-integral average needed in Eq. 
(\ref{P gesamt}) can be evaluated easily by introducing 
center-of-mass and relative coordinates
\be
R \Def \frac{m_1 x_1 + m_2 x_2}{M} \> \> , \> \> r \Def x_1 - x_2 \> \> , 
\> \> M \EQ  m_1 + m_2 \> . 
\ee
It is well-known that the Lagrange fuction (\ref{zwei Teilchen}) then separates
into a center-of-mass part $ \> M\dot R^2/2 \> $ and a relative part 
$ \> m_1 m_2 \dot r^2/(2 M ) - V(r) \> $, which leads to a factorization
of the corresponding path integrals.
Because of $  m_1 \dot x_1 + m_2 \dot x_2  = M \dot R $ 
the path integral over the relative coordinate  -- which, in general, cannot be performed --
cancels in the averaging procedure. Therefore we can employ the result for the free particle
and obtain
\be
 P^{\> \rm total} \E M \, \frac{R_b - R_a}{T} \> . 
\ee
\renewcommand{\baselinestretch}{1.2}
\normalsize
\vspace{0.2cm}

\noindent
If the Jacobian contributes, i.e. if the "measure" is not invariant,
additional terms appear which are called an \textcolor{blue}{\bf anomaly}: 
The quantum theory does not have the symmetry of the classical theory. 
This plays an important role in field theory (see {\bf chapter} {\bf \ref{sec3: Anomal}}).


\subsection{\textcolor{blue}{Numerical Treatment of Path Integrals}}
\label{sec1: Numerik}

\vspace{0.3cm}
If  neither an exact analytic evaluation of the path integral is possible
nor perturbation theory or semi-classical expansions are applicable
one has to try to calculate the functional integral numerically.
In practice, this is only possible in {\bf Euclidean time} since in real 
time the oscillations in the integrand are not under control numerically
 \footnote{There are some attempts to treat path integrals numerically in real
 time, e.g. in Ref. \cite{SSBa}.}:
 One changes to \textcolor{blue}{\bf imaginary times}
\be
T \E - i \beta \> , \hspace{0.5cm} t \E - i \tau
\label{euklid Zeit}
\ee
and investigates the \textcolor{blue}{\bf partition function}
\be
\boxed{
\qquad Z(\beta) \E {\rm tr} \left ( e^{- \beta \hat H/\hbar} \right ) \> . \quad 
} 
\ee
We can write down immediately a path-integral representation
for the partition function of a particle utilizing the one derived
for time-evolution operators. If in 
\be
{\rm tr} \left (\hat U(T,0) \right ) = {\rm tr} \left (e^{- i T\hat H/\hbar}
\right ) = \int dx \int_{x(0)=x}^{x(T)=x} {\cal D}x \> \exp \left [
\frac{i}{\hbar} \int_0^T dt \left( \frac{m}{2} \dot x^2 - V(x) 
\right ) \> \right ]
\ee
we perform the transformation (\ref{euklid Zeit}), then we obtain the required result
\vspace{0.2cm}

\fcolorbox{blue}{white}{\parbox{14cm}
{
\bea
Z(\beta) \EA  \int dx \int_{x(0)=x}^{x(\beta)=x} {\cal D}x \> \exp \left [
 - \frac{1}{\hbar} \int_0^{\beta} d\tau \left( \frac{m}{2} \dot x^2 + V(x) 
\right ) \>
\right ] \EQ \oint\limits_{x(0)=x(\beta)}\! {\cal D}x  \>
e^{- S_E[x]/\hbar} \> . \no
\eea
}}
\vspace{-2cm}

\bea
 \label{Z Pfad}
\eea
\vspace{0.8cm}


\noindent
Here
\be
S_E[x] \E \int_0^{\beta} d\tau \> \left [ \, \frac{m}{2} \dot x^2 + V(x)
\, \right ]
\label{S_E}
\ee
is the \textcolor{blue}{\bf Euclidean} action \footnote{The name derives from the
fact that this transformation changes the relativistic indefinite metric
into a four-dimensional Euclidean one: $x_{\mu}x^{\mu} = c^2 t^2
- \fx^2 = - \left ( \fx^2 + c^2 \tau^2 \right )$.}.
Notice the change of sign in Eq. \eqref{S_E} between
kinetic and potential energy compared to the normal action!

\noindent
Similar as the ground state is projected out in the expression for the Green function
the \textcolor{blue}{\bf ground-state energy} of the system
$ \> E_0 = < 0 | \, \hat H \, | 0 > \> $ can be filtered out from the partition function 
by considering large Euclidean times: 
\bea
E_0 \EA \lim_{\beta \to \infty} \> \frac{ {\rm tr} \left ( \> \hat H 
e^{- \beta \hat H/\hbar} \> \right )}{{\rm tr} \left ( \> 
e^{- \beta \hat H/\hbar} \right ) } \> =\>  \lim_{\beta \to \infty} \> 
\frac{ E_0 e^{- \beta E_0/\hbar} 
+ E_1 e^{- \beta E_1/\hbar} + \ldots}
{e^{- \beta E_0/\hbar} +  e^{- \beta E_1/\hbar} + \ldots} \non
\EA\lim_{\beta \to \infty} \> \left [ \> - \hbar
\frac{\partial}{\partial \beta} \, \ln Z \> \right ] \> .
\eea
If we consider the discretized form of the path integral (\ref{Z Pfad}) 
\be
Z \E  \lim_{N \to \infty} \> A^N_{\epsilon} \, \int dx_0 \ldots dx_{N-1}
\> \exp \left [ \> -  S_E(x_0 \ldots x_N)/\hbar \> \right ]
\ee
with $ \> x_0 = x_N \> , \> \>
A_{\epsilon} = \left [ m/(2 \pi \epsilon \hbar) \right ]^{1/2} \> $ and
\be 
S_E(x_0 \ldots x_N) \E 
\epsilon \sum_{j=1}^N \left [ 
\, \frac{m}{2} \left ( \frac{x_j - x_{j-1}}{\epsilon} \right )^2 + V(x_j) \,
\right ] 
\ee
then we obtain
\be
E_0 \E \lim_{\latop{N \to \infty}{\epsilon \to 0}} \> 
\frac{ \int dx_0 \ldots dx_{N-1} \> \frac{1}{N} \sum_{j=1}^N \left [
\, \frac{m}{2} \left ( \frac{x_j - x_{j-1}}{\epsilon} \right )^2 + V(x_j)
- \frac{\hbar}{2 \epsilon} \, \right ] \> \exp \left ( - S_E/\hbar \right ) }
{\int dx_0 \ldots dx_{N-1} \>\exp \left ( - S_E/\hbar \right ) } \> .
\ee
This is not well suited for nmerical purposes because terms inverse in
 $\epsilon$ have to cancel to produce a finite result in the limit $ \epsilon \to 0 $.
As can be seen the
culprit is the operator of the kinetic energy which is non-local and 
causes these difficulties when
  taking expectation values.
  
\noindent
One can circumvent this difficulty by using the \textcolor{blue}{\bf virial theorem} 
\footnote{Proof: Evaluate the commutator
$ \> [ \, \hat x \hat p \, , \, \hat p^2/(2 m) + V(\hat x) \, ] \> $ and 
take the ground-state expectation value on both sides.}
\be
\left < 0 \left | \> \frac{ \hat p^2}{2 m} \> \right | 0 \right > \E 
\frac{1}{2} \left < 0 \left | \> \hat x V'(\hat x) \> \right | 0 \right > \> .
\label{virial}
\ee
By this we obtain
\bea
E_0 \EA \lim_{\latop{N \to \infty}{\epsilon \to 0}} \>
\frac{ \int dx_0 \ldots dx_{N-1} \> \frac{1}{N} \sum_{j=1}^N \left [
\,  \frac{1}{2} x_j V'(x_j) + V(x_j) \, \right ] \> 
\exp \left ( - S_E/\hbar \right ) }
{\int dx_0 \ldots dx_{N-1} \>\exp \left ( - S_E/\hbar \right ) } \non
&\EQ & \left < \> \frac{1}{N} \sum_{j=1}^N \left [
\,  \frac{1}{2} x_j V'(x_j) + V(x_j) \, \right ] \> \right > \> .
\label{E0}
\eea
\vspace{0.2cm}

\textcolor{blue}{\bf Excitation energies} can be obtained from the euclideam
Green functions (also called correlation functions) if the expectation value
of the corresponding operator vanishes in the ground state
\bea
G_{AA}(t=-i \beta) \EA  \sum_n \> \left | < 0 \, | \hat A | \, n > \right |^2 
\, e^{ - (E_n - E_0) \beta/\hbar} \non 
&\stackrel{\beta \to \infty}{\longrightarrow}& 
\underbrace{\left | < 0 \, | \hat A | \, 0 > \right |^2}_{=0} \> + \> 
\left | < 0 \, | \hat A | \, 1 > \right |^2\, e^{ - (E_1 - E_0) \beta/\hbar} 
\> \ldots
\label{Abfall Korr funk}
\eea
In other words, the correlation functions fall off exponentially in
Euclidean time and their decay constant is determined by the excitation energy
of the first state which has an overlap with the ground state by means of the 
operator $ \hat A $. For the numerical calculation
it is crucial that the expectation value of the operator in the ground state
vanishes {\it exactly} because any contribution of the first term in Eq. 
(\ref{Abfall Korr funk}), no matter how small, would overwhelm 
the exponentially vanishing signal of the excited state.

\noindent
The simplest operator whose expectation value vanishes in the ground state
due to the parity selection rule, is 
$ \> \hat A = \hat x \> $. Thus we have, for example
\be
E_1 - E_0 \E \lim_{\beta \to \infty} \left \{ \> - \frac{\hbar}{\beta}
\, \ln \, \frac{\int {\cal D}x \> x(\beta) x(0) \, \exp \left [ - S_E/\hbar
\right ] }{\int {\cal D}x \>\exp \left [ - S_E/\hbar\right ] } \> \right \} 
\ee
and the corresponding discrete form can be used for a numerical analysis. 
In the 3-dimensional case, for instance, the lowest-lying state for a given 
angular momentum $ \ell $ can be projected out by the operator
$ \> \hat A = Y_{\ell 0} (\hat r) \> $.

Similarly one proceeds in field theory in which the particles are considered as 
excitations above a "vacuum'' whose energy is set to $ E_0 = 0 $.  For example, 
in Quantum Chromo Dynamics (to be discussed in greater detail in \rot{\bf section 3}) 
the operator $ \> \hat A = \bar \psi \gamma_5 \psi \> $, 
where $ \psi $ are suitable quark field operators connects the vacuum with the pion, 
the lowest pseudoscalar state. In this way, one can (in principle) determine 
the masses of the lowest-lying hadrons numerically from the fundamental field theory 
(more on this in  {\bf chapter \ref{sec3: Gitter}}).

\vspace{0.3cm}

\noindent
Up to now we only have discussed the formal tools to extract energies or masses
from the Euclidean path integral. How does one calculate in reality those
functional integrals numerically?
Let's take again as a concrete example the case of a particle moving in one dimension
under the influence of a given potential. In order to get the ground-state energy 
we see from Eq. (\ref{E0}) that we have to 
evaluate a $N$-dimensional integral numerically  
and that $ N $ has to be very large.
Direct integration by means of methods for one- or low-dimensional integrals
(for example Simpson's rule or Gaussian integration) is clearly not feasable
as the required computing time would increase like (integration points) $^N$ ,
a feature sometimes called "the curse of dimensions".
Fortunately, one can use statistical (stochastic) methods which depend 
much less on the dimensions of the integrals.
However, such  
\textcolor{blue}{\bf Monte-Carlo methods} have the disadvantage that 
the accuracy of the result only increases like the square root of the number 
of "thrown dices". Nevertheless, at present they are the only methods to treat
high-dimensional integrals (like the discretized path integral).
\vspace{0.1cm}

\noindent
If, for example, one wants to evaluate the one-dimensional
integral
\be
I \E \int_0^1 dx \> f(x)
\ee
over an arbitrary function $ f(x) $ by "rolling the dices",
one simply generates $ M $ uniformly distributed random numbers
$ \> x_i \in [0,1] \> $ and takes the average
\be
I \> \simeq \> \frac{1}{M} \sum_{i=1}^M f(x_i) \EQ  \bar f \> .
\ee
The error made by this procedure can be estimated by
\be
\left ( \Delta I \right )^2 \E \frac{1}{M} \left [ \> \frac{1}{M}
\sum_{i=1}^M f^2(x_i) \> - \left ( \frac{1}{M} \sum_{i=1}^M f(x_i) \right )^2
\> \right ] \EQ \frac{1}{M} \left [ \> \overline{f^2} - (\bar f)^2 \> 
\right ] \> .
\ee
This is not very effective and a simple inprovement is obtained
by rolling the dice  more specifically, i.e by using
\textcolor{blue}{\bf ``importance sampling''}: Let
$ \> w(x) > 0 \> $ 
be a weight function which essentially mimicks the behaviour of the
function $f(x)$. Then we write
\be
I \E \int_0^1 dx \> \frac{f(x)}{w(x)} \, w(x)
\ee
and the transformation
\be
y(x) \E \int_0^x dz \> w(z) \hspace{0.5cm} \Rightarrow \> dy = w(x) dx
\ee
has the effect that in the integral
\be
I \E \int_0^{y(1)} dy \> \frac{f(x(y))}{w(x(y))} 
\ee
the integrand varies much less. If
\be
y(1) \E \int_0^1 dz \> w(z) \E 1
\ee
then we can calculate immediately
\be
I \simeq \frac{1}{M} \sum_{i=1}^M \frac{ f(x(y_i))}{w(x(y_i))}
\ee
where
 $ \> y_i \in [0,1] \> $ are uniformly distributed random numbers.
$ \> x_i = x(y_i) \> $ are now the integration points weighted by the function
 $ w(x) $, i.e. they are more concentrated at those points where
  $ \> w(x) \simeq f(x) \> $ is large. The
  disadvantage of this method is that an inversion
 $ x(y) $ is needed.

\noindent
There exist methods to generate weighted integration points 
$\> x_i \> $  {\it without} inversion. 
As the most prominent (and oldest) we now discuss the \textcolor{blue}{\bf Metropolis method}
although it is - as may be expected --  not the most efficient method anymore.
In general, one starts with an arbitrary point
$ x_0 $ in the region of integration
and generates a new one $ x_1 $ by a specific algorithm, from that a new one etc.
\be
x_0 \> \rightarrow \> x_1 \> \rightarrow \> x_2 \> \rightarrow \> \ldots
\ee
This is called  a "random walk'' through the integration region.
The Metropolis algorithm chooses a new point, for example by the scheme
\be
x_t \E x_n + \delta \cdot \left ( z_1 - 0.5 \right ) \> ,
\label{neuer Punkt}
\ee
where $\> \delta \> $ is a given interval and $ z_1 $ a uniformly
distributed random number between $ 0 $ and $ 1 $. Then one 
forms the ratio
\be
r \E \frac{w(x_t)}{w(x_n)}
\ee
and accepts the new point 
 $ x_t $ according to the following criteria
\bit
\item[a)] if $ r > 1 $ then one sets $ x_{n+1} = x_t $ ,
\item[b)] if $ r < 1 $ then  $ x_t $ is accepted with probability $ r $ ,
otherwise $ x_{n+1} = x_n $ remains. In practice, this is done by generating a
second, uniformly distributed random number 
 $ z_2 \in [0,1] $ and by setting  $ x_{n+1} = x_t $, if $ z_2 < r \> $,  but 
 $\>  x_{n+1} = x_n \> $ if $ z_2 > r $.
\eit

\vspace{0.6cm}
\noindent
{\bf Assertion :} If  $ \> n \> $ becomes large, then the 
distribution of points generated in such a way approaches 
more and more
accurately the one weighted by  $ w(x)$ .

\vspace{0.5cm}
\noindent
{\bf Proofs :} Consider a large collection of "random walks" with density
 $ N_n(x) $ in the  $n^{\rm th}$ step. At the $(n+1)^{\rm th}$ step the net effect
is
\be
\Delta N_n(x,y) \E N_n(x) \, P(x \to y) - N_n(y) \, P(y \to x) \> ,
\ee
where $ \> P(x \to y) \> $ is the transition probability from  $ x $ to
$ y $. If
\be
\frac{N_n(x)}{N_n(y)} \E \frac{P(x \to y)}{P(y \to x)} \> = 
\frac{N(x)}{N(y)}
\ee
then there will an equilibrium. In the Metropolis algorithm
the probability that a move is made from  $ x $ to $ y $ is given by 
\be
P(x \to y) \E T(x \to y) \, A(x \to y)
\ee
where  $ \, A(x \to y) \, $ is the probability of acceptance and
$ \, T(x \to y) \, $ 
the probabalility that a move leads from 
$ x $ to $ y $. It is crucial that in the prescription of  
Eq. (\ref{neuer Punkt}) to generate the point  $ x_t $ 
the probability to move from $ y $ to $ x $ is equally large
("{\bf detailed balance}''):
\be
T(x \to y) \E T(y \to x ) \hspace{0.5cm} \Rightarrow \> 
\frac{N_n(x)}{N_n(y)} \E \frac{A(y \to x)}{A(x \to y)} \> .
\ee
We now distinguish two cases:
\bit
\item[a)] $ w(x) > w(y)$ ; then $ A(y \to x) = 1$ since  $ r > 1 $ and
$ A(x \to y) = r = w(y)/w(x) $. From this it follows that
\be
\frac{N(x)}{N(y)} \E \frac{1}{w(y)/w(x)} \> = \frac{w(x)}{w(y)} \> .
\label{Gleichgew}
\ee
\item[b)] $  w(x) < w(y) $; then $ A(y \to x) = w(x)/w(y) $ and    
$ A(x \to y) = 1 $ since  $ r > 1 $. Again Eq. (\ref{Gleichgew}) follows.
\eit

\noindent
In both cases we therefore have the desired result that the
equilibrium distribution of the "random walks" generated by the
Metro\-polis-Algorithmus is proportional to the weight function $ w (x) $.

\vspace{1.3cm}

\renewcommand{\baselinestretch}{0.9}
\scriptsize
\refstepcounter{tief}
\noindent
\blau{\bf Detail \arabic{tief}:} {\bf Ground-State Energy of the Anharmonic Oscillator 
by Monte-Carlo Methods and
\vspace*{0.06cm}

\hspace*{0.8cm} FORTRAN Program}

\vspace{0.4cm}

\noindent
\begin{subequations}
From Eq. (\ref{E0}) we obtain for the potential $ \> V(x) = m \omega^2 \, x^2/2 
+ \lambda \, x^4 \> $
\be
E_0 \E \lim_{\latop{N \to \infty}{\epsilon \to 0}}
\> \left < \> \frac{1}{N} \sum_{i=1}^N \left [ \, x_i^2 + 3 \lambda x_i^4
\, \right ] \> \right > \> ,
\ee
with the weight function 
\bea
w \left (\fx\right ) \EA \frac{e^{-S_E(\fx)}}{\int dx_0 
\ldots dx_{N-1} \> e^{-S_E(\fx)}} \> \> \> , \> \> \fx \E \left (
x_0 \ldots x_{N-1} \right ) \\
S_E(\fx) \EA \sum_{i=1}^N \left [ \, \frac{(x_i-x_{i-1})^2}{2 \epsilon}
+ \frac{\epsilon}{2} x_i^2 + \epsilon \lambda x_i^4 \, \right ] \> .
\eea
It fulfills all requirements: $ \> w > 0 \>, \> \int dx_0
\ldots dx_{N-1} \> w(\fx) = 1 \> $.
According to the Metropolis algorithm one has to calculate
\be
r \E \frac{w(\fx_t)}{w(\fx)} \> = \exp \left [ - (S_E(\fx_t)
- S_E(\fx) \right ] \E e^{-\Delta S_E}   \> .                            
\ee
This change of action is only made at a single point of the "time lattice"; then 
one doesn't have to evaluate the full action but only 
\bea
\Delta S_E \EA S_E[ \ldots x_i^{\rm neu} \ldots ] - 
S_E[ \ldots x_i^{\rm alt} \ldots ] \E  \left ( x_i^{\rm neu} - 
x_i^{\rm alt}\right ) \non 
&& \cdot \Biggl \{ \> \left [
\frac{1}{\epsilon} + \frac{\epsilon}{2} + \epsilon \lambda \left (  
{x_i^{\rm neu}}^2 + {x_i^{\rm alt}}^2 \right ) \right ] \left (
x_i^{\rm neu} + x_i^{\rm alt} \right )  
- \frac {1}{\epsilon} \left (
x_{i+1}^{\rm alt} + x_{i-1}^{\rm alt} \right ) \> \Biggr \} ,
\eea
where  one has to take $ \> i = 0, \ldots N-1 \> , \> x_{-1} \EQ x_{N-1} \> , 
\> x_N = x_0 \> $.

\noindent
A simple FORTRAN program which performs this algorithm is given in the following. 
It only serves for illustration and is not really up to professional  rules
-- both for the statistical analysis as well for the random number
generator \footnote{Linear congruent generator: 
$ \> X_{j+1} = a X_j + b \> \> 
({\rm mod} \> c) \> , \>  z_j = X_j/c \> $ with 
$ \> a = 7^5, \, b = 0, \, c  = 2^{31} - 1 \> $. For a more
detailed discussion see, e.g. {\bf \{Num. Recipes\}}, ch. 7.}
-- even for this simple quantum mechanical example.

\vspace{0.3cm}

\color[rgb]{0.3,0,0.7}
\begin{verbatim}
C
C  Calculates the ground-state energy of the anharmonic oscillator
C  by the Metropolis method for the Euclidean path integral
C
C  Units : h bar = m = omega = 1
C
C  Parameter : ALA    = lambda (anharmonicity)
C              EPS    = time step
C              N      = number of (time) lattice points
C              NTH    = number of thermalisation sweeps
C              NHIT   = number of additional Monte-Carlo trials 
C                       at each lattice point
C              NSWEEP = total number of sweeps
C              NMESS  = number of sweeps after which a measurement is made
C              DELTA  = max. increase of x at each lattice point
C
      DOUBLE PRECISION DSEED
      DIMENSION X(-1:200)
      WRITE(*,*) 'Eingabe: ala,eps,n,nth,nhit,nsweep,nmess,delta'
      READ(5,*) ALA,EPS,N,NTH,NHIT,NSWEEP,NMESS,DELTA
      WRITE(6,1) ALA,EPS,N,NTH,NHIT,NSWEEP,NMESS,DELTA
1     FORMAT(/'   lambda = ',F6.2,2X,' EPS = ',F6.2,2X,'N = ',I3,2X,
     &    'NTH = ',I4,2X,'NHIT = ',I2// ' NSWEEP = ',I7,2X,'NMESS = ',
     &    I4,2X,'DELTA = ',F6.3/)
C
      DSEED = 3.72D2
      WRITE(6,2) DSEED
2     FORMAT(/'   SEED =',D15.8//)
C
C  auxiliary calculations
C
      HILF = 1./EPS
      HELF= HILF + 0.5*EPS
      HALF = ALA*EPS
      N1 = N - 1
C
C  setting of initial values
C
      NACC = 0
      SUM1 = 0.
      SUM2 = 0.
      DO 10 I = -1,N
10    X(I) = 0.5
C
C Sweep
C
      WRITE(6,5)
5     FORMAT(/'  sweep',5x,'energy',4x,'mean energy',5x,'error',
     &  5x,'acceptance'//)
      DO 20 I =1,NSWEEP
         DO 30 K = 0,N1
            DO 35 M = 1,NHIT
               Z = ZUFALL(DSEED)
               XT = X(K) + DELTA*(Z-0.5)
C
C  change of action
C
               DS = HALF*(XT*XT + X(K)*X(K))
               DS = (HELF + DS)*(XT+X(K))
               DS = DS - HILF*(X(K-1) + X(K+1))
               DS = (XT - X(K))*DS
C
C  Metropolis test
C
               IF(DS .GE. 0.) THEN
                  R = EXP(-DS)
                  Z = ZUFALL(DSEED)
                  IF(Z .GT. R) GO TO 35
               ENDIF
               X(K) = XT
               NACC = NACC + 1
35          CONTINUE
            IF(K .EQ. 0) X(N) = X(0) 
            IF(K .EQ. N1) X(-1) = X(N1)
30       CONTINUE
         IF(I .LE. NTH) GO TO 20
C
C  measurement
C
         II = I - NTH
         IC = II/NMESS
         IF(NMESS*IC .NE. II) GO TO 20
         E = 0.
         DO 25 K = 0,N1 
25       E = E + X(K)*X(K) + 3.*ALA*X(K)**4
         E = E/N
         SUM1 = SUM1 + E
         SUM2 = SUM2 + E*E
         EM = SUM1/IC
         EMM = SUM2/IC
         DE = SQRT((EMM-EM*EM)/IC)
         ACC = FLOAT(NACC)/(I*N*NHIT)
C
C  print out
C
         WRITE(6,3) II,E,EM,DE,ACC
3        FORMAT(I8,4(3X,F10.4))
20    CONTINUE
      STOP
      END

C++++++++++++++++++++++++ subprogram ZUFALL ++++++++++++++++++

       FUNCTION ZUFALL(DSEED)
C
C  generates uniformly distributed random numbers in the interval [0,1]
C
      DOUBLE PRECISION DSEED,A,C
      DATA A,C /16807.D0,2147483647.D0/
      DSEED = DMOD(A*DSEED,C)
      ZUFALL = DSEED/C
      RETURN
      END
\end{verbatim}

\color{black}
\vspace{0.3cm}

\noindent
With this program we produce a new $ x_i $-value at any point of the "time lattice''
and then calculate the change in the action. According to the 
Metropolis method the new value is then accepted or not. It is
crucial for the correct description of quantum fluctuations that
also configurations with a larger action as the foregoing ones
have to be accepted with a certain probability, otherwise
one would end up inevitably in the state of deepest Euclidean action (= energy),
i.e. in the classical ground state. Once one has gone through the lattice 
one has performed a "{\bf sweep}".
Since the random walk is not completely random (the new points generated
are more or less in the neighborhood of the old ones) one rolls the dice several times
at each point of the lattice (NHIT = 3 in our example),
before moving on and calculates
the ground-state energy after each ``sweep''. The $ \delta $-parameter
is adjusted so that approximately $ 50 \% $ of all trials are accepted;
with too small $ \delta $ more is accepted  but one remains too close to the
old configuration; at large  $ \delta $ one indeed has more independent
configurations, but most have such a large action
that they are not accepted. The time step $ \epsilon $ should be small
compared to the period of oscillation ($ 2 \pi / \omega = 2 \pi $ for the 
pure harmonic oscillator) and $ N $ should be so large that the first
excited state is sufficiently suppressed, that is,
$ \exp (- \hbar (E_1-E_0) N \epsilon) = \exp (-N \epsilon) \ll 1 $ for 
the pure harmonic oscillator.

\noindent
Initially all $x_i$-values are set arbitrarly to the fixed value
$ x_i = 0.5 $ -- this is called a "{\bf cold start}" (as the system 
in Euclidean time is equivalent to a statistical system at a given temperature,  
a thermodynamic nomenclature
is frequently used: At low temperatures all spins in a spin system 
point in the same direction). A "{\bf hot start}" would have the 
 $x_i$-values randomly distributed at the beginning. 
 Of course, the result of the calculation must be independent 
 from the initial preparations.
Therefore one disregards the first
NTH (= 1000 in our example)  "sweeps'' until "{\bf thermalisation}"
has been achieved and after 
NMESS (= 500 in the example) ``sweeps'' one then "measures" the ground state energy.
For example, for $ \lambda = 0 $ (i.e. the harmonic oscillator)
one obtains the following printout

\vspace{0.5cm}
\color[rgb]{0.3,0,0.7}

  lambda =   0.00\hspace{0.2cm}   EPS = 0.10  \hspace{0.2cm}N = 100  
\hspace{0.2cm} NTH =  1000  \hspace{0.2cm}NHIT =  3 \\

  NSWEEP =    110000  \hspace{0.2cm} NMESS =   500  \hspace{0.2cm}
 DELTA =  1.300\\

  SEED = 0.37200000D+03\\
\vspace{0.2cm}

\noindent
sweep \hspace{0.6cm}energy \hspace{0.3cm}mean energy \hspace{0.5cm}
    error \hspace{0.5cm} acceptance 
  
\vspace{0.2cm}

\hspace*{-4mm}500\hspace{9.2mm}0.5407\hspace{9mm}0.5407\hspace{9mm}0.0000
\hspace{8mm}0.5045\\
1000\hspace{9mm}0.3529\hspace{9mm}0.4468\hspace{9mm}0.0664\hspace{9mm}0.5039\\
1500\hspace{9mm}0.2990\hspace{9mm}0.3976\hspace{9mm}0.0598\hspace{9mm}0.5037\\
2000\hspace{9mm}0.2488\hspace{9mm}0.3604\hspace{9mm}0.0552\hspace{9mm}0.5037\\
2500\hspace{9mm}0.4176\hspace{9mm}0.3718\hspace{9mm}0.0454\hspace{9mm}0.5037\\
3000\hspace{9mm}0.4441\hspace{9mm}0.3839\hspace{9mm}0.0394\hspace{9mm}0.5033\\
3500\hspace{9mm}0.5483\hspace{9mm}0.4073\hspace{9mm}0.0401\hspace{9mm}0.5035\\
4000\hspace{9mm}0.4838\hspace{9mm}0.4169\hspace{9mm}0.0362\hspace{9mm}0.5033\\
4500\hspace{9mm}0.4182\hspace{9mm}0.4171\hspace{9mm}0.0322\hspace{9mm}0.5033\\
5000\hspace{9mm}0.5140\hspace{9mm}0.4267\hspace{9mm}0.0304\hspace{9mm}0.5032\\
5500\hspace{9mm}0.4527\hspace{9mm}0.4291\hspace{9mm}0.0277\hspace{9mm}0.5033\\
6000\hspace{9mm}0.1740\hspace{9mm}0.4078\hspace{9mm}0.0326\hspace{9mm}0.5031\\
6500\hspace{9mm}1.0234\hspace{9mm}0.4552\hspace{9mm}0.0545\hspace{9mm}0.5030\\
7000\hspace{9mm}0.3397\hspace{9mm}0.4469\hspace{9mm}0.0513\hspace{9mm}0.5031\\
7500\hspace{9mm}0.8015\hspace{9mm}0.4706\hspace{9mm}0.0530\hspace{9mm}0.5031\\
8000\hspace{9mm}0.4014\hspace{9mm}0.4663\hspace{9mm}0.0499\hspace{9mm}0.5030\\
8500\hspace{9mm}0.6237\hspace{9mm}0.4755\hspace{9mm}0.0478\hspace{9mm}0.5031\\
9000\hspace{9mm}0.3360\hspace{9mm}0.4678\hspace{9mm}0.0458\hspace{9mm}0.5030\\
9500\hspace{9mm}0.4740\hspace{9mm}0.4681\hspace{9mm}0.0434\hspace{9mm}0.5031\\
\hspace*{-1.5mm}10000\hspace{9mm}0.5682\hspace{9mm}0.4731\hspace{9mm}0.0415\hspace{9mm}0.5031\\

\color{black}
\vspace{0.3cm}

\noindent
i.e. after 10000 sweeps one obtains
\be
E_0^{\rm h.o.}  \E 0.473 \pm 0.042 \>,
\ee
in good agreemeent with the exact value $ E_0 = 0.5 $. As already mentioned 
the error estimate is much too naive and should take into account the
correlations which obviously still exist between the different configurations
(see the printout).

\end{subequations}
\renewcommand{\baselinestretch}{1.2}
\normalsize

\vspace{1cm}

\noindent
For the anharmonic case we obtain with the same stochastic parameters the
values collected in Table 1:

\begin{table}[htbp]
\vspace{0.1cm}
\bce
\begin{tabular}{|r|c|l|l|} \hline
~~$\lambda$~~ & ~perturbation theory~ & ~Monte-Carlo~ & ~exact~
\\ \hline
 0.      & 0.50           & 0.473 $\pm$ 0.042 & 0.500000    \\
 1.      & 1.25           & 0.809 $\pm$ 0.067 & 0.803771  \\ 
 10.     & 8.00           & 1.386 $\pm$ 0.122 & 1.504972 \\ \hline
\end{tabular}
\label{tab:1.8.1}
\ece
\vspace{0.1cm}
{\bf Tab.} 1 :  Ground-state energy of the anharmonic oscillator for several 
values of the anharmonicity  parameter\\
\hspace*{1.4cm} $ \lambda $. The column titled "perturbation theory"
gives the values obtained in 1$^{\rm st} $ order perturbation\\ 
\hspace*{1.4cm} theory (see Eq. (\ref{Stoerung})), the column "Monte-Carlo'' the results of the
stochastic calculation and\\ 
\hspace*{1.4cm} the last  column the exact values (see, e.g. Ref.  \cite{HsCh}).
\vspace{0.3cm}

\end{table}

As one can see the stochastic calculation of the ground-state energy
even works well for large anharmonicities  while 1$^{\rm st} $-order
perturbation theory fails completely as expected. There is no much help
when including higher terms of the perturbative series  since this 
is an asymptotic (i.e. in the mathematical sense divergent) series.


\vspace{0.5cm}

\subsection{\textcolor{blue}{Tunneling and Instanton Solutions}}
\label{sec1: Tunneln}

\begin{figure}[hbtp]
\bce
\vspace{-0.5cm}
\includegraphics[angle=0,scale=0.6]{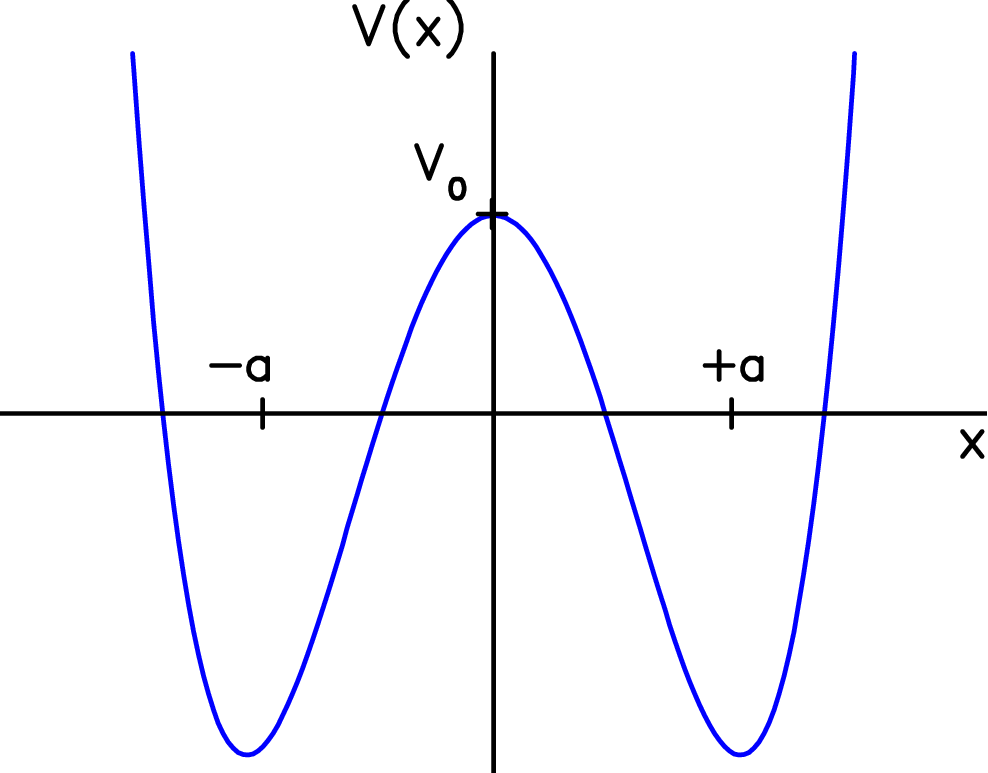}
\label{abb:1.8.1}
\ece
\bce
{\bf Fig. \arabic{abb}} : Double-well potential.
\ece
\end{figure}

Again we are considering the anharmonic oscillator with the potential
\be
V(x)  \E  V_0 + \frac{m}{2} \Omega^2 x^2 + \lambda x^4 \> ,
\ee
but this time for $ \Omega^2 < 0 $. In this "double-well potential"
(see Fig. \ref{abb:1.8.1}) new effects arise by the fact that there are 
two degenerate minima of the potential lying at
\be
x_{\pm} \E \pm \sqrt{- \frac{ m \Omega^2}{4 \lambda}} \deF \pm \, a
\label{def omega}
\ee
For convenience we choose $ V_0 $ such that 
$\, V(x_{\pm}) = V_0 - m^2 \Omega^4/(16 \lambda)= 0 \, $ --
this we can do as  a constant term in 
the potential does not have any influence on the dynamics of the system but only
fixes the scale of the energy. By this choice the potential becomes
\be
V(x)  \E  - \frac{m \Omega^2}{4 a^2} \, \left ( x^2 - a^2 \right )^2 \deF 
\frac{m \omega^2}{8 a^2} \, \left ( x - a \right )^2 \, \left ( x + 
a \right )^2 \> , \hspace{0.5cm} \omega^2 = - 2 \, \Omega^2 \> .
\label{doppelmulden pot}
\ee
\vspace{0.3cm}

\refstepcounter{abb}

\noindent
In the vicinity of $ x = \pm \, a $ we have approximately an
oscillator potential
\be
V(x) \> \stackrel{x \to \pm a}{\longrightarrow} \> \frac{m \omega^2}{2} \, 
\left ( x \mp a \right )^2
\ee
and therefore we obtain for the case of far-away minima separated by 
high barriers approximately as ground-state energy
\be
E_0 \> \simeq \>  \frac{1}{2} \, \hbar \omega\> .
\ee
The corresponding ground-state wave functions are superpositions of the 
ground-state wave functions 
of the oscillators localized at  $x = \pm a $ 
\be
\psi_0 (x) \> \simeq \> \frac{1}{\sqrt{2}} \, \Bigl [ \> \phi_0(x-a) \pm  
\phi_0(x+a) \> 
\Bigr ] \> .
\label{Grund0 pm}
\ee
This is because in quantum mechanics the wave functions must 
exhibit the symmetry of the Hamiltonian (there is no 
"spontaneous symmetry breaking"). Here this means that one can
classify the wave functions according to their behaviour under parity transformations
(the Hamiltonian is invariant under $x \to -x$); obviously 
Eq. (\ref{Grund0 pm}) describes wave functions of positive and negative parity
which are degenerate in energy.

Since for $ \lambda \ne 0 $ the barrier at $ x = 0 $ 
\be
V(0) \E \frac{m \omega^2}{8} a^2 \E \frac{m^2 \Omega^4}{16 \lambda}
\ee
is not infinitely high (and thick) the particle can "tunnel" through the
wall and lift the degenarcy: We expect an energy splitting between 
the state with positive parity (which is the lowest-lying state) and 
the state of negative parity.
This energy splitting can be calculated for small anharmonicities 
by semi-classical methods -- best in the path-integral formalism
(and not in the usual WKB approximation for Schr\"odinger's equation)
because this can be extended with not too much effort to systems
with many, even infinite many degrees of freedom.

Therefore we consider the matrix element of the 
\textcolor{blue}{\bf Euclidean} 
time-evolution operator which brings the particle from the 
minimum at  $ x = - a $ to the minimum at $ x = + a $ 
\bea
U \left ( a, \frac{\beta}{2}; -a, - \frac{\beta}{2} \right ) & \EQ & 
\left < +a \, | \, e^{-\beta \hat H/\hbar} \, | - a \right > 
\E \sum_n \psi_n(a) 
\psi_n(-a) \, e^{- \beta E_n /\hbar} \non
\EA \int\limits_{x(-\beta/2)=-a}^{x(\beta/2)=+a} {\cal D} x 
\> e^{-S_E[x]/\hbar}
\label{Pfad}\\
&\stackrel{\beta \to \infty}{\longrightarrow}& \psi_0(a) \psi_0(-a) 
e^{-\beta E_0/\hbar} 
+ \psi_1(a) \psi_1(-a) e^{-\beta E_1/\hbar} + \ldots
\label{spektral eukl}
\eea 
Thus we can determine both the ground-state energy as well as 
the energy of the first excited state from the behaviour of the 
time-evolution operators at large Euclidean times. Moreover,
the sign of $  \psi_0(a) \psi_0(-a)
= \pm | \psi_0(a) |^2 $ tells us which parity state is the lowest.
\vspace{0.3cm}

\noindent
We will proceed in several steps:
\vspace{0.5cm}

\bdes
\item{\bf (i) Determine the classical path}

\begin{figure}[hbtp]
\bce
\vspace{-0.5cm}
\includegraphics[angle=0,scale=0.6]{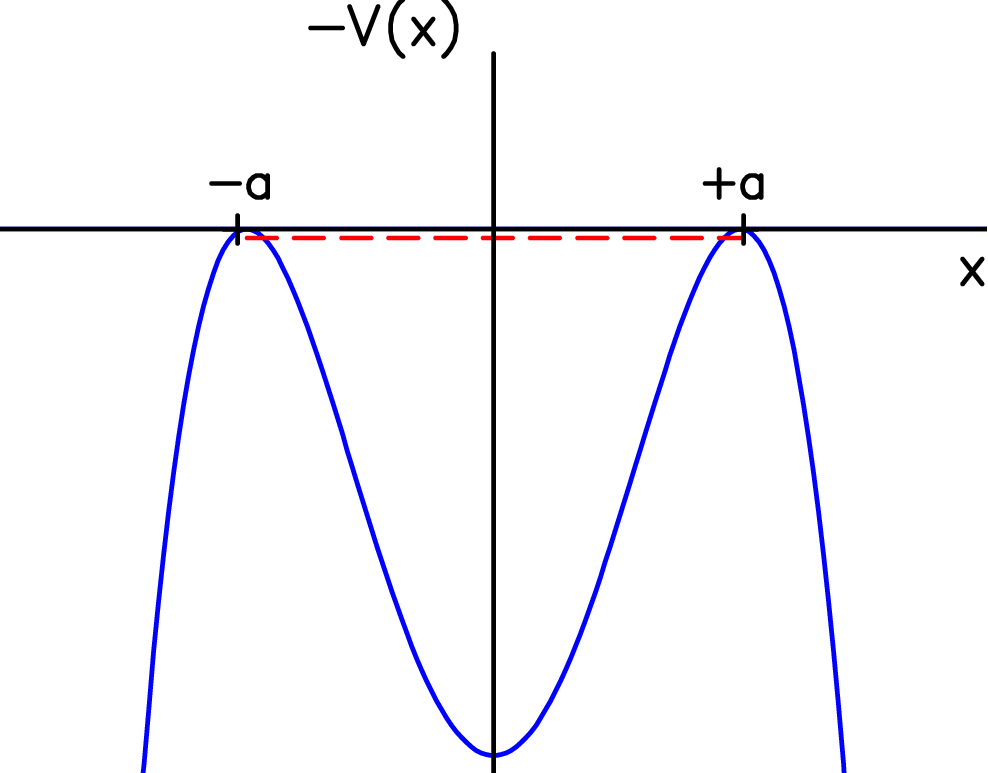}
\label{abb:1.8.2}
\ece
\bce
{\bf Fig. \arabic{abb}} : Motion in the inverted  potential
for Euclidean "energy'' $ = 0$.
\ece
\end{figure}

In the semi-classical approximation we first determine the path
which makes the action stationary. Since we are working in Euclidean
time this is a path in the \textcolor{blue}{\bf inverted} potential 
(see Fig. \ref{abb:1.8.2}):
\be
\frac{\delta S_E[x]}{\delta x(\tau)}  \E  0 \> \> \Longrightarrow \> \> 
m \frac{d^2 x_{\rm cl}}{d\tau^2} - V'\left ( x_{\rm cl} \right ) \E 0 \> ,
\hspace{0.3cm}  {\rm with } \> \> \> x_{\rm cl}(-\beta/2) \E - a \> , 
\> x_{\rm cl}(\beta/2) \E + a \> .
\label{klass Bewegungsgl}
\ee

\refstepcounter{abb}


\noindent
From the path-integral representation (\ref {Pfad}) of the Euclidean
 time-evolution operator it follows that those paths determine the
ground-state energy in the limit $ \beta \to \infty $ which have 
an Euclidean action as low as possible. Since $ V(x) \ge $ 0 we have
\be 
S_E [x] \E  \int_{-\beta/2}^{+\beta/2} d\tau \left [ \frac{m}{2} \left (
\frac{dx}{d \tau} \right )^2 + V(x) \right ] \> \ge \>  
\int_{-\beta/2}^{+\beta/2} d\tau \frac{m}{2} \left (
\frac{dx}{d \tau} \right )^2 \>,
\ee
and therefore all paths whose speed is finite  at $\tau = \pm \beta/2 $
will give an infinite Euclidean action 
 for  $ \beta \to \infty $. Only that path which reaches the points 
$ \pm a $ with zero velocity at $ \tau \to \infty $ 
is excluded from this, i.e. a trajectory with Euclidean "energy"  $ = 0 $ 
where the particle "crawls" up and down the hills in the inverted 
potential. This path can be determined easily by integrating
the classical equation of motion (\ref{klass Bewegungsgl}):
The Euclidean "energy" is
\be
\frac{m}{2}  \left (
\frac{dx_1}{d \tau} \right )^2 - V(x_1)  \> \stackrel{!}{=}\> 0
\label{eukl Energie}
\ee
and another integration by separating the variables gives
\be
\tau \E \tau_0 + \int_{x_1(\tau_0)}^{x_1(\tau)} dx \> 
\sqrt{ \frac{m}{2 V(x)}} \> .
\label{class sep}
\ee
For the double-well potential (\ref{doppelmulden pot}) it is possible to do
 the integral analytically and also to solve the implicit equation for this 
 special path. One obtains (\purpur{\bf Problem \ref{Null Mode} a)})
\be
x_1(\tau) \EQ x_{\rm cl}(\tau - \tau_0)  \E a \tanh \left [ \, \frac{\omega}{2} \left ( \tau - \tau_0 
\right ) \, \right ] \> ,
\label{kink}
\ee
where the time $ \tau_0 $ has been fixed (arbitrarily) by demanding 
that the origin of the potential is traversed at this time.

This solution is called an \textcolor{blue}{\bf instanton} (or also a 
\blau{\bf kink}) because the particle spends most of its time 
to crawl up and down the potential hills but then relatively fast
-- localized in time around $\tau = \tau_0$ --  crosses over from the 
negative $x$-axis  to the positive $x$-axis. The solution
$ -x_1(\tau) $ which connects the  $(x =+a)$-minimum of the potential 
with the one  at $x = - a $ is called an \textcolor{blue}{\bf anti-instanton}.
\vspace{0.2cm}

\refstepcounter{abb}
\begin{figure}[hbtp]
\bce
\vspace{-0.5cm}
\includegraphics[angle=0,scale=0.6]{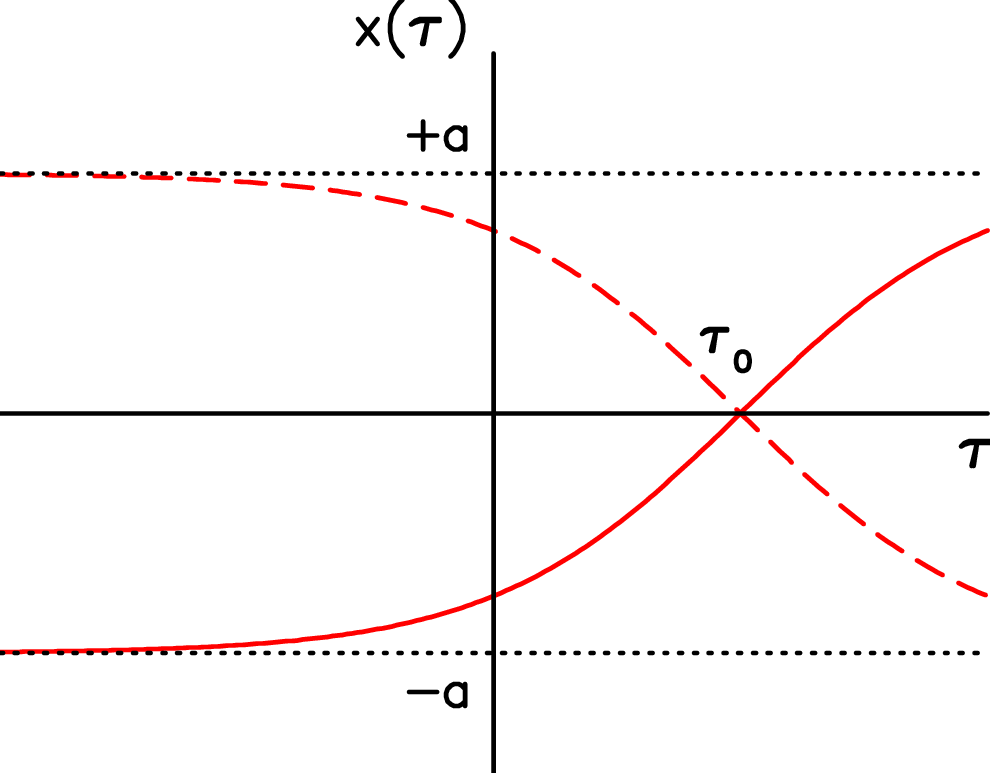}
\label{abb:1.8.3}
\ece
\bce
{\bf Fig. \arabic{abb}} : The instanton solution  and the
anti-instanton solution (dashed line).
\ece
\end{figure}

\noindent
The action of this instanton solution is most easily calculated 
by using the "energy" conservation (\ref{eukl Energie}) 
\bea
S_1 \EA \int_{-\beta/2}^{\beta/2} d\tau \> m \left ( 
\frac{dx_1(\tau)}{d\tau} \right )^2 \> 
\stackrel{\beta \to \infty}{\longrightarrow} \> 
 m \int_{-a}^{+a} dx_1 \> 
\frac{dx_1(\tau)}{d\tau} \non
\EA  \int_{-a}^{+a} dx_1 \> \sqrt{2 m V(x_1)} \E \int_{-a}^{+a} dx_1 \> 
\frac{m \omega}{2 a} \, \left ( a^2 - x_1^2 \right ) \E \frac{2}{3} m 
\omega a^2
\label{S1}
\eea
and, indeed, a finite action is found.
%
%
\vspace*{0.4cm}

\item{\bf (ii) Evaluate the quadratic fluctuations and treat the zero mode}

Although this (one-)instanton solution is not sufficient to obtain the energy splitting
in the double-well potential, we first concentrate on calculating
the quadratic fluctuations around this solution in the
euclidean path integral. If we set
\be
x(\tau) \E x_{\rm cl}(\tau - \tau_0) + \eta(\tau ,\tau_0)
\label{class+fluct}
\ee
the functional Taylor expansion of the action gives
\be
S_E[x_1+\eta] \E S_n + \frac{1}{2} \int_{-\beta/2}^{+\beta/2}
d\tau \, d\tau' \> \frac{\delta^2 S}{\delta x(\tau) \delta x(\tau')} 
\Biggr |_{x = x_{\rm cl}} \, \eta(\tau) \, \eta(\tau') + \ldots 
\ee
since the first functional derivative vanishes. Here we have
\be 
\frac{\delta^2 S}{\delta x(\tau) \delta x(\tau')} \Biggr |_{x = x_{\rm cl}}  \E 
\lsp - m
\frac{d^2 }{d \tau^2}  + \, V''\lrp x_{\rm cl}(\tau) \rrp \, \rsp \, 
\delta(\tau - \tau') \deF {\cal O}_{V''} \, \delta(\tau - \tau') \> .
\ee  
As usual quadratic actions can be integrated out exactly in the path integral:
Expand the fluctuations in a complete, orthonormal system 
(preferently in eigenfunctions of the operator $ {\cal O}_{V''} $)
\be
\eta(\tau) \E \sum_{n=0}\, c_n \, y_n(\tau)\> , \hspace{0.5cm}
{\cal O}_{V''}\,  y_n(\tau) \E e_n \, y_i(\tau) \> , \>  \la y_m \big | y_n \ra \Def 
\int_{-\beta/2}^{+\beta/2} d\tau \> 
y_m(\tau) \, y_n(\tau) \E \delta_{m n}
\ee
which turns the functional integral into a product of Gaussian integrals
over the expansion coefficients $ c_n. $ 
If the eigenvalues are positive  $ e_n > 0$, then the result of the 
integrations simply is
\be 
U \lrp a,\frac{\beta}{2}; -a, - \frac{\beta}{2} \rrp  \simeq  {\cal N} \> e^{-S_1/\hbar} \, 
\prod_{n=0} \lsp \int_{-\infty}^{+\infty} dc_n \, \exp \lrp - \frac{1}{2 \hbar} e_n c_n^2 \rrp  
\rsp \deF F^{\rm d.w.}_1(\beta) \> e^{-S_1/\hbar} 
\ee
where 
\be
F^{\rm d.w.}_1(\beta) \E {\cal N} \, \prod_{n=0} \lrp \frac{2 \pi \hbar}{e_n} \rrp^{1/2} \E 
\sqrt{ \frac{m}{ 2 \pi \hbar \, \prod_n e_n}} \E \sqrt{ \frac{m}{ 2 \pi \hbar \,
\left [ \, \fdet \, {\cal O}_{V''} \, \right ]}}
\ee
is the one-instanton prefactor of the semi-classical path integral for the double well (d.w.)
potential due to the fluctuations around the classical path. Note that from our previous
derivation of the semi-classical approximation (see Eqs. \eqref{HO propagator}
and \eqref{prefactor} transformed to Euclidean times) we know the correct normalization factor
to be
\be 
{\cal N} \E \sqrt{\frac{m}{2 \pi \hbar}} \,  \prod_{n=0} \lrp 2 \pi \hbar \rrp^{-1/2} \> .
\label{normalization}
\ee
However, as should be investigated in more detail in \purpur{\bf Problem \ref{Null Mode} b)}, there 
is an eigenfunction of $ {\cal O}_{V''} $ with eigenvalue $ 0 $
\be 
\lsp - m\frac{d^2 }{d \tau^2}  + \, V''\left (x_1(\tau) \right ) \rsp  \, 
 \frac{\partial x_1(\tau-\tau_0)}{\partial \tau_0} \E 
\lsp - m\frac{d^2 }{d \tau^2}  + m \omega^2 - \frac{3}{2} 
\frac{m \omega^2}{\cosh^2(\omega (\tau - \tau_0)/2)} \rsp \, 
\frac{\partial x_1(\tau-\tau_0)}{\partial \tau_0} \E 0 \> ,
\label{zero Mod}
\ee
a {\bf zero mode} or {\bf Goldstone mode}! 
This zero mode is a consequence of the translational invariance of the solution in the limit
$ \beta \to \infty $: The position of the center of the instanton is then irrelevant and it doesn't ``cost'' 
to shift the center by an infinitesimal amount. Therefore, the zero-mode
wavefunction must be proportional to the derivative of the classical (instanton) solution
\be
y_0(\tau-\tau_0) \E A_0 \, \lim_{\epsilon \to 0} \, 
\frac{x_{\rm cl}(\tau-(\tau_0 + \epsilon) - x_{\rm cl}(\tau - \tau_0 )}{\epsilon}   
 \E A_0 \frac{\partial x_{\rm cl}(\tau - \tau_0)}{\partial \tau_0} \EQ - A_0 \, 
\dot x_{\rm cl}(\tau - \tau_0)
\label{0 mode}
\ee
(see \purpur{\bf Problem \ref{Null Mode} b)}).
%
%
Hence, the integration over $ c_0 $ is undamped and the euclidean path integral diverges. This problem
can be remedied by the {\bf method of collective co-ordinates}: The integration over $c_0$ is replaced by 
an integration over "the collective co-ordinate" $\tau_0$. This is achieved by applying the Faddeev-Popov trick 
(as in {\bf chapter} \ref{sec1: Potstreu}),
i.e. by multiplying the path integral by the following "one"
\be 
1 \E \int_{-\beta/2}^{+\beta/2} d\tau_0 \> \delta \lsp f(\tau_0) \rsp \, \Bigg | \frac{\partial f}{\partial \tau_0} \Bigg |
\> .
\label{FP null mode}
\ee
This means that one first fixes $\tau_0$ by requiring that an arbitrary function of it should vanish (due to the $\delta$-function) and then\,  integrates over all possible $\tau_0$. 
Choosing 
\be 
f(\tau_0) \E \la \eta \, \big | y_0 \ra \E A_0 \int_{-\beta/2}^{+\beta/2}  d\tau \> \Big [ \,  x(\tau) - x_{\rm cl}(\tau,\tau_0) \, \Big ]
\, \frac{\partial x_{\rm cl}(\tau,\tau_0)}{\partial \tau_0}
\label{FP for zero mode}
\ee
demands the fluctuations to be orthogonal to the zero mode. 
If this is inserted one finds that the expansion coefficient $ \> c_0 \> $ for the zero mode is set
to zero and replaced by the integration over the "center" $ \> \tau_0 \> $ of the instanton.
More precisely, 
one obtains for the one-instanton prefactor of the semi-classical approximation to the path integral for the double-well potential
\be 
F^{\rm d.w.}_1 (\beta) \E  
{\cal N} \> \int_{-\beta/2}^{+\beta/2} d\tau_0 \> \frac{1}{A_0} \,  \prod_{n=1} \lrp
\frac{2 \pi \hbar}{e_n} \rrp^{1/2} \EQ \beta \, \sqrt{\frac{m}{2 \pi \hbar}} \, \sqrt{\frac{1}{ 
2 \pi \hbar \, A_0^2 \, \fdet' \, {\cal O}_{V''} }} \> .
\ee
where $ \fdet' $ indicates the functional determinant with the zero mode removed.
Thus the usual factor for the quadratic fluctuations is replaced by
\be 
\frac{1}{\sqrt{\fdet \, {\cal O}_{\rm V''}}} \> \longrightarrow \> 
\sqrt{\frac{1}{ 2 \pi \hbar \, A_0^2 \, \fdet' \, {\cal O}_{V''} }} \, \int_{-\beta/2}^{+\beta/2} d \tau_0\> ,
\ee
where 
\be 
\frac{1}{A_0^2} \E  \la \frac{\partial x_{\rm cl}}{\partial \tau_0} \, \Big | \, \frac{\partial x_{\rm cl}}{\partial \tau_0} \ra \E \frac{S_1}{m}
\label{A0}
\ee
is from the normalization of the zero mode (see \purpur{\bf Problem \ref{Null Mode} b)} ).
The factor $ \> (2 \pi \hbar)^{-1/2} \> $ is a left-over from the $\> (n=0)$-term in the product of Eq. \eqref{normalization} for the normalization factor $ \> {\cal N} $. 
Note that we are only interested in the limit $ \> \beta \to \infty \> $.
\vspace*{0.4cm}

\renewcommand{\baselinestretch}{0.9}
\scriptsize
\refstepcounter{tief}
\noindent
\blau{\bf Detail \arabic{tief}:} {\bf Path integral with Zero Mode}\\
\vspace{0.1cm}

\noindent
As the original path-integral variable $ \> x(\tau) = x_{\rm cl}(\tau,\tau_0) + \eta(\tau,\tau_0) \> $ does not depend on the collective variable$ \> \tau_0 \> $ we find from Eqs. \eqref{FP null mode}, \eqref{FP for zero mode}
\begin{subequations}
\be 
\frac{\partial f(\tau_0)}{\partial \tau_0} \E - \frac{1}{A_0} \la y_0 \big | y_0 \ra + \la \eta \bigg | 
\frac{\partial y_0}{\partial \tau_0}\ra \> .
\ee
Thus with $ \> \la y_0 \big | y_0 \ra = 1 \> $ and the usual
integral representation of the $\delta$-function
the one-instanton prefactor of the semi-classical path integral for the double well potential  becomes (suppressing the subscript 
of the fluctuation operator  $ {\cal O} $ )
\be 
F^{\rm d.w.}_1(\beta) \E \int_{-\beta/2}^{+\beta/2} d\tau_0 \int \frac{d\alpha}{2 \pi \hbar} 
\int_{\eta(-\beta/2)=0}^{\eta(+\beta/2)=0} {\cal D}\eta \> 
\exp \lsp - \frac{1}{2\hbar}\la \eta \big | {\cal O} \big |\eta \ra + \frac{i}{\hbar} \alpha \la \eta \big | y_0 \ra \rsp \,  
\Big | - \frac{1}{A_0} + \la \eta \bigg | \frac{\partial y_0}{\partial \tau_0} \ra \, \Big | \> .
\ee 
The exchange $ \> \eta \to -\eta \> $ and $ \> \alpha \to - \alpha \> $ leaves the exponent invariant but
changes the sign of the last term (linear in $ \> \eta \> $) which therefore vanishes. 
As $ \> {\cal O} \> $ has a zero mode one cannot perform the
Gaussian functional integral over $ \> \eta \> $ immediately. Following \meingruen{\bf Negele \& Orland}
p. 214 - 217 we regularize the operator, i.e. we replace
\be 
{\cal O} \> \longrightarrow \> {\cal O}_{\epsilon} \Def \epsilon  | \, y_0\!\left.\ra\!\la\right.\!y_0  | + {\cal O}_{\perp}
\ee 
where $ \epsilon > 0 $ at the end is set to zero. $ \> {\cal O}_{\perp} \> $ is the projection of $ {\cal O}$  on the subspace orthogonal to the zero mode $ \> y_0 $ . Thus
\be 
{\cal O}_{\epsilon} \E 
\begin{pmatrix} \epsilon & 0 \\
                    0    & {\cal O}_{\perp}
\end{pmatrix} \> , \qquad  \fdet \> {\cal O}_{\epsilon} \E \epsilon \, \fdet \> {\cal O}_{\perp} \> .
\ee
With the proper normalization of the fluctuation path integral we then  obtain 
\be 
F^{\rm d.w.}_1(\beta) \E \int_{-\beta/2}^{+\beta/2} d\tau_0 \int \frac{d\alpha}{2 \pi \hbar} 
\, \left ( \frac{m}{2 \pi \hbar A_0^2 \, \fdet {\cal O}_{\epsilon}} \right )^{1/2} \, \exp \lsp - \frac{\alpha^2}{2 \hbar} \la y_0 \big |
{\cal O}_{\epsilon}^{-1} \big | y_0 \ra \rsp \> .
\ee
Since $ \> y_0 \> $ is the zero mode, one has 
$ \> {\cal O}_{\epsilon}^{-1} y_0 = \frac{1}{\epsilon} \, y_0 \> $ and after performing the Gaussian 
$\alpha$-integration and performing the limit $ \> \epsilon \to 0 \> $ one arrives at
\be 
F^{\rm d.w.}_1(\beta) \E \lim_{\epsilon \to 0} \int_{-\beta/2}^{+\beta/2} d\tau_0  \, 
\sqrt{\frac{m}{ 2 \pi \hbar}}
\, \left ( 2 \pi \hbar A_0^2 \epsilon \, \fdet \, {\cal O}_{\perp} \right )^{-1/2} \sqrt{\epsilon} 
\EQ \beta \, \sqrt{\frac{m}{2 \pi \hbar } \, \frac{1}{2 \pi \hbar A_0^2 \fdet' {\cal O}}} \> .
\label{F dw 0 mode}
\ee
More generally:
If there is a continous symmetry for which the classical equations of motion 
( $ \> \delta S/\delta x(t) \big |_{\rm cl} = 0 \> $ ) 
have degenerate solutions
$ \> x_{\rm cl}(t, \tau_i) \> , i = 1 \ldots n \> $, then there exist $ n $ Goldstone modes 
$ \> \partial x_{\rm cl}(t,\vec{\tau}/(\partial \tau_i) \> $ which are annihilated by 
$ \> \delta^2 S/\delta x(t) \delta x(t') \big |_{\rm cl} \> $
and Eq. \eqref{F dw 0 mode} generalizes to
\be 
F_n(\beta) \E \sqrt{\frac{m}{2 \pi \hbar \, \fdet {\cal O}_{\perp}}} \, \prod_{i=1}^n \lrp \int
\frac{d \tau_i}{\sqrt{2 \pi \hbar}} \rrp \, \lsp \det_{i,j} \lrp \la \frac{\partial x_{\rm cl}}{\partial \tau_i}
\Big | \frac{\partial x_{\rm cl}}{\partial \tau_j} \ra \rrp \rsp^{1/2}
\label{pre zero gener}
\ee 
(Problem 4.6 in \meingruen{\bf Negele \& Orland}, p. 228 - 230).
\end{subequations}
\renewcommand{\baselinestretch}{1.1}
\normalsize
\vspace*{0.4cm}

\item{\bf (iii) Calculate the functional determinant without the zero mode}

How do we calculate the reduced determinant $ \> \fdet' {\cal O}_{V''} \> $ ? The straight-forward way would be to determine
the other eigenvalues of the operator $ {\cal O}_{V''} $, i.e. solving the Schr\"odinger-like 
equation (after evaluating $ \> V''(x_1(\tau)) \> $)
\be
\lcp -m \frac{d^2}{d\tau^2} + m \omega^2 - \frac{3 m \omega^2}{2} \frac{1}{\cosh^2 [ 
\omega(\tau-\tau_0)/2 ] }  \rcp 
y_n(\tau-\tau_0) \E e_n \, y_n(\tau-\tau_0)
\label{eigen}
\ee
with {\bf boundary conditions} $ \> y_n(-\beta/2) = y_n(+\beta/2) = 0 \> $ as the fluctuations vanish 
at the boundaries. Then
\be 
\fdet'  {\cal O}_{V''}  \E \prod_{n=1} e_n \EQ \frac{ \fdet  {\cal O}_{V''} }{e_0}
\ee
This is possible since these are potentials of the 
Rosen-Morse type for which
analytic solutions are available (see \meingruen{\bf Kleinert}, ch. 17.3 ). However, this is very cumbersome, in part because one also has to consider the continous spectrum. Fortunately, there is a simpler method \footnote{This derivation follows \meingruen{\bf Kleinert}, ch. 17.5.} based on the {\bf Gel'fand-Yaglom formula}~:

We have to find the solution $ f_{\rm GY}(\tau) $ of the differential equation \eqref{eigen} 
with the r.h.s. vanishing (i.e. for eigenvalue zero) subject to the {\bf initial condition} 
$ \> f_{\rm GY}(-\beta/2) = 0 \> , \dot f_{\rm GY}(-\beta/2) = 1 \> $
and the determinant is then given by 
\be
\fdet  \> {\cal O}_{V''} \E f_{\rm GY}(\beta/2) \> .
\ee
We already know one solution for the 
full Gel'fand-Yaglom equation: It is (proportional to) the zero-mode solution 
\be 
f^{(1)}(\tau) \EQ y_0(\tau) \E  - A_0 \dot x_{\rm cl}(\tau) \E - A_0 a \, \frac{\omega/2}{\cosh^2 [ \omega (\tau -\tau_0)/2 ] }
\> \stackrel{\tau \to \pm \infty}{\longrightarrow} \> - 2 \omega a A_0 \, e^{-\omega |\tau|} \> ,
\ee
which, however, does not fulfill the initial conditions.
To achieve that we need a second independent solution $ f^{(2)}(\tau) $ which behaves at infinity like
\be 
f^{(2)}(\tau) \> \stackrel{\tau \to \pm \infty}{\longrightarrow} \> \mp  2 \omega a A_0 \, e^{\omega |\tau|} \> 
\ee
(it should have asymptotic behaviour and parity opposite to the first solution). Then we may form the linear combination
\be 
f_{\rm GY}(\tau) \E C \,  \lsp f^{(1)} \left (-\beta/2 \right )  \, f^{(2)}(\tau) - f^{(2)} \left (- \beta/2 \right )  \, f^{(1)}(\tau) \rsp
\ee
which obviously vanishes at $ \> \tau = - \beta/2 \> $. From the requirement that 
$ \> \dot f_{\rm GY}(-\beta/2) \stackrel{!}{=} 1 \> $ one determines the constant as
\be 
C \E \frac{1}{W(f^{(1)},f^{(2)})} \E \frac{1}{8 \omega^3 a^2 A_0^2}
\ee
where $ \>  W(f^{(1)},f^{(2)}) \equiv  f^{(1)} \dot f^{(2)} - \dot f^{(1)} f^{(2)} \> $ is the Wronskian of the two solutions evaluated at $ \> \tau = - \beta/2 \> $. As is well known the Wronskian of a linear 
second-order differential equation 
is a constant (\purpur{\bf Problem \ref{Null Mode} c)}).
Thus 
\be 
f_{\rm GY} \left (+\beta/2 \right) \E C \, \lsp f^{(1)} \left (-\beta/2 \right )  \,
 f^{(2)} \left (+\beta/2 \right) - f^{(2)} \left (-\beta/2 \right )  \, f^{(1)}
  \left (+\beta/2 \right ) \rsp \E C \, \cdot \, 8 \omega^2 a^2 A_0^2 \E \frac{1}{\omega} \> .
\ee
It remains to evaluate the  would-be zero eigenvalue $ \, e_0 \, $ at finite $ \, \beta \, $ 
to obtain
\be 
\fdet' {\cal O}_{V''} \E \frac{f_{\rm GY}(+\beta/2)}{e_0} \> .
\ee
At  finite $\> \beta  \> $ there is no exactly vanishing eigenvalue -- instead we expect that
$ e_0 \sim \exp (- \omega \beta) $. 
Indeed (as to be shown in \purpur{\bf Problem \ref{Null Mode} d)})
the leading behaviour of the eigenvalue $ e_0 $ at large $ \> \beta \> $ can be found in 
perturbation theory as
\be
e_0 \E 24 \, m \omega^2 \, e^{-\omega \beta} \> .
\ee
Thus
\be 
F^{\rm d.w.}_1(\beta) \> \stackrel{\beta \to \infty}{\longrightarrow} \> \beta \, 
\sqrt{\frac{m}{2 \pi \hbar}} \, e^{- \omega \beta/2} \, \sqrt{\frac{24 \omega^3 S_1}{2 \pi \hbar}}
\deF \beta \, \sqrt{\frac{m \omega}{ \pi \hbar}} \, e^{- \omega \beta/2} \> \cdot \sqrt{K_1}
\label{vorfaktor 1 inst}
\ee
where
\be 
K_1 \E \frac{4 m a^2 \omega^3}{\pi \hbar}
\label{K1}
\ee
is the single-instanton correction factor to the result which would be obtained for individual harmonic oscillators, i.e.
if the inverse $ \cosh^2$- term would be neglected in the Gel'fand-Yaglom equation.
\vspace*{0.4cm}

%
%

\item{\bf (iv) Consider $n$ instantons}

However, this 1-instanton solution is not sufficient to 
calculate the energy splitting in the double-well potential;
one has to take into account addtional solutions with finite action.
These solutions (called $ x_n(\tau) )$   consist of  $ 3,5 \ldots $ instantons/anti-instantons
which connect the classical minimum at  $x = - a $ with the one
at $ x = + a $ (see Fig. 11).

\refstepcounter{abb}
\begin{figure}[hbtp]
\bce
\vspace{-0.5cm}
\includegraphics[angle=0,scale=0.6]{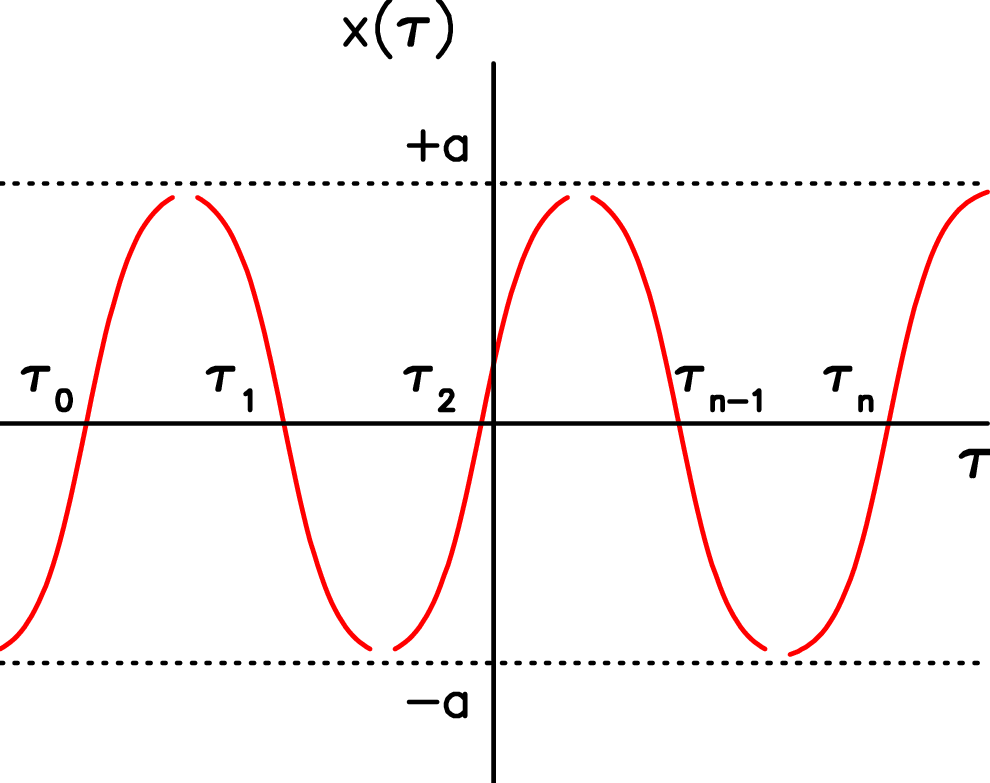}
\label{abb:1.8.4}
\ece
\bce 
{\bf Fig. \arabic{abb}} : Multi-instanton-anti-instanton solution.
\ece
\end{figure}

\vspace{0.2cm}

\noindent
Strictly speaking these are not exact solutions anymore since, of course, 
the superposition principle doesn't hold in the non-linear
equation of motion (\ref{klass Bewegungsgl}). However,  
$\delta S_E/\delta x \big |_{x=x_n} $ is very small if the mutual
 distance (in Euclidean time) fulfills  $\tau_k - \tau_{k-1} \gg 1/\omega $.
 This approximation is called the "{\bf dilute instanton gas}". Under these 
conditions it is also allowed to set
\be
S_n \simeq n \, S_1 \> \> , \> n \E 3,5 \ldots
\ee
for the corresponding actions. When we calculate the quadratic fluctuations 
we encounter $n$ zero modes which are treated in a completely 
analogous way as in the case of a single instanton: Again the method of collective co-ordinates
replaces the integration over the undamped modes by an integration over the 
centers $\tau_k , \, k = 1 \ldots n $ of the multi-instantons, under the condition
\be
-\frac{\beta}{2} \le \tau_1 \le \tau_2  \ldots \le \tau_{n-1} \le \tau_n 
\le \frac{\beta}{2} \> .
\ee
With Eqs. \eqref{pre zero gener}, \eqref{A0} this gives
\be 
\int dc_0^{(1)} \, dc_0^{(2)} \ldots dc_0^{(n)} \> \longrightarrow \> \int_{-\beta/2}^{+\beta/2} d\tau_n \, \int_{-\beta/2}^{\tau_n} d\tau_{n-1} \ldots \int_{-\beta/2}^{\tau_2} d\tau_1 \> \lrp 
\frac{1}{\sqrt{2 \pi \hbar} A_0} \rrp^n
\E \lrp \frac{\beta}{\sqrt{2 \pi \hbar} A_0} \rrp^n \, \frac{1}{n!}
\ee 
and therefore an $n$-instanton correction factor of  $ \> K_n \simeq K_1^n \>$.
\edes
\vspace{0.3cm}

\noindent
The Euclidean time-evolution operator then is 
\bea
U \left ( a, \frac{\beta}{2}; -a, - \frac{\beta}{2} \right ) &\simeq& \! \! 
\sum_{n=1,3 \ldots}  \, \sqrt{\frac{m \omega}{\pi \hbar}} e^{-\omega \beta/2}
\, \frac{1}{n!} \left ( \sqrt{K_1} \beta e^{-S_1/\hbar} \right )^n
\E \sqrt{\frac{m \omega}{\pi \hbar}} \,  e^{-\omega \beta/2} \, 
\sinh \left ( \sqrt{K_1} \beta e^{-S_1/\hbar} \right ) \non
\EA \frac{1}{2}  \sqrt{\frac{m \omega}{\pi \hbar}} \,  e^{-\omega \beta/2}  
\Bigg \{  
\exp \left [ \beta \sqrt{K_1} e^{-S_1/\hbar}\right ] - 
\exp \left [ - \beta \sqrt{K_1} e^{-S_1/\hbar} \right ]  \Bigg \} \, .
\eea
Comparing with the spectral representation (\ref{spektral eukl}) 
we thus conclude that
\bea
E_0 \EA \frac{1}{2} \hbar \omega - \hbar  \sqrt{K_1} e^{-S_1/\hbar} \> ,
\hspace{0.5cm} \psi_0(a) \psi_0(-a) = \frac{1}{2}  
\sqrt{\frac{m \omega}{\pi \hbar}} \\
E_1 \EA \frac{1}{2} \hbar \omega + \hbar  \sqrt{K_1} e^{-S_1/\hbar} \> ,
\hspace{0.5cm} \psi_1(a) \psi_1(-a) = - \frac{1}{2}  
\sqrt{\frac{m \omega}{\pi \hbar}} \> ,
\eea
i.e. the ground state has positive parity and is lowered while
the first excited state has negative parity and has been raised
in energy by tunneling. The wave function at $ x = \pm a $ agrees with 
Eq. (\ref{Grund0 pm}) if one neglects the (exponentially suppressed)
tail of wave function from the other well and recalls that
the wave function of the harmonic oscillator has
$\phi_0(0) = \left ( m \omega/ \pi \hbar \right )^{1/4} $ 
(see Eq. (\ref{psi0})).
Taking the value \eqref{K1} for the constant $ K_1 $ we obtain 
the (leading term for the) splitting of the energy levels
\be
\boxed{
\qquad \Delta E \E E_1 - E_0 \E 2 \hbar  \sqrt{K_1} e^{-S_1/\hbar} \E 4 \hbar \omega
\sqrt{\frac{m \omega a^2}{\pi \hbar}} \, 
\exp \left [ - \frac{2}{3} m \omega a^2/\hbar \right ] \> . \quad
}
\label{Aufspaltung}
\ee
\vspace{1cm}

\noindent
{\bf Remarks}
\ben
\item The factor  $ e^{-S_1/\hbar} $ is the expected suppression factor
for the tunneling probability. From Eq. (\ref{S1}) it can be seen that 
this is exactly the usual WKB result.

\item In terms of the usual oscillator length $ \> b = \sqrt{\hbar/(m \omega)} \> $ the result
\eqref{Aufspaltung} reads
\be 
\Delta E \E \frac{4}{\sqrt{\pi}} \, \hbar \omega \, \frac{a}{b} \, \exp \lrp - \frac{2}{3} \frac{a^2}{b^2} \rrp \> .
\ee

\item Since $ 1/a^2 \sim \lambda $ essentially is the anharmonicity parameter
(cf. Eq. (\ref{def omega})) one sees that the result (\ref{Aufspaltung}) for 
the splitting
\be
\Delta E \> \sim \> \frac{{\rm const}}{\sqrt{\lambda}} \, e^{-{{\rm const}_2}/\lambda}
\ee
depends non-analytically on $ \lambda $ and hence cannot be obtained in perturbation
theory.

\item Higher-order terms for the splitting have been calculated (see, e.g.  Ref.
\cite{Turb} and references therein).
%
%
%
%
%
%

\een

\newpage
\pagestyle{myheadings}
\markboth{\textcolor{green}{Section 2 : Many-Body Physics}}{\textcolor{green}{R. Rosenfelder :
 Path Integrals in Quantum Physics}}

\section{\textcolor{red}{Path Integrals in Statistical Mechanics  \hspace{-0.5mm}and   
\hspace{-0.5mm}Many-Body Physics}}

\renewcommand{\thesubsection}{\textcolor{blue}{2.\arabic{subsection}}}
\renewcommand{\theequation}{2.\arabic{equation}}

\setcounter{equation}{0}

\subsection{\textcolor{blue}{Partition Function}}
\label{sec2: Zustandssum}

The partition function
\be
\boxed{
\qquad Z(\beta) \E  {\rm tr} \left ( e^{- \beta \hat H} \right )\> , \> \> \> \beta \E 
\frac{1}{k_B T} \qquad
}
\ee
($k_B$ : Boltzmann's constant, $T$ : temperature) is a central quantity
in statistiscal mechanics as the following thermodynamical quantities
can be derived from it
\bit
\item free energy $ F = - \frac{1}{\beta} \ln Z $, i.e.
\be
\hspace{-2cm} Z \E  \exp( - \beta F)
\label{def freie En}
\ee
\item pressure $ P = - \partial F / \partial V $ \quad ($V$ : volume)
\item entropy $S = - \partial F / \partial T $ \hspace{1cm} etc.
\eit
To calculate expectation values of arbitrary observables 
$\hat A$ in the thermodynamical equilibrium one needs the 
(equilibrium) density matrix 
\be
\hat \rho_{\beta} \E \frac{1}{Z} \, e^{-\beta \hat H} \> \> , \> \>  
\left < \hat A \right > \E 
{\rm tr}  \left (  \, \hat A \, \hat \rho_{\beta} \, \right ) .
\ee

\noindent
We already have derived a path-integral representation for the partition function 
of a particle in a potential in {\bf chapter} {\bf \ref{sec1: Numerik}} and only have to
replace there $\beta \to \beta \hbar$ :
\vspace{0.2cm}

\fcolorbox{black}{white}{\parbox{15cm}
{
\bea
Z(\beta) \EA \int dx \int_{x(0)=x}^{x(\beta \hbar)=x} {\cal D}x \, \exp \left [
 - \frac{1}{\hbar} \int_0^{\beta \hbar} d\tau \left( \frac{m}{2} \dot x^2 
+ V(x) \right ) \,
\right ] \EQ \oint\limits_{x(0)=x(\beta \hbar)}\! {\cal D}x  \,
e^{- S_E[x]/\hbar}  \> , \non 
S_E[x] \EA \int_0^{\beta\hbar} d\tau \> \left (\frac{m}{2} \dot x^2 + V(x) 
\right ) \> . \no
\eea
}}
\vspace{-2.3cm}

\bea
\label{Z Pfad2}
\eea
\vspace{1cm}


\noindent
For example, when we consider a particle in a harmonic potential we obtain from
Eqs.  (\ref{HO propagator}, \ref{S kl fuer HO}, \ref{HO Vorfaktor}) after transformation 
to Euclidean times \footnote{Set $t_a = 0 \> , \> 
t_b = T = - i \beta \hbar $ and $\sin(iz) = i \sinh(z) \> , \> \cos(iz) = \cosh(z) $.}
the following expression for the density matrix
\bea
\rho_{\beta}^{\rm h.o.} (x,x') &:=& \frac{1}{Z} \, \left < x \left | \, 
\exp ( -\beta \hat H^{\rm h.o.} ) \, \right | x' \right > \non
\EA \frac{1}{Z} \, \sqrt{ \frac{m \omega}{2 \pi \hbar \sinh (\beta \hbar \omega)} }
\, \exp \left [ \, - \frac{m \omega}{2 \hbar} \, 
\frac{ (x^2 +x'^2) \cosh(\beta \hbar \omega) - 2 x x'}{ \sinh (\beta \hbar \omega)} 
\, \right ] \> .
\label{HO Dichtematrix}
\eea
After performing the trace the corresponding partition function reads
\be
Z^{\rm h.o.}(\beta) \E \int_{-\infty}^{+\infty} dx \, \left < x \left | \, 
\exp ( -\beta \hat H^{\rm h.o.} ) \, \right | x \right > \E 
\frac{1}{2 \sinh (\beta \hbar \omega/2 )} \> ,
\label{Z hO}
\ee
where we have used  $\cosh z - 1 = 2 \sinh^2(z/2)$.
Of course, this agrees with the sum over all energy levels of the
harmonic oscillator:
\be
Z^{\rm h.o.}(\beta) \E \sum_n \, \exp ( - \beta E_n ) \> \stackrel{{\rm h.o.}}{=} \>   
\sum_{n=0}^{\infty} \, 
\exp \left [  \, - \beta \hbar \omega \left ( n + \frac{1}{2} \right ) \, \right ] 
\> .
\ee
\vspace{0.2cm}

\noindent
For many applications one needs the
{\bf partition function of the forced harmonic oscillator}:
\be
Z^{\rm forced \, h.o.}(\beta) \E Z^{\rm h.o.}(\beta) \, \exp \left \{ \, 
\frac{1}{2 m \omega \hbar} \, \int_0^{\beta \hbar} d\tau \int_0^{\tau} d\tau' \, 
e(\tau) \, K \left (\omega (\tau - \tau'),\omega \beta \hbar \right )
e(\tau') \right \}
\label{Z erzw. hO}
\ee
with the integral kernel
\be
K(x,y) \E \frac{\cosh(y/2 - x)}{\sinh(y/2)} \> .
\label{Kern erzw. hO}
\ee
\vspace{0.5cm}

\renewcommand{\baselinestretch}{0.9}
\scriptsize
\refstepcounter{tief}
\noindent
\blau{\bf Detail \arabic{tief}:} {\bf  Partition Function of the Forced Harmonic Oscillator}\\

\noindent
\begin{subequations}
The derivation of this result can be made by direct calculation but is a little bit 
cumbersme. One starts from the result (\ref{erzwung HO}) in real time and performs,
in addition to the transformation
$t_a = 0, t_b = T = -i\beta \hbar$, the substitution $ t = - i\tau , t' = - i \tau' $
in the integrals over the external force. This gives
\be
Z^{\rm forced \, h.o.}(\beta) \E 
\sqrt{ \frac{m \omega}{2 \pi \hbar \sinh (\omega \beta \hbar)}} \, 
\int_{-\infty}^{+\infty}dx \, \exp \left ( - a x^2 - b x + c \right ) 
\ee
with
\bea
a \EA \frac{m \omega}{\hbar \sinh (\omega \beta \hbar)} \> \> , \hspace{0.5cm}
b \E \frac{1}{\hbar \, \sinh (\omega \beta \hbar)} \non
c \EA \frac{1}{m \omega \hbar \, \sinh (\omega \beta \hbar)} \, 
\int_0^{\beta \hbar} 
d\tau \int_0^{\tau} d\tau' \, e(\tau) \, e(\tau') \, 
\sinh \omega (\beta \hbar - \tau) \, \sinh (\omega\tau') \> .
\eea
The Gaussian integral can be done immediately and gives
\be
Z^{\rm forced \, h.o.}(\beta) \E Z^{\rm h.o.}(\beta) \, \exp \left ( \frac{b^2}{4 a} + c
\right ) \> .
\ee
If one evaluates the exponent, uses the relation (\ref{Intpotenz})
with $j = 2$ for combining the double integrals of $b^2$ and $c$
and employs the addition theorems for the hyperbolic functions, one obtains
Eq. (\ref{Z erzw. hO}) with the kernel (\ref{Kern erzw. hO}).\\



\end{subequations}
\renewcommand{\baselinestretch}{1.1}
\normalsize
\vspace{0.5cm}

We may take over the result (\ref{Z Pfad2}) immediately to the case
where we have a system of $ N $ {\bf distinguishable} quantum mechanical 
particles moving in an exterior (or mean) potential:
\be
\hat H = \sum_{i=1}^N \left [ \frac{\hat p_i^2}{2 m} \> + \> V(\hat x_i)
\right ] = \sum_{i=1}^N \hat H_i \> .
\ee
Then
\be 
Z = {\rm tr} \left ( e^{- \beta \sum_{i=1}^N \hat H_i} \right )
\E \left [ \> {\rm tr} \left ( e^{- \beta \hat H} \right ) \> \right ]^N
\> ,
\ee
since the particles are moving independently. For each of the
sub-partition functions the path-integral representation
(\ref{Z Pfad2}) is valid.

\vspace{0.3cm}

As another application we will derive the \textcolor{blue}{\bf high-temperature expansion} 
of the partition function from the path integral. This is also known as
{\bf Wigner-Kirkwood expansion}. For this purpose we write the
path which leads from the point  $ x $ at "time"  $ \> \tau = 0 \> $ to  $ x $ at
"time" $ \> \tau = \beta \hbar\> $ as
\be
x(\tau) = x + \xi(\tau)
\ee
and assume that in the high-temperature case $ \> \xi \> $ is small compared to $\> x \> $.
Then we can expand
\bea
V(x + \xi) \EA V(x) + \xi V'(x) + \frac{1}{2} \xi^2 V''(x) + \ldots 
\non
\EA V - \frac{V'^2}{2 V''} + \frac{1}{2} V'' \left (\xi + \frac{V'}{V''} 
\right )^2 + {\cal O}(\xi^3)
\eea
and obtain similar as in the semi-classical expansion of the propagator
in {\bf chapter} {\bf \ref{sec1: halbklass}}
\be
Z^{1/N} \simeq \int dx \> 
\exp \left [ -\beta \left ( V - V'^2/(2 V'') \right ) \right ]
\int {\cal D}\eta \> \exp \left [ - \frac{1}{\hbar} \int_0^{\beta \hbar} 
d\tau \left ( \frac{m}{2}
\dot \eta^2 + \frac{1}{2} V'' \eta^2 \right ) \right ] \> .
\label{Z1}
\ee
The boundary conditions for the shifted  integration variable
$ \eta(\tau) = \xi(\tau) + V'/V''$ are  $ \eta(0) = \eta(\beta \hbar) = V'/V''$
and all potential values have to be taken at the fixed point $ x $ over which
finally one also has to integrate. The path integral (\ref{Z1})
is that of a harmonic oscillator in Euclidean time with $ m \omega^2 = V''(x)$
and therefore we can employ immediately the result given in
Eq. (\ref{HO Dichtematrix}). In addition, one has to expand the
hyperbolic functions for small  $\beta$ in order to remain consistent 
with the assumption that  $\xi(\tau) = {\cal O}(\tau^2)$ is small.
In this way one obtains
\bea
Z^{1/N} &\simeq& \int dx \> \exp \left [ -\beta \left ( V - V'^2/(2 V'') 
\right ) \right ] \> \sqrt{\frac{m}{2 \pi \hbar^2 \beta \> ( 1 
+ \omega^2 \beta^2 \hbar^2/6 + \ldots )}} \non
&& \hspace{2cm} \cdot \exp \left [ - \frac{m \omega}{\hbar} \left ( 
\frac{V'}{V''} \right )^2
\frac{\omega}{2} \beta \hbar \left ( 1 - \frac{1}{12} \omega^2 \beta^2 \hbar^2
 + \ldots \right ) \right ] \non
&\simeq& \int dx \> \exp \left [ -\beta V(x) \right ] \> \frac{1}{\hbar} 
\sqrt{\frac{m}{2 \pi  \beta}} \> \left [ 1 + 
\frac{\beta^2 \hbar^2}{24 m} \left ( \beta V'^2 - 2 V'' \right ) 
+ \ldots \right ] \> .
\eea
Realizing that
\be
\frac{d^2}{d x^2} e^{- \beta V(x) } = \beta \left ( \beta V'^2 - 2V'' \right )
e^{- \beta V(x) }
\ee
one can perform an integration by parts to obtain the result
\be
\boxed{
\qquad Z^{1/N} \simeq \frac{1}{\hbar} \sqrt{\frac{m}{2 \pi  \beta}}
\int dx \> \exp \lsp -\beta V(x) \rsp
\> \left [ 1  - \frac{\beta^3 \hbar^2}{24 m} V'^2(x) + \ldots \right ] \> . \quad
}
\label{Wigner-Kirkwood}
\ee
The first term
\be 
Z_{\rm class.}^{1/N} = \frac{1}{\hbar} \sqrt{\frac{m}{2 \pi  \beta}}
\int dx \> \exp \left [ -\beta V(x) \right ] \E \int 
\frac{ dx \, dp}{2 \pi \hbar} \> \exp \left [ - \beta \left ( \frac{p^2}{2 m}
 + V(x) \right ) \right] 
\label{Z klass}
\ee
is the classical result as can be seen  most clearly in the phase-space
representation (the second expression in Eq. (\ref{Z klass})):
Apart from the division of phase space into cells of size
 $ \> h = 2 \pi \hbar \> $ this form does not contain any dependence on
 Planck's elementary quantum but only the classical Boltzmann factor
 $ \exp(-\beta H(p,x))$. The second term in the square bracket of 
Eq. (\ref{Wigner-Kirkwood}) is the quantum correction for high temperatures.
It is obvious that higher corrections may be calculated by additional terms in the 
Taylor expansion of the potential.

\vspace{0.5cm}

\subsection{\textcolor{blue}{The Polaron}}
\label{sec2: Polaron}

A celebrated application of the path-integral method in statistical mechanics or
solid-state physics is the motion of electrons in an ionic crystal, e.g. NaCl.
Here Feynman's path-integral method \cite{Feyn2} gives results which are clearly superior over
those obtained in conventional approaches.

The electron interacts with the ions which are not tightly bound and thus generates
a distortion of the crystal which it tugs along during its motion. This "quasiparticle"
is called a  \textcolor{blue}{\bf polaron}. A simple model Hamiltonian describing
this effect has been given by H. Fr\"ohlich \footnote{This 
``Fr\"ohlich'' polaron is also called "large" polaron as its dimensions are so large
compared to the lattice constant of the solid that one can treat the  crystal as a
continuum. This is not the case for "small" or ``Holstein'' polarons. See, e.g. 
http://en.wikipedia.org/wiki/Polaron.}:
One solves the Poisson equation for the potential acting on the electron
under the assumption that the induced charge density is proportional
to the divergence of a longitudinal displacement wave which can be
expanded in modes
\be
{\bf P}(\fx) \>  \sim \> \sum_{\fk} \left [ \>  a_{\fk} 
\frac{\fk}{ |\fk |} e^{i \fk \cdot \fx}  \> + \> c.c. \> 
\right ] \> .
\ee
In the quantized theory one interpretes $ \, \hat a_{\fk} \, $ as a creation operator for a {\bf phonon} with 
 momentum $ \fk $ . For small momenta only the optical branch of the phonons
  contributes 
with a frequency  $ \omega $ which is constant, i.e. independent of  $ \fk $ 
 in the simplest approximation. If one adds a free part 
 for the phonons then the Hamiltonian of the system reads
(in units  $ \> \hbar = \omega = m = 1 \> $)
\be
\hat H = \frac{1}{2} \hat \fp^2 + \sum_{\fk} \> \hat a^{\dagger}_{\fk}
\hat a_{\fk} +  i \left (2 \sqrt 2 \pi \alpha \right )^{1/2}
\frac{1}{\sqrt{V}} \sum_{\fk} \> \frac{1}{ |\fk|} \> 
\left [ \> \hat a^{\dagger}_{\fk} e^{- i {\fk} \cdot \hat \fx}  -  {\rm h.c.} \> 
\right ] \EQ  \hat H_0 + \hat H_1 \> .
\label{Froehlich H}
\ee
Here $ V $ denotes the volume in which the total system is enclosed
 \footnote{This is to make the spectrum discrete and countable:
$ k_i = n_i \pi/L \> , V = L^3 $. Summation over  $\fk$ is a shorthand
for the summation over $  n_i = 1, 2 \ldots \> , i = 1, 2 , 3 \> $. 
For $ L \to \infty $ one can replace this summation by an integral so that
 $ (1/V) \sum_{\fk} \to \int d^3k/(2\pi)^3 $.}
and $ \alpha$ is the dimensionless electron-phonon coupling constant. 
In actual applications it takes values between  
 1 and 10 -- and therefore is not necessarily small.

First, however, we assume that perturbation theory is applicable.
In lowest order the energy of the electron is
$ \> E^{(0)} = \fp^2/2 \> $. The next two orders of perturbation 
theory give $ \> \Delta E^{(1)} = 0 \> $ and
\be
\Delta E^{(2)} \E  \sum_n \frac{ <0 | \hat H_1 | n >
< n | \hat H_1 | 0 > }{E^{(0)}-E^{(n)}} =  - \frac{2 \sqrt 2 \pi \alpha}
{V} \sum_{\fk} \> \frac{1}{\fk^2} \> \left [ \frac{1}{2}
( \fp - {\fk} )^2 + 1 - \frac{1}{2}\fp^2 \right ]^{-1} \> .
\ee
The first order of perturbation theory vanishes since there is no
phonon in initial and final state. As usual, the second order
leads to a lowering of the (ground-state) energy and can be
visualized as the emission and absorption of a phonon with momentum $ {\fk} $ 
from the (bare) electron:

\refstepcounter{abb}
\begin{figure}[hbtp]
\bce
\vspace*{-9.8cm}
\hspace*{3cm}\includegraphics[angle=0,scale=0.8]{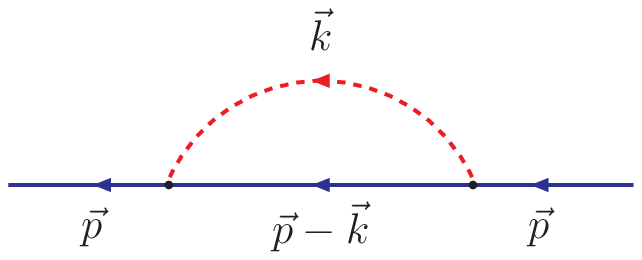}
\label{abb:2.2.1}
\ece
\vspace{-11.5cm}
\bce
{\bf Fig. \arabic{abb}} : Second-order perturbation theory for the
self-energy of the polaron.
\ece
\end{figure}
 
\vspace{0.5cm}
\noindent
In the continuum limit we obtain a (loop) integral
whose calculation gives 
\be
\Delta E^{(2)} \E - \alpha \frac{\sqrt 2}{p} \arcsin \left ( 
\frac{p}{\sqrt 2} \right ) \> .
\label{PT}
\ee
For small momenta $ p = |\fp| $ of the electron we therefore have
\be
E \E \frac{1}{2} \fp^2 - \alpha - \frac{\alpha}{12} \fp^2
+ {\cal O}(\alpha^2, \fp^4) \simeq - \alpha +
\frac{\fp^2}{2 ( 1 + \alpha/6)} + \ldots \> .
\ee
In other words: The electron gets an effective mass
$ m_{\rm eff}/m = 1 + \alpha/6 $ and has a rest energy  \footnote{The numerical factors 
in front of the interaction term in the 
Fr\"ohlich Hamiltonian \eqref{Froehlich H} look rather arbitrary but have
been chosen intentionally such that the term linear in $ \alpha $ for the energy has the "nice" 
coefficient $-1$... This legitimate cosmetics is quite common and leads to a simpler and, 
in some sense, more "elegant" final result.
However, I am far away from supporting Dirac's statement
\textsf{"It is more important to have {\it beauty} in one's equation than to have 
them fit experiment"}, in {\bf \{Farmelo\}}, p.376 .
Rather I prefer the attitude of J. W. Tukey \textsf{"An approximate solution of the exact problem
is often more {\it useful} than the exact solution of an approximate problem"}, in 
"Physics Today", July 2001, p. 80.
}
\be
E_0  \E - \alpha + {\cal O}(\alpha^2) \> .
\label{E0 1the ord}
\ee
Higher orders of perturbation theory have been calculated (over the years/decades)  
\cite{HoMu},\cite{Smon},\cite{Ros3} giving the result
\be 
E_0(\alpha) \E -\alpha - 1.59196 \> \left ( \frac{\alpha}{10} \right )^2 - 0.8061 \> 
\left ( \frac{\alpha}{10} \right )^3 - 0.533 \> \left ( \frac{\alpha}{10} \right )^4
- 0.38  \> \left ( \frac{\alpha}{10} \right )^5  
+ {\cal O} (\alpha^6) \> .
\label{E alpha klein}
\ee
It is obvious that this series becomes useless for large coupling constants.
Here the path-integral method comes to rescue : First, Feynman realized that 
the phonon coordinates appear at most quadratically and therefore can 
be \textcolor{blue}{\bf integrated out exactly}. 

\vspace{0.8cm}

\renewcommand{\baselinestretch}{0.9}
\scriptsize
\refstepcounter{tief}
\noindent
\blau{\bf Detail \arabic{tief}:} {\bf Integrating out the Phonons}\\

\noindent
\begin{subequations}
This is done most easily by introducing the (reversed) coordinate and momentum operators
instead of the creation and annihilation operators in the Fr\"ohlich Hamiltonian (\ref{Froehlich H})
\bea
\hat q_{\fk} \E  \frac{i}{\sqrt 2} \left ( \hat a_{-{\fk}}^{\dagger}
- \hat a_{\fk} \right ) \> , \hspace{1cm} 
\hat p_{\fk} \E - \frac{1}{\sqrt 2} \left ( \hat a_{{\fk}}^{\dagger}
+ \hat a_{-{\fk}} \right ) \> .
\label{kanon Tr}
\eea
As one can check easily, these operators obey the commutation relations $ [ \hat q_{{\fk}}, \hat p_{{\fk}'}] = i \delta_{{\fk} {\fk}'} $, i.e. Eq. (\ref{kanon Tr}) is a canonical transformation.
Then the Fr\"ohlich Hamiltonian reads 
\be
\hat H \E \frac{1}{2} \hat \fp^2 + \frac{1}{2}\sum_{\fk} \> \left ( \hat p_{\fk} \, 
\hat p_{-\fk} + \hat q_{\fk}\, \hat q_{-\fk} \right ) - \sum_{\fk} \frac{1}{2}
+ 2 \left (\sqrt 2 \pi \alpha \right )^{1/2}
\frac{1}{\sqrt{V}} \sum_{\fk} \> \frac{1}{ |\fk|} \, \hat q_{\fk} \,  
e^{i {\fk} \cdot \hat \fx}  \> , 
\label{Froehlich H2}
\ee 
where the $3^{\rm rd} $ term subtracts the (infinite) zero-point energy of all modes
 $ \fk $ . The path integral for the partition function now has the following form
\bea
Z(\beta) \EA \oint\limits_{\fx(0)=\fx(\beta)} \! {\cal D}^3x(\tau) 
\> \exp \left ( \, - \int_0^{\beta} d\tau \, \frac{1}{2} \dot \fx^2(\tau) \, \right ) 
\, \prod_{\fk} \Biggl \{ \,  \exp ( \beta/2 ) \! \! \! 
\oint\limits_{q_k(0)=q_k(\beta)} \! \!  {\cal D}q_k(\tau) \non 
&& \cdot \, \exp \left [  \, - \int_0^{\beta} d\tau \, \sum_{\fk}
\left ( \, \frac{1}{2} \dot q_{\fk}(\tau) \, \dot q_{-\fk}(\tau) + 
\frac{1}{2} q_{\fk}(\tau) \,  q_{-\fk}(\tau)  + e_{\fk}(\tau) \, q_{\fk}(\tau)
\right ) \, \right ] \, \Biggr \}
\eea
with
\be
e_{\fk}(\tau) \E 2 \left ( \sqrt{2}\pi \alpha \right )^{1/2} 
\frac{1}{\sqrt{V}} \,           
\frac{1}{ |\fk| } e^{i \fk \cdot {\bf x}(\tau)}  \> .
\label{def ek}
\ee
Due to the canonical transformation (\ref{kanon Tr}) we have 
$ \> q_{\fk}^{\star}(\tau) \E q_{-\fk}(\tau) \> $ which means
$ \gamma_{-\fk} = \gamma_{\fk} $ for the real part and 
$\eta_{-\fk} = - \eta_{\fk}$ for the imaginary part. Therefore 
we only have to integrate over the positive modes with the actions
\be
\sum_{\fk \ge 0} \, \int_0^{\beta} d\tau \left [ \, \dot \gamma_{\fk}^2 +  
\gamma_{\fk}^2 + \left ( e_{\fk} + e_{-\fk} \right ) \, \gamma_{\fk} 
\, \right ] \> , \> {\rm or} \quad
\sum_{\fk > 0} \, \int_0^{\beta} d\tau \left [ \, \dot \eta_{\fk}^2 +  
\eta_{\fk}^2 + i \left ( e_{\fk} - e_{-\fk} \right )  \, \eta_{\fk} \, 
\right ] \> .
\ee
In both cases we can use the result (\ref{Z erzw. hO}) 
(by setting there $m = 2, \omega = \hbar = 1 $ ). Then we obtain
for each mode the factor
\be
Z^{\rm h.o.}(\beta) \, \cdot \, \exp \left \{  \, \frac{1}{2} 
\int_0^{\beta} d\tau \int_0^{\tau} 
d\tau' \, K \left ( \tau - \tau',\beta \right ) \, 
\left [ \, e_{\fk}(\tau) e_{-\fk}(\tau')
+  e_{-\fk}(\tau) e_{\fk}(\tau') \, \right ] \, \right \} \> .
\ee
By this the partition function of the electron becomes
\be
Z(\beta) \E \prod_{\fk} \left ( \frac{\exp(\beta/2)}{2 \sinh(\beta/2)} \right ) \, 
\oint\limits_{\fx(0)=\fx(\beta)}  {\cal D}^3x(\tau) \> e^{ - S_{\rm eff}[\fx(\tau)] } 
\> , 
\label{Z polaron allg.}
\ee
with the effective (Euclidean) action
\bea 
S_{\rm eff}[\fx(\tau)] \E \int_0^{\beta} d\tau \> \frac{1}{2} \dot \fx^2
\> - \frac{1}{2}
\int_0^{\beta} d\tau \int_0^{\tau} d\tau' \, \frac{\cosh \left ( \beta/2 - (\tau - \tau')
\right )}{\sinh(\beta/2)} \, \sum_{\fk} e_{\fk}(\tau) e_{-\fk}(\tau') \> . \nonumber
\eea
Inserting Eq. (\ref{def ek}) and taking the limit of infinite quantization volume
gives the result
\bea
S_{\rm eff}[\fx(\tau)] \E \int_0^{\beta} d\tau \> \frac{1}{2} \dot \fx^2
\> - 2 \pi \sqrt{2} \alpha \, 
\int_0^{\beta} d\tau \int_0^{\tau} d\tau' \, \frac{\cosh \left ( \beta/2 - (\tau - \tau')
\right )}{\sinh(\beta/2)} \,  \int \frac{d^3 k}{(2 \pi)^3} \, \frac{1}{\fk^2} \, 
\exp \left [ i \fk \cdot \left (\fx(\tau) - \fx(\tau') \right ) \right ] \> .
\label{S eff 1}
\eea
The momentum integral is that for a Coulomb potential  
\be
\int \frac{d^3 k}{(2 \pi)^3} \, \frac{1}{\fk^2} \, 
\exp \left ( i \fk \cdot \fy \right ) \E \frac{1}{4 \pi} \, \frac{1}{|\fy|} \> ,
\ee
so that the final result reads
\bea 
S_{\rm eff}[\fx(\tau)] \E \int_0^{\beta} d\tau \> \frac{1}{2} \dot \fx^2
\> - \> \alpha \, \sqrt 2 \int_0^{\beta} d\tau \int_0^{\tau} d\tau' \> 
G_{\beta}(\tau - \tau') \,  \frac{1}{| \fx(\tau) - \fx(\tau') |}
\label{S eff 2}
\eea
Here the retardation function is defined by
\be
G_{\beta}(t) \Def \frac{\cosh \left ( \beta/2 - |t| \right )}{2 \sinh(\beta/2)} \> 
\stackrel{\beta \gg |t|}{\longrightarrow} \> \frac{1}{2} \, e^{- |t|} \> .
\label{polaron ret fkt}
\ee

\end{subequations}
\renewcommand{\baselinestretch}{1.1}
\normalsize
\vspace{1cm}

\noindent
By integrating out the phonons the polaron problem has been reduced to an 
one-body problem for the electron! If we are only interested in the ground-state
energy we can also (partially) perform the limit $\beta \to \infty $ :
The zero-point energy of the individual phonon oscillators cancels in the
pre-factor of Eq. (\ref{Z polaron allg.}) and the integral kernel can be
simplified by assuming $ \tau, \tau' \ll \beta $ . We then have
\be
Z(\beta) \E \oint\limits_{\fx(0)=\fx(\beta)} \, {\cal D}^3x(\tau) 
\> e^{-S_{\rm eff}[\fx(\tau)]}
\ee
with the  \textcolor{blue}{\bf effective action}
\be
\boxed{
\qquad S_{\rm eff}[\fx(\tau)] \E \int_0^{\beta} d\tau \> \frac{1}{2} \dot \fx^2
\> - \> \frac{\alpha}{\sqrt{2}} \int_0^{\beta} d\tau
\int_0^{\tau} d\tau' \> \frac{e^{- (\tau-\tau')}}{| \fx(\tau) - \fx(\tau') |} \> . \quad
}
\label{S eff 3}
\ee 
However, the remaining path integral cannot be solved exactly anymore.
Feynman, in a second step, therefore applied a \textcolor{blue}{\bf variational principle} .
It is based on the identity (remember: The path integral works with ordinary, commuting
 numbers!)
\be
\int {\cal D} x \> e^{-S} \E \int {\cal D} x \> e^{-S_t} \cdot
\frac{ \int {\cal D} x \> e^{-(S-S_t)} e^{-S_t}}{ \int {\cal D} x \> 
e^{-S_t}} \EQ \int {\cal D} x \> e^{-S_t} \cdot \left < e^{-(S-S_t)} \right >  
\label{Pfad trial}
\ee
and \textcolor{blue}{\bf Jensen's inequality} for convex functions
\be
\boxed{
\qquad \left < e^{-\Delta S} \right > \geq e^{- \left < \Delta S \right > } \qquad
}
\label{Jensen}
\ee
which holds for averaging with positive weight functions which is the case 
for  $ \exp(-S_t) $. Here $ S_t $ is a trial action which should approximate
the true action  $ S $ as well as possible but also allows the path integral to be performed.
This technical requirement restricts our choice to the class of 
\textcolor{blue}{\bf quadratic trial actions}. Since the 
effective action (\ref{S eff 3}) shows a 
\textcolor{blue}{\bf retardation} which takes into account that 
phonons emitted at an earlier time are absorbed at a later time, it is crucial
to have a retardation also built into the trial action. Therefore one makes the
{\it ansatz}
\be
S_t = \int_0^{\beta} d\tau \> \frac{1}{2} \dot \fx^2 + 
\int_0^{\beta} d\tau \int_0^{\tau} d\tau' \> f\left ( \tau - \tau' \right ) \> 
\left [ \> \fx(\tau) - \fx(\tau') \> \right ]^2
\label{S trial}
\ee
where $ f(\sigma) $ is a free, undetermined retardation function.
The ground-state energy of the electron in the crystal is obtained from
\be
Z(\beta) \E e^{- \beta F} \>  \> \stackrel{\beta \to \infty}{\longrightarrow} 
\> \> e^{-\beta E_0} \> ,
\ee
i.e. with the help of Eqs. (\ref{Pfad trial}) and (\ref{Jensen}) 
\be
\boxed{
\qquad E_0 \> \leq \> E_t  + \lim_{\beta \to \infty} \frac{1}{\beta} 
\la S_{\rm eff} - S_t \ra  \> . \quad
}
\label{E var allg}
\ee
This expression is a generalization of the upper limit of energy
from Hamiltonians (i.e. the well-known Rayleigh-Ritz variational principle) to general actions.
The calculation of  $E_t$ and $ \left < S_{\rm eff} - S_t \right > $ with
the {\it ansatz} (\ref{S trial}) is  slightly involved but
may be done best by Fourier expanding the electron path as 
in {\bf chapter} {\bf \ref{sec1: Lagr,Ham}}. The result is
\cite{RoSch} 
\be
E_0 \> \leq \> \Omega - \frac{\alpha}{\sqrt \pi} \int_0^{\infty}
d\sigma \> \frac{e^{-\sigma}}{\mu(\sigma)} \> ,
\label{E var quad}
\ee
with
\bea
\Omega \EA \frac{3}{2 \pi} \int_0^{\infty} dE \left [ \> \ln A(E) + 
\frac{1}{A(E)} - 1 \> \right ] \\
\mu^2(\sigma) \EA \frac{4}{\pi} \int_0^{\infty} dE \> 
\frac{\sin^2(\sigma E/2)}{E^2 A(E)} \> .
\eea
Here, the function  $ A(E) $ is obtained from the retardation function
by
\be
A(E) \E 1 + \frac{8}{E^2} \int_0^{\infty} d\sigma \> f(\sigma) \> 
\sin^2 \left ( \frac{\sigma E}{2} \right ) \> .
\ee
Feynman chose
\be
f(\sigma) = f_F(\sigma) = C \cdot e^{-w \sigma}
\ee
with two variational parameters $ C $ (strength of the retardation) and $ w $
(retardation time). Then one obtains
\be
\Omega_F = \frac{3}{4 v} (v - w)^2 \> , \hspace{0.4cm}
A_F(E) = \frac{v^2 + E^2}{w^2 + E^2} \> , \hspace{0.4cm}
\mu_F(\sigma) = 
\left [ \frac{w^2}{v^2} \sigma + \frac {v^2 - w^2}{v^3} \>
(1 - e^{-v \sigma } )
\right ]^{1/2} \> ,
\ee
where one uses -- as is customary -- the quantity
$ \> v = \sqrt{w^2 + 4 C/w} \> $ ($ v \geq w $)
instead of the strength $ C $ .
For small coupling constants
$ \alpha $ one can evaluate the variation w.r.t. the parameters
 $ v, w $ (\purpur{\bf Problem \ref{Feyn Polaron}}) analytically and one obtains
\be
E_{F} = - \alpha - 0.012346 \> \alpha^2  - 0.6344 \cdot 10^{-3} \> \alpha^3 
-0.4643 \cdot 10^{-4}  \> \alpha^4 - 0.3957  \cdot 10^{-5} \> \alpha^5
+ {\cal O}(\alpha^6) \> .
\ee
This is only slightly worse than the perturbative expansion
(\ref{E alpha klein}). For very large coupling constants one also can work out 
Feynman's energy analytically and finds
\be
E_{F} \E - \underbrace{\frac{1}{3\pi}}_{=0.1061} \! \! \! \alpha^2  - 3 \ln 2 - \frac{3}{4}
  + {\cal O}(\alpha^{-2}) \E 
- 0.1061 \, \alpha^2 - 2.83 + {\cal O}(\alpha^{-2}) \> ,
\ee
while the exact result in this limit has been derived as
\be
E_0(\alpha) \E  -0.10851 \, \alpha ^2 - 2.84 + {\cal O}\left (\frac{1}{\alpha^2} \right ) \> ,
\ee
(the leading term will be determined in {\bf chapter} {\bf \ref{sec2: Luttinger-Pekar}}). 
For arbitray $ \alpha $ one has to calculate the remaining integral
in Eq. (\ref{E var quad}) numerically. When doing that it turns out that
Feynman's result is the best upper limit~\footnote{There also exist (in principle exact) 
 Monte-Carlo calculations \cite{AlRo}, \cite{TPC} with which one can compare.}
obtained when different approximation schemes are compared for 
{\it all} values of $ \alpha $ . While perturbation theory is better for 
 $ \alpha \ll 1$, it completely fails for large coupling constants where Feynman's result
only deviates by at most 2.2~\%. This success is mainly due to the fact that
the infinite phonon degreees of freedom have been integrated out {\bf exactly}
so that -- irrespective how the electronic part is approximated --
the lattice distortion accompanying the electron is treated correctly.
Note that although the effective action (\ref{S eff 3}) is an one-body action,
there is no Hamiltonian description available anymore and therefore
also no  Schr\"odinger equation to be solved.

\vspace{1cm}

\renewcommand{\baselinestretch}{0.9}
\scriptsize
\refstepcounter{tief}
\begin{subequations}
\noindent
\blau{\bf Detail \arabic{tief}:} {\bf Mean Number of Phonons in the Polaron}\\          

\noindent
It is easy to calculate the mean number of phonons
\be 
\bar N \EQ \la \hat N \ra \E \la \, \sum_{\fk} \hat a_{\fk}^{\dagger} a_{\fk} \ra \E \lim_{\beta \to \infty} 
\, \frac{{\rm tr} \lrp \hat N \, 
e^{-\beta \hat H} \rrp}{{\rm tr} \lrp e^{-\beta \hat H} \rrp}
\ee
which are present in the polaron ground-state -- the "dressed electron". This is because
one can generate the number operator $ \hat N $ by differentiating a modified Hamiltonian
w.r.t. an artificial parameter  $ \lambda $ which then is set to the value ``1'' :
\be
\bar N \E - \frac{\partial}{\partial \lambda} \, \lim_{\beta \to \infty} \, \frac{1}{\beta}
\, \ln {\rm tr} \lrp e^{-\beta \hat H_{\lambda}} \rrp \, \Biggr |_{\lambda = 1} \> .
\label{mittleres N 1} 
\ee
Here
\be
\hat H_{\lambda} \E  \frac{1}{2} \hat \fp^2 + \lambda \, \sum_{\fk} \> \hat a^{\dagger}_{\fk}
\hat a_{\fk} +  i \left (2 \sqrt 2 \pi \alpha \right )^{1/2}
\frac{1}{\sqrt{V}} \sum_{\fk} \> \frac{1}{ |\fk|} \> 
\left [ \> \hat a^{\dagger}_{\fk} e^{- i {\fk} \cdot \hat \fx}  -  {\rm h.c.} \> 
\right ]  \> .
\label{H lambda}
\ee
Since for large $ \beta $ the modified partition function fulfills
\be
{\rm tr} \lrp e^{-\beta \hat H_{\lambda}} \rrp \> \stackrel{\beta \to \infty}{\longrightarrow} \> 
e^{- \beta \, E_0(\lambda,\alpha)}
\ee
one obtains from Eq. \eqref{mittleres N 1}
\be
\bar N \E \frac{\partial}{\partial \lambda} \, E_0(\lambda,\alpha)  \, \Bigr |_{\lambda = 1} 
\label{mittleres N 2}
\ee
and one only has to determine the lowest eigenvalue of the Hamiltonian $ \, \hat H_{\lambda} \, $. 
However, this is easily done by relating it to the original Hamiltonian $ \, \hat H \, $:
First, one divides by $ \lambda $
\be
\frac{1}{\lambda} \, \hat H_{\lambda} \E \frac{1}{2 \lambda} \hat \fp^2 + \sum_{\fk} \> \hat a^{\dagger}_{\fk}
\hat a_{\fk} +  i \left (2 \sqrt 2 \pi \alpha \right )^{1/2}
\frac{1}{ \lambda \sqrt{V}} \sum_{\fk} \> \frac{1}{ |\fk|} \> 
\left [ \> \hat a^{\dagger}_{\fk} e^{- i {\fk} \cdot \hat \fx}  -  {\rm h.c.} \> 
\right ]  \>,
\ee
and then scales momentum and position operators of the electron
(in opposite ways to preserve the commutation relation)
\be
\hat \fp \E \sqrt{\lambda} \, \tilde \fp \> , \quad \hat \fx \E \frac{1}{\sqrt{\lambda}} \, \tilde \fx
\> .
\ee
In the interaction term we then set $ \> \fk = \sqrt{\lambda} \, \tilde \fk \> $  in the summation
and obtain
\bea
\frac{1}{\lambda} \, \hat H_{\lambda} \EA \frac{1}{2} \tilde \fp^2 + \sum_{\fk} \> \hat a^{\dagger}_{\fk}
\hat a_{\fk} +  i \left (2 \sqrt 2 \pi \alpha \right )^{1/2}
\frac{\lambda^{3/4}}{\lambda \sqrt{\tilde V}} \sum_{\tilde \fk} \> \frac{1}{\sqrt{\lambda} |\tilde \fk|} \> 
\left [ \> \hat a^{\dagger}_{\tilde \fk} e^{- i {\tilde \fk} \cdot \tilde \fx}  -  {\rm h.c.} \> 
\right ]  \non
\EA \hat H_0 +  i \left (2 \sqrt 2 \pi \frac{\alpha}{\lambda^{3/2}} \right )^{1/2}
\frac{1}{\sqrt{\tilde V}} \sum_{\tilde \fk} \> \frac{1}{|\tilde \fk|} \> 
\left [ \> \hat a^{\dagger}_{\tilde \fk} e^{- i {\tilde \fk} \cdot \tilde \fx}  -  {\rm h.c.} \right ] 
\> .
\eea
Therefore we can read off
\be
E_0(\lambda,\alpha) \E \lambda \, E_0 \lrp \frac{\alpha}{\lambda^{3/2}} \rrp
\ee
and Eq. \eqref{mittleres N 2} gives
\be
\boxed{
\bar N \E E_0(\alpha) - \frac{3}{2} \, \alpha \, \frac{\partial}{\partial \alpha} \, E_0(\alpha) \> .
}
\ee
\noindent
This means that for large $ \alpha $ the mean number of phonons grows like $ 0.217 \, \alpha^2 $, i.e.
at $ \alpha = 10 $ a cloud of roughly $ 22 $ phonons surrounds a bare electron.

\end{subequations}
\renewcommand{\baselinestretch}{1.2}
\normalsize


\vspace{2cm}

\subsection{\textcolor{blue}{Dissipative Quantum Systems}}
\label{sec2: Dissip}
Aside from the rather special polaron problem, Feynman's treatment of the phonons
has become important for the description of dissipative systems
where the heat bath (the rest of the system which is not treated explicitly) 
is also modelled by a collection of oscillators. This is the topic of the present
chapter which essentially follows Ref. \cite{Ing}.
\vspace{0.2cm}

In classical physics dissipation is frequently described in a phenomenological way
by adding a velocity-dependent term in the equations of motion.
In quantum mechanics this is not possible anymore as a description
with a time-independent Hamiltonian implies energy- and probability conservation.

\noindent
The simple example of a damped pendulum shows how one can get
a better, more physical model for a dissipative system: The degree of freedom 
we are interested in -- the elongation of the pendulum -- is damped because
it interacts with other degrees of freedom (the molecules of the air, the suspension 
mechanism etc.) We may describe
the pendulum, the molecules of the air and the suspension
as a large system which conserves the total energy
(if isolated well enough from other degrees of freedom). 
The energy of the pendulum, however, is not conserved in general but
will be shared with the environment, i.e. the bath. A popular model due to Caldeira and 
Leggett \cite{CaLeg} therefore assumes a total Hamiltonian of the form
\be
\boxed{
\qquad H \E H_{\rm system} + H_{\rm bath} + H_{\rm interaction} \qquad
}
\label{CL Modell}
\ee
where
\be
 H_{\rm system} \E \frac{p^2}{2 m} + V(x)
\label{H System}
\ee
describes  a particle with mass  $ m $ moving in a potential $ V $. 
The model becomes tractable by assuming that the bath degrees of freedom 
can be described by a collection of  $ N $ harmonic oscillators
\be
\boxed{
\qquad H_{\rm bath} \E \sum_{n=1}^N \, \left ( \, \frac{p_n^2}{2 m_n} + \frac{1}{2} 
m_n \omega_n^2 q_n^2 \, \right ) \qquad
} 
\label{H Umgebung}
\ee
and the interaction by a bilinear coupling of particle
and bath coordinates
\be
\boxed{
\qquad H_{\rm interaction} \E - x  \, \sum_{n=1}^N c_n q_n + x^2 \,  \sum_{n=1}^N \, 
\frac{c_n^2}{2 m_n \omega_n^2} \qquad \> .
}
\label{H Wechselwirk}
\ee
The last term in Eq. (\ref{H Wechselwirk}) is really a potential term for the particle,
i.e. should be  part of $ V $, 
but it will turn out to be useful if written in this way.


First we want to show that the model describes the
expected damping already on the \textcolor{blue}{\bf classical level}.
For this purpose we derive the equations of motion for the bath degrees of freedom
\be
\dot p_n \E - m_n \omega_n^2 q_n + c_n x \> , \hspace{1cm} \dot q_n \E 
\frac{p_n}{m_n}
\label{Beweggl Umgeb}
\ee
and for those of the system
\be
\dot p \E - \frac{\partial V}{\partial x} +  \sum_{n=1}^N \, c_n q_n - 
x \sum_{n=1}^N \, \frac{c_n^2}{2 m_n \omega_n^2} \> , \hspace{1cm}
\dot x \E \frac{p}{m} \> .
\label{Beweggl System}
\ee
Next, we solve Eq. (\ref{Beweggl Umgeb}) for the bath coordinates
in such a way that we consider the coordinate  $ x(t )$ 
of the particle as a given function of time. Then the inhomogeneous differential equation
has the solution
\be
q_n(t) \E q_n(0) \, \cos (\omega_n t) + \frac{p_n(0)}{m_n \omega_n} \, 
\sin (\omega_n t) + \frac{c_n}{m_n \omega_n} \, \int_0^t ds \> \sin \left [ 
\omega_n (t-s) \right ] \, x(s) \> ,
\ee
and inserting that result into Eq. (\ref{Beweggl System}) gives
\bea
&& m \, \ddot x(t) -  \int_0^t ds \> \sum_{n=1}^N \,  \frac{c_n^2}{m_n \omega_n^2} \, 
\sin \left [ \omega_n (t-s) \right ] \, x(s) +    \frac{\partial V}{\partial x} 
+ x(t) \sum_{n=1}^N \, \frac{c_n^2}{m_n \omega_n^2} \non
\hspace{1cm} \EA  \sum_{n=1}^N c_n \left [ \, q_n(0) \, \cos (\omega_n t) + 
\frac{p_n(0)}{m_n \omega_n} \, \sin (\omega_n t) \, \right ] \> .
\eea
With the help of an integration by parts in the second term on the l.h.s.
the equation of motion of the particle can be brought into the final form
\be
m \, \ddot x(t)  + m \int_0^t ds \> \gamma(t-s) \> \dot x(s) + 
\frac{\partial V}{\partial x} \E \xi(t) \> .
\label{gedaempfte Beweggl} 
\ee
Note that the damping term needs information about the particle coordinate
at earlier times ("{\bf memory effect}"). On the r.h.s. we have a fluctuating force
\be
\xi(t) \E \sum_{n=1}^N c_n \left [ \, \left ( q_n(0) - \frac{c_n}{m_n \omega_n^2} 
x(0) \right ) \, \cos (\omega_n t) + 
\frac{p_n(0)}{m_n \omega_n} \, \sin (\omega_n t) \, \right ] \> ,
\ee
which vanishes if averaged over the bath degrees of freedom.
The damping kernel
\be
\gamma(t) \E \frac{1}{m} \, \sum_{n=1}^N \, \frac{c_n^2}{m_n \omega_n^2} \, 
\cos (\omega_n t)
\ee
can be expressed by the spectral density of the bath oscillators
\be
J(\omega) \Def \pi \sum_{n=1}^N \, \frac{c_n^2}{2 m_n \omega_n} \, \delta 
\left ( \omega - \omega_n \right ) 
\label{def Spektraldichte}
\ee
as is easily verified:
\be
\gamma(t) \E \frac{2}{m \pi} \, \int_0^{\infty} d\omega \> 
\frac{J(\omega)}{\omega} \, \cos (\omega t) \> .
\label{gamma durch J}
\ee
Thus, for practical applications there is no need to quantify all parameters $ m_n, 
\omega_n $ and $ c_n $ which enter Eqs. (\ref{H Umgebung},
\ref{H Wechselwirk}); it is sufficient to specify the spectral density $ J(\omega) $ .
The form most frequently used is called
{\bf ``Ohmic damping''}
\be
J^{\rm Ohm} (\omega) \E m \gamma \, \omega \> ,
\label{Ohmsche Daempf}
\ee
which leads to the damping term $ \gamma(t) = 2 \gamma \, \delta(t) $ and thus to
a classical damping (without "memory"). However, a realistic spectral density 
cannot grow without limit and thus Eq. 
(\ref{Ohmsche Daempf}) has to be modified at high frequencies. One possibility
is the suppression of frequencies above a characteristic 
{\bf Drude frequency} $\omega_D$: 
\be
J^{\rm Drude}(\omega) \E m \gamma \, \omega \, \frac{\omega_D^2}{\omega^2 + 
\omega_D^2} \> \> \Longrightarrow \> \> \gamma(t) \E \gamma \, \omega_D \, 
\exp \left (-\omega_D |t| \right ) \> .
\label{Drude Daempf}
\ee

At first sight,
the description of the environment by a  "heat bath" of harmonic
oscillators and the bilinear coupling of the particle to it may appear problematic
because the system can return to its original state after a sufficient long time
and therefore no real, irreversible damping ever happens. However, if the number 
$ N $ of oscillators goes to infinity this so-called \textcolor{blue}{\bf Poincar\'e recurrence time}
also goes to infinity and the system  {\bf always} loses energy to the environment.
This is also exactly the limiting case, in which the spectral density $ J(\omega) $ 
becomes a continous function rather than a sum of $ \delta $-functions.

\vspace{0.5cm}

Now we want to consider the \textcolor{blue}{\bf quantum mechanical} description 
of the system by means of Euclidean path integrals. Thus we consider
the full partition function of the total system at a temperature
$ T = 1/(k_B \beta)$ 
(and we omit the index "E" for "Euclidean" in the following)
\bea
\hspace{-0.5cm} Z(\beta) \EA  {\rm tr}_{\rm \> system + bath + interaction} 
\left ( \, \exp ( - \beta \hat H) \, \right ) \non
\hspace{-0.5cm}\EA \! \! \oint {\cal D}x  \left ( \prod_{n=1}^N {\cal D} q_n \! 
\right ) \exp \left [  - \frac{1}{\hbar} \left (  S_{\rm system}[x] 
+  S_{\rm bath}[q_n] +  S_{\rm interaction}[x,q_n] \right ) \right ] ,
\eea
with
\bea
S_{\rm system}\, [x] \EA \int_0^{\beta \hbar} d\tau \> \left [ \, \frac{m}{2} 
\dot x^2 + V(x) \, \right ] \\
S_{\rm bath} \, [q_n] \EA  \int_0^{\beta \hbar} d\tau \> \sum_{n=1}^N 
\, \frac{m_n}{2} \left ( \dot q_n^2 + \omega_n^2 q_n^2 \right ) \\
S_{\rm interaction} \, [x,q_n] \EA \int_0^{\beta \hbar} d\tau \> \sum_{n=1}^N
\, \left ( \, - x \, c_n q_n + x^2 \, \frac{c_n^2}{2 m_n \omega_n^2} \, \right ) 
\> .
\eea
The trace in the definition of the partition function again requires periodic
boundary conditions for all paths and integration over the endpoints.
We can write the partition function in the following form
\be
\boxed{
\qquad Z(\beta) \E \E \oint {\cal D}x  \> \exp \left ( \,   - \frac{1}{\hbar} 
 S_{\rm system} [x] \, \right ) \, {\cal F} [x] \qquad
}
\label{def Einfluss}
\ee
where the \textcolor{blue}{\bf influence functional} $  \> {\cal F} [x]\>  $ 
is given as a product of path integrals for each environmental oscillator:
\bea 
{\cal F} [x] \EA  \prod_{n=1}^N \,  {\cal F}_n [x]  \\
{\cal F}_n [x] \EA  \oint {\cal D} q_n \> 
\exp \left \{ \, - \frac{1}{\hbar} \int_0^{\beta \hbar} d\tau \, 
\frac{m_n}{2} \left [ \dot q_n^2(\tau) + 
\omega_n^2 \left (q_n(\tau) - \frac{c_n}{m_n \omega_n^2} x(\tau) \right )^2 \right ]
\, \right \} \> .
\eea
Similar as in the polaron case we can perform the 
path integral over the bath oscillators exactly as each mode~$n$ is a forced harmonic 
oscillator.
Indeed, the application of  Eq. (\ref{Z erzw. hO}) immediately gives
\bea
{\cal F}_n [x] \EA Z_n^{\rm h.o.} \, \exp \left  [ \, \frac{1}{\hbar}
\int_0^{\beta \hbar} d\tau \int_0^{\tau} d\tau' \> 
K_n(\tau - \tau') \, x(\tau) x(\tau') \, - \frac{c_n^2}{2m_n \omega_n^2 \hbar} \, 
\int_0^{\beta \hbar} d\tau \> x^2(\tau) \right  ]
\> ,
\label{Einfluss n}
\eea
where the kernel reads
\bea
K_n(s) \EA \frac{c_n^2}{2m_n \omega_n}  \, \frac{\cosh \left ( \omega_n 
(\beta \hbar/2 - s) \right )}{\sinh \left ( \omega_n \beta \hbar/2 \right )} 
\> \> , \quad s > 0 
\eea
(see Eq. (\ref{Kern erzw. hO})). Different from the polaron problem where we only
were interested in the ground-state energy (i.e. $ \beta \to \infty $), it is not 
possible to simplify it for finite temperatures.
Note that the kernel is symmetric around  $ s = \beta \hbar/2 $ . Since we only need it 
in the interval $ [0, \beta \hbar] $ , we may assume periodicity outside this
interval and expand it in a Fourier series
\be
K_n(s) \E \sum_{l=-\infty}^{\infty} c_l^{(n)} \, e^{i \nu_l s} \> , 
\label{Kern Fourier}
\ee
where
\be 
\nu_l \Def \frac{2 \pi l}{\beta \hbar}
\label{def Matsubara Freq}
\ee
are the so-called \textcolor{blue}{\bf Matsubara frequencies}. The coefficients are given by
\be
c_l^{(n)} \E \frac{1}{\beta \hbar} \, \int_0^{\beta \hbar} ds \> K_n(s) \,  
e^{- i \nu_l s} \E \frac{c_n^2}{m_n \beta \hbar} \, \frac{1}{\omega_n^2 + \nu_l^2} \> .
\label{Fourier Koeff Kern}
\ee
Using the result (\ref{Z hO}) we obtain
\be
{\cal F}_n [x] \E \frac{1}{2 \sinh (\beta \hbar \omega_n/2 )} \, 
\exp \left [ \, - \frac{1}{\hbar} \int_0^{\beta \hbar} d\tau \int_0^{\tau} 
d\tau' \>  k_n(\tau - \tau') \, x(\tau) x(\tau') \, \right ] \> .
\ee
Here we have split up the Fourier expansion of the kernel $ K_n(s) $ (see Eqs. 
(\ref{Kern Fourier}, \ref{Fourier Koeff Kern})) into
\bea
K_n(s) \EA \frac{c_n^2}{m_n \omega_n^2 \beta \hbar} \, \sum_{l=-\infty}^{\infty} \, 
e^{i \nu_l s} - \frac{c_n^2}{m_n \omega_n^2 \beta \hbar} \, 
\sum_{l=-\infty}^{\infty} \, \frac{\nu_l^2}{\omega_n^2 + \nu_l^2} \, e^{i \nu_l s} 
\non
&=:& \frac{c_n^2}{m_n \omega_n^2 } \, \sum_{j=-\infty}^{\infty} \, 
\delta \left ( s - j \beta \hbar \right ) - k_n(s) \> .
\eea
In the interval $ \> [0, \beta \hbar] \>  $ only the  $ j = 0 $-term contributes to the sum
of $\delta$-functions and exactly cancels the last term in Eq. (\ref{Einfluss n}), 
i.e. the explicit "potential"-term in the interaction between system and
environment (caution: Due to the integration limits the 
$\delta$-functions contribute only half of their strength!) 
It can be shown that the reduced kernel $ k_n(s) $ does not contain a local (i.e. one-time)
contribution anymore.
\vspace{0.2cm}

\noindent
The full influence functional therefore is
\be
\boxed{
\qquad {\cal F} [x] \E \prod_{n=1}^N \left ( \frac{1}{2 \sinh (\beta \hbar \omega_n)} 
\right ) \, \exp \left [ \, - \frac{1}{\hbar} \int_0^{\beta \hbar} 
d\tau \int_0^{\tau} d\tau' \> 
k(\tau - \tau') \, x(\tau) x(\tau') \, \right ] \> , \quad
}
\label{Einfluss}
\ee
with
\be 
k(s) \E \sum_{n=1}^N \frac{c_n^2}{m_n \omega_n^2} \, \frac{1}{\beta \hbar} \, 
\sum_{l=-\infty}^{+\infty} \, \frac{\nu_l^2}{\nu_l^2 + \omega_n^2} \, e^{i \nu_l s} 
\EQ  \frac{2}{\pi \beta \hbar} \, \int_0^{\infty} d\omega \> 
\frac{J(\omega)}{\omega} \, \sum_{l=-\infty}^{+\infty} \, 
\frac{\nu_l^2}{\nu_l^2 + \omega^2} \, e^{i \nu_l s}  \> .
\label{voller red Kern 1}
\ee
In the second line we have used the definition (\ref{def Spektraldichte}) of
the spectral density of the bath oscillators. As the damping kernel
is determined as well by the spectral density via 
Eq. (\ref{gamma durch J}) one can express the integrand in Eq. (\ref{voller red Kern 1}) 
also by the Laplace transform of $\gamma(t)$ 
\be
\tilde \gamma(z) \Def \int_0^{\infty} dt \> \gamma(t) \, e^{-z t} \E
\frac{2}{\pi m} \int_0^{\infty} d\omega \> 
\frac{J(\omega)}{\omega} \, 
\frac{z}{z^2 + \omega} \> .
\ee
Thus
\be
k(s) \E \frac{m}{\beta \hbar} \,
\sum_{l=-\infty}^{+\infty} \, |\nu_l| \, \tilde \gamma \left ( |\nu_l| \right )  
\, e^{i \nu_l s}  \> .
\label{voller red Kern 2}
\ee
The Laplace transform of the Drude damping (\ref{Drude Daempf}) is
\be
\tilde \gamma^{\rm Drude} (z) \E \gamma \, \frac{\omega_D}{\omega_D + z} \> ,
\ee
which reduces to $  \gamma^{\rm Ohm} (z) = \gamma $ for pure Ohmic damping.

As in the polaron case we have seen that the influence of the environment
can be taken into account by adding a non-local
(\textcolor{blue}{\bf two-time}) term to the original action.
The properties of the infinitely many bath oscillators can be modelled 
by simple {\it ans\"atze}. 
\vspace{1cm}

\noindent
{\bf Examples} : 
\vspace{0.6cm}

\noindent
{\bf a) The Damped Harmonic Oscillator}
\vspace{0.2cm}

\noindent
If the particle moves in a harmonic potential
\be
V(x) \E \frac{1}{2} m \, \omega_0^2 \, x^2
\ee
all further steps can be done analytically.
The partition function of the undamped harmonic oscillator is given by
Eq. (\ref{Z hO}), or in the product representation of Eq. (\ref{HO Vor prod})
\be
Z^{\, \rm h.o.} \E \frac{1}{\beta \hbar \omega_0} \, \prod_{k=1}^{\infty} \, 
\frac{\nu_k^2}{\nu_k^2+\omega_0^2} \> .
\ee
If the fluctuations for the damped harmonic oscillator are also expanded in a Fourier series,
we may perform the path integral (\ref{def Einfluss}) with the influence functional
 (\ref{Einfluss}) and obtain a result which is modified by damping
\be
Z^{\, \rm damped \, h.o.} \E \frac{1}{\beta \hbar \omega_0} \, \prod_{k=1}^{\infty} \, 
\frac{\nu_k^2}{\nu_k^2+ \nu_k \tilde \gamma(\nu_k) + \omega_0^2} \> .
\label{Z gedaempft hO}
\ee
Which information is contained in Eq. (\ref{Z gedaempft hO})? First, we can determine
the free energy (\ref{def freie En}) at inverse temperature $\beta$
\be
F^{\, \rm damped \, h.o.} \E \frac{1}{\beta} \, \ln \left (\beta \hbar \omega_0 \right ) 
+ \frac{1}{\beta} \, \sum_{k=1}^{\infty} \, \ln \left ( 1 + 
\frac{\tilde \gamma(\nu_k)}{\nu_k} + \frac{\omega_0^2}{\nu_k^2} \, \right )
\ee
In the limit $\beta \to \infty $  the ground-state energy of
the damped harmonic oscillator follows from that
\be
E_0^{\, \rm damped \, h.o.} \E \frac{\hbar}{2 \pi} \, \int_0^{\infty} d\nu \> 
\ln \left ( 1 + 
\frac{\tilde \gamma(\nu)}{\nu} + \frac{\omega_0^2}{\nu^2} \, \right ) \> ,
\ee
since the difference between the Matsubara frequencies shrinks to zero and
the sum can then be replaced by an integral.
For Ohmic damping with Drude regularization the partition function as well as the 
ground-state energy can be given in closed analytical form {\bf \{Weiss\}}. 
However, it is instructive to consider the results for weak damping:
\bea
E_0^{\, \rm damped \, h.o.} & \to & \frac{\hbar}{2 \pi} \, \int_0^{\infty} d\nu \> 
\ln \left ( 1 + \frac{\omega_0^2}{\nu^2} \, \right ) + 
\frac{\hbar}{2 \pi} \, \int_0^{\infty} d\nu \> \frac{\nu}{\nu^2 + \omega_0^2} \, 
\tilde \gamma(\nu) + \ldots \non
\EA \frac{1}{2} \hbar \omega_0 + \frac{\hbar}{2 \pi m} \, \int_0^{\infty} d\omega
\> \frac{J(\omega)}{\omega (\omega + \omega_0)} \> .
\eea
With the spectral density  (\ref{Drude Daempf}) one obtains
\be 
E_0^{\, \rm damped \, h.o.} \E \frac{1}{2} \hbar \omega_0 + \gamma \, \left [\,  
\frac{\hbar}{2 \pi} \, \ln \left ( \frac{\omega_D}{\omega_0} \right ) + 
{\cal O} \left ( \omega_D^{-1} \right ) \, \right ] 
+  {\cal O} \left ( \gamma^2 \right ) \> ,
\label{E0 Ord gamma}
\ee
which is in agreement with second-order perturbation theory for
the interaction term (\ref{H Wechselwirk}) (note that Eq. 
(\ref{E0 Ord gamma}) diverges if the cut-off frequency $ \omega_D $ 
goes to infinity).
Thus, the coupling to the environmental degrees of freedom 
leads to an energy shift, similar as in the polaron problem
or in the interaction of atoms with the radiation field (``Lamb shift''). 
Actually, the similarity with the latter system is even more pronounced since
the excited states also get a finite lifetime by the damping.
This can be determined, e.g., from the density of states of the damped system
\be
Z(\beta) \deF  \int_0^{\infty} dE \> \rho(E) \, e^{-\beta E}
\> \> \Longrightarrow \> \> \rho(E) \E \frac{1}{2 \pi i} \, 
\int_{c-i\infty}^{c-i\infty}
d\beta \> Z(\beta) \, e^{\beta E} \>. 
\ee
In the inverse Laplace transform one has to choose the constant $ c $ such 
that the path of integration is to the right of all
poles of the integrand. The explicit calculations \cite{HaZw}
show that instead of a sum of $ \delta $- functions for the
undamped harmonic oscillator, the density of state of the
damped oscillator consists of  one $ \delta $-function for the ground state and
(for small damping) a series of peaks which are slightly shifted from the 
excitation energies of the pure harmonic oscillator while they become 
broader and broader for increasing excitation energy until
they cannot be resolved as individual peaks anymore.
This is in qualitative and quantitative agreement with Fermi's "Golden Rule'' 
according to which the width of the $ n^{\rm th} $ level is given by
\be
\Gamma_n \E \frac{2 \pi}{\hbar^2} \, \sum_{j=1}^{\infty} \> 
\Bigl | \left < n+1,1_j \, | c_j q x_j | \, n, 0 \right > \Bigr |^2 \, \delta 
\left ( \omega_0 - \omega_j \right ) \> .
\ee 
Here we have used that the dipole interaction
(\ref{H Wechselwirk}) can only connect neighbouring states and that the environment 
cannot give energy but only can take it.
With the help of the matrix element
\be
\left < n+1,1_j \, | c_j q \, x_j | \, n, 0 \right > \E 
\frac{\hbar}{2 \sqrt{m m_j \omega_0 \omega_j}} \, n^{1/2}
\ee
one finds
\be
\Gamma_n \E \frac{n}{m \omega_0} \, J(\omega_0) \> \stackrel{\rm Ohm}{=} \> 
n \, \gamma \> ,
\label{Gamma n}
\ee
which is particularly simple for Ohmic damping. As expected the width increases 
with increasing damping  $\gamma$ and level number  $ n $ .
\vspace{1cm}

\noindent
{\bf b) Structure Function of the Damped Harmonic Oscillator}
\vspace{0.2cm}

\noindent
This can also be seen in the structure function \eqref{def Struk} which we have studied for
a harmonically bound particle in {\bf chapter} {\bf \ref{sec1: greensche Funk}} .
It is possible to apply the formalism used for dissipative quantum systems also to this 
process either by an analytic continuation of the temperature result to real times
 \cite{Ros2} or by direct calculation of the Green function
\be
G_{{\mathcal {O}} {\mathcal {O}}^{\dagger}}^{\> {\rm damped \, h.o.}}(T) \E {\rm const.} \, 
\int {\cal D}x \> 
\exp\lsp i \int_{-\infty}^{+\infty} dt \lrp \frac{m}{2} \dot x^2 - \frac{m}{2} \omega_0^2 x^2 - 
b(t) \cdot x \rrp \rsp \> \prod_{n=1}^N {\cal F}_n[x] \> ,
\ee
where the "impulse function" $ b(t) $  has been defined in Eq. \eqref{def b(t)}.
The influence functional in real time is now
\bea
{\cal F}_n[x] \EA \int {\cal D}q_n \> \exp\lcp i \int_{-\infty}^{+\infty} dt \lsp \frac{m}{2} \dot q_n^2 
- \frac{m_n}{2} \omega_n^2  \lrp q_n - \frac{c_n}{m_n \omega_n^2} \, x \rrp^2 \rsp \rcp \non
\EA {\rm const'.} \, \exp \lcp - i \frac{c_n^2}{2 m_n} \int_{-\infty}^{+\infty} dt \, dt' \> x(t)
\lrp t \lvl \frac{1}{-\partial^2_t - \omega_n^2 + i 0^+} + \frac{1}{\omega_n^2} \rvl t' \rrp \, x(t') 
\rcp \> .
\eea
Not unexpected the
functional integration over the particle coordinate  $ x(t)$ can be done as a Gaussian integral 
and with the correct normalization at $ q = 0 $ one obtains instead of Eq. \eqref{G HO}
\be
G_{{\mathcal {O}} {\mathcal {O}}^{\dagger}}^{\> {\rm damped \, h.o.}}(T) \E 
\exp \lsp - i \frac{q^2}{2m} \int_{-\infty}^{+\infty} \frac{dE}{\pi}  \frac{ 1 - \cos (ET)}{E^2 - \omega_0^2 
- \Sigma(E)} \rsp \> .
\label{G ged HO}
\ee
The effect of coupling the particle to additional degrees of freedom shows up as a complex
"self-energy"
\be
\Sigma(E) \E \sum_{n=1}^N \frac{c_n^2}{m m_n} \lrp \frac{1}{E^2 - \omega_n^2 + i 0^+} + \frac{1}{\omega_n^2} 
\rrp \> \stackrel{N \to \infty}{\longrightarrow}\>
\frac{2}{\pi} \, \int_0^{\infty} d\omega \> \frac{J(\omega)}{m \omega} \, \frac{E^2}{E^2 - \omega^2 + i 0^+} \> ,
\ee
where $ J(\omega) $  is the (continous)  spectral density of the environmental oscillators 
defined in Eq. \eqref{def Spektraldichte}. One sees that 
 $ \Sigma(E) $ is even and vanishes at $ E = 0 $ : Thus the ground state remains a sharp line.
Due to the decomposition \eqref{Zerleg Green} the imaginary part is always negative, i.e. excitations
now have a {\bf finite width}. For Ohmic damping \eqref{Ohmsche Daempf} one finds
\be
\Sigma^{\rm Ohm}(E) \E - i \gamma \, |E| \> ,
\ee
which explains the growing width  \eqref{Gamma n} because $ E \simeq n \omega_0 $ . 
\vspace{1cm}


\renewcommand{\baselinestretch}{0.9}
\scriptsize
\noindent
\begin{subequations}

\noindent
As in Eq. \eqref{elast form HO} the square of the elastic form factor is given by the 
$T$-independent part of Eq. \eqref{G ged HO}. A simple calculation gives again a Gaussian 
dependence on the momnetum transfer
\be
\lrp F_{00}^{\> {\rm  damped \, h.o.}}(q) \rrp^2  \E \exp \lsp - i \frac{q^2}{2m} \, \int_{-\infty}^{+\infty} 
\frac{dE}{\pi} \> \frac{1}{E^2 - \omega_0^2 \, + i \gamma |E|}\rsp \E 
\exp \lsp - \frac{q^2}{2m \Omega} \, \frac{2}{\pi} \arctan \lrp \frac{2 \Omega}{\gamma} \rrp \rsp
\> ,
\ee
but with the shifted frequency
\be
\Omega \Def \sqrt{\omega_0^2 - \frac{\gamma^2}{4}} \le \omega_0 
\ee
(here we only consider the so-called oscillatory case $ \gamma \le 2 \omega_0 $). The coupling to the
bath degrees of freedom thus reduces the target size:
The mean square radius which can be measured in the elastic form factor
($ \> F_{00}(q) \to 1 - q^2 \la r^2 \ra/6 + \ldots $ for $ q \to 0 \> $) becomes for small damping
\be 
\la r^2 \ra \E \frac{3}{2 m \omega_0} \, \lsp 1 - \frac{\gamma}{\pi \omega_0} + {\cal O} \lrp \gamma^2\rrp \rsp \> .
\ee
\end{subequations}

\renewcommand{\baselinestretch}{1.2}
\normalsize

\newpage

\vspace{0.3cm}

\refstepcounter{abb}
\begin{figure}[hbtp]
\bce
\includegraphics[angle=90,scale=0.45]{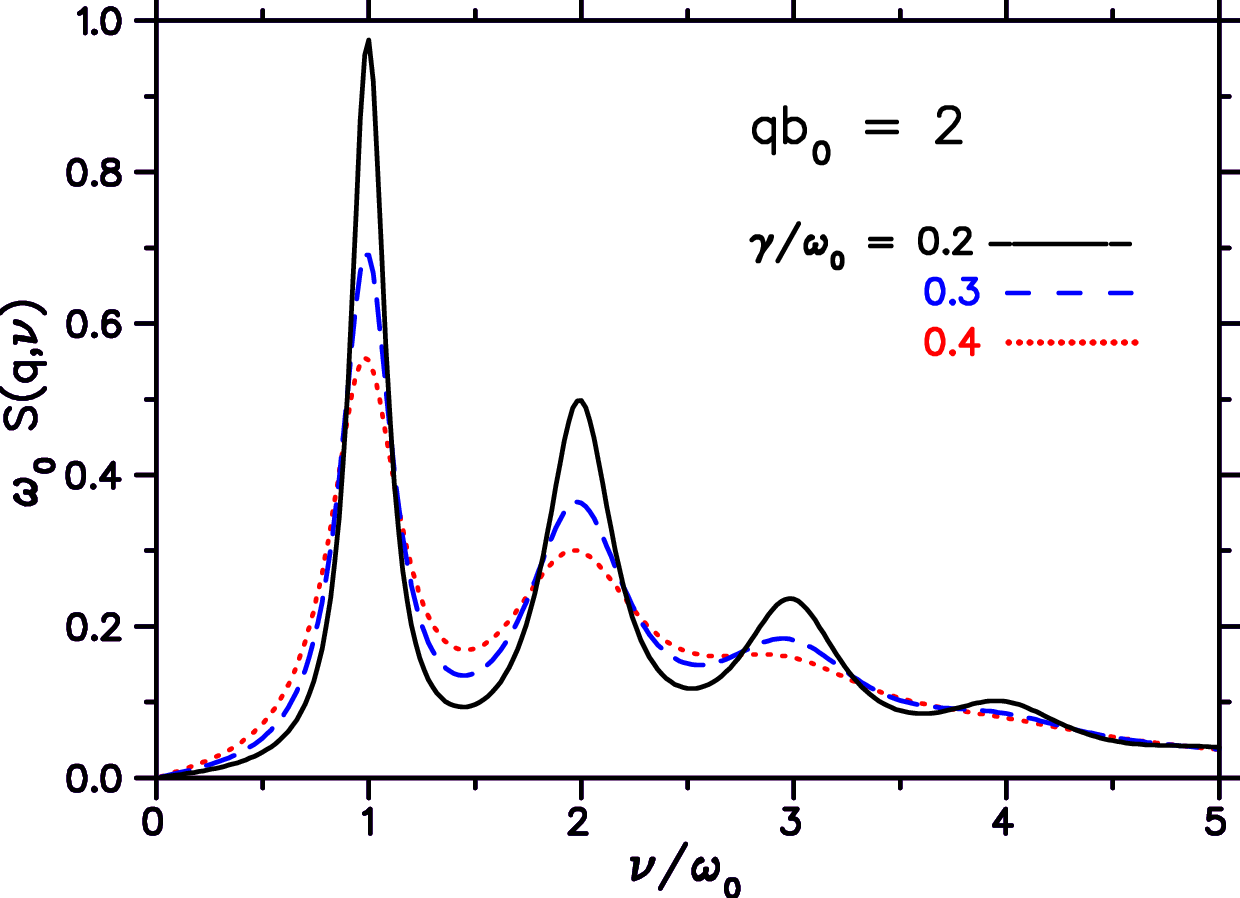} \\
\vspace{0.5cm}

\includegraphics[angle=90,scale=0.45]{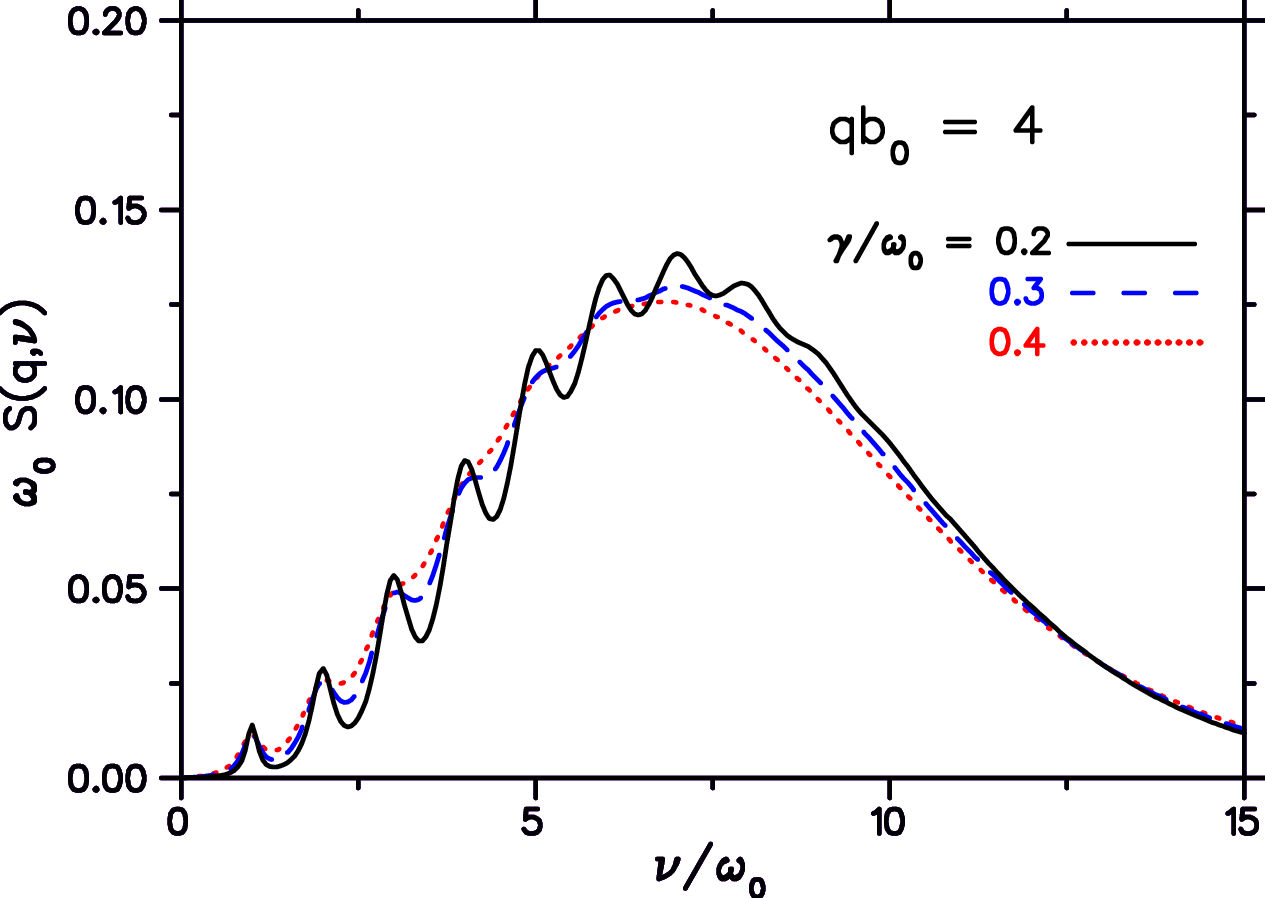}
\ece
\vspace*{0.5cm}
{\bf Fig. \arabic{abb}} : Inelastic structure function of the damped
harmonic oscillator for several momentum transfers $ q $ \\
\hspace*{1.5cm}  and Ohmic damping parameters $\gamma \>  $ \cite{Ros2}.
The oscillator frequency of the undamped system is denoted\\
\hspace*{1.5cm} by  $ \> \omega_0 $ , $ b_0 = (m \omega_0)^{-1/2} $ 
is the oscillator length and serves as unit of length.
\label{abb:2.3.1}
\end{figure}

\vspace{0.6cm}

\noindent
Fig. \ref{abb:2.3.1} shows the result of a calculation for two different 
momentum transfers and several values of the Ohmic damping parameter 
 $\gamma$. This damping may simulate the coupling to more complicated 
states like the continuum in which parts of the target have been ejected or the 
conversion of the struck quark inside the proton into observable hadrons.
As can be seen, at high momentum transfer the individual excitation lines 
merge into a continous curve, the "quasielastic peak" which has its maximum
at $ \nu \simeq q^2/(2m) $ and whose width reflects the momentum distribution
of the bound particle. Starting from a consistent Hamiltonian for system + environment
there are no unphysical excitations with $\nu < 0$ : The individual broadened lines
are therefore not exact Lorentz- or Breit-Wigner curves which also would extend
to negative excitation energies. Therefore the sum rule obtained by
integration over  $\nu$ from Eq. (\ref{def Struk}) 
\be
\int_0^{\infty} d\nu \> S^{\> {\rm inelast.}} \left (q, \nu \right ) +  F_{00}^2(q)  \E 
 \Bigl <  \,  0 \,  \Bigl | \, e^{- i q \cdot \hat x} \> \sum_n \, \Bigr | \, n \, \Bigr >   \Bigl <  n \,\left |  \> 
e^{i q \cdot \hat x} \, \right | \, 0 \Bigr > \E \la 0\,  | \, 0 \ra  \> \E 1
\ee
is exactly conserved. It is based on the completeness of all target states
and is independent of the special form of the Hamiltonian and its eigenstates.
Thus, by coupling to other degrees of freedom the spectrum is shifted and distorted but the 
total strength remains the unchanged.

\vspace{1.5cm}

\subsection{\textcolor{blue}{Particle-Number Representation \hspace{-0.5mm}and 
\hspace{-0.5mm}Path Integrals
\hspace{-1mm}over Coherent \hspace{-1mm} 
States}}
\label{sec2: Pfad ueber kohaer Zust}

While in the first chapter of this section the particles always moved in an external potential
we now consider a genuine many-body problem: A non-relativistic system of
 $ N $ \textcolor{blue}{\bf identical} particles interacting via
 two-body forces:
\be
\hat H \E \sum_{i=1}^N \frac{ \hat p_i^2}{2 m} + \sum_{i < j} \hat V_{ij}
\E \sum_{i=1}^N \frac{ \hat p_i^2}{2 m} + \frac{1}{2} \sum_{i \neq j} 
\hat V_{ij} \> .
\label{H 1. Quant}
\ee
In systems of identical particles it is a fundamental principle of quantum mechanics 
\footnote{Also called the spin-statistics theorem and proven for local, relativistic 
 quantum field theory by Fierz and Pauli in 1939/40.} 
that the wave function for bosons (particles with integer spin) should be
\textcolor{blue}{\bf symmetric}  and the one  for fermions
(particles with half-integer spin)  \textcolor{blue}{\bf antisymmetric} under the exchange
of particles
\bea
\hat H | \> \Psi > \EA E | \> \Psi > \non
\Psi (  ... i ... j ...) \EA \zeta \> \Psi (  ... j ... i ...) \>, 
\hspace{0.5cm}
\zeta = \left \{ \begin{array}{r@{\quad:\quad}l} 
                 +1 & {\rm bosons} \\ -1 & {\rm fermions}
                 \end{array} \right. \> \> .
\eea
In the usual formulation of quantum mechanics it is cumbersome or difficult
to realize this symmetry requirement in every step of the calculation.
It is much more convenient to have an formalism at hand which incorporates 
this requirement automatically. This is the case for the
\textcolor{blue}{\bf particle-number representation} 
(also misleadingly and confusingly called \textcolor{blue}{\bf "second quantization"}), 
which is already known from treating the harmonic oscillator.
Instead of describing the state of the system by a wave function
 $ \Psi $, which depends on the coordinates and momenta of the particles, 
 one specifies how many particles are in a given state $ | \> \alpha > $ 
 of a chosen complete basis. One then defines annihilation and creation operators,
$ \> \hat a_{\alpha} \> $,
and $ \> \hat a^{\dagger}_{\alpha} \> $, respectively,
which annihilate or create a particle in the state  
$ \> | \> \alpha > $. The "vacuum" is that state of lowest energy in which there are no
particles at all:
\be
\hat a_{\alpha} | \> 0 > \E 0 \> \> \> \> \> \forall \> \alpha \> .
\ee
An one-particle operator,  e.g. the kinetic energy, then has the
following representation
\be
\hat T \E \sum_{i=1}^N \hat T_i \To
\sum_{\alpha,\beta} < \alpha \> | \> \hat T \> | \> \beta > \> 
\hat a^{\dagger}_{\alpha} \, \hat a_{\beta} \> ,
\label{kin}
\ee
and a two-particle operator, e.g. the potential energy,
\be
\hat V \E \frac{1}{2}\sum_{i \neq j}^N \hat V_{ij} \> \longrightarrow \>
\frac{1}{2} \sum_{\alpha,\beta,\gamma,\delta} 
< \alpha \beta \> | \> \hat V \> | \> \gamma \delta > \> \hat 
a^{\dagger}_{\alpha} \, \hat a^{\dagger}_{\beta} \, \hat a_{\delta} \,
\hat a_{\gamma}
\> .
\label{pot}
\ee

Note the different order of the indices in the 
two-particle matrix element and the annihilation operators. 
It is also noteworthy that the operators in the particle-number representation 
do not contain any information about the particle number
 -- the summations are unrestricted and run over all quantum numbers of the basis, 
 not over the number of particles! This is because the operators act in
 the \textcolor{blue}{\bf Fock space} which is a a direct sum of Hilbert spaces with
particle number 0, 1, 2 ... In each subspace $ {\cal H}_N $ they then have 
the same matrix elements as the operators with which
we started. Since both in the kinetic as well as in the potential
energy always pairs of $ \> \hat a^{\dagger} \, \hat a \> $
occur (see Eqs. (\ref{kin}, \ref{pot})), the number of particles remains
conserved and the dynamics of the system never carries us outside the 
chosen Hilbert space.
In contrast to this we see single  $ \> \hat a^{\dagger} \> $'s and $\> \hat a\> $'s 
in the polaron Hamiltonian (\ref{Froehlich H}) -- a typical feature of a field theory is 
that particle (here: phonon) number may not be conserved.
\vspace{0.2cm}

The particle-number representation is extremely advantageous in many-body physics because 
the symmetry requirement is automatically built into the commutation relations:
\bce
\vspace{0.2cm}

\fcolorbox{blue}{white}{\parbox{8cm}
{
\bea
 \left [ \hat a_{\alpha}, \hat a^{\dagger}_{\beta} \right ]_{-\zeta}  \Def
\hat a_{\alpha} \> \hat a^{\dagger}_{\beta} \> - \zeta 
\hat a^{\dagger}_{\beta}
\> \hat a_{\alpha}   \EA\delta_{\alpha \beta} \non
\left [ \hat a_{\alpha}, \hat a_{\beta} \right ]_{-\zeta} \E 
 \left [ \hat a^{\dagger}_{\alpha}, \hat a^{\dagger}_{\beta} 
\right ]_{-\zeta} \EA 0 \> , \no
\eea
}}
\ece
\vspace{-2cm}

\bea
\label{Kommutatoren}
\eea
\vspace{0.4cm}

\noindent
i.e. for bosons one has to use commutators and for fermions anti-commutators.
The latter realize the \textcolor{blue}{\bf  Pauli principle} in the following way
\be
\hat a^{\dagger \> 2}_{\alpha} | \> 0 > \E - 
\hat a^{\dagger \> 2}_{\alpha} |\>  0 > \E 0 \> ,
\ee
which means that two fermions cannot exist in the same state.

\noindent
The one-particle basis must be orthonormal and complete
\be
< \alpha | \> \beta > \E \delta_{\alpha \beta} \> ,
\hspace{0.4cm} \sum_{\alpha} | \> \alpha > < \alpha \> | \E \hat 1 \> ,
\ee
but otherwise can be arbitrary. For an electron, e.g., one could use the 
states of a 3-dimensional harmonic oscillator characterized by
$ \> n \> $ = prinipal quantum number, $ \> l \> $ = orbital angular momentum,
$ \> j \> $ = total angular momentum, $ \> m_j \> $ = magnetic quantum number.
For nucleons one would have to add the isospin, for quarks the color etc.
Going over to another basis, e.g. to the position basis, is possible by the transformation
\be
\hat a_\fr \EQ \hat \phi(\fr) \E \sum_{\alpha} < \fr | 
\> \alpha > \hat a_{\alpha} \hspace{1cm}  {\rm etc.} \> .
\ee
The \textcolor{blue}{\bf field operators} $ \> \hat \phi(\fr) , 
\hat \phi^{\dagger}(\fr) \> $ annihilate or create a particle at the point $ \> \fr \>$
(with the other quantum numbers not written out explicitly).
 It is easily found that they obey the 
commutation relations
\be
\left [ \hat \phi(\fr), \hat \phi^{\dagger}(\fr') \right]_{-\zeta}
\E \delta \left ( \fr - \fr' \right ) \> .
\ee
For a local potential $ \hat V_{ij} = V(\hat x_i - \hat x_j) $ the Hamiltonian
then reads
\be
\hat H \E \int d^3r \> \hat \phi^{\dagger}(\fr) \left ( -
\frac{\hbar^2}{2 m} \Delta\right ) \hat \phi(\fr) + \frac{1}{2} \int d^3r \, 
d^3r' \> \hat \phi^{\dagger}(\fr) \,  \hat \phi^{\dagger}(\fr') \> 
V(\fr - \fr') \> \hat \phi(\fr') \, \hat \phi(\fr) \> .
\label{H op}
\ee
If one writes $ \> \hat H = \int d^3 r \> \hat {\cal H}(\fr) \> $ 
then the form of the Hamilton density  $ {\cal H} $ is very similar to that of 
a $\phi^4$- field theory the only difference being that the interaction is
non-local here: The fields are not taken at the same position
which is not necessary in a non-relativistic theory with an instantaneous potental.
\vspace{0.2cm}


Due to the great importance of the particle-number representation it is 
advantageous to choose a basis where
the states are no longer eigenstates of the position operator but
of the annihilation operator when deriving the path integral \footnote{\label{Pfad-Bezeich}
Although the visualization as "sum over all paths" doesn't exist anymore,
we will still speak of  \blau{\bf path integrals} for convenience. The name
"functional integral" is more general but less physical.}.
These are the \textcolor{blue}{\bf coherent states} 
which are also important in quantum optics (Glauber 1963). 
This representation goes also under the name "{\bf holomorphic}" or "{\bf Bargmann}" representation.
In the following some properties are collected for the case of 
a particle in a harmonic potential
\bea
\hat H = \frac{\hat p^2}{2 m} + \frac{1}{2}m \omega^2 \hat x^2  \EA \> 
\hbar \omega \left ( \hat a^{\dagger} \hat a + \frac{1}{2} \right ) 
\label{H oszill}\\
\hat x = \sqrt{\frac{\hbar}{2 m \omega}} \left ( \hat a^{\dagger} + 
\hat a \right ) \> , \hspace{0.4cm} \hat p \EA i 
\sqrt{ \frac{m \hbar \omega}{2} } \left ( \hat a^{\dagger} - \hat a \right )
 \> , \hspace{0.4cm} \left [ \hat a , \hat a^{\dagger} \right ] \E 1 \> ,
\eea
namely

\vspace{0.5cm}

\bdes

\item[a)] Definition: Coherent states are eigenstates of the annihilation
operator \footnote{One can easily see that the creation operator doesn't have
eigenstates since it raises the minimal particle number by one.}
\bce
\vspace{0.3cm}

\fcolorbox{blue}{white}{\parbox{6cm}
{
\bea
\hat a \> | \> z > \EA z \> | \> z > \quad ,  \> \> z \quad
{\rm complex} \> . \no
\eea
}}
\ece
\vspace{-2cm}

\bea
\label{def kohaerent}
\eea
\vspace{0.3cm}


\noindent
Since  $ \hat a $ is not hermitean, the eigenvalue is necessarily complex.

\item[b)] Explicit form:
\be
\boxed{ 
\qquad | \> z > \E \exp \left ( z \hat a^{\dagger} \right ) \> | \> 0 >
\E \sum_{n=0} \frac{z^n}{\sqrt{ n!}} \> | \> n  > \> , \quad
}
\ee
where $ \> | \> n > = \left ( a^{\dagger} \right )^n | \> 0 >/\sqrt{ n!} \> $
is the normalized eigenstate of the Hamiltonian (\ref{H oszill}).

\item[c)] Coherent states are states of minimal uncertainty
(\purpur{\bf Problem \ref{kohaer Zust}}), i.e. particularly "classical".

\item[d)] Overlap:
\be
< z_2 \> | \> z_1 > \E \exp \left (  \> z_2^{\ast} z_1 \> \right ) \> ,
\label{Ueberlapp}
\ee 
i.e. coherent states are not orthogonal.

\item[e)] Representation of unity (closure relation):
\be
\boxed{ 
\qquad \hat 1 \E \frac{1}{2 \pi i} \int dz^{\ast} dz \> | z >  < z | \> 
e^{- |z|^2} \> . \quad 
}
\label{Einheit}
\ee
\vspace{0.3cm}

\renewcommand{\baselinestretch}{0.9}
\scriptsize

\begin{subequations}
\noindent
{\bf Proof :} Using polar coordinates for the complex variable
$ \> z \> $ one obtains (the Jacobian is $ 2 i \> $)
\bea
\frac{1}{2 \pi i} \int dz^{\ast} dz \> | \> z >  < z \> | \> e^{- |z|^2} 
\EA \frac{1}{\pi} \int d ( {\rm Re} \, z) \> d ( {\rm Im} \, z) \sum_{n,m} 
\frac{ z^{\ast \> n} z^m}{\sqrt{n!\>  m! }} \> | \> m > < n \> | \> 
e^{- |z|^2} \non
\EA \frac{1}{\pi} \sum_{n,m} \frac{ | \> m > < n \> |}{\sqrt{n!\>  m! }} 
\> \int_0^{\infty} dr \> r e^{-r^2} 
r^{m+n} \int_0^{2 \pi} d\phi \> e^{-i n \phi + i m \phi} \non
\EA \sum_n \frac{| \> n > < n \> |}{n!} \> 2 \int_0^{\infty} \! \! dr \> 
r^{2 n + 1} e^{-r^2}  =  \sum_n | \> n > < n\>  | \E \hat 1 \>  .
\label{1 aus kohaerenten Zu}
\eea
\end{subequations}
\renewcommand{\baselinestretch}{1.2}
\normalsize
\vspace{0.2cm}

\noindent
Because of these properties coherent states are also called 
``\blau{\bf overcomplete}'': Every state in Fock space can be 
represented but the basis vectors are linearly dependent:
\be
| \> \psi > \E \int \frac{dz^{\ast} dz}{2 \pi i} \> e^{- |z|^2} \> 
\psi(z^{\ast}) \> 
| \> z > \> \>, \> \> {\rm with} \> \> \psi(z^{\ast}) \E 
< z \> | \> \psi > \> .
\ee

\item[f)] Action of $ \hat a$ and $ \hat a^{\dagger} $:
\be
\hat a = \frac{\partial}{\partial z^{\ast}} \>, \hspace{0.4cm}
\hat a^{\dagger} = z^{\ast} \> ,
\label{a,ak kohaerent}
\ee
since $ < \!z \> | \hat a | \> \psi > = < 0 \> | \exp(z^{\ast} \hat a) \hat a 
| \> \psi > = \partial \psi(z^{\ast})/ \partial z^{\ast} $ and 
$ < \!z\> | \hat a^{\dagger} | \> \psi > = \left ( \hat a \> | \> z > 
\right )^{\dagger} | \> \psi > = z^{\ast} \psi(z^{\ast}).$  
In the holomorphic representation the  Schr\"odinger equation for 
a Hamiltonian $ H \left ( \hat a^{\dagger}, \hat a \right ) $ 
therefore is
$ \> H \left (z^{\ast}, \partial /\partial z^{\ast} \right ) \psi(z^{\ast})
= E \> \psi(z^{\ast}) \> $ . Indeed, Eq. (\ref{a,ak kohaerent}) is an
explicit representation of the bosonic commutation relations.

\item[g)] Matrix elements of operators: 
These are particularly simple if the operators are
\textcolor{blue}{\bf normal ordered}, i.e. if all annihilation operators
$\> \hat a \> $ are to the right of the creation operators
$ \> \hat a^{\dagger} \> $ . Let
$ \> \hat A \left (\hat a^{\dagger}, \hat a \right ) \> $ be such an
operator. From the definition (\ref{def kohaerent}) it then  follows immediately
\be
< z \> | \> \hat A \left (\hat a^{\dagger}, \hat a \right ) \> | \> z' >
\E A \left ( z^{\ast}, z' \right ) < z \> | \> z' > \E 
A \left ( z^{\ast}, z' \right ) \> e^{ \> z^{\ast} z'} \> .
\label{ME normal}
\ee
By inserting the unity operator  (\ref{Einheit}) we obtain 
for the trace of the operator 
\bea
{\rm tr} \> \hat A \EA \sum_n < n \> | \hat A | \> n > \E \sum_n \int
\frac{ dz^{\ast} dz}{2 \pi i} < n \> | \> z > < z \> | \hat A | \> n > 
\> e^{-|z|^2} \non
\EA \! \! \int\frac{ dz^{\ast} dz}{2 \pi i}   <\! z \> |  \hat A  \sum_n 
| \> n \!>
<\! n \> | \> z \!> e^{-|z|^2} \! = \! \int\frac{ dz^{\ast} dz}{2 \pi i}  
<\! z \> | \hat A  | \> z \!> e^{-|z|^2} \! .
\label{Spur kohaerent}
\eea

\edes
\vspace{1cm}

\noindent
Going over to \textcolor{blue}{\bf many-body systems} is easy for bosons
(fermions will be treated in the next chapter): One only has to replace
the complex eigenvalue $ z $ by a set of complex numbers
 $ \> z \To z_{\alpha} \> $ where $\alpha$ denotes the occupied 
 one-particle state and one has to sum over the states. The bosonic
 coherent state therefore is defined by
 
\be
\boxed{
\qquad | \> z > \E \exp \left ( \> \sum_{\alpha} z_{\alpha} \, \hat 
a^{\dagger}_{\alpha} \> \right ) \> | \> 0 > \qquad \quad ,
}
\ee
where here and in the following we always will characterize the state
simply by $ z \EQ  \{ z_{\alpha} \} $ .
The unit operator in bosonic Fock space is represented by
\be
\hat 1 \E \int \prod_{\alpha} 
\frac{ dz^{\ast}_{\alpha} \, dz_{\alpha}}{ 2 \pi i } \> 
e^{- \sum_{\alpha} |z_{\alpha}|^2} \> | \> z > < z \> |
\label{Einheit mehr}
\ee
and all other relations are to be rewritten correspondingly.
It should be clear that bosonic coherent states
\textcolor{blue}{\bf do not have a fixed particle number} as they are a 
superposition of states from different Hilbert spaces. Indeed, one 
finds for the mean particle number
\be
\bar N \E \frac{< z \> | \hat N | \> z >}{< z \> | \> z >} \E 
\frac{ \sum_{\alpha} < z \> | \hat a^{\dagger}_{\alpha} \, \hat a_{\alpha} 
|\> z >}{< z \> |\>  z >} \E \sum_{\alpha} | z_{\alpha}|^2 \> ,
\ee
but for the mean-square deviation
\be
\left (\Delta N \right )^2 \E \frac{< z \> | \hat N^2 | \> z >}
{< z \> | \> z >} \> 
- \> \bar N^2 \E \sum_{\alpha} | z_{\alpha}|^2 \E \bar N  \> .
\ee
Only in the thermodynamic limit in which $ \bar N  \to \infty $, the
relative deviation  $ \Delta N /\bar N = 1/\sqrt{\bar N} $ goes to zero; then
the coherent states are sharply centered around the mean particle number.

\vspace{0.8cm}

After this preliminaries we may begin to derive 
the path-integral representation of the time-evolution
operator for a bosonic many-particle sytem
\be
\hat U\left(t_f,t_i \right ) \E \exp \left ( - \frac{i}{\hbar} 
\hat H (t_f - t_i) \right )
\ee
in the coherent basis. As in the one-particle case we achieve
that goal by splitting the time interval $\> t_f - t_i\> $ into $ M $ 
sub-intervals of length  $ \epsilon $ and by inserting the 
representation (\ref{Einheit mehr}) at  $ M - 1$ places. Then we obtain
\bea
<z_f | \> \hat U\left(t_f,t_i \right ) \> | z_i > &\EQ & U \left (
z^{\ast}_f,t_f ; z_i, t_i \right ) \E \lim_{M \to \infty} {\cal N}
\int \left ( \prod_{k=1}^{M-1} \>  \prod_{\alpha} dz^{\ast}_{\alpha,k} \>  
dz_{\alpha,k} \right ) \non
&& \cdot \> \exp \left ( - \sum_{k=1}^{M-1} \sum_{\alpha} | z_{\alpha,k}|^2 
\right ) \> \prod_{k=1}^{M} \left < z_k \> \left | 
e^{-i \epsilon \hat H /\hbar} \right | \>  z_{k-1} \right > \> .
\label{U kohaerent 1}
\eea
Here $ z_0 = z_i, z_M = z_f $ and $ {\cal N} $ is a normalization factor
which is of no concern for us. For small time steps $ \epsilon $
we expand the last factor in Eq. (\ref{U kohaerent 1}) 
and rewrite it then again as exponential function \footnote{A more detailed derivation
can be found in Ref. \cite{SuZh}.}
\bea
\left < z_k \> \left | e^{-i \epsilon \hat H /\hbar} \right | \> 
z_{k-1} \right > \EA  < z_k \> | \> z_{k-1} > - \frac{i \epsilon}{\hbar}
< z_k \> | \hat H | \> z_{k-1} > + {\cal O} (\epsilon^2) \non 
\EA < z_k \> | \> z_{k-1} > \> \exp \left ( \> - \frac{i \epsilon}{\hbar} 
\frac{< z_k \> | \hat H | \> z_{k-1} >}{< z_k \> | \> z_{k-1} > } \> \right )
+ {\cal O}(\epsilon^2) \> .
\eea
Assuming that the Hamiltonian is normal-ordered as, e.g. in Eq. \eqref{H op}, 
we then obtain by means of Eqs. (\ref{Ueberlapp}, \ref{ME normal}) 
\bea
U \left (z^{\ast}_f,t_f ; z_i, t_i \right ) \EA \lim_{M \to \infty} 
{\cal N} \int \left ( \prod_{k=1}^{M-1} \> \prod_{\alpha} 
dz^{\ast}_{\alpha,k} \> dz_{\alpha,k} \right ) \> 
\exp \left ( - \sum_{k=1}^{M-1} \sum_{\alpha} | z_{\alpha,k}|^2 \right )
\non
&& \hspace{1cm} \cdot \exp \left [ \> \sum_{k=1}^M \left ( 
\sum_{\alpha} z_{\alpha,k}^{\ast} z_{\alpha,k-1}
- \frac{i \epsilon}{\hbar} H(z^{\ast}_k,z_{k-1}) \right ) \> \right ] \> .
\label{U kohaerent 2}
\eea
Now we adopt a continous (and, of course, more symbolic) notation:
\be
\left \{ \> z_{\alpha,1} , z_{\alpha,2}, \ldots z_{\alpha,M} \> \right \}
\To z_{\alpha}(t)
\ee
and
\be
z^{\ast}_{\alpha,k} \> \> \frac{ z_{\alpha,k} - z_{\alpha,k-1}}{\epsilon}
\To z_{\alpha}^{\ast}(t) \> \frac{\partial}{\partial t} 
z_{\alpha}(t) 
\label{Ableitung 1}
\ee
\be
H \left ( z^{\ast}_k,z_{k-1} \right ) \To H \left ( 
z^{\ast}(t),z(t) \right ) \> .
\ee
For velocity-dependent interactions the last equation 
has to be modified as in the one-particle case. In the continous notation
the argument of all exponential functions in Eq. (\ref{U kohaerent 2}) becomes
\bea
&& \! \! \! \sum_{\alpha} z^{\ast}_{\alpha,M} \> z_{\alpha,M-1} - 
\frac{i \epsilon}{\hbar}
H \left (z^{\ast}_M,z_{M-1} \right ) + \frac{i \epsilon}{\hbar} 
\sum_{k=1}^{M-1}
\left [  i \hbar \sum_{\alpha} z^{\ast}_{\alpha,k} \left ( \frac{z_{\alpha,k}
-z_{\alpha,k-1}}{\epsilon} \right ) - H \left ( z^{\ast}_k, z_{k-1} \right ) 
\right ] \non
&& \longrightarrow \> \sum_{\alpha} z_{\alpha}^{\ast}(t_f) \, z_{\alpha}(t_f)
+ \frac{i}{\hbar} \int_{t_i}^{t_f} dt \> \left [ \> i \hbar \sum_{\alpha} 
z^{\ast}_{\alpha}(t) \, \frac{\partial z_{\alpha}(t)}{\partial t} 
- H \left ( z^{\ast}(t),z(t) \right ) \> \right ] \non
&& \EQ  \sum_{\alpha} z_{\alpha}^{\ast}(t_f) \, z_{\alpha}(t_f) + 
\frac{i}{\hbar} \int_{t_i}^{t_f} \! dt \, L \left (z^{\ast}(t),z(t) \right )
\EQ \sum_{\alpha} z_{\alpha}^{\ast}(t_f) \, z_{\alpha}(t_f) + 
\frac{i}{\hbar} S[z^{\ast}(t),z(t)] \, .
\eea
Here
\be 
\hat L \E  i \hbar \frac{\partial}{\partial t} - \hat H 
\label{L Schroedinger}
\ee 
is the  Lagrange operator of the  Schr\"odinger theory \footnote{Formally Schr\"odinger's 
equation is the Euler-Lagrange equation of the action
$ S[\psi^{\star},\psi] = \int dt
\, \psi^{\star} \lrp i \hbar \frac{\partial}{\partial t} - \hat H \rrp \psi $.
Thus, the operator in the integrand can be considered as Lagrange operator.}            
and $\> S \>  $ the corresponding classical action. If we omit the index $ \alpha $
for the modes or take 
$ z(t) = \left \{z_{\alpha}(t) \right \} $ as column vector, 
$ \> \bar z = \left \{z^{\ast}_{\alpha}(t) \right \} $ as row vector,
then the path-integral representation for the
time-evolution operator in the coherent basis reads
\be
U \left (
\bar z_f,t_f ; z_i, t_i \right ) \E 
\int_{z(t_i)=z_i}^{\bar z(t_f)=\bar z_f} {\cal D}\bar z(t) \> {\cal D}z(t) 
\> \exp \left ( \> \bar z(t_f) z(t_f)  + \frac{i}{\hbar} S[\bar z(t),z(t)]
\> \right ) \> .
\label{U kohaerent 3}
\ee
The term $ \> \bar z(t_f) z(t_f) \> $ in the first exponent is a 
remainder from collecting the individual terms for the derivative in Eq. 
(\ref{Ableitung 1}). If we had collected
\be
\frac{ - z^{\ast}_{\alpha,k+1} + z^{\ast}_{\alpha,k}}{\epsilon} \> 
z_{\alpha,k} \To\left ( - \frac{\partial}{\partial t} 
z^{\ast}_{\alpha}(t) \right ) z_{\alpha}(t) 
\label{Ableitung 2}
\ee
then  $ \> \bar z_1 z_0 \to \bar z(t_i) z(t_i) \> $ would have been left over.
Both results are from the same discrete expression and are thus equivalent.
If one wants to have an expression which is symmetric in initial and final time
one may average both.
\vspace{0.3cm}

Now we are able to give the path-integral representation of the bosonic
partition function. By means of Eq. (\ref{Spur kohaerent}) (the remainder is canceled!)
and after going over to Euclidean time we obtain
\bce
\vspace{0.1cm}

\fcolorbox{blue}{white}{\parbox{13cm}
{
\bea
Z \E {\rm tr} \left ( e^{- \beta \hat H} \right ) \EA 
\oint\limits_{z(0)=z(\beta \hbar)} \! {\cal D}\bar z(\tau) \> {\cal D} z(\tau) 
\> \exp \Bigl( \> - \frac{1}{\hbar} S_E [ \bar z(\tau),z(\tau) ] \> \Bigr ) \> , \no
\eea
}}
\ece
\vspace{-2cm}

\bea
\label{Z Pfad boson}
\eea
\vspace{0.4cm}


\noindent
where the Euclidean action is given by
\bce
\vspace{0.3cm}

\fcolorbox{blue}{white}{\parbox{10cm}
{
\bea
S_E[\bar z(\tau),z(\tau)] \EA \int_0^{\beta \hbar} d\tau \> \left [ \> \hbar \,   
\bar z(\tau) 
\frac{\partial z(\tau)}{\partial \tau} + H ( \bar z(\tau), z(\tau) ) \> 
\right ] \no \> .
\eea
}}
\ece
\vspace{-2cm}

\bea
\label{S Mehr boson}
\eea
\vspace{0.2cm}


\noindent
\vspace{0.4cm}

What we can treat analytically are systems where the particles move independently
in an external or mean potential: In such a case the Hamiltonian (and thus the action) is 
quadratic in the variables $ \, \bar z(\tau) , \, z(\tau) $. 
We then need to evaluate complex multidimensional Gaussian integrals with hermitean
matrices, i.e. need an extension of our toolbox used up to now \ldots . 
\vspace{0.2cm}

\noindent
Indeed one finds for the
\blau{\bf complex Gaussian integral with a hermitean and positive definite  
$ n \times n $ matrix} $ \He $ 
\bce
\vspace{0.2cm}

\fcolorbox{blue}{white}{\parbox{8cm}
{
\bea
\quad G_{2n}(\He) \Def \int d^nx \, d^ny \> \exp \left ( -\fz^{\dagger}   \He  \, 
\fz \right ) \E \frac{\pi^n}{{\det}_n \,  \He} \> ,  \no
\eea
}}
\ece
\vspace{-2cm}

\bea
\label{Gauss  hermit}
\eea
\vspace{0.4cm}


\noindent
where one integrates over real and imaginary part of the complex vector
\be
\fz \E \left ( \begin{array}{c} z_1 \\ z_2 \\ \vdots \\ z_n \end{array} \right) \> , \quad 
\fz^{\dagger}\E \left ( z_1^*, z_2^* \ldots z_n^* \right ) \> , \quad z_j \E x_j + i \, y_j \> .
\ee
This result is very plausible: If one integrates over the real as well as the imaginary part
of the complex vector  $ \fz $ then one obtains the square of the corresponding real-symmetric
Gaussian integral \eqref{Gauss 3}.
\vspace{1.5cm}

\renewcommand{\baselinestretch}{0.9}
\scriptsize
\refstepcounter{tief}

\noindent
\blau{\bf Detail \arabic{tief}:} {\bf Gaussian Complex Integral for Hermitean Matrices}\\
\vspace{0.2cm}

\noindent
\begin{subequations}
As in Eq. \eqref{1 aus kohaerenten Zu} one writes the integral \eqref{Gauss hermit} 
also as
\be
G_{2n}(\He) \E \int d^nz \, d^nz^* \> J_{2n} \>  \exp \left ( -\fz^{\dagger}  \, \He \, \fz 
\right ) \> ,
\label{herm G2n b}
\ee
where one treats $ \fz, \fz^* $ as independent  (real) variables. 
Because of $ \fx = (\fz + \fz^*)/2 \> , \> \fy = (\fz - \fz^*)/(2i) $ the associated
Jacobian is
\be 
J_{2n} \E {\det}_{2n} \left ( \begin{array}{cc}
\frac{\partial x_i}{\partial z_j} & \frac{\partial x_i}{\partial z_j^*} \\
\frac{\partial y_i}{\partial z_j} & \frac{\partial y_i}{\partial z_j^*} \end{array}
\right ) \E {\det}_{2n} \left ( \begin{array}{cc} 1/2 &  1/2 \\
 1/(2i) & - 1/(2i)\end{array} \right ) 
\E \left ( \frac{i}{2} \right)^n \> .
\ee
In the last line we have used the identity (see, e.g. http://en.wikipedia.org/wiki/Determinant)
\be
{\det}_{2n} \left ( \begin{array}{cc} A &  B \\
C & D\end{array} \right ) 
\E {\det}_n  \, A \, \cdot \, {\det}_n \lrp D - C A^{-1} B \rrp
\label{Block Det}
\ee
which holds for block matrices  ($ A $ must be invertible which is trivially fulfilled
for the diagonal matrices encountered here).

\noindent
The integral is real since $ \He$ is  hermitean  ( $ H_{ij}^* = H_{ji}$ )
\be
\left ( \fz^{\dagger}  \, \He  \, \fz \right )^* \E \left ( z_i^*  \, H_{ij}  \, z_j 
\right )^* \E z_j^*  \, \left (H_{ij} \right )^*  \, z_i \E  z_j^*  \, H_{ji}  \, z_i 
\> \stackrel{i \leftrightarrow j}{=} \> 
z_i^*  \, H_{ij}  \, z_j \> \equiv \>   \fz^{\dagger}  \, \He \, \fz  \> .
\ee
However, for the convergence of the integral we have to require additionally
that $ \He $ is positive definite, i.e. 
\be
z_i \, H_{ij} \, z_j \> > \> 0 \> , \quad \mbox{for all} \quad \fz\in \C \> .
\ee
This can be achieved easily by not considering  $ \, \He \, $ but
$ \, \He - E_0 \, \hat 1 \, $
where $ E_0 $ is the lowest eigenvalue (the ground-state energy).
\vspace{0.1cm}

In many texbooks the calculation of this complex Gaussian integral is performed
in close analogy to the real-symmetric case: One uses the fact that a 
hermitean matrix can be diagonalized by an {\bf unitary transformation}
\be
\He \E U^{\dagger} D U \> , \quad {\rm with} \quad U^{\dagger} U \E 1 \quad {\rm and} \quad 
D \E \left (\lambda_1 \ldots \lambda_n \right ) \> , \> \> \lambda_i > 0
\ee
and defines 
\be
z_i' \Def U_{ij} \, z_j \> \Longrightarrow \> z_i \E \left( U^{\dagger} \right )_{ij} \, 
z_j' \> . 
\label{unitary trans}
\ee
The complex Gaussian integral \eqref{herm G2n b} then becomes
\be
G_{2n}(\He) \E \left ( \frac{i}{2} \right )^n \, \int d^nz' \, d^nz'^* \> J_U \> 
\exp \left ( - \fz'^{\dagger} \, D \, \fz' \right ) 
\E \int d^nx' \, d^ny'\> J_U \> 
\exp \left [ \,  - \sum_{i=1}^n \lambda_i \left ( x_i'^2 + y_i'^2  \right ) \, \right ] \> ,
\ee
where the Jacobi determinant is
\be
J_U \E {\det}_{2n} \left ( \begin{array}{cc}
\frac{\partial z_i}{\partial z_j'} & \frac{\partial z_i}{\partial z_j'^*} \\
\frac{\partial z_i^*}{\partial z_j'} & \frac{\partial z_i^*}{\partial z_j'^*} \end{array}
\right ) \E {\det}_{2n} \left ( \begin{array}{cc} U^{\dagger}  &  0 \\
 0 & U^T  \end{array} \right ) \E 
{\det}_n \,  U^{\dagger} \cdot {\det}_n \,  U^T  \E {\det}_n \,  U^{\dagger} \cdot 
{\det}_n \,  U \E {\det}_n \,  \left ( U^{\dagger} U \right ) \E 1 \> .
\ee
The remaining integrations over  $\fx', \fy' $ are then trivial and one obtains
\be
G_{2n}(\He) \E \left ( \prod_{i=1}^n \, \sqrt{ \frac{\pi}{\lambda_i}} \right ) \cdot 
\left ( \prod_{i=1}^n \, \sqrt{ \frac{\pi}{\lambda_i}} \right ) \E \frac{\pi^n}{{\det}_n \,
  \He} \> ,
\label{Ergeb G2n}
\ee
i.e. the result \eqref{Gauss hermit}.
\vspace{0.2cm}

\noindent
However, there is "snag" in this derivation: The unitary transformation
\eqref{unitary trans} distorts the integration path into the complex plane 
and it is not obvious whether and how one can bring it back to the
real axis for integrating over the coordinates  $x_i',y_i'$  (even in the 1-dimensional
Fresnel case \eqref{Fresnel} this requires some effort).

\noindent
Therefore we will evaluate the integral \eqref{Gauss hermit} by {\bf real} methods.
If we set
\be
\He \deF \A + i \B \> , \quad \He^{\dagger} \E \He \> \Longrightarrow \> \A^T = \A \> , \> \> 
\B^T = - \B
\ee
then
\be
\fz^{\dagger}  \, \He \ \, fz \E \left (\fx^T - i \fy^T \right ) \, \left ( \A + i \B \right ) 
\, 
\left ( \fx + i \fy \right ) \deF \left (\fx^T, \fy^T \right ) \, \tilde \He \, \left ( 
\begin{array}{c}
                                                                       \fx \\ \fy 
\end{array} \right ) \> ,
\ee
where
\be
\tilde \He \Def \left ( \begin{array}{cc}
                     \A & -\B \\
                     \B & \A 
                           \end{array} \right )
\label{H tilde}
\ee
is a {\bf symmetric}  $2n \times 2n $ matrix. 
Hence we can utilize the result \eqref{Gauss 3} and obtain
\be
G_{2n}(\He) \E \frac{\pi^{2n/2}}{\sqrt{{\det_{2n}} \tilde \He}} \> .
\ee
There only remains the task to calculate the determinant of
 $ \tilde \He$ . Due to its block form that is easily done by means of the relation
\eqref{Block Det} and we obtain
\be
{\det}_{2n} \tilde \He \E {\det}_{2n} \left ( \begin{array}{cc}
                                    \A & -\B \\
                                    \B & \A 
                                   \end{array} \right ) \E {\det}_n \,  \A \cdot {\det}_n \,  
\left ( \A - \B \A^{-1} (-\B) \right ) \E \left ( {\det}_n \,  \A \right )^2 \cdot {\det}_n \,  
\left ( 1 + \A^{-1} \B \A^{-1} \B \right ) \> .
\ee
This can be related to the determinant of the original hermitean
matrix $ \He $ which is real and positive (as are the eigenvalues 
of $ \He $ by assumption). Hence we get
\bea
{\det}_n \,  \He \EA {\det}_n \,  \A \cdot {\det}_n \,  \left ( 1 + i \A^{-1} \B \right ) \> , 
\quad
{\det}_n \,  \He^T \E {\det}_n \,  \He \E  {\det}_n \,  \A \cdot {\det}_n \,  
\left ( 1 - i \A^{-1} \B \right ) \non
\Longrightarrow && \left ( {\det}_n \,  \He \right )^2 \E {\det}_n \,  \He \cdot {\det}_n \,  
\He^T \E \left ( {\det}_n \,  \A \right )^2 
\cdot {\det}_n \,  \left ( 1 + \A^{-1} \B \A^{-1} \B \right ) \equiv {\det}_{2n} \tilde \He \> .
\eea
As the integral is real and poitive we have to take the positive root and thereby obtain
the same result \eqref{Ergeb G2n} as was derived before (by somehow doubtful methods).

\end{subequations}

\renewcommand{\baselinestretch}{1.2}
\normalsize
\vspace{0.6cm}

\noindent
By completing the square one obtains for the extended integral
\bce
\vspace{0.2cm}

\fcolorbox{blue}{white}{\parbox{10cm}
{
\bea
\int d^nx \, d^ny \> \exp \left ( \, -\fz^{\dagger}   \He  \, 
\fz  + z^{\dagger} b + b^{\dagger} z \, \right ) \E \frac{\pi^n}{{\det}_n \,  \He} \> \exp \lrp b^{\dagger} \, \He^{-1} \, b \rrp
,  \no
\eea
}}
\ece
\vspace{-2cm}

\bea
\label{Gauss  hermit erweit}
\eea
\vspace{0.4cm}

\noindent
where  $ \, b, b^{\dagger} \, $ are  $n$-dimensional complex vectors.


\vspace{0.5cm}

\subsection{\textcolor{blue}{Description of Fermions : Grassmann Variables}}
\label{sec2: Fermionen}

If we want to describe fermions (obeying anticommutation relations) by 
coherent states which are eigenstates of the annihilation operator
\be
\boxed{
\qquad \hat a_{\alpha} \> | \> \cy{\xi} > \E \cy{\xi_{\alpha}} \> | \> \cy{\xi} >  \qquad ,
}
\label{def fermion kohaerent}
\ee
then we see that the eigenvalues  $ \,\cy{\xi_{\alpha}} \, $ cannot be
ordinary numbers:
$ \> \left ( \hat a_{\alpha} \hat a_{\beta} + \hat a_{\beta}  \hat a_{\alpha} 
\right ) | \, \cy{\xi} > = 0 \> $ requires that the eigenvalues 
 $ \, \cy{\xi_{\alpha}} \, $ should be
\textcolor{blue}{\bf anticommuting} quantities. Algebras of anticommuting numbers
are known as \textcolor{blue}{\bf Grassmann} algebras and are defined by a set of 
generators which fulfill
\bce
\vspace{0.2cm}

\fcolorbox{blue}{white}{\parbox{7cm}
{
\bea
\cy{\xi_{\alpha}} \>\cy{\xi_{\beta}} \> + \> \cy{\xi_{\beta}}\> \cy{\xi_{\alpha}} \EA 0 \> \> ,
\hspace{0.3cm} \alpha \> , \> \beta \> = 1, 2 , \ldots m  \> . \no
\eea
}}
\ece
\vspace{-1.5cm}

\bea
\label{Grassmann antikomm.}
\eea

\vspace{0.4cm}

\noindent
(Here and below Grassmann-valued quantities are always denoted by a \cy{blue-green}
color).
If we have an even number of generators $ m = 2 n $ one can define a conjugation
(also called an involution): We choose a set of  $ n $ generators
and to each Generator $ \cy{\xi_{\alpha}} $ we assign a generator
which we call  $ \cy{\bar{\xi}_{\alpha}} $. If
$ \lambda $ is a complex number, then one has
\be
\overline{ \left ( \> \lambda \> \cy{\xi_{\alpha}} \> \right ) } \E 
\lambda^{\ast} \cy{\bar{\xi}_{\alpha}} \> ,
\ee
and for a product of generators
\be
\overline{\left ( \cy{\xi}_{\alpha_1} \> \ldots \> \cy{\xi}_{\alpha_n} \right ) } \E
\cy{\bar{\xi}}_{\alpha_n}  \> \ldots \> \cy{\bar{\xi}}_{\alpha_1} \> .
\ee
Similar as for ordinary complex functions one can define a differentiation 
(acting to the right) for functions of Grassmann variables. It is defined 
as for ordinary functions with the difference that the variable $ \cy{\xi_{\alpha}} $ 
in the function in question has to be anticommuted until it is in front of the 
differential operator.
For example,
\bea
\frac{\partial}{\partial \cy{\xi_{\beta}}} \> \cy{\bar{\xi}_{\alpha}} \, \cy{\xi_{\beta}}
\E \frac{\partial}{\partial \cy{\xi_{\beta}}} \left ( \> - \cy{\xi_{\beta}} \> 
\cy{\bar{\xi}_{\alpha}} \> \right ) \E - \cy{\bar{\xi}_{\alpha}} \nonumber \> .
\eea
More uncommon is integration over Grassmann variables:
Berezin has shown {\bf \{Berezin\}} that
\bce
\vspace{0.2cm}

\fcolorbox{red}{white}{\parbox{9cm}
{
\bea
\int d\cy{\xi_{\alpha}} \> 1 \EA 0\> , \hspace{0.4cm}
\int d\cy{\xi_{\alpha}} \> \cy{\xi_{\alpha}} \E 1 \non 
\int d\cy{\bar{\xi}_{\alpha}} \> 1 \EA 0\> , \hspace{0.4cm}
\int d\cy{\bar{\xi}_{\alpha}} \> \cy{\bar{\xi}_{\alpha}} \E 1 \no
\eea
}}
\ece
\vspace{-3cm}

\bea
\label{Berezin xi}
\eea
\vspace{-1cm}

\bea
\label{Berezin xi bar}
\eea
\vspace{0.4cm}


\noindent
leads to a consistent description.
This "integration" actually behaves  more like a differentiation -- the power of a variable in
integrand is decreased by one.
For our purposes it is sufficient
to consider these definitions as a clever design to obtain the
the diverse minus signs which are connected with the 
anti-symmetry of fermionic systems without trying to inject too much physical intuition
in it.
Since Gaussian integrals played a dominant role in the previous treatment of the path integrals, 
it is quite natural to
investigate the corresponding Gaussian Grassmann integral.
Let us do that, first, for the case $ n = 1 $,
i.e. for two generators of  a Grassmann algebra: 
Because the square of a Grassmann variable vanishes
the exponential series for the Gaussian integrand terminates after 
the second term; after anticommutating and applying the
 integration rules (\ref{Berezin xi}, \ref {Berezin xi bar})
one finds
\be
\int d \cy{\bar{\xi}} \> d \cy{\xi} \> \exp \left ( \> - \lambda \> \cy{\bar{\xi}} \, \cy{\xi} 
\> \right ) \E \int d \cy{\bar{\xi}} \> d \cy{\xi} \> \left ( \> 1 - \lambda \> 
\cy{\bar{\xi}} \, \cy{\xi} \> 
\right ) \E \lambda \int d \cy{\bar{\xi}} \> d \cy{\xi} \> \cy{\xi} \, \cy{\bar{\xi}} \E 
\lambda \> .
\label{ferm Gauss n=2}
\ee
This \blau{\bf Gaussian Grassmann integral} can be easily generalized to 
an arbitrary even number of Grassmann generators 
and for a \blau{\bf hermitean matrix} $ \He $ :
\bce
\vspace{0.2cm}

\fcolorbox{blue}{white}{\parbox{10cm}
{
\bea
\int \left ( \prod_{\alpha=1}^{n} d \cy{\bar{\xi}_{\alpha}} \> d \cy{\xi_{\alpha}} 
\right ) \> \exp \left ( \> - \sum_{\alpha,\beta} \cy{\bar{\xi}_{\alpha}} \> 
\He_{\alpha, \beta} \> \cy{\xi_{\beta}} \> \right ) \EA \det \He \> . \no
\eea
}}
\ece
\vspace{-2cm}

\bea
\label{Gauss fermionisch}
\eea
\vspace{1cm}


\renewcommand{\baselinestretch}{0.9}
\scriptsize
\refstepcounter{tief}
\noindent
\blau{\bf Detail \arabic{tief}:} {\bf Gaussian Grassmann Integral}\\

\noindent
\begin{subequations}
The proof of Eq. (\ref{Gauss fermionisch}) is straightforward as we do not have to
consider boundary conditions: We diagonalize
 $ \He $ by an unitary matrix so that
\be
U^{\dagger} \He \, U \E {\rm diag} \> \left (\lambda_1, \ldots 
\lambda_n \right ) \> .
\ee
A crucial difference compared to ordinary integrals is that in integrals over
Grassmann variables the inverse of the absolute value of
the Jacobian is to be used if a linear transformation of the variables is performed.
This can already be seen in Eq. (\ref{ferm Gauss n=2}) by setting
$\cy{\xi} = \cy{\xi}'/\sqrt{\lambda} , \cy{\bar{\xi}} = \cy{\bar{\xi}}'/\sqrt{\lambda}$~:
The Jacobian is
\be
{\cal J} \E \det \left (
\begin{array}{rr}
 \frac{\partial \cy{\xi}}{\partial \cy{\xi}'} & \frac{\partial \cy{\xi}}{\partial \cy{\bar{\xi}}'} \\
\frac{\partial \cy{\bar{\xi}}}{\partial \cy{\xi}'} & \frac{\partial \cy{\bar{\xi}}}{\partial 
\cy{\bar{\xi}}'} 
\end{array} \right ) \E
\det \left (
\begin{array}{rr}
1/\sqrt{\lambda} & 0 \\
0 & 1/\sqrt{\lambda}
\end{array} \right ) \E \frac{1}{\lambda} \> ,
\ee
but the integral has the value
\be
\lambda \> \stackrel{!}{=} \> 
\int d \cy{\bar{\xi}}' \> d \cy{\xi}' \> C \, \exp \left ( \> -\cy{\bar{\xi}}' \, \cy{\xi}' 
\> \right ) \E C \> \> \Longrightarrow \> C \E \lambda \EQ {\cal J}^{-1} \> .
\ee
However, in the case under discussion the Jacobian is one and hence one obtains
for the Gaussian Grassmann integral directly
\bea
\int \left ( \prod_{\alpha=1}^{n} d \cy{\bar{\xi}_{\alpha}} \> d \cy{\xi_{\alpha}} 
\right ) \> \exp \left ( \> - \sum_{\alpha,\beta} \cy{\bar{\xi}_{\alpha}} \> 
\He_{\alpha, \beta} \> \cy{\xi_{\beta}} \> \right ) \EA  
\int \left ( \prod_{\alpha=1}^{n} d \cy{\bar{\xi}}'_{\alpha} \> d \cy{\xi}'_{\alpha} 
\right )
\> \exp \left ( \> - \sum_{\alpha} \cy{\bar{\xi}}'_{\alpha} \> \lambda_{\alpha}
\> \cy{\xi}'_{\alpha} \> \right ) \non
\EA \int \left ( \prod_{\alpha=1}^{n} d \cy{\bar{\xi}}'_{\alpha} \> d \cy{\xi}'_{\alpha}
\> \Bigl [ \> 1 - \cy{\bar{\xi}}'_{\alpha} \> \lambda_{\alpha} \> \cy{\xi}'_{\alpha}  
\> \Bigr ] \> \right ) \E
\prod_{\alpha=1}^{n} \> \lambda_{\alpha} \E \det \He \> .
\eea

\end{subequations}
\renewcommand{\baselinestretch}{1.2}
\normalsize
\vspace{0.5cm}

\noindent
In most applications the Gaussian integral
(\ref{Gauss fermionisch}) is the only result which is needed
when integrating over Grassmann variables. Slightly exaggerating 
one could say: In general, Grassmann variables are introduced only
to get rid of them as fast as possible...

\noindent
Comparison with the bosonic case in Eq. \eqref{Gauss hermit} shows
that -- apart from the irrelevant normalization constants --
fermionic Gaussian integrals just give the inverse of the bosonic integrals.
Using $ \> \det \He = \exp ( {\rm tr} \ln \He ) \> $ 
(\purpur{\bf Problem \ref{Det Spur}}) one can combine the bosonic and fermionic
cases for the extended Gaussian integral                  
\bce
\vspace{0.2cm}

\fcolorbox{red}{white}{\parbox{12cm}
{
\bea
 \int d \bar z \> d z \> \exp \lrp - \bar z \, \He \, z + \bar z \, b + \bar b \, z \rrp 
\EA {\rm const} \> \cdot 
\> \exp \lrp  \, - \zeta \> {\rm tr} \, \ln \He  +   \bar b \, \He^{-1} \, b \rrp 
\quad \hspace{2.7cm}
\label{Gauss ferm/boson}
\eea
}}
\ece

\vspace{0.4cm}


\noindent
where  $ z, b $ and $ \bar z, \bar b $ are complex or Grassmann-valued variables
depending on whether bosonic or fermionic systems are considered.
The different sign in the exponent of
Eq. (\ref{Gauss ferm/boson}) leads to the important property that in perturbation theory
fermion loops get a minus sign relative to boson loops.

\vspace{0.3cm}
\noindent
In close analogy to the bosonic case, fermionic coherent states
can now be defined by
\be
\boxed{
\qquad | \> \cy{\xi} > \>   = \> \exp \left ( \> - \sum_{\alpha} \cy{\xi_{\alpha}} \, \hat 
a^{\dagger}_{\alpha} \>  \right ) \> | \> 0 > \qquad .
}
\ee
It is convenient to require (i.e. to define) that
Grassmann variables also anticommute with creation and annihilation operators.
Then $ \> \cy{\xi_{\alpha} \, \hat a^{\dagger}_{\alpha}} \> $  is
\blau{Grassmann even}, i.e. it commutes with all other 
$ \>  \cy{\xi_{\beta} \, \hat a^{\dagger}_{\beta}} \> $ 
in the expansion of the exponential function and one obtains
\be
| \> \cy{\xi} > 
\E \prod_{\alpha} \left ( \> 1 - \cy{\xi_{\alpha} \, \hat a^{\dagger}_{\alpha}}
\>  \right ) \> | \> 0 >  \> .
\label{fermion kohaerent}
\ee
Fermionic coherent states practically have the same properties as the
bosonic coherent states studied in the previous chapter  -- with some small, but
crucial differences. For example, the overlap is
\be
< \cy{\xi} \> | \> \cy{\xi'} > \E \exp\left ( \> \sum_{\alpha} \cy{\bar{\xi}_{\alpha}} \,
\cy{\xi'_{\alpha}} \>  \right ) \> ,
\ee
and the unit operator in fermionic Fock space is represented by
\be
\hat 1 \E \int \left ( \prod_{\alpha} d \cy{\bar{\xi}_{\alpha}} \> d \cy{\xi_{\alpha}}
\right ) \> \exp\left ( \> - \sum_{\alpha} \cy{\bar{\xi}_{\alpha}} \,
\cy{\xi_{\alpha}} \>  \right ) \> | \> \cy{\xi} > < \cy{\xi} \> |\> .
\ee
As in the bosonic case the trace of a operator is given by
\be
{\rm tr} \> \hat A \E \int \left ( \prod_{\alpha} d \cy{\bar{\xi}_{\alpha}} \> 
d \cy{\xi_{\alpha}} \right ) \> \exp\left ( \> - \sum_{\alpha} \cy{\bar{\xi}_{\alpha}} \,
\cy{\xi_{\alpha}} \>  \right ) \> \sum_n < n \> | \> \cy{\xi} > < \cy{\xi} \> | \hat A |
\> n > \> .
\label{Spur ferm 1}
\ee
However, since the matrix elements  $  \> < n \> | \> \cy{\xi} > \> $ and 
$ \> < \cy{\xi} \> | \> m > \> $ between states  $ \> | \> n > , | \> m > \> $
in Fock space and coherent states also contain Grassmann numbers linearly
 (cf. Eq. (\ref{fermion kohaerent})), it follows from the anticommutation rules that
\be
< n \> | \> \cy{\xi} > \> < \cy{\xi} \> | \> m > \E < -\cy{\xi} \> | \> m > \> 
< n \> | \> \cy{\xi} > 
\ee
holds. With that Eq. (\ref{Spur ferm 1}) becomes
\bea
{\rm tr} \> \hat A \EA \int \left ( \prod_{\alpha} d \cy{\bar{\xi}_{\alpha}} \>
d \cy{\xi_{\alpha}} \right ) \> \exp\left ( \> - \sum_{\alpha} \cy{\bar{\xi}_{\alpha}}
\cy{\xi_{\alpha}} \>  \right ) \>  < - \cy{\xi} \> | \hat A \> \sum_n | \> n >
< n \> | \> \cy{\xi}  > \non
\EA \int \left ( \prod_{\alpha} d \cy{\bar{\xi}_{\alpha}} \>
d \cy{\xi_{\alpha}} \right ) \> \exp\left ( \> - \sum_{\alpha} \cy{\bar{\xi}_{\alpha}}
\cy{\xi_{\alpha}} \>  \right ) \> < - \cy{\xi} \> | \hat A  | \> \cy{\xi} > \> .
\label{Spur ferm 2}
\eea
This change compared to the bosonic form (\ref{Spur kohaerent}) entails
that in the path-integral representation of the fermionic partition function
the boundary condition is
\be 
\boxed{
\qquad \cy{\xi}(0) \E - \> \cy{\xi}(\beta \hbar) \qquad \> ,
}
\label{antiperiod Rand}
\ee
i.e. that fermions have to obey
\textcolor{blue}{\bf antiperiodic} boundary conditions.
\vspace{0.2cm}

It is now possible to give an unified path-integral representation
for the partition function of a many-body system
\vspace{0.2cm}

\fcolorbox{red}{white}{\parbox{14cm}
{
\bea
Z(\beta) \EA \int\limits_{z(\beta \hbar)=\zeta z(0)} \! {\cal D} \bar z(\tau)
\> {\cal D} z(\tau) \> \exp \left \{ \> - \frac{1}{\hbar} \int_0^{\beta \hbar} 
d\tau \> \left [ \, \hbar   \,            
\bar z(\tau) \> \frac{\partial z(\tau)}{\partial \tau} + H \left ( \bar 
z(\tau), z(\tau)\right ) \, \right ] \> \right \} \> . \no
\eea
}}
\vspace{-2cm}

\bea
\label{Z boson/ferm}
\eea
\vspace{0.4cm}



\noindent
For the two-particle interaction (\ref{pot}) the Hamiltonian 
contains terms up to $ \> \> \bar z \, \bar z \, z \, z \> \> $  and hence 
this path integral is not a Gaussian functional integral anymore
which can be solved exactly. However, 
Eq. (\ref{Z boson/ferm}) serves as a starting point for an unified  perturbation theory in powers
of these terms, both for bosons and fermions.

\vspace{0.5cm}

\subsection{\textcolor{blue}{Perturbation Theory and Diagrams}}
\label{sec2: Stoerung + Diagr}

We now want to calculate the partition function for a bosonic or fermionic
system in a systematic way in order to deduce or predict its macroscopic properties from the
microscopic Hamiltonian. For simplicity of notation we will use a system of units in which
 $ \hbar = 1$ .
When using coherent states without fixed particle number it seems best to adopt
the \textcolor{blue}{\bf grand-canonical ensemble } for the description of the system.
where it is in contact with a particle and heat bath. The probability to observe it with energy 
 $ E $ and particle number $ N $ is proportional to $ \> \exp \left [ - (E - \mu N)/(k_B T)
\right ] \> $ (the chemical potential  $ \mu $ controls the mean particle number). Therefore we will use the partition function
\be
\boxed{
\qquad Z \E {\rm tr} \, \left [ \> e^{- \beta (\hat H - \mu \hat N )} \> \right ]
\deF e^{ - \beta \Omega} \qquad
}
\ee
to describe the macroscopic thermodynamics of the system.

Similarly as for the free energy one obtains the pressure or the entropy
from the {\bf thermodynamic} (grand-canonical) 
{\bf potential} $ \> \Omega(\mu,V,T) \> $
by differentiation w.r.t. the volume $ V $ or the temperature 
$ T $ , respectively,  and by differentiation w.r.t. 
the chemical potential the mean particle number
\be
\bar N \E - \frac{\partial \Omega}{\partial \mu}  \E \frac{1}{Z} \, {\rm tr} \> 
\left [ \> \hat N \, e^{- \beta (\hat H - \mu \hat N )} \> \right ]\> .
\label{N mittel}
\ee
More generally, the thermodynamic average of an operator $ \> \hat {\cal O} \> $ 
is given by
\be
\left < \hat {\cal O} \right >_{\beta} \E \frac{1}{Z} \, {\rm tr} \> 
\left [ \> \hat {\cal O} \, e^{- \beta (\hat H - \mu \hat N )} \> \right ] \> .
\label{Mittelwert}
\ee
This can be expressed by the corresponding $n$-particle Green function
\bea
G^{(n)} \left ( \alpha_1 \tau_1 \ldots \alpha_n \tau_n \, \big | \, 
\alpha_{n+1} \tau_{n+1} \ldots \alpha_{2n}\tau_{2n} \right )
\EA \frac{1}{Z} {\rm tr} \> \Biggl \{ \,{\cal T} \> e^{- \beta (\hat H
- \mu \hat N)} \, \hat a_{{\alpha}_1}^{(H)}(\tau_1) \ldots 
\hat a_{{\alpha}_n}^{(H)}(\tau_n) \non 
&& \hspace{1cm} \cdot \hat a_{{\alpha}_{n+1}}^{(H) \> \dagger}(\tau_{n+1}) 
\ldots \hat a_{{\alpha}_{2n}}^{(H) \> \dagger}(\tau_{2n}) \> \Biggr \} \> ,
\eea
where
\bea
\hat a_{{\alpha}}^{(H)}(\tau) \EA e^{ \tau (\hat H - \mu \hat N)} \,
\hat a_{\alpha} \, e^{ - \tau (\hat H - \mu \hat N)} \non
\hat a_{{\alpha}}^{(H) \> \dagger }(\tau) \EA e^{ \tau (\hat H - \mu \hat N)}
 \, \hat a_{\alpha}^{\dagger} \, e^{ - \tau (\hat H - \mu \hat N)} 
\eea
are the Heisenberg operators in imaginary time and $ \> {\cal T} \> $ orders in 
imaginary time (note that $ \> \hat a_{\alpha} \> $
and $ \> \hat a_{\alpha}^{\dagger} \> $ are not hermitean adjungated anymore).

The path-integral representation of the grand-canonical partition function
is a simple generalization of  Eq. (\ref{Z boson/ferm}):
\vspace{0.2cm}

\fcolorbox{blue}{white}{\parbox{14cm}
{
\bea
Z \EA \int\limits_{z(\beta)=\zeta z(0)} \! {\cal D} ( \bar z(\tau) \,
z(\tau)) \> \exp \left \{\>   -  \int_0^{\beta} d\tau \> \left [ \,
\bar z(\tau) \> \left ( \frac{\partial }{\partial \tau} - \mu \right )z(\tau) 
+ H \left ( \bar z(\tau),
z(\tau)\right ) \, \right ] \> \right \} \> , \no
\eea
}}
\vspace{-2cm}

\bea
\label{Z gross-kanon}
\eea
\vspace{0.4cm}


\noindent
as one only has to replace the Hamiltonian by $ \> \hat H - \mu \sum_{\alpha}
\hat a^{\dagger}_\alpha \hat a_{\alpha} \> $ . Similarly the
 $(2n)$-point function is given by
\vspace{0.2cm}

\fcolorbox{black}{white}{\parbox{14cm}
{
\bea
&& \hspace{-2cm} G^{(n)} \left ( \alpha_1 \tau_1 \ldots \alpha_n \tau_n \, \big | \, 
\alpha_{n+1}
\tau_{n+1} \ldots \alpha_{2n}\tau_{2n} \right ) \non
\EA  \frac{1}{Z} \>  \int\limits_{z(\beta)=\zeta z(0)} \! {\cal D} 
( \bar z(\tau) \, z(\tau) ) \>  \> 
z_{{\alpha}_1}(\tau_1) \ldots
z_{{\alpha}_n}(\tau_n) \>
\bar z_{{\alpha}_{n+1}}(\tau_{n+1}) \ldots
\bar z_{{\alpha}_{2n}}(\tau_{2n}) \non
&& \hspace{1cm} \cdot 
\> \exp \left \{\>  -  \int_0^{\beta} d\tau \left [ \,
\bar z(\tau) \> \left ( \frac{\partial }{\partial \tau} - \mu \right )z(\tau)
+ H \left ( \bar z(\tau),
z(\tau)\right ) \, \right ] \> \right \} \> . \no
\eea
}}
\vspace{-2cm}

\bea
\label{Green gross-kanon}
\eea
\vspace{0.4cm}


Since we cannot perform the path integrals exactly, we will combine the single-particle part
 of $ \> H (\bar z, z) \> $ together with the other
quadratic terms and develop the many-body part of the
Hamiltonian in a Taylor series. This will lead to functional integrals over
Gaussian functions times polynomials which can be evaluated directly.
The resulting perturbative series will be discusssed for \textcolor{blue}{\bf finite temperature} 
as this is somehow easier to handle than the direct treatment of the system at
zero temperature.

\begin{figure}[hbtp]
\bce
\includegraphics[angle=0,scale=0.3]{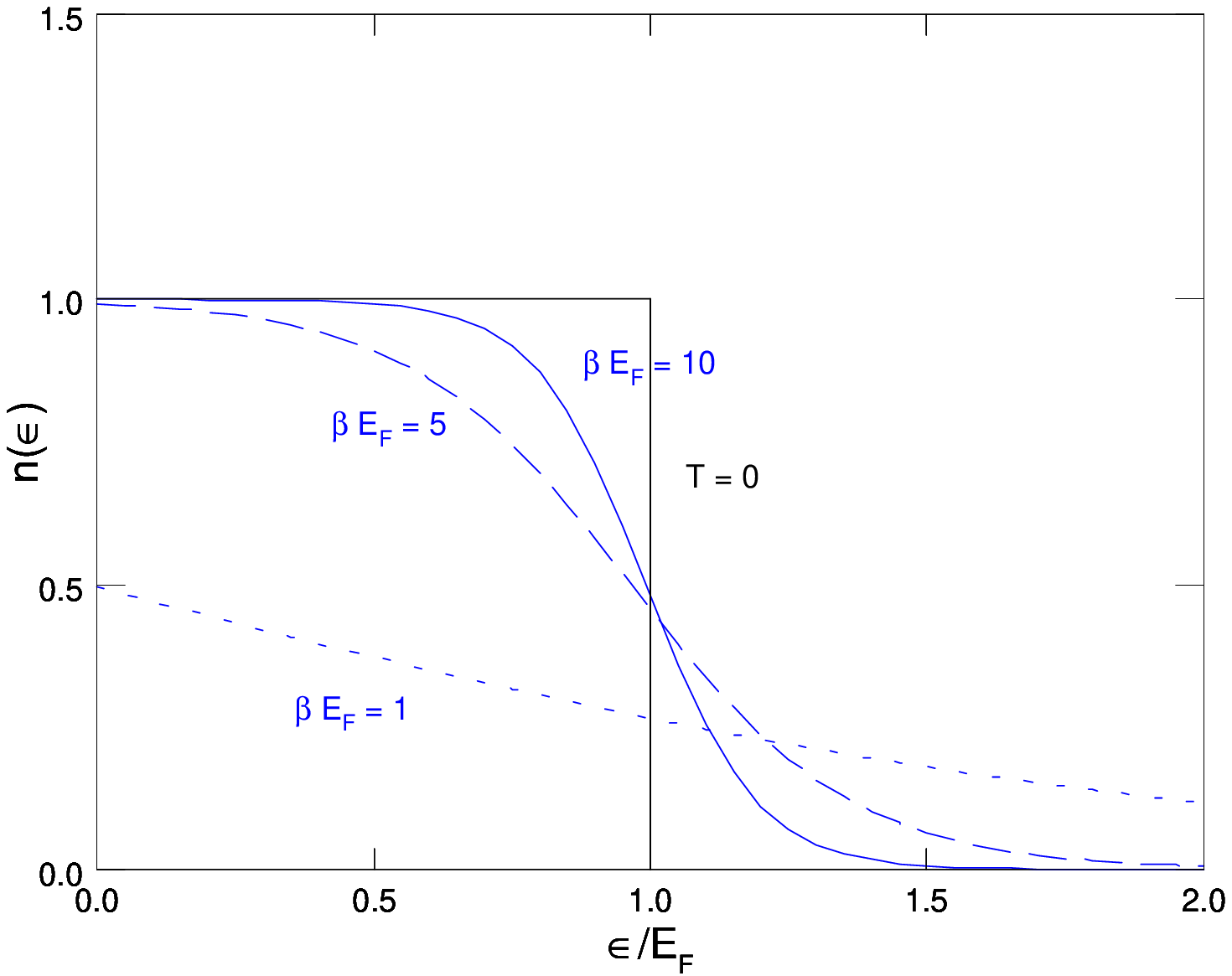}
\label{abb: Fermiverteil}
\ece
{\bf Fig. \arabic{abb}} : Distribution functions of an ideal Fermi gas at different temperatures
and constant density
\hspace*{1.9cm } $ \, N/V \, $. The Fermi energy $ \, E_F = k_F^2/(2m) \, $ 
from Eq. \eqref{Dichte FG} is used as an energy scale. For\\
\hspace*{1.9cm} $ \> \beta E_F = 10 , \,  5 , \,  \,  1 \> $  the corresponding chemical potential
has been calculated numerically as\\ 
\hspace*{2cm}$ \> \mu/E_F = 9.916 , \,  4.823 , \, -0.0214  $.  
The occupation probablity for $  T = 0  $ , i.e. 
$ \, \beta = \infty \, $ is also depicted.
\vspace{0.4cm}
\end{figure}

As a warm-up (pun intended) we will calculate the partition function and
the one-particle Green function for a system of \textcolor{blue}{\bf non-interacting} 
particles which is described by a one-body Hamiltonian
\be
\hat H_0 \E \sum_{\alpha} \> \epsilon_{\alpha} \, \hat a_{\alpha}^{\dagger}
\hat a_{\alpha} \> .
\label{1Teilchen H}
\ee
In Eq. (\ref{1Teilchen H}) we have chosen a basis in which 
$ \> \hat H_0 \> $ is diagonal -- including the case of an external potential
as well as the case of a mean potential acting between the particles.
The  ``free'' partition function is
\bea
Z_0 \EA \int\limits_{z(\beta)=\zeta z(0)} \! {\cal D} ( \bar z(\tau) \, 
z(\tau) ) \> \exp \left \{ \>  - \int_0^{\beta} d\tau \> 
\sum_{\alpha} \left [ \,
\bar z_{\alpha}(\tau) \> \left ( \frac{\partial}{\partial \tau} - \mu \right )
z_{\alpha} (\tau) + \epsilon_{\alpha} \bar z_{\alpha}(\tau) z_{\alpha} (\tau)
 \, \right ] \> \right \} \non
\EA {\rm const.} \prod_{\alpha} \> \fdet^{-\zeta} \> \left [ \,  
\frac{\partial}{\partial \tau} - \mu + \epsilon_{\alpha} \, \right ]\> .
\eea
As usual we calculate the determinant as product of the eigenvalues
of the corresponding operator with the given boundary conditions.
In this way one obtains for the eigenvalues $ \> \epsilon_{\alpha} - \mu + 
i \omega_n \> $ with  $ \> n = 0, \pm 1, \pm 2 \ldots \> $,  where
\be
\omega_n \>= \> \left \{ \begin{array}{r@{\quad:\quad}l}
                \frac{2 n \pi}{\beta} & {\rm bosons} \> \> (\zeta = 1) \\ 
                \frac{(2 n + 1) \pi}{\beta} & {\rm fermions} \> \> 
                                                            (\zeta = -1)
                \end{array} \right .
\ee
are the {\bf Matsubara frequencies}. As in the treatment of the harmonic oscillator in 
{\bf chapter} {\bf \ref{sec1: Lagr,Ham}} we use Euler's formula
$ \> \prod_{n=1}^{\infty} \left [ 1 + 
x^2/(n^2 \pi^2) \right ] = \sinh x/x \> $ to obtain the well-known expression
\be
\boxed{
\qquad Z_0 \E \prod_{\alpha} \left [ \> 1 - \zeta e^{- \beta (\epsilon_{\alpha}
- \mu)} \> \right ]^{-\zeta}  \> .\qquad
}
\label{Z0}
\ee
By means of Eq. (\ref{N mittel}) one then obtains $ N = \prod_{\alpha} n_{\alpha} $
where
\be
\boxed
{
\qquad n_{\alpha} \E \frac{1}{ \exp \left [ \, \beta (\epsilon_{\alpha} - \mu \, 
\right ] - \zeta } \qquad 
}
\label{Besetz}
\ee
is the occupation probability of the state $ \alpha $ .
%
%
For free fermions  ( $  \zeta = - 1 $ )  the occpation probability is depicted schematically  in Fig.\ref{abb: Fermiverteil}  as a function of  the energy.

\refstepcounter{abb}
\vspace{0.5cm}

Although the expressions for bosons and fermions differ "only" by a
sign this has far-reaching consequences:
For example, the expression in the square bracket of 
Eq. (\ref{Z0}) can vanish for $ \> \mu \le 0 \> $ if we consider bosons but 
not for fermions. This is the phenomen of {\bf Bose-Einstein condensation}
which already shows up for non-interacting particles \footnote{See, e.g., 
 {\bf \{Fetter-Walecka\}}, p. 39 - 44 .}.

\noindent
Similarly we calculate the one-particle Green function: 
\be
G \left ( \alpha \tau \, \big | \, \alpha' \tau' \right ) \E    
\frac{\delta^2}{\delta J_{\alpha}(\tau) \, \delta \bar J_{\alpha'}(\tau')}
\> \ln \, \int {\cal D} ( \bar z \,  z)  \> \exp \left \{ \> -  
S[\bar z, z]  + (\bar J,z) + (\bar z,J) \> \right \} \Biggr |_{\bar J = J = 0} 
\> ,
\ee
where we use again a compact notation as in {\bf chapter}
{\bf \ref{sec1: greensche Funk}}. For instance, the "free" action 
may be written as   
\be
S_0[\bar z, z] \E \int_0^{\beta} dt \> \sum_{\gamma} \, 
\bar z_{\gamma}(t) \left ( \, \frac{\partial}{\partial t} + \epsilon_{\gamma}
 - \mu \, \right ) z_{\gamma}(t) \EQ \left ( \bar z, \, \lrp 
\frac{\partial}{\partial t} + \epsilon - \mu \rrp \, z \right ) \> .
\ee
Completing the square in the generating functional gives
\bea
G_0 \left ( \alpha \tau \, \big | \, \alpha' \tau' \right ) \EA
\frac{\delta^2}{\delta J_{\alpha}(\tau) \, \delta \bar J_{\alpha'}(\tau')}
\> \ln \, \exp \left \{ \> \left ( \bar J, \, \frac{1}{\partial/\partial t 
+ \epsilon - \mu} \,  J \right ) \> \right \} \non
\EA \delta_{\alpha \alpha'} \> \left < \tau \left | 
\frac{1}{\partial/\partial t
+ \epsilon_{\alpha} - \mu} \right | \tau' \right > \> ,
\eea
as the operator to be inverted is diagonal in the chosen basis.
As can be seen the ``free'' one-particle Green function is solution
of the equation
\be
 \left ( \, \frac{\partial}{\partial \tau} + \epsilon_{\alpha} - \mu \,
\right ) \, G_0 \left ( \alpha \tau \, \big | \, \alpha' \tau' \right )
\E \delta_{\alpha \alpha'} \, \delta ( \tau - \tau')
\ee
with boundary condition
\be
G_0 \left ( \alpha \beta \, \big | \, \alpha' \tau' \right ) \E 
 \zeta \, G_0 \left ( \alpha 0\, \big | \, \alpha' \tau' \right ) \> .
\ee
In the discretized formulation the calculation is more cumbersome
but this is compensated by the advantage of getting an unambigous result for 
equal times $ \> \tau = \tau' \> $ which is not the case in the continous 
(and more symbolic) notation. The final result is
\be      
G_0 \left ( \alpha \tau \, \big | \, \alpha' \tau' \right ) \E
\delta_{\alpha \alpha'} \, e^{- (\epsilon_{\alpha} - \mu) ( \tau - \tau')}
\> \Bigl [ \, \Theta(\tau - \tau' - 0^+) \, \left ( 1 + \zeta \, 
n_{\alpha} \right ) 
\, + \, \zeta \Theta(\tau'-\tau + 0^+) \, n_{\alpha} \> \Bigr ] \> .
\label{G0}
\ee
\noindent
From the one-particle Green function the thermal expectation
value of an arbitrary one-particle operator
\be
\hat A \E \sum_{\alpha, \alpha'}  \> \left < \alpha' \left | 
\hat A \right | 
\alpha \right > \> \hat a_{\alpha'}^{\dagger} \hat a_{\alpha} \> ,
\label{1-Teil Op}
\ee
can be calculated easily with the help of Eq. (\ref{Mittelwert}) 
\be
\left < \hat A \right >_{\beta} \E \zeta \lim_{\tau' \to \tau^{+}}
\, \sum_{\alpha, \alpha'}  \left < \alpha' \left | \hat A \right |
\alpha \right > \> G \left ( \alpha \tau \big | \alpha' \tau' \right ) \> .
\label{1-Teil Mittel}
\ee
This is because the infinitesimally greater time $ \tau' $ in the Green function
exactly gives the correct ordering of creation and annihilation operators
required in Eq. (\ref{1-Teil Op}).

\vspace{1cm}

\noindent
{\bf Example : Mean Particle Number and Energy of a Free Fermi Gas}
\vspace{0.2cm}

\noindent
The basis is made up of plane waves  $ \> \exp ( i {\bf k} \cdot 
\fx)/\sqrt{V} \> $  normalized to one in the volume $V$ .
From Eq. (\ref{G0}) we find for $ \zeta = - 1, \> \> 
\lim_{\tau' \to \tau^{+}} G_0 = - n_{\bf k} $ 
and hence
\bea
\left < N \right >_{\beta} \EA 2 \, \sum_{\bf k} \> \frac{1}{ \exp \left [ 
\beta
\left ( k^2/(2 m) - \mu \right ) \right ] + 1 } \non
\left < E \right >_{\beta} \EA 2 \, \sum_{\bf k} \> \frac{ k^2}{2m}  
\frac{1}{ \exp \left [ \beta
\left (  k^2/(2 m) - \mu \right ) \right ] + 1 } \> ,
\eea
where the factor of 2 comes from the summation over the spin $ 1/2 $ of the particles.
In the continous limit  $  \> \sum_{\bf k} \> $  turns again into
$ \> V \, \int d^3k/(2 \pi)^3 \> $ . The integrals are particularly simple 
for $ T = 0 $ since the occupation probability then becomes a step function cutting off
the momenta at the Fermi momentum $ k_F $  (~$ \mu = E_F = k_F^2/(2 m) $ ), see Fig. \ref{abb: Fermiverteil}. We then obtain
\bea
N &\EQ & < \, N \, >_{\beta = \infty} \E \frac{ 2 V}{(2 \pi)^3} 
\frac{4 \pi}{3} k_F^3 
\label{Dichte FG}\\
\bar E &\EQ & < \, E \, >_{\beta = \infty} \> = \>  \frac{ 2 V}{(2 \pi)^3} 
\frac{4 \pi}{5} \,              
\frac{ k_F^5}{2m} \E N \, \frac{3}{5} E_F \> .
\label{E mittel FG}
\eea
Eq. (\ref{Dichte FG}) provides the relation between density $ N/V $
and Fermi momentum and
Eq. (\ref{E mittel FG}) expresses the mean energy of the particles by the
Fermi energy.


\vspace{0.5cm}

\noindent
Let us now return to the problem of calculating the partition function 
for a system whose Hamiltonian contains many-body interactions:
\be
\hat H \E \hat H_0 + V \left (\hat a^{\dagger}_{\alpha}, 
\hat a^{\dagger}_{\beta} \ldots \, ; \hat a_{\gamma}, \hat a_{\delta} \ldots
\right ) \> .
\ee
Obviously one can expand
\bea
Z  \EA Z_0 \> \left < \, \exp \left\{\> - \int_0^{\beta} d \tau \> 
V(\bar z \ldots; z \ldots )  \> \right \} \, \right >_{0, \beta} \non
\EA Z_0 \> \sum_{n=0}^{\infty}  \frac{(-)^n}{n !} 
\> \int_0^{\beta}
d\tau_1 \ldots d\tau_n \> \left < \, V( \bar z(\tau_1) \ldots ) \> \ldots
V( \bar z(\tau_n) \ldots ) \, \right >_{0, \beta} 
\label{Stoer1}
\eea
in a series containing powers of the perturbative potential. In each term of this series
the expectation value is to be taken w.r.t. the "free" action  $ \> S_0 \> $ 
and at inverse temperature $ \beta $. Let us assume that the perturbation potential
is a polynomial in creation and annihilation operators (e.g. a two-body potential).
Then in Eq. (\ref{Stoer1}) one has to evaluate functional integrals over polynomials
multiplied by Gaussian functions. This is done with the help of 
\textcolor{blue}{\bf Wick's theorem} which is based on the following identity


\be
\boxed{
\qquad \frac{ \int {\cal D} (\bar z z) \> z_{{i_1}} \ldots z_{{i_n}} \, 
\bar z_{{j_1}} \ldots \bar z_{{j_n}} \> \exp \left [ - \sum_{i,j} \bar z_i 
M_{i j} z_j \right ] }{\int {\cal D} (\bar z z) \> \exp \left [ - \sum_{i,j} 
\bar z_i M_{i j} z_j \right ] } 
\E  \sum_{\rm permutations} \eta^P
\> M^{-1}_{{i_{P n}},{j_{P n}}} \ldots M^{-1}_{{i_{P 1}},{j_{P 1}}} \> . \quad
}
\label{Wick}
\ee
Here the index $i$ denotes the time and state and 
 $ \eta $ is the parity of the permutation, i.e.  $ \pm 1 $ for 
 an even/odd number of permutations.
 
\vspace{0.5cm}
\renewcommand{\baselinestretch}{0.9}
\scriptsize
\refstepcounter{tief}
\noindent
\blau{\bf Detail \arabic{tief}:} {\bf Proof of  Wick's Theorem}\\

\begin{subequations}
\noindent
This can be verified by differentiating the corresponding generating functional
\bea
W(\bar J, J) \EA  \frac{ \int {\cal D} (\bar z z) \> 
\exp \left [ - \sum_{i,j} \bar z_i M_{i j} z_j + \sum_i \left ( \bar J_i z_i
+ \bar z_i J_i \right ) \right ] }
{\int {\cal D} (\bar z z) \> \exp \left [ - \sum_{i,j}
\bar z_i M_{i j} z_j \right ] } \non 
\EA  \exp \left [ \> \sum_{i,j} \bar J_i M^{-1}_{i j} J_j
\right ]
\label{Beweis Wick}
\eea
w.r.t. to the external sources:
\be
\frac{\delta^{2n} W}{\delta \bar J_{{i_1}} \ldots \delta \bar J_{{i_n}}
\delta \bar J_{{j_n}} \ldots \delta \bar J_{{j_1}} } \Biggr |_{\bar J = J = 0}
\E  \eta^n \> 
\frac{ \int {\cal D} (\bar z z) \> z_{{i_1}} \ldots z_{{i_n}} \,
\bar z_{{j_1}} \ldots \bar z_{{j_n}} \> \exp \left [ - \sum_{i,j} \bar z_i
M_{i j} z_j \right ] }{\int {\cal D} (\bar z z) \> \exp \left [ - \sum_{i,j}
\bar z_i M_{i j} z_j \right ] }  \> .
\ee
Indeed, if one performs the derivatives in  Eq. (\ref{Beweis Wick}) 
carefully (taking into account order and sign), one obtains Wick's theorem (\ref{Wick}).
 
\end{subequations}
\renewcommand{\baselinestretch}{1.2}
\normalsize
\vspace{0.5cm}

\noindent
The usual formulation is obtained by defining a 
\textcolor{blue}{\bf contraction} 
\be
\underbrace{ \hat a_{\alpha}^{(H)}(\tau) \, 
\hat a_{\alpha'}^{(H) \> \dagger}(\tau') } \Def \left < \, {\cal T}
\hat a_{\alpha}^{(H)}(\tau) \,\hat a_{\alpha'}^{(H)\> \dagger}(\tau') \,
\right >_{0 \beta} \E G_0 (\alpha \tau \big | \alpha' \tau' )
\EQ \delta_{\alpha \alpha'} \> g_{\alpha}(\tau - \tau') \> .
\ee
Contractions of $ \> \hat a \hat a \> $ and $ \> \hat a^{\dagger} \hat 
a^{\dagger} \> $ are defined to be zero. Then the identity  
(\ref{Wick}) for the Gaussian integrals is equivalent to
\be
\left < \> {\cal T} \,
\hat b_{\alpha_1}^{(H)}(\tau_1) \ldots \hat b_{\alpha_n}^{(H)}(\tau_n) \>
\right >_{0 \beta} \E \sum_{\hat b \E \hat a, \,\hat a^{\dagger}} 
\> \left ( \> {\rm all \> \> complete \> \> contractions} 
\right )\> . 
\label{Wick ueblich}
\ee
For a two-body interaction
\be 
V\left ( \bar z(\tau), z(\tau) \right ) \E \frac{1}{2} \,
\sum_{\alpha \beta \gamma \delta} \left < \alpha \beta \left | V \right |
\gamma \delta \right > \, \bar z_{\alpha}(\tau) \bar z_{\beta}(\tau) \,
z_{\delta}(\tau) z_{\gamma}(\tau)
\ee
the  $n^{\rm th}$ term in the perturbative expansion (\ref{Stoer1}) reads
\vspace{0.2cm}

\fcolorbox{black}{white}{\parbox{13cm}
{
\bea
\left ( \frac{Z}{Z_0} \right )^{(n)} \EA \frac{(-)^n}{n ! \, 2^n} \> 
\sum_{\latop{\alpha_1 \beta_1}{\gamma_1 \delta_1}} \ldots 
\sum_{\latop{\alpha_n \beta_n}{\gamma_n \delta_n}}
\left < \alpha_1 \beta_1 \left | V \right |
\gamma_1 \delta_1 \right >  \ldots 
\left < \alpha_n \beta_n \left | V \right |
\gamma_n \delta_n \right >  \non
&& \cdot \int_0^{\beta} d\tau_1 \ldots d\tau_n \> \Bigl < \>
\bar z_{\alpha_1}(\tau_1) \bar z_{\beta_1}(\tau_1) \,
z_{\delta_1}(\tau_1) z_{\gamma_1}(\tau_1) \> \ldots \non
&& \hspace{3cm} \ldots \bar z_{\alpha_n}(\tau_n) \bar z_{\beta_n}(\tau_n) \,
z_{\delta_n}(\tau_n) z_{\gamma_n}(\tau_n) \> \Bigr >_{0 \beta} \> , \no
\eea
}}
\vspace{-0.8cm}

\bea
\label{Zn Stoer}
\eea


\noindent
for which Wick's theorem  (\ref{Wick ueblich}) has to be applied when evaluating 
its expectation value. 
\vspace{0.1cm}

\noindent
For "book keeping" it is very helpful to have a graphical
representation for the individual terms:

\refstepcounter{abb}
\begin{figure}[hbtp]
\vspace*{-4cm}
\bce
\includegraphics[angle=0,scale=0.6]{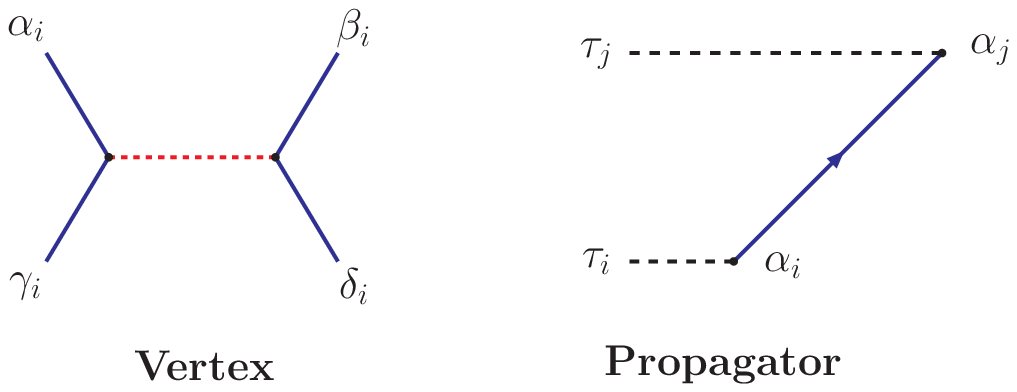}
\label{abb:2.5.1}
\ece
\vspace*{-8cm}
{\bf Fig. \arabic{abb}} : Graphical representation of a vertex for a two-body interaction
and of a propagator \\
\hspace*{1.5cm}  in "labeled" Feynman diagrams.
\end{figure}
\vspace{1cm}


\noindent
A \textcolor{blue}{\bf vertex} represents
\be
{\rm vertex} \> \hat = \> \left < \alpha_i \beta_i \left | V \right |\gamma_i 
\delta_i \right >  \> ,
\label{Vertex}
\ee
and a \textcolor{blue}{\bf propagator} 
\bea
{\rm propagator} &\hat =&
\delta_{\alpha_i \alpha_j} \, g_{\alpha_i} \left (\tau_j - \tau_i \right )
\non
\EA \delta_{\alpha_i \alpha_j} \, e^{ - (\epsilon_{{\alpha_i}} -\mu)
(\tau_j - \tau_i)} \, \Bigl [ \> \left ( 1 + \zeta n_{{\alpha_i}} \right )
\Theta(\tau_j - \tau_i) + \zeta n_{{\alpha_i}} \Theta(\tau_i - \tau_j) \> 
\Bigr ] \> .
\label{Propagator}
\eea
\vspace{0.1cm}

\noindent
In $ n^{\rm th} $ order the rules for these ``labeled'' 
Feynman diagrams are then as follows:

\bdes
\item[1.]: 
Draw all \blau{\bf distinct} labeled diagrams with $n$ vertices (diagrams
are distinct if they cannot be deformed in such a way that they are identical
-- including all time labels, left-right-labels and 
the direction of arrows).

\item[2.]: Assign a single-particle index to each directed line and give it the factor (\ref{Propagator}).

\item[3.]: Give each vertex the factor (\ref{Vertex}).
\item[4.]: Sum over all single-particle indices and integrate over all times in the
interval $ \> [0,\beta] \> $.

\item[5.]: Multiply the result by
 $ \> (-)^n \zeta^{n_L}/(n ! \, 2^n)  \> $ where
 $ n_L $ is the number of closed loops of single-particle propagators\\
 \hspace{0.3cm} in the diagram.

\edes

\newpage
\noindent
{\bf Example : $1^{\rm st} $-Order Perturbation Theory}\\

\begin{figure}[hbtp]
\vspace*{-7.5cm}
\bce
\includegraphics[angle=0,scale=0.9]{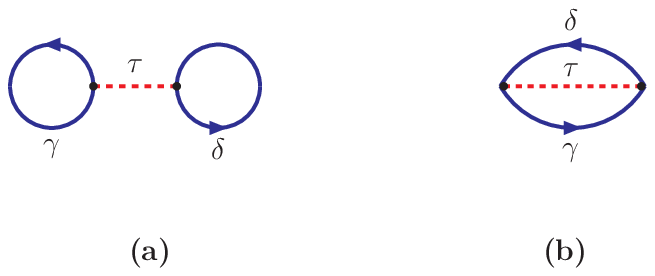}
\label{abb:2.5.2}
\ece
\vspace*{-14cm}
{\bf Fig. \arabic{abb}} : Diagrams for the partition function in 
$1^{\rm st} $-order perturbation theory at finite temperature.
\vspace*{1cm}
\end{figure}

\noindent
For $ n = 1 $ there are two diagrams (see Fig. \ref{abb:2.5.2}):

\refstepcounter{abb}

\vspace{0.5cm}

\noindent
Diagram (a) is given by
\bea
(a) \EA - \frac{1}{2} \int_0^{\beta} d\tau \> \sum_{\alpha \beta \gamma 
\delta}
\underbrace {\bar z_{\alpha}(\tau) \overbrace {\bar z_{\beta}(\tau) 
z_{\delta}(\tau)} z_{\gamma}(\tau)} \, \left < \gamma \delta | V | \gamma 
\delta \right > \non
\EA - \frac{1}{2} \int_0^{\beta} d\tau \> \sum_{\gamma \delta}
\> g_{\gamma}(0) g_{\delta}(0) \, \left < \gamma \delta | V | \gamma \delta 
\right > \> .
\eea
Diagram (b) results from contracting $ \bar z_{\alpha}(\tau) $ with
$ z_{\delta}(\tau) $ and $ \bar z_{\beta}(\tau) $ with
$ z_{\gamma}(\tau)$. It has the value
\be
(b) \E - \frac{1}{2}  \zeta \int_0^{\beta} d\tau \> \sum_{\gamma \delta}
\> g_{\gamma}(0) g_{\delta}(0) \, \left <  \delta \gamma | V | \gamma \delta
\right > \> .
\ee
By means of the explicit expression (\ref{Propagator}) we thus obtain
\be
\frac{Z}{Z_0} \E 1 - \frac{\beta}{2} \sum_{\gamma \delta}
\> n_{\gamma} n_{\delta} \Bigl [ \> \left <  \gamma \delta | V | \gamma \delta
\right >  + \zeta \left <  \delta \gamma | V | \gamma \delta
\right > \> \Bigr ] + {\cal O}(V^2)
\ee
and from that for the thermodynamic potential
\be
\frac{Z}{Z_0} \E e^{ - \beta ( \Omega - \Omega_0 ) } \E 
1 - \beta ( \Omega_1 - \Omega_0 ) + \ldots
\ee
\be
\Omega_1 \E \Omega_0 
+ \frac{1}{2} \sum_{\gamma \delta}
\> n_{\gamma} n_{\delta} \Bigl [ \> \left <  \gamma \delta | V | \gamma \delta
\right >  + \zeta \left <  \delta \gamma | V | \gamma \delta
\right > \> \Bigr ]  \> .
\ee
For fermions ($ \zeta = - 1 $) we can determine the ground-state energy by the limit
 $ T \to 0 $, i.e. $ \beta \to \infty $ 
\be
\boxed{
\qquad E_1 \E E_0 + \frac{1}{2} \sum_{\gamma \delta < F} \> \Bigl [ \> \left <  
\gamma \delta | V | \gamma \delta
\right >  - \left <  \delta \gamma | V | \gamma \delta
\right > \> \Bigr ]  \>. \quad
}
\label{E 1}
\ee
In this expression the summation over single-particle states extends 
up to the Fermi energy since at zero temperature the occupation number
becomes a step-function.
The first term is the direct or \textcolor{blue}{\bf ``Hartree''} term, the second one 
the exchange or \textcolor{blue}{\bf ``Fock''} term for the 
energy \footnote{In general this designation is not fully correct as the basis here is arbitrary 
and not optimal (selfconsistent), see the next chapter.
However, for homogeneous systems plane waves are also the selfconsistent states.}. 

\vspace{0.3cm}

\noindent
The number of "labeled" Feynman diagrams grows 
dramatically with the order of perturbation theory.
As many have the same numerical value, it is useful
to introduce "unlabeled" diagrams which are weighted by a {\bf symmetry factor}.
The modified Feynman rules for these types of diagrams are derived in the book 
of \meingruen{Negele \& Orland}  and are not presented here.
Likewise, it is obvious that in {\bf homogeneous systems} 
further simplifications of the rules can be achieved by going over to energy and/or
momentum representation.

\vspace{0.5cm}

\subsection{\textcolor{blue}{Auxiliary Fields and Hartree Approximation}}
\label{sec2: Hilfsfelder}

Perturbation theory, of course, fails at strong coupling and then the famous question 
arises \blau{\textsf{``What Is to Be Done?''}}{\bf \{Lenin\}}.
Apart from a numerical evaluation of the (Euclidean) path integral
there also exist analytical methods which are helpful in this case.
These are based on the introduction of a \blau{\bf mean field} which is generated by the particles 
themselves and in which they move independently in lowest approximation.
This approximation is physically plausible and the fundament for the successful description
of atoms and nuclei by shell models.

\vspace{0.2cm}

In the path-integral formalism one derives these approaches most easily by transforming
Eq. (\ref{U kohaerent 3}) for the
time-evolution operator in the coherent basis. 
For a local (spinless) two-body interaction
 $ V(\fx  - \fx') $ this formula reads in position representation
\be
U \left ( \Phi^*_f t_f; \Phi_i t_i \right ) \E \int_{\Phi(t_i) = \Phi_i}^
{\Phi^*(t_f) = \Phi^*_f} {\cal D} \Phi^*(\fx,t) {\cal D} 
\Phi(\fx,t) \> 
\exp \left ( \, \int d^3 x \> \Phi^*_f(\fx) \Phi_f(\fx) + \frac{i}{\hbar} \, 
S [ \Phi^*, \Phi ] \, \right )
\label{U phi(x)}
\ee
with the action (see Eq. (\ref{H op}))  
\bea 
&& S [ \Phi^*, \Phi ] \E \int_{t_i}^{t_f} dt \, \int d^3 x \> \Biggl [ \, 
\Phi^*(\fx,t) \left ( i \hbar \, \frac{\partial}{\partial t} 
+ \frac{\hbar^2 \Delta}{2 m} \right )  \Phi(\fx,t) \non 
&& \hspace{2cm} - \frac{1}{2}\int d^3 x'  \> \Phi^*(\fx,t)  \Phi^*(\fx',t) \, 
V(\fx - \fx') \, \Phi(\fx',t) \Phi(\fx,t) \, \Biggr ] \> .
\label{wirk phi}
\eea 
As is well-known the last (quadri-linear) term prevents an evaluation of the path integral 
by analytical means. However, it is possible to express this interaction term formally 
by a term which is quadratically in the fields.
This is based on the following identity for the (one-dimensional) Gaussian integral which we 
already have used in Eq. \eqref{aufheb quad} (replace there 
$ a \to - V/2 , \> x \to ix $)
\be
\exp \left ( - \frac{i}{2} V \, x^2 \right ) \E \frac{1}{\sqrt{2 \pi i V}} \, 
\int_{-\infty}^{\infty} dy \> \exp \left ( \frac{i}{2 V} y^2 - i y \, x \right )
\> .
\label{undoing the square}
\ee
This is called \blau{\bf "undoing the square"} because on the r.h.s. the variable 
 $x$ now only appears linearly. As always such Gaussian integrals can be generalized to 
 the multi-dimensional, even infinite-dimensional case -- in many-body physics 
 the corresponding identity is usually called the 
\textcolor{blue}{\bf Hubbard-Stratonovich transformation}. If we introduce the density
\be
\rho(\fx,t) \Def \Phi^*(\fx,t) \, \Phi(\fx,t)
\label{Dichte}
\ee
then we may write the interaction term for  bosons as well as for fermions 
(anticommute twice) in the form \footnote{In the treatment by \meingruen{\bf Negele \& Orland}, ch. 7 
unphysical selfenergy terms arise which are canceled in higher orders.} 
\be
\frac{1}{2} \int d^3 x \, d^3 x'  \> \Phi^*(\fx,t)  \Phi^*(\fx',t) \, 
V(\fx - \fx') \, \Phi(\fx',t) \Phi(\fx,t) \E \frac{1}{2} \int d^3 x \, d^3 x'  
\> \rho(\fx,t) \, V(\fx - \fx') \, \rho(\fx',t) \> .
\label{def dichte}
\ee
The functional generalization of Eq. (\ref{undoing the square}) 
is then (we now set $\hbar = 1$)
\vspace{0.2cm}

\fcolorbox{black}{white}{\parbox{14cm}
{
\bea
&& \exp \left [ \, - \frac{i}{2} \int dt \int d^3 x \, d^3 x' \> 
\rho(\fx,t)\,  V(\fx -\fx') 
 \rho(\fx',t) \, \right ] \E {\rm const.} \> \fdet^{-1/2} (V) \, 
\int {\cal D} \sigma(\fx,t) \non
&& \cdot \exp \left \{  i\int dt  \left [ \frac{1}{2} \int d^3 x d^3 x' \, 
\sigma(\fx,t) \left ( V^{-1} 
\right )(\fx,\fx')  \, \sigma(\fx',t) -  \int d^3 x \, \rho(\fx,t) \, 
\sigma(\fx,t) \, \right ] \right \} \> .\no
\eea
}}
\vspace{-2cm}

\bea
\label{HS transf}
\eea
\vspace{0.6cm}


\noindent
With that the matrix element (\ref{U phi(x)}) of the  time-evolution operator
assumes the form
\bea
U \left ( \Phi^*_f t_f; \Phi_i t_i \right ) \EA  \exp \left (\, \int d^3 x \, 
\Phi^*_f(\fx) \Phi_f(\fx) \, \right ) \, 
\int {\cal D}
\sigma(\fx,t)  \, W [ \sigma ] \, \int_{\Phi(t_i) = \Phi_i}^
{\Phi^*(t_f) = \Phi^*_f} \! \! {\cal D} \Phi^*(\fx,t) {\cal D} 
\Phi(\fx,t) \non 
&& \hspace{1cm} \cdot \exp \left [ \, i \int dt \int d^3 x 
\left ( i \Phi^*(\fx,t) 
\frac{\partial\Phi(\fx,t)}{\partial t} - {\cal H}_{\sigma}(\Phi^*,\Phi) 
\right ) \, \right ] \> ,
\label{U sigma}
\eea
where
\be 
{\cal H}_{\sigma}(\Phi^*,\Phi) \E \Phi^*(\fx,t) \left ( - 
\frac{\Delta}{2m}\right )  \Phi(\fx,t) + \sigma(\fx,t) \, \Phi^*(\fx,t) 
\EQ \Phi^*(\fx,t) \, \hat H_{\sigma} \, \Phi(\fx,t)
\Phi(\fx,t)
\ee
is the Hamilton density of particles moving independently in the auxiliary field
$\sigma(\fx,t)$ and thus a 
\textcolor{blue}{\bf single-particle} operator (c.f. Eq. (\ref{kin})).
However, one has to integrate functionally over this auxiliary field with the weight
\be
W [ \sigma ] \E  {\rm const.} \> \exp \left [ \, \frac{i}{2} \int dt 
\int d^3 x \, d^3 x' \, \sigma(\fx,t) 
\, \left ( V^{-1} \right )(\fx,\fx')  \, \sigma(\fx',t) \, \right ] 
\ee
which contains the information about the actual two-body interaction  $ V(\fx -\fx')$ .
Note that this weight does not have a kinetic term for the auxiliary field
 $\sigma(\fx,t)$ with the consequence that the $\sigma$-configurations can be 
very "rough", i.e. discontinous, similar as in the phase-space path integral.

\vspace{0.2cm}

The Hubbard-Stratonovich transform gives the exact expression (\ref{U sigma})
in which we can calculate the single-particle problem in the auxiliary field
but not the subsequent functional integral over $ \sigma(\fx, t) $ --
which illustrates the well-known principle of conservation of difficulties ...
However, if we apply the \textcolor{blue}{\bf stationary phase approximation} to
the $ \sigma $ - functional integral we get a meaningful approximation which
can be improved systematically. The auxiliary field which makes the action in
Eq. (\ref{U sigma}) stationary is obtained by variation of the 
$\sigma$-dependent exponents  w.r.t. $\sigma(\fx,t)$:
\be
\int d^3x \> \left ( V^{-1} \right ) (\fx,\fx') \, \sigma(\fx',t) - \rho(\fx,t)
\stackrel{!}{=} \> 0 \> ,
\ee
i.e. the stationary auxiliary field is the average of the two-body interaction
over the density of the particles:
\be
\boxed{
\qquad \sigma_0(\fx,t) \E \int d^3 x' \> V(\fx - \fx') \>  \rho(\fx,t)
\> , \quad
}
\label{sigma0}
\ee
and thus the expected mean field.
If we consider a fermionic system with  $N$ particles we may expand w.r.t.
eigenfunctions $\, \varphi_k(\fx,t) \, $ of the single-particle time-evolution operator.
The ground state wave function in the (fixed) auxiliary field
$\sigma_0(\fx,t)$ is a 
{\bf Slater determinant}
\be 
\Psi_H \lrp \fx_1 \ldots \fx_N, t \rrp \E \frac{1}{\sqrt{N!}} \, \det_{j, \, k} \> \Big [ \varphi_k(\fx_j,t) \, \Big ]
\ee
built from this complete, orthonormal
single-particle basis and the density is given by 
the sum over all occupied single-particle states 
\be
\rho(\fx,t) \E \sum_{k=1}^N \, \left | \varphi_{k} (\fx,t) \right |^2 \> .
\ee
Due to the Hubbard-Stratonovich transformation the remaining
functional integral over $\varphi^*, \varphi$ is a Gaussian integral
to which we may apply the stationary-phase method as well (which gives the
exact result in this case). By variation w.r.t.
$\varphi_k^*(\fx,t)$ we then obtain the Schr\"odinger equation
\be
i \frac{\partial\varphi_k(\fx,t)}{\partial t} \E - \frac{\Delta}{2m}  
\varphi_k(\fx,t) + \sigma_0(\fx,t) \, \varphi_k(\fx,t)   
\label{Hartree}
\ee
for the motion in the mean field  $\sigma_0(\fx,t)$. Eqs.
 (\ref{sigma0}-\ref{Hartree}) are coupled equations and define the well-known 
\textcolor{blue}{\bf Hartree approximation}. 

If we assume a time-independent mean field $ \sigma(\fx) $ 
the time dependence of the single-particle wave functions can be separated, i.e.
we can set
$ \> \varphi_k(\fx,t) = \varphi_k(\fx) \, \exp(-i \epsilon_k t) \> $ and then
has to solve the non-linear Schr\"o\-dinger equation
\be
\left [ \,  - \frac{\Delta}{2m}  + \sum_{l=1}^N \int d^3 x' \, V(\fx -\fx') 
\left | \varphi_l(\fx') \right |^2 \,  \right ] \, \varphi_k(\fx) 
\E \epsilon_k \, \varphi_k(\fx) \> .
\label{einteil zeitunab}
\ee
If this would be a normal Schr\"odinger equation then the total energy would be
\bea
\sum_{k=1}^N \epsilon_k \E \sum_{k=1}^N \left < \varphi_k \left | 
-\frac{\Delta}{2m}  \right | \, \varphi_k \right >  + 
\sum_{k,l} \left < \varphi_k \varphi_l | \hat V | 
\, \varphi_k \varphi_l \right >  \nonumber \> ,
\eea
where the last relation follows from Eq. (\ref{einteil zeitunab}) after 
multiplying with $\varphi^*_k(\fx)$, integrating over $\fx$  and 
using the orthonormality of the eigenfunctions. Obviously, the potential 
energy has been counted twice in this procedure.
The correct relation is obtained from the resolvent
\be        
G(E) \E {\rm tr} \left ( \frac{1}{E - \hat H + i 0^+} \right ) \E 
- i \, \int_0^{\infty} dT \> e^{i E T } \, {\rm tr} \> \hat U(T,0) \> ,
\label{resolvent}
\ee
which in the $N$-particle sector  in a fixed basis of Slater determinants $ \> \Psi_{k} \> $ 
takes the form
\be 
G_N(E) \E  -i \, \sum_{k_1 ... k_N} \int_0^{\infty} dT \, e^{i E T } \, \int {\cal D}\sigma \, 
\exp \left [ \frac{i}{2} \int_0^T \int d^3 x \, d^3 x' \sigma(\fx) 
V^{-1}(\fx,\fx') \sigma(\fx' )+ \ln \la \Psi_k \big | \hat U_{\sigma}(T,0) \big | \Psi_k \ra  
\right ] 
\label{resolvent 2}
\ee
where $ \> \hat U_{\sigma}(T,0) = {\cal T} \, \exp \lrp - i \int_0^T dt \, \hat H_{\sigma} \rrp \> $ is the time-evolution operator for independent particles in the potential $ \> \sigma(\fx,t) \> $.
In static Hartree approximation this becomes
\be 
G_N(E) \> \simeq \> - i \, \sum_{k_1 ... k_N} \int_0^{\infty} dT \> e^{i E T } 
\exp \left [ \, \frac{i T}{2} \int d^3 x \, d^3 x' \sigma_0(\fx) 
V^{-1}(\fx,\fx') \sigma_0(\fx' ) - i T \sum_{i=1}^{N} \epsilon_{k_i} \, \right ] \> .
\ee
The pole of the resolvent gives us the total energy
\be 
\boxed{
\quad E_H \E  \sum_{k=1}^N \epsilon_k  - \frac{1}{2} \int d^3 x \, d^3 x' 
\sigma_0(\fx) V^{-1}(\fx,\fx') \sigma_0(\fx' ) \E 
\sum_{k=1}^N \left < \varphi_k \lvl -\frac{\Delta}{2m} \rvl \, \varphi_k \right >  + 
\frac{1}{2} \sum_{k,l} \left < \varphi_k \varphi_l | \hat V | 
\, \varphi_k \varphi_l \right >  \> , \quad
}
\label{E Hartree}
\ee
in which the potential energy between the particles has been counted correctly.
In contrast to the perturbation theory considered in the last chapter, now the
single-particle wave functions are not specified anymore but have to be determined 
\textcolor{blue}{\bf self-consistently} from Eq. (\ref{einteil zeitunab}) which 
contains the interaction potential as well. This is essential for finite systems 
(electrons in an atom, nucleons in a nucleus).

An obvious disadvantage of this approximation is the violation 
of translational invariance: While the two-body potential
 $ \> V(\fx - \fx')\> $ is unchanged by a shift of coordinates, the mean field
 $\> \sigma_0(\fx) \> $ has a fixed origin (for which one usually takes the center
 of mass of the system).

\vspace{0.2cm}
It is clear that one can calculate systematic corrections to the lowest order
stationary-phase approximation. Taking into account the quadratic fluctuations 
leads to the so-called \textcolor{blue}{\bf RPA} \footnote{for 
\textcolor{blue}{\bf ``Random Phase Approximation ''}, the name traditionally given for
this approximation although it is not very illuminating in the present context ...}
corrections to the Hartree approximation. 
\vspace{1cm}
 
\renewcommand{\baselinestretch}{0.9}
\scriptsize
\refstepcounter{tief}
\noindent
\blau{\bf Detail \arabic{tief}:} {\bf RPA  Corrections}\\
\vspace{0.2cm}

\noindent
\begin{subequations}
Following \meingruen{\bf Negele \& Orland}, p. 344 - 348 (beware of misprints!) we make the change of variables
$ \> \sigma(\fx,t) \E \sigma_0(\fx) + \eta(\fx,t)\> $
where $ \> \sigma_0(\fx) \> $ is the Hartree solution and expand 
the exponential in the resolvent of Eq. \eqref{resolvent 2} up to second order in $ \eta $. This gives 
\be
G_N(E) = {\cal N}  \sum_{k_1 ... k_N} (-i)\int_0^{\infty} dT \, e^{i (E-E_H) T } 
\int {\cal D} \eta \, \exp \Bigg \{ - \frac{1}{2} \int d1 \, d2 \> \eta(1) \Bigg [ \underbrace{- i v^{-1}(1,2)- \frac{\delta^2   \ln \la \Psi_k \big | U_{\sigma} \big | \Psi_k \ra}{\delta \sigma(1) \delta \sigma(2)} \Big |_{\sigma = \sigma_0} }_{\deF \Gamma^{-1}(1,2)} \Bigg ] \eta (2) + {\cal O}\lrp \eta^3 \rrp \Bigg \}
\ee
because the terms independent of $ \> \eta \> $ generate the Hartree energy $ \> E_H \> $ and the first-order term vanishes due to the stationary-phase approximation. To simplify the notation we abbreviate
$ \> 1 \EQ (\fx_1,t_1) \> $ etc. and set $ \> v^{-1}(1,2) \Def \delta(t_1-t_2) (V^{-1})(\fx_1,\fx_2) \> $.
Time integrations run from zero to $ T $. In quadratic  approximation the $ \eta$-integral gives
\be 
{\cal Z} \E {\cal N} \, \lsp \fdet (\Gamma^{-1}) \rsp^{-1/2} \E \lsp \frac{\fdet (\Gamma^{-1})}{\fdet (-i v^{-1})} \rsp^{-1/2} \E \lsp \fdet( i v \Gamma^{-1}) \rsp^{-1/2} 
\ee
where we normalize to the case of first-order interaction.
Evaluation of the derivative appearing in $ \Gamma $ leads to 
\be 
\frac{\delta^2   \ln \la \Psi_H \big | U_{\sigma} \big | \Psi_H \ra}{\delta \sigma(1) \delta \sigma(2)} \Bigg |_{\sigma = \sigma_0}  \E - \la \Psi_H \big | {\cal T} \lrp \hat \psi^{\dagger}(1) \hat \psi(2) \rrp \big  | \Psi_H \ra \, \la \Psi_H \big | {\cal T} \lrp \hat \psi(1) \hat \psi^{\dagger}(2) \rrp \big  | \Psi_H \ra 
\ee
where 
\be 
\hat \psi^{\dagger}(1) \E \exp \lrp i h_{\sigma_0} t_1 \rrp \, \hat \psi^{\dagger}(\fx_1) \, 
\exp \lrp -i h_{\sigma_0} t_1 \rrp 
\ee
is the creation operator for a particle at space-time "1"
in the Heisenberg representation for the Hartree Hamiltonian. Introducing
the Hartree Green function  $ \> G(1,2) \Def \, -i \la \Psi_H \big | {\cal T} \lrp \hat \psi(1) \hat \psi^{\dagger}(2) \rrp \big  | \Psi_H \ra \> $
and the Hartree particle-hole propagator (depicted graphically in Fig. \ref{abb:2.7.1} (a))
\be 
i D_0(1,2) \Def G(1,2) \, G(2,1)
\ee
one obtains for the quadratic fluctuations 
\be 
{\cal Z} \E \Big [ \fdet \lrp 1 - v D_0 \rrp \Big ]^{-1/2} \E \exp \lsp \frac{1}{2} \sum_{n=1}^{\infty} 
\frac{1}{n} {\rm tr} \lrp (v D_0)^n \rrp \rsp \> .
\label{RPA corr}
\ee
The $n$th term in the exponent reads
\be 
\frac{1}{2 n} \int_0^T dt_1 d^3 x_1 d^3 x_1' \ldots dt_n d^3x_n d^3x_n' \> V(\fx_1 - \fx_2) D_0(\fx_1't_1,\fx_2 t_2 ) \ldots V(\fx_n'-\fx_n) D_0(\fx_n't_n, \fx_1,t_1)
\ee 
and can be represented by a {\bf ring} diagram as shown in Fig.  \ref{abb:2.7.1} (b).

\refstepcounter{abb}
\begin{figure}[hbtp]
\vspace*{-8.5cm}
\bce
\includegraphics[angle=0,scale=1.05]{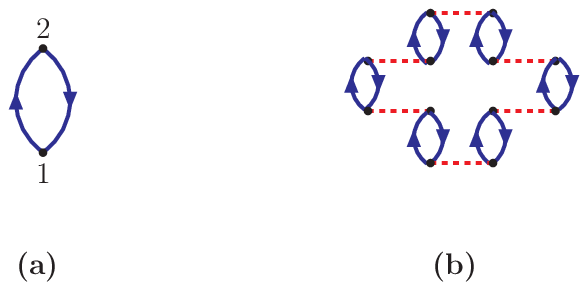}
\label{abb:2.7.1}
\ece
\vspace*{-16.3cm}
\hspace*{2cm} {\scriptsize\bf Fig. \arabic{abb}} : {\scriptsize (a) Graphical representation of the particle-hole propagator 
$ \> D_0(1,2) \> $. \\ \hspace*{3.4cm} (b) Graphical representation of the $ \, n = 6 \, $ term in the 
expansion \eqref{RPA corr} for the RPA corrections.}
\vspace*{1cm}
\end{figure}

\noindent
Writing $ \> {\cal Z} \sim \exp ( - i E T ) \> $ for $ \> T \to \infty \> $ one can read off the 
contributions to the energies from the quadratic fluctuations: 
The $n=1$-term provides the exchange energy which was absent in the Hartree approximation and the 
higher terms give the RPA-energy 
\be 
E_{\rm RPA} \E  \lim_{T \to \infty} \frac{i}{2 T} \, \sum_{n=2}^{\infty} \frac{1}{n} \, {\rm tr} 
\lrp D_0 V \rrp^n \> .
\ee
\end{subequations}
\renewcommand{\baselinestretch}{1.1}
\normalsize

\vspace*{1cm}

As already mentioned the RPA correction also contains the
exchange ( "Fock") term (\ref {E 1}) for the energy. One may
wonder why this term does not appear already in the lowest order;
in the case of nuclear forces, for example, it is
much more attractive than the direct ( "Hartree") term which does not even
bind the nucleons if realistic interctions are used.
As a matter of fact one has considerable freedom to modify the lowest order result
by a different grouping of fields than used in Eq. \eqref{def dichte}
and it is indeed  possible to obtain the Hartree-Fock approximation or more advanced
approximations as lowest-order result of the stationary-phase
method \footnote{For example, the nuclear pairing force can be included via 
the so-called 
Hartree-Fock-Bogoliubov approximation \cite{ToSev}. }.                            
This is due to a lack of a systematic expansion parameter: Since Planck's 
elementary quantum  $ \hbar $ also shows up in the action (\ref{wirk phi}) 
the stationary-phase method is not a semi-classical approximation anymore.
Different choices of the zeroth approximation for which one then calculates corrections
correspond to an arbitrary split-up of the Hamiltonian
$ \hat H = \hat H_0 + \hat H_1 $ and must be justified by the physics of the particular
problem.

\vspace{0.5cm}

\subsection{\textcolor{blue}{Asymptotic Expansion of a Class of Path Integrals}}
\label{sec2: Luttinger-Pekar}

In contrast to perturbation theory for weak coupling 
a general expansion for strong coupling is not available.
This is due to the (much deplored) inability to calculate non-gaussian functional 
integrals by analytic methods. However, following a work of Luttinger \cite{Lutt} 
it is possible to evaluate a particular class of path integrals asymptotically.
This will be presented in this chapter and applied to the polaron problem. 
Although only the  single-particle problem is treated here, methods of many-body physics
are used and the polaron problem (a single-particle problem after integrating out the phonons)
really belongs to "infinite-particle" physics, i.e. to field theory.

\vspace{0.3cm}

We begin with the expression \eqref{Z Pfad2} for the partition function of a particle
moving (in 3 dimensions) under the influence of an external potential
(in the following we set  $\hbar = 1 $)
\be
Z(\beta) \E {\rm tr} \lrp  e^{- \beta \hat H} \rrp
\E \oint_{\fx(0)=\fx(\beta)} {\cal D}^3x  \> 
\exp \lsp - \int_0^{\beta} d\tau \> \lrp \frac{m}{2} \dot \fx^2 + V(\fx)
\rrp \rsp \> ,
\label{Z pot 2}
\ee
From the partition function we may determine the ground-state energy 
by taking the limit  $ \beta \to \infty $ :
\be
E_0 \E \lim_{\beta \to \infty} \, \lcp - \frac{1}{\beta} \, \ln Z(\beta) \rcp \> .
\label{E grund}
\ee
It can be shown that the partition function \eqref{Z pot 2} can also be represented by the
following path integral over bosonic fields $ \Phi(\fx), \Phi^{\star}(\fx) $ and fermionic
 (Grassmann-valued) fields $ \cy{\eta}(\fx), \cy{\bar \eta}(\fx) $ :
\bce
\vspace{0.2cm}

\fcolorbox{black}{white}{\parbox{12cm}{
\bea
Z(\beta) \EA \int {\cal D} \Phi^{\star} \, {\cal D} \Phi \, {\cal D} \cy{\bar \eta} \, {\cal D} \cy{\eta}
\> \int d^3x \> \> \cy{\bar \eta}(\fx) \, \cy{\eta}(\fx) \> \> \delta \Bigl ( \, (\Phi^{\star},\Phi) + 
(\cy{\bar \eta}, \cy{\eta}) - 1 \, \Bigr ) \non
&& \hspace{3.5cm} \cdot \exp \Bigl \{  - \beta \, \bigl [ \,  (\Phi^{\star},H \Phi) + 
(\cy{\bar \eta}, H \cy{\eta}) \, \bigr ] \, \Bigr \} \> . \no
\eea
}}
\ece
\vspace{-2cm}

\bea
\label{Z Lutt}
\eea
\vspace{0.6cm}

\noindent
Here we have used the abbreviations
\be
(\Phi^{\star}, {\cal O} \, \Phi) \Def \int d^3x \, d^3y   \> \Phi^{\star}(\fx) \, \la \fx \left | 
\hat {\cal O}
\right | \fy \ra \, \Phi(\fy) \ , \quad   (\cy{\bar \eta}, \cy{\eta}) \Def
\int d^3x  \, d^3y \> \, \cy{\bar \eta}(\fx) \, \la \fx \left | \hat {\cal O}
\right | \fy \ra \, \cy{\eta}(\fy) \\
\ee
for the operators $ \hat {\cal O} = \hat 1 , \> \hat H $ and the bosonic path integral has been
normalized to
\be
\int {\cal D} \Phi^{\star} \, {\cal D} \Phi \> \exp \lsp - \lrp \Phi^{\star}, \Phi \rrp \rsp \E 1 \> .
\ee

\vspace{0.8cm}

\renewcommand{\baselinestretch}{0.9}
\scriptsize
\refstepcounter{tief}
\noindent
\blau{\bf Detail \arabic{tief}:} {\bf Luttinger's Path-Integral Relation}\\
\vspace{0.2cm}

\noindent
\begin{subequations}
Due to the constraint by the $\delta$-function we can also write 
Eq. \eqref{Z Lutt} as
\be
Z(\beta) \E e^{-\beta E_0} \, \int {\cal D} \Phi^{\star} \, {\cal D} \Phi \, {\cal D} \cy{\bar \eta} \, 
{\cal D} \cy{\eta}
\> \int d^3x \> \> \cy{\bar \eta}(\fx) \, \cy{\eta}(\fx) \> \> \delta \Bigl ( \, \Phi^{\star},\Phi) + 
(\cy{\bar \eta}, \cy{\eta}) - 1 \, 
\Bigr ) \cdot \exp \Bigl \{  - \beta \, \bigl [ \,  (\Phi^{\star},h \Phi) + (\cy{\bar \eta}, h \cy{\eta}) 
\, \bigr ] \, \Bigr \} 
\ee
where $ h \Def H - E_0 $ has only  positive eigenvalues.    
If we use the Fourier representation of the
$\delta$-function 
\be
 \delta \Bigl ( \, \Phi^{\star},\Phi) + (\cy{\bar \eta}, \cy{\eta}) - 1 \, \Bigr ) \E \frac{1}{2 \pi} \, 
\int_{-\infty}^{+\infty}
d\omega \> \exp \Bigl \{ - \, i \omega \lsp  \, (\Phi^{\star},\Phi) + (\cy{\bar \eta}, \cy{\eta}) \rsp + 
i \omega \, \Bigr \} 
\ee
then we can perform the integrations over bosonic and fermionic auxiliary fields by means of
Eq. \eqref{Gauss ferm/boson} :
\bea
\int {\cal D} \Phi^{\star} \, {\cal D} \Phi \> \exp \lsp - \lrp \Phi^{\star}, ( i \omega + \beta h ) 
\Phi \rrp \rsp
\EA  \frac{1}{{\cal D}{\rm et} \lrp  i \omega + \beta \hat{h} \rrp } \\
 \int  {\cal D} \cy{\bar \eta} \, {\cal D} \cy{\eta} \> \underbrace{\int d^3x \> \cy{\bar \eta}(\fx) \, 
\cy{\eta}(\fx)}_{\equiv (\cy{\bar \eta},\cy{\eta})} \> 
 \exp \lsp - \lrp \cy{\bar \eta}, ( i \omega + \beta h) \cy{\eta} \rrp \rsp \EA 
i \frac{\partial}{\partial \omega} \> 
 \int  {\cal D} \cy{\bar \eta} \, {\cal D} \cy{\eta} \>  
 \exp \lsp - \lrp \cy{\bar \eta}, ( i \omega + \beta h) \cy{\eta} \rrp \rsp \non
 \EA i \frac{\partial}{\partial \omega} \, {\cal D}{\rm et} \lrp i \omega + \beta \hat{h} \, \rrp  \non
\EA - {\rm tr} \lrp
 (\, i \omega + \beta \hat{h} \, )^{-1} \rrp \,     
 {\cal D}{\rm et} (\, i \omega + \beta \hat{h} \, ) 
\> .
\eea
In the last line the result of \purpur{\bf Problem \ref{Det Spur} b)} for the differentiation 
of a determinant w.r.t. a parameter has been used. We see that the inverse determinant from the
bosonic and the determinant from the fermionic integration cancel.
With the theorem of residues we then indeed obtain
\be 
Z(\beta) \E - e^{-\beta E_0} \, \frac{1}{2 \pi} \, \int_{-\infty}^{+\infty} 
d\omega \> e^{i \omega} \, {\rm tr} \lrp
 \frac{1}{i \omega + \beta  \hat{h} } \rrp \E  e^{- \beta E_0} \, \frac{i}{2 \pi} \, 
\int_{-\infty}^{+\infty} d\omega \> e^{i \omega} \, {\rm tr} \lrp
 \frac{1}{\omega -i  \beta (\hat H -E_0 )} \rrp \E   {\rm tr} \lrp e^{-\beta \hat H} \rrp \> , 
\ee
as we have to close the integration contour in the upper half plane where all the poles are located.
\end{subequations}
\renewcommand{\baselinestretch}{1.2}
\normalsize
\vspace{0.8cm}

\noindent
On first sight the representation \eqref{Z Lutt} looks like an unnecessary complication of the ``simple''
path integral over \textcolor{blue}{\bf trajectories}: Now one integrates over bosonic and fermionic
\textcolor{blue}{\bf fields}! However, as we will see shortly, this representation has some advantages:

\noindent
If we express on the l.h.s. the term
\be
\int_0^{\beta} d\tau \> V(\fx(\tau)) \E \beta \, \int d^3x \> L_{\beta}(\fx) \, V(\fx)
\ee
by the ``local time'' 
\be
L_{\beta}(\fx) \Def \frac{1}{\beta} \, \int_0^{\beta} d\tau \> \delta \lrp \fx - \fx(\tau) \rrp
\label{lokale Zeit}
\ee
then the interaction term has the same form as on the r.h.s. where 
\bea
\lrp \Phi^{\star}, H \Phi \rrp + \lrp \cy{\bar \eta}, H \cy{\eta} \rrp 
\EA \underbrace{\int d^3x \> \Phi^{\star}(\fx) \lrp - \frac{\Delta}{2m} \rrp \, 
\Phi(\fx)}_{= (\Phi^{\star},H_0 \Phi)}
+ \int d^3x \> \Phi^{\star}(\fx) \, V(\fx) \, \Phi(\fx) \non
&& + \lrp \cy{\bar \eta}, H_0 \cy{\eta} \rrp 
+ \int d^3x \> \cy{\bar \eta} (\fx) \, V(\fx) \, \cy{\eta} (\fx)
\eea
show up. We thus see that going from a description by paths to a description 
by fields amounts to the replacement
\be
L_{\beta}(\fx) \> \longrightarrow \> \Phi^{\star}(\fx) \, \Phi(\fx)
+  \cy{\bar \eta} (\fx) \, \cy{\eta} (\fx) \> .
\ee
Now one can multiply both sides by an arbitrary functional of $ V $ (let's call it
$ \Gamma[V] $) and functionally integrate over both sides; in this way we extend the class
of simple path integrals over one-time actions to more general forms.
If the functional $ F_{\beta}[L_{\beta}] $ is defined by
\be
\exp \lrp - \beta F_{\beta} \lsp L_{\beta} \rsp \rrp \Def \int {\cal D} V \> \Gamma[V] \, 
\exp \lrp - \beta \int d^3x \> L_{\beta}(\fx) \, V(\fx) \rrp
\ee
then the relation
\bea
\oint {\cal D}^3x \> \exp \lcp - S_0[x] -  \beta \, F_{\beta} \lsp L_{\beta} \rsp \rcp
\EA \int {\cal D} \Phi^{\star} \, {\cal D} \Phi \, {\cal D} \cy{\bar \eta} \, {\cal D} \cy{\eta}
\> \int d^3x \> \> \cy{\bar \eta}(\fx) \, \cy{\eta}(\fx) \> \> \delta \Bigl ( \, \Phi^{\star},\Phi) + 
(\cy{\bar \eta}, \cy{\eta}) - 1 \, 
\Bigr ) \non
&& \cdot \exp \Bigl \{  - \beta \lsp \lrp \Phi^{\star},H_0 \Phi \rrp + \lrp  \cy{\bar \eta}, H_0 \, 
\cy{\eta} \rrp 
\rsp -  \beta F_{\beta} \, \lsp \Phi^{\star} \Phi  + \cy{\bar \eta} \cy{\eta} \rsp \, \Bigr \}  
\label{Z Lutt allg}
\eea
follows.
This means that we can ``translate'' path integrals over more general functionals into
functional integrals over fields -- but unfortunately, in general we also cannot ``do'' the latter ones...
However, in the limit $ \, \beta \to \infty \, $ the r.h.s. of Eq. \eqref{Z Lutt allg} simplifies: If one
scales $ \cy{\eta} \to \cy{\eta}/\beta \> , \> \> \cy{\bar \eta} \to  \cy{\bar \eta}/\beta $ then we 
know that the fermionic ``measure'' transforms with the inverse Jacobian, i.e. all
$\cy{\bar \eta}, \cy{\eta}$-integrals remain invariant. This means that the fermionic
parts in the integrand are suppressed and that aymptotically the partition function
is determined solely by the bosonic integral
\bea
\lim_{\beta \to \infty} \, \oint {\cal D}^3x \> \exp \lcp - S_0[x] -  
\beta \, F_{\beta} \lsp L_{\beta} \rsp \rcp
\EA \int {\cal D} \Phi^{\star} \, {\cal D} \Phi 
\> \int d^3x \> \> \cy{\bar \eta}(\fx) \, \cy{\eta}(\fx) \> \> \delta \Bigl ( \, \Phi^{\star},\Phi) - 1 \, 
\Bigr ) \non
&& \cdot \exp \Bigl \{  - \beta \lrp \Phi^{\star},H_0 \Phi \rrp  
-  \beta F_{\beta \to \infty} \, \lsp \Phi^{\star} \Phi  \rsp \, \Bigr \} \> . 
\label{Z Lutt beta gross}
\eea
The leading term of this integral may be determined by
\textcolor{blue}{\bf Laplace's method} (see Eq. \eqref{Int a la Laplace}): Recall that for large values of 
a parameter in the exponential function  of the integral only the minimum of the function (here: of the 
functional) contributes which is multiplied by this parameter.
Therefore we obtain from Eq. \eqref{E grund}
\vspace{0.2cm}
\be
\boxed{
E_0 \E {\rm min}_{(\Phi^{\star},\Phi) = 1} \, \Biggl \{ \>   \lrp \Phi^{\star}, H_0 \Phi \rrp  
+  F_{\infty} \, \lsp \Phi^{\star} \Phi  \rsp \> \Biggr \}  \> .
}
\label{E0 Pekar}
\ee
\newpage 
\noindent
{\bf Applications}:
\bdes
\item[ \textcolor{blue}{a) Quantum Mechanics}]: If
$ \> F_{\beta}[L_{\beta}] = \int d^3x \> L_{\beta}(\fx) \, U(\fx) \> $,
then Eq. \eqref{E0 Pekar} is nothing else than the Rayleigh-Ritz variational principle
for the motion of a non-relativistic particle in the potential $ \> U (\fx) $.
Unconstraint variation w.r.t. the normalized (wave) function $\Phi^{\star}(\fx)$
gives the usual time-independent \textcolor{blue}{\bf linear Schr\"odinger equation}.

\item[\textcolor{blue}{b) Polaron in the Limit of Strong Coupling}]: For arbitrary coupling constants 
the interaction term in the effective action of the polaron (after integrating out the phonons) 
does {\bf not} have a form so that it can be expressed solely in terms of the
``local time'' \eqref{lokale Zeit}. However, this is the case for large coupling constants
as can be seen by scaling \cite{AdGeLe}
\bea
\hspace{-3cm} && \bar \tau \E \lambda  \tau \> , \qquad \bar \fx(\bar \tau) \E \sqrt{\lambda} \, 
\fx(\tau) \qquad \lambda > 0 \non 
\Longrightarrow &&
\frac{1}{2} \, \int_0^{\beta} d\tau \> \lrp \frac{\fx(\tau)}{d \tau} \rrp^2
\> \longrightarrow \> \frac{1}{2} \, \int_0^{\lambda \beta} d \bar \tau \> \lrp 
\frac{\bar \fx(\bar \tau)}{d \bar \tau} \rrp^2 \non
&& - \frac{\alpha}{\sqrt 2} \, \int_0^{\beta} d\tau \, d\tau' \> \frac{G_{\beta}\ (\tau-\tau')}
{ |\fx(\tau) - \fx(\tau') |} \, \longrightarrow \, - \frac{\alpha}{\sqrt 2} \, 
\frac{1}{\lambda^{3/2}} \, \int_0^{\lambda \beta} d \bar \tau \, d \bar \tau' \> 
\frac{G_{\beta}\lrp (\tau - \tau') /\lambda \rrp}
{ | \bar \fx(\bar \tau) - \bar \fx(\bar \tau') |} \> .
\eea
Here (cf. Eq. \eqref{polaron ret fkt})
\be
G_{\beta}(t) \E \frac{\cosh(\beta/2 - |t|)}{2 \sinh(\beta/2)} \no
\ee
is the polaron retardation function for finite  $ \beta$. If we choose
\be
\lambda \E \lrp \kappa \, \alpha \beta \rrp^2 
\label{Wahl lambda}
\ee
(with an arbitrary numerical factor $ \kappa $), then we see that for fixed $ \beta$ 
the retardation vanishes for $ \alpha \to \infty $ and that in this case the polaron self-interaction 
can indeed be written as a functional of the local time. Replacing the primed quantities
by unprimed ones we get
\bea
 - \frac{\alpha}{\sqrt 2} \, \frac{1}{\lambda} \, \frac{G_{\beta}(0)}{\kappa \alpha \, \beta} \, 
\int_0^{\lambda \beta} d\tau \, d\tau' \> 
\frac{1}{ | \fx(\tau) - \fx(\tau') |} \EA - \frac{G_{\beta}(0)}{\kappa \sqrt 2} \, \frac{1}{T} \, 
\int_0^{T} d\tau \, d\tau' \> \frac{1}{ | \fx(\tau) - \fx(\tau') |} \non
&& \hspace{-2cm} \equiv \>  T \, \int d^3x \, d^3y \> L_T(\fx) \, 
\lrp - \frac{G_{\beta}(0)}{ \kappa \sqrt 2} 
\frac{1}{ \, |\fx - \fy | } \rrp \, L_T(\fy) \> .
\eea
Instead of the Euclidean time $ \beta $  we now have the quantity
\be
T \E \lambda \, \beta 
\ee
which also becomes asymptically large in the limit of strong coupling and finite $\beta $. 
Hence we can apply Luttinger's method and from Eq. \eqref{E0 Pekar} and the choice \eqref{Wahl lambda}
we obtain for the free energy
\be
F(\beta,\alpha)  \> \stackrel{\alpha \to \infty}{\longrightarrow} \>  \frac{\alpha^2 \beta^2}{\beta}
\, G_{\beta}(0) \> \gamma_P \E \beta \, \frac{1}{2} \coth (\beta/2) \> \gamma_P \> \alpha^2
\label{freie En}
\ee
with
\be
\gamma_P \E \kappa^2 \, {\rm min}_{(\Phi^{\star},\Phi) = 1} \, \Biggl \{ \>   
\int d^3x \> \Phi^{\star}(\fx) \, 
\lrp - \frac{\Delta}{2} \rrp \, \Phi(\fx) - \frac{1}{\kappa \sqrt 2} \, \int d^3x \, d^3y \> 
\frac{\Phi^{\star \> 2 }(\fx) \, \Phi^2(\fy)}{|\fx - \fy |} \> \Biggr \} \> .
\label{Pek 1}
\ee
The free energy should tend to the ground-state energy for small temperature (large $ \beta $).
Unfortunately, we cannot perform the limit $ \beta \to \infty $ in 
Eq. \eqref{freie En} anymore as essential parts of the action
have been removed by the $ (\alpha \to \infty) $ - limit -- obviously, both limits do not commute.
The general property that the free energy is always smaller than the ground-state energy and grows
monotonically with $\beta$ (\purpur{\bf Problem \ref{freie Energie}}) comes to our rescue.
In the limit $\beta \to 0 $ we have
\be
\beta \, \frac{1}{2} \, \coth (\beta/2) \> \stackrel{\beta \to 0}{\longrightarrow} \> 
\beta \frac{1}{\beta} \E 1 ,
\ee
so that $ F(0) = \gamma_P \,  \alpha^2 $ is a  {\bf lower bound} for the ground-state energy
of the strong-coupling polaron. Exactly the same result
had been derived much earlier by Pekar by a quantum-mechanical {\it ansatz} for the
total wave function of electron + phonon system \cite{Pekar} in a Rayleigh-Ritz variational calculation.
Since it is basic knowledge that the variational principle leads to an {\bf upper bound} for the true
ground-state energy, the equality of these two bounds allows the conclusion that indeed
the ground-state energy of the polaron at strong coupling is given by  \footnote{Rigorously proven by Lieb and Thomas \cite{LiebThom}.}
\be
\boxed{
E_0  \> \stackrel{\alpha \to \infty}{\longrightarrow} \>  \gamma_P \> \alpha^2 \> .
} 
\label{polaron gross alpha}
\ee
Independent variation of the functional \eqref{Pek 1} w.r.t. $ \Phi^{\star} $ and $ \Phi $ gives
\bea 
 - \frac{\Delta}{2} \, \Phi(\fr) + \, \int d^3y \> \Phi^2(\fy) \, V(\fr,\fy) \, \cdot \,  
\Phi^{\star}(\fr) - \epsilon_0 \, \Phi(\fr) \EA 0 \\
 - \frac{\Delta}{2} \, \Phi^{\star} (\fr) + \, \int d^3x \> \Phi^{\star \> 2}(\fx) \, 
V(\fx,\fr) \, \cdot \,  \Phi(\fr) - \epsilon_0 \, \Phi^{\star} (\fr) \EA 0 \> , 
\eea
where the constraint  $(\Phi^{\star},\Phi) = 1 $ has been incorporated into the functional by a 
Lagrange multiplier $ \epsilon_0  \lrp (\Phi^{\star},\Phi) - 1 \rrp $.
We see that $ \Phi^{\star} = \Phi$ is a solution since the interaction has the property
\be
V(\fx,\fy) \Def - \frac{\sqrt 2}{\kappa} \, \frac{1}{ | \fx - \fy |} \E V(\fy,\fx) \> .
\label{pot polaron}
\ee
Therefore we can write Eq. \eqref{Pek 1} also as
\be
\gamma_P \E \kappa^2 \, {\rm min}_{(\Phi,\Phi) = 1} \, \Biggl \{ \>   \int d^3x \> \Phi(\fx) \, 
\lrp - \frac{\Delta}{2} \rrp \, \Phi(\fx) + \frac{1}{2} \, \int d^3x \, d^3y \> 
\Phi^2(\fx) \, V(\fx, \fy) \, \Phi^2(\fy) \,  \Biggr \} \> .
\label{Pek 2}
\ee
The numerical value of $ \kappa > 0 $ is irrelevant because the scaling behaviour of the 
Coulomb-like interaction \eqref{pot polaron} compensates other values. Usually one chooses
$ \kappa = 1 \> $  \cite{AdGeLe} but one also could take the value $ \kappa = \sqrt 2 $ 
which simplifies the interaction \eqref{pot polaron}.
\vspace{0.3cm}

Pekar's {\it ansatz} which leads to the exact result for strong coupling, is built on the intuitive 
picture that in this limit the ``bare' electron moves so fast \footnote{In contrast, 
at fixed momentum the ``dressed'' electron = polaron moves slower and slower since its effective mass grows 
$ \propto \alpha^4 $.} that the phonons only feel a mean field $ \Phi^2(\fx) $. 
Actually Eq. \eqref{Pek 2} is identical with the \textcolor{blue}{\bf Hartree approximation} of the previous 
chapter and the variational equation
\be
- \frac{\Delta}{2} \, \Phi(\fr) + \, \int d^3x \> V(\fr,\fx) \, \Phi^2(\fx) \, \cdot \, \Phi(\fr)  \E 
\epsilon_0 \, \Phi(\fr) \> , 
\label{Hartree 1}
\ee
a \textcolor{blue}{\bf non-linear} Schr\"odinger equation, is identical with the Hartree equation
\eqref{einteil zeitunab}. 
If we multiply Eq. \eqref{Hartree 1} with $\Phi(\fr) $ and integrate over $ \fr $, then we obtain 
by means of the normalization and by comparison with Eq. \eqref{Pek 2}
\be
\epsilon_0 \E \lrp \Phi,H_0 \Phi \rrp + \lrp \Phi^2, V \Phi^2 \rrp
\E \frac{E_0}{\kappa^2 \, \alpha^2} + \frac{1}{2} \, \lrp \Phi^2, V \Phi^2 \rrp \> ,
\ee 
in agreement with the result \eqref{E Hartree} of the Hartree approximation.

Choosing simple {\it ans\"atze} for $ \Phi(\fx) $ the minimal principle \eqref{Pek 2} can be evaluated 
and one finds (\purpur{\bf Problem \ref{Pekar} a) b)})
\bea
\gamma_P \EA -  \frac{25}{256} \E -0.0977 \quad  \mbox{for} \quad \Phi(\fx) \E C \, e^{-r/a} 
\label{ansatz exp}\\
\gamma_P \EA - \frac{1}{3 \pi} \E - 0.1061  \quad  \mbox{for} \quad \> \Phi(\fx) \E C \, e^{-r^2/a^2} 
\label{ansatz gauss} \> .
\eea
The result \eqref{ansatz gauss} is exactly the same as obtained by Feynman's quadratic retarded 
trial action at large coupling (see \purpur{\bf Problem \ref{Feyn Polaron} b)} ). This is not surprising 
since we have seen that in this limit the retardation does not play a role and since we know
that quadratic (local) actions correspond to oscillator-like wave functions.
\vspace{0.3cm}

How can we determine the exact value for the Pekar constant $ \gamma_P $, i.e. for the polaron 
ground-state energy at large coupling?
%
%
One option is to use better and better variational functions in the minimum principle 
\eqref{Pek 2}, another possibility is to determine $ \gamma_P $ numerically from the Hartree equation
\eqref{Hartree 1}. Since this equation is a non-linear one, we have to solve it iteratively
until the wave function, which also determines the interaction, is obtained {\bf selfconsistently}.

Here we choose the first option and determine the Pekar coefficient $ \gamma_P $ by variation
of the coefficient in a suitable {\it ansatz} for the S-wave \footnote{We assume that the ground-state 
wave function is spherically symmetric which is always the case for rotationally symmetric problems in quantum mechanics.}
\be
\Phi(\fr) \E \frac{y(r)}{r} \, \frac{1}{\sqrt{4 \pi}} \> , \quad r \equiv |\fr| \> .
\ee
This projects the monopole part out of the Coulomb interaction
\be
\int d\Omega_x \, d\Omega_y \> \frac{1}{|\fx - \fy|} \E \frac{(4 \pi)^2}{{\rm max}\lrp |\fx|,|\fy| 
\rrp} \> ,
\ee
so that the minimal functional reads (we choose $ \kappa = 1 $)
\be
\gamma_P \E {\rm min}_{(y,y)=1} \> \Bigl [ \,  \la T \ra + \la V \ra \Bigr ]
\ee
with
\be
\int_0^{\infty} dr \> y^2(r) \E 1 \> , \quad \la T \ra \E \frac{1}{2} \int_0^{\infty} dr \> y'^2(r) 
\> , \quad
 \la V \E \ra - \frac{1}{\sqrt{2}} \, \int_0^{\infty} dr \> y^2(r) \, \int_0^{\infty} ds \> 
\frac{y^2(s)}{{\rm max}(r,s)} \> .
\ee
\vspace{0.8cm} 

\renewcommand{\baselinestretch}{0.9}
\scriptsize
\refstepcounter{tief}
\noindent
\blau{\bf Detail \arabic{tief}:} {\bf Numerical Treatment and FORTRAN Program}\\

\noindent
\begin{subequations} 

\noindent
The simplest {\it ansatz} which generalizes the trial wave function in \purpur{\bf Problem \ref{Pekar} b)}
and represents a complete function system for $ N \to \infty $, is
\be
y(r) \E C \, r \sum_{i=0}^N c_i \, \lrp \frac{r}{a} \rrp^i \, e^{-r/a} \> , \quad c_0 \EQ 1 \> .
\label{ansatz}
\ee
With the help of the integral
\be
\int_x^{\infty} dt \> t^n e^{-\lambda t} \E \frac{n!}{\lambda^{n+1}} \, e^{-\lambda x} \, 
\sum_{j=0}^n 
\frac{(\lambda x)^j}{j!} \> , \quad \mbox{in particular:} \> \> \int_0^{\infty} dt \>t^n e^{-\lambda t} 
\E  \frac{n!}{\lambda^{n+1}} 
\ee
(which can be obtained by subsequent integrations by parts) one easily finds the 
normalization of the trial wave function \eqref{ansatz}
\be
C^{-2} \E a^3 \, \sum_{i,j=0}^N c_i c_j \, d_{i+j+2} \> , \quad {\rm with} \> \> \>  d_n 
\Def \frac{n!}{2^{n+1}} \> ,
\ee
the expectation value of the kinetic energy
\be
\la T \ra \E \frac{1}{2} C^2 a \,  \sum_{i,j=0}^N c_i c_j \, \Bigl [ \>  (i+1) (j+1) d_{i+j} - 
d_{i+j+2} \> \Bigr ] \> ,
\ee
and of the potential (self-)energy
\be
\la V \ra \E - \sqrt{2}\,  C^4 a^5 \,  \sum_{i,j,k,l=0}^N c_i c_j \> \frac{1}{2^{i+j+3}} \>  
c_k c_l \> d_{k+l+1} \, \sum_{m=0}^{k+l+1} \frac{d_{i+j+m+2}}{m!} \> .
\ee
Due to the simple scaling properties ($C^2 \sim a^{-3} , \la T \ra \sim a^{-2} ,  
\la V \ra \sim a^{-1} $ ) the variation w.r.t. the scaling parameter $ a $ can be performed
analytically:
\be 
\frac{\partial}{\partial a} \lsp \frac{1}{a^2} \, \la T \ra_{a=1} + \frac{1}{a} \la V \ra_{a=1} 
\rsp \E 0 \quad
\Longrightarrow \> a_{\rm optimal} \E - 2 \frac{\la T \ra}{\la V \ra} \Biggr |_{a=1} \> , \quad 
\Bigl [ \,  \la T \ra + \la V \ra \Bigr ]_{\rm optimal} \E - \frac{1}{4} \, 
\frac{ \la V \ra^2}{\la T \ra } \Biggr |_{a=1}
\ee
and one ``only'' has to vary the energy functional (which is now given completely analytically)
w.r.t. the remaining coefficients $  c_1, c_2 \ldots c_N $. 
However, "there is no free lunch", as non-linear multi-dimensional minimalization is a difficult numerical problem in need of an efficient, robust method.
\vspace{0.2cm}

We will use the AMOEBA program in the (freely available) Fortran77 version of the book
{\bf \{Num. Recipes\}}
to perform numerically the minimalization over the coefficients $  c_1, c_2 \ldots c_N $.
This procedure may be slow as the ``amoeba'' only crawls down the hills but is applicable
to general functions. It is based on the ``downhill''- simplex method and not only needs a starting 
value but $N+1$ points defining an initial simplex and the corresponding
function values.
A simplex is a geometrical figure consisting of $ N + 1 $ points (or vertices) in $ N $ dimensions.
We determine these points from an initial point
$ {\bf P}_1 = (c_1,c_2 \ldots c_N) $ and  generate additional points simply by
\be 
{\bf P}_i \E {\bf P}_1 + \lrp c_1, c_2 \ldots ,\underbrace{c_j + \lambda^j}_{j=i-1}, 
\ldots c_N \rrp
\quad i \E 2, 3 \ldots N + 1 \> ,
\ee
where $ \lambda $ is a suitably chosen constant (in the example below we take $ \lambda = 0.3 $).
Of course, with $ N = 0 $ one just obtains the result \eqref{ansatz exp} from
\purpur{\bf Problem \ref{Pekar} b)} and we increase step-by-step the number of coefficients to be varied.
As initial point we take the minimalization result from the previous run with $ (N - 1) $ coefficients
and set the additional coefficient $c_N = 0 $ at the point $ {\bf P}_1 $. By this procedure it is 
guaranteed that at each $ N $ we obtain at least the previous result. 
The big risk is that during iteration one gets stuck in a local minimum; therefore it is
recommended (at each $ N $ ) to start again with the obtained minimum as new starting point.

The AMOEBA program stops if a prescribed accuracy (with single precision we take FTOL = $10^{-6} $ ) 
has been achieved and presents $ N + 1 $ points as result which lie inside the precision bound 
around the discovered minimum. In the program presented below we average over the function values
of this final simplex.
\vspace{0.6cm}

\color[rgb]{0.3,0,0.7}
\begin{verbatim}

c++++++++++++++++++++++++++ MAIN PROGRAM +++++++++++++++++++++++++++++++
c 
c   Program minimizes the Pekar functional for the polaron problem
c   in the limit of strong coupling
c   using the ansatz for the S-wave function
c 
c   y(r) = C r [ 1 + sum_{i=1}^N c_i r^i exp(-r/a) ] ;  
c                      int_0^{infinity} dr y^2(r) = 1 .
c 
c   The scale a is determined analytically by variation.
c   Minimization by "downhill simplex method" for the coeffizients c_i;   
c   program from NUMERICAL RECIPES, ch. 10.4   
c 
c   Loop over N = 0,1,  ... NMAX-1 = 9 coefficients
c             
      EXTERNAL ENERGIE        
      PARAMETER (NMAX = 10,FTOL=1.E-6,ALAM=0.3)) 
      COMMON N,D(0:4*NMAX+3) 
      DIMENSION P(NMAX+1,NMAX),X(NMAX),Y(NMAX),XNEU(NMAX),XALT(NMAX+1) 
C 
C   tabellation of  n!/2^(n+1) 
C  
      NHOCH = 4*NMAX + 3 
      D(0) = 0.5 
      DO 10 I = 1,NHOCH 
         D(I) = I*D(I-1)*0.5 
 10   CONTINUE 
C  
      WRITE(6,9) FTOL 
 9    FORMAT(//' rel. accuracy = ',e10.3/) 

      DO 100 NN = 1,NMAX                ! loop over different number of coefficients

            N = NN - 1 

            ISTART = 0                  ! initial start
C 
C   N+1 start simplices for minimization routine AMOEBA 
C 
 50         CONTINUE 

            DO 20 I = 1,NN 
               DO 25 J = 1,N 
               IF(ISTART .EQ. 0) P(I,J) = XALT(J) 
               IF(ISTART .EQ. 1) P(I,J) = XNEU(J)               
               IF(J. EQ. (I-1)) P(I,J) = P(I,J) + ALAM**J 
 25         CONTINUE 
 20      CONTINUE 
C 
         DO 30 I = 1,NN 
            DO 35 J = 1,N 
               X(J) = P(I,J) 
 35         CONTINUE 
            Y(I) = ENERGIE(X) 
 30      CONTINUE 
C 
         CALL AMOEBA(P,Y,NMAX+1,NMAX,N,FTOL,ENERGIE,ITER) 
C 
         DO 38 J = 1,N                  ! store first simplex for restart  
            XNEU(J) = P(1,J) 
            XALT(J) = XNEU(J)           ! store first simplex for next N 
 38      CONTINUE                
         XALT(NN) = 0.                  ! and add c(N+1)=0 as last coefficient 
C 
C   minimal value averaged over the simplices 
C 
         YMIT = 0. 
         DO 40 I = 1,NN 
            YMIT = YMIT + Y(I) 
 40      CONTINUE 
         YMIT = YMIT/NN 
C 
         IF(ISTART .EQ. 0) WRITE(6,98) N 
 98      FORMAT(//' N = ',I3/) 
         IF(ISTART .EQ. 1) WRITE (*,*) 'restart:' 
         WRITE(6,99) ITER,YMIT 
 99      FORMAT(' iterations = ',I4,10X, 
     &      'mean minimal value = ',F10.7) 
C 
         ISTART = ISTART + 1 
         IF(ISTART .EQ. 1) GO TO 50 
 100  CONTINUE 
      STOP 
      END  

C++++++++++++++++++++++++ SUBPROGRAM ENERGIE  ++++++++++++++++++++++++

      FUNCTION ENERGIE(X) 
C 
      PARAMETER (NMAX=10) 
      COMMON N,D(0:4*NMAX+3) 
      DIMENSION X(NMAX),C(0:NMAX) 
      DATA WU2 /1.414213562/                      ! square root from  2 
C 
      C(0) = 1.                                   ! fixed 
      DO 10 I = 1,N 
         C(I) = X(I) 
 10   CONTINUE 
C 
C   normalization
C 
      ANORM = 0. 
      DO 20 I = 0,N 
         DO 25 J = 0,N 
            ANORM = ANORM + C(I)*C(J)*D(I+J+2) 
 25      CONTINUE 
 20   CONTINUE 
      ANORM = 1./ANORM 
C 
C   kinetic energy 
C 
      T = 0. 
      DO 30 I = 0,N 
         DO 35 J = 0,N 
            DT = (I+1.)*(J+1.)*D(I+J) - D(I+J+2) 
            T = T + DT*C(I)*C(J) 
 35      CONTINUE 
 30   CONTINUE 
      T = 0.5*ANORM*T 
C 
C   potential energy
C 
      V = 0. 
      DO 40 I = 0,N 
         DO 45 J = 0,N 
            ISUM = I + J + 2 
            DO 50 K = 0,N 
               DO 55 L = 0,N 
                  KSUM = K + L + 1 
                  HILF = C(I)*C(J)/(2**(ISUM+1)) 
                  HILF = HILF*C(K)*C(L)*D(KSUM) 
                  V1 = 0. 
                  FAK = 1.                        !   0! 
                  DO 60 M = 0,KSUM                      
                     MSUM = ISUM + M  
                     V1 = V1 + D(MSUM)/FAK 
                     FAK = (M+1.)*FAK             !  (M+1)! 
 60               CONTINUE 
                  V = V + HILF*V1 
 55            CONTINUE 
 50         CONTINUE 
 45      CONTINUE 
 40   CONTINUE 
      V = -WU2*V*ANORM*ANORM 
C 
C   energy after variation of scale parameter   
C 
      ENERGIE = -0.25*V*V/T 
C 
      RETURN 
      END 

C++++++++++++++++++++++++++ SUB PROGRAM AMOEBA ++++++++++++++++++++++++


      SUBROUTINE AMOEBA(P,Y,MP,NP,NDIM,FTOL,FUNK,ITER) 
C 
C 
C ***********************************************************************
C
C   from: W. H.Press et al. "Numerical Recipes in Fortran 77" , ch. 10.4 
C  
C         Cambridge University Press, 2nd ed. (1992) 
C
C ***********************************************************************
C
C   Multidimensional minimization of the function FUNK(X) where X is an  
C   NDIM-dimensional vector, by the downhill simplex method of Nelder and Mead.
C  
C   Input is a matrix P whose NDIM+1 rows are NDIM-dimensional vectors which  
C   are the vertices of the starting simplex. 
C   [Logical dimensions of P are P(NDIM+1,NDIM); physical dimensions are input as P(MP,NP)]. 
C   Also input is the vector Y of length  NDIM+1, whose components must be  
C   pre-initialized to the values of FUNK evaluated at the NDIM+1 vertices (rows) of P; 
C   and FTOL the fractional convergence tolerance to be achieved in the  
C   function value (n.b.!). 
C   On output, P and Y will have been reset to NDIM+1  
C   new points all within FTOL of a minimum function value, 
C   and ITER gives the number of iterations taken. 
C 
      PARAMETER (NMAX=10,ALPHA=1.0,BETA=0.5,GAMMA=2.0,ITMAX=1000) 
C 
C   Expected maximum number of dimensions, three parameters which define  
C   the expansions and contractions, and maximum allowed number of iterations. 
C 
      DIMENSION P(MP,NP),Y(MP),PR(NMAX),PRR(NMAX),PBAR(NMAX) 
C 
      MPTS = NDIM + 1                   ! Note that MP is the physical dimension 
                                        ! corresponding to the logical dimension  
                                        ! MPTS, NP to NDIM 
      ITER = 0 
1     ILO = 1                           ! First we must determine which point is the 
                                        ! highest (worst), next-highest, and 
                                        ! lowest (best). 
      IF(Y(1) .GT. Y(2)) THEN 
         IHI = 1 
         INHI = 2 
      ELSE 
         IHI = 2 
         INHI = 1 
      ENDIF 
C 
      DO 11 I = 1,MPTS                  ! by looping over the points in the simplex 
         IF(Y(I) .LT. Y(ILO)) ILO = I 
         IF(Y(I) .GT. Y(IHI)) THEN 
            INHI = IHI 
            IHI = I 
         ELSE IF(Y(I) .GT. Y(INHI)) THEN 
            IF(I .NE. IHI) INHI = I 
            ENDIF 
11    CONTINUE 
C 
C   Compute the fractional range from highest to lowest and return if satisfactory 
C 
      RTOL = 2.*ABS(Y(IHI)-Y(ILO))/(ABS(Y(IHI))+ABS(Y(ILO))) 
      IF(RTOL .LT. FTOL) RETURN 
      IF(ITER .EQ. ITMAX) PAUSE 'Amoeba exceeding maximum iterations.' 
      ITER = ITER + 1 
      DO 12 J = 1,NDIM 
         PBAR(J) = 0.  
12    CONTINUE                  ! Begin a new iteration. Compute the vector average 
                                ! of all points except the highest, i.e. the center 
                                ! of the "face'' of the simplex across from the high  
                                ! point. We will subsequently explore along the ray 
                                ! from the high point through that center. 
      DO 14 I = 1,MPTS 
         IF(I .NE. IHI) THEN 
            DO 13 J = 1,NDIM 
               PBAR(J) = PBAR(J) + P(I,J) 
13          CONTINUE 
         ENDIF 
14    CONTINUE 
      DO 15 J = 1,NDIM                  ! Extrapolate by a factor ALPHA through the  
                                        ! face, i.e. reflect the simplex from the 
                                        ! high point. 
         PBAR(J) = PBAR(J)/NDIM 
         PR(J) = (1.+ ALPHA)*PBAR(J) - ALPHA*P(IHI,J) 
15    CONTINUE  
      YPR = FUNK(PR)                    ! Evaluate the function at the reflected point. 
      IF(YPR .LE. Y(ILO)) THEN          ! Gives a result better than the best point, 
                                        ! so try an additional extrapolation by a 
                                        ! factor GAMMA, 
         DO 16 J = 1,NDIM 
            PRR(J) = GAMMA*PR(J) + (1.-GAMMA)*PBAR(J) 
16       CONTINUE 
         YPRR = FUNK(PRR)               ! check out the function there. 
         IF(YPRR .LT. Y(ILO)) THEN      ! The additional extrapolation succeeded, 
                                        ! and replaces the high point. 
            DO 17 J = 1,NDIM 
               P(IHI,J) = PRR(J) 
17          CONTINUE 
         Y(IHI) = YPRR 
      ELSE                              ! The additional extrapolation failed, 
                                        ! but we can still use the reflected point 
         DO 18 J = 1,NDIM 
            P(IHI,J) = PR(J) 
18       CONTINUE 
         Y(IHI) = YPR 
      ENDIF 
      ELSE IF(YPR .GE. Y(INHI)) THEN    ! The reflected point is worse then  
                                        ! the second-highest 
         IF(YPR .LT. Y(IHI)) THEN       ! If it's better than the highest,  
                                        ! then replace the highest, 
            DO 19 J = 1,NDIM 
               P(IHI,J) = PR(J) 
19          CONTINUE 
            Y(IHI) = YPR 
         ENDIF 
         DO 21 J = 1,NDIM               ! but look for an intermediate lower point 

            PRR(J) = BETA*P(IHI,J) + (1.-BETA)*PBAR(J) 

21       CONTINUE                       ! in other words, perform  a contraction  
                                        ! of the simplex along one dimension.  
                                        ! Then evaluate the function. 
         YPRR = FUNK(PRR)
         IF(YPRR .LT. Y(IHI)) THEN      ! Contraction gives an improvement, 
            DO 22 J = 1,NDIM            ! so accept it. 
               P(IHI,J) = PRR(J) 
22          CONTINUE 
            Y(IHI) = YPRR 
         ELSE                           ! Can't seem to get rid of that high point. 
                                        ! Better contract aroand the lowest (best) 
                                        ! point. 
            DO 24 I = 1,MPTS 
               IF(I .NE. ILO) THEN 
                  DO 23 J = 1,NDIM 
                     PR(J) = 0.5*(P(I,J) + P(ILO,J)) 
                     P(I,J) = PR(J) 
23                CONTINUE 
                  Y(I) = FUNK(PR) 
               ENDIF 
24          CONTINUE 
         ENDIF 
      ELSE                              ! We arrive here if the original reflection gives
                                        ! a middling point.
                                        ! Replace the old high point and continue 
         DO 25 J = 1,NDIM 
            P(IHI,J) = PR(J) 
25       CONTINUE 
         Y(IHI) = YPR 
      ENDIF 
      GO TO 1                           ! for the test of doneness and the next  
                                        ! iteration. 
      END 

\end{verbatim}
\vspace{0.5cm}

\color{black}

After running this program we obtain the following print-out:

\vspace{0.6cm}

\color[rgb]{0.3,0,0.7}

 rel. accuracy =   .100E-05\\

 N =   0

 iterations =    1   \hspace{2.1cm}       mean minimal value =  -.0976562\\
 restart:\\
 iterations =    0   \hspace{2.1cm}       mean minimal value =  -.0976562\\

 N =   1

 iterations =    8   \hspace{2.1cm}       mean minimal value =  -.1080244\\
 restart:\\
 iterations =    4   \hspace{2.1cm}       mean minimal value =  -.1080244\\

 N =   2

 iterations =   16    \hspace{2cm}      mean minimal value =  -.1085069\\
 restart:\\
 iterations =   12    \hspace{2cm}      mean minimal value =  -.1085069\\

 N =   3

 iterations =   12   \hspace{2cm}       mean minimal value =  -.1085069\\
 restart:\\
 iterations =   12   \hspace{2cm}       mean minimal value =  -.1085069\\

 N =   4

 iterations =   21   \hspace{2cm}       mean minimal value =  -.1085092\\
 restart:\\
 iterations =   65   \hspace{2cm}       mean minimal value =  -.1085098\\

 N =   5

 iterations =   81   \hspace{2cm}       mean minimal value =  -.1085106\\
 restart:\\
 iterations =   82   \hspace{2cm}       mean minimal value =  -.1085114\\

 N =   6

 iterations =  117   \hspace{1.9cm}       mean minimal value =  -.1085121\\
 restart:\\
 iterations =   96   \hspace{2cm}       mean minimal value =  -.1085124\\

 N =   7

 iterations =  136   \hspace{2cm}       mean minimal value =  -.1085123\\
 restart:\\
 iterations =  134   \hspace{2cm}       mean minimal value =  -.1085125\\

 N =   8

 iterations =  154   \hspace{2cm}       mean minimal value =  -.1085124\\
 restart:\\
 iterations =  154   \hspace{2cm}       mean minimal value =  -.1085124\\

 N =   9

 iterations =  171   \hspace{2cm}       mean minimal value =  -.1085125\\
 restart:\\
 iterations =  152   \hspace{2cm}       mean minimal value =  -.1085126\\

\end{subequations}
\color{black}

\renewcommand{\baselinestretch}{1.2}
\normalsize

\vspace{0.5cm}

\noindent
The numerical value of Pekar's coefficient for the polaron energy at strong coupling therefore is
\be
\gamma_P \E - 0.108513(1) 
\ee
with an estimated error of (1) in the last digit. A more precise value can be obtained
by taking a smaller accuracy parameter (FTOL) and a larger number of coefficients (NMAX)
but then double-precision arithmetic is required  (\purpur{\bf Problem \ref{Pekar} c)} ).

\edes

\newpage

\section{\textcolor{red}{Path Integrals in Field Theory}}

\pagestyle{myheadings}
\markboth{\textcolor{green}{Section 3 : Field Theory}}{\textcolor{green}{R. Rosenfelder : 
Path Integrals in Quantum Physics}}
\renewcommand{\thesubsection}{\textcolor{blue}{3.\arabic{subsection}}}
\renewcommand{\theequation}{3.\arabic{equation}}

\setcounter{equation}{0}

\subsection{\textcolor{blue}{Generating Functionals and Perturbation Theory} }     
\label{sec3: erzeug Funk}

The transition from many-body quantum physics to field theory
is a transition from a finite number of degrees of freedom
to an infinite number. We will perform this transition {\bf formally}
thus omitting in a first step all subtleties like renormalization, even ignoring
the question of the sheer existence of an interacting theory.

In the beginning we consider for simplicity a system of particles of mass $ \> m \> $ 
which is described by a neutral scalar field $ \> \Phi(x) \> $.
Let the Lagrange density (Lagrangian) be \footnote{Charged scalar particles
are represented by complex fields in whose Lagrangian the factor  $1/2$ is missing, see
\purpur{\bf Problem \ref{N skalare Teil}}.}
\be
\boxed{
\qquad {\cal L} \E  \frac{1}{2} \left ( \> \partial_{\mu} \Phi \, \partial^{\mu} \Phi
- m^2 \Phi^2 \> \right ) - V(\Phi) \> , \quad
}
\label{skalare Lagrangedichte}
\ee
where  $ \> V(\Phi) \> $ denotes the self-interaction of the field \footnote{In this chapter we 
set $ \> \hbar = 1 \> $. Here and in the following we always choose a system of units with
$ c = 1$, employ the metric $(+,- - -)$, i.e. $ \,a \cdot b = a_0 b_0 - {\bf a} \cdot \fb \,  $,
and sum over identical indices.}.

If one requires the theory to be renormalizable (in 1 time and 3 space dimensions)
this interaction can only be a polynomial of the field
up to degree four; a typical example is
\be
\boxed{
\qquad V(\Phi) \E \frac{\lambda}{4 !} \Phi^4 \> , \quad 
}
\label{Phi4 Theorie}
\ee
which is of great importance in superconductivity (Ginzburg-Landau theory) and 
particle physics    
(Higgs mechanism). 

As in the quantum-mechanical case we consider the transition matrix element
\be
\left < \Phi_f \> \left | \> e^{- i \hat H (t_f - t_i)} \> \right | \> 
\Phi_i \right >
\ee
of the time-evolution operator between field configurations $ \> \Phi_i \> $
and $ \> \Phi_f \> $. $ \> \hat H = \int d^3x \, \hat {\cal H} \> \> $ 
ist the Hamilton operator of the systems. The classical Hamilton density 
follows from  Eq. (\ref{skalare Lagrangedichte}) by the usual
Legendre trans\-for\-mation
\be
{\cal H} = \pi \dot \Phi - {\cal L} \E \frac{1}{2} \pi^2 + 
\frac{1}{2} \left ( \nabla \Phi \right )^2
+ \frac{1}{2} m^2 \Phi^2 + V(\Phi) \> ,
\ee
where $ \> \pi = \partial {\cal L}/\partial \dot \Phi = \dot \Phi \> $
is the canonically conjugated momentum (density) of the field.
We divide the (finite) space volume $ L^3 $ into $ N $ small cells with volume 
 $ \> v = L^3/N \> $ and the time interval  $ \> t_f - t_i \> $ in $ M $ intervals
 of length  $ \epsilon $. If we use the completeness for the fields
$ \, \Phi(t,\fx) = \Phi^l_j , \> l = 1 \ldots M-1 , \> j = 1 \ldots N \,  $ 
at each time and space point
\be
\int d\Phi^l_j \> \left | \> \Phi^l_j \left > \right < \Phi^l_j \> \right |
\E 1
\ee
then we obtain
\bea
\left < \Phi_f \> \left | \> e^{- i \hat H (t_f - t_i)} \> \right | \> 
\Phi_i \right >
\EA  \lim_{N, M \to \infty} \int \prod_{j=1}^N \> \Bigl \{ \> 
d \Phi^{M-1}_j \ldots d \Phi^1_j \> \>
\left < \Phi_j^M \> | \> 
e^{-i \epsilon v \hat {\cal H}} | \> \Phi_j^{M-1} \right > \non
&& \hspace{5cm} \ldots \> \left < \Phi_j^1 \> | \>
e^{-i \epsilon v \hat {\cal H}} | \> \Phi_j^0 \right > \> \Bigr \} \> .
\eea
As before we have set 
$ \> \Phi_j^M = \Phi_f(\fx_j) \> $ and
$ \> \Phi_j^0 = \Phi_i(\fx_j) \> $ as abbreviation.
Exactly as in the quantum-mechanical case one finds for small times
\be
\left < \Phi_j^{l+1} \> \left | \>
e^{-i \epsilon v \hat {\cal H}} \> \right | \> \Phi_j^l \right >  \> \simeq 
\> \int \frac{d p_j^l}{2 \pi} \> \exp \left [ \> i p_j^l \> ( \Phi_j^{l+1} 
- \Phi_j^l) - i \epsilon v {\cal H}_j^l \> \right ] \> .
\ee
If we write $ \> p_j^l = v \pi_j^l \> $, we obtain
\bea
\left < \Phi_f \> | \> e^{- i \hat H (t_f - t_i)} \> | \> \Phi_i\right >
\EA \lim_{N, M \to \infty} \int \prod_{j=1}^N \> \left (
\prod_{l=1}^{M-1} \frac{d \Phi_j^l v d \pi_j^{l}}{2 \pi}
\right ) \int \frac{v d \pi_j^{M}}{2 \pi} \non
&& \cdot \exp \left \{ \> i \epsilon \sum_{l=0}^{M-1} v
\sum_{j=1}^N  \left ( \pi_j^l \frac{\Phi_j^{l+1} - \Phi_j^l}{\epsilon}
- {\cal H}_j^l \right )  \> \right \} \non
&\EQ &   \int \frac{ {\cal D}\Phi(x) {\cal D} \pi(x)}{2 \pi} 
\> \exp \left \{ \> i \int_{t_i}^{t_f} \! dt \int d^3x \> \left [ \pi
\dot \Phi - {\cal H}(\pi,\Phi) \right ] \> \right \} \> .
\eea
This is the path integral in Hamilton form
(note footnote \ref{Pfad-Bezeich}). By completing the square\\
($ \> \pi \dot \Phi - \pi^2/2 = - (\pi - \dot \Phi)^2/2 
+ \dot \Phi^2/2 \> $ ) we can transform the $\pi$-integral into a Gaussian integral
and we then obtain the path integral in Lagrange form
\be
\left < \Phi_f \> \left | \> e^{- i \hat H (t_f - t_i)} \> \right | \> 
\Phi_i \right >
\E {\rm const.} \cdot \int {\cal D}\Phi(x) \> 
e^{i S[\Phi] }\> .
\label{Lagrange}
\ee
Again we have the (classical) action in the exponent because
\bea
&& \! \! \int_{t_i}^{t_f} dt \> \int d^3x \> \left [
\frac{1}{2} \dot \Phi^2 - \frac{1}{2} ( \nabla \Phi )^2  - \frac{1}{2} m^2
\Phi^2 - V(\Phi) \right ] \non
\EA \int_{t_i}^{t_f} dt \> \int d^3x \> \left [
\frac{1}{2} \partial_{\mu} \Phi \partial^{\mu} \Phi  - \frac{1}{2} m^2 \Phi^2
- V(\Phi) \right ] 
=  \int_{t_i}^{t_f} dt \> \int d^3x \> {\cal L}
\left ( \Phi(x),\partial_{\mu} \Phi(x) \right ) \EQ S[\Phi] \> .
\eea
However, this representation of the transition matrix element
between field configurations is only an intermediate step as this is not
the relevant quantity in field theory.
Actually, all observables in field theory can be obtained from the 
\textcolor{blue}{\bf n-point functions}
or \textcolor{blue}{\bf Green functions} 
\be
G_n(x_1 \ldots x_n) \E \left < 0 \> \left | \> {\cal T} \left [         
\hat \Phi(x_1)  \ldots \hat \Phi(x_n) \right ] \> \right | \> 0 \right >
\ee
(as can be seen below in the example of scattering of two scalar particles).
Here  $ \> | \> 0 > \> $ is the exact ground state of the theory
(the ``vacuum''), $ \> {\cal T} \> $ the time-ordering operator
and $ \> \hat \Phi(x) \> $ are the exact field operators in the Heisenberg picture.
What we need therefore is a path-integral represention for the Green functions.

Fortunately, this already has been done in quantum mechanics
(in {\bf chapter} {\bf \ref{sec1: greensche Funk}}): The ground state
can be projected out by the unphysical limit \footnote{As a reminder: \quad     
$e^{- i \hat H (t_f - t_i)} \E |\, 0><0 \,| \,
e^{- i E_0 (t_f - t_i) } + |\,1><1\,| \, e^{- i E_1 (t_f - t_i) } + \ldots $
Also, one doesn't have to take purely imaginary times but can go to infinity along
a suitable ray in the complex plane -- the essential point is that the
excited states are sufficiently damped in this limit.}
 $ \> t_i \to i \infty \> , t_f \to - i \infty \> $ 
 and one has
\bce
\vspace{0.2cm}

\fcolorbox{blue}{white}{\parbox{10cm}
{
\bea
G_n(x_1 \ldots x_n) \EA \lim_{\latop{t_i \to i \infty}{t_f \to - i \infty}}
\frac{ \int {\cal D} \Phi \> \Phi(x_1) \ldots \Phi(x_n) \> \exp ( \> i 
S[\Phi]\> ) }
{\int {\cal D} \Phi \>\exp ( \> i S[\Phi] \> )} \> . \no
\eea
}}
\ece
\vspace{-2cm}

\bea
\label{GF Pfad}
\eea
\vspace{0.2cm}


\noindent
The set of all $n$-point functions can be obtained from the \textcolor{blue}{\bf generating 
functional}
\bce

\fcolorbox{red}{white}{\parbox{10cm}
{
\bea
Z[J] \EA \int {\cal D} \Phi \> \exp \left ( \> i S[\Phi] \>
+ i \int d^4 x \> J(x) \, \Phi(x) \> \right ) 
\hspace{5.5cm}\label{def erzeug functional}
\eea
}}
\ece

\vspace{0.4cm}


\noindent
by functional differentiation:
\bce
\vspace{0.2cm}

\fcolorbox{red}{white}{\parbox{10cm}
{
\bea
G_n(x_1 \ldots x_n) \EA \left (\frac{1}{i}\right )^n \> 
\frac{ \delta^n}{\delta J(x_1) \ldots                        
\delta J(x_n)} \> \frac{Z[J]}{Z[0]} \Biggr |_{J=0} \> . \no
\eea
}}
\ece
\vspace{-2cm}

\bea
\label{Gn}
\eea
\vspace{0.5cm}


\noindent
As one can see the normalization of the functional integral cancels.
Eq. \eqref{def erzeug functional} can also be considered as functional Fourier transform of
$ \> \exp(i S[\Phi]) \> $  and thus the artificially introduced source  $ \> J(x) \> $ can be seen
as variable conjugate to the  field $ \> \Phi(x) \> $ .


In most cases the generating functional can be calculated only for
vanishing self-interaction $ \> V \> $ :
\bea
Z_0[J] \EA \int {\cal D} \Phi \> \exp \left [ \> i \int d^4 x \> \left ( 
\frac{1}{2} \partial_{\mu} \Phi \partial^{\mu} \Phi - \frac{1}{2} m^2 
\Phi^2 + J \Phi \> \right ) \right ] \non
\EA \int {\cal D} \Phi \> \exp \left [ \> \frac{i}{2} \int d^4 x \> 
\Phi(x) \left ( - \Box - m^2 \right ) \Phi(x) + i \int d^4 x \> J(x) \Phi(x)
\> \right ] \> .
\label{Z0 1}
\eea
In the second line an integration by parts has been performed under the assumption
that boundary terms vanish. The functional integration 
in Eq. (\ref{Z0 1}) can be done since the exponent
\bea 
&& \hspace{-1.5cm}
\frac{i}{2} \int d^4 x \> d^4y \> \Phi(x) K_0(x,y) \Phi(y) + 
i \int d^4 x \> J(x) \Phi(x) \non
&& \hspace{-1.5cm} =  \frac{i}{2} \int d^4x \> d^4y \> \left ( \Phi 
K_0 + J \right )(x) 
K_0^{-1}(x,y) \left ( K_0 \Phi + J \right )(y) - 
 \frac{i}{2} \int d^4x \> d^4y \> J(x) K_0^{-1}(x,y) J(y)
\eea
is quadratic in the fields. The Gaussian integral over  $\Phi$ thus gives
\be
\boxed{
\qquad Z_0[J] \E {\rm const.} \cdot \exp \lsp - \> \frac{i}{2} 
\int d^4 x \, d^4y \> J(x) \, K_0^{-1}(x,y) \, J(y) \rsp  \> \equiv  \> Z_0[0] \> \exp \lsp -
\frac{i}{2} \, \lrp J, K_0^{-1} J \rrp \rsp \> , \quad
}
\label{Z0 2}
\ee
where the constant does not depend on the source. In addition, we have used a
convenient short-hand notation. The inverse of the kernel
\be
K_0(x,y) = \left ( - \Box_x - m^2 \right ) \,  \delta^4 (x - y )
\label{Kern}
\ee
is calculated most easily in momentum space as -- due to translational invariance -- 
$ \> K_0 \> $ only depends on the difference  $ \> x - y \> $:
\be
\qquad \Delta(x,y) \EQ K_0^{-1}(x,y) = \int \frac{d^4 k}{(2 \pi)^4} \> 
\tilde \Delta(k) \, e^{ i k \cdot (x - y)} \> . \quad   
\label{Delta(x,y)}
\ee
The equation $ \> \left ( -\Box_x - m^2 \right ) \Delta(x,y)
= \delta^4 (x - y ) \> $ then just becomes an algebraic equation \\
$ \> (k^2 - m^2) \tilde \Delta(k) = 1 \> $ with the solution
\bce
\vspace{0.2cm}

\fcolorbox{blue}{white}{\parbox{5cm}
{
\bea
\tilde \Delta_F(k) \EA \frac{1}{k^2 - m^2 + i \, 0^+} \> . \no
\eea
}}
\ece
\vspace{-2cm}

\bea
\label{skalarer Propagator}
\eea
\vspace{0.4cm}


\noindent

This is the \textcolor{blue}{\bf Feynman propagator} in momentum space.
Of course, one has to specify how to treat the pole at $ \> k^2 = m^2 \> $ .


As in \blau{\bf Detail \ref{tief}} this can be achieved by a more careful 
treatment of the limit $ \> t_i \to i \infty, t_f \to - i \infty \> $ in the
path-integral representation (\ref{GF Pfad}) for the Green functions, or 
simpler by introducing a {\bf convergence factor}
\be 
\exp \lsp  - \frac{\epsilon}{2} \,  \int d^4x \> \Phi^2(x) \rsp \> , \quad \epsilon > 0
\ee
into the path integral. By combining this term with the mass term in the Lagrangian
this leads to the replacement
\be
\boxed{
m^2 \To m^2 - i \, 0^+
}
\ee
analogous to the rule \eqref{feyn regel} in the case of the harmonic oscillator.
($ i \, 0^+ $ is a short-hand for a small, postive imaginary part
 $ i \epsilon , \, \epsilon > 0 $ which at the end of the calculation is set to zero.).

\noindent
From Eq. (\ref{Z0 2}) one thus finds that the 2-point function
(or the \blau{\bf propagator}) in lowest order is given by
\be
G_2^{(0)}(x_1,x_2) \E i \Delta_F(x_1,x_2) \> .
\label{prop 0}
\ee

The generating functional \textcolor{blue}{\bf with} interaction can be obtained formally 
from $ \> Z_0[J] \> $ by functional differentiation. As functional differentiation w.r.t.
the source  $ \> J(x) \> $ generates the field
$\> \Phi(x) \> $ we thus have
\vspace{0.2cm}

\fcolorbox{black}{white}{\parbox{14cm}
{
\bea
Z[J] \EA \int {\cal D} \Phi \> \exp \left [ \> - i \int d^4x \> 
V \left ( \frac{1}{i} \frac{\delta}{\delta J(x)} \right ) \right ]
\> \exp \left [ \> i \int d^4y \> \left ( {\cal L}_0 + J(y) \Phi(y) \right )
\right ] \non
\EA \exp \left [ \> - i \int d^4x \>
V \left ( \frac{1}{i} \frac{\delta}{\delta J(x)} \right ) \right ]
\> Z_0 [ J ] \> .\no
\eea
}}
\vspace{-2cm}

\bea
\label{Z aus Z0}
\eea

\vspace{1.6cm}

\renewcommand{\baselinestretch}{0.9}
\scriptsize
\refstepcounter{tief}
\noindent
\blau{\bf Detail \arabic{tief}:} {\bf A Functional Equation for the Generating Functional}\\

\noindent
\begin{subequations}
The representation of the generating functional as a functional integal
offers many possibilities for transformations known from usual integral calculus.
For instance, we may perform an infinitesimal shift of the integration variable
\be
\Phi(x) \E \varphi(x) + \epsilon \, f(x)
\ee
in the path integral \eqref{def erzeug functional} (\meingruen{\bf Itzykson \& Zuber},
 p. 447). As the Jacobian obviously is "1", we have
\be
Z[J] \E \int {\cal D} \varphi \> \lcp 1 + i \epsilon \int d^4x \> f(x) \, \frac{\delta}{\delta \varphi(x)} \> \exp \lrp i S[\varphi] 
+ i \int d^4x \, J(x) \, \varphi(x) \rrp \rcp\> .
\ee
Since the function $ f(x )$ is totally arbitrary, the (functional) integral over a total (functional) derivative must vanish (as in ordinary integral calculus):
\be 
0 \E  \int {\cal D} \varphi \> \frac{\delta}{\delta \varphi(x)} \> \exp \lrp i S[\varphi] 
+ i \int d^4x \, J(x) \, \varphi(x) \rrp  \E i \, \int {\cal D} \varphi \> \lcp \frac{\delta S[\varphi]}{\delta \varphi(x)} + J(x) \rcp \, \exp \lrp i S[\varphi] 
+ i \int d^4x \, J(x) \, \varphi(x) \rrp \> .
\ee
We may pull out the terms, which contain  $ \varphi $ in the curly bracket, from 
the functional integral if we replace $ \varphi(x) $ by $ \delta/(i \delta J(x)) $ (acting on
the generating functional) in these terms. Then we obtain
\be
\lcp \frac{\delta S}{\delta \varphi(x)} \lsp \frac{\delta}{i \delta J(x)} \rsp + J(x) \rcp \, 
Z[J] \E 0 \> ,
\ee
which for the scalar theory \eqref{skalare Lagrangedichte} explicitly reads
\be
\lcp \lrp \Box + m^2 \rrp \,  \frac{\delta}{i \delta J(x)} + V' \lrp  \frac{\delta}{i \delta J(x)} \rrp
- J(x) \rcp \, Z[J] \E 0 \> .
\ee
This so-called {\bf Schwinger equation} therefore connects different $ n $-point functions.

\end{subequations}
\renewcommand{\baselinestretch}{1.1}
\normalsize
\vspace{0.5cm}

\renewcommand{\thesubsubsection}{\textcolor{blue}{3.\arabic{subsection}.\arabic{subsubsection}}}

\subsubsection{\blau {Perturbation Theory}}

\begin{figure}[hbtp]
\vspace*{-5cm}
\bce
\includegraphics[angle=0,scale=0.8]{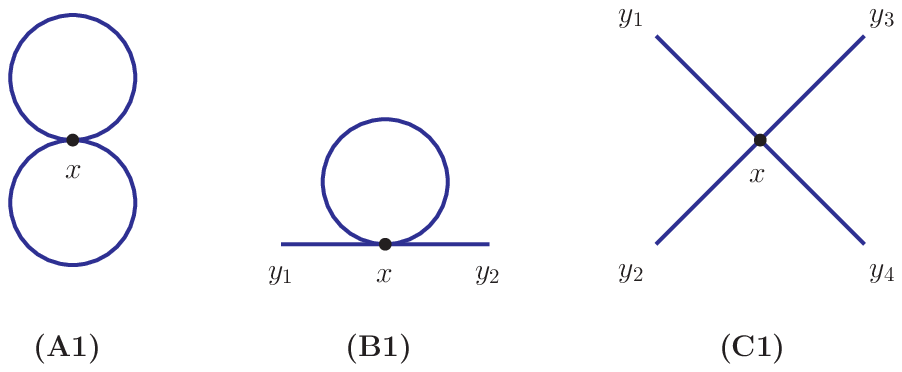}
\ece
\vspace*{-12cm}
{\bf Fig. \arabic{abb}} : Graphical representation of first-order perturbation
theory  $ \> \omega_1 \> $ for the generating \\
\hspace*{1.7cm} functional in $ \>\Phi^4$-theory.
\label{abb: omega1}
\end{figure}

The connection \eqref{Z aus Z0} between the functional with and without interaction is a purely
formal relation which only can be made concrete by an expansion in powers of the interaction.
This will be illustrated in the case of the $ \> \> \Phi^4 $-Theorie (\ref{Phi4 Theorie}).
In lowest orders of a power series expansion w.r.t. the coupling constant 
$ \> \lambda \> $ we obtain
\be
Z[J] \E Z_0 [ J ] \> \Bigl \{ \> 1 + \lambda \omega_1[J]
+ \lambda^2 \omega_2[J] + \ldots \> \Bigr \}
\ee
with
\be
\omega_1[J] \E  - \frac{i}{4 !} \, \frac{1}{ Z_0 [ J ]} \> 
\int d^4x \> \left (\frac{\delta}{\delta J(x)} \right )^4 Z_0 [ J ] \> .
\ee
Doing the differentiations gives
\bea
\omega_1[J]  \EA - \frac{i}{4 !}  \, \int d^4x \>  \Biggr \{ \>
3 \, (-i)^2 \, \Delta_F(x,x) \, \Delta_F(x,x) \hspace{4.2cm} (\rm{A}1) \non
&& + 6 \, (-i)^3 \int d^4 y_1 d^4 y_2 \> \, \Delta_F(x,y_1) \, J(y_1) \, 
\Delta_F(x,x) \, \Delta_F(x,y_2) \, J(y_2) \hspace{0.7cm}  ({\rm B}1) 
\label{omega1} \\
&& + (-i)^4 \int d^4 y_1 d^4 y_2 d^4 y_3 d^4 y_4 \>  \prod_{j=1}^4 \, 
\Delta_F(x,y_j) \, J(y_j)  \> \Biggr \} \> . \hspace{2.3cm} ({\rm C}1) \nonumber  
\eea
\noindent
The different terms can be ordered according to the number of sources $J$ and represented
graphically: Each propagator  $ \> \Delta_F(x,y) \> $ is symbolized by a line leading from 
$ y $ to $ x $ and each interaction vertex by a point into which 4 lines are going in or out
(see Fig. \ref{abb: omega1}).

\refstepcounter{abb}
\vspace{0.4cm}

\noindent
The first line of Eq. (\ref{omega1}) -- the term (A1) --   
describes so-called "vacuum graphs" which do not have any sources and which cancel
by division with $ Z[0] $ in Eq. (\ref{Gn}) for the Green function \footnote{
If normalized to the free generating functional $ Z_0[0] $, this term determines
the vacuum-energy density produced in $1^{\rm st} $ order perturbation 
theory by the self-interaction.}.
Note that the different terms carry different numerical factors ("multiplicities")
making up the {\bf symmetry factors} -- these are the numbes with which one has to
divide the particular diagram according to the Feynman rules. Considering that 
functional differentation w.r.t.  $ \, 2 j \, $ sources gives an additional 
factor $ \> (2 j)! \> $ for the particular diagram, Eq. \eqref{omega1} gives multiplicities
for the graphs (A1), (B1), (C1) which are in agreement
with the ones listed in tables I -- III  of Ref. \cite{KPKB}.

\begin{figure}[hbtp]
\vspace*{-5cm}
\bce
\includegraphics[angle=0,scale=0.8]{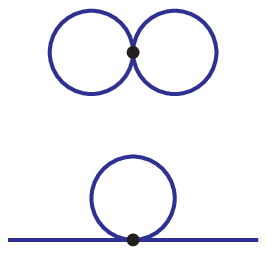}
\ece
\vspace*{-15.3cm}
{\bf Fig. \arabic{abb}} : A disconnected graph in $ 2^{\rm nd} $ order 
perturbation theory for the  $ \>\Phi^4$-theory.
\label{abb: unverbund}
\end{figure}

In second-order perturbation theory one finds
\be
\omega_2[J] \E \frac{1}{2} \omega_1^2[J] \> + \> \tilde \omega_2[J]
\> ,
\label{omega2}
\ee
where the first contribution on the r.h.s. contains
\textcolor{blue}{\bf disconnected graphs} in which two parts are not connected
by a common line (see, e.g. Fig.\ref{abb: unverbund} where the first-order graphs
 (A1) and (B1) are taken together without connection).

\refstepcounter{abb}

\refstepcounter{abb}
\begin{figure}[hbtp]
\vspace*{-3.5cm}
\bce
\includegraphics[angle=0,scale=0.56]{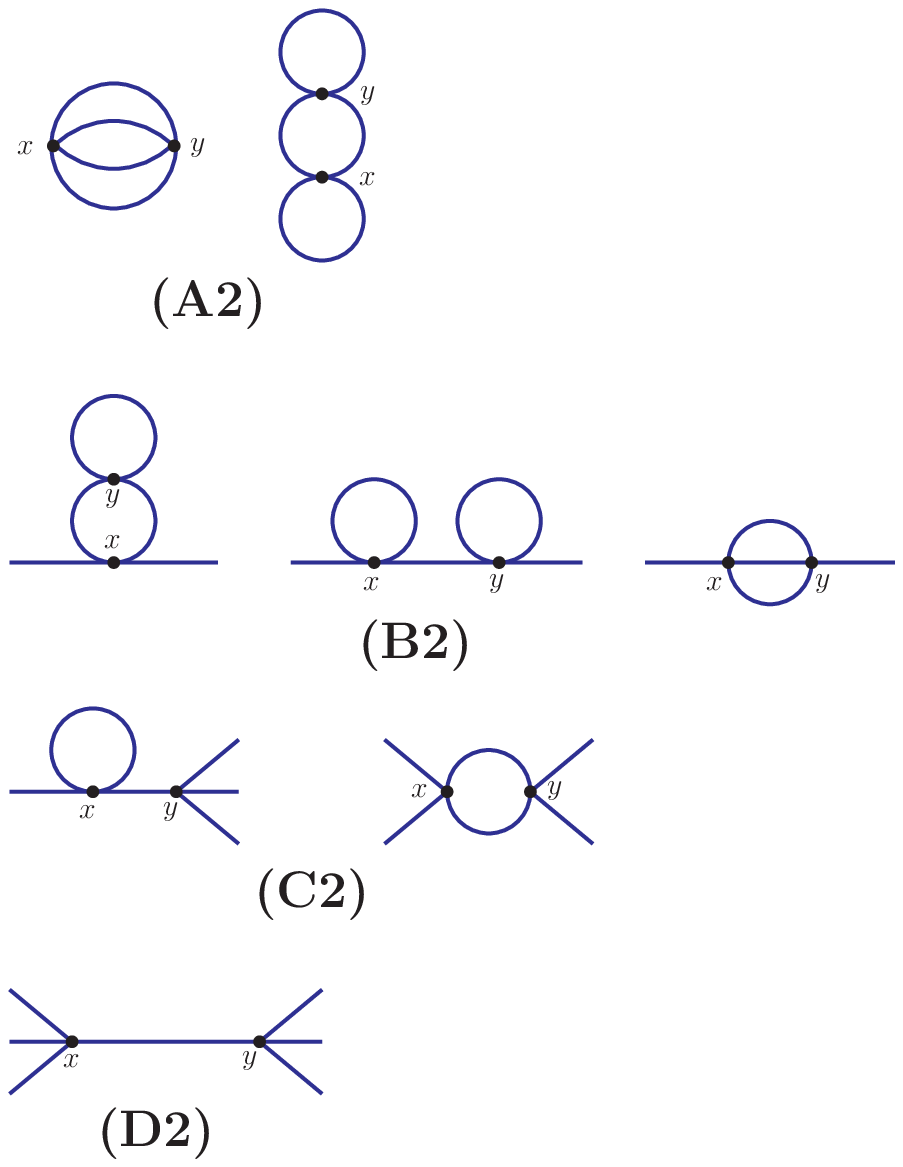}
\ece
\vspace*{-2.7cm}
{\bf Fig. \arabic{abb}} : Graphical representation of the 
$ 2^{\rm nd} $ order perturbation theory $ \> \tilde \omega_2 \> $ for the generating \\
\hspace*{1.5cm} functional of the  $ \>\Phi^4$-theory (see Eq. \eqref{tilde omega2}).
\label{abb: omega2}
\end{figure}
\vspace{1.5cm}

\renewcommand{\baselinestretch}{0.9}
\scriptsize
\refstepcounter{tief}

\noindent
\blau{\bf Detail \arabic{tief}:} {\bf Generating Functional of the  $\Phi^4$-Theory  in $ 2^{\rm nd} $ Order}\\
\vspace{0.3cm}

\noindent
\begin{subequations}
The direct calculation of
\bea
\omega_2[J] \E \frac{(-i)^2}{2 (4!)^2} \, \frac{1}{Z_0[J]} \, \int d^4x \, d^4y \> 
\left ( \frac{\delta}{\delta J(y)} \right )^4 \, 
\left ( \frac{\delta}{\delta J(x)} \right )^4 \> Z_0[J] 
\deF \frac{1}{2} \omega_1^2[J] + 
\frac{(-i)^2}{2 (4!)^2} \, \sum_{j=0}^3 \, (-i)^{4+j} \, 
\tilde \omega_2^{(2 j)}[J] 
\eea
is a little bit cumbersome but feasible when using Leibniz's rule. It gives the
multiplicities for the individual graphs without combinatorics. In doing so
it is recommended to use a condensed notation: $\, \Delta_{12} $ for
$\, \Delta_F(y_1,y_2)$ , $J_1$ for $J(y_1)$ etc. and integration over repeated indices:
$ \, \Delta_{xx} \EQ  \int d^4x \, \Delta_F(x,x)$ , $ \Delta_{xy}^4 \EQ \int d^4x d^4y \, 
\lrp \Delta_F(x,y) \rrp^4 $  
etc. With
$\, \Delta_F(x,y) = \Delta_F(y,x) $ one obtains the following
contributions with $ 2 j $ external sources
\bea
\tilde \omega_2^{(0)}[J] \EA 24 \, \Delta_{xy}^4 + 72 \,  \Delta_{xx} \Delta_{xy}^2
\Delta_{yy} \hspace{7.95cm} (\rm{A}2) \non
\tilde \omega_2^{(2)}[J] \EA 144 \, \Delta_{xy}^2  \Delta_{yy}
\left (\Delta_{x1} J_1 \right )^2 + 144 \, \Delta_{xx}   \Delta_{xy} \Delta_{yy} 
\Delta_{x1} J_1 \Delta_{y2} J_2 + 96 \, \Delta_{xy}^3 \Delta_{x1} J_1 
\Delta_{y2} J_2 \hspace{1.6cm} (\rm{B}2) \non
&& \label{tilde omega2}\\                                               
\tilde \omega_2^{(4)}[J] \EA 96 \,  \Delta_{xx} \Delta_{xy}  \left (\Delta_{y2} J_2
\right )^3  \Delta_{x1} J_1 + 72 \, \left ( \Delta_{x1} J_1 
\right )^2  \Delta_{xy}^2 \left ( \Delta_{y2} J_2 \right )^2 \hspace{4.0cm} (\rm{C}2)\non
\tilde \omega_2^{(6)}[J] \EA 16 \,  \left (\Delta_{x1} J_1
\right )^3  \Delta_{xy} \left (\Delta_{y2} J_2 \right )^3 \> . \hspace{7.6cm} (\rm{D}2) 
\nonumber
\eea
After multiplication with $ \> (2j)! \> $ the numerical factors agree with those
given in Ref. \cite{KPKB}. In this reference and elsewhere \cite{HHL}, one can find
general procedures and computer programs which give the symmetry factors for all graphs in arbitrary
order and for different interactions.

\end{subequations}
\renewcommand{\baselinestretch}{1.1}
\normalsize
\vspace{1cm}

\noindent
The second part $ \> \tilde \omega_2[J] \> $ in Eq. \eqref{omega2} is graphically 
represented in Fig. \ref{abb: omega2}.
\vspace{1cm}

\subsubsection{\blau{Connected and Amputated Green Functions}}

It is obvious that the disconnected graphs from the first term on the r.h.s. of
Eq. \eqref{omega2} do not contribute to physical processes. They are canceled
if we define
\bce
\vspace{0.5cm}

\fcolorbox{blue}{white}{\parbox{10cm}
{
\bea
G_c(x_1 \ldots x_n) \EA \left ( \frac{1}{i} \right )^n \> \frac{ \delta^n}{\delta J(x_1) 
\ldots \delta J(x_n)} \> \ln Z[J] \Biggr |_{J=0} \no
\eea
}}
\ece
\vspace{-2cm}

\bea
\label{verbundene GF}
\eea
\vspace{0.4cm}


\noindent
as \textcolor{blue}{\bf ``connected''} Green functions. Indeed, one finds
\bea
\ln Z[J] \EA \ln Z_0[J] + \ln \left \{  1 + \frac{Z[J]-Z_0[J]}{Z_0[J]}
\right \} \E \ln Z_0[J] + \ln \left \{ 1 + \lambda \omega_1[J]
+ \lambda^2 \omega_2[J] + \ldots \right \} \non
\EA - \frac{i}{2} \int d^4x d^4y \> J(x) \Delta(x,y) J(y)
+  \lambda \omega_1[J] + \lambda^2 \omega_2[J]  - \frac{1}{2}
\lambda^2 \omega_1^2[J] + {\cal O}(\lambda^3) \> ,
\eea
so that the disconnected part in $ \> \omega_2[J]  \> $ is taken away exactly.
One can show that this happens in all orders and that therefore

\be
\boxed{
\qquad W[J] \Def - i \> \ln Z[J] \qquad
}
\label{def W[J]}
\ee
is the generating functional for the connected Green functions.
The expansion of $ Z[J] $ w.r.t. these connected functions is a        
\textcolor{blue}{\bf cumulant expansion}, which is also important 
in other areas, for example in statistics (see \purpur{\bf Problem \ref{Kumulante}}).

If the system exhibits translation invariance it is more convenient to employ
(connected) Green functions in momentum space where a  $  \delta $-function can be split off
which expresses the total energy-momentum conservation:  
\be
\boxed{
\quad G_n(p_1 \ldots p_n) \E  \prod_{i=1}^n \lrp \int d^4x_i \> e^{-i p_i \cdot x_i} \rrp \> 
G_n(x_1 \ldots x_n)               
\deF (2 \pi)^4 \delta^4 \left ( \sum_{i=1}^n p_i \right ) \> \bar G_n(p_1 \ldots p_n)  
\> . \quad
}
\ee
\vspace{0.3cm}    

\noindent
{\bf Example: $\Phi^4$-Theory}
\vspace{0.2cm}

\noindent
The first correction to the propagator (\ref{prop 0}) follows from the second line
of Eq. (\ref{omega1}), i.e. the term (B1)
\be
G_2^{(1)}(x_1,x_2) \E - \frac{\lambda}{2} \, \int d^4x \> \Delta_F(x_1,x) \,
 \Delta_F(x,x) \,  \Delta_F(x,x_2) \> .
\label{prop 1}
\ee
This is a loop correction which after Fourier transform and use of the expression
 (\ref{Delta(x,y)}, \ref{skalarer Propagator}) takes the following form
\bea
G_2^{(1)}\left ( p_1,p_2 \right ) \EA - \frac{\lambda}{2} \, \int d^4x \, d^4x_1 \, d^4x_2 \> 
\prod_{j=1}^3 \left ( \int \frac{d^4k_j}{(2 \pi)^4}  \frac{1}{k_j^2 - m^2 + i \, 0^+} 
\right ) \> e^{-i p_1 \cdot x_1 - i p_2 \cdot x_2} \, \cdot \, 
e^{i k_1 \cdot (x_1-x) +i k_3 \cdot (x-x_2)} \non
\EA - \frac{\lambda}{2} \, (2 \pi)^4 \, \delta \left ( p_1 + p_2 \right ) \>
\Delta_F(p_1) \, \Delta_F(p_2) \> \int \frac{d^4 k_2}{(2 \pi)^4} \> 
\frac{1}{k_2^2 - m^2 + i \, 0^+} \> .
\eea
After separation  of the four-momentum $ \delta $-function the total propagator (which, by translation invariance only depends on $ p^2 $ ) therefore reads
\bea
\bar G_2(p) \EA  \frac{i}{p^2 - m^2 + i \, 0^+}  +  
i \left ( \frac{1}{p^2 - m^2 + i \, 0^+} \right )^2 \> 
\left ( i \frac{\lambda}{2} \, \int \frac{d^4 k}{(2 \pi)^4} \> \frac{1}{k^2 - m^2 
+ i \, 0^+} \right )  + {\cal O}(\lambda^2) \non
&& \simeq \frac{i}{p^2 - m^2 + i \, 0^+ - \Sigma^{(1)}} + {\cal O}(\lambda^2) \> ,
\label{prop 1a}
\eea
where
\be
\Sigma^{(1)} \Def i \frac{\lambda}{2} \, \int \frac{d^4 k}{(2 \pi)^4} \> 
\frac{1}{k^2 - m^2 + i \, 0^+} 
\ee
is called (first-order) {\bf ``selfenergy''}.
\newcommand{\mph}{m_{\rm phys}}
In the present case it simply is a (divergent) constant -- independent of the 
external momentum  $ p $ -- which can be combined with the parameter $m^2$. This gives the 
physical mass which can be defined as pole of the propagator:
\be
\left [ \, \bar G_2(p^2 = \mph^2) \, \right ]^{-1} \E 0 
\quad \Longrightarrow \quad  \mph^2  \E m^2 + i \, \frac{\lambda}{2} \, 
\int \frac{d^4 k}{(2 \pi)^4} \> \frac{1}{k^2 - m^2 + i \, 0^+} + {\cal O}(\lambda^2) 
\> .
\label{Massenverschieb1}
\ee
In \purpur{\bf Problem \ref{Selbstenergie phi4}} the divergence of the first-order mass shift
is investigated in more detail.
\vspace{0.2cm}

\noindent
In the general case, the exact 2-point function can be written in the same manner as in Eq. \eqref{prop 1a}
but the selfenergy is a function of the square of the external momentum and can be expanded around
 $ p^2 = \mph^2 $ 
\be
\Sigma(p^2) \E \Sigma \lrp \mph^2 \rrp + \lrp p^2 - \mph^2 \rrp \,\Sigma' \lrp \mph^2 \rrp + \ldots
\> ,
\ee
where the prime indicates the derivative w.r.t. $ p^2 $ . With that we have in the
vicinity of the pole 
\bea
\bar G_2(p) \lrp p^2 \rrp \EA \frac{i}{p^2 - m^2 - \Sigma(p^2) + i0^+} 
\quad \stackrel{p^2 \to \mph^2}{\longrightarrow} \quad
\frac{i}{p^2 - m^2 - \Sigma(\mph^2) - (p^2 - \mph^2) \, \Sigma'(\mph^2)} \non
& \equiv &
\frac{i Z_{\Phi} }{p^2 - \mph^2 + i0^+} \> ,
\eea
with
\be
\mph^2 \E m^2 + \Sigma(\mph^2) \> , \quad Z_{\Phi} \E \lsp 1 -  \Sigma'(\mph^2) \rsp^{-1} \> .
\ee
Since in $ \Phi^4 $-theory the first-order selfenergy is a constant it follows that
$ \>  Z_{\Phi} = 1 + {\cal O}(\lambda^2) $. 

\noindent
By means of the geometric series in  Eq. \eqref{prop 1a} one has summed up a lot of 
recurring corrections for the full propagator; hence the selfenergy is the sum of a 
particular (smaller) class of diagrams, the 
\blau{\bf proper} or \blau{\bf one-particle irreducible} graphs, which we will
consider in the next {\bf chapter} {\bf \ref{sec3: eff Wirk}}.

\vspace{0.2cm}

In contrast to the propagator the connected 4-point function only gets its first contribution 
in first-order perturbation theory, simply because it needs interaction. From the third line
-- the term (C1)  --  of Eq. \eqref{omega1} one obtains 
\be
G_4^{(1)}(x_1,x_2,x_3,x_4) \E - i \lambda \, \int d^4x \, \prod_{j=1}^4 \Bigl [ \,  
\Delta(x,x_j) \, \Bigr ]
\ee
and thereby after Fourier transformation
\be
G_4^{(1)}(p_1,p_2,p_3,p_4) \E (2 \pi)^4 \delta \left (p_1 + p_2 + p_3 + p_4 \right ) \, 
\left ( - i \lambda \right ) \, \prod_{j=1}^4 \left ( \frac{1}{p_j^2 - m^2 + i \, 0^+} 
\right )\> .
\label{G4 1}
\ee
The term (C1) describes a  {\bf tree graph} while (C2) denotes the first 
loop correction for the 4-point function. Correspondingly, the perturbation theory 
for the 6-point function starts with the tree graph (D2) of second order.

\vspace{0.3cm}

As \textcolor{blue}{\bf amputated} Green functions one denotes those $n$-point functions
where the external "legs" have been removed
\bce
\vspace{0.2cm}

\fcolorbox{blue}{white}{\parbox{12cm}
{
\bea
G_n^{\> \rm amp}(p_1 \ldots p_n) \EA \prod_{i=1}^n \, \lsp
\frac{1}{i \tilde \Delta_F(p_i)} \rsp G_n(p_1 \ldots p_n) \no
\eea
}}
\ece
\vspace{-2cm}

\bea
\label{def amp Green}
\eea
\vspace{0.4cm}


\noindent
with          
\be
\tilde \Delta_F (p) \Def \frac{1}{p^2 - \mph^2 + i \, 0^+} \>.
\ee
Note that here the \textcolor{blue}{\bf physical}, i.e. measured \blau{\bf mass} 
of the in- and out-going particles enters, but not the parameter $m$ in the Lagrangian
(\ref{skalare Lagrangedichte}). This is because the "bare" mass is modified by quantum corrections
as we already have seen in the above example in first-order perturbation theory.
This is an essential difference to a non-relativistic theory with finite particle number
in which the particle properties are {\bf not} modified by the interaction \footnote{Due 
to a  ``superselection rule'' in Galilei-invariant quantum mechanics found by Bargmann in 1954.}.
\vspace{0.5cm}
 
The amputated Green functions are of particular importance since they are
directly proportional to the scattering amplitudes and thus to the
physical observables. This relation is provided by the 
\textcolor{blue}{\bf reduction formulas} \footnote{See, e.g.
z. B., \meingruen{\bf Itzykson \& Zuber},
ch. 5-1, or \meingruen{\bf Peskin \& Schroeder} ch. 7.2.
In honor of the authors Lehmann, Symanzik, Zimmermann they are also called
LSZ formulas.}.            
For instance the $S$-matrix element for the scattering of 2 scalar particles 
of mass $\mph$ with initial momenta  $k_1, k_2$ and final momenta
$k'_1, k'_2 $ is given by
\vspace{0.4cm}

\fcolorbox{blue}{white}{\parbox{15.2cm}
{
\bea
&& \hspace{-0.8cm}
\left < k_1' k_2' | \, \hat S \, | k_1 k_2 \right > \E {\rm disconnected} \> {\rm terms} 
+ \left ( \frac{i}{\sqrt{Z_{\Phi}}} \right )^4 \int d^4 x_1 \, d^4 x_2 \,
d^4 y_1 \, d^4y_2 \> \exp \lsp i \lrp k_1' \cdot y_1 + k_2' \cdot y_2 \rrp \rsp \non
&& \hspace{3cm} \cdot \> \exp \lsp - i  \lrp k_1 \cdot x_1 + k_2 \cdot x_2  \rrp \rsp \> 
\cdot \left [ \Box_{y_1} + \mph^2 \right ] \, \left [ \Box_{y_2} + \mph^2 \right ] \non
&& \hspace{3cm} \cdot \> \left [ \Box_{x_1} + \mph^2 \right ] \, \left [ \Box_{x_2} + \mph^2 \right ] 
\> \left < 0 | {\cal T} \left ( \hat \Phi(y_1) \hat \Phi(y_2) 
\hat \Phi(x_1) \hat \Phi(x_2) \right ) | 0 \right >_c \>,\no
\eea
}}
\vspace{-1.1cm}

\bea
\label{LSZ fuer 4Punkt}
\eea
\vspace{0.4cm}


\noindent
where the index ``c'' shall indicate the connected parts and $ Z_{\Phi} $ is the so-called
"wave function renormalization constant" which in lowest order has the value one.
Possible bound states of the two-particle system show up as poles of the
$S$ matrix below the square of the total energy $ \> (k_1 + k_2)^2 = 4 \mph^2 \> $.

\vspace{1cm}

\renewcommand{\baselinestretch}{0.9}
\newcommand{\partialboth}{{\stackrel{\leftrightarrow}{\partial}}}
\scriptsize
\refstepcounter{tief}
\noindent
\blau{\bf Detail \arabic{tief}:} {\bf More about the Reduction Formulas}\\

\noindent
\begin{subequations}
The $S$-matrix element for the scattering of two scalar particles is given by
\be
S_{12 \to 1'2'} \E \la f \> {\rm out} \, | i \> {\rm in} \ra 
\ee
where  $  \Phi_{\latop{\rm out}{in}}(x) $ is the field operator for a free state with 
the quantum numbers of the out/in-going particles. We will now show that this 
$S$-matrix element can be expressed by the full Green functions.
For that purpose we use
\bea
\left | \, k_1, k_2, {\rm in} \ra \EA \hat a_{\rm in}(k_1)^{\dagger} \,  
\hat a_{\rm in}(k_2)^{\dagger} 
\, | \, 0 > \non
\la  k_1', k_2', {\rm out}  \, \right | \EA < 0 \,  |  \, \hat a_{\rm out}(k_1') \,  
\hat a_{\rm out}(k_2')
\eea
where $\hat a^{\dagger}(k_n)$ generates a particle with four-momentum
$k_n$  ( $ k_n^2 \E k_n^{' 2} \E \mph^2 $).

\noindent
We can "take out" the initial state $ |\, k_1, {\rm in} \, > \>  $ from the matrix element.
This is possible since the "in"-field is a free field and has the expansion
\be 
\hat \phi_{\rm free}(x) \E \int  \frac{d^3k}{(2 \pi)^3 2 E_k}
\> \left [ \, \hat a(k) \, e^{-i k \cdot x} 
+ \hat a^{\dagger}(k) \, e^{i k \cdot x}\, \right ] \> , \quad E_k \E \sqrt{\fk^2 + \mph^2} \> .
\label{phi expansion}
\ee
The inverse relations read
\be
\hat a(k) \E i \int d^3 x \> e^{i k \cdot x} \, \partialboth_0 \, \hat \phi_{\rm free}(x)  
\Bigr |_t \> , \qquad 
\hat a^{\dagger}(k) \E - i \int d^3 x \> e^{- i k \cdot x} \, \partialboth_0 \, 
\hat \phi_{\rm free}(x)  \Bigr |_t 
\label{a, a dag from phi}
\ee
and are valid for any time $ t \equiv x_0 $ . Here the operator
$\partialboth_0$ is defined in such a way that it acts like
\be
f(t) \, \partialboth_t \, g(t) \Def f(t) \, \left ( \frac{\partial g(t)}{\partial t} \right ) 
- \left ( \frac{\partial f(t)}{\partial t} \right ) \, g(t) \> .
\ee
We can use Eq. \eqref{a, a dag from phi} to get
\be
S_{1 1' \to 2 2'} \E \int d^3 x_1 \> e^{-i k_1 \cdot x_1} \,  (-i) \partialboth_{t_1} 
\la k_1', k_2' \, | \, \hat \phi_{\rm in}(x_1) \Bigr |_{t_1} \, | \, k_2 \ra \> .
\ee
The {\bf ``adiabatic hypothesis''}  is the assumption that the fully-interacting field operator
agrees with the "in"-field in the distant past (where the particle is prepared for 
the scattering process) up to a constant
\be
\hat \phi(x_1) \quad \stackrel{t_1 \to - \infty}{\longrightarrow} \quad \sqrt{Z_{\Phi}} \> 
\hat \phi_{\rm in}(x_1) \> .
\ee
As discussed in all field-theory textbooks this cannot be understood as an
operator equation but is only valid for matrix elements. Here $ Z_{\Phi} $ 
is a constant which takes into account that acting with
$\hat \phi$ on the vacuum not only generates one-particle states but also 
states with additional particles and antiparticles (if the Lagrangian is even in
$ \phi $). According to this argument one expects that  $ Z_{\Phi} $ has a value
between zero and one (the case without interaction) but it turns out that 
$ Z_{\Phi} $ diverges in higher orders ...

\noindent
Disregarding this (general field-theoretical) problem we can replace at $ t_1 = - \infty $ 
\be
\hat \phi_{\rm in}(x_1) \To  \lim_{t_1 \to - \infty} \, \frac{1}{\sqrt{Z_{\Phi}}} \, \hat \phi(x_1)
\ee
so that we obtain
\be
S_{1 1' \to 2 2'}\E \frac{-i}{\sqrt{Z_{\Phi}}} \, \lim_{t_1 \to - \infty} \int d^3 x_1 \> 
e^{-i k_1 \cdot x_1} 
\, \partialboth_{t_1} \, \la k_1', k_2', {\rm out} \, \left  | \, \hat \phi(x_1) \, 
\right | \, k_2, {\rm in} \ra 
\ee
This process can be repeated for the remaining "in"-state of the second particle and 
for both "out"-states. In the latter case one relies on the hypothesis that in the far future
\be
\hat \phi_{\rm out}(x_n') \To  \lim_{t_n' \to + \infty} \, \frac{1}{\sqrt{Z_{\Phi}}} \, 
\hat \phi(x_n')
\> , \quad n = 1,2 \> .
\ee
The final result is
\be
S_{1 1' \to 2 2'} \E  \frac{1}{Z_{\Phi}^2} \prod_{n=1}^2  \, \lcp  
\> \> 
\lim_{\latop{t_n \to -\infty}{t_n' \to + \infty}} \> \> \int d^3x_n \, d^3x_n' \>  
e^{-i k_n \cdot x_n} \, e^{+i k_n' \cdot x_n'}
\> \partialboth_{t_n} \,  \partialboth_{t_n'} \rcp \> 
\cdot \> \la 0 \, \left | \, {\cal T} \lrp \hat \phi(x_1) \hat \phi(x_2) 
\hat \phi(x_1') \hat \phi(x_2') \rrp \, \right | \, 0 \ra 
\label{reduc 1}
\ee
where ${\cal T}$ is the time-ordering operator. One sees that the last factor is nothing
but the 4-point Green function.

\noindent
The last step is to convert the limits 
$t_n, t_n' \to \pm \infty$ into integrals over a total time derivative:
\bea
\int d^4 x \> \partial_0 \, \lsp  e^{\pm i p \cdot x} \, \partialboth_0 \, f(x) \rsp  \EA 
\int d^4 x \> \lsp  \partial_0^2 \, f(x) + p_0^2 \, f(x) \rsp \, e^{\pm i p \cdot x} 
\E \int d^4 x \> \lsp  \partial_0^2 \, f(x) + f(x) \, \underbrace{\left ( m^2 + 
\fp^2 \right ) }_{= (\mph^2 - \Delta )}\, 
\rsp \, e^{\pm i p \cdot x} \non
\EA \int d^4 x \>  e^{\pm i p \cdot x} \, \lsp  \partial_0^2 - \Delta + \mph^2 \rsp \, f(x) \> , 
\label{trick for covar form}
\eea
where  $x$ stands for each $x_n, x_n'$, $p$ stands for each  $k_n, k_n'$  and $ \Delta $ 
is Laplace's operator.
In the last line we have performed an integration by parts in the spatial coordinates 
which does not produce boundary terms if wave packets instead of plane waves
would have been used in the very first beginning. 
This gives Eq. \eqref{LSZ fuer 4Punkt}.
\vspace{0.1cm}

\noindent
After an integration by parts (assuming that the fields vanish at infinity)
one obtains for the r.h.s. of this equation
\bea
&& \!\! \! \frac{1}{Z_{\Phi}^2} \,  \left (\mph^2 -k_1^2  \right )  \left (\mph^2 -k_2^2  \right)  
\left (\mph^2 -k^{' \, 2}_1 \right )  \left (\mph^2 -k^{' \, 2}_2  \right )
 \int d^4 x_1  d^4 x_2  d^4 y_1  d^4y_2   
\, \exp \left [  i \left ( k_1 \cdot y_1 + k_2 \cdot y_2 
- k'_1 \cdot x_1 - k'_2 \cdot x_2 \right )  \right ] \non
&& \hspace{6cm} \cdot  \left < 0 | {\cal T} \left ( \hat \Phi(y_1) 
\hat \Phi(y_2) 
\hat \Phi(x_1) \hat \Phi(x_2) \right )| 0 \right >_c \EQ Z_{\Phi}^{-2} \, 
G^{\> \rm amp}_4 \left ( -k_1, -k_2, k'_1, k'_2 \right ) \> , 
\label{LSZ mit amput G}
\eea
i.e., essentially the amputated, connected 4-point function
 \footnote{Roughly speaking, the reduction formula "fishes" for poles 
 of the Green function corresponding to the external particles and 
 uses the residues of these poles as physical observables.}.
The $T$ matrix is defined via
$ \hat S = 1 + i \hat T $ and we can split off the four-momentum conservation
\be
\la k_1', k_2' | \, \hat T \, | k_1, k_2 \ra \E \left (2 \pi \right )^4 \, \delta^{(4)} \lrp 
k_1 + k_2 -  k_1' - k_2' \rrp \, \la k_1', k_2' | \, {\cal M}\, | k_1, k_2 \ra \> .
\label{M from T}
\ee
In the center-of-mass (CM) system
the scattering cross section for equal-mass particles is then simply given by
\be 
\lrp \frac{d \sigma}{d\Omega} \rrp_{CM} \E \frac{1}{64 \pi^2  
E_{CM}^2 } \,  \left |  \, {\cal M} \, \right |^2 \>.
\ee
(see, e.g., \meingruen{\bf Peskin \& Schroeder}, eq. (4.85)).

\end{subequations}
\renewcommand{\baselinestretch}{1.2}
\normalsize
\vspace{1cm} 

The perturbative calculation of the amputated $n$-point functions
can be summarized by a set of \textcolor{blue}{\bf Feynman rules}:

\bit
\item[1.] Draw all possible, connected, topologically distinctive diagrams with
 $n$ vertices.

\item[2.] To each internal line belongs a propagator
$ \> i \tilde \Delta_F(k) = i/(k^2 - \mph^2 + i \, 0^+) \> $, to each vertex
(in $\Phi^4$-theory) a factor $ \> - i \lambda \> $.

\item[3.] For each internal momentum $ k $ , which is not fixed by momentum conservation
perform an integration $ \> \int d^4k /(2 \pi)^4 \> $.

\item[4.] Each graph has to be divided by a symmetry factor 
 $ S $ which corresponds to the number of permutations of internal lines which
 leave the diagram unchanged  with fixed vertices.

\item[5.] There is an additional vertex connecting two lines which corresponds 
to the mass-shift factor $ \frac{1}{2} \delta m^2 =  \frac{1}{2} (\mph^2 - m^2 ) $.
\eit
\vspace{0.5cm} 

As the measured mass enters the amputated Green functions, it is convenient to do that
also for all quantities which are altered ("renormalized" as one says) by the
interaction. To these belong the coupling constant and the normalization
of the field operator. Therefore one defines (here for the $\Phi^4$-theory)
\bea
\Phi_{\rm phys} &:=& Z_{\Phi}^{-1/2} \, \Phi \\
\mph^2  &:=& m^2 + \delta m^2 \> , \quad {\rm or} \quad \mph^2 = Z_m^{-1} \,Z_{\Phi} \, m^2 \\
\lambda_{\rm phys}  &:=&  Z_{\lambda}^{-1} \, Z^2_{\Phi} \, \lambda
\label{renorm}
\eea
and writes the Lagrangian \eqref{skalare Lagrangedichte} with the self-interaction 
\eqref{Phi4 Theorie} as
\bea
{\cal L} \EA \frac{1}{2} \, \lsp \lrp \partial_{\mu}  \Phi_{\rm phys} \rrp^2 - \mph^2
\,  \Phi_{\rm phys}^2 \rsp 
- \frac{\lambda_{\rm phys}}{4!} \, \Phi_{\rm phys}^4 \non
&& \hspace{-0.2cm} +  ( Z_{\Phi} - 1 )  \frac{1}{2} \lsp \lrp \partial_{\mu}  \Phi_{\rm phys} \rrp^2 
- \mph^2
\,  \Phi_{\rm phys}^2 \rsp - \lrp Z_{\lambda} - 1 \rrp 
\frac{\lambda_{\rm phys}}{4!} \, \Phi_{\rm phys}^4 -  \frac{1}{2}  ( Z_m - 1 ) \mph^2 \, 
\Phi_{\rm phys}^2 \, .
\eea 
The additional terms which have been generated by splitting
 $ \> Z_i = 1 \, + \, (Z_i - 1) \> , i = \Phi, \lambda, m \> $ are called ``{\bf counterterms}''. 
Having the same form as those in the original Lagrangian, they are determined such that in each 
order of perturbation theory the emerging divergences are compensated. In a renormalizable theory
their number is finite.
This {\bf ``renormalized perturbation theory''} is just a convenient
re-organization of the original perturbation theory which used  "bare" parameters. In any case, one 
has to specify  how (or at which energy) the physical parameters are measured (renormalization conditions). One has considerable freedom to do so, which, however, only leads to different results
in higher orders of perturbation theory than considered.
\vspace{0.3cm}

\noindent
That {\bf all} divergences in {\bf all} orders of perturbation theory can be subsumed
in the constants 
$ Z_{\Phi}, Z_{\lambda}, Z_m$ is highly non-trivial and expresses the fact that the
\blau{\bf $\Phi^4$-theory} is (perturbatively) \blau{\bf renormalizable}.
A necessary condition for that is the property that the coupling constant in 4 space-time 
dimensions is dimensionless or, more generally in arbitrary dimension
$ \, d \> $, has mass dimension $ \ge 0 $  ({\bf \{Le Bellac\}}, ch. 6.1).
As the action has to be dimensionless (we have $ \exp (i S) $ in the path integral) one finds
\bea
{\rm dim} \, \lsp \int d^d x \>  m^2 \, \Phi^2 \rsp \EA 
- d + 2 + 2 \,  {\rm dim} \, \Phi \E 0 \> , \quad \Longrightarrow \> {\rm dim} \,  \Phi 
\E \frac{d}{2} - 1 \\
{\rm dim} \, \lsp \int d^d x \>  \lambda\, \Phi^n \rsp \EA - d + {\rm dim} \, \lambda + n \frac{d}{2} - n \E 0         
\> , \quad \Longrightarrow \> {\rm dim} \, \lambda \E n - d \lrp \frac{n}{2} -1 \rrp \> .
\eea
While in quantum mechanics the potential can be arbitrary to a large extent (e.g. 
$ \> V(x) = g \, x^6 $ ), in quantum field theory the requirement of renormalizability 
does not allow  $ g \, \Phi^6 $-terms in the four-dimensional Lagrangian: The coupling constant 
 $ g $ would then have the mass dimension $ 6 - 2d = -2 $.
\vspace{0.2cm}

How can one prove that a particular theory is renormalizable? For this lecture we adopt
the attitude \\
 ``\blau{\textsf{ ... das ist ein {\it zu} weites Feld}}'' {\bf \{Fontane\}} \footnote{"\textsf{... that is too wide a field}".}
 and refer to more specialized monographs, e.g.  {\bf \{Collins\}} or {\bf \{Muta\}}.

\vspace{2cm}

\renewcommand{\baselinestretch}{0.9}
\scriptsize
\refstepcounter{tief}
\noindent
\blau{\bf Detail \arabic{tief}:} {\bf Additional Generating Functionals} \\

\begin{subequations}
\noindent
Instead of the formal representation \eqref{Z aus Z0} one can derive the (equally formal) 
expression
\be
Z[J] \E {\rm const.} \> \exp \lsp - \frac{i}{2} \lrp J, \Delta_F J \rrp \rsp \> \lcp \exp \lsp 
\frac{i}{2} \lrp \frac{\delta}{\delta \Phi}, \Delta_F \frac{\delta}{\delta \Phi} \rrp \rsp  \> 
\exp \lrp - i \int d^4x \> V(\Phi) \rrp \rcp_{\Phi = \Delta_F J} 
\ee
(\purpur{\bf Problem \ref{Ableit erzeug Funk}$^{\star}$}) which sometimes is more convenient.\\

\noindent
Instead of coupling the source $ J(x) $ to {\bf one} field one may introduce 
{\bf bi-linear} sources, e.g.
\be
\tilde Z[K] \Def  {\rm const.} \> \int {\cal D} \Phi \> \exp \lsp i \int d^4x \> \lrp {\cal L}
- \frac{1}{2} K(x)  \, \Phi^2(x) \rrp \rsp 
\ee
for a single scalar field. In \purpur{\bf Problem \ref{Bi-Quelle}} this is generalized to 
a  $N$-component field with $ O(N) $-symmetry.\\

\noindent
More interesting is to write down a {\bf generating functional for amputated Green functions}  \footnote{See, e.g., the appendix of Ref. \cite{Kop} where it is
normalized to $ Z_0[0] $ for determining the vaccum-energy density.}:
Starting from the definition \eqref{def amp Green}
and from  Eq. \eqref{Gn} we define
\be
\varphi \Def - \Delta_F \, J \> , \quad {\rm d.h.} \quad \varphi(x) \E - \int d^4y \> 
\Delta_F(x,y) \, J(y)\> .
\ee
As the free part of the action can be written as
\be
\frac{1}{2} \, \lrp \Phi, ( -\Box - m^2 ) \Phi \rrp \E \frac{1}{2} \, \lrp \Phi, ( -\Box - \mph^2 )
 \, \Phi \rrp
+  \frac{1}{2} \, \delta m^2 (\Phi, \Phi) \EQ  \frac{1}{2} \, \lrp \Phi, \Delta_F^{-1} \Phi \rrp 
+  \frac{1}{2} \, \delta m^2 (\Phi, \Phi)
\ee
we have
\be
G^{\rm amp}(1  \ldots n) \E \frac{1}{Z[0]}  \, \frac{\delta^n}{\delta \varphi_1 \ldots 
\delta \varphi_n} \, 
\int {\cal D} \Phi \> \exp \lsp \frac{i}{2} \lrp \Phi,\Delta_F^{-1} \Phi \rrp - i 
\lrp \varphi,  \Delta_F^{-1} \Phi \rrp + i \, S_{\rm int}[\Phi] \rsp \> \Biggr |_{\varphi = 0} \> ,
\ee
with an additional term  $ \delta m^2 \, (\Phi,\Phi)/2 $)
in the interacting part $ \, S_{\rm int}[\Phi] $ . If we shift the integration variable
$ \> \Phi = \Phi' + \varphi \> $ we obtain
\be 
\boxed{
\quad G^{\rm amp}(1  \ldots n) \E    \frac{\delta^n}{\delta \varphi_1 \ldots 
\delta \varphi_n} \, \frac{1}{Z[0]} \, \exp\lsp -\frac{i}{2} (\varphi, \Delta_F^{-1} \, \varphi) 
\rsp \, \int {\cal D} \Phi' \> \exp \lrp i \, S_0[\Phi'] + i \, S_{\rm int}[\Phi' + \varphi] \rrp 
\> \Bigr |_{\varphi = 0} \> ,
}
\label{erzeug G amp}
\ee
where the normalization is given by
\be
Z[0] \E  \int {\cal D} \Phi' \> \exp \lsp i \, S_0[\Phi'] + i \, S_{\rm int}[\Phi'] \rsp \> .
\ee
The connected, amputated  $n$-point functions are again determined by the logarithm of the 
generating functionals, i.e.
\be
G^{\rm amp}_c(1  \ldots n) \E   \, \frac{\delta^n}{\delta \varphi_1 \ldots 
\delta \varphi_n} \, \lsp -\frac{i}{2} (\varphi, \Delta_F^{-1} \, \varphi) + 
 \ln \, \lcp \frac{\int {\cal D} \Phi' \> \exp \lrp i \, S_0[\Phi'] + i \, 
S_{\rm int}[\Phi' + \varphi] \rrp }{\int {\cal D} \Phi' \> \exp \lrp i \, S_0[\Phi'] + i \, 
S_{\rm int}[\Phi'] \rrp} \rcp \> \rsp_{\varphi = 0} \> .
\label{erzeug G amp zusammen}
\ee
Let's take as examples the 2- and 4-point functions in lowest order
perturbation theory: It is seen immediately that for vanishing interaction only the
connected, amputated 2-point function 
\be
G^{{\rm amp} \, (0)}_c \left (x_1,x_2 \right ) \E -i \Delta_F^{-1}(x_1,x_2)
\E i \lrp \Box_{x_1} + \mph^2 \rrp \, \delta^{(4)}\left (x_1 - x_2 \right )
\ee
exists. Obviously, a connected 4-point function shows up only with interaction:
Since $ \> V(\Phi' + \varphi) = {\cal O}(\varphi^0,\varphi^1,\varphi^2,\varphi^3) + 
\lambda \varphi^4/4! \> $ , one finds in first-order perturbation theory for this
amputated Green function
\bea
&&G_c^{{\rm amp} \, (1)} \left ( x_1,x_2,x_3,x_4 \right ) \E \frac{\delta^4}{\delta \varphi(x_1) \ldots 
\delta \varphi(x_4)} \, \lrp \frac{-i \lambda}{4!} \rrp  \, \int d^4x \> \varphi^4(x) 
\E \lrp - i \lambda \rrp \, \delta^{(4)} \left ( x_1 - x_2 \right )   \delta^{(4)} \left ( x_2 - x_3 
\right )
 \delta^{(4)} \left ( x_3 - x_4 \right ) \non
\Longrightarrow && G_c^{{\rm amp} \, (1)} \left ( p_1,p_2,p_3,p_4 \right ) \E ( 2 \pi )^4 \, \delta^{(4)} 
\left ( p_1 + p_2 + p_3 + p_4 \right ) \> ( - i \lambda ) \> ,
\eea
which agrees with the Feynman rules for graph  (C1) in Fig. \ref{abb: omega1} or the amputation of
Eq. \eqref{G4 1}.
\vspace{0.2cm}

\noindent
Since the connected, amputated Green functions directly give the $S$ matrix (see, e.g.
Eq. \eqref{LSZ mit amput G}) one can define a {\bf $S$-matrix functional}
\be
{\cal F}[\varphi] \Def {\cal N} \, \int {\cal D} \Phi' \> \exp \Bigl ( \, i \,   S_0[\Phi'] 
+ i \, S_{\rm int}[\Phi' + \varphi] \> \Bigr )
\ee
(c.f. \meingruen{\bf Nair}, eq. (8.49)), which after functional differentiation and Fourier transformation gives the $S$ matrix 
\be
S_{k_1,\ldots k_m \to k'_1,\ldots k'_n} =  \lrp \frac{i}{\sqrt{Z_{\Phi}}} \rrp^{m+n}
 \, \prod_i^m \lrp \int d^4x_i \, e^{-i k_i \cdot x_i} \rrp \, 
\prod_j^n \lrp \int d^4y_j \, e^{i k'_j \cdot y_j} \rrp 
\>  \frac{\delta^{n+m}}{\delta \varphi(x_1) \ldots \delta \varphi(x_m) \>
\delta \varphi(y_1) \ldots \delta \varphi(y_n)} \>  \ln \, {\cal F}[\varphi] \, 
\Biggr |_{\varphi = 0}
\label{S von S Funk}
\ee
(\meingruen{\bf Nair}, eq. (5.24)). The non-interacting part
$ \> - i (\varphi, \Delta_F^{-1} \varphi )/2  \> $ 
in Eq. \eqref{erzeug G amp zusammen}, which only contributes to the 2-point function
can be omitted in this expression.
With the help of the chain rule (our conventions for Fourier transforms 
are given in  Eq. \eqref{Delta(x,y)})
\be      
\frac{\delta}{\delta \tilde \varphi(k)} \E \int d^4x \> \frac{\delta \varphi(x)}{\delta \tilde \varphi(k)}\, \frac{\delta}{\delta \varphi(x)} \E \int \frac{d^4x}{(2 \pi)^4} \> e^{i k \cdot x} \, \frac{\delta}{\delta \varphi(x)}
\ee
this can written more compactly as
\be
\boxed{
S_{k_1,\ldots k_m \to k'_1,\ldots k'_n} =  \lrp \frac{(2 \pi)^4 i}{\sqrt{Z_{\Phi}}} \rrp^{m+n}
\>  \frac{\delta^{n+m}}{\delta \tilde \varphi(-k_1) \ldots \delta \tilde \varphi(-k_m) \>
\delta \tilde \varphi(k'_1) \ldots \delta \tilde \varphi(k'_n)} \>  \ln \, {\cal F}[\varphi] \, \Biggr |_{\varphi = 0}
}
\label{S von tilde phi}
\ee
Though this has been derived only for a scalar theory the formulas 
\eqref{S von S Funk} or \eqref{S von tilde phi} can be generalized (with obvious modifications) to
other theories.\\

\noindent
For the generating functional of the  {\bf proper} or {\bf one-particle-irreducible diagrams} 
see the next  {\bf chapter} {\bf \ref{sec3: eff Wirk}}.
\end{subequations}
\renewcommand{\baselinestretch}{1.2}
\normalsize
\vspace{2cm}


%
%
\noindent
{\bf Free Propagators of Other Theories:}
\bdes
\item[{\bf (a)}] Charged scalar particles are described by two fields with the same mass
or equivalently by a complex field $ \Phi(x) $ with the free Lagrangian
\be 
\boxed{
{\cal L}_0 \E \lrp \partial_{\mu} \Phi^{\star} \rrp \, \lrp \partial^{\mu} \Phi \rrp - m^2 \, 
\Phi^{\star} \Phi
\label{L0 geladen skalar}
}
\ee
(note the missing factor  $1/2$ compared to the neutral Lagrangian \eqref{skalare Lagrangedichte} ).
This is the topic in \purpur{\bf Problem \ref{N skalare Teil}} where one also should show that
the 2-point function invokes the same Feynmann propagator as in the neutral case
\be 
G_2(x-y) \EQ \la \Phi^{\star}(y) \, \Phi(x) \ra \E i \, \Delta_F(x-y) \> .
\ee

\item[{\bf (b)}] A massive spin-1 particle (also called a vector particle) is described by
the field $ V_{\mu}(x) $ and the free Lagrangian
\be 
\boxed{
{\cal L}_0^V \E - \frac{1}{4} \, F_{\mu \nu} F^{\mu \nu} + \frac{1}{2} m^2_V \, V_{\mu} V^{\mu} \> , \quad
F_{\mu \nu} \Def \partial_{\mu} V_{\nu} - \partial_{\nu} V_{\mu} 
\label{vec L0}
}
\ee
(notice the different sign of the mass term compared to the scalar case!).
As can be derived in \purpur{\bf Problem \ref{Vektor-Teilchen} a)} its propagator is 
\be 
\boxed{
\tilde D_V^{\mu \nu}(k) \E - \frac{1}{k^2 - m_V^2 + i 0^+} \, \lsp g^{\mu \nu} 
- \frac{k^{\mu} k^{\nu}}{m_V^2} \rsp 
\label{vec prop} \> .
}
\ee
\item[{\bf (c)}] In an analogous way one can treat \textcolor{blue}{\bf free
spin-$\frac{1}{2}$ particles}, i.e. fermions, with the Lagrange density
\be
\boxed{
\qquad {\cal L}_0 \E \cy{\bar{\psi}} \left ( i \dslash - m \right ) \cy{\psi}
\> , \hspace{1cm} \dslash \EQ \gamma^{\mu} \partial_{\mu} \> ,
\hspace{0.5cm} \left [\gamma^{\mu}, \gamma^{\nu} \right ]_{+} = 2 g^{\mu \nu} \> .\qquad 
}
\label{L0 fermion}
\ee
To derive the equation of motions (the {\bf Dirac equations}) for 
$ \> \cy{\psi}, \, \cy{\bar{\psi}} \> $ one should use 
\be
{\cal L}_0 \E \frac{i}{2} \, \lsp \cy{\bar{\psi}} \, \gamma^{\mu} \lrp \partial_{\mu} \cy{\psi} \rrp
- \lrp \partial_{\mu} \cy{\bar{\psi}} \rrp \, \gamma^{\mu} \, \cy{\psi} \rsp - m \, \cy{\bar{\psi}} 
\cy{\psi}
\ee
but for the action both forms are equally valid (after an integration by part).
To obtain a generating functional for fermionic Green functions, one has to introduce 
\textcolor{blue}{\bf anticommuting external sources}
$ \> \cy{\eta}(x), \cy{\bar{\eta}}(x)\> $ and to integrate functionally over Grassmann-valued 
fields \footnote{\label{g gerade} As one wants to work with commuting numbers  in the path integral,
the individual terms in the exponent (action, Lagrangian, source terms) have to be
{\bf Grassmann-even}, i.e. they should contain an even number of Grassmann-valued factors.}:
\be
Z_0[\cy{\eta}, \cy{\bar{\eta}}] \E \int {\cal D} \cy{\bar{\psi}}(x) {\cal D} \cy{\psi}(x)
\> \exp \left [ \> i \int d^4x \> \left ( {\cal L}_0(\cy{\psi},\cy{\bar{\psi}}) 
+ \cy{\bar{\psi}} \cy{\eta} + \cy{\bar{\eta}} \cy{\psi} \right ) \> \right ] \> .
\ee
As in the bosonic case one finds by completing the square
\be
\boxed{
\qquad Z_0[\cy{\eta}, \cy{\bar{\eta}}] \E {\rm const.} \cdot \exp \left [ \> - 
i \int d^4x \, 
d^4y \> \cy{\bar{\eta}}(x) \, S_F(x,y) \, \cy{\eta}(y) \> \right ] \> , \quad
}
\label{Z0 fermionisch}
\ee
where
\be
\boxed{
\qquad S_F(x,y) \E \left ( i \dslash - m + i \, 0^+\right )^{-1}(x,y)
\E \int \frac{d^4k}{(2 \pi)^4} \> e^{ i k \cdot (x-y)} \>
\frac{1}{\kslash - m + i \, 0^+} \qquad
}
\ee
is the Feynman propagator of the fermion.
\edes
\noindent
Interactions can be included -- in principle -- as in the bosonic case, unless
they follow from a gauge principle (see chapter \ref{sec3: Eichtheo}).

\vspace{0.8cm}

\renewcommand{\baselinestretch}{0.9}
\scriptsize
\refstepcounter{tief}
\noindent
\blau{\bf Detail \arabic{tief}:} {\bf $\sigma-\omega$  (or Walecka) Model}\\
\vspace{0.2cm}

\begin{subequations}
\noindent
Walecka has proposed a relativistic model for the description of nuclei \cite{Wale}, 
in which nucleons (represented  by the Dirac field $ \cy{\Psi} $ with mass $ M $) 
interact with each other by the exchange of scalar mesons ($ \sigma $,  $ m_{\sigma} $) and vector
mesons ($ \omega_{\mu} $, $m_{\omega} $) :
\be 
{\cal L}_W \E \cy{\bar \Psi} \left ( i \dslash - M \right ) \cy{\Psi}
+ \frac{1}{2} \lsp \left (\partial_{\mu} \sigma \right )^2 - m_{\sigma}^2 \sigma^2
\rsp  - \frac{1}{4} \, F_{\mu \nu} F^{\mu \nu}  + \frac{1}{2} m_{\omega}^2 \, \omega_{\mu} 
\, \omega^{\mu} - g_{\sigma} \cy{\bar \Psi \Psi} \, \sigma - g_{\omega}  \cy{\bar \Psi} \gamma_{\mu} 
\cy{\Psi} \, \omega^{\mu} \> , \> 
F_{\mu \nu} \EQ \partial_{\mu} \omega_{\nu} - \partial_{\nu} \omega_{\mu} \> .
\label{L Walecka}
\ee
Aiming at a description of the strong interaction inside nuclei, the corresponding dimensionless
coupling constants $ g_{\sigma} $ and $ g_{\omega} $ of the mesons to the nucleons
are not small: $g^2/(4 \pi) = {\cal O}(10) $, while, e.g. in quantum electrodynamics
the so-called fine-structure constant has the value $ e^2/(4 \pi) = 1/137 $.
Hence perturbation theory is not applicable which is, in general, also the case for all
{\bf bindung problems}
(another "\blau{\textsf{weites Feld}}" {\bf \{Fontane\}} which is not covered here ...)
\vspace{0.2cm}

Why is this Lagrangian a possible model for the mutual interaction of nucleons?
To answer this question we consider the non-relativistic limit for the
generating functional (without external mesons): After integrating out the meson fields
(Gaussian integral!) we have
\bea 
\hspace*{-0.7cm} Z[\cy{\bar \eta}, \cy{\eta}] \! \!\EA \! \!\int {\cal D} \cy{\bar \Psi}  {\cal D} \cy{\Psi}  
{\cal D} \sigma  \prod_{\mu} 
{\cal D} \omega_{\mu} \, e^{i \int d^4x \bigl [ {\cal  L} + \cy{\bar \eta \Psi} + 
\cy{\bar \Psi \eta} \bigr ] } 
= \int{\cal D} \cy{\bar \Psi} {\cal D} \cy{\Psi} \, \exp \biggl \{ i \int d^4x  \biggl [
\cy{\bar \Psi} \left ( i \dslash - M \right ) \cy{\Psi} + \cy{\bar \eta \Psi} + \cy{\bar \Psi \eta} \biggr ] 
\biggr \} \cdot 
e^{i S_{\sigma \, \omega}[\cy{\bar \Psi}, \cy{\Psi}]} 
\label{Z Walecka}\\
S_{\sigma \, \omega}  [ \cy{\bar \Psi}, \cy{\Psi} ] \EA - \frac{1}{2} \int d^4x \, d^4y \> 
\lsp g _{\sigma}^2 \, \rho_S(x) D_{\sigma}(x-y) \rho_S(y) + g_{\omega}^2 \, 
J_{\mu}(x) D^{\mu \nu}(x-y) J_{\nu}(y) \rsp  \> .
\label{S WW}
\eea
Here $ \> \rho_S(x) = \cy{\bar \Psi}(x) \, \cy{\Psi}(x) \> $ is the scalar density of the nucleons
and
\be 
J_{\mu}(x) \E \cy{\bar \Psi}(x) \gamma_{\mu} \cy{\Psi}(x)
\label{Strom Walecka}
\ee
their current while $ \> \tilde D_{\sigma}(k) = 1/(k^2 - m_{\sigma}^2 + i 0^+) \> $ denotes
the usual Feynman propagator of the scalar field and  $  D_{\omega}^{\mu \nu}(x-y) $
the one of the vector meson.
As derived in \purpur{\bf Problem \ref{Noether Feld}} the nucleon current is conserved:
 $ \partial^{\mu} J_{\mu} = 0 $ and thus the terms proportional to
$k^{\mu} k^{\nu} $ in Eq. \eqref{vec prop} do not contribute to the action \eqref{S WW} bei. 
Because of that the behaviour of the vector-meson propagator at high
 $ k $ is less "dangerous" and the model remains renormalizable.
\vspace{0.1cm}

The non-relativistic limit is most transparent when we allow the
"velocity of light  $ c \to \infty $"  so that it is recommended to 
re-introduce the natural constants $ c, \hbar $ . If we use
$ \Box = \partial^2/(c^2 \partial t^2)- \Delta $ 
then the scalar propagator in position space, e.g., turns into
\bea
D_{\sigma} (x-y) \EA \!
\int \! \frac{d^4 k}{(2 \pi \hbar)^4} \, 
\frac{\exp [ - i k \cdot (x-y)/\hbar]}{k_0^2/(c^2 
\hbar^2)  - {\bf k}^2/\hbar^2 - m_{\sigma}^2 c^2/\hbar^2 + i0^+} 
\, \stackrel{c \to \infty}{\longrightarrow}\,   - \hbar^2 \delta 
\left ( x_0-y_0 \right ) \, 
\int \! \frac{d^3 k}{(2 \pi \hbar)^3} \, \frac{\exp [  i {\bf k} \cdot ( {\bf x}-
{\bf y} )/\hbar ]}{{\bf k^2} + m_{\sigma}^2 c^2} \non
&& \hspace{6.1cm} = \> - \delta \left ( x_0-y_0 \right ) \,
\frac{1}{4 \pi |{\bf x} - {\bf y}| } \, 
\exp \left ( - \frac{m_{\sigma} c}{\hbar}  |{\bf x} - 
{\bf y}| \right ) \> ,
\eea
i.e. an instantaneous Yukawa term where the Compton wavelength $ \> \hbar/(m_{\sigma} c)\>  $ 
of the exchanged particle determines the range. The same is valid for the
vector-meson propagator \eqref{vec prop}, which, however has an opposite sign compared to that
of the scalar exchange.

The behaviour of the Dirac nucleons in the non-relativistic limit is standard
in relativistic quantum mechanics (see, e.g.   {\bf \{Bjorken-Drell\}}, ch. 4).
While the usual approach is an elimination of the "small" components
in the Dirac equation or the systematic Foldy-Wouthuysen transformation, in the path integral
one may simply integrate out the small component
 (see \purpur{\bf Problem \ref{Dirac nonrel}$^{\star}$} ).
 Then one obtains a non-relativistic action for the "large" component, the bi-spinor
 $ \cy{\phi}(\fx,t) $ with a dominant central potential and a sub-dominant spin-orbit potential.
 In addition, the spatial components of the nucleon current
\eqref{Strom Walecka} are suppressed since in this case the nucleons only move slowly compared 
to the velocity of light.

If we substitute all this simplifications into Eq. \eqref{Z Walecka} and in leading order replace
 $\rho_S(x) \simeq J_0(x) \simeq \cy{\phi^{\dagger}}(\fx,t) \cy{\phi}(\fx,t) $ , then all
$c$-factors cancel and -- returning to our standard system of units   $ \hbar = c = 1 $ -- 
we obtain 
\be 
S_{\sigma \, \omega}[\cy{\phi^{\dagger}},\cy{\phi}] \E - \frac{1}{2} \int dt \int d^3x \, d^3y \> 
\cy{\phi^{\dagger}}(\fx,t) \cy{\phi}(\fx,t) \, \bigg [ \> \underbrace{- \frac{g_{\sigma}^2}{4 \pi} \, 
\frac{e^{- m_{\sigma} r}}{r} + \frac{g_{\omega}^2}{4 \pi} \, 
\frac{e^{- m_{\omega} r}}{r}}_{=: V_{\sigma \omega}(r = |\fx - \fy|)} \> \bigg ] \, \cy{\phi^{\dagger}}(\fy,t) 
\cy{\phi}(\fy,t) \> .
\label{pot Walecka}
\ee
Comparison with Eqs. \eqref{L Schroedinger} and \eqref{H op} identifies 
$ V_{\sigma \omega}(r) $ as two-particle potential between nucleons in a fermionic
Schr\"odinger theory. In other words: The 
{\bf exchange of scalar mesons} leads to an {\bf attractive} Yukawa potential,  
the {\bf exchange of vector mesons} to a {\bf repulsive} interaction which dominates 
at small distances if  $ g_{\omega} > g_{\sigma} $. For  $ m_{\omega} > m_{\sigma} $ 
the scalar exchange dominates at larger distances and together this generates a potential
as depicted in  Fig. \ref{abb:Walecka}. This potential reflects the empirical 
properties of the nucleon-nucleon interaction: Attractive at larger distances to bind the
nucleons in the nucleus and strongly repulsive at small distances so that the density of 
heavy nuclei is practically constant.

\refstepcounter{abb}
\begin{figure}[hbtp]
\bce
\vspace{-0.5cm}
\includegraphics[angle=0,scale=0.3]{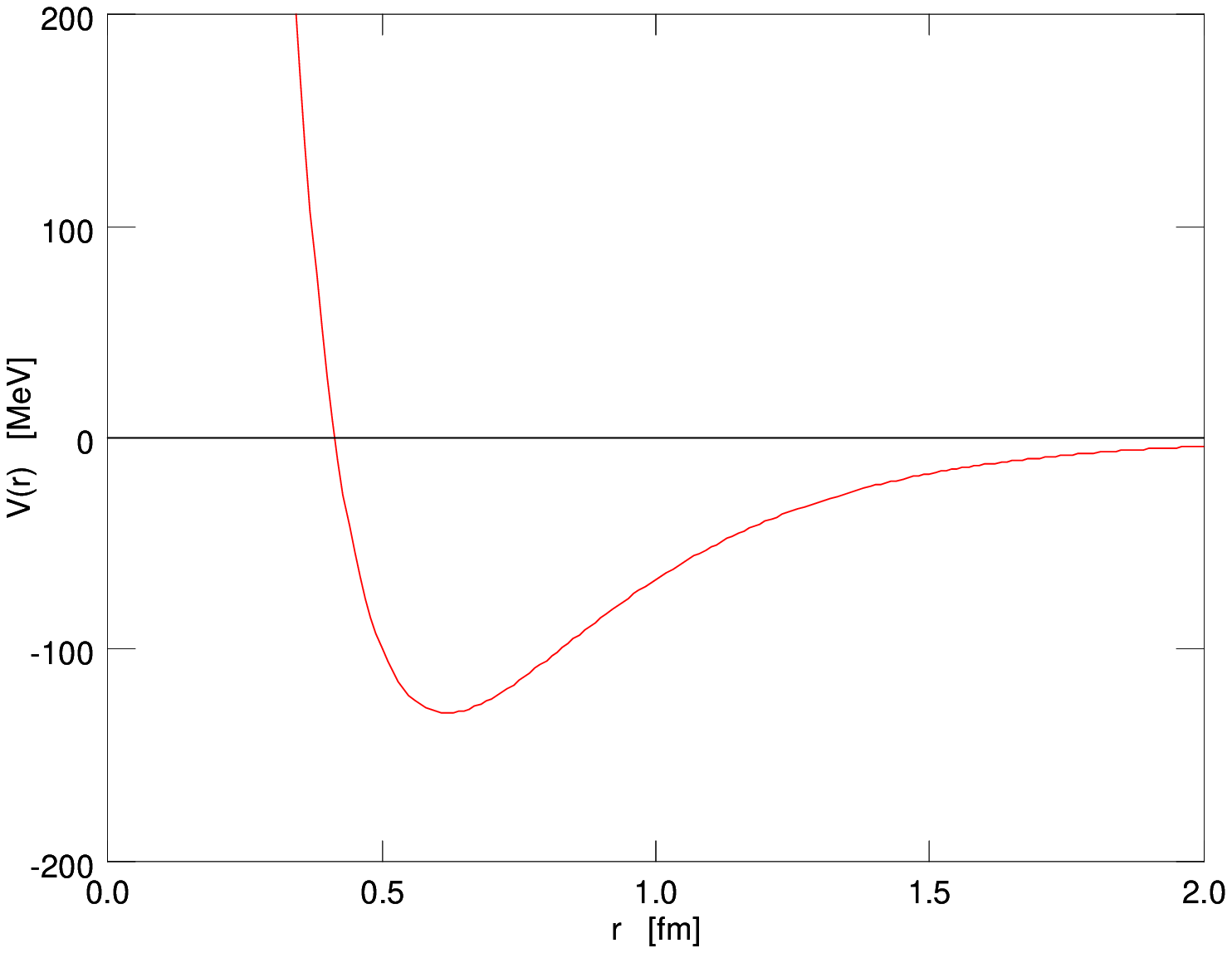}
\label{abb:Walecka}
\ece
\renewcommand{\baselinestretch}{0.9}
{\scriptsize \bf Fig. \arabic{abb}} : {\scriptsize The non-relativistic potential
\eqref{pot Walecka} between nucleons generated in the Walecka model by the exchange of
scalar  $\sigma$- and  \\
\hspace*{1.4cm} vectorial  $\omega$-mesons. The meson masses are taken as 
 $ \> m_{\sigma} = 520 \> {\rm MeV} , \> m_{\omega} = 783 \> {\rm MeV} $  and the coupling
 constants as \\ 
\hspace*{1.3cm}
$ \> g^2_{\sigma}/(4 \pi) = 8.72 , \> g^2_{\omega}/(4 \pi) = 15. 15 $ \cite{HoSe}.}
\vspace{0.5cm}
\end{figure}

If one also considers the  $1/M^2$-corrections in  \purpur{\bf Problem \ref{Dirac nonrel}$^{\star}$} ,
then one sees that also a {\bf strong spin-orbit interaction} is generated by the $ \sigma-\omega$-exchange:
Although  being suppressed by the large nucleon mass the large scalar and vector potentials add in 
absolute value in this term while they nearly cancel in 
the central potential. This strong spin-orbit interaction has been observed empirically 
long ago in the excitation spectra of atomic nuclei. Therefore the Walecka model offers
a relativistic description of nuclei: By fitting its few parameters to experimental
data it gives (in a relativistic mean-field Hartree approximation,
see {\bf chapter} {\bf \ref{sec2: Hilfsfelder}}) 
an equally good account of closed-shell nuclei as sophisticated non-relativistic potential models.
In addition, as mentioned before, it is renormalizable which, however, is not very
relevant in practice  as the exchange of scalar and vector mesons can only be an effective 
or schematic theory of nucleon-nucleon interactions; for instance, it is well-known 
that the longest range of this interaction is mediated by the exchange of the lightest hadrons, the pions, 
which are not contained in this model at all.

\end{subequations}

\renewcommand{\baselinestretch}{1.2}
\normalsize
\vspace{0.5cm}

However, as mentioned above, the quantization of fermionic theories whose interaction with 
bosonic fields is governed by a \blau{\bf gauge principle} -- as, e.g.
\rot{\bf Q}\blau{uantum-}\rot{\bf E}\blau{lectro-}\rot{\bf D}\blau{ynamics} 
(\rot{\bf QED}) or                                           
\rot{\bf Q}\blau{uantum-}\rot{\bf C}\blau{hromo-}\rot{\bf D}\blau{ynamics} 
(\rot{\bf QCD}) -- needs a special treatment. This is due to the fact
that the gauge fields are necessarily massless and that in consequence the inversion of the
bosonic kernel is not possible if the gauge degrees of freedom
have not been separated. We will address that problem in 
{\bf chapter} {\bf \ref{sec3: Eichtheo}}.


\vspace{0.2cm}

\subsection{\textcolor{blue}{Effective Action}}
\label{sec3: eff Wirk}

\begin{figure}[hbtp]
\vspace*{-6cm}
\bce
\includegraphics[angle=0,scale=0.9]{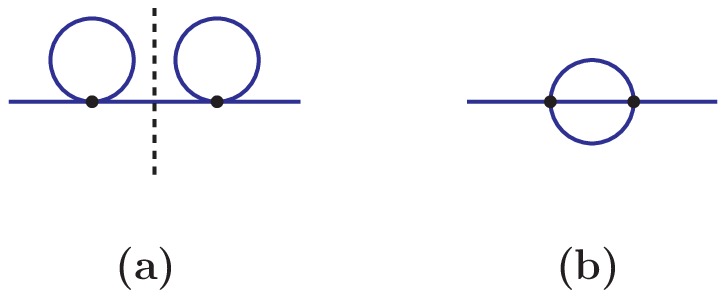}
\ece
\vspace*{-15cm}
{\bf Fig. \arabic{abb}} : (a) An one-particle-reducible graph which can be cut along the dashed line\\
\hspace*{1.6cm} (b) An one-particle-irreducible graph (in $2^{\rm nd}$ order perturbation theory for the 
$\Phi^4$-theory).
\label{abb: eigentliche Dia}
\end{figure}

We now consider again a scalar theory (e.g. that of Eq. (\ref{Phi4 Theorie}))
and denote explicitly the dependence on Planck's elementary quantum:
\be
Z[J] \E \exp \left ( \> \frac{i}{\hbar} W [J] \> \right ) \> .
\ee
One can define a \textcolor{blue}{\bf ``classical field''} as vacuum expectation value
of the field operator in presence of the source
\be
\boxed{
\qquad \Phi_{\rm cl}(x) \Def \frac{\delta W[J]}{\delta J(x)} \E
\frac{1}{Z} \int {\cal D}\Phi \> \Phi(x) \> \exp \left \{ \> 
\frac{i}{\hbar} \int d^4y \> \lsp {\cal L} + J(y) \, \Phi(y) \rsp \> 
\right \} \> . \quad
}
\label{def Phi_cl}
\ee
Of course, this field depends on the source $ \> J \> $ since otherwise
the functional integral over an odd integrand would vanish.
We assume that this relation can be inverted (which for weak sources is feasible
at least perturbatively), i.e. 
\be
J \E J(\Phi_{\rm cl}) \> .
\ee
Then one defines the quantity
\be
\boxed{
\qquad \Gamma[\Phi_{\rm cl}] \Def \left \{ \> W[J] - \int d^4x \> J(x)\, \Phi_{\rm cl}(x)
\> \right \}_{J = J(\Phi_{\rm cl})} \>. \quad
}
\label{def eff Wirk}
\ee
as  \textcolor{blue}{\bf effective action}. Note that this
is a \textcolor{blue}{\bf Legendre transformation} 
similar as in classical mechanics by which one goes from the Lagrange to the Hamilton function.
One can show (for the 2-point function: See \purpur{\bf Problem \ref{eff Wirk}}) that the 
effective action is the generating functional for the \textcolor{blue}{\bf proper}
or \textcolor{blue}{\bf one-particle-irreducible diagrams}
\be
\Gamma[\Phi_{\rm cl}] \E \sum_{n=0}^{\infty} \> 
\frac{1}{n !} \int d^4x_1 \ldots d^4x_n \> \Gamma^{(n)}(x_1 \ldots x_n)
\> \Phi_{\rm cl}(x_1) \ldots \Phi_{\rm cl}(x_n) \> .
\ee
These are those diagrams which do not collapse into two disconnected lower-order diagrams 
when one cuts a single internal line  (see Fig. \ref{abb: eigentliche Dia}).

\refstepcounter{abb}
\vspace{0.4cm}

\noindent
The effective actions gets an additional meaning from a relation
which is obtained by functional differentiation of the definition  (\ref{def eff Wirk}) 
w.r.t. classical field (see \purpur{\bf Problem \ref{eff Wirk}})      
\be 
\frac{\delta \Gamma [\Phi_{\rm cl}]}{\delta \Phi_{\rm cl}(x)} \E - J(x) \> .
\label{Ableit eff Wirk}
\ee
If the external source vanishes, we then have
\be
\boxed{ 
\qquad \frac{\delta \Gamma [\Phi_{\rm cl}]}{\delta \Phi_{\rm cl}} \Biggr |_{J=0} \E 0
\label{Extremum eff Wirk} \> , \quad
}
\ee
which is the exact counterpart to the equation of motion for a classical field.
The main difference is, however, that  $ \Gamma [\Phi_{\rm cl}] $ contains \textcolor{blue}{\bf all
quantum corrections} explaining the name ``effective action''.
Borrowing from thermodynamics one can interprete $ - W[J] $ as vacuum energy
induced by the external source and understand the effective action
 $\Gamma [\Phi_{\rm cl}] $ as analogon to Gibbs' free energy whose minimum
determines the dynamically most stable state of the system \footnote{See, e.g.
 \meingruen{\bf Peskin \& Schroeder}, ch. 11.3.}. Studying the effective action, therefore
allows us to determine the ground state of the system including all quantum fluctuations.
This is particularly simple if the solutions of Eq. (\ref{Extremum eff Wirk}) are constant
because then one has only to determine the minimum of a {\it function} and not of a functional:
\be
\Phi_{\rm cl}\E {\rm const.} \hspace{0.3cm} \Longrightarrow \hspace{0.3cm}
\Gamma [\Phi_{\rm cl}] \E - {\cal V} \, \cdot \, V_{\rm eff} \left ( \Phi_{\rm cl}
\right ) \> .
\ee
Here $ {\cal V}$ is the space-time volume and $ V_{\rm eff} $ the
\textcolor{blue}{\bf effective potential}.

How can one calculate the effective action? Apart from perturbation theory we also have 
to our disposal the 
\textcolor{blue}{\bf semi-classical expansion} which we already have used several times
in the first two main parts of this lecture.
Also here it is based on the application of the stationary-phase method to the 
functional integral
\be
Z[J] \E \int {\cal D}\Phi \> \exp \left \{ \> \frac{i}{\hbar}
\int d^4x \> \left [ \frac{1}{2} \partial_{\mu} \Phi \partial^{\mu} \Phi
- \frac{1}{2} m^2 \Phi^2 - V(\Phi) + J \Phi \right ] \> \right \} 
\EQ \int {\cal D}\Phi \> e^{i S[\Phi,J]/\hbar} \> .
\label{Z als Pfad}
\ee
The field configuration which is stationary fulfills $ \> \delta S[\Phi,J]/\delta \Phi = 0 \> $,
i.e.
\be
\left ( \Box + m^2 \right ) \Phi_0(x) + V'(\Phi_0) \E J(x) \> ,
\ee
and we assume that for $ \> J = 0 \> $ the solution is $ \> \Phi_0 = 0 \> $ .
In leading order of an expansion in $ \hbar$ the functional integral
(\ref{Z als Pfad}) is given by the value of the integrand for the stationary configuration:     
\be
Z^{(0)}[J] \E \exp \left ( \> \frac{i}{\hbar} S[\Phi_0] + 
\frac{i}{\hbar} \int d^4x J(x) \Phi_0(x) \> \right ) \E 
\exp \left ( \> \frac{i}{\hbar} W^{(0)} \> \right ) \> .
\ee
According to Eq. (\ref{def Phi_cl}) the ``classical field''
is given in this order by
\be
\Phi^{(0)}_{\rm cl}(x) \E \Phi_0(x) + \int d^4y \> J(y) 
\frac{\delta \Phi_0(y)}{\delta J(x)} + \frac{\delta S[\Phi_0]}{\delta J(x)}
\E \Phi_0(x) + \int d^4y \> \frac{\delta S[\Phi_0,J]}{\delta \Phi_0(y)}
\frac{\delta \Phi_0(y)}{\delta J(x)} \> = \Phi_0(x) \> .
\ee
Here the chain rule has been used and the fact that
$ \> S[\Phi,J] \> $ is stationary at $ \> \Phi_0 \> $ . Consequently, we have the
expected result that for $ \> \hbar = 0 \> $ the effective action is identical with
the classical action:
\be
\boxed{
\qquad \Gamma^{(0)}[\Phi_{\rm cl}] \E S[\Phi_{\rm cl}] \> . \quad
}
\ee
As usual, the quantum corrections are generated by fluctuations around the stationary
configuration. Setting 
\be
\Phi(x) \E \Phi_0(x) + \sqrt \hbar \> \phi(x)
\ee
we obtain
\bea
Z[J] \EA Z^{(0)}[J] \cdot \int {\cal D}\phi \> \exp \Biggl \{ \> 
i \int d^4x \> \Bigl [ \frac{1}{2} \partial_{\mu}\phi \partial^{\mu}\phi
- \frac{1}{2} \left ( m^2 + V''(\Phi_0)\right) \phi^2 \non
&& \hspace{3cm} - \sum_{m\ge 3} \hbar^{m/2 -1} \frac{1}{m!} 
V^{(m)}(\Phi_0) \phi^m \> \Bigr ] \> \Biggr \} \> .
\label{hbar Entwickl}
\eea
Obviously the quadratic terms in $ \> \phi \> $ are the leading ones in a systematic 
expansion in powers of $ \> \hbar \> $. We will see that the quadratic fluctuations
lead to one-loop diagrams and it is not difficult to derive that in general the
$ \> \hbar$-expansion is an \textcolor{blue}{\bf expansion in loops}.
\vspace{0.2cm}

\begin{figure}[hbtp]
\vspace*{-7cm}
\bce
\includegraphics[angle=0,scale=0.8]{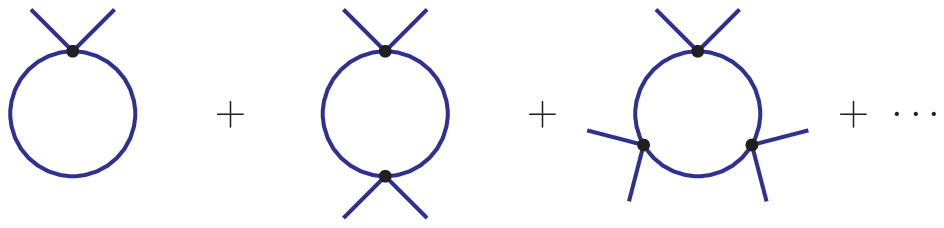}
\label{abb:3.2.1}
\ece
\vspace*{-13cm}
\bce
{\bf Fig. \arabic{abb}} : One-loop correction for the effective action
in the $ \> \Phi^4$-theory.
\ece
\end{figure}

Let us now calculate the one-loop correction to the classical action. From
Eq. (\ref{hbar Entwickl}) we obtain by means of the usual Gaussian integral 
\be
\int {\cal D}\phi \> \exp \Biggl \{ \> - \frac{i}{2}
\int d^4x \> \phi(x) \left [ \> \Box + m^2 + V''(\Phi_0) \> \right] \phi(x)
 \> \Biggr \} \E \frac{\rm const.}{
 \fdet^{1/2} \left ( \> \Box + m^2 + V''(\Phi_0) \> \right)} \> .
\ee
We may write the constant as value of the integral for $ \> V = 0 \> $  as we do not need the
normalization of the generating functional. In this way the quantum correction to the effective action 
also vanishes automatically for vanishing interaction. Employing the trace-representation 
of the determinant (\purpur{\bf Problem \ref{Det Spur}}) we obtain 
\be
Z^{(1)}[J] \E Z^{(0)}[J] \> \cdot \> \exp \left \{ \> -\frac{1}{2} 
{\rm tr} \ln \left [ \> 1 +  \frac{1}{\Box + m^2 - i \, 0^+} 
V''(\Phi_0) \> \right ] \> \right \} \> ,
\ee
or
\be
W^{(1)}[J] \E S[\Phi_0,J] 
+ \frac{1}{2} i \hbar \> {\rm tr} \ln \left [ \> 1 - \Delta_F V''(\Phi_0) \>
\right ] \> .
\ee
For evaluating the effective action we use
$ \> \Phi_{\rm cl} = \Phi_0 + {\cal O}(\hbar) \> $. Moreover, since 
$ \> S[\Phi,J] \> $ is stationary at  $ \> \Phi_0 \> $ we also have
$ \> S[\Phi_{\rm cl},J] = S[\Phi_0,J] +  {\cal O}(\hbar^2) \> $.
By this it follows that
\be
\boxed{
\qquad \Gamma[\Phi_{\rm cl}] \E S[\Phi_{\rm cl}] + \frac{1}{2} i \hbar 
\> {\rm tr} \ln 
\left [ \> 1 - \Delta_F V''(\Phi_{\rm cl}) \> \right ] + 
{\cal O}\left (\hbar^2 \right ) \>. \quad
}
\label{1 Schleif eff W}
\ee
The perturbative expansion of
\bea
&& \frac{1}{2} i \hbar \> {\rm tr} \ln
\left [ \> 1 - \Delta_F V''(\Phi_0) \> \right ]
\E  -i \hbar \sum_{n=1}^{\infty} \frac{1}{2 n} 
{\rm tr} \left [ \Delta_F V''(\Phi_0) \right ]^n \non
\EA -i \hbar \sum_{n=1}^{\infty} \frac{1}{2 n} 
\, \int d^4z_1  \ldots d^4z_n 
\> \Delta_F(z_1-z_2) V''(\Phi_0(z_2)) 
\ldots V''(\Phi_0(z_n)) \Delta_F(z_n-z_1) V''(\Phi_0(z_1)) 
\eea
shows that the additional term includes the contributions of all
one-loop diagrams made up of $n$ propagators \\
$ \> i \Delta_F(z_i-z_{i+1}) \> $ and $n$ vertices 
$ \> - i V''(\Phi_0(z_{i+1})) = - i \lambda \Phi_0^2(z_{i+1})/2 \> $ 
(see Fig. \ref{abb:3.2.1}).

\refstepcounter{abb}

In many cases one only is interested in the effective potential
which we already have introduced above for constant fields. However, it
also can be seen as the first term in an expansion of the effective action
in higher and higher derivatives of the field
\be
\Gamma[\Phi_{\rm cl}] \E \int d^4x \> \left [ \> - V_{\rm eff}
(\Phi_{\rm cl})
+ \frac{1}{2} Z_{\rm eff}(\Phi_{\rm cl}) \>  \partial_{\mu} \Phi_{\rm cl} 
\partial^{\mu} \Phi_{\rm cl} + \ldots \> \right ] \> .
\label{def eff pot}
\ee
For the $ \> \Phi^4$-theory we obtain with $ \Phi_{\rm cl} = $ const. 
\be
V_{\rm eff}(\Phi_{\rm cl}) \E \frac{1}{2} m^2 \Phi_{\rm cl}^2
+ \frac{\lambda}{4 !} \Phi_{\rm cl}^4 - \frac{i \hbar}{2} \int 
\frac{d^4k}{(2 \pi)^4} \> \ln \left [ \> 1 - \frac{\lambda}{2} 
\frac{\Phi_{\rm cl}^2}{k^2 - m^2 + i \, 0^+}
\> \right ] + {\cal O} \left ( \hbar^2 \right ) \> .
\ee
The loop integral is divergent and must be made finite by a counter term
of the form
\be
V_{\rm counter term}(\Phi_{\rm cl}) \E \frac{A}{2} \Phi_{\rm cl}^2    
+ \frac{B}{4 !}\Phi_{\rm cl}^4 \> .
\ee
The constants $ \> A, B \> $ can be determined by requiring that the physical mass
and the physical coupling constant are measured at $ \> \Phi_{\rm cl}=0 \> $ :
\be
\frac{\partial^2 V_{\rm eff}}{\partial \Phi_{\rm cl}^2}
\Biggr|_{\Phi_{\rm cl}=0} \E m_{\rm phys}^2 \> , \hspace{0.5cm}
\frac{\partial^4 V_{\rm eff}}{\partial \Phi_{\rm cl}^4}
\Biggr|_{\Phi_{\rm cl}=0} \E \lambda_{\rm phys} \> .
\ee
A particularly interesting case happens if one starts from an originally massless theory:
$ \> m^2 = 0 \> $. However, one then has to impose the renormalization condition that the
coupling constant has a fixed value at, say  $ \> \Phi_{\rm cl} = M \> $
\be 
\lambda_M \E \frac{\partial^4 V_{\rm eff}}
{\partial \Phi_{\rm cl}^4}\Biggr|_{\Phi_{\rm cl}=M} \> .
\ee
Then one obtains \cite{CoWei}
\be 
\boxed{
\qquad V_{\rm eff}(\Phi_{\rm cl}) \E \frac{\lambda_M}{4 !} \Phi_{\rm cl}^4
+ \frac{\hbar \lambda_M^2}{(16 \pi)^2}  \Phi_{\rm cl}^4 \> \left ( \> 
\ln \frac{\Phi_{\rm cl}^2}{M^2} - \frac{25}{16} \> \right ) + \ldots \> . \quad
}
\label{CW eff pot}
\ee
As the last term is negative for small $ \> \Phi_{\rm cl} \> $ the minimum is shifted from
the origin  $ \>\Phi_{\rm cl} = 0 \> $, i.e. the quantum corrections of the one-loop correction
have generated a mass \textcolor{blue}{ \bf ``spontanously''}. Due to Eq. (\ref{Ableit eff Wirk}) 
this means that a vacuum expectation value of the field $ \> \Phi^{\rm min}_{\rm cl} \> $ 
exists even for vanishing external source $J(x)$. However, as one can verify, this value 
lies outside the validity of the semi-classical expansion (the correction is as large as the classical
contribution) so that this mechanism of {\bf dynamical mass generation} remains unsettled.
\vspace{0.5cm}

\subsection{\textcolor{blue}{Quantization of Gauge Theories}}
\label{sec3: Eichtheo}

The free Lagrangian (\ref{L0 fermion}) for electrons is invariant under the 
\textcolor{blue}{\bf global gauge transformation}
\be
\cy{\psi}(x) \To e^{- i e \Theta} \> \cy{\psi}(x) \> ,
\hspace{1cm} \cy{\bar{\psi}}(x)\To\cy{\bar{\psi}}(x) \> 
 e^{i e \Theta} \> .
\label{Eich global}
\ee
Here $ \> \Theta \> $ is a constant parameter and the electric charge
$ \> e \> $ of the particles has been taken out explicitly for convenience.
One can extend the gauge transformation
(\ref{Eich global}) to a \textcolor{blue}{\bf local} one
\be
\boxed{
\qquad \cy{\psi}(x) \To e^{- i e \Theta(x)} \> \cy{\psi}(x) \> ,
\hspace{1cm} \cy{\bar{\psi}}(x)\To\cy{\bar{\psi}}(x) \> 
 e^{i e \Theta(x)} \> , \quad 
}
\label{Eich lokal}
\ee
if one introduces a ``\textcolor{blue}{\bf gauge field}'' $ \> A_{\mu}(x) \> $ which compensates
the additional term from the derivative in $ \> {\cal L}_0 \> $ :
\be
{\cal L}_0 \To{\cal L} \E 
\cy{\bar{\psi}}(x) \left ( \> i \dslash - m 
- e \Aslash(x) \> \right ) \cy{\psi}(x) \> .
\ee
To achieve that  $ \> A_{\mu}(x) \> $ obviously has to transform like
\be
A_{\mu}(x) \To A_{\mu}(x) + \partial_{\mu} \Theta(x) \> .
\label{Eich A abelsch}
\ee
\vspace{0.2cm}

If one adds an invariant kinetic energy term for the ``photon field'' 
$ \> A_{\mu} \> $ which at most is quadratic in the derivatives, then the Lagrangian for
\rot{\bf Q}\blau{uantum}\rot{\bf E}\blau{lectro-}\rot{\bf D}\blau{ynamics} 
(\rot{\bf QED}) is complete:
\bce

\fcolorbox{blue}{white}{\parbox{12cm}
{
\bea
{\cal L}_{\rm QED} \EA - \frac{1}{4} F_{\mu \nu} F^{\mu \nu} + 
\cy{\bar{\psi}} \left ( \> i \diffslash - m
- e \Aslash \> \right ) \cy{\psi} \> , \hspace{1cm} F_{\mu \nu} \Def \partial_{\mu}
A_{\nu} - \partial_{\nu} A_{\mu} \> . \hspace{2.8cm}
\label{L QED}
\eea
}}
\ece

\vspace{0.1cm}

\noindent
Note that the photons have to be \textcolor{blue}{ massless} -- a mass term of the form
$ \> A_{\mu} A^{\mu} \> $ would violate the gauge invariance. The interaction between electrons 
and photons has thus been generated from ``\blau{\bf gauging}'' the phase transformation
(\ref{Eich global}). As mathematically the group of these transformations constitutes an U(1) group 
one says that QED is an (abelian) U(1) gauge theory.

\vspace{0.2cm}

This remarkable gauge principle can be extended to
\textcolor{blue}{ non-abelian gauge groups}: Let the Lie algebra of a (compact, semi-simple) Lie group
$ G \> $ \footnote{For an explanation of these concepts see, e.g., 
http://de.wikipedia.org/wiki/Liealgebra or introductions to group theory for physicists like 
 {\bf \{Georgi\}}. Here all this mostly is too ``heavy artillery'' as we only will consider the 
group $ SU(N) $  -- the group of unitary transformations $ U $ of $ N $-dimensional vectors with 
$ \det U = 1 $ (i.e. with traceless generators) -- in the following.}
be generated by the (hermitean) generators $ \> T^a \> $ satisfying
\be
\boxed{
\quad \left [ \> T^a, T^b \> \right ] \E i f^{a b c } \, T^c \> , \qquad a = 1 \ldots n \>
}
\ee
(identical indices are to be summed over). The constants
$ \> f^{a b c } \> $ are the structure constants which characterize the Lie algebra of the group G.
As known from the treatment of angular momentum in quantum mechanics the generators can be realized in different (matrix) representations: Most
important for physical applications are the \blau{\bf fundamental representation} as smallest possible
representation (according to which the fermions transform) and the \blau{\bf adjoint representation} which is defined by
\be
\lrp T^a \rrp_{bc} \E - i \, f^{abc} \> .
\label{adjung}
\ee
Both the field tensor $ F_{\mu \nu} $ as well as the gauge field
must transform under the adjoint representation of the gauge group
(see, e.g. \meingruen{\bf Das}, ch. 12.2).
We have
\be
{\rm tr} \, \lrp T^a T^b \rrp \E C_2 \, \delta_{ab} \> ,
\ee
where the constant $ C_2 $ only depends on the representation (usually
$ C_2 = 1/2 $ is chosen for the fundamental representation).

\vspace{0.8cm}

\noindent
{\bf Examples :}
\bit
\item[a)] $ G = SU(2) \> : \> f^{a b c } \EQ \epsilon^{a b c} \> ,
\> $
(total antisymmetric tensor, $ \epsilon^{1 2 3} = 1 $). In the fundamental
representation the generators are given by the {\bf Pauli matrices}: $ \> T^a = \tau^a/2 \> $
with
\be
\tau_1 \E \begin{pmatrix} 0 & 1 \\ 1 & 0 \end{pmatrix} \> , \qquad 
\tau_2 \E \begin{pmatrix} 0 & -i \\ i & 0 \end{pmatrix} \> ,\qquad 
\tau_3 \E \begin{pmatrix} 1 & 0 \\ 0 & -1 \end{pmatrix} \> .
\ee

\item[b)] $ G = SU(3) \> $ : The gauging of this group leads to 
\rot{\bf Q}\blau{uantum}\rot{\bf C}\blau{hromo}\rot{\bf D}\blau{ynamics} 
(\rot{\bf QCD}), the theory of strong interactions of quarks and gluons.
In the fundamental representation the generators are given by the
{\bf Gell-Mann matrices} : $ \> T^a = \lambda^a/2 \> \> , 
\> a = 1 \ldots 8 \> $ with
\bea
\lambda_1 \EA \begin{pmatrix} 0 & 1 & 0 \\ 1 & 0 & 0 \\ 0 & 0 & 0\end{pmatrix} \, ,\> 
\lambda_2 \E  \begin{pmatrix} 0 & -i & 0 \\ i & 0 & 0 \\ 0 & 0 & 0 \end{pmatrix} \, ,\>
\lambda_3 \E  \begin{pmatrix} 1 & 0 & 0 \\ 0 & -1 & 0 \\ 0 & 0 & 0\end{pmatrix} \, ,\>
\lambda_4 \E  \begin{pmatrix} 0 & 0 & 1 \\ 0 & 0 & 0 \\ 1 & 0 & 0\end{pmatrix} \non
\lambda_5 \EA \begin{pmatrix} 0 & 0 & -i \\ 0 & 0 & 0 \\ i & 0 & 0\end{pmatrix} \, ,\>
\lambda_6 \E  \begin{pmatrix} 0 & 0 & 0 \\ 0 & 0 & 1 \\ 0 & 1 & 0 \end{pmatrix} \, ,\>
\lambda_7 \E  \begin{pmatrix} 0 & 0 & 0 \\ 0 & 0 & -i \\ 0 & i & 0 \end{pmatrix} \, ,\>
\lambda_8 =  \frac{1}{\sqrt{3}} \begin{pmatrix} 1 & 0 & 0 \\ 0 & 1 & 0 \\ 0 & 0 & -2  
\end{pmatrix}.
\label{Gellmann}
\eea
The structure constants $ f^{abc} $ are totally antisymmetric and the non-vanishing 
elements have the values (as can be seen from the explicit representation \eqref{Gellmann}) 
\be
f^{123} \E 1 , \>   f^{147} \E f^{246} \E  f^{257} \E 
f^{345} \E \frac{1}{2} , \>  f^{156} \E f^{367} \E -\frac{1}{2} , \> f^{458} \E f^{678} \E 
\sqrt{\frac{3}{2}} \> .
\ee

\eit 
\vspace{0.3cm}

\noindent
In general one has
\be
n \E N^2 - 1 \hspace{1.5cm} \mbox{generators for} \> \>  SU(N) \> .
\ee
The fermion field now appears in $ N $ species, for instance,  there exist 2 states of the nucleon 
-- proton and neutron -- for isospin SU(2) and 3 colored quarks for color SU(3).
It transforms under the fundamental representation
\be 
\boxed{                                                    
\qquad \cy{\psi}(x) \E \exp \left ( - i g \Theta^a T^a \right ) \> \cy{\psi}'(x)\> , \quad  
}
\hspace{1cm} \cy{\psi}(x) \E \left (\begin{array}{c}
                                              \cy{\psi_1}(x) \\
                                              \vdots \\
                                              \cy{\psi_N}(x) \end{array} \right)
\> ,
\label{Eich nichtabelsch global}
\ee
where $ \> T^a \> $ is a matrix representation of the generators. The free Lagrangian
(\ref{L0 fermion}) (in which $ m $ now is  a mass matrix)   
is again invariant under the global transformation
(\ref{Eich nichtabelsch global}) but no longer under the  \blau{\bf local
gauge transformation}
\be
\cy{\psi}(x) \E \exp \left [ \>  - i g \Theta^a(x) T^a \> \right ] \>  
\cy{\psi}'(x) \> .
\label{Eich nichtabelsch lokal}
\ee
For compensation we need a gauge field $ \> A_{\mu}^a(x) \> ,
\> a = 1 \ldots n \> $ which transforms like
\be
A_{\mu}^a(x) \E A_{\mu}^{a '}(x) + g \, f^{a b c} \Theta^b(x)
A_{\mu}^{c '}(x) + \partial_{\mu} \Theta^a(x) + {\cal O}(\Theta^2)
\label{Eich A nichtabelsch}
\ee
under an infinitesimal gauge transformation.
\vspace{0.6cm}

\renewcommand{\baselinestretch}{0.9}
\scriptsize
\begin{subequations}
\noindent
{\bf Proof:} We write the gauge transformation \eqref{Eich nichtabelsch lokal} as
\be
\cy{\psi}(x) \E U(x) \, \cy{\psi}'(x) \qquad  {\rm with} \quad U(x) \E e^{-i g \Theta^a(x) T^a} \> ,
\quad U^{\dagger}(x) \, U(x) \E U(x) \, U^{\dagger}(x) \E 1 
\label{Eich U}
\ee
and introduce again a (matrix-valued) gauge field $ \> {\cal A}_{\mu}(x) \> $ which 
should compensate the additional terms which have appeared by the local gauge transformation
of the free fermionic Lagrangian
\bea
{\cal L} \lrp \cy{\bar \psi}, \cy{\psi},{\cal A}\rrp  \EA \cy{\bar \psi} \lsp \gamma^{\mu} \, 
\lrp i \partial_{\mu} - g  {\cal A}_{\mu} \rrp - m \rsp \cy{\psi} 
\E \cy{\bar \psi}' U^{\dagger}(x) \Bigl \{ \gamma^{\mu} \, \lsp i U(x) \partial_{\mu} + i \lrp 
\partial_{\mu} U(x) \rrp
- g  {\cal A}_{\mu} U(x) \rsp - m U(x) \Bigr \} \cy{\psi}' \non
\EA {\cal L}\lrp \cy{\bar \psi}', \cy{\psi}',{\cal A}'\rrp +  \cy{\bar \psi}'  \gamma^{\mu} \lsp i \lrp
U^{\dagger}\partial_{\mu} U(x) \rrp + g  {\cal A}_{\mu}' - g  U^{\dagger}(x)  {\cal A}_{\mu} U(x) 
\rsp \cy{\psi}'\> .
\eea
To cancel the unwanted terms the vector potential $ \>  {\cal A}_{\mu} \> $ must thus transform as
\be
{\cal A}_{\mu}  \E U(x) {\cal A}_{\mu}' U^{\dagger}(x) +  \frac{i}{g} \, 
\lrp \partial_{\mu} U(x) \rrp \,U^{\dagger}(x) \> .
\label{Eich A nichtabelsch 2}
\ee
For the abelian theory this agrees with Eq. \eqref{Eich A abelsch} (there the coupling constant
has been denoted by $e$ ). In an non-abelian theory we write
\be
{\cal A}_{\mu}(x) \deF A_{\mu}^a(x) \, T^a
\ee
and restrict ourselves to infinitesimal transformations (which is always possible in Lie algebras)
$ \> U(x) \E 1 - i g \Theta^a(x) \, T^a + {\cal O}\lrp \Theta^2 \rrp \> $ . Then we obtain
from Eq. \eqref{Eich A nichtabelsch 2}
\bea
A_{\mu}^{a } T^a \EA A_{\mu}^{a '} T^a - i g \Theta^b(x) \, T^b \,  A_{\mu}^{a '} T^a
+ i g A_{\mu}^{a '} T^a \,  \Theta^b(x) \, T^b + \partial_{\mu} \Theta^a(x) \, T^a \non
\EA  A_{\mu}^{a '} T^a  + \partial_{\mu} \Theta^a(x) \, T^a - i g  \,  \Theta^b(x) \, 
\underbrace{\lsp T^b, T^a\rsp}_{=if^{bac} T^c} \, A_{\mu}^{a '} \E
A_{\mu}^{a '} T^a  + \partial_{\mu} \Theta^a(x) \, T^a + g  f^{bac} \Theta^b(x) \, A_{\mu}^{a '} \, T^c \> .
\eea
If we exchange $ \> a \leftrightarrow c \> $ in the last term and use $\> f^{bca} = f^{abc} \>  $, 
we get Eq. \eqref{Eich A nichtabelsch}.

\end{subequations}
\renewcommand{\baselinestretch}{1.2}
\normalsize
\vspace{0.5cm}

\noindent
Hence, 
if we add an invariant, kinetic term for the gauge fields then the Lagrange density of a non-abelian
gauge theory is given by
\bce

\fcolorbox{blue}{white}{\parbox{8cm}
{
\bea
\qquad {\cal L}_{\rm non-abelian} \E - \frac{1}{4} F^a_{\mu \nu} F^{a \> \mu \nu}
+ \cy{\bar{\psi}} \left ( i \ddslash - m \right ) \cy{\psi} \no
\eea
}}
\ece
\vspace*{-2cm}

\bea
\label{L nichtabelsch}
\eea
\vspace{0.2cm}


\noindent
where
\be
\boxed{
\qquad D_{\mu}(x) \E \partial_{\mu} + i g \, T^a  A_{\mu}^a(x)
}
\label{kov ableit}
\ee
is the  \blau{\bf  covariant derivative}. In components it reads
\be
\lrp D_{\mu}(x)\rrp_{jk}  \E \partial_{\mu} \, \delta_{jk} + i g \lrp T^a \rrp_{jk} A_{\mu}^a(x)\> ,
\ee
where $ \, \lrp T^a \rrp_{ij} \, $ is the matrix represention of the generator $ T^a $ in the 
fundamentalen representation. Note that e.g. in $SU(N): j,k = 1, \ldots N $ while
$ \, a,b \, $ run from $ 1 $ to $ n = N^2-1 $. In the adjoint representation we have
\be
\lrp D_{\mu}(x) \rrp^{ab} \E \partial_{\mu} \, \delta^{ab} + g f^{acb} \, A_{\mu}^c(x)    
\label{kov ableit adjung}
\ee
and therefore we can write the behaviour of the potential \eqref{Eich A nichtabelsch}
under a gauge transformation simply as
\be
A_{\mu}^a(x) \To A_{\mu}^a(x) + \delta  A_{\mu}^a(x)\> , \quad  \delta  A_{\mu}^a(x) \E 
\lrp D_{\mu}(x) \rrp^{ab}\, \Theta^b \EQ \lrp D_{\mu}(x)\, \Theta \rrp^a  \> .
\label{delta A}
\ee
\vspace{0.1cm}


\noindent
The field tensor can be determined from the relation
\be
\lsp D_{\mu}, D_{\nu} \rsp \E i g \, {\cal F}_{\mu \nu} \E i g \, F_{\mu \nu}^a T^a
\label{F aus D}
\ee
with the result
\bce
\vspace{0.2cm}

\fcolorbox{blue}{white}{\parbox{7cm}
{
\bea 
F^a_{\mu \nu} \EA \partial_{\mu} A_{\nu}^a - \partial_{\nu} A_{\mu}^a
- g f^{a b c} A_{\mu}^b A_{\nu}^c \no \> .
\eea
}}
\ece
\vspace{-3cm}

\bea
\label{feldstaerke nichtabelsch}
\eea
\vspace{1.5cm}


\noindent
It is now {\bf non-linear} in the vector potentials -- therefore the gauge bosons carry 
"charges" (e.g. color) and interact with each other, in contrast to the abelian case.
\vspace{0.6cm}

\renewcommand{\baselinestretch}{0.9}
\scriptsize
\begin{subequations}
\noindent
That the kinetic term for the gauge bosons is gauge invariant can be seen
from the behaviour of the covariant derivative under gauge transformations:
This quantity is constructed in such a way that 
$ \> \cy{\bar \psi} ( i \dddslash - m ) \cy{\psi} \> $ is invariant under the gauge trnasformation
\eqref{Eich U} , i.e.
\be
D_{\mu}(x) \To U(x) \, D_{\mu}(x) \, U^{\dagger}(x) \> . 
\ee
Then Eq. \eqref{F aus D} transforms as
\be 
{\cal F}_{\mu \nu}(x) \To - \frac{i}{g} \, \lsp U(x) \, D_{\mu}(x) \, U^{\dagger}(x) \, , \, 
U(x) \, D_{\nu}(x) \, U^{\dagger} (x) \rsp \E  - \frac{i}{g} \,  U(x) \, \Bigl [ \,   D_{\mu}(x) ,  
D_{\nu}(x) \, \Bigr ]
\, U^{\dagger}(x) \EQ  U(x) \, {\cal F}_{\mu\ nu}(x) \, U^{\dagger}(x)
\ee
and, indeed, the kinetic term for the gauge bosons
\be
- \frac{1}{2} \, {\rm tr} \lrp {\cal F}_{\mu \nu} \, {\cal F}^{\mu \nu} \rrp \To
- \frac{1}{2} \, {\rm tr} \lrp U(x) \,  {\cal F}_{\mu \nu} \, {\cal F}^{\mu \nu} \, U^{\dagger}(x) \rrp
\E - \frac{1}{2} \, {\rm tr} \lrp {\cal F}_{\mu \nu} \, {\cal F}^{\mu \nu} \rrp \E - \frac{1}{4}
\, F_{\mu \nu}^a \,  F^{\mu \nu \, a}
\ee
does not change (in the final step it was used that in a trace one can move factors cyclically).
\end{subequations}
\renewcommand{\baselinestretch}{1.2}
\normalsize

\vspace{0.5cm}
\noindent
As already mentioned we encounter difficulties when trying to quantize gauge
theories because gauge fields are physically equivalent if they
are connected by a gauge transformation (\ref{Eich A nichtabelsch}). As this is 
not tied to the existence of fermions we will first consider the pure 
``\textcolor{blue}{\bf Yang-Mills theory}'' without fermions.
Naively the generating functional would be
\be
Z^{\rm YM}[J] \E \int {\cal D}A_{\mu}^a(x) \> \exp \left [ i \int d^4 x 
\left ( {\cal L} + J_{\mu}^a A^{a \> \mu} \right ) \right ] \> .
\ee
In particular, for the free case ($ \> g = 0 \> $) the field tensor
equals the abelian one and after an integration by parts we have
\be
Z^{\rm YM}_0[J] \E \int {\cal D}A_{\mu}^a(x) \> \exp \left \{ \> i \int d^4 x
\left [ \> \frac{1}{2} A_{\mu}^a \left (g^{\mu \nu} \Box - \partial^{\mu}
\partial^{\nu} \right ) A_{\nu}^a + J_{\mu}^a A^{a \> \mu} \> \right ] 
\> \right \} \> .
\ee
Following the usual scheme of completing the square we would obtain
\be
Z^{\rm YM}_0[J] \E \frac{\rm const.}{\fdet^{1/2} K} \> \exp \left \{
- \frac{i}{2} \int d^4x \> d^4y \> J^{a \> \mu}(x) \left ( K^{-1} 
\right )^{a b}_{\mu \nu} (x,y) J^{b \> \nu}(y) \> \right \} 
\ee
where
\be
K^{a b}_{\mu \nu} (x,y) \E \delta^{a b} \> \delta^4(x-y) \> \left ( 
g_{\mu \nu} \Box - \partial_{\mu} \partial_{\nu} \right ) 
\ee
has to be inverted. However,  $ K $ does  {\bf not} have an inverse: 
There exist eigenfunctions $ \> k_{\nu} \exp ( i k \cdot x) \> $ with eigenvalue
$ 0 $ ! (Equivalent aspects of this fact are that $ \> K \> $ is a projection
operator which projects out the transverse degrees of freedom of the gauge field
or that the spin-1 propagator \eqref{vec prop} diverges for $ m = 0 $).          
Therefore $ \> \fdet \, K \> $ vanishes and the previous approach to derive 
the Feynman rules is not applicable.

This difficulty arises from the fact that we have summed over all gauge-field 
configurations in the path integral, also over those which are connected by gauge 
transformations. These are redundant, i.e. unphysical. We have to take out the
(infinite) contribution of such configurations from the path integral, i.e.
in order to quantize a gauge theory one has to \blau{\bf fix the gauge}.
This can be done by a condition
\be
{\cal H}^a \left ( A_{\mu}^{\Theta} \right ) \E h^a(x)    
\label{Eichfixierung}
\ee
where the index $ \> \Theta \> $ shall indicate the chosen gauge with gauge parameter 
$ \> \Theta \> $ . The decisive step now is to multiply the 
path integral for the generating functional with a `` 1 ''
in the form
\be
\boxed{
 \qquad 1 \E \Delta_{FP} (A) \int{\cal D} \Theta(x) \> \delta \left [ \> 
{\cal H} \left ( A_{\mu}^{\Theta}(x) \right ) - h(x) \> \right ] \qquad
}
\ee
where 
\be
{\cal D} \Theta(x) \> = \prod_{a=1}^n \prod_j d \Theta^a(x_j)
\ee
is the invariant functional measure in the group space (for $SU(2)$ e.g.
this means functional integration over the 3 Euler angles by which the group
can be parametrized) and                          
\be
\delta \lsp {\cal H}(x) - h(x) \rsp \E  \prod_{a=1}^n \prod_j \, \delta \lrp {\cal H}^a(x_j) - h^a(x_j) \rrp 
\ee
is the functional $\> \delta$-function. The
\textcolor{blue}{\bf Faddeev-Popov} factor $ \> \Delta_{FP} \> $  is given by
\be
\Delta_{FP} (A) \E \fdet \left [ \> \frac{\delta {\cal H}
(A_{\mu}^{\Theta})}{\delta \Theta} \> \right ] \> ,
\ee 
where the determinant also has to be taken in group space.
\vspace{0.8cm}

\renewcommand{\baselinestretch}{0.9}
\scriptsize
\refstepcounter{tief}
\noindent
\blau{\bf Detail \arabic{tief}:} {\bf A Two-dimensional Example 
\footnote{\meingruen{\bf Cheng \& Li, p. 250 - 252}}}\\

\begin{subequations}
\vspace{0.1cm}

\noindent
Let us assume that in the integral
\bea
Z \E \int d^2 r \> \exp [i S(\fr) ]
\label{Z Beispiel}
\eea
the "action"  $S$ only dependes on the radius  $r$ of the vector $\fr = (r,\varphi) $.
Then the integrand is invariant under a rotation  $ \fr \to 
\fr^{\Theta} = (r, \varphi + \Theta) $ by the fixed angle $\Theta$. 
Of course, normally one would use this invariance immediately to perform
the $\varphi$-integration which gives a factor $2 \pi $. But let's assume
that we are unable to identify the "relevant degree of freedom"
(here the radius $r$) -- nevertheless it is possible to factor out the contribution
of the irrelevant degree of freedom  ($\varphi$) in the integral: We fix the rotation
angle  $\Theta$ by the constraint
\be
{\cal H}\left ( \fr^{\Theta} \right ) \E h \E {\rm const.}   
\ee
and integrate over all rotation angles. Due to the relation
$\delta (F(x) - h) = \sum_i \delta (x - x_i)/|F'(x_i)| \> $ where
$F(x_i) = h $, we thus have
\be
1 \E \Delta(\fr) \, \int d \Theta \> \delta \left ( {\cal H} 
\left ( \fr^{\Theta} \right ) - h \right ) \> , \>  \> {\rm with} \> \> \> \> 
\Delta(\fr) \E \frac{ \partial {\cal H} \left ( \fr^{\Theta} \right )}
{\partial \Theta} \Biggr |_{\Theta = \Theta_0} \> , 
\ee
provided that the equation ${\cal H}\left ( \fr^{\Theta} \right ) = h \> $ has just
{\bf one} solution $\Theta_0$. Hence we obtain for the integral 
in Eq. (\ref{Z Beispiel})
\be
Z \E \int d\Theta \> \underbrace{ \int d^2 r \> 
e^{i S(r)} \, \Delta(\fr) \, \delta \left ( {\cal H} 
\left ( \fr^{\Theta} \right ) - h \right )}_{\Def Z^{\Theta}} \>,
\ee 
where  $Z^{\Theta}$ now is independent of $\Theta $. This is shown by rotation with another
angle $\Theta'$:
\be
Z^{\Theta'} \E \int d^2 r \> 
e^{i S(r)} \, \Delta(\fr) \, \delta \left ( {\cal H} 
\left ( \fr^{\Theta'} \right ) - h \right ) \> .
\ee
By construction the Faddeev-Popov factor is independent of the rotation angle
as is the "action" $S$. Thus, if we choose 
$\fr^{\Theta'} = (r,\varphi + \Theta') \deF (r,\varphi') = \fr' $ 
as new integration variable, we get
\be 
Z^{\Theta'} \E \int d^2 r' \> 
e^{i S(r')} \, \Delta({\bf r'}) \, \delta \left ( {\cal H} 
\left ( {\bf r'} \right ) - g \right ) \EQ 
Z^{\Theta = 0} 
\ee
and we can do the angle integration, i.e. we can factorize the contribution 
of the irrelevant degrees of freedom
\be
Z \E 2 \pi \, \cdot \, Z^{\Theta = 0}  \> . 
\ee
The choice of the constraint  ${\cal H}$ is as arbitrary as is the choice
of the constant $ h $; we only have to make sure that the rotation angle $\Theta$
is fixed unambigously. A (trivial) example is 
${\cal H}(\fr^{\Theta}) = \varphi + \Theta \stackrel{!}{=} h $, which
fixes the polar angle for all $r$ and leads to $\Delta(\fr) = 1 $ .

\end{subequations}
\renewcommand{\baselinestretch}{1.2}
\normalsize

\vspace{0.4cm}

\noindent
One therefore has assumed additionally that for each  $ \> \Theta^a \> $ there exists 
precisely one gauge field  $ \> A^a \> $ which fulfills the gauge condition (\ref{Eichfixierung}) 
\footnote{Non-perturbatively this is not the case in general, a phenomenon called 
{\bf Gribov ambiguity.}}.   
If we now perform a gauge transformation in the path-integral representation of the generating
functionals
\be
Z^{\rm YM}_0[J] \E \int D \Theta \int {\cal D}A^{\Theta} \> 
\Delta_{FP}(A^{\Theta}) \>
\exp \left [ \>  i S_0[A^{\Theta}] + i (J,A^{\Theta}) \> \right ] 
\delta \left [ \> {\cal H}(A^{\Theta}) -h(x) \> \right ]
\ee
then the action $ \> S_0[A] \> $ is invariant 
as well as the Faddeev-Popov factor $ \> \Delta_{FP} \> $ and the integration measure.
If we only consider gauge-invariant quantities then also in the source term
\be
\left ( J, A^{\Theta} \right ) \EQ \int d^4x \> J^{a \> \mu}(x) 
A_{\mu}^{\Theta \> a}(x)
\ee
we can replace $ \> A^{\Theta} \> $ by $ \> A^{\Theta'} \> $ \footnote{$(J,A) $ is
not gauge-invariant. Consequently, also Green functions are not gauge-independent but
only $S$-matrix elements.}. By re-gauging we thus obtain
\bea
Z^{\rm YM}_0[J] \EA \int D \Theta \int {\cal D}A^{\Theta'} \> 
\Delta_{FP}(A^{\Theta'}) \>
\exp \left ( i S_0[A^{\Theta'}] + i (J,A^{\Theta'}) \right ) 
\delta \left [ {\cal H}(A^{\Theta'}) - h(x) \right ] \non
\EA {\rm const.} 
\int {\cal D}A^{\Theta'} \>
\Delta_{FP}(A^{\Theta'}) \>
\exp \left ( i S_0[A^{\Theta'}] + i (J,A^{\Theta'}) \right )
\delta \left [ {\cal H}(A^{\Theta'}) - h(x) \right ] \> ,
\eea
as now the integration over the gauge parametrs can be performed. This only produces 
an (infinite) factor which cancels when calculating Green functions.
In this way we have integrated over all configurations which only 
are produced by a gauge transformation.

As the functions $ \> h^a(x) \> $ in the gauge fixing (\ref{Eichfixierung})
are arbitrary, one may functionally integrate over them with the weight
\be
\exp \left ( - \frac{i}{2 \lambda} \int d^4x \> h^a(x) \, h^a(x)\right )
\ee
where  $ \lambda$ is an arbitrary parameter. In this way the gauge fixing 
becomes a part of the Lagrangian and one obtains

\be
\boxed{
\qquad Z^{\rm YM}_0[J] \E {\rm const.} \int {\cal D} A \> \fdet \left [ \> 
\frac{\delta {\cal H}_a(A)}{\delta \Theta} \> \right ]\> \exp \left [
\> i \int d^4x \> \left ( {\cal L}_0 - \frac{1}{2 \lambda} {\cal H}^2(A)
+ J^{a \> \mu} A_{\mu}^a \right ) \> \right ] \> . \quad
}
\ee
\vspace{0.2cm}

\noindent
Exactly the same purpose serves the 
representation of the determinant (of the Faddeev-Popov factor) 
as an integral over \textcolor{blue}{\bf fictitious, anticommuting}
fields $ \> \cy{\chi^a}(x), \cy{\bar{\chi}^a}(x) \> $ :
\be
\fdet \left [ \>\frac{\delta {\cal H}_(A)}{\delta \Theta} \> 
\right ]\E \int {\cal D}\cy{\bar{\chi}}(x) \> {\cal D}\cy{\chi}(x) \> 
\exp \left [ \> i \int d^4x \, d^4y \> \cy{\bar{\chi}^a}(x) K_{a b}(x,y)   
\cy{\chi^b}(y) \> \right ]
\> ,
\label{FP Geister}
\ee
where
\be
K_{a b}(x,y) \E \frac{\delta {\cal H}_a(A(x))}{\delta \Theta_b(y)} \> .
\label{FP Kern}
\ee
These are the``\textcolor{blue}{\bf Faddeev-Popov ghosts}'' -- scalar fields anticommuting 
like fermionic fields.

How does one determine the kernel $ \> K \> $ in Eq. (\ref{FP Kern}) ? 
We know how the gauge field transforms under an infinitesimal gauge transformation (see Eq. 
(\ref{Eich A nichtabelsch})). If, for simplicity, we only consider {\bf linear 
gauge fixings}
\be
{\cal H}^a(A) \E {\cal H}^{\mu} \> A_{\mu}^a(x)
\ee
(where $ {\cal H}^{\mu} $ could also be an operator), we obtain
\be
K_{a b}(x,y) \E {\cal H}^{\mu} \left ( f^{a b c} A_{\mu}^c(x)
+ \delta_{a b} \frac{1}{g} \partial_{\mu} \right ) \, \delta^{(4)} (x-y) \E \frac{1}{g} \,  {\cal H}^{\mu}
\, \left (D_{\mu}(x) \right )^{ab} \,  \delta^{(4)} (x-y)
\> ,
\ee
where  $ D_{\mu}(x) $ is the covariant derivative -- now in the adjoint representation
\eqref{kov ableit adjung}. We can take out the factor $ \>g^{-1} \> $ from  $ \> K_{a b} \> $ 
by a re-definition of the ghost fields and thus find
\bce
\vspace{0.2cm}

\fcolorbox{blue}{white}{\parbox{14cm}
{
\bea
Z^{\rm YM}_0[J] \EA {\rm const.} \int {\cal D}A \> {\cal D} \cy{\bar{\chi}} \> 
{\cal D}\cy{\chi} \> \exp \left \{ \> i \int d^4 x \left [ {\cal L}_0 
- \frac{1}{2 \lambda} \left ( {\cal H}_a^{\mu} A_{\mu}^a \right )^2 \> 
\right ]  \> \right \} \non
&& \hspace{0.5cm} \cdot \>\exp \left \{ \> i \int d^4 x \left [ \> 
J^{a \> \mu} A_{\mu}^a +                         
\cy{\bar{\chi}^a} {\cal H}^{\mu} \, \left (D_{\mu}(x) \right )^{ab} \, 
\cy{\chi^b}\> \right ] \> \right \} \> . \no
\eea
}}
\ece
\vspace{-1.8cm}

\bea
\label{Z0 mit Geistern}
\eea
\vspace{0.4cm}


\noindent
The covariant derivative in the last exponential function also contains
a term of order  $g$ which rather is a part of the interaction and therefore
(for getting the free propagators) can be replaced by $ \partial_{\mu} \, \delta_{ab} $.
\vspace{1cm}

\noindent
{\bf Notes :}
\bit

\item[a)] In \textcolor{blue}{\bf abelian  theories} (like \rot{\bf QED}), in which the
structure constants  are $ \> f^{a b c} = 0 \> $ the ghosts do not couple to the gauge
field and therefore are not relevant.

\item[b)] In a \textcolor{blue}{\bf non-abelian theory} (like
\rot{\bf QCD}) they couple in general to the gauge bosons (the gluons)
but only in $ \> {\cal O}(g) \> $.

\item[c)] In a covariant gauge
\be
{\cal H}_{\mu} \E \partial_{\mu}
\ee
we now obtain after an integration by part for the {\bf free generating      
functional} (the ghosts can be integrated out immediately and only contribute
to the irrelevant constant in front of the path integral)      
\bea
Z_0[J] \EA {\rm const.} \int {\cal D} A \> \exp \left \{ 
i \int d^4x \> \left [ \frac{1}{2} A_{\mu} \left ( g^{\mu \nu} \Box - 
\partial^{\mu} \partial^{\nu} + \frac{1}{\lambda} \partial^{\mu}
\partial^{\nu} \right ) A_{\nu}  + J^{\mu} A_{\mu} \right ] \> \right \} 
\non
\EA {\rm const.}' \exp \left \{ - \frac{i}{2}  \int d^4x \> d^4y \> 
J^{\mu}(x) \Delta_{\mu \nu}(x,y) J^{\nu}(y) \> \right \},
\eea
where the kernel fulfills
\be
\left [ \> g^{\mu \nu} \Box_x - \left ( 1 - \frac{1}{\lambda} \right ) 
\> \partial^{\mu}_x
\partial^{\nu}_x \> \right ] \Delta_{\nu \rho}(x,y) \E 
g^{\mu}_{\rho} \> \delta^4(x-y) \> .
\ee
This can now be inverted in Fourier space and one obtains the following
propagator for the gauge bosons (\purpur{\bf Problem \ref{Vektor-Teilchen} b)})    
\be
\boxed{
\qquad \Delta_{\mu \nu}(k) \E - \frac{1}{k^2 + i \, 0^+} \> 
\left [ \> g_{\mu \nu}  - ( 1 - \lambda) 
\frac{k_{\mu} k_{\nu}}{k^2 + i \, 0^+}\> \right ] \> . \quad
}
\label{Eichboson Prop}
\ee
If one chooses the gauge parameter  $ \> \lambda = 1 \> $, then one works 
 in \textcolor{blue}{\bf Feynman gauge}, for $ \> \lambda = 0 \> $
in \textcolor{blue}{\bf Landau gauge}. In non-abelian gauge theories the gauge-boson propagator
(\ref{Eichboson Prop}) gets an additional factor $ \> \delta_{a b} \> $.

Physical observables, as cross sections, masses, decay rates must be independent of the 
gauge parameter $ \, \lambda \, $ which gives a necessary criterion for a correct calculation.
On the other hand Green functions are gauge dependent (see, e.g. Eqs.
\eqref{BN prop} and \eqref{BN prop exponent}).

\item[d)] One can obtain the ghost propagator in a covariant gauge directly from 
the corresponding part of the free generating functional if 
$ \> \partial_{\mu} \, \left ( D_{\mu}\right )^{ab} = \delta_{ab} \, \Box + {\cal O}(g) \> $ 
is used. From
\be
Z_0 \E \ldots \int {\cal D} \cy{\bar{\chi}} \> 
{\cal D}\cy{\chi} \> \exp \left [ \> \int d^4 x  \> \cy{\bar{\chi}^a}(x)    
\delta_{a b} \Box \cy{\chi^B}(x) \> \right ]
\ee
one can read off
\be
\boxed{
\qquad \Delta_{FP}^{a b}(k) \E - \delta_{a b} \> \frac{1}{k^2 + i \, 0^+} \> .
 \quad
}
\ee

\item[e)] There exist gauges in which the Faddeev-Popov factor is independent of
the gauge field and in which therefore no ghosts appear, e.g. the {\bf axial gauge} 
$ \> n^{\mu} A_{\mu}^a =  0, \> \> \> (
 \> n^{\mu}\> $ is a spacelike vector) or  temporal gauge
$ \> A_0^a = \> 0 \> $. However, these gauges result in a complicated gluon propagator and
are not explicitly covariant.
\eit
\vspace{1cm}

\renewcommand{\baselinestretch}{1.}
\scriptsize
\refstepcounter{tief}
\noindent
\blau{\bf Detail \arabic{tief}:} {\bf Infrared Problem}\\
\vspace*{0.4cm}

\begin{subequations}

\noindent
Quantized gauge theories not only have a divergence problem at high energies when one is calculating
loops ({\bf ``ultraviolet problem''}) which is treated 
(maybe one even can say "solved") by renormalization but also at very small energies 
({\bf ``infrared problem''}) because the gauge bosons are massless.
Among other things this shows up in the full fermionic Green functions which 
in reality do not exhibit a simple pole but a branch point at $ p^2 = m^2 $ .
One can study that in \rot{\bf QED} by invoking the  {\bf Bloch-Nordsieck approximation} 
which simplifies drastically the fermion spin degrees of freedom but correctly describes
the effect of low-energy ("soft") photons \footnote{In this approximation
the Dirac matrices for the electron are replaced by a constant four-vector, essentially
the velocity, since it doesn't change practically in processes with low-energy
emitted and absorbed photons.}
and one obtains
(\purpur{\bf Problem \ref{Bloch-Nordsieck}$^{\star}$})
\be
G_2(p) \E \frac{i \, m^{1 + 2 \kappa} Z_2}{(p^2 - m^2 + i0)^{1 + \kappa}} \> ,   
\label{BN prop}
\ee
with the gauge-dependent exponent
\be
\kappa \E \frac{e^2}{8 \pi^2} \, \lrp 3 - \lambda \rrp \> .
\label{BN prop exponent}
\ee
A physical electron, thus, is always surrounded by a cloud of (very soft) photons
(similar as in the polaron problem the bare electron is surrounded by a
cloud of phonons, which, howver, have a fixed frequency = mass)
 and the LSZ formulas are strictly not applicable. 
As a remedy one is giving the photons a small mass $ \mu $  and is including
processes with ultra-soft photons (summed incoherently) which cannot be distinguihed 
from the actual final state due to the finite energy resolution of the detector.
The Bloch-Nordseck theorem then guarantees that the limit 
$ \mu \to 0 $ can be performed in the final result and that essentially the photon mass
is replaced by the energy resolution.

\noindent
One can do that also in non-abelian theories like 
 \rot{\bf QCD} but that doesn't circumvent the main problem, viz. the "{\bf confinement}" 
 of quarks and gluons which do neither appear in the initial nor the final state as free particles.
For the treatment of infrared divergencies in this case see  {\bf \{Muta\}}, ch. 6.
\end{subequations}

\renewcommand{\baselinestretch}{1.2}
\normalsize
\vspace{1cm}

\begin{figure}[hbtp]
\vspace*{-6cm}
\bce
\includegraphics[angle=0,scale=0.8]{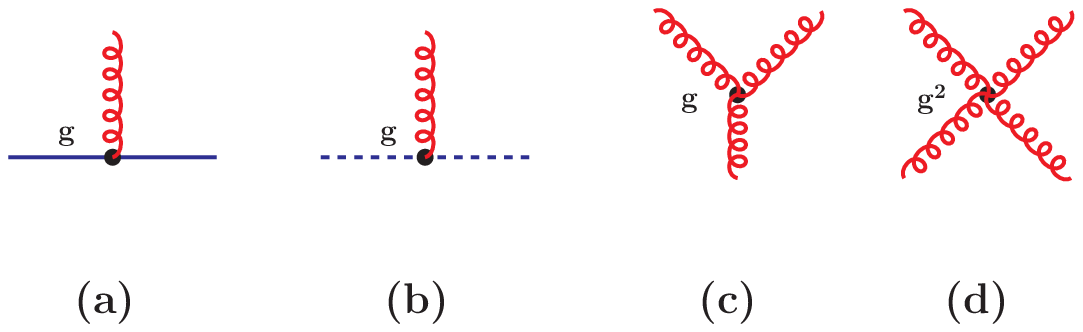}
\label{qcd_vertices}
\vspace*{-12cm}
\ece
{\bf Fig. \arabic{abb}} : Vertices of a non-abelian gauge theory (like \rot{\bf QCD}) according to Eq. 
\eqref{Vertices}. The full blue line \\
\hspace*{1.7cm} represents the fermions (quarks), the dashed blue line the
Faddeev-Popov ghosts and the curly red\\
\hspace*{1.7cm} lines the gauge bosons (gluons).
\vspace*{0.8cm}

\end{figure}

\noindent
The full generating functional -- including the fermions -- and with interactions 
reads
\be
Z[J,\cy{\eta},\cy{\bar{\eta}}] \E \int {\cal D}A \> {\cal D} \cy{\bar{\psi}} \> 
{\cal D} \cy{\psi} \> {\cal D} \cy{\bar{\chi}} \> {\cal D} \cy{\chi}  \> 
\exp \left \{ \> i \int d^4x \> \left [
{\cal L}_0 + {\cal L}' + J^a_{\mu} A^{a \mu} + \cy{\bar{\psi}} \cy{\eta} + \cy{\bar{\eta}} \cy{\psi} \right ]
\> \right \} \> ,
\ee
where
\be
{\cal L}_0 = \frac{1}{2} A_{\mu}^a \left [ \> g^{\mu \nu} \Box - 
\left ( 1 - \frac{1}{\lambda} \right ) \partial^{\mu}
\partial^{\nu} \> \right ] A_{\nu}^a + \cy{\bar{\psi}} \left ( i \dslash - m
\right ) \cy{\psi}
+ \cy{\bar{\chi}^a} \Box \cy{\chi^a}                 
\ee
is the free Lagrangian density (i.e. everything quadratic in the fields)
while $ \> {\cal L}' \> $ contains the couplings
\bea
\hspace*{-5cm}{\cal L}'\!\!  &=& \! \! - g \cy{\bar{\psi}}  T^a  \Aslash^a  \cy{\psi} + g \, f^{abc}  
\cy{\bar{\chi}^a}  
\cy{\chi^b}  \partial \cdot A^c + \frac{g}{2}  f^{abc} \left (\partial_{\mu} A_{\nu}^a -     
\partial_{\nu} A_{\mu}^a \right)   A^{\mu \, b}  A^{\nu \, c}-\frac{g^2}{4} f^{abc} f^{ade}  
A_{\mu}^b  A_{\nu}^c  A^{\mu \, d}  A^{\nu \, e} .
\label{Vertices}\\
&&\hspace{0.8cm}{\bf (a)}\hspace{2cm}{\bf (b)}\hspace{3cm}{\bf (c)}\hspace{4.5cm}{\bf (d)} \no
\eea
These determine the vertices of the theory displayed in 
Fig.  \ref{qcd_vertices}. Note that the different vertices are all fixed by 
\blau{\bf one coupling constant} $ \> g \> $.
\vspace{0.3cm}

\refstepcounter{abb}

\noindent
With that (and the propagators following from the quadratic terms) it is easy to derive 
the Feynman rules, say for \rot{\bf QCD}, \footnote{See, e.g. \meingruen{\bf Cheng \& Li}, ch. 9.2 .}.   
\vspace{1cm}

\renewcommand{\baselinestretch}{1.}
\scriptsize
\refstepcounter{tief}
\noindent
\blau{\bf Detail \arabic{tief}:} {\bf BRST symmetry}\\
\vspace*{0.2cm}

\noindent
\begin{subequations}
\noindent
That this gives a consistent, renormalizable theory in all orders is due to a symmetry which
the gauge-fixed, non-abelian Lagrangian with ghost fields possesses: Instead 
of the classical local gauge invariance the full Lagrangian in a covariant gauge
\bea
{\cal L} \EA {\cal L}_f +  {\cal L}_g +  {\cal L}_{gauge} + {\cal L}_{FP} \non
{\cal L}_f \EA \cy{\bar{\psi}}(x) \, \lrp i \dddslash(x) - m \rrp \, \cy{\psi}(x) \> , \quad
D_{\mu}(x) \E \partial_{\mu} + i g \, T^a A_{\mu}^a(x) \non
{\cal L}_g \EA - \frac{1}{4} F_{\mu \nu}^a \, F^{\mu \nu \, a} \> , \quad F_{\mu \nu}^a \E 
\partial_{\mu}  A_{\nu}^a - \partial_{\nu}  A_{\mu}^a - g \, f^{abc} \, A_{\mu}^b A_{\nu}^c \non
{\cal L}_{gauge} \EA - \frac{1}{2 \lambda} \, \lrp \partial^{\mu} A_{\mu}^a \rrp^2 \> , \quad
{\cal L}_{FP} \E \cy{\bar{\chi}^a} \, \partial^{\mu} \, \lrp D_{\mu}(x)\rrp^{ab} \, \cy{\chi^b} ,
\quad D_{\mu}^{ab} = \partial_{\mu} \delta^{ab} + g f^{acb} A_{\mu}^c(x)          
\label{volle Lagrangefunk}   
\eea
is invariant under the \textcolor{blue}{\bf Becchi-Rouet-Stora-Tyutin (BRST) transformation}:
\bea
\delta A_{\mu}^a \EA \cy{\omega} \lrp D^{\mu} \, \cy{\chi} \rrp^a \> , \quad 
\delta \cy{\psi} \E  i g \,  \cy{\omega} \cy{\chi^a} \, T^a \, \cy{\psi}  \non               
\delta \cy{\bar{\chi}^a} \EA - \frac{1}{\lambda} \, \cy{\omega} \partial^{\mu} A_{\mu}^a \> , \quad 
\delta \cy{\chi^a}\E - \frac{g}{2} \, \cy{\omega} f^{abc} \, \cy{\chi^b} \cy{\chi^c} \> , 
\label{BRST 1}
\eea
where $ \, \cy{\omega} \, $ is a Grassmann valued constant parameter.
One finds (\purpur{\bf Problem \ref{BRST}$^{\star}$}) 
\be
\delta \lrp D_{\mu} \cy{\chi} \rrp^a \E  \cy{0} \> , \quad \delta \lrp f^{abc} \cy{\chi^b} \cy{\chi^c} 
\rrp \E 0 
\ee
and 
\be 
\delta \lrp \partial_{\mu} A^{\mu \, a} \rrp \E 0 \> ,
\ee  
if the equation of motion for the ghost field $ \, \cy{\chi^a} \, $ is used.
This means that applying the BRST transformation twice on gauge and ghost fields
yields zero
\be
\delta_2 \, \delta_1 \, \Phi^a \E 0 \> ,
\ee
where $ \, \Phi^a = A_{\mu}^a $ or $ \cy{\chi^a} $ or $ \cy{\bar{\chi}^a} $ and the two transformations could have 
different parameters $ \cy{\omega_{1/2}} $. These transformations induce the following change of the 
Lagrangian
\be
\delta {\cal L} \E - \partial^{\mu} \, \lsp \frac{\cy{\omega}}{\lambda} \lrp \partial^{\nu} A_{\nu}^a \rrp
\, D_{\mu} \cy{\chi^a} \rsp \> , 
\ee
i.e. a total derivative and therefore demonstrate the invariance of the action.

\noindent
It is advantageous to represent the gauge-fixing term in the Lagrangian by an auxiliary field
\be
B^a(x) \E \frac{1}{\lambda} \, \partial^{\mu} A_{\mu}^a(x)
\ee
(again an application of the ``undoing the square''-trick or the Hubbard-Stratonovich transformation!)
\be
\exp \lcp - \frac{i}{2 \lambda}\int d^4x \> \lsp \partial_{\mu} A_{\mu}^a(x) \rsp^2 \rcp \E 
{\rm const.} \, \int {\cal D} B(x) \> \exp \lcp i \int d^4x \> \lsp \frac{\lambda}{2} B^a(x) B^a(x)
+ A_{\mu}(x) \, \partial^{\mu} B^a(x) \rsp \rcp \> .
\ee
This allows to choose directly the Landau gauge ($ \lambda = 0 $) and simplifies the 
BRST transformation \eqref{BRST 1}:
\bea
\delta A_{\mu}^a \EA \cy{\omega} \lrp D^{\mu} \, \cy{\chi} \rrp^a \> , \quad
\delta \cy{\psi} \E  i g \,  \cy{\chi^a} \, T^a \, \cy{\psi} \,  \cy{\omega} \, \non    
\delta \cy{\bar{\chi}^a} \EA - B^a \,  \cy{\omega} \> , \quad
\delta \cy{\chi^a} \E - \frac{g}{2} \, f^{abc} \, \cy{\chi^b} \cy{\chi^c} \>  \cy{\omega} \>, \quad 
\delta B^a \E 0 \> .
\label{BRST 2}
\eea
allowing an easier proof of the invariance (see \purpur{\bf Problem \ref{BRST}}). 

\noindent
According to Noether's theorem (see {\bf chapter} {\bf \ref{sec1: Symm}}) a conserved current 
\be
\cy{J}_{\mu}^{\rm BRST} \E F_{\mu \nu} ^a  D^{\nu} \, \cy{\chi^a} + B^a \, D_{\mu} \, \cy{\chi^a} 
- \frac{1}{2} f^{abc} \, \lrp \partial_{\mu} \cy{\bar{\chi}^a} \rrp \, \cy{\chi^b} \, \cy{\chi^c}
\ee
is associated with the BRST invariance and the spatial integral over the 0$^{\rm th}$ component
defines a conserved ``charge''$\cy{Q}^{\rm BRST}$ which in the quantized theory is the generator
of the BRST transformations. We assume that the BRST symmetry is not ``broken'', i.e. that not only the
Lagrangian (or the Hamiltonian) remain unchanged under the transformation but also the physical states:
\bea
&& e^{-i \cy{\omega} \hat{\cy{Q}}^{\rm BRST}} \, \bigl | \, {\rm phys} \, \bigr > \E \lrp 1 - i 
\cy{\omega} \hat{\cy{Q}}^{\rm BRST} \rrp
\, \bigl | \, {\rm phys} \, \bigr > \E \bigl | \, {\rm phys} \, \bigr > \non
&& \Longrightarrow \> \cy{Q}^{\rm BRST} \>  \bigl | \, {\rm phys} \, \bigr > \E 0 \> .
\eea
(This is a necessary constraint for the states in the Hilbert space of the gauge theory
and obviously requires the language and formalism of canonical quantization -- the path integral
doesn''t know anything about states!)
\vspace{0.1cm}

This can be used to show that the gauge-fixing and ghost terms, which have been added to the
Lagrangian, do not give a contribution to physical matrix elements of the theory
(\meingruen{\bf Nair}, ch. 12.4). To achieve that we define the Grassmann-valued quantity 
\be 
\cy{\Xi} \Def - \partial^{\mu} \cy{\bar{\chi}^a} - \frac{\lambda}{2} \, \cy{\bar{\chi}^a} \, B^a 
\ee
and by applying the BRST- transformationen \eqref{BRST 2} find that
\bea
\delta \, \cy{\Xi} &\Def& \delta \, \lsp  - \partial^{\mu} \cy{\bar{\chi}^a} - \frac{\lambda}{2} \, 
\cy{\bar{\chi}^a} \, B^a \rsp 
\E \cy{\omega} \partial^{\mu} B^a A_{\mu}^a - \partial^{\mu}  \cy{\omega} \lrp D_{\mu} \cy{\chi} \rrp^a - 
\frac{\lambda}{2} \lrp - \cy{\omega} B^a \rrp B^a
\E \cy{\omega} \, \lrp {\cal L}_{\rm gauge} + {\cal L}_{FP} \rrp \non
&\widehat{=} & i \, \cy{\omega} \, \lsp \hat{\cy{Q}}^{\rm BRST} ,  \hat{\cy{\Xi}}  \rsp_+ \> , 
\label{delta Xi}
\eea
i.e. that the additional terms added to the original Lagrangian may be written as a BRST variation.
The equivalence displayed in the last term of Eq. \eqref{delta Xi}  is due to the fact that in the quantized
theory the variation of a Grassmann-valued operator is given by the anti-commutator with the generator
\be
e^{i \cy{\omega}   \hat{\cy{Q}}^{\rm BRST}} \, \hat{\cy{\Xi}} \, e^{-i \cy{\omega}   
\hat{\cy{Q}}^{\rm BRST}} \E  
\hat{\cy{\Xi}} + i \cy{\omega} \, 
\hat{\cy{Q}}^{\rm BRST} \, \hat{\cy{\Xi}} - i  \hat{\cy{\Xi}} \, \cy{\omega} \, 
\hat{\cy{Q}}^{\rm BRST} \E \hat{\cy{\Xi}} 
+ i \cy{\omega} \, 
\hat{\cy{Q}}^{\rm BRST} \, \hat{\cy{\Xi}} + i  \cy{\omega} \, \hat{\cy{\Xi}} \, \hat{\cy{Q}}^{\rm BRST} 
\EQ  \hat{\cy{\Xi}} + i \cy{\omega} \, 
\lsp \hat{\cy{Q}}^{\rm BRST} , \hat{\cy{\Xi}} \rsp_+ \deF  \hat{\cy{\Xi}} + \delta \, \cy{\Xi} \, .
\ee
By this it follows that
\be 
\bigl < {\rm phys} \bigl | \> {\cal L}_{\rm gauge} + {\cal L}_{FP} \> \bigr | \, {\rm phys}' \bigr >
\E  i \, \la {\rm phys} \left | \, \lsp \hat{\cy{Q}}^{\rm BRST} , \hat{\cy{\Xi}} \rsp_+ \,  \right | \, 
{\rm phys}' \ra \E 0
\ee
between arbitrary physical states which implies that the added terms do not contribute to
physical processes,
\vspace{0.1cm}

The BRST invariance also gives relations for Green functions which are essential for the
renormalizability of the theory. These relations are known as {\bf Ward} or {\bf Slavnov-Taylor identities} and 
sketched in \meingruen{\bf Das}, ch. 12.5.

\end{subequations}
\renewcommand{\baselinestretch}{1.2}
\normalsize

\vspace{0.8cm}

\subsection{\textcolor{blue}{Worldline Formalism and Spin in the Path Integral}}
\label{sec3: Weltlinien}

In this chapter we will study the \textcolor{blue}{effective action} in one-loop approximation
by means of a quantum-mechanical formalism which has (re)gained significance in recent years.
Surprisingly the motivation for this development came from "String Theory" where new methods for perturbative expansions had been found which -- in special cases -- could be transferred to quantum field theory \footnote{See, e.g.., Ref. \cite{Schub}.  Further examples for the application of the worldline technique can be found in Ref. \cite{RSS}.}.

We start again with a scalar theory and recall that the one-loop correction to the effective action is given by
\be
\Gamma^{(1)} \left [ \Phi_{\rm cl} \right ] \E \frac{i \hbar}{2} 
\ln \frac{ \fdet \left ( \> \Box + m^2 + V''(\Phi_{\rm cl}) \> \right)} 
{\fdet \left ( \> \Box + m^2\> \right)} \E 
\frac{i \hbar}{2} {\rm tr} \ln \left [ \frac{ \Box + m^2 + 
V''(\Phi_{\rm cl}) - i \, 0^+}{ \Box + m^2 - i \, 0^+} \right ]
\label{Gamma 1a}
\ee
(see Eq. (\ref{1 Schleif eff W})). Using the integral representation of the logarithm 
\be
\ln \frac{a}{b} \E \int_0^{\infty} dT \> \frac{1}{T} \left ( \, 
e^{-b T} - e^{-a T} \, \right ) \> , \hspace{0.5cm} {\rm Re} \, a \> , 
{\rm Re} \, b  > 0 \> \> ,
\label{Integraldarst log}
\ee
we write that as
\be
\Gamma^{(1)} \left [ \Phi_{\rm cl} \right ] \E {\rm const.} - 
\frac{i \hbar}{2} \int_0^{\infty} dT \> \frac{1}{T} \> {\rm tr} \, 
\exp \Bigl \{ \, 
- i T \left [ \> -  (i\partial_{\mu}) (i\partial^{\mu})
+ m^2 + V''(\Phi_{\rm cl}) - i \, 0^+\right ] \, \Bigr \}  \> .
\label{Gamma 1b}
\ee
Here we have put the interaction-independent part in an (infinite) constant which will 
be omitted in the following (a constant term in the action is irrelevant).
However, the trace to be taken in Eq. (\ref{Gamma 1b}) over all degrees of freedom
cannot be evaluated simply in momentum space (as was done for the effective potential)
as  $ V''(\Phi_{\rm cl}) $ now depends on the space-time point  $x$ . By writing it as
\be
{\rm tr} \exp \left \{ \> \ldots  \> \right \} \E 
\int d^4 x \> \left < x  \left | \> \exp \left \{ - i T \left [ \, -
\hat p_{\mu} \hat p^{\mu} + m^2 + V''(\Phi_{\rm cl}(\hat x))  \, \right] \> 
\right \} \> \right | x \right >
\label{Spur} 
\ee
we see that formally this is a matrix element of the time-evolution operator for 
a quantum-mechanical particle with "mass" $ \> - 1/2 \> $ which moves under the
influence of the "potential"
 $ \> V''(\Phi_{\rm cl}(x)) \> $ in four-dimensional Minkowski space.
 The particle starts at "time" $ \> 0 \> $ 
at the space-time point $ x $ and returns to that point at "time" 
$\> T \> $ . Based on this quantum-mechanical analogy we therefore can give immediately
a path-integral representation for  Eq. (\ref{Spur}) 
(see \purpur{\bf Problem \ref{Weltlin Pot}}) and we obtain the following expression for the one-loop correction to the effective action
\be
\boxed{
\qquad \Gamma^{(1)} \left [ \Phi_{\rm cl} \right ] \E -
\frac{i \hbar}{2} \int_0^{\infty} dT \> \frac{e^{ - i  m^2 T} }{T} \! \!  \! 
\oint\limits_{x(0)=x(T)} \! \! \! {\cal D} x(t) \>
\exp \left \{ \> i \int_0^T dt
\, \left [ - \frac{1}{4} \dot x^2 + V''\left ( \Phi_{\rm cl}(x(t)) \right )
\, \right ] \> \right \} \> . \quad 
}
\label{Gamma 1c}
\ee
This is  referred to as \textcolor{blue}{\bf worldline} or particle representation
as the system is no longer described by quantized fields but by particles with trajectories
 $ x_{\mu}(t) $ parametrized by the \textcolor{blue}{proper time} $ t $. Note that this
 also holds for the usual time $ x_0 $ and that one has to integrate over the final proper time
  $ T $ at the very end of the calculation. 
Eq. (\ref{Gamma 1c}) offers the possiblity to perform one-loop calculations
with arbitrary many external lines in a fast and efficient way.
This is because the expression (\ref{Gamma 1c}) contains many Feynman diagrams
which only arise by a permutation of the external lines.
To see that one may expand the interacting part of the exponent, say for the $\Phi^4$-theory
\be
\exp \left \{ i \int_0^T dt
\, V''\left ( \Phi_{\rm cl}(x(t)) \right )  \right \} \E 
\sum_{n=0} \frac{(i \lambda)^n}{2^n \, n!} \, \int_0^T \! dt_1 \, dt_2 \ldots
dt_n \> \Phi_{\rm cl}^2(x(t_1)) \, \Phi_{\rm cl}^2(x(t_2)) \,
\ldots \Phi_{\rm cl}^2(x(t_n)) \> ,
\ee
and realize that {\bf all} time orderings  of the respective interactions are included.
In addition, one does not have to perform four-dimensional momentum integrations
but only one-dimensional time integrations.

\begin{figure}[hbtp]
\vspace{-2cm}
\bce
\includegraphics[angle=0,scale=0.6]{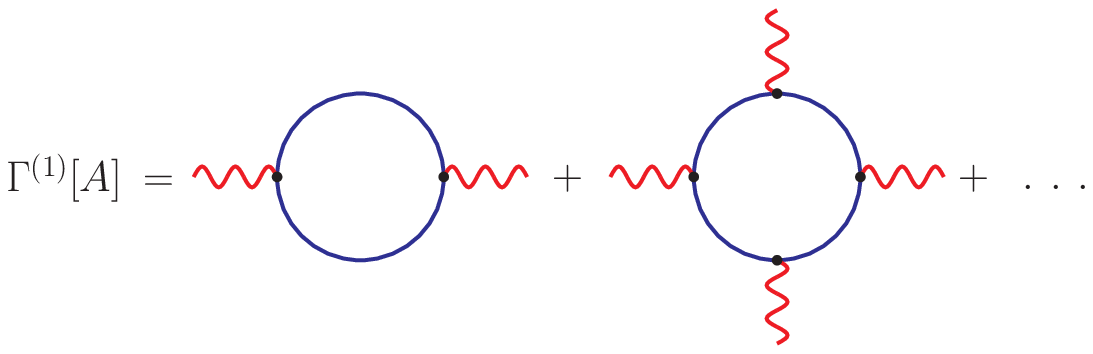}
\label{abb:3.4.1}
\ece
\vspace{-11cm}
{\bf Fig. \arabic{abb}} : One-loop approximation for the effective action of Quantum Electrodynamics.
The wavy lines \\ \hspace*{1.5cm} represent the photons.

\end{figure}

\vspace{0.3cm}
Can this formalism also be applied to \textcolor{blue}{\bf fermions} ?
The answer is ``{\bf Yes}'', but it requires an explicit description of the 
spin degrees of freedom in the quantum-mechanical path integral.
This is an old problem with many suggestions for a solution.
Here we will present a description by Grassmann-valued trajectories for the case
of quantum electrodynamics with  the Lagrangian
(\ref{L QED}). We only consider processes without external fermions, i.e. the
generating functional
\be
Z[J] \E \int {\cal D}A_{\mu}(x) \, {\cal D} \cy{\bar{\psi}}(x)\,  
{\cal D} \cy{\psi}(x)
\> \exp \left \{ \> \frac{i}{\hbar} \int d^4 x
\left [ {\cal L}_{\rm QED}(\cy{\psi},\cy{\bar{\psi}}, A) + J_{\mu} A^{\mu} \right ] \> 
\right \} \> .
\ee
Since the Lagrangian is bilinear in the fermion fields we may integrate them out
immediately with the result
\bea
Z[J] \EA \int {\cal D}A_{\mu}(x) \> \fdet \left ( \> i \dslash - m
- e \Aslash \> \right ) \> \exp \left \{ \>  \frac{i}{\hbar} 
\int d^4 x \> \left [ \, 
{\cal L}_0 (A) + J_{\mu} A^{\mu} \right ] \>\right \} 
\non
\EA \! \! \int {\cal D}A_{\mu}(x) \, \exp \left \{ \,  \frac{i}{\hbar} 
\left [ \, -i  \hbar \, {\rm tr}
\ln \left ( \, i \dslash - m - e \Aslash \, \right ) + 
\int d^4 x \, \left ( {\cal L}_0 (A) + J_{\mu} A^{\mu} \right ) \, \right ] 
\, \right \} .
\eea
Here $ {\cal L}_0 (A) $ is the free Lagrangian of the photons (we do not display the gauge-fixing term explicitly). As this part does not contain any interaction we immediately
can write down the effective action in zeroth and first order semi-classical approximation:
\bea
\Gamma^{(0)} [A_{\rm cl}] \EA \int d^4 x \>{\cal L}_0 (A_{\rm cl})  \E \int d^4x \> 
\left [ \> - \frac{1}{4} F_{\mu \nu} F^{\mu \nu} \> \right ] \E 
\int d^4x \> \frac{1}{2} \left ( {\bf E}^2 - {\bf B}^2 \right ) 
\label{Gammaf 0} \\
\Gamma^{(1)} [A_{\rm cl}] \EA - i \, \hbar \, {\rm tr} \ln \frac{i \ddslash - m + i \, 0^+}
{i \dslash - m + i \, 0^+} \E -i  \, \hbar \, {\rm tr} \ln \left [ 
1 - e S_F \Aslash_{\rm cl} \right ] \> .
\label{Gammaf 1a}
\eea
In the following we will set again $ \hbar = 1$ and omit the index "cl" for the photon field
 $ A_{\mu} $ to simplify the notation. $ {\bf E},
{\bf B} \> $ are the corresponding electric and magnetic field strengths.
In addition we have normalized $ \Gamma^{(1)}$ such that it vanishes for 
 $ e = 0 $ .
 As can be seen, the effective action in first semi-classical (or one-loop) approximation 
 is given by the fermionic determinant, i.e. by the pair production of charged electrons and
 positrons. When expanding the logarithm in powers of the coupling constant (or in powers of the
 photon field)
\be
\Gamma^{(1)}[A] \E  i \sum_{n=1} \, \frac{e^{2n}}{2n} \, {\rm tr} \left [
S_F \Aslash \right]^{2n} 
\ee
only even powers appear which can be proved by using the charge conjugation properties of
Feynman propagator and vertex
({\bf Furry-Theorem}) \footnote{See, e.g., \meingruen{\bf Itzykson \& Zuber}, p. 276.}. 
This is depicted in Fig. \ref{abb:3.4.1}.

\refstepcounter{abb}

\vspace{0.5cm}

\noindent
Note that the fermionic effective action has a different sign compared to the 
scalar theory in  Eq. (\ref{Gamma 1a}) which has its origin in the integration over the anticommuting
fields $ \cy{\psi}, \cy{\bar{\psi}} $.
Moreover, there is a factor of $ 2 $ as two charged particles run in the loop here.

\par
We now want to derive a worldline representation of the fermionic effective action in one-loop approximation. To achieve that we first iterate the Dirac operator
 $ \> i \ddslash - m \> $ in Eq. (\ref{Gammaf 1a})
\be
\Gamma^{(1)}[A] \E  - \frac{i}{2} \ln \left [ \fdet \left( 
i \ddslash - m \right ) \left ( - i \ddslash - m \right ) \right ] \E 
- \frac{i}{2} \ln \left [ \fdet \left( \ddslash^2 + m^2 \right )
\right ] \> , 
\ee
which is possible due to the relation \footnote{$\gamma_5 = 
i \gamma_0\gamma_1\gamma_2 \gamma_3$ anticommutes with all Dirac matrices
and fulfills $ \gamma_5^2 = 1$.} $ \> \fdet (\ddslash - m ) = 
\fdet(\gamma_5 \ddslash \gamma_5 - m)
= \fdet (- \ddslash - m ) \> $. Now we have
\bea
\ddslash^2 \EA \gamma^{\mu} \gamma^{\nu} D_{\mu} D_{\nu} \E 
\left ( g^{\mu \nu} + \frac{1}{2} \left [ \gamma^{\mu}, \gamma^{\nu} \right ]
\right ) D_{\mu} D_{\nu} \E 
D^2 + \frac{1}{4} \left [ \gamma^{\mu}, \gamma^{\nu} \right ]
\left [ D_{\mu}, D_{\nu} \right ] \non
\EA \left ( \partial + i e A \right )^2 
+ \frac{1}{4} \, \left [ \gamma^{\mu}, \gamma^{\nu} \right ] \, ie
\left ( \partial_{\mu} A_{\nu} - \partial_{\nu} A_{\mu} \right ) 
\E  \left ( \partial +
i e A \right )^2 + \frac{ie}{2} \gamma^{\mu} \gamma^{\nu} F_{\mu \nu} \> .
\eea
If we use the integral representation (\ref{Integraldarst log}) and define
\be
H(\hat x, \hat p, \gamma ) \Def - \left( \hat p + e A(\hat x) 
\right )^2
+ \frac{ie}{2} \gamma^{\mu} \gamma^{\nu} F_{\mu \nu} (\hat x) 
\label{H spin}
\ee
as ``Hamiltonian'' of the system, then we may again express 
\be
\boxed{
\qquad \Gamma^{(1)}[A] \E \frac{i}{2} \int_0^{\infty} dT \> 
\frac{e^{-i m^2 T}}{T} \> {\rm tr} \int d^4 x \>  \left < x \left | \, 
e^{- i T H(\hat x, \hat p, \gamma )} \, \right | x \right > \qquad
}
\ee
formally as matrix element of a quantum-mechanical operator
 $ \hat U(T,0) $ . The symbol "tr" indicates the trace over the Dirac matrices.
 As usual we split the proper-time evolution into 
$N$ steps and obtain
\bea
{\rm tr} \> \hat U(T,0) &\EQ & {\rm tr}
 \int d^4 x \> \left < x \left | \, e^{- i T H(\hat x, \hat p,
\gamma )} \, \right | x \right > \non
\EA  {\rm tr} \lim_{N \to \infty}
\int d^4 x_1 \ldots d^4 x_N \> \frac{d^4 p_1}{(2 \pi)^4} \ldots
\frac{d^4 p_N}{(2 \pi)^4} 
 \> \exp \left [ - i \sum_{i=1}^N p_i \cdot
(x_i - x_{i-1} ) \right ] \non
&& \cdot \> \exp
\left [ \> - i H_W(x_N,p_N,\gamma_N) \Delta t \> \right ] \> \ldots
\> \cdot \> \exp \left [\>  - i  H_W(x_1,p_1,\gamma_1) \Delta t \>
\right ] \> ,
\label{time-sliced phase space PI}
\eea
where $ x_0 = x_N $ has to be taken. Here
\be
H_W(x,p,\gamma) = \int d^4 y \> \la x -\frac{y}{2} \> | \> \hat H
\> | \>
x + \frac{y}{2} \ra e^{-i p \cdot y}
\ee
is the  \blau{Wigner transform (or the Weyl symbol) of the Hamiltonian},
which -- as we know -- is the appropriate classical analogon for the (Weyl-ordered)
quantum operator. In the following we will omit the index "W".

\noindent
We need two essential steps to derive a path integral with spin from
Eq. (\ref{time-sliced phase space PI}) \cite{FrGi}, \cite{ARS}:

\ben
\item[1.]
As the Dirac matrices do not commute, the ordering of the factors
is all-important and it is not possible to combine the individual exponents.
By giving the Dirac matrices an artificial time dependence
we can write the (proper) time evolution operator as a time-ordered
path integral
\bea
 {\rm tr} \, \hat U(T,0) \EA {\rm tr}
\oint\limits_{x(0)=x(T)} \frac{{\cal D}'x \, {\cal D}p}{(2 \pi)^4} 
\> \> {\cal T}
\exp \left \{ - i \int_0^T dt \left [ \> p \cdot \dot x  +
H \left (x(t),p(t),\gamma(t)\right ) \> \right ] \> \right \}
\non
\EA {\rm tr}
\oint\frac{{\cal D}'x {\cal D}p}{(2 \pi)^4}
\> \exp \left \{ - i \int_0^T dt \left [ \> p \cdot \dot x  +
H\left ( x(t),p(t),\frac{\delta}{\delta \cy{\rho}(t)} \right ) \>
\right ] \> \right \}
\non
&& \hspace{5cm} \cdot \> {\cal T} \exp \left [ \int_0^T dt \> \cy{\rho^{\mu}}(t)
\cy{\gamma_{\mu}}(t) \right ]_{\cy{\rho^{\mu}}=\cy{0}} \> .
\label{G hat with diff}
\eea
Here  $\,  \cy{\rho^{\mu}}(t) \, $ are Grassmann sources which we assume to anticommute with
the Dirac matrices \footnote{This is the reason why we write
$\cy{\gamma_{\mu}}$, see footnote \ref{g gerade}.}.
The time-ordering symbol $ {\cal T} $ would be devastating for all subsequent
manipulations in the path integral since we want it to be based on numbers and not on operators.
Fortunately, in this special case the time ordering can be eliminated by the relation
\bea
Y(T) &\EQ & {\cal T} \exp \left \{ \int_0^T dt \> \cy{\rho^{\mu}}(t)
\cy{\gamma_{\mu}}(t) \>
\right \}  \non
\EA \exp \left \{-\int_0^T dt_1 \int_0^{t_1} dt_2 \> \cy{\rho^{\mu}}(t_1)
\cy{\rho_{\mu}}(t_2)\> \right \} \> \cdot
        \> \exp \left \{\int_0^T dt \> \cy{\rho^{\mu}}(t)\cy{\gamma_{\mu}} \>
\right \}  \> .
\label{time ordering eliminated}
\eea
This can be proved by solving the evolution equation
\be
\frac{\partial  Y(T)}{\partial  T} \E \cy{\rho^{\mu}}(T)\cy{\gamma_{\mu}} \> Y(T)
\> \> , \> \> \> \> Y(0) \E 1
\ee
by means of the Magnus expansion \footnote{This is the continous analogon to the
better-known Baker-Campbell-Haussdorff formula, see, e.g. Ref. \cite {Wil}.}
\be
Y(T) \E \exp \Biggl \{ \>  \int_0^T dt \> \cy{\rho^{\mu}}(t) \cy{\gamma_{\mu}}
\> + \> \frac{1}{2} \int_0^T dt_1 \int_0^{t_1} dt_2 \>
\left [ \cy{\rho_{\mu}}(t_1) \cy{\gamma^{\mu}} , \cy{\rho_{\nu}}(t_2) \cy{\gamma^{\nu}} \right ]
\> + \> \ldots \> \Biggr \} \> .
\ee
The commutator gives $ - 2 \cy{\rho_{\mu}}(t_1) \cy{\rho^{\mu}}(t_2) $ which is a commuting number.
Therefore all higher terms in the expansion, expressed by multiple commutators, vanish.
Due to that result we now may omit the artificial time-dependence of the Dirac matrices.

\item[2.]
The differentiations w.r.t.  $\cy{\rho^{\mu}}(t)$ required in Eq. 
(\ref{G hat with diff}), can only be performed analytically in all orders if the variables 
appear  \textcolor{blue}{\it linearly} in the exponent. This can be achieved by using
the trick of "undoing the square" which we already have employed in
{\bf chapter} {\bf \ref{sec1: Potstreu}}. 
Since $\cy{\rho^{\mu}(t)}$ are anticommuting and we want to have a Grassmann-even object
in the exponent of the time-evolution operator we have to undo the square by a Grassmann path
integral. We therefore use the identity
\bea
\exp \left \{- \int_0^T \! dt_1 \int_0^{t_1} \! dt_2 \> \cy{\rho^{\mu}}(t_1)
\cy{\rho_{\mu}}(t_2) \right \} \! \EA \!
\int {\cal D}\cy{\xi} \> \exp \left \{\int_0^T
dt \left[ - \frac{1}{4} \cy{\xi_{\mu}}(t)\dot{\cy{\xi}}^{\cy \mu}(t) +
\cy{\rho^{\mu}}(t)\cy{\xi_{\mu}}(t)\right] \right \}
\non
&&  \cdot \>
\left [ \int {\cal D} \cy{\xi} \>\> \exp \left ( - \frac{1}{4} \int_0^T
dt \> \cy{\xi_{\mu}}(t)\dot{\cy{\xi}}^{\cy\mu}(t)\right ) \right ]^{-1} \> .
\label{Berezin integral}
\eea
The boundary conditions for the Grassmann path integral are
\be
\cy{\xi_{\mu}}(0) + \cy{\xi_{\mu}}(T) \E  0 \>.
\label{boandary cond xi}
\ee
Eq. (\ref{Berezin integral}) can be verified by the stationary phase method which is 
exact for quadratic actions.
\een

\noindent
Finally we use the representation
\be
\exp \left \{\int_0^T dt \> \cy{\rho^{\mu}}(t)\cy{\gamma_{\mu}} \right \} \E
\exp \left \{ \cy{\gamma_{\mu}}\frac{\partial}{\partial \cy{\Gamma_{\mu}}}\right \}
\> \> \exp \left \{\int_0^T
dt \> \cy{\rho^{\mu}}(t)\cy{\Gamma_{\mu}} \right \}\Biggr|_{\cy{\Gamma_{\mu}} = \cy{0}}
\label{Weyl symbol}
\ee
and obtain
\bea
{\rm tr} \, \hat U(T,0) \EA {\rm tr} \, \exp \left ( \cy{\gamma} \cdot 
\frac{\partial}{\partial \cy{\Gamma}} \right ) \, 
\oint \frac{{\cal D}'x  \> {\cal D}p}{(2 \pi)^4}  \> {\cal D}\cy{\xi} \> 
{\cal N}^{\> \rm spin}_0 \non
&& \hspace{2cm}\cdot \exp \Biggl \{ \> - i\int_0^T dt \> 
\biggl [  \, p \cdot \dot x - \frac{i}{4} \cy{\xi} \cdot \dot{\cy{\xi}} 
+ H(x,p,\cy{\xi} + \cy{\Gamma}) \,  \biggr ] \>   \Biggr \}
_{\cy{\Gamma}=\cy{0}}  .
\label{phase space PI a}
\eea
Here
\be
{\cal N}^{\> \rm spin}_0 \E \left [ \int {\cal D}\cy{\xi} \> \exp \left ( - \frac{1}{4} 
\int_0^T dt \> \cy{\xi_{\mu}} \dot{\cy{\xi}}^{\cy\mu} \right )
\right ]^{-1}
\label{N0 spin}
\ee
is a normalization factor for the spin integral.
Note that, in general, the procedure required in Eq. (\ref{Weyl symbol}) is
{\bf not} only a replacement of the variable $ \cy{\Gamma} $ by the corresponding
Dirac matrix $\cy{\gamma}$ but also involves an antisymmetrization.
For example, we have
\be
\exp \left \{ \cy{\gamma} \cdot\frac{\partial}{\partial \cy{\Gamma}} \right \}
\cy{\Gamma_{\mu}} \Biggr|_{\cy{\Gamma} = \cy{0}} \E  \cy{\gamma_{\mu}} \> ,
\ee
but
\bea
\exp \left \{ \cy{\gamma} \cdot \frac{\partial}{\partial \cy{\Gamma}}\right \}
\cy{\Gamma_{\mu}} \cy{\Gamma_{\nu}}\Biggr|_{\cy{\Gamma} = \cy{0}} \EA\frac{1}{2}
\left ( \cy{\gamma} \cdot \frac{\partial}{\partial \cy{\Gamma}} \right )^2
\cy{\Gamma_{\mu}} \cy{\Gamma_{\nu}}\Biggr|_{\cy{\Gamma} = \cy{0}} \E \frac{1}{2}
\left ( \cy{\gamma} \cdot \frac{\partial}{\partial \cy{\Gamma}} \right ) \left ( \>
\cy{\gamma_{\mu}} \cy{\Gamma_{\nu}} + \cy{\Gamma_{\mu}} \cy{\gamma_{\nu}} \> \right )
\Biggr|_{\cy{\Gamma} = \cy{0}} \non
\EA \frac{1}{2} \left ( \> - \cy{\gamma_{\nu}} \cy{\gamma_{\mu}} + \cy{\gamma_{\mu}}
\cy{\gamma_{\nu}} \> \right ) \> .
\label{zwei gamma}
\eea
Taking the trace over the Dirac matrices entails the simplification that only the "1" in
the series expansion of
$ \exp(i \cy{\gamma} \cdot \partial/ \partial \cy{\Gamma}) $ remains: At most terms with up to
four $\cy{\Gamma}$'s can show up, the trace over an odd number of 
 $\cy{\gamma}$-matrices vanishes as well as the trace over 
Eq. (\ref{zwei gamma}); finally we have
$ \> \> {\rm tr} \, \cy{\gamma_0} \cy{\gamma_1} \cy{\gamma_2} \cy{\gamma_3} = - i \, {\rm tr}\,  
\cy{\gamma_5} = \cy{0}  \> $. 
By that Eq. (\ref{phase space PI a}) simply becomes
\be
\qquad {\rm tr} \, \hat U(T,0) \E  4 \, 
\oint\limits_p \frac{{\cal D}'x  {\cal D}p}{(2 \pi)^4}  \oint\limits_{ap} 
{\cal D}\cy{\xi} \> 
{\cal N}^{\> \rm spin}_0 \> 
\exp \left \{ \> - i\int_0^T dt \> 
\biggl [  \, p \cdot \dot x - \frac{i}{4} \cy{\xi} \cdot \dot{\cy{\xi}} +
H( p,x,\cy{\xi}) \,  \biggr ] \>   \right \} \> , \quad
\label{phase space PI b}
\ee
where we indicate by the subscrips "p" or "ap" that periodic or antiperiodic boundary conditions 
have to be taken for the  $x$ or $\cy{\xi}$ integral. 
From Eq. (\ref{H spin}) we see that the Hamilton function
\be
H( p,x,\cy{\xi}) \E - \Pi^2 + \frac{i e}{2} \cy{\xi_{\mu}} \cy{\xi_{\nu}}
F^{\mu \nu}
\ee
is a quadratic function of the kinematic momentum
\be
\Pi_{\mu} \E p_{\mu} - e A_{\mu} \> .
\ee
By shifting the integration variable in the phase-space path integral
(\ref{phase space PI b}) we may perform the (functional) momentum integration and obtain
\be
\boxed{
\qquad {\rm tr} \, \hat U(T,0) \E
\oint\limits_p {\cal D}'x  \, \oint\limits_{ap} {\cal D}\cy{\xi} \> 
{\cal N}_0(T) \, 
\exp \left \{ \> i \int_0^T dt \> L(x,\dot x, \cy{\xi}, \dot{\cy{\xi}}) \> \right \} \qquad
}
\label{PI a}
\ee
with the normalization
\be
{\cal N}_0(T) \E \left [ \int {\cal D}\cy{\xi} \> \exp \left ( 
- \frac{1}{4}\int_0^T dt \> \cy{\xi} \cdot  \dot{\cy{\xi}} \right ) \right ]^{-1}
 \> \cdot \> \int \frac{{\cal D}\Pi}{(2 \pi)^4} ~\exp \left ( i \int_0^T dt \>
\Pi^2 \right )
\label{norm}
\ee
and the Lagrange function 
\be
\boxed{
\qquad L \lrp x,\dot x,\cy{\xi},\dot{\cy{\xi}} \rrp \E  -\frac{1}{4} \dot{x}^2 +
\frac{i}{4} \cy{\xi} \cdot \dot{\cy{\xi}} - e \, \dot{x} \cdot A(x) - \frac{ie}{2}
F_{\mu\nu} (x)\cy{\xi^{\mu}}\cy{\xi^{\nu}} \> . \quad 
}
\label{L spin}
\ee
As can be seen, the spin of the relativistic particle is
described by a Grassmann trajectory  $ \cy{\xi_{\mu}}(t) $ over which one has     
to integrate functionally with antiperiodic boundary conditions. The Lagrange function  
(\ref{L spin}) has a simple interpretation: It contains kinetic terms for the orbital and
spin motion of a relativistic particle and a coupling of the spin current 
$ \, \cy{\xi^{\mu}}\cy{\xi^{\nu}}\,  $ to the
electromagnetic field strength (non-relativistically this is the well-known 
$ {\bfx \sigma} \cdot {\bf B} $ term). In addition, 
$ \, \dot{x} \cdot A(x) $ describes the usual interaction of the convection current with the vector potential.

\noindent
The effective action of \rot{\bf QED} therefore is given in one-loop approximation
by
\be
\Gamma^{(1)}[A] \E 2i  \int_0^{\infty} dT \> 
\frac{e^{-i m^2 T}}{T}  \! \!    
\oint\limits_{x(0)=x(T)} \! \! {\cal D}'x \! \! 
\oint\limits_{\cy{\xi}(0)=-\cy{\xi}(T)} \! \! {\cal D}\cy{\xi} \> {\cal N}_0(T) \> 
\exp \left \{ \> i \int_0^T dt \> L(x,\dot x, \cy{\xi}, \dot{\cy{\xi}})
\> \right \} \> .
\label{Gammaf 1b}
\ee
It is advantageous to split the trajectory into
\be
x(t) \E x_0 + y(t) \> , \hspace{0.5cm} {\rm with} \> \> \> 
\int_0^T dt \> y(t) \E 0.
\label{Null Modus}
\ee
and to integrate separately over the "zero mode" $ x_0 $ . Then we obtain
\bea
\Gamma^{(1)}[A] \EA 2i  \int_0^{\infty} dT \>
\frac{e^{-i m^2 T}}{T}  \int d^4x_0 \> 
\underbrace{ \frac{ \int {\cal D}y {\cal D}\cy{\xi}
\> \exp \left ( i S[y,\cy{\xi}] \right )}{\int {\cal D}y {\cal D}\cy{\xi}
\> \exp \left (i S_0[y,\cy{\xi}] \right )}}_{:= \left < \,  \exp (i(S - S_0) \, 
\right >} \non
&& \cdot 
\int\limits_{y(0)=y(T)} \! \! {\cal D}'y \! \!
\oint\limits_{\cy{\xi}(0)=-\cy{\xi}(T)} \! \! {\cal D}\cy{\xi} \> N_0(T) \>
\exp \left \{ \> i \int_0^T dt \> L_0(x,\dot x, \cy{\xi}, \dot{\cy{\xi}}) \> \right \} 
\> ,
\label{Mittelung}
\eea 
where  $ L_0 $ is the free Lagrange density ($ e = 0 $ ). The free path integral can
be performed immediately: Due to the normalization factor (\ref{norm}) the spin integral 
is one and the functional  $y$-integral is that of a free particle with mass
 $ -1/2 $ ($ 1/2$) for the temporal (spatial) component. From  Eq. (\ref{freier Prop 2})
 one therefore obtains
\be
\int\limits_{y(0)=y(T)} \! \! {\cal D}'y \! \!
\oint\limits_{\cy{\xi}(0)=-\cy{\xi}(T)} \! \! {\cal D}\cy{\xi} \> {\cal N}_0(T) \>
\exp \left \{ \> i \int_0^T dt \> L_0(x,\dot x, \cy{\xi}, \dot{\cy{\xi}}) \> \right \} 
\E 
\sqrt{ \frac{-1/2}{2 \pi i T}} \sqrt{ \frac{1/2}{2 \pi i T}}^3 \E 
\frac{i}{(4 \pi i T)^2} \> .
\ee
If we write the effective action as 
\be
\Gamma^{(1)}[A] \E \int d^4 x_0 \> \delta {\cal L}[A]
\ee
then the first quantum correction to the classical Lagrangian is given by
\be
\delta {\cal L}[A] \E \frac{1}{8 \pi^2} \int_0^{\infty} dT \> 
\frac{e^{-i m^2 T}}{T^3}
\> \left < \> 
\exp \left \{ \> - i e \int_0^T dt \> \left [\, \dot y \cdot A +
\frac{i}{2} F_{\mu \nu} \cy{\xi^{\mu}} \cy{\xi^{\nu}} \, \right ] \, \right \}
\> \right > \> .
\label{Gammaf 1c}
\ee
Here the averaging is defined as in Eq. (\ref{Mittelung}) 
as ratio of the path integral with interaction to the free path integral.

\vspace{1.4cm}

\noindent
{\bf Example: The Euler-Heisenberg Effective Lagrangian}\\

\noindent
Let us consider the case of a {\bf constant} electromagnetic field $F_{\mu \nu}$, for which
the \blau{\bf Fock-Schwinger gauge}
\be 
A_{\mu}(y) \E \frac{1}{2} y^{\nu} F_{\nu \mu}
\ee
is most convenient.  Then the action is quadratic in the bosonic (b) as well as in the fermionic
(f) components and can be written as
\be
S[y,\cy{\xi}] \E \Bigl (x_{\mu}, {\cal O}_b^{\mu \nu} \, x_{\nu} \Bigr )
+ \left (\cy{\xi_{\mu}}, {\cal O}_f^{\mu \nu} \, \cy{\xi_{\nu}} \right )
\ee
with
\bea
{\cal O}_b^{\mu \nu}(t-t') \EA \left [ \> \frac{1}{4} g^{\mu \nu} 
\frac{\partial^2}{\partial t^2} 
- \frac{e}{2} F^{\mu \nu} \frac{\partial}{\partial t} \> \right ] \, 
\delta \left ( t - t' \right ) \\
{\cal O}_f^{\mu \nu}(t-t') \EA \left [ \> \frac{i}{4} g^{\mu \nu}
\frac{\partial}{\partial t} 
- \frac{ie}{2} F^{\mu \nu} \> \right ] \, \delta \left (t - t' \right )
\> .
\eea
Note that $ \> {\cal O}_b = - i \partial \, {\cal O}_f/ \partial t 
\> $ which is  an expression of a  \textcolor{blue}{\bf super-symmetry} between bosonic
and fermionic components \cite{BDZVH}.

As the path integral (\ref{Gammaf 1c}) is now a Gaussian we can perform all
functional integrations immediately and obtain 
\bea
\delta {\cal L}[A] \EA \frac{1}{8 \pi^2} \int_0^{\infty} dT \> 
\frac{e^{-i m^2 T}}{T^3} \> \left ( 
\frac{\fdet_{ap} {\cal O}_f}
{\fdet'_p {\cal O}_b} \right )^{1/2}  \> \cdot \left (
\frac{\fdet'_{p} {\cal O}_b^{(0)}}
{\fdet_{ap} {\cal O}_f^{(0)}} \right )^{1/2}  \non
\EA \frac{1}{8 \pi^2} \int_0^{\infty} dT \>
\frac{e^{-i m^2 T}}{T^3} \> 
\fdet^{' \> -1/2}_{p} \left [ 1 - 2e 
\frac{1}{\partial_t^2} \, F \, \partial_t \right ] \> 
\fdet^{1/2}_{ap} \left [ 1 - 2e \frac{1}{\partial_t} \, F \right ]
\> .
\label{delta L}
\eea
Due to the super-symmetry the bosonic and the fermionic
determinant would cancel (up to a constant) but the boundary conditions are different.
Another difference is that the zero mode  $x_0$ has been eliminated in the bosonic determinant which
is indicated by a prime:  $ \fdet'$. ``Det'' here means a determinant both in functional space
as in the Lorentz indices; hence only a power $ 1/2 $ (and not $ d/2 = 2 $) of the determinants
appear. The inverse operators (or worldline Green functions) $ 1/\partial_t^2 $ and $ 1/\partial_t$
can be calculated by expanding their eigenfunctions in Fourier modes 
\be
\sum_{\latop{k = -\infty}{k \ne 0}}^{+\infty} b_k \> 
e^{2 \pi i k t/T} \> , \hspace{0.5cm} {\rm or}\hspace{0.5cm}
\sum_{k = -\infty}^{+\infty} f_k \> e^{2 \pi i (k+ 1/2) t/T} \qquad  {\rm respectively}    
\ee
in the bosonic and fermionic case, respectively. These expansions obey the boundary conditions,
eliminate the zero mode and diagonalize the corresponding operator. Applying the $ (\ln \det = {\rm tr} \ln) $-rule then gives for the bosonic determinant
\bea
\fdet'_p \EA \exp \left \{ \> {\rm tr} \sum_{\latop{k = -\infty}{k \ne 0}}
^{+\infty} \ln \left [ \, 1 - \frac{2e}{2 \pi i k/T} F \, \right ] \> 
\right \} \E 
\exp \left \{ \> {\rm tr} \sum_{k =1}^{\infty} \ln \left [ \, 1 + 
\frac{e^2 T^2}{k^2 \pi^2} F^2 \, \right ] \> \right \} \non
\EA \exp \left \{ \> - \sum_{n=1}^{\infty} 
\left ( \frac{- e^2 T^2}{\pi^2} \right )^n
\frac{1}{n} \> \zeta(2n) \> {\rm tr} \, F^{2n} \> \right \} \> .
\eea
In the last line we have expanded the logarithm:
~$ \ln (1 + x) = x - x^2/2 + x^3/3 + \ldots $ and used the Riemann Zeta-function 
 \footnote{See, e.g. , {\bf \{Gradshteyn-Ryzhik\}}, eq. 0.233.3 .}
\be
\zeta(2n) \E \sum_{k=1}^{\infty}\> \frac{1}{k^{2n}} \E 
\frac{2^{2n-1} \pi^{2n}}{(2 n) !} \left | B_{2n} \right | \> .
\ee
$ B_0 = 1, \> B_2 = 1/6, \> B_4 = - 1/30 $ etc.
are the Bernoulli numbers  and ``tr'' denotes the trace over Lorentz indices.
The fermionic determinant is calculated in a similar way:
\bea
\fdet_{ap} \EA \exp \left \{ {\rm tr} \sum_{k = -\infty}
^{+\infty} \ln \left [ \, 1 - \frac{2e}{2 \pi i (k+1/2)/T} F \, \right ] \>
\right \} \non
\EA \exp \left \{ - \sum_{n=1}^{\infty} \left ( \frac{e T}{i \pi} \right )^n
\frac{1}{n} \, {\rm tr} \, F^n  \sum_{k=-\infty}^{+\infty} \! \left ( 
k + \frac{1}{2} \right )^{-n}
\right \} \>.
\eea
The trace over odd powers of the antisymmetric field strength tensor 
$ F_{\mu \nu} $ vanishes and if we use the formula \footnote{{\bf \{Gradshteyn-Ryzhik\}}, eq. 0.233.5 .}
\be
\sum_{k=-\infty}^{+\infty} \> \frac{1}{(k + 1/2)^{2n}} \E 2^{2n+1}
\sum_{k=1}^{\infty} \> \frac{1}{(2k-1)^{2n}} \E 
2^{2n} \frac{2^{2n}-1}{(2n)!} \pi^{2n} \left | B_{2n} \right |
\ee
we obtain for the fermionic determinant
\be
\fdet_{ap} \E \exp \left \{ \> - \sum_{n=1}^{\infty} 
\left ( - 4 e^2 T^2 \right )^n \, \frac{2^{2n} -1}{2 n (2n)!} 
\left | B_{2n} \right | \> {\rm tr} \, F^{2n} \> \right \} \> .
\ee
Inserted into Eq. (\ref{delta L}) this gives
\be
\delta {\cal L}[A] \E \frac{1}{8 \pi^2}  \int_0^{\infty} \frac{dT}{T^3} \>
e^{- i m^2 T}  \exp \left \{ \> - \sum_{n=1}^{\infty}
 \, \frac{(-4 e^2 T^2)^n}{2 n (2n)!} \left ( 2^{2n-1} - 1 \right ) 
\left | B_{2n} \right | \> 
{\rm tr} \, F^{2n} \, \right \} \> .
\ee
Although one can do the summation analytically \footnote{Schwinger's classical result
can be found , for example, in \meingruen{\bf Itzykson \& Zuber}, p. 196.}, it is more instructive
to expand the individual terms and to integrate term-by-term over  $ T $ :
The $ F $-independent term gives a (divergent) constant, the quadratic one 
renormalizes the classical Lagrangian  (\ref{Gammaf 0}) while the term quartic in $ F $
gives an additional piece
\be
\delta {\cal L}^{(4)} \E \frac{1}{8 \pi^2} 
\int_0^{\infty} \frac{dT}{T^3} \>e^{- i m^2 T}  \left \{ \> - \frac{7}{180}
e^4 T^4 \> {\rm tr} \, F^4 \> + \> \frac{1}{72} e^4 T^4 \> 
\left ( {\rm tr} \, F^2 \right )^2 \> \right \} \> .
\ee
Because of 
\bea
{\rm tr} \, F^2  \EA 2 \left ( \, {\bf E}^2 - {\bf B}^2 \, \right ) \\
{\rm tr} \, F^4  \EA 2 \left ( \, {\bf E}^2 - {\bf B}^2 \, \right )^2 + 
4 \left (\, {\bf E} \cdot {\bf B} \, \right )^2
\eea
we finally obtain
\be
\boxed{
\qquad \delta {\cal L}^{(4)} \E \frac{2 \alpha^2}{45 m^4} \left [ \> 
({\bf E}^2 - {\bf B}^2)^2 + 7 ({\bf E} \cdot {\bf B})^2 \> \right ] \> , \quad 
}
\ee
\vspace{0.1cm}

\noindent
where $ \alpha = e^2/(4 \pi) \simeq 1/137.036 $ denotes the fine-structure constant.
This is the effective Lagrangian of 
\blau{H. Euler} \footnote{Ann. Phys. {\bf 26} (1936) 398. Of course, the great
L. Euler whose name is attached to the  $\Gamma$-function and many other results, lived much earlier.
Also  at that time ``Ann. Phys.'' was the abbreviation for  ``Annalen der Physik''...}
and \blau{Heisenberg} describing non-linear effects in electrodynamics due to quantum corrections.
\vspace{2cm}

\renewcommand{\baselinestretch}{0.9}
\scriptsize
\refstepcounter{tief}
\noindent
\blau{\bf Detail \arabic{tief}:} {\bf Color in the Path Integral}\\
\vspace{0.6cm}

\begin{subequations}
\noindent
For the treatment of spin in the world-line path integral we have used  special properties of the Dirac matrices.
How can that be taken over to other inner degrees of freedom of fermions, for example the color of quarks? One possible method using fermionic auxiliary variables has been developed by 
 D'Hoker and Gagn\'{e} \cite{DHGa} which we will follow here. 
 We consider that part in Eq. \eqref{G hat with diff} which describes the dynamics of the inner degrees of freedom and leave out the orbital motion which is treated as usual.
 This means: We want to find a path-integral representation for
\be 
Z[M] \Def  {\rm tr} \lrp {\cal T} \, e^{i \int_0^T d\tau \, M(\tau)} \rrp \deF {\rm tr} \, U(T,0)
\label{Z DHG}
\ee
where $ M (\tau) $ is a traceless, hermitean $ N \times N $ matrix which describes the inner degrees of freedom -- e.g., a combination of Gell-Mann matrices. Again the task is to replace the time-ordering symbol (which is necessary because $ M(\tau) $ doesn't commute at different times)
by a suitable functional integral over a (Grassmann even) action made up by auxiliary fields.
\vspace{0.1cm}

\noindent
{\bf Assertion}:  Eq. \eqref{Z DHG} can be represented by the worldline path integral
\be 
Z[M] \E {\cal N} \, \int_0^{2 \pi} d\phi \, \int_{\rm ap}
{\cal D} \cy{\bar \lambda} \,  {\cal D} \cy{\lambda} \> \exp \lcp i \phi \lrp \cy{\bar \lambda} \cy{\lambda} + 
\frac{N}{2} - 1 \rrp \rcp \> \exp \lcp - \int_0^T d\tau \, \lsp \cy{\bar \lambda} \cy{\dot \lambda} 
- i \cy{\bar \lambda} M(\tau) \cy{\lambda} \rsp \rcp \> .
\label{Z Farbe}
\ee
Here  $ \cy{\bar \lambda}, \cy{\lambda} \> \> $ are $N$-component  Grassmann variables over which
one has to integrate functionally with
anti-periodic ("ap") boundary conditions $ \> \cy{\bar \lambda}(T) = - \cy{\bar \lambda}(0) , \> \cy{\lambda}(T) = - \cy{\lambda}(0) \> $ . In the first exponential factor they can be taken at an arbitrary time between $0$ and $T$ , i.e. one is allowed to write there 
$ \frac{1}{T} \int_0^T d\tau \cy{\bar \lambda} \, \cy{\lambda} $ . $ {\cal N}$  is a normalization
factor to be determined.
\vspace{0.1cm}

\noindent
{\bf Proof}: Grassmann integration over $ \cy{\bar \lambda}, \cy{\lambda} $ (with the help of  Eq. \eqref{Gauss fermionisch}) gives
\be 
Z[M] \E {\cal N} \, \int_0^{2 \pi} d\phi \, \fdet \lsp \frac{d}{d\tau}
- i M - \frac{i \phi}{T} \rsp \, e^{i \phi (N/2 - 1)} \> .
\ee
As usual the functional determinant is calculated as infinite product of the eigenvalues
of the corresponding operator. The eigenfunctions $ f(\tau) $ are solutions of the equation
$ d f/d\tau \E \lrp  i M + i \phi/T + \kappa \rrp \, f $, i.e.
\be
f(\tau) \E \exp \lrp \kappa \tau + i \frac{\phi \tau}{T} \rrp \, U(\tau,0) \, f(0) \> .
\ee
Since $ M $ is hermitean, $ U(\tau,0) $ in Eq. \eqref{Z DHG} is unitary at every time and we can write
\be 
U(\tau,0) \EQ {\cal T} \, \exp \lsp i \int_0^{\tau} d\tau' M(\tau') \rsp \E \exp \Big [ \, i \, {\rm diag}(m_1(\tau), \ldots m_N(\tau) \, \Big ] \> , 
\ee
where
\be 
\sum_{n=1}^N m_n(\tau) \E 0
\label{spurlos}
\ee
 since  $ M $ is assumed to be traceless. The eigenvalues $ \kappa $ are determined by the boundary condition $ f(T) = - f(0) $ as
\be 
\kappa_n^{(k)} \E \frac{i}{T} \lsp (2 k + 1) \pi - \phi - m_n(T)  \rsp \> , \quad k = 0, \pm 1, \pm 2 \ldots\> .
\ee
With that the functional determinant is found to be
\be  
\fdet \lsp \frac{d}{d\tau}
-iM - \frac{i \phi}{T} \rsp \E \prod_{n=1}^N \, \prod_{k=0}^{\infty} \lsp \frac{\pi^2 (2 k + 1)^2}{T^2}
\rsp \, \prod_{l=0}^{\infty} \lsp 1 - \frac{(\phi + m_n(T) )^2}{(2 l + 1)^2 \pi^2} \rsp 
\E \fdet \lsp \frac{d}{d \tau}\rsp  \, \cdot \prod_{n=1}^N \cos \lrp \frac{\phi + m_n(T)}{2} \rrp \> .
\ee
In the last step the product representation of the cosine function (see, e.g.  eq. 1.431.3 in {\bf \{Gradshteyn-Ryzhik\}}) has been used. After inserting that and splitting the cosine into exponential functions we have
\bea 
Z[M] \EA \frac{{\cal N}}{2^N} \, \fdet \lsp \frac{d}{d \tau}\rsp \, \cdot \, \int_0^{2 \pi} d\phi \> 
e^{i (N/2-1) \phi} \exp \lrp - i \phi N/2 -
\frac{i}{2} \sum_{n=1}^N m_n(T) \rrp \, \prod_{n=1}^N \lsp 1 + e^{i (\phi + m_n(T))} \rsp \non
\EA \frac{{\cal N}}{2^N} \, \fdet \lsp \frac{d}{d \tau}\rsp \, \cdot \, \int_0^{2 \pi} d\phi \> e^{-i \phi} \, \lcp 1 + \sum_n  e^{i (\phi + m_n(T))} + \sum_{n, m < n}  e^{i (\phi + m_n(T))} 
\,  e^{i (\phi + m_m(T))} + \ldots \rcp
\eea
where Eq. \eqref{spurlos} has been used. Due to the orthogonality relation
 $ \int_0^{2 \pi} d\phi \exp ( i k \phi ) = 2 \pi \, \delta_{k0} $, obviously only the term which contains just one factor  $ e^{i\phi} $ contributes in the expansion of the product. 
Hence the desired result 
\be 
Z[M] \E \frac{ 2 \pi {\cal N} }{2^N} \, \fdet \lsp \frac{d}{d \tau}\rsp \cdot \sum_{n=1}^N 
e^{i m_n(T)} \E {\rm tr} \lrp {\cal T} e^{i \int_0^T d\tau M(\tau)} \rrp \> ,
\ee
follows if the normalization factor is chosen as $ {\cal N}^{-1} = 2 \pi \, \fdet \lsp \frac{d}{d \tau}\rsp/2^N $ .\\
\noindent
A method which employs bosonic auxiliary variables can be found in Ref. \cite{Lun}.

\end{subequations}

\renewcommand{\baselinestretch}{1.2}
\normalsize

\vspace{1.2cm}

\subsection{\textcolor{blue}{Anomalies}}
\label{sec3: Anomal}
Due to the global symmetry of the  \rot{\bf QCD} Lagrangian under the transformation
\be
\cy{\psi}(x) \To e^{ i \alpha } \, \cy{\psi}(x) \> , \hspace{0.5cm}
\cy{\bar{\psi}}(x) \>\longrightarrow\>   \cy{\bar{\psi}}(x) \, e^{ - i \alpha } 
\label{Vektor trans}
\ee
the \textcolor{blue}{\bf vector current}
\be
\boxed{
\qquad V_{\mu}(x) \E \cy{\bar{\psi}}(x) \gamma_{\mu} \cy{\psi}(x) \qquad
}
\label{Vektorstrom}
\ee 
is conserved in  \rot{\bf QCD}. As the masses of $u$-, $d$- and 
$s$-quarks  in nature are small compared to a typical hadronic scale
(estimates are  $m_{u/d} < 10$ MeV, $m_s \simeq 130$ MeV),
they can be neglected in many cases. Then the resulting Lagrangian
\be
{\cal L}_{m=0} \E - \frac{1}{4} F^{a \> \mu \nu} F^a_{\mu \nu} 
 + \cy{\bar{\psi}}(x) \left ( i \dslash - g \Aslash^a T^a \right )
\cy{\psi}
\ee
is also invariant under an \textcolor{blue}{ axial} transformation
\be
\cy{\psi}(x) \To e^{ i \alpha \gamma_5} \, \cy{\psi}(x) \> , 
\hspace{0.5cm}
\>  \cy{\bar{\psi}}(x) \To  \cy{\bar{\psi}}(x) \, 
e^{ i \alpha \gamma_5} \> ,
\label{Axial trans}
\ee
where $ \> \gamma_5 = i \gamma_0 \gamma_1 \gamma_2 \gamma_3 \> $ anticommutes with all
Dirac matrices. According to the classical Noether theorem the 
\textcolor{blue}{\bf axial current}
\be
\boxed{
\qquad A_{\mu}(x) \E \cy{\bar{\psi}}(x) \gamma_{\mu} \gamma_5 \cy{\psi}(x) \qquad
}
\label{Axialvektorstrom}
\ee
should also be conserved in this limit:
\be
\partial^{\mu} A_{\mu}(x) \> \stackrel{?}{ = } \> 0 \> .
\label{falsche A-Strom Erh.}
\ee
The fact that this does {\bf not} hold for the quanzized theory is called
an "\blau{anomaly}".
Besides the theoretical interest for this phenomenon there are also practical reasons to study anomalies: For example, the decay $ \> \pi^0 \to 2 \gamma \> $ can only be obtained
in the strength observed experimentally when the axial anomaly is taken into account!

\vspace{0.2cm}
How does the axial (or Adler-Bell-Jackiw (ABJ)) anomaly show up in the
path-integral representation? To answer this question we recall
{\bf chapter} {\bf \ref{sec1: Symm}} where we derived Noether's theorem in quantum mechanics.
As seen in \purpur{\bf Problem \ref{Noether Feld}} this applies to field theory
in a completely analogous way: We consider the full generating functional (gauge fixing and ghosts are irrelevant and are omitted to simplify the notation)
\be 
Z[\cy{\eta}, \cy{\bar{\eta}}] \E \int {\cal D} \cy{\bar{\psi}}(x) \, {\cal D} \cy{\psi}(x) \,
{\cal D} A(x)
\> \exp \left [ \> i \int d^4x \> \left ( {\cal L}_{m=0}(\cy{\psi},\cy{\bar{\psi}}, A)
+ \cy{\bar{\psi}} \cy{\eta} + \cy{\bar{\eta}} \cy{\psi} \right ) \> \right ] \> .
\ee
and perform, for instance  the vector transformation (\ref{Vektor trans})
as a \blau{local} substitution of integration variables, i.e we set
\be
\cy{\psi}(x) \E  e^{ - i \alpha(x) } \, \cy{\psi}'(x) \> , \hspace{0.5cm}
\cy{\bar{\psi}}(x) \E \cy{\bar{\psi}}'(x) \, e^{ i \alpha(x) } \> .
\label{vektor lokal}
\ee
As the numerical value of the integral does not change we have
\bea
&& \hspace{-1cm}\int {\cal D} \cy{\bar{\psi}}(x) \, {\cal D} \cy{\psi}(x) \, 
{\cal D} A(x)
\> \exp \left [ \> i \int d^4x \> \left ( {\cal L}_{m=0}(\cy{\psi},\cy{\bar{\psi}}, A)
+ \cy{\bar{\psi}} \cy{\eta} + \cy{\bar{\eta}} \cy{\psi} \right ) \> \right ] \non
&&\hspace{-1cm} = \> \int {\cal D} \cy{\bar{\psi}}'(x) \, {\cal D} \cy{\psi}'(x) \, 
{\cal J}^{-1} \, {\cal D} A(x)
\> \exp \left [ \> i \int d^4x \> \left ( {\cal L}_{m=0}(\cy{\psi},\cy{\bar{\psi}}, A)
+ \cy{\bar{\psi}} \cy{\eta} + \cy{\bar{\eta}} \cy{\psi} \right ) \> 
\right ]_{\latop{\cy{\psi} = \exp (-i \alpha(x))\cy{\psi}'}{
\cy{\bar{\psi}} = \cy{\bar{\psi}}' \exp (i \alpha(x))}} \> . 
\label{Pfadint trans}
\eea
Here
\be
{\cal J} \E \fdet _{x x'} \left ( \frac{\partial \cy{\psi}(x)}
{\partial \cy{\psi}'(x')} \right ) \, \cdot \, \fdet _{x x'} \left ( 
\frac{\partial \cy{\bar{\psi}}(x)}
{\partial \cy{\bar{\psi}}'(x')}\right )
\ee
is the \textcolor{blue}{ Jacobi determinant} of the transformation
(\ref{vektor lokal}) which -- as we know from {\bf chapter} {\bf \ref{sec2: Fermionen}} -- 
appears inversely when integrating over Grassmann variables.
Calculating the Jacobi determinant of the vector transformation
(\ref{vektor lokal}) we obtain
\be
{\cal J}_V \E \fdet _{x x'} \left (  \, e^{-i \alpha(x)} 
\delta (x - x') \, \right ) \, \fdet _{x x'} \left (  \, e^{i \alpha(x)}
\delta (x - x') \, \right ) \E 1  \> ,
\label{Jacobi vektor}
\ee
since this is an unitary (phase) transformation. Restricting ourselves to
\textcolor{blue}{ infini\-te\-simal} transformations we expand 
Eq. (\ref{Pfadint trans}) up to first order in $\alpha(x) $ to get
\bea
0 \> \EA \> i \,
\int {\cal D} \cy{\bar{\psi}}(x) \, {\cal D} \cy{\psi}(x) \,
{\cal D} A(x) \> \left \{ \> i \delta {\cal J} + 
\,\int d^4x \> \left [ \, \delta {\cal L}_V 
- i \alpha(x) \, \cy{\bar{\psi}} \cy{\eta} + 
i \alpha (x)  \, \cy{\bar{\eta}} \cy{\psi} \, \right ] \> \right \}  \non
&& \hspace{5cm} \cdot \exp \left [ \> i \int d^4x \> \left ( 
{\cal L}_{m=0}(\cy{\psi},\cy{\bar{\psi}}, A)
+ \cy{\bar{\psi}} \cy{\eta} + \cy{\bar{\eta}} \cy{\psi} \right ) \> \right ] \> . 
\label{Pfad diff}
\eea
Here $ \delta {\cal J}_V = 0 $  is the contribution of the  Jacobian
and
\be
\delta {\cal L}_V \E  \cy{\bar{\psi}}  \gamma^{\mu} \cy{\psi} \, \partial_{\mu} 
\alpha(x)
\ee
the infinitesimal change of the Lagrangian caused by the vector transformation (\ref{vektor lokal}).
As we only have made a local transformation of the fermion fields
it doesn't vanish  \footnote{Recall that in a local \blau{gauge} transformation the gauge field has to transform in a very specific way to compensate that change. Here, however, the gauge field remains 
unchanged.}.
With an integration by parts we can get rid of the derivative acting on  $\alpha(x)$ and obtain
\bea
0 \> \EA \> i \, 
\int {\cal D} \cy{\bar{\psi}}(x) \, {\cal D} \cy{\psi}(x) \, 
{\cal D} A(x)
\> \int d^4x \> \alpha(x) \left \{ \, - \partial_{\mu} V^{\mu}(x)
- i \, \cy{\bar{\psi}} \cy{\eta} + i \, \cy{\bar{\eta}} \cy{\psi} \, \right \}  \non
&& \hspace{4cm}\cdot \exp \left [ \> i \int d^4x \> \left ( 
{\cal L}_{m=0}(\cy{\psi},\cy{\bar{\psi}}, A)
+ \cy{\bar{\psi}} \cy{\eta} + \cy{\bar{\eta}} \cy{\psi} \right ) \> \right ] \> . 
\eea
Here  $V^{\mu}(x) $ exactly is the vector current (\ref{Vektorstrom}).
Since  $\alpha(x)$ is arbitrary the whole functional integral over the 
curly bracket must vanish and this for arbitrary values of the external source
 $ \> \cy{\eta}(x), \, \cy{\bar{\eta}}(x)  \> $. Using again the notation
\be
\left < {\cal O} \right > \EQ
\int {\cal D} \cy{\bar{\psi}}  \, {\cal D} \cy{\psi} \, {\cal D} A \> 
{\cal O}(\cy{\psi}, \cy{\bar{\psi}}, A) \> e^{ i \int d^4 x {\cal L}_{m=0} (\cy{\psi}, \cy{\bar{\psi}},A)}
\label{def Mittelung}
\ee
we therefore see that for $ \cy{\bar{\eta}} = \cy{\eta} = 0 $ Noether's theorem
for the vector transformation in the path-integral formulation simply reads
\be 
\boxed{
\qquad \left < \> \partial_{\mu} V^{\mu} \> \right > \E 0 \> .\qquad
}
\ee
By differentiation w.r.t. the external sources
$ \> \cy{\bar{\eta}}, \, \cy{\eta}  \> $ additional 
\textcolor{blue}{ Ward identities} can be obtained which are exact relations
between $n$-point functions into which the conserved Noether current has been inserted. 

\vspace{0.2cm}
If we would carry out this derivation as well for the axial vector transformations 
(\ref{Axial trans}) we would obtain Ward identities for the axial current, in particular
also $ \, \left < \partial_\mu A^{\mu} \right > = 0 \, $. However, this {\bf not correct}
because in this case the Jacobian gives a contribution as first recognized by 
\blau{\bf Fujikawa} \cite{Fuji}. The Jacobi determinant of the axial transformation
namely is
\be
{\cal J}_A \E \fdet \left ( e^{-i \alpha \, \gamma_5} \right ) \, \cdot \,
\fdet \left ( e^{-i \alpha \, \gamma_5} \right ) \E e^{ \> - 2 i \, 
{\rm tr} \,  \alpha \, \gamma_5 } \> ,
\ee
where we have used again the formal relation between determinant and the trace of 
the logarithm of a matrix (see \purpur{\bf Problem \ref{Det Spur}}).            
For infinite matrices this expression (in general) is divergent
and Fujikawa has proposed to use a \blau{gauge invariant} regularization
which suppresses the high-energy modes of the Dirac field:
\be
{\cal J}_A \E \lim_{M \to \infty} \exp \left [ \> - 2 i \, 
{\rm tr} \,  \left ( \alpha \, \gamma_5 \, e^{-\dddslash^2/M^2} \right ) 
\> \right ] 
\> ,
\ee
where $D_{\mu}$ is the covariant derivative. Other regularizations yield the same
result provided they conserve the vector current.
Recall the relation
\be
\ddslash^2 \E D^2 + \frac{1}{4} \left [ \gamma^{\mu}, \gamma^{\nu} \right ]
 \, \left [ D_{\mu}, D_{\nu} \right ] \E D^2 + \frac{g T^a}{2}
\, \sigma^{\mu \nu} \, F_{\mu \nu}^a
\ee
which we already have encountered in the abelian case in 
{\bf chapter} {\bf \ref{sec3: Weltlinien}} \footnote{
$ \sigma_{\mu \nu} = i [\gamma_{\mu}, \gamma_{\nu}]/2 \> $ ,
$T^a$ are the generators of the $SU(N)$ Lie algebra.}. We now have to calculate
\be
\lim_{M \to \infty} {\rm tr} \,  \left ( \gamma_5 \, e^{-\dddslash^2/M^2} 
\right ) \E  \lim_{M \to \infty} {\rm tr} \, \int d^4x \> \left < x \, 
\left |
\gamma_5 \exp \left [ \, - \frac{1}{M^2} \left ( D^2 + \frac{g T^a}{2}
\, \sigma^{\mu \nu} \, F_{\mu \nu}^a \right ) \, \right ] \right | \, 
x \right > \> .
\ee
As the regulator mass $M$ should go to infinity we can focus our attention to the 
asymptotic part of the spectrum in which the momentum of the Dirac field is large 
while the gauge field remains bounded. If we expand in powers of 
 $ F_{\mu \nu}^a $ we have to "bring down" four Dirac matrices from the exponential
 function in order to obtain a non-vanishing trace with
$ \gamma_5$. This is because  tr $\gamma_5$ = tr $ \gamma_5 \gamma_{\mu} \gamma_{\nu} = 0 $.
Then the leading term of this \textcolor{blue}{`` heat kernel''} expansion
is the one in which we expand the exponential function up to order
$ (\sigma \cdot F)^2 $ and neglect the gauge field in all other terms. This gives
\be
\lim_{M \to \infty} {\rm tr} \,  \left ( \gamma_5 \, e^{-\dddslash^2/M^2}
\right ) \E  \lim_{M \to \infty} \, \int d^4x \> 
{\rm tr} \left [ \, \gamma_5 \, \frac{1}{2} \left ( \frac{g T^a}{2 M^2}
\sigma_{\mu \nu} F_{\mu \nu}^a \right )^2 \, \right ] \, \left < x \, \left |
e^{-\partial^2/M^2} \right | \, x \right > \> .
\label{Fujikawa 1}
\ee
The matrix element in Eq. (\ref{Fujikawa 1}) has the value
\be
\left < x \, \left |
e^{-\partial^2/M^2} \right | \, x \right > \E \int \frac{d^4 k}{(2 \pi)^4}
\> e^{k^2/M^2} \E \frac{1}{(2 \pi)^4} \left ( - \pi M^2 \right )^{1/2}
\left ( \pi M^2 \right )^{3/2} \E \frac{i M^4}{16 \pi^2} \> .
\ee
The trace over the Dirac indices can also be calculated easily
\be
{\rm tr} \left ( \, \gamma_5 \sigma^{\mu \nu} \sigma^{\alpha \beta} \, 
\right ) \E - 4 i \epsilon^{\alpha \beta \mu \nu} \> ,
\ee
where $\epsilon^{\alpha \beta \mu \nu} $ is the total antisymmetric tensor
in four dimensions. Hence
\be
\lim_{M \to \infty} {\rm tr} \,  \left ( \gamma_5 \, e^{-\dddslash^2/M^2}
\right ) \E   
- \frac{g^2}{32 \pi^2} \, \epsilon^{\alpha \beta \mu \nu} 
\, {\rm tr} \left ( T^a T^b \right ) \, 
\int d^4 x \, F_{\alpha \beta}^a(x) \, F_{\mu \nu}^b(x) \> .
\ee
Using $ \> {\rm tr} \, T^a T^b = \delta_{a b}/2 \>  $ we thereby have for the Jacobian of the axial
transformation
\be
{\cal J}_A \E \exp \left [ \> - i \int d^4 x \, \alpha(x)  \, N_f \, 
\frac{g^2}{32 \pi^2} \, \epsilon^{\alpha \beta \mu \nu} F_{\mu \nu}^a(x)
F_{\alpha \beta}^a(x) \> \right ] \> .
\ee
The factor $ N_f = 3 $ originates from the trace over those individual quarks
which can be considered massless, i.e. over their 
``{\bf flavor}''.

By this somewhat lengthy calculation we have found how the the
integrand of the generating functional changes under an axial transformation.
For infinitesimal $\alpha(x)$ we then obtain with the help of Eq. (\ref{Pfad diff})
\bea
0 \> \EA \> i \,
\int {\cal D} \cy{\bar{\psi}}(x) \, {\cal D} \cy{\psi}(x) \,
{\cal D} A(x)
\> \int d^4x \> \alpha(x) \Biggl \{ \>  - \partial_{\mu} A^{\mu}(x)  +
N_f \frac{g^2}{32 \pi^2} \epsilon^{\alpha \beta \mu \nu} F_{\mu \nu}^a(x)
F_{\alpha \beta}^a(x) \non
&& \hspace{1cm} - i \alpha(x) \, \cy{\bar{\psi}} \cy{\eta} + 
i \alpha (x)  \, \cy{\bar{\eta}} \cy{\psi} \, \Biggl \} \,
\, \cdot \exp \left [ \> i \int d^4x \> \left ( {\cal L}_{m=0}(\cy{\psi},\cy{\bar{\psi}}, A)
+ \cy{\bar{\psi}} \cy{\eta} + \cy{\bar{\eta}} \cy{\psi} \right ) \> \right ] \> . 
\eea
In particular, for vanishing sources we have
\bce
\vspace{0.2cm}

\fcolorbox{blue}{white}{\parbox{7cm}
{
\bea
\left < \, \partial_{\mu} A^{\mu} \, \right> \E   N_f \frac{g^2}{32 \pi^2} 
\epsilon^{\alpha \beta \mu \nu} 
\left < \, F_{\mu \nu}^a F_{\alpha \beta}^a \,  \right > \> . \hspace*{5.5cm}
\label{Anomalie}
\eea
}}
\ece

\vspace{0.4cm}

\begin{figure}[hbtp]
\vspace*{-3cm}
\bce
\includegraphics[angle=0,scale=0.7]{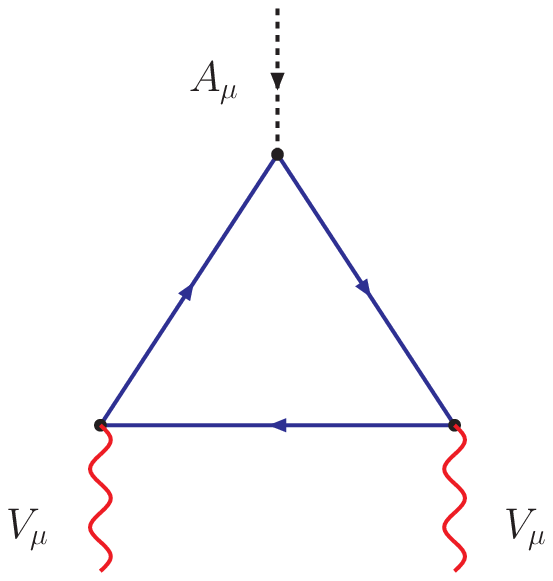}
\label{abb:3.5.1}
\ece
\vspace*{-10cm}
{\bf Fig. \arabic{abb}} : Triangle diagram for the axial current 
$A_{\mu}$ and two vector currents  $V_{\mu}$ which gives an anomalous\\
\hspace*{1.7cm} contribution for the divergence of the axial current.
\vspace*{0.4cm}

\end{figure}

\noindent
This is 
in direct correspondence with the ABJ relation in operator notation
which was first found by a perturbative analysis of diagrams like the one 
shown in Fig. \ref{abb:3.5.1}. 
The path-integral derivation has the advantage that it holds in all orders and 
thereby is in agreeement with the so-called Bardeen theorem which says that the anomaly 
is not changed in higher orders. The regularization of the divergent trace used above may 
look somehow {\it ad hoc} but it can justified by a more careful treatment (for instance, 
by a zetafunction regularization \cite{Reu}).

\refstepcounter{abb}

\vspace{0.3cm}

Finally we want to sketch how the anomaly determines the decay $ \> \pi^0 \to 2 \gamma \> $ :
Consider the $S$ matrix element
\be
\langle \gamma(k_1,\epsilon_1) \, \gamma(k_2,\epsilon_2) \, \left | \, S
\right |
\pi^0(q) \rangle \E i  (2 \pi)^4 \, \delta \left ( q - k_1 - k_2 \right)
\epsilon^{\mu}(k_1) \epsilon^{\nu}(k_2) \, \Gamma_{\mu \nu} \left ( k_1, k_2,
 q \right ) \> ,
\ee
where $ k_i, \epsilon_i , i = 1,2$ denote the momenta and polarisations of the photons. 
In the "soft pion limit" (i.e.  $ q_{\mu} \to 0 $) the amplitude
\be
\Gamma_{\mu \nu} \left ( k_1, k_2, q \right ) \E e^2 
\left (q^2 - m_{\pi}^2 \right ) \int d^4 x_1 \,
d^4 x_2 \> e^{i k_1 \cdot x_1 + i k_2 \cdot x_2} \, \langle 0 \left |
{\cal T} J_{\mu}^{\rm em}(x_1) J_{\nu}^{\rm em}(x_2) 
 \Phi_{\pi}(0) \right |  \rangle
\ee
can be related to the anomalous divergence of the axial current from the coupling of
the quarks (here not to the gluons but) to the electromagnetic current  $ J_{\mu}^{\rm em} $.
This can be done by means of the PCAC relation \footnote{``Partially conserved axial current''.
$ f_{\pi} \simeq 93 $ MeV is the pion decay constant.}
\be
\partial_{\mu} A_{\mu}^a \E f_{\pi} m_{\pi}^2 \Phi_{\pi}^a \> .
\hspace{0.5cm} a = 1, 2 ,3 \> .
\label{PCAC}
\ee
In Eq. (\ref{Anomalie}) we therefore have to replace $g^2$ by the electromagnetic coupling constant
$e^2 \simeq 4 \pi/137 $ and to modify the trace terms:
\be
\partial^{\mu} A_{\mu}^a \E \frac{e^2}{32 \pi^2} 
\epsilon^{\alpha \beta \mu \nu} F_{\mu \nu} F_{\alpha \beta} \>
\underbrace{{\rm tr} \left ( \tau^a Q^2 \right) }_{=: S^a} \> .
\ee
Here $ \tau^a $ is the isospin matrix of the pion and 
$Q$ the matrix of quark charges. The trace factor only contributes for
$ a = 3$ (i.e. for neutral pions) and has the value
\footnote{It is amusing that already in 1949 Steinberger has evaluated the triangle diagram with
{\it nucleons} in the loop \cite{Stein} for which one also obtains the value 1 .}
\be
S^{(a=3)} \E  N_c \left ( Q_d^2 - Q_u^2 \right ) = 
3 \cdot \left (\frac{4}{9} - \frac{1}{9} \right ) = 1 \> .
\ee
The further calculation for the decay rate \footnote{See, e.g., \meingruen{\bf Peskin \& Schroeder}, p. 674 - 676 .} then gives
\be
\boxed{
\qquad \Gamma (\pi^0 \to 2 \gamma) \E \frac{\alpha^2}{64 \pi^3} 
\frac{m_{\pi}^3}{f_{\pi}^2} \> ,\qquad
}
\ee
in good agreement with the experimental result $ \> \Gamma = 7.8 \pm 0.6 \> $ eV. Without the
color of the quarks running in the loop \footnote{A different interpretation is advocated in Ref. \cite{BaeWie}.} it would be smaller by a factor $ N_c^2 = 9 $. \\
(Much) more about anomalies in quantum field theory can be found in {\bf \{Bertlmann\}}.

\vspace{0.5cm}

\subsection{\textcolor{blue}{Lattice Field Theories}}
\label{sec3: Gitter}
Up to now 
in this field-theoretical part we have used mostly methods of perturbation theory
to evaluate path integrals. This is may be sufficient  for theories
with small coupling constants (like \rot{\bf QED} for which  $ \alpha \simeq 1/137 $)
but clearly is  not sufficient in the region of strong coupling which occur in 
\rot{\bf QCD} at low energies \footnote{The property of \blau{\bf asymptotic freedom}
of \rot{\bf QCD} allows application of perturbation theory also at high energies.}.
This is exactly the region in which quarks and gluons do not show up as 
asymptotically observable particles but combine to form the observed hadrons.
Moreover, perturbation theory is not complete: Phenomena like tunneling through a barrier
or non-linear soliton solutions of classical field equations and probably also 
the "\blau{\bf confinement}" of quarks and gluons in \rot{\bf QCD}
depend non-analytically on the coupling constant, i.e. are non-perturbative effects.

\noindent
The exact treatment of non-perturbative effects in the path integral
requires a (numerical) calculation of the functional integral as we already have done
in {\bf chapter} {\bf \ref{sec1: Numerik}} for quantum-mechanical problems.
We recall that we have discretized the path integral there, in accordance 
with its original, heuristic introduction, and evaluated in Euclidean time.
Similarly, the numerical treatment now will be performed on a finite 
\blau{\bf Euclidean space-time lattice} with lattice constant $ a $.
In field theory this has the additional advantage that all momentum integrals
are cut off at $\sim \> 1/a $ and thus the ultraviolet divergences are regularized automatically.
There is an obvious disadvantage that translation and rotation invariance are violated
and will only be restored if the lattice is made larger and finer.
\vspace{0.3cm}

We start our introduction in this vast theoretical field by first treating a scalar theory.
Consider the partition function
\be
Z \E \oint {\cal D} \Phi(x) \> e^{- S_E[\Phi]} \> ,
\ee
where
\be
S_E[\Phi] \E \int d^4x \> \left [ \, \frac{1}{2}
\left ( \partial \Phi \right )^2  + \frac{1}{2} m^2 \Phi^2
+ \frac{\lambda}{4 !} \Phi^4 \, \right ] \E \int d^4x \> \lsp  \frac{1}{2} \, \Phi(x) 
\lrp - \, \Box + m^2 \rrp \, \Phi(x) + \frac{\lambda}{4 !} \Phi^4(x) \rsp
\ee
is the Euclidean action (for instance of the  $ \Phi^4 $-theory). In the following
we will omit the index "E" as we will work exclusively in Euclidean space:
\be
x_0 \E - i x_4 \> ,
\ee
so that, for example,  $ \> \Box \Phi   \E \sum_{\mu=1}^4  \partial_{\mu}^2 \, \Phi \> $.

Let's apply a naive discretization: We introduce a 4-dimensional space-time lattice with 
 $ N $ lattice points and lattice distance $a$ in such a way that 
 each lattice point is determined by a 4-dimensional vector
 $n$: 
\be
x  \EQ a \, (n_1,n_2,n_3,n_4)
\E a \sum_{\mu=1}^4 \> n_{\mu} \, e_{\mu} \> ,
\ee
where $ e_{\mu} $ is a unit vector in direction  $\mu$. 

The integers $ n_{\mu} $ are between $ - N/2 $ and $ N/2 $ and it is advisable to extend them
outside that range:  $ \, n \, \hat = \, n + N$ (periodic lattice).
In the discrete case the 4-dimensional integration is replaced by
\be
\int d^4x \> \ldots \To a^4 \sum_n \> \ldots
\ee
and the scalar field exists at each lattice point
\be
\Phi(x) \To\Phi(n) \E \Phi(n_1,n_2,n_3,n_4)
\EQ \Phi_n \> .
\ee
Finally we replace the  Laplace operator by the simplest, symmetric form   
\be
\Box \, \Phi(x) \To\frac{1}{a^2} \, \sum_{\mu=1}^4 \lsp 
\Phi(n + e_{\mu}) + \Phi(n - e_{\mu}) - 2 \Phi(n) \rsp
\EQ \frac{1}{a^2} \,  \sum_{\mu=1}^4 \left ( \, \Phi_{n+\mu} + \Phi_{n-\mu} - 2 \Phi_n \, \right )
\> ,
\ee
which tends to the l.h.s. in the continuum limit $ \, a \to 0 \, $. With that
the lattice action (which is now a {\it function} of the fields  $ \Phi_n$) reads
\bea
S\{\Phi_n\} \EA \, \sum_{n} \lcp \frac{a^2}{2}                                      
\sum_{\mu=1}^4 \lsp - \Phi_n \Phi_{n+\mu} - \Phi_n \Phi_{n-\mu} + 2 \Phi_n^2 \rsp \, + 
a^4 \lrp \frac{m^2}{2} \Phi_n^2 + \frac{\lambda}{4 !} \Phi_n^4 \rrp  \rcp \non
\EA \sum_n \lcp - a^2 \, \sum_{\mu=1}^4 \Phi_n \Phi_{n+\mu} + \frac{a^2}{2} \lrp 8 + m^2 a^2 
\rrp \, 
\Phi_n^2 + a^4 \frac{\lambda}{4 !} \Phi_n^4 \rcp
\label{Gitter skalar}
\eea
and the partition function becomes
\be
Z \E \int \left ( \prod_n d \Phi_n \right ) \> e^{- S\{\Phi_n\}} \> .
\ee
It is obvious that the lattice action is not unique: It is possible to add arbitrary terms 
which vanish for $ \, a \to 0 \, .$  In particular, one may try to improve the discretization 
such that the errors are of higher order than the ones from the given "naive" form.
This is the basic philosophy behind the \blau{"improved actions"} which are used more and
more frequently \footnote{See, e.g. Ref. \cite{Lep}.}.
Similarly we calculate correlation functions
\be
\left < \Phi_{n_1} \, \Phi_{n_2} \ldots \Phi_{n_k} \right > \E 
\frac{1}{Z} \int \left ( \prod_n d \Phi_n \right ) \> 
\Phi_{n_1} \, \Phi_{n_2} \ldots \Phi_{n_k} \> e^{- S\{\Phi_n\}} \> .
\ee
As an example let's consider the 2-point function in the free discretized theory. We 
obtain (\purpur{\bf Problem \ref{Gitter 2-Punkt} a)}~)
\be
\boxed{
\qquad \left < \Phi_n \Phi_{n'} \right > \E \int_{-\pi/a}^{\pi/a} 
\frac{d^4 p}{(2 \pi)^4} \> \frac{1}{2 \sum_{\mu} \left [ \, 1 - 
\cos(a p_{\mu}) \right ]/a^2 + m^2} \> e^{i p \cdot x} \> , \quad    
}
\label{Gitter 2-Punkt-Funktion}
\ee
where  $ x_{\mu} = a (n_{\mu} - n'_{\mu}) $ is the distance on the lattice.
From this result we can see the following points:
\ben
\item[(1)] The momentum integration is cut off: $ |\> p_{\mu}| \le \pi/a 
\> $ and only extends over the first \textcolor{blue}{ ``Brillouin zone''}.
\item[(2)] For $ a \to 0 $ we obtain from the expansion
$ \> 1 - a \cos(a p_{\mu}) \to a^2 p_{\mu}^2 /2 + \ldots $ the usual
(Euclidean) Feynman propagator
\be
\left < \Phi_n \Phi_{n'} \right > \To
\int\limits_{-\pi/a \to -\infty}
^{\pi/a \to +\infty}
\frac{d^4 p}{(2 \pi)^4} \> \frac{1}{m^2 + p^2 + {\cal O}(a^2)} \> 
e^{i p \cdot x} \> .
\ee
\item[(3)] For $ a \ne 0 $ the dispersion law in the continuum is changed
\be
m^2 + p^2 \To m^2 + \frac{4}{a^2} \sum_{\mu} \sin^2 \left (
\frac{a p_{\mu}}{2} \right ) \> .
\ee
However, there are no additional zeros which would indicate (additional) asymptotically propagating  
particles. This is not the case if \blau{\bf fermions} are put on the lattice: As 
can be derived in 
\purpur{\bf Problem \ref{Gitter 2-Punkt} b)} in a naive discretization one obtains
\be
\boxed{
\quad \la \, \cy{\psi}_{\alpha}(n) \cy{\bar{\psi}}_{\beta}(n') \, \ra \E \int_{-\pi/a}^{\pi/a} 
\frac{d^4 p}{(2 \pi)^4} \> \lsp - i \sum_{\mu} \gamma^E_{\mu} \tilde p_{\mu}  + m
 \rsp_{\alpha \beta} \> 
\frac{\exp \lrp i p \cdot x \rrp }{m^2 + \sum_{\mu} \tilde p_{\mu}^2}
\> , \quad \tilde p_{\mu} \Def \frac{\sin \lrp a p_{\mu} \rrp }{a} \quad
} \> ,
\label{Gitter 2-Punkt fermion}
\ee
where the Euclidean Dirac matrices fulfill
$ [\gamma_{\mu}^E, \gamma_{\nu}^E ]_+ = 2 \delta_{\mu \nu} $ . Although 
$ \, \tilde p_{\mu} \to p_{\mu} \, $ for $ a \to 0 $,  the propagator
\eqref{Gitter 2-Punkt fermion} "repeats" itself at the end of the Brillouin zone since
$ \, \tilde p_{\mu} =  \sin ( \pi - p_{\mu}a )/a \to \pi/a - p_{\mu} $ for 
$ p_{\mu} \to \pi/a $.
This means that one doesn't have propagation of a single species of particles
but $2^4 -1 = 15 $ additional copies ... Euphemistically one calls that 
\blau{\bf ``fermion doubling''}.

\een
\vspace{0.3cm}

\noindent
It is possible to remove the coupling constant $ \lambda $ from the action by rescaling of the field:
\be 
\Phi_n' \E \sqrt{\lambda} \, \Phi_n \> .
\ee
Then we obtain
\be
S\{\Phi_n\} \E \frac{1}{\lambda} \, S'\{\Phi'_n\}
\ee
with
\be
S'\{\Phi_n'\} \E \sum_{n} \lcp \frac{a^2}{2}                                       
\sum_{\mu=1}^4 \lsp - \Phi'_n \Phi'_{n+\mu} - \Phi'_n \Phi'_{n-\mu} + 2 \Phi_n'^2 \rsp \, + 
a^4 \lrp \frac{m^2}{2} \Phi_n'^2 + \frac{1}{4 !} \Phi_n'^4 \rrp  \rcp 
\ee
and for the partition function (constant factors are irrelevant)
\be
Z \E \int \left ( \prod_n d \Phi'_n \right ) \> e^{- \frac{1}{\lambda} 
S'\{\Phi'_n\}} \> .
\ee
If we compare that with the partition function in statistical mechanics we see the
correspondence
\be
\qquad \frac{1}{\lambda} \> \longleftrightarrow \> \frac{1}{k_B T} \> ,\qquad 
\ee
i.e.  an \blau{expansion for strong couplings} corresponds to a \blau{high-temperature expansion}
of statistical systems.
\vspace{0.2cm}

\renewcommand{\thesubsubsection}{\textcolor{blue}{3.\arabic{subsection}.\arabic{subsubsection}}}
\subsubsection{\textcolor{blue}{Gauge Theories on the Lattice}}

Gauge theories on a space-time lattice require a particular treatment in order
to maintain gauge invariance in each step. In 1974
\blau{K. Wilson} has proposed a formulation where the gauge fields are not defined 
on the lattice points but on {\bf ``links''}:
\be
U\left ( n,\mu \right ) \E \exp \left [ \> i g a {\cal A}_{\mu}(n)
\> \right ] \> , \hspace{0.5cm}
{\cal A}_{\mu}  = A_{\mu}^a T^a \in SU(N) \>, \> \left ( \> T^a = 
\frac{\lambda_a}{2} \> \> \mbox{for} \> \> SU(3) \> \right )
\ee
is the link which points from the point $ n $ in positive direction $ \mu $.
Wilson has also proposed an action which reduces to the Euclidean Yang-Mills action
\be
S_{YM} \E \frac{1}{2} \int d^4x \>  {\rm tr} {\cal F}_{\mu \nu}
{\cal F}_{\mu \nu} \E   \frac{1}{4} \int d^4x \>   {\cal F}_{\mu \nu}^a
{\cal F}_{\mu \nu}^a \> , \quad   {\cal F}_{\mu \nu} \EQ F_{\mu \nu}^a \, T^a
\label{S Kont}
\ee
in the continuum limit:
\be
\boxed{
\quad S_{\rm Wilson} \E  \beta \sum_{n, \mu , \nu} \, \Biggl \{ \> 
1 - \frac{1}{N} {\rm Re} \> \>  {\rm tr} \Bigl [ \, U \left ( n,\mu \right )
\, U \left ( n + \mu,\nu \right ) 
 \, U \left ( n + \mu + \nu,-\mu \right ) \, U \left ( n + \nu,-\nu \right ) 
\, \Bigr ] \> \Biggr \}  \> . \quad
}
\label{S Wilson}
\ee
Here $ \beta $ is a yet undetermined factor and the summation has to be performed over all
elementary squares (``{\bf plaquettes}'') of the lattice:

\vspace{0.5cm}
\refstepcounter{abb}
\begin{figure}[hbtp]
\vspace*{-6cm}
\bce
\hspace*{3cm}\includegraphics[angle=0,scale=1]{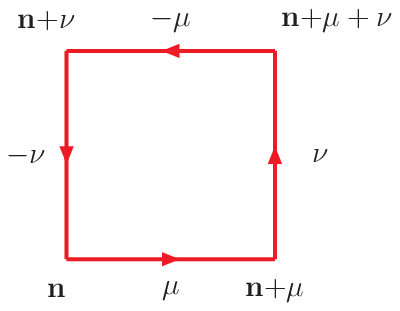}
\label{abb:3.6.1}
\ece
\vspace*{-18cm}
\bce
{\bf Fig. \arabic{abb}} : The elementary plaquette in lattice gauge theory.
\ece
\end{figure}
\vspace{0.5cm}

\renewcommand{\baselinestretch}{0.9}
\scriptsize
\refstepcounter{tief}
\noindent
\blau{\bf Detail \arabic{tief}:} {\bf Wilon's Action at Small Lattice Distances}\\
\vspace*{0.4cm}

\noindent
\begin{subequations}
It is relatively easy to show that the Wilson action goes over into the continuum action for $ a \to 0 $. 
For that  we use 
$ \> U \left ( n,\mu \right ) \, U \left ( n + \mu,-\mu \right ) = 1 \, , $ i.e. 
\be
U \left ( n,\mu \right ) \E U^{-1} \left ( n + \mu,-\mu \right )  \> .
\ee
Then we obtain
\be
S_{\rm Wilson} \E \beta \sum_{\rm plaquettes} \, \Biggl \{ \> 1 - 
\frac{1}{N} {\rm Re} \> {\rm tr} \Bigl [ \, e^{i g a {\cal A}_{\mu}(x)} \, 
e^{i g a {\cal A}_{\nu}(x + a e_{\mu})} 
 \, e^{-i g a {\cal A}_{\mu}(x + a e_{\nu})} 
\, e^{-i g a {\cal A}_{\nu}(x)} \, \Bigr ] \> \Biggr \}  \> .
\ee
To combine the exponential functions we utilize the Baker-Campbell-Haussdorff formula
\be
e^{a \hat A} \, e^{a \hat B} \E \exp \left ( \, a \hat A + a \hat B
+ \frac{a^2}{2} [\hat A, \hat B] + {\cal O}(a^3) \, \right )   \> .
\ee
Then the result is
\bea
S_{\rm Wilson} \EA \beta \sum_P \, \Biggl \{ \> 1 - \frac{1}{N}
{\rm Re} \> {\rm tr} \exp \Bigl \{ \> i g a \left ( \, {\cal A}_{\mu}(x)
+ {\cal A}_{\nu}(x + a e_{\mu}) - {\cal A}_{\mu}(x + a e_{\nu}) - 
{\cal A}_{\nu}(x) \, \right ) \non
&& \hspace{2cm} + \frac{(iga)^2}{2} \Bigl ( \, 
\left [ {\cal A}_{\mu}(x),{\cal A}_{\nu}(x + a e_{\mu}) \right ] 
- \left [ {\cal A}_{\mu}(x) + {\cal A}_{\nu}(x + a e_{\mu}), 
{\cal A}_{\mu}(x + a e_{\nu}) \right ] \non
&& \hspace{3cm} - \left [ {\cal A}_{\mu}(x)
+ {\cal A}_{\nu}(x + a e_{\mu}) - {\cal A}_{\mu}(x + a e_{\nu}), 
{\cal A}_{\nu}(x)
\right ] \, \Bigr ) \> \Bigr \} \> \Biggr \}   \> .
\eea
By a Taylor expansion of the vector fields we obtain
\bea
S_{\rm Wilson} \EA \beta \sum_P \, \Biggl \{  1 - \frac{1}{N}
{\rm Re} \> {\rm tr} \exp \Bigl\{ \, i g a \left ( a \partial_{\mu} 
{\cal A}_{\nu}(x)
- a \partial_{\nu} {\cal A}_{\mu}(x) \right ) 
+ (iga)^2 \Bigl [ {\cal A}_{\mu}(x) , {\cal A}_{\nu}(x) \Bigr ] \, \Bigl \}
 \Biggr \}  \non
& \stackrel{ a \to 0 }{\longrightarrow}& \beta \sum_P \, 
\Biggl \{ \> 1 - \frac{1}{N}
{\rm Re} \> {\rm tr} \Bigl( \, 1 + i g a^2 {\cal F}_{\mu \nu} + \frac{1}{2}
(i g a^2)^2 {\cal F}_{\mu \nu} {\cal F}_{\mu \nu} + \ldots \Bigl ) \> 
\Biggr \}  \> .
\eea
In the limit $ a \to 0 $ only the 2$^{\rm nd}$ order term contributes since
$ {\rm tr} {\cal F}_{\mu \nu} = 0 $  (the $SU(N)$ generators are traceless)
and we have
\be
S_{\rm Wilson} \> \stackrel{ a \to 0 }{\longrightarrow} \> 
\beta \frac{1}{2}\sum_{n,\mu \nu} \, \frac{1}{N} \frac{1}{2} (g a^2)^2
{\rm tr} \, {\cal F}_{\mu \nu} {\cal F}_{\mu \nu} \E  
\frac{ \beta g^2}{2 N}
\int d^4 x \> {\rm tr} \sum_{\mu \nu} {\cal F}_{\mu \nu}(x) {\cal F}_{\mu \nu}
(x)   \> .
\ee
The factor $\frac{1}{2}$ has its origin in the relation
$\sum_P = \frac{1}{2} \sum_{\mu \nu} $. 
\end{subequations}
\renewcommand{\baselinestretch}{1.2}
\normalsize

\vspace{0.5cm}
\noindent
Thus we have to choose
\be
\boxed{
\qquad \beta \E \frac{2 N}{g^2} \qquad
}
\ee
to obtain the action (\ref{S Kont}) of the continuum theory; 
for $ N = 1$ one has to take $ \beta = 1/g^2 $. 
The Wilson action is distinguished by the property that it is \textcolor{blue}{\bf invariant
under a local gauge transformation}
\be
U \left ( n,\mu \right ) \To V ( n ) \, 
U \left ( n,\mu \right ) V^{-1} \left ( n + \mu \right )
\ee
with an arbitrary matrix $ V \in SU(N) $ .

\noindent 
\begin{subequations}  
\renewcommand{\baselinestretch}{0.9}
\scriptsize
\bea
{\bf {\rm Proof}:} \qquad \qquad \quad S_{\rm Wilson} &\longrightarrow& \beta \sum_P \, \Biggl \{ \> 1 - \frac{1}{N}
{\rm Re} {\rm tr} \Bigl [ \,  V ( n ) U \left ( n,\mu \right )
\underbrace{  V^{-1} \left ( n + \mu \right )
V \left ( n + \mu \right )}_{=1} U \left ( n+ \mu
, \nu \right ) \non
&& \cdot \underbrace{ V^{-1} \left ( n + \mu + \nu 
\right ) V \left ( n + \mu + \nu \right ) }_{=1}
U \left ( n+ \mu + \nu,-\mu \right ) \non
&& \cdot \underbrace{ V^{-1} \left ( n + \mu + \nu
-\mu \right )  V \left ( n + \nu \right ) }_{=1}
U \left ( n + \nu,-\nu \right ) V^{-1} \left ( n + 
\nu - \nu \right ) \, \Bigr ] \> \Biggr \} \non
\EA \beta \sum_P \, \Biggl \{ \> 1 - \frac{1}{N}{\rm Re} \> {\rm tr} 
\Bigl [ \,  V ( n ) U U U U V^{-1}(n) \, \Bigr ] \> \Biggr \}
\E \beta \sum_P \, \Biggl \{ \> 1 - \frac{1}{N}{\rm Re}\>  {\rm tr}
\Bigl [ \,  U U U U  \, \Bigr ] \> \Biggr \} \non
\EA S_{\rm Wilson} \> , \nonumber
\eea
\hspace{1cm} as $ {\rm tr} ( \hat A \hat B ) = {\rm tr} ( \hat B \hat A ) $  \qquad
{\bf q.e.d.}

\end{subequations}
\renewcommand{\baselinestretch}{1.2}
\normalsize
\vspace{0.2cm}

\noindent
In this formulation
the lattice gauge theory therefore respects the gauge invariance 
at the expense of translation and thus Lorentz invariance. The latter ones should be restored in 
the continuum limit.
Note that unlike the continuum theory no gauge fixing is necessary
since the gauge fields are represented by a finite unitary matrix $ U $ (``compact'' action). One does not
sum over infinite degrees of freedom which are connected by gauge transformations. 

\vspace{0.2cm}

\subsubsection{\textcolor{blue}{Wilson loops and Confinement}}

To obtain a criterion for the postulated ``{\bf confinement}'' of the quarks
we will study now the energy of a system made up of a quark at  
$ x' = (t,{\bf 0})$ and an antiquark at $ x = (t,{\bf R})$ .
Here the quarks are assumed to be so {\bf heavy} that they do not move and only
constitute a source for the gluon fields. If the quarks are not confined we expect
that
\be
E(R) \To 2 m \hspace{0.5cm} \mbox{for} \> R \to \infty \> ,
\ee
where $m$ is the quark mass. \textcolor{blue}{ Confinement} means that the potential energy of the quarks
grows without bound:
\be
E(R) \To \infty \hspace{0.5cm} \mbox{for} \> R \to \infty \> . 
\ee
The quark-antiquark system cannot be represented simply by the state
\be
\bar q(x') \, q(x) \, | \, 0 \rangle \> ,
\ee
since this is not gauge invariant. The correct expression
\be
\Gamma(x',x,C) \E \bar q(x') \, {\cal P} \exp \left ( i g \int_x^{x'} 
{\cal A}_{\mu} (y) d y^{\mu} \right ) \, q(x) \, | \, 0 \rangle \E 
\bar q(x') \, U(x',x,C) \, q(x) \, | \, 0 \rangle \> 
\ee
contains a phase factor between the field operators creating a quark/antiquark at different 
space-time points. It is obvious that this factor also depends on the path $C$ which 
connects $ x $ and $ x' $.
${\cal P}$ denotes the \textcolor{blue}{ path ordering} which is necessary because the gauge field 
does not commute at different points ~\footnote{Analogously to the time ordering the path ordering 
is defined by the following procedure: Parametrize the path from $ x $ to $ x' $ by a continously increasing
parameter $s \in [0,1] $ and expand the exponential function
$ \exp \left ( i g \int_0^1 ds \, \frac{dx_{\mu}}{ds} {\cal A}_{\mu} (s) ds \right ) $ into a power series.
Then the  matrices  ${\cal A}_{\mu} (s)$ have to ordered such that the largest value of $s$ is on the l.h.s.}.

\begin{figure}[hbtp]
\vspace{-6cm}
\bce
\includegraphics[angle=0,scale=0.6]{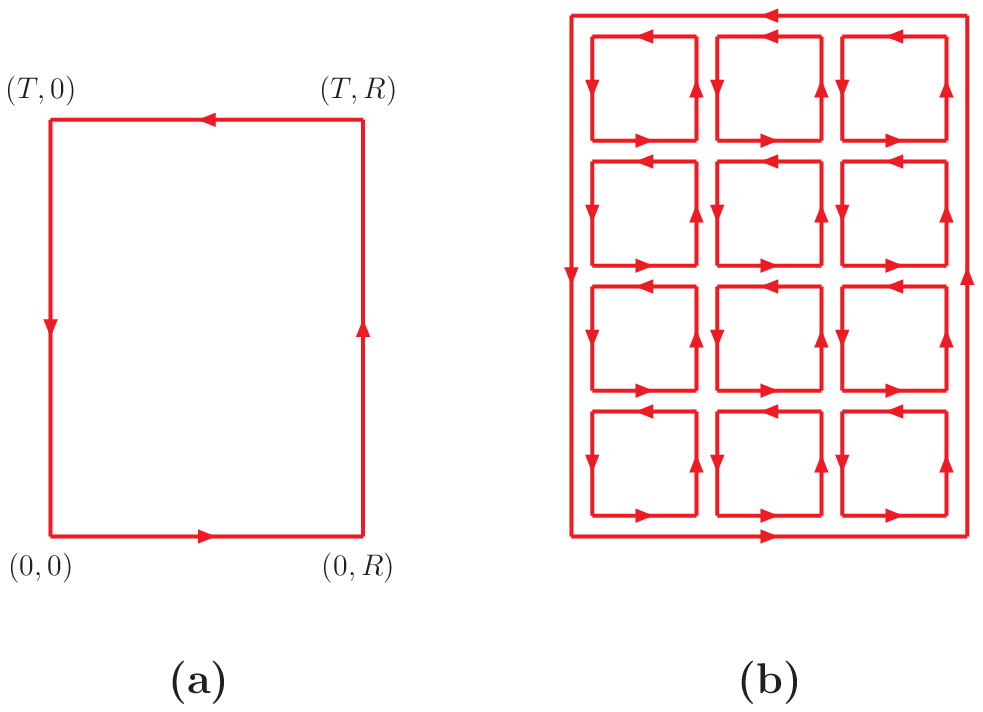}
\label{abb:3.6.2}
\ece
\vspace{-4.5cm}
{\bf Fig. \arabic{abb}} : (a) Wilson loop with rectangular path. \quad (b) A Wilson loop completely covered by plaquettes.
\end{figure}

We now want to calculate the overlap between the $q \bar q$-Zustand at $ t = 0 $ and the same state at
$ t = T $ :
\be
\Omega(T,R) \E \langle 0 | \, \Gamma^{\dagger} \left [ (0,{\bf 0}) ,
(0,{\bf R}), C \right ] \, \Gamma \left [ (T,{\bf 0}), (T,{\bf R}), C \right ]
\, | 0 \rangle \> .
\ee
As usual we insert a complete set of intermediate states and investigate 
the behaviour for large $ T $. Then we see that the intermediate state dominates which has 
the smallest potential energy of the $q \bar q$ system at distance  $R$ :
\be
\Omega(T,R) \> \stackrel{T \to \infty}{\simeq} \> {\rm const.}
\, e^{- E(R) \, T} \> .
\label{Omega 1}
\ee
On the other hand we may express the overlap by means of the quark field operators:
\be
\Omega(T,R) \E \la 0 | \> \bar q(0,{\bf R}) \, U \left [ (0,{\bf R}),
(0,{\bf 0}); C  \right ] \, q(0,{\bf 0}) \, \bar q(T,{\bf 0}) U \left [ (T,{\bf 0}),
(T,{\bf R}); C \right ] \, q(T,{\bf R}) \> | 0 \ra \> .
\label{Omega 2}
\ee
For a heavy, non-moving quark the Euclidean Dirac equation reads
\be
\gamma_4 \left ( \partial_4 - i g {\cal A}_4 \right ) q(x) = - m q(x) \> .
\ee
In this extreme non-relativistic limit it is also allowed to set
$ \gamma_4 = 1 $  and there is no quark pair production by the gauge field which acts like an 
external field. The solution then simply is
\be 
q(x_4 = t,\fx) \E {\rm const.} \, e^{-m t} \, {\cal P}
\exp \left (i g \int_0^t dt' \> {\cal A}_4(t') \right ) \> ,
\ee
and hence the quark propagator is
\be
 \langle 0 | \, q_{\beta}(t',\fx) \bar q_{\alpha}(t,\fx) \, | 0 
\rangle \E U \left [ (t',\fx), (t,\fx); C \right ] \, 
\delta_{\alpha \beta} \, e^{-m |t-t'|} \> .
\label{schweres quark}
\ee
If we contract the quark field operators in Eq. (\ref{Omega 2}) and use
Eq. (\ref{schweres quark}) we obtain
\bea
\Omega(T,R) &\simeq& {\rm const.} \> e^{- 2 m T} \> \langle 0 | \, 
U \left [ (T,{\bf R}), (0,{\bf R}) \right ] \,
U \left [ (0,{\bf R}), (0,{\bf 0}) \right ] \non
&& \hspace{2cm} \cdot U \left [ (0,{\bf 0}), (T,{\bf 0}) \right ] \, 
U \left [ (T,{\bf 0}), (T,{\bf R}) \right ] \, | 0 \rangle \non
\EA {\rm const.} \> e^{- 2 m T} \> \underbrace{ \> \langle 0 | \, 
{\rm tr} \, U \left [
(0,{\bf 0}), (0,{\bf 0}); C \right ] \,  0 \rangle \> }_{=: W(C)} \> .
\label{Wilson Schleife}
\eea
Here  $ C $ is the rectangular path shown in Fig. \ref{abb:3.6.2} (a) .
From the 
\textcolor{blue}{\bf Wilson loop} $ \> W(C)$ one can determine the
potential energy between a heavy quark and a heavy antiquark: 
\be
\lim_{T \to \infty} W(C) \E {\rm const.} \> e^{\, - 
\, \left [ \, E(R) - 2m \, \right ] \, T} \> .
\ee
As will be shown the Wilson loop obeys an \blau{area law} in
the strong-coupling limit
\be
W(C) \> \stackrel{g^2 \to \infty}{\simeq} \> e^{- K A(C) } \> ,
\ee
where $ K $ is a constant and $ A(C) $ is the area enclosed by the path $ C $.
For the rectangular path we have $ \> A(C) = T R \> $ and therefore a 
\textcolor{blue}{ linearly increasing potential} between heavy quark and antiquark.
\be
E(R) - 2 m \E K R \> .
\ee
The graphic picture one can associate with such an increasing potential is that of a \blau{``{\bf string}''}
of gluon fields which is generated between the static color sources. Therefore the constant $ K $ is also
called \blau{``string tension''}.

\vspace{0.5cm}
\refstepcounter{abb}
\unitlength1pt

\vspace{0.7cm}

\noindent
The Wilson loop can be calculated directly on the lattice:
\bea
W(C) \EA \langle 0 | \, {\rm tr} \, U(x,x;C) \, | 0 \rangle \E 
\frac{1}{Z} \int {\cal D} {\cal A}_{\mu}(x) \> {\rm tr} \, U(x,x;C) \,
e^{- S_E[{\cal A}]} \non
& \stackrel{\mbox{lattice}}{\longrightarrow} &
\frac{1}{Z} \int \prod_{n,\mu} dU \left ( n, n + \mu \right ) 
\> {\rm tr} \, U(n,n;C) \, e^{- S_{\rm Wilson}(U)} \> , 
\eea
where
\be 
Z \E \int \prod_{n,\mu} dU \left ( n, n + \mu \right )  \>  
e^{- S_{\rm Wilson}(U)} 
\ee
is the partition function. Here we do not integrate directly over the gauge fields but over the links $ U $
which are proportional to them for small lattice distances.

We now try to calculate the Wilson loop \textcolor{blue}{ for large coupling constants analytically}.
For that purpose we write the Wilson action (\ref{S Wilson}) as
\be
S_{\rm Wilson} \E \beta \sum_P \left (1 - \frac{1}{N} {\rm Re} \> {\rm tr}
\, U_P \right ) \> ,
\ee
omit the irrelevant constant in the action and expand in powers of $ \beta = N/(2 g^2) $.
This gives
\be
W(C) \E \frac{1}{Z} \int dU \> {\rm tr} \, U(x,x;C) \,
\left \{ \> 1 + \frac{1}{2 g^2} \sum_P {\rm sp } \, U_P + \frac{1}{2 !} 
\left ( \frac{1}{2 g^2} \right )^2 \sum_{P,P'} {\rm tr} \, U_P \, 
{\rm tr} \, U_{P'} \, + \ldots \right \} \> .
\ee
To evaluate that result we should know how to integrate over
the group elements of $ SU(N)$. This is easy in the $U(1)$ case since
there the group elements are phase factors parametrized by
\mbox{$ e^{i \Theta}, -\pi < \Theta \le \pi $} :
\be
\frac{1}{2 \pi} \int_{-\pi}^{+\pi} d \Theta \E 1 \> \> , \hspace{1cm}
\frac{1}{2 \pi} \int_{-\pi}^{+\pi} d \Theta \> e^{i \Theta} \E  0 \> .
\ee
In the general case the integration over group elements is done with the help of the
{\bf Haar measure} \footnote{See, e.g., {\bf \{Creutz\}}, ch. 8.}.
However, for the present purposes it is sufficient to list the integrals
\bea
\int dU \EA 1 \\
\int dU \> U_{i j} \EA 0 \\
\int dU \> U_{i j} \, U^{\dagger}_{k l} \EA \frac{1}{N} \delta_{il} \delta_{jk} \> .
\eea
The second rule says that no ``links'' should remain after integration while the third rule
means that a non-vanishing result is only obtained if one
integrates over two ``links'' in opposite direction.
This has the consequence that a Wilson loop has to be covered completely with plaquettes whose
adjacent ``links'' have opposite direction in order to yield a  contribution (Fig. \ref{abb:3.6.2} (b)).
The lowest (non-vanishing in $1/g^2$) contribution to $ \> W(C) \> $ therefore gives the term with
\be
W(C) \> \sim \> \left ( \frac{1}{2 g^2} \right )^{N_P} \> ,
\ee
where  $N_P$ is the minimal number of plaquettes required to cover the area enclosed
by the path $ C $. Since
$ A(C) = a^2 N_P $, this is equivalent to the area law
\be
W(C) \> \sim \> \left ( g^2 \right )^{- A(C)/a^2} \E 
\exp \left ( - \frac{T R}{a^2} \ln g^2 \right ) \> . 
\ee
Thus in the limit of strong coupling we indeed find
\be
\boxed{
\qquad E(R) - 2 m\> \sim \> \frac{\ln g^2}{a^2} \, R \EQ K \, R 
\>  \> , \hspace{0.8cm} g \gg 1 \qquad
}
\label{lineares Pot}
\ee
a \textcolor{blue}{ linearly increasing (confining) potential} between heavy quark and antiquark.

However, this result is obviously independent of $ N $, i.e. also an abelian $U(1)$ theory like
\rot{\bf QED} leads to confinement in the strong-coupling limit! This seems unreasonable but can be brought into agreement with reality by the following argument: In the continuum limit $ a \to 0 $ 
one has to change 
the (bare) coupling constant $ g = g(a) $ in such a way that the physical observables (e.g. the string tension) 
do not change. As can be seen from Eq. (\ref{lineares Pot}) this implies that the (bare) coupling constant 
has to decrease in the continuum limit. It is generally believed and supported by  numerical data -- 
as we will see in the next chapter -- that \rot{\bf QCD} and \rot{\bf QED} behave totally different in the limit
of weak coupling. While in \rot{\bf QCD} the area law still holds and the quarks remain confined there is a 
{\bf phase transition} in \rot{\bf QED}. Instead of an area law one finds a \blau{perimeter law} at weak (bare) coupling:
The potential between $e^{+}$ and $e^-$ approaches a constant at large distances which means that the constituents
may appear as free particles at sufficient energy in agreement with the experimental observation.
\vspace{0.5cm}

\subsubsection{\textcolor{blue}{Numerical Calculation of Observables on the Lattice}}

Perturbation theory and the strong-coupling expansion are insufficient to describe
non-perturbative phenomena in the continuum. This is evident for non-abelian theories
like \rot{\bf QCD} but should also hold for the postulated transition of confinement to Coulomb phase
in \rot{\bf QED}. In such cases (and with the present
state of theoretical tools) there only remains to perform a numerical evaluation of the path integral.

We already have covered that extensively in the {\bf chapter} {\bf \ref{sec1: Numerik}} 
for the anharmonic oscillator in quantum mechanics. Our aim is to extend that to a 4-dimensional field theory.
To simplify the programming effort we illustrate the numerical simulation
not for the (much more interesting) $SU(3)$ theory of \rot{\bf QCD} but ``only'' for the abelian $U(1)$ theory.
In this case the ``links'' are simply given by the phase factors
\be
U(n,\mu) \E e^{i g a A_{\mu}(n)} \E e^{i \Theta_{\mu}(n)} \> ,
\hspace{0.4cm} 0 < \Theta_{\mu}(n) \le 2 \pi
\ee
with
\be
U(n+\mu,-\mu) \E U^{-1}(n,\mu) \E U^{\dagger}(n,\mu) \E 
U^{\star}(n,\mu) \> .
\ee
By means of the Metropolis algorithm we again generate ``configurations'' 
(i.e. all possible $U$'s on the lattice) distributed according to
\be
w(U) \E \frac{e^{-S_{\rm Wilson}(U)}}{\int dU \, e^{-S_{\rm Wilson}(U)}} \> .
\ee
We recall that we have to generate trial configurations $ (U_t) $
and then -- depending on the value of 
\be
r \E e^{- \left [ \, S(U_t) - S(U) \, \right ]} \EQ e^{-\Delta S}
\ee
-- have to decide whether we accept these or not.
It is reasonable to change the configuration only at one ``link'' 
$U(n,\mu)$ because then one only has to calculate the change of the action
\bea
\Delta S \EA - \beta \sum_{\nu \neq \mu} \, \Delta U(n,\mu) \, \Bigl \{ \>
U(n+\mu,\nu) \, U^{\ast}(n+\nu,\mu) \, U^{\ast}(n,\nu) \non
&& \hspace{3cm} + \, 
U(n+\mu,-\nu) \, U^{\ast}(n-\nu,\mu) \, U^{\ast}(n,-\nu)  \> \Bigr \}
\eea
which in 4 dimensions precisely affects 6 ``links''.
We take
\be
\Delta U(n,\mu) \E  U(n,\mu) \, \cdot \, \left [ \, 1 - 
\frac{\exp(2 \pi i z_1) 
+ \delta}{|\exp(2 \pi i z_1) + \delta|} \, \right ] \> ,
\ee
as change where $\delta$ is a given parameter and $z_1$ a random number (uniformly distributed betwee
$0$ and $1$) 
After sufficient ``sweeps'' through the lattice (so that thermalization has occurred) one measures the 
expectation value of an arbitrary variable ${\cal O}(U)$ by
\be
\langle {\cal O} \rangle \E \frac{ \int dU \, {\cal O}(U) e^{-S(U)}}
{\int dU \, e^{-S(U)}} \> \simeq \> \frac{1}{M} \sum_{i=1}^M {\cal O}(U_i) 
\> .
\ee
A popular observable is the \blau{\bf mean plaquette} 
\be
\langle P \rangle \E \left <  \, \left ( 1 - \frac{1}{N} \sum_P 
{\rm tr} \, U_P \right ) \, \right >_{N=1}  \> ,
\ee
since it is easy to calculate and as an \blau{\bf order parameter} signalizes the transition 
to another phase.
With a small lattice this easily can be done also on a PC.

\newpage

\renewcommand{\baselinestretch}{0.9}
\scriptsize
\refstepcounter{tief}
\noindent
\blau{\bf Detail \arabic{tief}:} {\bf  FORTRAN Program for the Calculation of the Mean Plaquette
\footnote{Thanks to Manfred Kremer (Mainz/J\"ulich) who provided the first version of this program.}}\\
\vspace{0.2cm}

\color[rgb]{0.3,0,0.7}
\begin{verbatim}
C
C  Calculates mean plaquette in 4-dimensional U(1) theory on a 8^4 lattice
C  as function of beta = 1/g^2
C
C  Parameter: 
C  NTH = number of thermalization sweeps
C  NHIT = number of additional Monte Carlo calls at each lattice point
C  NSWEEP = number of sweeps (<500) 
C  DELTA = parameter for a new configuration
C  BETA0 = initial value of beta
C  DBETA = beta step size
C  NBETA = number of beta values
C
      PARAMETER(LS=8,LT=8,NN=LS**3*LT)
      COMPLEX U(NN,4),V,ZPI
      DOUBLE PRECISION DSEED
      DIMENSION NNF(NN,4),NNB(NN,4),ACT(500)
      COMMON/ZUF/ DSEED,ZPI,DELTA
      COMMON /PAR/ NAKZ
      DATA PI /3.1415926/
      WRITE(*,*) 'Input: ntherm,nhit,nsweep,delta'
      READ (*,*) NTHERM,NHIT,NSWEEP,DELTA
      WRITE(*,*) 'Input: beta0,dbeta,nbeta'
      READ (*,*) BETA0,DBETA,NBETA
C
C  auxiliary calculations
C
      LS2 = LS*LS
      LS3 = LS2*LS
      DSEED = 12365.D0
      ZPI = 2.*PI*CMPLX(0.,1.)
      NAKZ = 0
      WRITE (6,100) LS,LT
100   FORMAT(// '  U(1) theory on a  (',I2,'**3)*',I2,'   lattice'//)
      WRITE(6,102) NTHERM,NHIT,NSWEEP,DELTA
102   FORMAT('  NTHERM =',I3,'  NHIT =',I2,'  NSWEEP =',I3,
     &   '  DELTA = ',F6.3//)
C
c  calculate next-neighbor addresses
C
      DO 10 IX = 1,LS
         DO 10 IY = 1,LS
            DO 10 IZ = 1,LS
               DO 10 IT = 1,LT
                  I = IX + LS*(IY-1+LS*(IZ-1+LS*(IT-1)))
                  NNF(I,1) = I - IX + 1 + MOD(IX,LS)
                  NNF(I,2) = I - (IY-1-MOD(IY,LS))*LS
                  NNF(I,3) = I - (IZ-1-MOD(IZ,LS))*LS2
                  NNF(I,4) = I - (IT-1-MOD(IT,LT))*LS3
                  NNB(I,1) = I - IX + 1 + MOD(IX+LS-2,LS)
                  NNB(I,2) = I - (IY-1-MOD(IY+LS-2,LS))*LS
                  NNB(I,3) = I - (IZ-1-MOD(IZ+LS-2,LS))*LS2
                  NNB(I,4) = I - (IT-1-MOD(IT+LT-2,LT))*LS3
10    CONTINUE
C
C  initialization and thermalization
C
      WRITE(6,*) ('  cold start'//)
      DO 12 I = 1,NN
         DO 12 I1 = 1,4
12    U(I,I1) = 1.
      DO 15 I = 1,NTHERM
15    CALL UPDATE(U,NN,NHIT,BETA,NNF,NNB)
C
C  beta loop
C
      DO 90 IBETA = 1,NBETA
         BETA = BETA0 + (IBETA - 1)*DBETA
C  sweeps
         NAKZ = 0
         DO 20 L = 1,NSWEEP
            CALL UPDATE(U,NN,NHIT,BETA,NNF,NNB)
C
C  measurement
C
            ACT(L) = 0.
            DO 30 I = 1,NN
               DO 30 K = 1,3
               KP1 = K + 1
               V = 0.
               DO 40 K1 = KP1,4
                  I1 = NNF(I,K)
                  I2 = NNF(I,K1)
                  V = V + U(I1,K1)*CONJG(U(I2,K))*CONJG(U(I,K1))
40             CONTINUE
               ACT(L) = ACT(L) + U(I,K)*V
30          CONTINUE
            ACT(L) = (1. - ACT(L)/(6.*NN))
20       CONTINUE
         SACT1 = 0.
         SACT2 = 0.
         DO 50 L = 1,NSWEEP
            SACT1 = SACT1 + ACT(L)
50       SACT2 = SACT2 + ACT(L)**2
         SACT = SACT1/NSWEEP
         SACT2 = SACT2/NSWEEP
         SACT2 = SACT2 - SACT*SACT
         SACT2 = SQRT(SACT2)
         XAKZ = NAKZ/(4.*NHIT*NSWEEP*NN)
C
C print out
C
         WRITE(6,104) BETA,SACT,SACT2,XAKZ
104      FORMAT(' beta =',F7.3,5X,'mean plaquette =',F9.3,
     &   ' +/-',F6.3,5X,'acceptance =',F7.3/)
90    CONTINUE
      STOP
      END

C++++++++++++++++++++ SUBPROGRAM UPDATE ++++++++++++++++++++++++++++++

      SUBROUTINE UPDATE (U,NN,NHIT,BETA,NNF,NNB)
C
C  New trial configuration (update)
C
      COMPLEX U(NN,4),V,W,W1,ZPI
      DIMENSION NNF(NN,4),NNB(NN,4)
      DOUBLE PRECISION DSEED
      COMMON /ZUF/ DSEED,ZPI,DELTA
      COMMON /PAR/ NAKZ
      DO 10 I = 1,NN
         DO 10 K = 1,4
            V= 0.
            DO 20 K1 = 1,4
               IF (K1 .EQ. K) GO TO 20
               I1 = NNF(I,K)
               I2 = NNF(I,K1)
               I4 = NNB(I,K1)
               I5 = NNF(I4,K)
               V = V + U(I1,K1)*CONJG(U(I2,K))*CONJG(U(I,K1))
     &               + U(I4,K1)*CONJG(U(I5,K))*CONJG(U(I4,K1))    ! Birbaumer-
                                                                  ! correction
20          CONTINUE
C
C  Metropolis algorithm
C
            DO 30 KIND = 1,NHIT
               CALL CREATE(W1)
               W = W1*U(I,K)
               SNEU = V*W
               SALT = V*U(I,K)
               DACT = (SNEU - SALT)*BETA
               IF (DACT .LT. 0.) THEN
                  P = EXP(DACT)
                  XR = ZUFALL(DSEED)
               END IF
               IF (DACT .GE. 0.) GO TO 7
               IF(XR .GT. P) GO TO 30
7              U(I,K) = W
               NAKZ = NAKZ + 1
30          CONTINUE
10    CONTINUE
      RETURN
      END


C++++++++++++++++++++++++  SUBPROGRAMM CREATE +++++++++++++++++++++++++

      SUBROUTINE CREATE(U)
C
C  creates new exp(i theta)
C
      DOUBLE PRECISION DSEED
      COMPLEX U,ZPI,A0
      COMMON /ZUF/ DSEED,ZPI,DELTA
      A0 = ZUFALL(DSEED)*ZPI
      U = CEXP(A0) + DELTA
      U = U/CABS(U)
      RETURN
      END

C++++++++++++++++++++++ SUBPROGRAMM ZUFALL ++++++++++++++++++++++++++++

      FUNCTION ZUFALL(DSEED)
C
C  generates uniformly distributed random numbers in the interval [0,1]
C
      DOUBLE PRECISION A,C
      DATA A,C /16807.D0,2147483647.D0/
      DSEED = DMOD(A*DSEED,C)
      ZUFALL = DSEED/C
      RETURN
      END
\end{verbatim}
\color{black}
\vspace{0.5cm}

\noindent
If we make a "cold" start with this program on a  $8^4$-Gitter  and let it run 
with $100$ ``sweeps'', throwing the Monte-Carlo dice 5 times at each lattice point and demanding 20
different $\beta$ values \footnote{That takes about 1.5 min on a 2 GHz PC, for the
$12^4$ lattice  about 7.5 min.} we obtain the following printout:

\renewcommand{\baselinestretch}{0.9}
\footnotesize
\vspace{1cm}

\color[rgb]{0.3,0,0.7}

  U(1) theory on a ( 8**3)*8   lattice\\

  NTHERM = 20  \hspace{5mm} NHIT = 5  \hspace{5mm} NSWEEP = 100 \hspace{5mm} DELTA = 1.5 \\

  cold start

\noindent
\hspace*{-1mm} beta =  0.100 \hspace{4mm}mean plaquette =    0.950 +- 0.004 
\hspace{5mm}    acceptance =  0.969\\
beta =  0.200\hspace{5mm}mean plaquette =    0.901 +- 0.004
\hspace{5mm}     acceptance =  0.938\\
beta =  0.300\hspace{5mm}mean plaquette =    0.852 +- 0.004 
\hspace{5mm}    acceptance =  0.908\\
beta =  0.400\hspace{5mm}mean plaquette =    0.804 +- 0.004
\hspace{5mm}     acceptance =  0.878\\
beta =  0.500\hspace{5mm}mean plaquette =    0.754 +- 0.004
\hspace{5mm}     acceptance =  0.848\\
beta =  0.600\hspace{5mm}mean plaquette =    0.706 +- 0.005
\hspace{5mm}     acceptance =  0.819\\
beta =  0.700\hspace{5mm}mean plaquette =    0.654 +- 0.004 
\hspace{5mm}    acceptance =  0.788\\
beta =  0.800\hspace{5mm}mean plaquette =    0.599 +- 0.006
\hspace{5mm}     acceptance =  0.758\\
beta =  0.900\hspace{5mm}mean plaquette =    0.536 +- 0.007
\hspace{5mm}    acceptance =  0.725\\
beta =  1.000\hspace{5mm}mean plaquette =    0.432 +- 0.022
\hspace{5mm}    acceptance =  0.682\\
beta =  1.100\hspace{5mm}mean plaquette =    0.311 +- 0.024
\hspace{5mm}     acceptance =  0.636\\
beta =  1.200\hspace{5mm}mean plaquette =    0.251 +- 0.004
\hspace{5mm}     acceptance =  0.608\\
beta =  1.300\hspace{5mm}mean plaquette =    0.224 +- 0.003
\hspace{5mm}     acceptance =  0.589\\
beta =  1.400\hspace{5mm}mean plaquette =    0.203 +- 0.003
\hspace{5mm}     acceptance =  0.572\\
beta =  1.500\hspace{5mm}mean plaquette =    0.188 +- 0.003 
\hspace{5mm}    acceptance =  0.556\\
beta =  1.600\hspace{5mm}mean plaquette =    0.174 +- 0.002
\hspace{5mm}     acceptance =  0.542\\
beta =  1.700\hspace{5mm}mean plaquette =    0.162 +- 0.002
\hspace{5mm}     acceptance =  0.529\\
beta =  1.800\hspace{5mm}mean plaquette =    0.152 +- 0.002
\hspace{5mm}     acceptance =  0.516\\
beta =  1.900\hspace{5mm}mean plaquette =    0.143 +- 0.002
\hspace{5mm}     acceptance =  0.504\\
beta =  2.000\hspace{5mm}mean plaquette =    0.135 +- 0.002
\hspace{5mm}     acceptance =  0.493\\
\color{black}

\renewcommand{\baselinestretch}{1.2}
\normalsize
\vspace{0.3cm}

\begin{figure}[hbtp]
\vspace*{0.5cm}
\bce
\includegraphics[angle=0,scale=0.55]{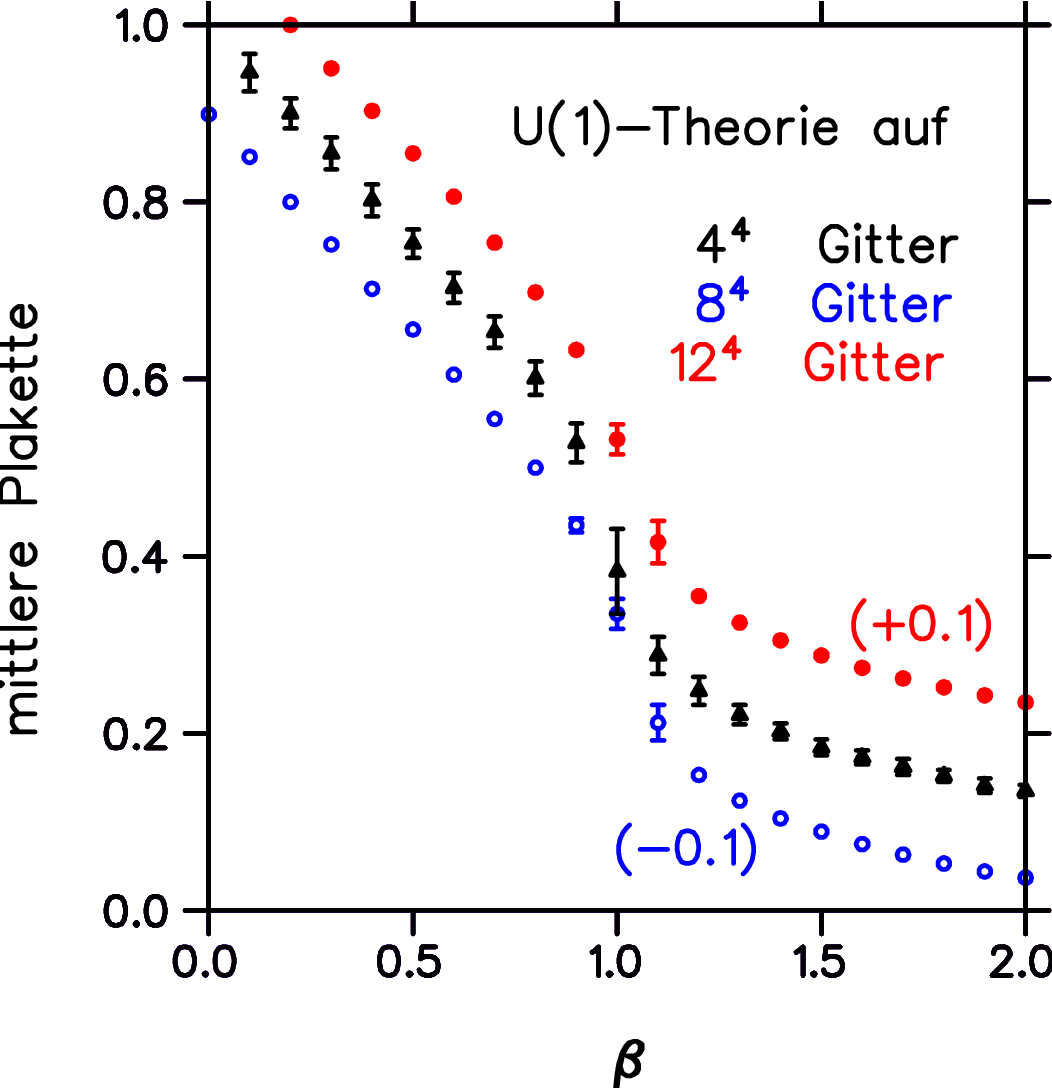}
\label{abb:3.6.3}
\ece
{\bf Fig. \arabic{abb}} : Mean plaquette $ \langle P \rangle $ as a function of
$\beta = 1/g^2$ for different lattice sizes. To make the results better 
\\ \hspace*{1.5cm} visible,
the results for the  $8^4$ lattice have been shifted downwards by the constant value  $ 0.1 $
while the   \\ \hspace*{1.5cm} ones for the $ 12^4$ lattice by $ 0.1 $ upwards.
\end{figure}

\noindent
If one plots the results for different lattice sizes 
(as in Fig. \ref{abb:3.6.3}; sorry, the legend is in German...))  then one sees that a rather rapid change of the mean plaquette 
values occurs at $ \beta \simeq 0.9 - 1.2 $, i.e.  the suspected phase transition
from the confining phase (large $g^2
\Rightarrow $ small $\beta$ ) to the Coulomb phase ( small $g^2 \Rightarrow $
large $\beta$) seems to take place.

\noindent
Since we are working on a very small, finite lattice this transition
is not so abrupt as in the infinite system where the order parameter 
vanishes in the phase where electrons and positrons are free.
Nevertheless our result agrees reasonably well with the best calculations
which localize the phase transition at $\beta = 1.011131(6) $ \cite{ABLS}.

\vspace{-0.4cm}

\refstepcounter{abb}
\unitlength1pt

 
\pagestyle{myheadings}
\markboth{\textcolor{green}{Appendix}}{\textcolor{green}{R. Rosenfelder : Path Integrals
for Quantum Physics}}

\newpage

\noindent
{\Large\bf \textcolor{red}{Additional Literature}} (in alphabetic order)

\vspace{1cm}

\bdes
\item[{\bf \{Bateman Proj. 2\}}  ]   {\tt A. Erdely (ed.)}, {\it Higher Transcendental 
Functions, based, in part, on notes left by Harry Bateman}, vol. 2 , McGraw-Hill, New York 
(1953).

\item[{\bf \{Berezin\}}  ] {\tt F. A. Berezin}, {\it The Method of Second 
Quantization}, Academic Press, New York (1965).

\item[\bf \{Bertlmann\} ] {\tt R. A. Bertlmann}, {\it Anomalies in Quantum Field Theory}, 
Oxford University Press, Oxford (2000).

\item[\bf \{Bjorken-Drell\}] {\tt J. D. Bjorken, S. Drell}, {\it Relativistic Quantum Mechanics},
McGraw-Hill, New York (1964)       

\item[{\bf \{Bronstein-Semendjajew\}}   ] {\tt I. N. Bronstein and K. A. Semendjajew}, 
{\it Taschenbuch der Mathematik}, \\ Harri Deutsch, Z\"urich and Frankfurt (1965). 

\item[{\bf \{Collins\}} ] {\tt J. C. Collins}, {\it Renormalization. An introduction to 
renormalization, the renormalization group, and the operator-product expansion}, 
Cambridge University Press, Cambridge (1984).

\item[{\bf \{Creutz\}}   ] {\tt M. Creutz}, \textit{Quarks, Gluons and Lattices}, 
Cambridge University Press, Cambridge (1983).


\item[{\bf \{Dennery-Krzywicki\}}   ] {\tt Ph. Dennery and A. Krzywicki}, \textit{Mathematics 
for Physicists}, Harper \& Row, \\ New York (1967).

\item[{\bf \{Eisenberg-Koltun\}} ] {\tt J. M. Eisenberg and D. S. Koltun}, \textit{Theory of
Meson Interactions with Nuclei}, \\John Wiley, New York (1980). 

\item[{\bf \{Euler\}}    ] {\tt L. Euler}, \textit{Introductio in Analysin Infinitorum}, 
Lausanne (1748). 

\item[{\bf \{Farmelo\}} ] {\tt G. Farmelo}, \textit{The Strangest Man: The Hidden Life of Paul Dirac, Quantum Genius}, Faber and Faber, London (2009).


\item[{\bf \{Fetter-Walecka\}}   ] {\tt A. Fetter and J. D. Walecka}, \textit{ Quantum Theory of 
Many Particle Systems}, McGraw-Hill, New York (1971).

\item[{\bf \{Fontane\}} ]{\tt Theodor Fontane}, \textit{Effi Briest} (1894), last sentence; 
insel taschenbuch 2131, Inselverlag (1997).

\item[{\bf \{Georgi\}} ] {\tt H. Georgi}, \textit{Lie Algebras in Particle Physics: From Isospin 
to Unified Theories}, Frontiers in Physics, vol. 54, Benjamin/Cummings, Reading (1982).
 
\item[{\bf \{Gradshteyn-Ryzhik\}}   ] {\tt I. S. Gradshteyn and I. M. Ryzhik}, 
\textit{Table of Integrals, Series and Products}, 4th edition, Academic Press, New York 
(1980).

\item[{\bf \{Handbook\}}  ] {\tt M. Abramowitz and I. Stegun (eds.)}, \textit{Handbook of 
Mathematical Functions}, Dover, New York (1965).

\item[{\bf \{Horn-Johnson\}}   ] {\tt R. A. Horn and C. R. Johnson}, \textit{Matrix Analysis},
 Cambridge University Press, Cambridge (1985).

\item[{\bf \{Landau-Lifschitz 1\}}  ] {\tt L. D. Landau and E. M. Lifschitz}, 
\textit{Course of theoretical physics}, vol. 1, Pergamon Press, Oxford (1978).

\item[{\bf \{Le Bellac\}} ] {\tt M. Le Bellac}, \textit{Quantum and Statistical Field Theory}, 
Oxford University Press, Oxford (1991).

\item[{\bf \{Lenin\}}  ] {\tt W. I. Lenin}, \textit{What Is to Be Done? Burning Questions of Our Movement (Russian tr. 
 Chto delat'?)} (1901). https://www.marxists.org/archive/lenin/works/1901/witbd/index.htm 

\item[{\bf \{Lighthill\}} ] {\tt M. J. Lighthill}, \textit{Introduction to Fourier Analysis and Generalized Functions} (Cambridge Monographs on Mechanics and Applied Mathematics),
Cambridge University Press, Cambridge (1958).

\item[{\bf \{Messiah\}}] {\tt Albert Messiah}, \textit{M\'ecanique Quantique}, Dunod, Paris (1959).

\item[{\bf \{Messiah 1/2\}}] {\tt A. Messiah}, \textit{Quantum Mechanics}, vol. I/II, North Holland, 
Amsterdam (1965).

\item[{\bf \{Muta\}}] {\tt T. Muta}, \textit{Foundations of Quantum Chromodynamics -- An Introduction 
  to Perturbative Methods in Gauge Theories}, World Scientific, Singapore (1988).

\item[{\bf \{Num. Recipes\}}   ] {\tt W. H. Press, S. A. Teukolsky, W.T. Vetterling and B. P. Flannery}, 
\textit{Numerical \\Recipes in Fortran 77}, 2nd ed.,  vol. 1, 
\textit{The Art of Scientific Computing} , Cambridge University Press, Cambridge (1992).
{\bf Free}, in Empanel~ format (http://www.nr.com/oldverswitcher.html).

\item[{\bf \{Oz\}} ] {\tt Amos Oz}, \textit{A Tale of Love and Darkness}, translator: Nicolas de Lange,
Harcourt (2004)

\item[{\bf \{Scheck\}} ] {\tt F. Scheck}, \textit{Leptons, Hadrons and Nuclei}, North-Holland, 
Amsterdam (1983).
  
\item[{\bf \{Schiller\}} ] {\tt Friedrich Schiller}, \textit{Die Verschw\"orung des Fiesco zu Genua. Ein 
republikanisches Trauerspiel} (1783).
 
\item[{\bf \{Weiss\}}  ] {\tt U. Weiss}, \textit{Quantum Dissipative Systems}, 
World Scientific (1999).

\edes

\vspace{2cm}

\noindent
{\Large\bf \textcolor{red}{Original Publications}}

\newpage

\pagestyle{myheadings}
\markboth{\textcolor{green}{Exercises}}{\textcolor{green}{R. Rosenfelder : Path Integrals in  
Quantum Physics}}

\bce
{\large \purpur{\bf Exercises for the Lecture "Path Integrals in Quantum Physics''}}

\vspace{0.5 cm}

{\bf Exercise 1}
\ece
\vspace{ 0.3cm}

\refstepcounter{ueb}     
\setcounter{equation}{0}
\renewcommand{\theequation}{\footnotesize\mbox{P }\arabic{ueb}.\arabic{equation}}    
\bdes
\item[ \purpur{\bf Problem \arabic{ueb}} :] 
\label{nr frei Prop}   
Show that
\par\noindent
\bdes
\item[\purpur{\bf a)}] the  {\bf free propagator} (the matrix element of the 
time-evolution operator) \eqref{freier Prop 2}   fulfills the Schr\"o\-din\-ger 
equation
\bea
i \hbar \> \frac{\partial U_0(b,a)}{\partial t_b} = - \frac{\hbar^2}{2 m} \>
\frac{\partial^2 U_0(b,a)}{\partial x_b^2}  
\eea
\par\noindent
\item[\purpur{\bf b)}] the {\bf free Green function}
\be 
 G_0(b,a) \E - \frac{i}{\hbar} \> \Theta (t_b - t_a) \> U_0(b,a) 
 \label{def G0}
\ee
obeys the equation
\be 
\left ( i \hbar \frac{\partial}{\partial t_b} \> + \frac{\hbar^2}{2 m}
\> \frac{\partial^2}{\partial x_b^2} \right ) \> G_0(b,a ) = \delta (t_a - t_b)
\> \delta(x_a - x_b) \> .
\ee
Show by Fourier transforming Eq. \eqref{def G0} that
\be 
G_0(b,a) \E \int_{-\infty}^{+\infty} \frac{dE}{2 \pi i \hbar} \, \int_{-\infty}^{+\infty}
\frac{dp}{2 \pi i \hbar} \> 
\frac{1}{E - p^2/(2m) + i 0^+ } \> e^{i p \cdot (x_b - x_a)/\hbar - i E (t_b - t_a)/\hbar} 
\ee
where $ i 0^+ $ denotes a small positive imaginary part which at the end of the calculation is set to
zero. Which Green function is obtained when a small {\it negative} imaginary part is taken?
\edes
\vspace{0.2cm}

\refstepcounter{ueb}     
\setcounter{equation}{0}
\renewcommand{\theequation}{\footnotesize\mbox{P }\arabic{ueb}.\arabic{equation}}    
\item[ \purpur{\bf Problem  \arabic{ueb}} :]
\label{klass Wirk}    
Determine the {\bf classical action} for
\par\noindent
\bdes
\item[\purpur{\bf a)}] a free particle (Lagrange function $ \> L = m \dot x^2/2 \> $), 
\item[\purpur{\bf b)}] a harmonic oscillator ($ \> L = m \dot x^2/2 - m \omega^2 x^2/2 \> $),
\item[\purpur{\bf c)}] a harmonic oscillator with an external, time-dependent
force  ($  \> L = m \dot x^2/2 - m \omega^2 x^2/2 - e(t) x \>  $).
\edes
\vspace{0.2cm}

\refstepcounter{ueb}     
\setcounter{equation}{0}
\renewcommand{\theequation}{\footnotesize\mbox{P }\arabic{ueb}.\arabic{equation}}    
\item[ \purpur{\bf Problem  \arabic{ueb}} :] 
\label{U fuer zeitabh. Pot}   
For a {\bf time-dependent potential} $ V(x,t) $
the Schr\"o\-din\-ger equation  for the   
time-evolution operator reads
\bea
\frac{\partial \hat U(t,t_0)}{\partial t} =  - \frac{i}{\hbar}
\> \hat H(t) \> \hat U(t,t_0) \quad {\rm with} \quad \hat U(t_0, t_0) = \hat 1  
\label{eq for U}
\eea
and the formal solution is not anymore $ \> \>
\exp\left( - i \hat H (t-t_0)/\hbar \right) \> \> $ but
\bea
\hat U(t,t_0) = {\cal T} \left \{ \exp \left(- i \int_{t_0}^t d\tau \>
\hat H(\tau) /\hbar)\right ) \right \} \> , 
\eea
where ${\cal T} $ is the time-ordering symbol.
\bdes
\item[\purpur{\bf a)}] Show that  $ \hat U $ is unitary if 
$ \hat H(t) $ is hermitean.
\item[\purpur{\bf b)}] Prove the composition law
$ \hat U(t,t_0) = \hat U(t,t_1) \, \hat U(t_1,t_0) $.
\item[\purpur{\bf c)}] Show with that und the ususal time-slicing method
that the path-integral representation of the matrix\\
\hspace{1.5cm} elements of
 $\hat U$ retains the given form, i.e. that no time-ordering symbol is necessary.
\edes
\edes
\newpage


\bce
{\large \purpur{\bf Problems for the Lecture ``Path Integrals in Quantum Physics"}}

\vspace{1 cm}

{\bf Exercise 2}
\ece
\vspace{ 0.4cm}

\bdes
\refstepcounter{ueb}     
\setcounter{equation}{0}
\renewcommand{\theequation}{\footnotesize\mbox{P }\arabic{ueb}.\arabic{equation}}    
\item[ \purpur{\bf Problem  \arabic{ueb}} :] 
\label{functionalableit}   
Calculate
\bdes
\item[\purpur{\bf a)}] the {\bf functional derivative}
$ \quad  \frac{\delta S}{\delta x(\sigma)} \quad $ for
$ \quad S[x] \E  \int_{t_a}^{t_b} dt \> \left [ \> 
\frac{m}{2} \dot x^2(t) - V(x(t)) \> \right ] $ ,
\item[\purpur{\bf b)}]
the second functional derivative
$ \quad  \frac{\delta^2 S}{\delta x(\sigma) \, \delta x(\sigma')} \quad $ 
of the action in \purpur{\bf a)} and 
derive the functional Taylor \\
\hspace{1.5cm} expansion of the action $ S $ around the classical trajectory.
\item[\purpur{\bf c)}]
Evaluate the functional derivative of the generating functional
\bea
Z[J] \E \int {\cal D}x \> \exp \left ( \, \frac{i}{\hbar} S[x] + \frac{i}{
\hbar}
\int_{-\infty}^{+\infty} dt \> 
x(t) \, J(t) \, \right ) 
\eea
\hspace{-0.3cm} w.r.t. the external source  $J(\sigma)$.
\edes 

\vspace{0.4cm}

\refstepcounter{ueb}     
\setcounter{equation}{0}
\renewcommand{\theequation}{\footnotesize\mbox{P }\arabic{ueb}.\arabic{equation}}    
\item[ \purpur{\bf Problem  \arabic{ueb}} :] 
\label{Wigner Trans}   
In 1932
{\bf Wigner} introduced the {\bf transformation} named after him
\bea
A_W(x,p) \E \int_{-\infty}^{+\infty} dy \> \left < \> x - \frac{y}{2} \>
\left | \> \hat A \> \right | \>  x + \frac{y}{2} \> \right >
\> e^{i p \cdot y/ \hbar}  
\eea
for an arbitrary quantum-mechanical operator $ \> \hat A \> $.
The Wigner transform $ \> A_W(x,p) \> $ is the form of the quantum-mechanical
operator which comes closest to the classical description.
\bdes        
\item[\purpur{\bf a)}] Calculate the Wigner transform of  a (local) potential operator $ V(\hat x) $
and of the kinetic energy $ T(\hat p) $.
\item[\purpur{\bf b)}] 
By using the inverse transformation show that for the classical phase space function
 $ \> A_W(x,p) \>  = \> p^m \> x^n \> $ Weyl's quantization prescription follows
 for the quantum-mechanical operator $ \> \hat A \> $.
\item[\purpur{\bf a)}] 
 Calculate the Wigner transform of the density matrix
$ \hat A = \left  |\psi_n \right > \left < \psi_n \right | $  
in the ground state ($n = 0$) and first excited state ($n = 1$) of
the harmonic oscillator and examine whether these functions are everywhere positive
so that they can be interpreted as genuine probabilities.
\edes

\vspace{0.8cm}

\refstepcounter{ueb}     
\setcounter{equation}{0}
\renewcommand{\theequation}{\footnotesize\mbox{P }\arabic{ueb}.\arabic{equation}}    
\item[ \purpur{\bf Problem  \arabic{ueb}} :] 
\label{klass Anfang/End}   
Show that for a {\bf classical motion} from
$\> (x_a,t_a) \> $ to $\> (x_b,t_b) \> $ the energy of a particle at the initial point
is given by
\bea
\frac{\partial S_{\rm cl}}{\partial t_a} 
\eea
and its momentum by
\bea
- \frac{\partial S_{\rm cl}}{\partial x_a}
\eea
where  $ S_{\rm cl}(x_b, t_b; x_a, t_a) \> $ is its classical action.
Check for the free case and for the harmonic oscillator case.

\edes

\newpage

\bce
{\large \purpur{\bf Problems for the Lecture``Path Integrals in 
Quantum Physics"}}

\vspace{1 cm}

{\bf Exercise 3}
\ece
\vspace{ 0.4cm}

\bdes
\refstepcounter{ueb}     
\setcounter{equation}{0}
\renewcommand{\theequation}{\footnotesize\mbox{P }\arabic{ueb}.\arabic{equation}}    
\item[ \purpur{\bf Problem  \arabic{ueb}} :] 
\label{Magnetfeld Eich}    
For a charged particle in a magnetic field
$\> {\bf B}(\fx) = {\rm rot} \> {\bf A}(\fx) \> $ 
the classical Hamilton function is
\bea
H(\fp,\fx) \E \frac{1}{2 m} \left ( \> \fp - \frac{e}{c}
{\bf A}(\fx) \> \right )^2  \> .
\eea

\bdes
\item[\purpur{\bf a)}] Starting from the  phase-space path integral
derive the Lagrange form of the  path-integral for the
time-evolution operator (propagator) of the particle.
\item[\purpur{\bf b)}] How does the propagator behave under a  
{\bf gauge transformation}
$ {\bf A}(\fx) \to {\bf A}(\fx) + {\rm grad} \>
\Lambda(\fx)$~?
\item[\purpur{\bf c)}] Show that only the mid-point rule in the discretized form of the path integral
gives the correct \\ \hspace{-0.3cm} Schr\"o\-dinger equation if 
$ \psi\left (\fx,t+\epsilon \right ) =
<\fx\> | \> \hat U(t+\epsilon,t) \> | \> \psi (t) > \> $ is evaluated for small $\epsilon$.
\edes
\vspace{1cm}

\refstepcounter{ueb}     
\setcounter{equation}{0}
\renewcommand{\theequation}{\footnotesize\mbox{P }\arabic{ueb}.\arabic{equation}}    
\item[ \purpur{\bf Problem  \arabic{ueb}} :] 
\label{Jacobi Eq.}    
Consider the family of (one-dimensional) classical paths
leaving the point $ x = a $ at time $t = 0$ with momentum $p$ . Show that
\bea
\hspace{-4cm} J(p,t) \E \frac{\partial x(p,t)}{\partial p} \E
\lim_{\epsilon \to 0} \frac{x(p+\epsilon,t) - x(p,t)}{\epsilon}
\eea
\hspace{9cm}\raisebox{2cm}{\includegraphics[angle=0,scale=0.6]{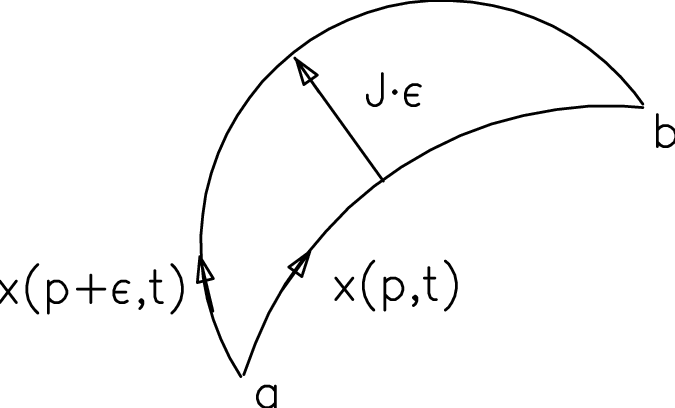}}
\bdes
\vspace{-3.8cm}
\item[\purpur{\bf a)}] fulfills the {\bf Jacobi equation} 
\vspace{0.5cm}
\bea
\hspace{-6.5cm}
\frac{d}{dt} \left ( \> \frac{\partial^2 L}{\partial \dot x^2} \dot J \>
\right ) + \left [ \> \frac{d}{dt} \left (
\frac{\partial^2 L}{\partial x \partial \dot x}
\right ) - \frac{\partial^2 L}{\partial x^2} \> \right ] J \E 0 
\eea
with $ J(p,0) = 0, \> \partial J(p,t)/\partial t \Bigr |_{t=0} = 1/m $ ;    
\item[\purpur{\bf b)}] this is identical with the Gel'fand-Yaglom equation
for a general quadratic Lagrange function;
\item[\purpur{\bf c)}] at the final point $J$ can be expressed by the classical action
$ \> S(x_b,t_b;x_a,t_a) \>$ (with the help of the result in 
\purpur{\bf Problem \ref{klass Anfang/End}}) as follows:
\bea
\frac{1}{J(p,t_b)} \E - \frac{\partial^2 S}{\partial x_a \partial x_b} \> .
\eea
\edes

\edes

\newpage
\bce
{\large  \purpur{\bf Problems for the Lecture ``Path Integrals in 
Quantum Physics"}}

\vspace{1 cm}

{\bf Exercise 4}
\ece
\vspace{ 0.4cm}

\bdes
\refstepcounter{ueb}     
\setcounter{equation}{0}
\renewcommand{\theequation}{\footnotesize\mbox{P }\arabic{ueb}.\arabic{equation}}    
\item[ \purpur{\bf Problem \arabic{ueb}} :] 
\label{Maslov}    
Consider the 1-dimensional motion of a particle in a harmonic potential.
Apply the composition law to (the explicit expression for)
the propagator of the harmonic oscillator
\bea
U^{\rm h.o.}\left (x_b,t_b;x_a,t_a \right ) \E \int_{-\infty}^{+\infty} dx_c \, 
U^{\rm h.o.}\left (x_b,t_b;x_c,t_c \right ) \, 
U^{\rm h.o.}\left (x_c,t_c;x_a,t_a \right ) 
\eea
and show that it indeed acquires a {\bf Maslov phase} factor 
\be 
- i \E e^{ -i \frac{\pi}{2} }  \> ,
\ee
if
$ \> T_1 = t_c - t_a < \pi/\omega \> , T_2 = t_b - t_c < \pi/\omega \> $ holds but
$T_1 + T_2 > \pi/\omega $, i.e. if the particle passes through a focal point
in the time between $t_a$ and $t_b$ .



\vspace{0.4cm}

\refstepcounter{ueb}     
\setcounter{equation}{0}
\renewcommand{\theequation}{\footnotesize\mbox{P }\arabic{ueb}.\arabic{equation}}    
\item[ \purpur{\bf Problem \arabic{ueb}$^{\star}$} :] 
\label{Null Mode}    
In the limit $ \beta \to \infty $ also
$ \> x_{\rm cl}(\tau - \tau_0) = a \tanh \frac{\omega}{2} \left ( \tau - 
\tau_0 \right ) \> $ is a solution of the classical equation of motion for an
{\bf instanton in the double-well potential}
$ \> V(x) = \frac{m\omega^2}{8a^2} \, \left ( x^2 - a^2 \right )^2 \> $ .
Due to time-translation invariance the "position" $\tau_0$ of the instanton is arbitrary.

\bdes
\item[\purpur{\bf a)}] Derive Eq. \eqref{kink} from Eq. \eqref{class sep} for the choice $ x_1(\tau_0) = 0 $.

\item[\purpur{\bf b)}] Show that the operator
\bea
{\cal O}_V \E -m \frac{d^2}{d \tau^2} + V'' \left ( x_{\rm cl}(\tau - \tau_0) \right )\> , 
\eea
which describes the quadratic fluctuations around the classical solution,
then possesses a {\bf zero eigenvalue (``zero mode'')} with eigenfunction 
\bea
y_0(\tau - \tau_0) \E \frac{\rm const.}{\cosh^2 \left [\omega \left ( \tau - \tau_0 
\right )/2 \right ] } \E A_0 \> \frac{\partial x_1(\tau)}{\partial \tau_0} \> .
\eea
Determine the normalization constant $ A_0 $ from $ \> \la y_0 \big | \, y_0 \ra = 1 \> $.


\item[\purpur{\bf c)}] Consider the {\bf Gelfand-Yaglom equation} $ \> {\cal O}_{\rm V''} f_{\rm GY} = 0 \> $ 
with $ \> f_{\rm GY}(-\beta/2) = 0 \> , \> \dot f_{\rm GY}(-\beta/2) = 1 \> $ 
and show that the Wronskian $ \> W = f^{(1)} \dot f^{(2)} -  f^{(2)} \dot f^{(1)} \> $ of two solutions 
is a constant.

\item[\purpur{\bf d)}] For finite, large $\beta$ determine the would-be zero eigenvalue $e_0$ from the 
equation
\be
Y_0(\tau) \E f_{\rm GY}(\tau) + e_0 \, \int_{-\beta/2}^{+\beta/2} d\tau' \> g(\tau,\tau') \, Y_0(\tau')
\label{eigen e0}
\ee
where
\be 
g(\tau, \tau') \E  \frac{1}{m W} \, \Theta(\tau-\tau') \, \lsp f^{(1)}(\tau) f^{(2)}(\tau') - f^{(2)}(\tau)
f^{(1)}(\tau') \rsp
\ee
is a Green function: $ \> {\cal O}_{\rm V''} \, g(\tau,\tau') = \delta(\tau-\tau') \> $  and $ \> f_{\rm GY} \> $ the solution of the Gel'fand-Yaglom equation with the correct initial conditions.
Evaluate $ e_0 $ to leading order in  $ \> \exp(-\omega \beta) \> $ by requiring that the l.h.s of Eq. 
\eqref{eigen e0} also vanishes for $ \> \tau = \beta/2 \> $ and by approximating $ Y_0(\tau') \simeq
f_{\rm GY}(\tau') \> $.

\edes

\edes

\newpage
\bce
{\large  \purpur{\bf Problems for the Lecture ``Path Integrals in Quantum Physics"}}

\vspace{1 cm}

{\bf Exercise 5}
\ece
\vspace{ 0.4cm}

\bdes
\refstepcounter{ueb}     
\setcounter{equation}{0}
\renewcommand{\theequation}{\footnotesize\mbox{P }\arabic{ueb}.\arabic{equation}}    
\item[ \purpur{\bf Problem \arabic{ueb}$^{\star}$} :] 
\label{Feyn Polaron}    
In {\bf Feynman's variational approach} for the polaron
one obtains for the ground-state energy
\bea
E_0 \le E_F \E  \frac{3}{4 v} (v-w)^2 - \frac{\alpha}{\sqrt \pi} \>
\int_0^{\infty} du \>  \frac{e^{-u}} {\mu(u)} 
\eea
with
\bea
\mu(u) \E \left [ \frac{w^2}{v^2} u + \frac {v^2 - w^2}{v^3} \> (1 - 
e^{-v u } ) \right ]^{1/2} \> .
\eea
Here $v$ and $w$ are two variational parameters for the strength
($ v^2 = w^2 + 4 C/w$) and the retardation of the trial action.
\bdes
\item[\purpur{\bf a)}] Determine the variational minimum for small coupling constants
$\alpha$ by the {\it ansatz} $ v = w( 1 + \epsilon )$,  where
$\epsilon = {\cal O}(\alpha) $. Show that the lowest energy
$E_F = - \alpha - \alpha^2/81 - {\cal O}(\alpha^3) $ is reached
for $ w = 3 $ and $\epsilon = 2 \alpha/27 $.
\item[\purpur{\bf b)}] Determine the variational minimum for large coupling constants 
$\alpha$ by the {\it ansatz} $ v \gg w $. Show that the lowest energy
$E_F = - \alpha^2/(3 \pi) $ is reached for $ v = 4 \alpha^2/(9 \pi) $.
\edes

\vspace{0.3cm}

\refstepcounter{ueb}     
\setcounter{equation}{0}
\renewcommand{\theequation}{\footnotesize\mbox{P }\arabic{ueb}.\arabic{equation}}    
\item[ \purpur{\bf Problem  \arabic{ueb}} :] 
\label{kohaer Zust}   
Show that the {\bf coherent states} of the harmonic oscillator,      
which can be defined by
\bea
\hat a^{\dagger} |\, z \, > \E  z \, | \, z \, > 
\eea
are states of minimal uncertainty: Calculate
\bea
\left ( \Delta x \right )^2 \EQ \frac{ < z \> | \> \hat x^2 \> | 
\> z > }{< z \> | \> z >} \>
- \> \left ( \frac{ < z \> | \> \hat x \> | \> z >}{< z \> | \> z >} 
\right )^2 \\
\left ( \Delta p \right )^2 \EQ \frac{ < z \> | \> \hat p^2 \> | 
\> z > }{< z \> | \> z >} \>
- \> \left ( \frac{< z \> | \> \hat p \> | \> z >}{< z \> | \> z >} \right )^2 \> .
\eea

\vspace{0.8cm}

\refstepcounter{ueb}     
\setcounter{equation}{0}
\renewcommand{\theequation}{\footnotesize\mbox{P }\arabic{ueb}.\arabic{equation}}    
\item[ \purpur{\bf Problem \arabic{ueb}} :] 
\label{Det Spur}   
Prove that
\vspace*{0.2cm}

\bdes
\item[\purpur{\bf a)}] $ \quad 
\det \A \> = \exp \lrp  {\rm tr} \ln \A \rrp \> ,$
\item[\purpur{\bf b)}] $ \quad
\frac{\partial}{\partial \lambda} \det \A(\lambda) \E \rm{Tr} \lrp \frac{\partial \A(\lambda)}
{\partial \lambda} \, \A^{-1}(\lambda) \rrp \, \cdot \, \det \A(\lambda)  \> , $ 
\item[\purpur{\bf c)}]  $ \quad
\det \lrp \A_0 + g \A_1 \rrp \E \det \lrp \A_0 \rrp \, \cdot \, \exp \lcp \int_0^g dg' \> 
\rm{Tr} \lsp \A_1 \, \lrp \A_0 + g' \A_1 \rrp^{-1} \rsp \rcp \> . $ 
\edes

\edes

\newpage

\bce
{\large  \purpur{\bf Problems for the Lecture ``Path Integrals in 
Quantum Physics"}}

\vspace{0.8 cm}

{\bf Exercise 6}
\ece
\vspace{ 0.4cm}

\bdes
\refstepcounter{ueb}     
\setcounter{equation}{0}
\renewcommand{\theequation}{\footnotesize\mbox{P }\arabic{ueb}.\arabic{equation}}    
\item[ \purpur{\bf Problem  \arabic{ueb}} :] 
\label{Grassmann}    
Consider a {\bf Grassmann algebra} with one generator   
$ \> \cy{\xi} \> $. Show that for all analytic functions
$ \> f( \cy{\xi}) \E f_0 + f_1 \cy{\xi}  \> $
the Grassmann-$\delta$ function is given by
\bea
\delta(\cy{\xi}-\cy{\xi}') \E \int d\cy{\eta} \> e^{- \cy{\eta} ( \cy{\xi} - \cy{\xi}')} \> ,
\eea
i.e. that
\bea
\int d\cy{\xi}' \> \delta(\cy{\xi}-\cy{\xi}') f(\cy{\xi}') \E f(\cy{\xi})  \> .
\eea
\vspace{1cm}

\refstepcounter{ueb} 
\setcounter{equation}{0}
\renewcommand{\theequation}{\footnotesize\mbox{P }\arabic{ueb}.\arabic{equation}}    
\item[\purpur{\bf Problem \arabic{ueb}} :] 
\label{fermion ho}   
Determine the partition function 
\bea 
Z_{\omega}(\beta) \EA \oint_{\rm{a.p.}} {\cal D} \cy {\bar \xi} \, {\cal D} \cy{\xi}  \, \exp \lcp - 
\int_0^{\hbar \beta} d\tau \> \lsp \cy {\bar \xi}(\tau) \, \frac{\partial \cy {\xi}(\tau)}{\partial \tau} + \omega \, \cy{\bar \xi}(\tau) \cy{\xi}(\tau) \rsp \rcp 
\label{Z omega}
\eea
for the {\bf fermionic harmonic oscillator}

\bdes
\item{\bf \purpur{a)}} by calculating the functional determinant
\bea
Z_{\omega}(\beta) \EA {\cal N} \, \cdot \,  \fdet \lsp \partial_{\tau} + \omega \rsp 
\eea
in the space of anti-periodic ("a.p.") eigenfunctions of the operator $ \> \partial_{\tau} + \omega $. 
Hint: Use the product representation (see, e.g. {\bf \{Gradshteyn-Ryzhik\}}, eq. 1.431.4)
\bea
\cosh x \E \prod_{k=0}^{\infty} \lrp 1 + \frac{4 x^2}{(2 k + 1)^2 \pi^2} \rrp  \>. 
\eea

\item{\bf \purpur{b)}} Determine the normalization constant $ {\cal N} $ (as in the bosonic case) 
by comparing with the special case $ \omega = 0 $ when using the discretized form
\bea
Z_{\omega=0}(\beta) \EA \lim_{N \to \infty} \prod_{n=0}^N \lrp 
\int d \cy{\bar \xi_n} d \cy{\xi_n} \rrp \> 
\exp \lcp - \sum_{n=1}^N \lsp \cy{\bar \xi_n} \lrp \cy{\xi_n}
- \cy{\xi_{n-1}} \rrp \rsp \rcp \> . 
\eea

\item{\bf \purpur{c)}} Show that the result implies a negative ground-state energy
\bea
E_0 \EA - \frac{1}{2} \, \hbar \omega 
\eea
of the fermionic harmonic oscillator which exactly cancels the positive ground-state energy
of the (bosonic) harmonic oscillator.
Does such a supersymmetry solve the problem of the (nearly) vanishing cosmological constant
which frequently is considered as vacuum energy of the universe?

\edes  
\edes
\newpage
\bce
{\large  \purpur{\bf Problems for the Lecture ``Path Integrals in 
Quantum Physics"}}

\vspace{0.5 cm}

{\bf Exercise 7}
\ece
\vspace{ 0.3cm}

\bdes
\refstepcounter{ueb}    
\setcounter{equation}{0}
\renewcommand{\theequation}{\footnotesize\mbox{P }\arabic{ueb}.\arabic{equation}}    
\item[\purpur{\bf Problem \arabic{ueb}} :]         
\label{freie Energie}    
Show that the {\bf free energy} $ F(\beta) $ of a system defined as 
\bea
F(\beta) \EA - \frac{1}{\beta} \, \ln \lcp \sum_{n=0} \, e^{-\beta E_n} \rcp 
\eea
is always smaller than the ground-state energy $ E_0$  and increases monotonically:
\bea
\vspace*{0.1cm}
F(\beta) \ge E_0 \> \> , \qquad F'(\beta) \ge 0 \qquad \forall \> \beta \ge 0 \> . 
\eea

\vspace{0.6cm}

\refstepcounter{ueb}     
\setcounter{equation}{0}
\renewcommand{\theequation}{\footnotesize\mbox{P }\arabic{ueb}.\arabic{equation}}    
\item[ \purpur{\bf Problem  \arabic{ueb}} :] 
\label{G_0 Vielteil}    
Derive the explicit form \eqref{G0} of the interaction-free {\bf one-particle Green function}
in the basis which diagonalizes $ H_0 $ by solving the differential equation
\bea
\lrp \partial_{\tau} + \epsilon_{\alpha} - \mu \rrp \, G_0 (\alpha, \tau | \alpha', \tau')
\EA \delta_{\alpha \alpha'} \, \delta(\tau - \tau') 
\eea
with the boundary condition
\bea
G_0 (\alpha, \beta | \alpha, 0) \EA \zeta  \, G_0 (\alpha, 0 | \alpha, 0) \> ,
\qquad  \zeta = \left \{ \begin{array}{r@{\quad:\quad}l} 
                 +1 & {\rm bosons} \\ -1 & {\rm fermions} \> .
                 \end{array} \right. 
\eea
Discuss why one has to set  $ \tau' \longrightarrow \tau' + 0^+ $, i.e. $\Theta(0) \longrightarrow \Theta(-0^+) = 0 $ in order to get the correct result for equal times.
\vspace{0.6cm}

\refstepcounter{ueb}  
\setcounter{equation}{0}
\renewcommand{\theequation}{\footnotesize\mbox{P }\arabic{ueb}.\arabic{equation}}    
\item[\purpur{\bf Problem \arabic{ueb}$^{\star}$} :]      
\label{Pekar}   
Calculate  {\bf Pekar's constant} for the polaron ground-state energy at large coupling constant~$\alpha$ ($  E_0 \to \gamma_P \, \alpha^2$ ) from the minimum of the functional
\bea
\gamma_P \EA \kappa^2 \, {\rm min}_{(y,y) = 1} \, \Big [ \la T\ra + \la V \ra \Big ] \>, \quad 
(y,y) \E \int_0^{\infty} \!dr \, y^2(r) \E 1 \> , \> \la T \ra \E \int_0^{\infty} \!dr \, \frac{1}{2} \, y'^2(r) \> , \non 
\la V \ra \EA - \frac{1}{\sqrt{2} \kappa} \, \int_0^{\infty} \!dr \> y^2(r) \, \int_0^{\infty} \, ds \>
\frac{y^2(s)}{{\rm max}(r,s)} \E - \frac{\sqrt 2}{\kappa} \, \int_0^{\infty} \! dr \> y^2(r) \, \int_r^{\infty} \!ds \, \frac{1}{s} y^2(s)
\eea

\bdes
\item[\purpur{\bf a)}] with the help of the {\it ansatz}
\bea
y(r) \E C_1 \,r \, e^{-r/a}
\eea
\item[\purpur{\bf b)}] with the help of the {\it ansatz}
\bea
y(r) \E C_2 \, r \, e^{-r^2/(2 b^2)} \> .
\eea
by variation w.r.t. the free parameters  $ a , b $ and show that the result is
independent of the arbitrary scale parameter  $ \kappa $ . 
$ \big [ $ Help with the integrals: Use Euler's integral representation of the $\Gamma$-function
\be 
\Gamma(x) \E  \int_0^{\infty}dt \> t^{x-1} \, e^{-t}
\label{Euler Gamma}
\ee
and the properties
$ \> \Gamma(x+1) = x \, \Gamma(x) \> , \quad \Gamma(1/2) = \sqrt{\pi} \> $. $ \big ] $
\item[\purpur{\bf c)}] Modify the given program for the numerical calculation of 
 $ \gamma_P $ to work with double-precision arithmetic and show that with
  FTOL = $ 10^{-8} $ and NMAX = $ 12 $ the more precise value
\be
\gamma_P \E - 0.10851197(2) 
\ee
is obtained.
\edes

\newpage

\bce
{\large  \purpur{\bf Problems for the Lecture ``Path Integrals in 
Quantum Physics"}}

\vspace{0.8 cm}

{\bf Exercise 8}
\ece
\vspace{ 0.4cm}

\bdes

\refstepcounter{ueb}     
\setcounter{equation}{0}
\renewcommand{\theequation}{\footnotesize\mbox{P }\arabic{ueb}.\arabic{equation}}    
\item[\purpur{\bf Problem \arabic{ueb}} :]          
\label{Selbstenergie phi4}   
Calculate the divergent integral which in first-order perturbation theory
determines the {\bf selfenergy in $ \Phi^4$-theory}
\bea
\Sigma^{(1)} \E i \ \frac{\lambda}{2} \, \mu^{4 - d}_0 \, \int \frac{d^dk}{(2 \pi)^d} \> 
\frac{1}{k^2 - m^2 + i \, 0^+} 
\eea
in $d$ dimensions ({\bf dimensional regularization}: $\> 1$ time and $d-1$ space dimensions). 
Here $\mu_0 $ is a mass parameter which one introduces to keep the coupling constant
$ \lambda $ dimensionless for all $ d $. In which dimension does this integral converge?

\noindent
Hint: Use the {\bf Fock-Schwinger representation} for the free propagator
\bea
\frac{1}{k^2 - m^2 + i \, 0^+} \E - i \, \int_0^{\infty} dT \> e^{i T (k^2 - m^2 + i \, 0^+)} 
\label{Fock-Schwinger}
\eea
and the usual Gaussian integral (caution: Minkowski metric!). Show that the result
\bea
\Sigma^{(1)} \E  \frac{\lambda}{32\pi^2} \, m^2 \, \left ( \frac{m^2}{4 \pi \mu_0^2} \right )^{d/2-2} \, 
\Gamma ( 1 - d/2)
\label{Sigma1 phi4}
\eea
is real. Set the dimension to  $ d = 4 - 2 \epsilon$ and
 investigate the limit  $\epsilon \to 0  $ by using properties of the Gamma function $ \Gamma(x)$. 
\vspace{0.5cm}

\refstepcounter{ueb}     
\setcounter{equation}{0}
\renewcommand{\theequation}{\footnotesize\mbox{P }\arabic{ueb}.\arabic{equation}}    
\item[\purpur{\bf Problem \arabic{ueb}} :]
\label{N skalare Teil}   
Consider a system of free scalar particles occurring in  $N$ species $\Phi_i$
 (with equal masses) so that their Lagrangian reads
\bea
{\cal L}_0^{(N)} \E \frac{1}{2} \sum_i^N \left [ \, \left ( \partial \Phi_i 
\right )^2 - m^2 \Phi_i^2 \, \right ] \> .
\eea
Show for the special case  $ N = 2 $ that for positively and negatively charged particles 
the free Lagrangian can be written as
\bea
{\cal L}_0^{(2)} \E \left | \partial \Phi \right |^2| - m^2 \left | \Phi 
\right |^2  
\eea
where $ \Phi $ is now a {\bf complex field}.
Evaluate the free generating functional 
\bea
Z_0 \left [ J^{\ast}, J \right ] \E \int {\cal D} \Phi^{\ast} \, {\cal D} \Phi
\> \exp \left [  i \int d^4 x \left ( {\cal L}_0^{(2)} + J^{\ast}(x) \Phi(x) + 
 \Phi^{\ast}(x) J(x) \right ) \, \right ] 
\eea
and the corresponding 2-point functions
$\left < \Phi^{\ast}(x_1) \Phi(x_2) \right > , 
\left < \Phi(x_1) \Phi(x_2) \right > $.

\edes

\newpage

\bce
{\large \purpur{\bf Problems for the Lecture ``Path Integrals in 
Quantum Physics"}}

\vspace{ 0.8 cm}

{\bf Exercise 9}
\ece
\vspace{0.4cm}

\bdes                 
\refstepcounter{ueb}                                      
\setcounter{equation}{0}
\renewcommand{\theequation}{\footnotesize\mbox{P }\arabic{ueb}.\arabic{equation}}    
\item[ \purpur{\bf Problem \arabic{ueb}} :] 
\label{Kumulante}   
Consider the Fourier transform of an ordinary function  $ f(x) $ expanded in moments
\bea
\tilde f(t) \Def \int dx \> f(x) \, e^{i t x} \E \sum_{n=0}^{\infty} \, m_n \, \frac{ (i t)^n}{n!} 
\> \> , 
\qquad m_n \> \equiv \ \int dx \> x^n \, f(x) \> \> ,
\label{mom entwick}
\eea
which all are supposed to exist.  For $ m_0 \neq 0 $ the {\bf cumulant expansion} of 
 $ \tilde f(t) $ is defined by
\bea
 \tilde f(t) \deF m_0 \> \exp \lcp \sum_{n=1}^{\infty} \, \lambda_n \,  \frac{ (i t)^n}{n!} \rcp \> .
\label{cum entwick}
\eea

\bdes
\item[\purpur{\bf a)}] Show that the cumulants $ \lambda_n $ can be determined recursively from  the moments $ m_n $ by
\bea
\lambda_{n+1} \E \frac{m_{n+1}}{m_0} - \sum_{k=0}^{n-1} \, {n \choose k} \>  \frac{m_{n-k}}{m_0} 
\> \lambda_{k+1}
\> \> \> , \quad n = 0 \> (\mbox{the sum is then empty})\> , 1, \ldots 
\label{cum from mom}
\eea
and give explicitly the first four (in statistics known under the names ``{\bf mean}'', 
``{\bf variance}'', ``{\bf skewness}'' and ``{\bf excess}''). 
Hint: Differentiate the moment and cumulant expansion w.r.t. the parameter  $t$ and compare
respective powers of $t$.

\item[\purpur{\bf b)}] Express the cumulants recursively in terms of the {\bf central moments}
\be 
c_n \Def \int dx \>  \lrp x - \frac{m_1}{m_0} \rrp^n \> f(x) \> .
\ee

\item[\purpur{\bf c)}] Generalize the cumulant expansion to the functional representation of the generating functional and list the first four connected Green functions when the action is even ( $ S[-\Phi] = S[\Phi] $).

\edes

\vspace{0.3cm}

\refstepcounter{ueb}                                      
\setcounter{equation}{0}
\renewcommand{\theequation}{\footnotesize\mbox{P }\arabic{ueb}.\arabic{equation}}    
\item[ \purpur{\bf Problem \arabic{ueb}$^{\star}$} :] 
\label{eff Wirk}   
Consider the  {\bf effective action}
\bea
\Gamma[\Phi] \E W - \int d^4x \> J(x) \Phi_{\rm cl}(x) 
\label{def Gamma}
\eea
of a scalar theory with the generating functional
\bea 
Z[J] \E \exp ( i W[J]/\hbar ) \hspace{0.5cm}  {\rm and}  \hspace{0.5cm} 
\Phi_{\rm cl}(x) \E \frac{\delta W[J]}{\delta J(x)} \> . 
\label{def Phi_kl}
\eea 
Prove that
\bea
\frac{\delta \Gamma[\Phi]}{\delta \Phi_{\rm cl}(x)} \E - J(x) 
\eea
and
\bea
\delta^4(x-y) \E - \int d^4z \> \frac{\delta^2 \Gamma}{\delta \Phi_{\rm cl}(z)
\delta \Phi_{\rm cl}(y)} \> \frac{\delta^2 W}{\delta J(z) \delta J(y)}  \> .
\eea
In momentum space the 2-point function can be written as
$ \> G^{(2)}(p) = i \left ( p^2 - m^2 - \Sigma(p) \right )^{-1} \> $.              
Show thereby  that $ \> \Gamma^{(2)}(p) \> $ gives the selfenergy
$ \> \Sigma(p) \> $ as sum of all one-particle irreducible contributions to the 
2-point function.

\edes

\newpage

\bce
{\large  \purpur{\bf Problems for the Lecture ``Path Integrals in 
Quantum Physics"}}

\vspace{ 0.8 cm}

{\bf Exercise 10}
\ece
\vspace{0.4cm}

\bdes
\refstepcounter{ueb}     
\setcounter{equation}{0}
\renewcommand{\theequation}{\footnotesize\mbox{P }\arabic{ueb}.\arabic{equation}}    
\item[\purpur{\bf Problem \arabic{ueb}$^{\star}$} :]
\label{Weltlin Pot}   
Consider charged scalar particles described by the Lagrangian
\bea
{\cal L}_1 \E  \, \left | \partial \Phi \right |^2 - 
\left ( m^2 + 2 m V(x) \right ) \left | \Phi \right |^2  \> ,
\eea 
where  $ \> V(x)$ is an external potential.

\bdes
\item[\purpur{\bf a)}] Show with the help of the {\it ansatz}
\bea
\Phi(x_0=t,{\bf x}) \E \frac{1}{\sqrt{2 m}} \, e^{-i m  t} \, \varphi(t,{\bf x}) 
 \> ,
\label{ansatz nr}
\eea
that in the limit $ m \to \infty $ the action
becomes the one of a non-relativistic system of particles in the external
potential $V(t,{\bf x})$ .

\vspace{0.3cm}

\item[\purpur{\bf b)}] Calculate the exact generating functional for this system
and show that the 2-point function (the propagator) is given by
\bea
\left < \, \Phi^{\ast}(x_2) \Phi(x_1) \, \right > \EQ G_2(x_2,x_1) \E i \, 
\left < x_2 \> \left | \> \frac{1}{ -\partial^2 - m^2 - 2 m V(x) + i \, 0^+} \>
\right | \> x_1 \right >  \> .
\eea

\vspace{0.3cm}

\item[\purpur{\bf c)}] By using the {\bf Fock-Schwinger representation}
\bea
\frac{1}{\hat A + i \, 0^+} \E - i \kappa \int_0^{\infty} dT \, 
\exp ( \, i \kappa T \, \hat A \, )  \> \> , \quad \kappa > 0 
\eea 
show that 
$ \> G(x_2,x_1) \> $ can be written like a non-relativistic path integral:
\bea
&& G_2(x_2,x_1) \E  \frac{1}{2m} \int_0^{\infty} dT \, e^{- i m T/2}
\int_{x(0)=x_1}^{x(T)=x_2} {\cal D}x(\tau) \, e^{i S[x(\tau)]}  
\label{G2 eigen}\\
&& S[x(t)] \E \int_0^T d\tau \, \left [ - \frac{m}{2} \dot x(\tau)^2 + V(x(\tau))
\right ]  \> .
\label{S eigen}
\eea
The paths $\> x_{\mu}(\tau) \> $ are now parametrized by the proper time
 $ \> \tau \> $ which runs from $0$ to $T$. Subsequently one has to integrate over
 $ T $  with the weight $ \exp (-im T/2) $ ({\bf worldline representation}).

\vspace{0.3cm}
%

\edes
\vspace*{0.5cm}

\refstepcounter{ueb} 
\setcounter{equation}{0}
\renewcommand{\theequation}{\footnotesize\mbox{P }\arabic{ueb}.\arabic{equation}}    
\item[\purpur{\bf Problem \arabic{ueb}$^{\star}$} :]  
\label{Ableit erzeug Funk}   
Verify
\bea
F \lrp \frac{1}{i} \frac{\partial}{\partial x} \rrp \> \exp \lrp - \frac{i}{2} a x^2 \rrp \E 
 \exp \lrp - \frac{i}{2} a x^2 \rrp \>  \exp \lrp - \frac{i}{2} a \frac{\partial^2}{\partial y^2} 
\rrp  \> F(y) \, \Bigr |_{y=ax} \> . 
\label{Ableit Bezieh}
\eea
Hint: Use the Fourier transform $ \> F(y) = \int_{-\infty}^{+\infty} dt \, \tilde F(t) \, 
\exp(-i t y)/(2 \pi) \>  $ together with the fact that
$ \exp ( t \partial/\partial x) \, f(x) = f(x + t ) $ gives a shift of argument.
\vspace*{0.1cm}

Generalizing that to functional derivatives show that the generating functional of a interacting scalar theory may be written as 
\bea
Z[J] \E {\rm const.} \> \exp \lsp - \frac{i}{2} \lrp J, \Delta_F J \rrp \rsp \> \lcp \exp \lsp 
\frac{i}{2} \lrp \frac{\delta}{\delta \varphi}, \Delta_F \frac{\delta}{\delta \varphi} \rrp \rsp  \> 
\exp \lrp - i \int d^4x \> V(\varphi) \rrp \rcp_{\varphi = \Delta_F J} 
\label{Kopitz result} 
\eea
where $ \> \Delta_F(k) = 1/(k^2 - m^2 + i0^+ ) \> $ is the free  propagator. Check in 
 $ \Phi^4 $-theory  whether the first correction to the free generating functional
$ \> \omega_1[J] \> $ (Eq. \eqref{omega1}) is obtained from this
representation.
\edes
\newpage

\bce
{\large  \purpur{\bf Problems for the Lecture ``Path Integrals in 
Quantum Physics"}}

\vspace{ 0.8 cm}

{\bf Exercise 11}
\ece
\vspace{0.3cm}

\bdes

\vspace*{0.3cm}

\refstepcounter{ueb}  
\setcounter{equation}{0}
\renewcommand{\theequation}{\footnotesize\mbox{P }\arabic{ueb}.\arabic{equation}}    
\item[\purpur{\bf Problem \arabic{ueb}} :]  
\label{Bi-Quelle}   
Consider, as in \purpur{\bf Problem \ref{N skalare Teil}}, scalar particles
which exist in  $ N $ species $ \, \Phi_i \, $ (with the same mass)
but now have a self-interaction
\bea
{\cal L}^{(N)} \E \frac{1}{2} \, \sum_i^N \lsp \lrp \partial \Phi_i \rrp^2 - m^2 \,  \Phi_i^2 \rsp
- \frac{\lambda}{4!} \, \lrp \sum_i^N \Phi_i^2 \rrp^2 \> . 
\eea
(Since the action is invariant under a orthogonal transformation in the $N$-dimensional 
space of fields. this is called a ``{\bf O(N)-symmetric $\Phi^4$-theory}'' in theoretical parlance).\\
In addition to the usual sources  $ \, J_i(x) \, $ introduce a source
 $ \, K(x) \, $ which couples to
$ \> \sum_i^N \Phi_i^2 \> $ :
\bea
Z[J_i,K] \Def \lrp \prod_{i=1}^N \, \int {\cal D} \Phi_i \rrp \>\exp \lcp i \int d^4x \> \lsp {\cal L}^{(N)}
- \frac{1}{2} K(x)  \sum_i^N \Phi_i^2(x) +  \sum_i^N J_i(x) \Phi_i(x)\rsp  \rcp \> . 
\eea
\bdes
\item[\purpur{\bf a)}] Show that the generating functional of the interacting theory is given
by
\bea
Z[J_i] \E \exp \lrp \frac{i \lambda}{6} \int d^4x \> \frac{\delta^2}{\delta K(x)^2} \rrp \> 
Z_0[J_i,K] \, \Biggr |_{K=0} \> .
\label{Z von Z0 bi}
\eea
\item[\purpur{\bf b)}] Prove that
\bea
Z_0[J_i,K] \E {\rm const.} \, \exp \lrp - \frac{i}{2} \sum_i^N \la J_i \left | {\cal O}_K^{-1} \right | 
J_i \ra - \frac{N}{2} \, {\rm tr} \ln {\cal O}_K \rrp \> \equiv \> e^{i W_0[J_i,K]}   
\eea
with  $ \> {\cal O}_K = -\partial^2 - m^2 - K  \> $.
\item[\purpur{\bf c)}] Show that the first-order correction for the generating functional
$ Z[J_i] $ now is
\bea
\omega_1 [J_i] \E \frac{i}{6} \int d^4x \> \lsp i \frac{\delta^2 W_0}{\delta K(x)^2}  
 - \lrp \frac{\delta W_0}{\delta K(x)} \rrp^2 \rsp_{K=0} \> .
\label{omega1 von W0}
\eea
Determine with that the symmetry factors for the individual first-order graphs 
depending on the number  $ N $ of components of the field $ \Phi_i$.
\edes

\edes

\newpage
\bce
{\large  \purpur{\bf Problems for the Lecture ``Path Integrals in 
Quantum Physics"}}

\vspace{0.8 cm}

{\bf Exercise 12}
\ece
\vspace{0.4cm}

%
%
\bdes
\refstepcounter{ueb}  
\setcounter{equation}{0}
\renewcommand{\theequation}{\footnotesize\mbox{P }\arabic{ueb}.\arabic{equation}}    
\item[\purpur{\bf Problem \arabic{ueb}} :] 
\label{Vektor-Teilchen}   
\bdes
\item[\purpur{\bf a)}]
Consider the free Lagrangian  \eqref{vec L0} for a {\bf massive spin-1 particle} (vector particle)
$ V_{\mu}(x) $. 
Evaluate the path integral for the free generating functional
\bea 
Z_0\lsp J_{\mu}\rsp \EA  \int \prod_{\mu} {\cal D}V_{\mu}  \exp \lcp i \int d^4x \lsp 
{\cal L}_0^V + J_{\mu}(x) V^{\mu}(x) \rsp \rcp \non
\EA {\rm const.}  \, \exp \lcp -\frac{i}{2} \int d^4x d^4y  \> J_{\mu}(x) \,  D_V^{\mu \nu}(x-y)  
\, J_{\nu}(y) \rcp  \> .
\eea
As in the scalar case determine the propagator by transforming into momentum space and thereby
verify Eq. \eqref{vec prop}. 
\item[\purpur{\bf b)}]
Do the same for the {\bf massless} photon propagator in a covariant gauge
with gauge fixing parameter $ \lambda $ given by        
\be
\Delta_{\mu \nu}(x,y) \E \la x \lvl \lsp g^{\mu \nu} \Box - \lrp 1 - \frac{1}{\lambda} \rrp 
\partial^{\mu} \partial^{\nu} \rsp^{-1} \rvl y \ra \> .
\ee
\edes

\vspace*{0.5cm}

\refstepcounter{ueb} 
\setcounter{equation}{0}
\renewcommand{\theequation}{\footnotesize\mbox{P }\arabic{ueb}.\arabic{equation}}    
\item[\purpur{\bf Problem \arabic{ueb}} :]            
\label{Noether Feld}   
Generalize {\bf Noether's theorem} derived in  {\bf chapter} {\bf \ref{sec1: Symm}} 
from quantum  mechanics to field theory:
Assume that the Lagrangian $ \> {\cal L}(\Phi,\partial_{\mu} \Phi)\>  $ 
of a theory is invariant under an infinitesimal transformation
\be
\Phi(x) \> \longrightarrow \> \Phi(x) + \alpha \, \Delta \Phi(x) \> , \quad \alpha = {\rm const.} 
\quad \Longrightarrow \quad 
{\cal L} \> \longrightarrow \> {\cal L} + \alpha \, \partial_{\mu} \, \Lambda^{\mu} 
\ee
up to a total derivative.
As in Eq. \eqref{Pfad Noether} show then by a transformation with a 
$x$-dependent parameter $ \> \alpha(x) \> $ that there exists a conserved current, i.e. that
\bea 
\partial_{\mu} \, \la J^{\mu} \ra \E {\cal N} \, \int {\cal D} \Phi \>  
\partial_{\mu}  \lcp  \frac{{\partial \cal L}}{\partial(\partial_{\mu} \Phi)} \Delta \Phi 
 - \Lambda^{\mu} \rcp \> e^{i S[\Phi] } \E 0 
\eea
holds. 
\bdes
\item[\purpur{\bf a)}] For complex fields, in particular, derive this current from the invariance
under the transformation
\bea
\Phi & \longrightarrow & e^{i \alpha} \, \Phi  \> , \qquad 
\Phi^{\star}  \longrightarrow  \Phi^{\star}e^{- i \alpha} 
\label{phase transf}
\eea
for the following theories:
\bdes
\item[\purpur{\bf a1)}] A charged scalar field with the Lagrangian \eqref{L0 geladen skalar}
\item[\purpur{\bf a2)}] A free Dirac field with the Lagrangian\eqref{L0 fermion}
\item[\purpur{\bf a3)$^{\star}$}] The nucleons in the Walecka model with Lagrangian \eqref{L Walecka}
\item[\purpur{\bf a4)}] A non-relativistic theory in which the particles interact via a {\it local}
two-particle potential \\ $ \la \fx_1',\fx_2' \lvl \hat V  \rvl \fx_1,\fx_2 \ra  = V(\fx_1-\fx_2) 
\delta(\fx_1-\fx_1') \delta (\fx_2 -\fx_2') $ (Lagrangian \eqref{L Schroedinger} 
and \eqref{H op}). 
\edes
\vspace{0.2cm}

\item[\purpur{\bf b)}] Consider for simplicity again a neutral scalar field $ \> \Phi(x) \> $. Under the shift $ \> x_{\mu} \longrightarrow x_{\mu} + a_{\mu} \> $
 ($ a_{\mu} $ constant) the action is invariant, i.e. the Lagrangian changes by a total derivative.
For an infinitesimal shift calculate this total derivative and $ \> \Delta {\cal L} \> $ from the field change $ \> \Delta \Phi(x) = a_{\mu} \partial^{\mu} \Phi(x) \> $. Determine 
from the invariance the conserved  energy-momentum tensor as 
\be 
T^{\mu \nu} \E \frac{\partial {\cal L}}{\partial \lrp \partial_{\nu} \Phi \rrp } \, \partial^{\mu} \Phi - g^{\mu \nu} {\cal L} \> .
\ee

\edes

\newpage

\bce
{\large  \purpur{\bf Problems for the Lecture ``Path Integrals in 
Quantum Physics"}}
\vspace{0.8cm}

{\bf Exercise 13}
\ece
\vspace{ 0.4cm}

\bdes
\refstepcounter{ueb}
\setcounter{equation}{0}
\renewcommand{\theequation}{\footnotesize\mbox{P }\arabic{ueb}.\arabic{equation}}    
\item[\purpur{\bf Problem \arabic{ueb}$^{\star}$} :]  
\label{Dirac nonrel}  
Consider a Dirac particle moving in external scalar and vector potentials
\be
{\cal L} \E \cy{\bar \Psi} \lrp i \ddslash - M^{\star} \rrp \, \cy{\Psi} \> , \quad D_{\mu} \E \partial_{\mu} + i A_{\mu}(x) \> , \quad M^{\star}(x) \Def M + U_S(x) \> .
\label{Dirac scalar+vector}
\ee
Given a constant four-velocity  $ \> v_{\mu} \>  (v^2 = 1 ) \> $ of the system one can decompose
\bea 
\cy{\phi}(x) & \Def &  \frac{1}{2} \lrp 1 + \vslash \rrp \, e^{i M v \cdot x} \, \cy{\Psi}(x) \> , \quad 
\cy{\chi}(x) \Def  \frac{1}{2} \lrp 1 - \vslash  \rrp \, e^{i M v \cdot x} \, \cy{\Psi}(x) \\
D_{\mu} \EA v_{\mu} \, \lrp v \cdot D \rrp + D_{\mu}^{\perp} \> , \quad  D_{\mu}^{\perp} \E \lrp 
g_{\mu \nu} - v_{\mu} v_{\nu} \rrp \, D^{\nu} \> .
\eea

\bdes
\item[\purpur{\bf a)}]  Prove
\be 
\vslash \phi \E \phi \> , \quad \vslash \chi \E - \chi \> ,\quad [\ddslash^{\perp}, \vslash ]_{+} 
\E 0 \> .
\ee
Hint: Use 
\be 
\gamma_{\mu} \gamma_{\nu} \E  \frac{1}{2} [\gamma_{\mu}, \gamma_{\nu}]_+ + 
 \frac{1}{2} [\gamma_{\mu}, \gamma_{\nu}]_- \E  g_{\mu \nu} - i \sigma_{\mu \nu}  
 \label{gamma gamma}
\ee
with $ \> \sigma_{\mu \nu} := \frac{i}{2} [\gamma_{\mu}, \gamma_{\nu} ]_{-}  $ \quad .
\item[\purpur{\bf b)}]  With these properties show that the action takes the form
\be 
S[\phi,\chi] \E \int d^4x \> \lsp \bar \phi \lrp i v\cdot D + M - M^{\star} \rrp \Phi - \bar \chi \lrp i v \cdot D +  M + M^{\star} \rrp \chi + \bar \chi \, i \ddslash^{\perp} \phi  + \bar \phi \, i \ddslash^{\perp} \chi \rsp \> .
\label{S phi chi}
\ee
In the heavy-mass (non-relativistic) limit integrate out the "small" component $  \> \chi \> $ and expand in inverse powers of $ M $. 
\item[\purpur{\bf c)}] Take the system at rest, i.e. $ \> v_{\mu} = (1,{\bf 0} ) \> $, and specialize to a situation where there is no preferred direction (e.g. a spherical or unpolarized system), i.e. 
$ \> A_{\mu}(x) = (U_V(\fx),{\bf 0}) \> $. Assume time-independent potentials depending only on
$ \> r = |\fx| \> $.
Show that in leading order one obtains a {\bf non-relativistic Hamiltonian} with a {\bf central potential}
$ \, V_c(r) \, $ 
and a {\bf spin- orbit potential} $ \, V_{LS} \, $ given by
\be
V_c(r) \E U_V(r) + U_S (r) \> , \quad 
V_{LS} \E \frac{1}{2 M^2} \, \frac{1}{r} \frac{d}{dr} \, \Big [ \,  U_S(r) + U_V(r) \, \Big ] \> 
{\bf L} \cdot {\bf S} 
\ee
where   $ \> {\bf L } = \fx \times \nabla/i \> $ is the orbital momentum
and $ \> {\bf S} =  \vec{\sigma}/2\>  $ the spin of the particle.\\
Hint: Use
\bea 
\gamma_0 \EA \begin{pmatrix} 1 & 0\\
                            0 & -1 \end{pmatrix} \> , \qquad 
\vec{\gamma} \E \begin{pmatrix} 0 & \vec{\sigma} \\
                                - \vec{\sigma} & 0 \end{pmatrix} 
\label{gamma explizit}
\eea                               
and the following relation for the Pauli matrices $ \> \vec{\sigma} \> $
\bea 
\vec{\sigma} \cdot  {\bf a} \> \> \vec{\sigma} \cdot {\bf b}\EA {\bf a} \cdot {\bf b} 
+  i \vec{\sigma} \cdot \lrp {\bf a} \times {\bf b} \rrp \> .
\label{Pauli sigma Rel}
\eea  

\edes
\edes

\newpage

\bce
{\large  \purpur{\bf Problems for the Lecture ``Path Integrals in 
Quantum Physics"}}

\vspace{ 0.8 cm}

{\bf Exercise 14}
\ece
\vspace{0.4cm}

\bdes
\refstepcounter{ueb}  
\setcounter{equation}{0}
\renewcommand{\theequation}{\footnotesize\mbox{P }\arabic{ueb}.\arabic{equation}}    
\item[\purpur{\bf Problem \arabic{ueb}} :]              
\label{x Prop}   
Evaluate --
by Fourier transformation  in $ d $ dimensions -- the {\bf free propagator in $x$-space}
\bdes
\item[\purpur{\bf a)}] for a scalar particle (use
Eq. \eqref{skalarer Propagator})
\item[\purpur{\bf b)}] for a photon in covariant gauge with gauge fixing parameter
$ \lambda $ (use Eq. \eqref{Eichboson Prop}).
\edes
As in \purpur{\bf Problem \ref{Selbstenergie phi4}}) one may utilize the Schwinger representation for the individual denominators and the Gaussian integral in Minkowski space.
How do these propagators behave if the relative space-time distance becomes small or large?

\vspace{0.3cm}

\refstepcounter{ueb}  
\setcounter{equation}{0}
\renewcommand{\theequation}{\footnotesize\mbox{P }\arabic{ueb}.\arabic{equation}}    
\item[\purpur{\bf Problem \arabic{ueb}$^{\star}$} :]               
\label{Bloch-Nordsieck}    
Derive the full (interacting) propagator of a electron
by functional differentiation of the generating functional of QED
and show that one obtains 
\bea
\bar G_2(p)=\!\!\int \!\!d^4(x-y) e^{-i p \cdot (x-y)} \! \! \int \!\!{\cal D}A 
\la x | {\cal O}^{-1} | y \ra  \, {\rm Det} \,{\cal O}  e^{i S_0[A]} , \,  
{\cal O}=i \dslash - e \Aslash(x) - m + i0^+ 
\eea
with  $S_0[A]$ as photon action.\\
The {\bf Bloch-Nordsieck approximation} for the interacting electron propagator
consists of neglecting the determinant in the path integral (which describes vacuum polarization)
and of replacing the Dirac matrices by a constant four-vector
\bea
\gamma_{\mu} \> \longrightarrow \> v_{\mu} \equiv \frac{p_{\mu}}{m_{\rm phys}} \>, 
\eea
such that the relation $ \, \pslash = m_{\rm phys}  $ is conserved on the mass shell.
\bdes
\item[\purpur{\bf a)}] Use again the Schwinger representation to show that
\bea
\hspace*{-1cm}\exp \Bigl [ i T \lrp i v \cdot \partial - e v \cdot A(x) - m \rrp  \Bigr ] =  
\exp \Bigl [ i T ( i v \cdot \partial - m ) \Bigr ] \cdot 
\exp \Biggl[ - i e \int\limits_0^T  \! d\tau \, u \cdot A(x + v \tau) \Biggr] .
\eea
With this result perform the functional integration over the photon field and prove 
\bea
\bar G_2(p) \>\simeq \> \bar G_2^{\rm BN}(p) \E \int_0^{\infty} dT \> \exp \lsp i \, 
\lrp v \cdot p - m \rrp \, T +  X(T) \rsp 
\eea
where 
\bea
X(T) \E -i \frac{e^2}{2} \, \int_0^T d\tau \, d\tau' \> v^{\mu} \, v^{\nu} \> 
\Delta_{\mu \nu} \lrp v \, ( \tau - \tau') \rrp 
\eea
is determined by the photon propagator $ \Delta_{\mu \nu}(x-y) $ in $x$-space.
\item[\purpur{\bf b)}] Calculate $ X(T) $ in dimensional regularization
($ e^2 \to e^2 \, \mu_0^{4-d}$) by using the result in
\purpur{\bf Problem \ref{x Prop}~b)} in $ d = 4 - 2 \epsilon $ dimensions.
Expand the result up to order $ \epsilon^0 $ and show thereby
\bea
\bar G_2^{\rm BN}(p) \E \int_0^{\infty} dT \> \exp \lsp i \frac{p^2 - m^2}{m} T + \kappa \, 
\ln \mu_0 T 
+ c \rsp \E i Z_2 \, \frac{m^{1+2\kappa}}{(p^2 - m^2 + i0^+)^{1 + \kappa}}  \> .    
\eea
Determine $ \, Z_2, m \, $ in this approximation and show that the exponent
$ \kappa $ depends on the gauge fixing parameter
$ \lambda $. Show that the  $ \ln T $-dependence in the proper-time integral and accordingly the essential singularity
of the electron propagator at $ p^2 = m_{\rm phys}^2 $ has its origin in the fact that the photons are massless.

\edes

\edes
\newpage

\bce
{\large  \purpur{\bf Problems for the Lecture ``Path Integrals in 
Quantum Physics"}}

\vspace{ 0.8 cm}

{\bf Exercise 15}
\ece
\vspace{0.4cm}

\bdes
\refstepcounter{ueb} 
\setcounter{equation}{0}
\renewcommand{\theequation}{\footnotesize\mbox{P }\arabic{ueb}.\arabic{equation}}    
\item[\purpur{\bf Problem \arabic{ueb}$^{\star}$} :]         
\label{BRST}    
To quantize a gauge theory one needs to fix the gauge and to introduce Faddeev-Popov ghosts 
which take away unphysical degrees of freedom.
Show that the Lagrangian \eqref{volle Lagrangefunk}
\bea
{\cal L} \EA {\cal L}_f +  {\cal L}_g +  {\cal L}_{gauge \, fix} + {\cal L}_{FP} \>, \quad 
 {\cal L}_{gauge \, fix} \E \frac{\lambda}{2} \, B^a B^a  -B^a \partial \cdot A^a 
\eea
of a non-abelian gauge theory with covariant gauge fixing, Faddeev-Popov ghosts and
auxiliary field $B^a$ is invariant under the
 {\bf Becchi-Rouet-Stora-Tyutin (BRST) transformation}:
\bea
\delta A_{\mu}^a \EA \cy{\omega} \lrp D^{\mu} \, \cy{\chi} \rrp^a \> , \quad 
\delta \cy{\psi} \E  i g \,  \cy{\omega} \, \cy{\chi^a} \, T^a \, \cy{\psi} \non
\delta \cy{\bar{\chi}^a}(x) \EA \cy{\omega} \, B^a \> , \quad 
\delta \cy{\chi^a} \E -\frac{g}{2} \, \cy{\omega} \, f^{abc} \, \cy{\chi^b} \cy{\chi^c} \> , 
\quad \delta B^a \E 0 
\eea
with a constant Grassmann-valued parameter  $ \, \cy{\omega} \, $.
Show first that the BRST transformation is a special, local gauge transformation
with the parameter $ \Theta^a(x) =   \cy{\omega} \, \cy{\chi^a}(x) \,$.   
Calculate then the variation of $ \, D_{\mu}^{ab} \cy{\chi^b} \, $ and with that
the variation of $ \> {\cal L}_{gauge} + {\cal L}_{FP} $.

Hint: Use the Jacobi identity for the structure constants
\bea
f^{ade} f^{bcd} + f^{bde} f^{cad} + f^{cde} f^{abd} \E 0 \> . 
\label{Jacobi ident}
\eea

\vspace{0.3cm}

\refstepcounter{ueb} 
\setcounter{equation}{0}
\renewcommand{\theequation}{\footnotesize\mbox{P }\arabic{ueb}.\arabic{equation}}    
\item[\purpur{\bf Problem \arabic{ueb}$^{\star}$ } :] 
\label{Gitter 2-Punkt}   
\vspace{0.1cm}

Calculate the  {\bf free 2-point function on a Euclidean space-time lattice}.
\bdes
\item[\purpur{\bf a)}] Verify Eq. \eqref{Gitter 2-Punkt-Funktion}) for a {\bf scalar field}.

\noindent
Hint: Calculate first (as in the continuum) the generating functional
\bea
Z_0(J_l) \E \prod_k \lrp \int d^4 \Phi_k \rrp \> \exp \lsp - S_E^{(0)}(\Phi) + \sum_l J_l \, 
\Phi_l \rsp 
\eea
with the $(\lambda = 0)$-action from Eq. \eqref{Gitter skalar} by Gaussian integration and 
then the 2-point function by differentiation w.r.t. the external source.
For the inversion of the kernel $ K $, which determines the quadratic part of the action,
use the Fourier representation
\be 
K(l-l') \E a^4 \, \int_{-\pi/a}^{+\pi/a} \frac{d^4k}{(2 \pi)^4} \> \tilde K(k) \> e^{i k \cdot 
(l-l') a} \> . 
\label{lattice FT}
\ee
\item[\purpur{\bf b)}] Derive Eq. \eqref{Gitter 2-Punkt fermion} for a {\bf fermionic field}.
In doing so, use the Euclidean action
\bea
S_E[\cy{\bar{\psi}}, \cy{\psi}] \E \int d^4x \> \cy{\bar{\psi}}(x) \, \lsp \sum_{\mu} 
\gamma_{\mu}^E \partial_{\mu} + m \rsp \, \cy{\psi}(x) \> , 
\eea
with the (Hermitean)  Euclidean Dirac matrices  $ \> \gamma_{\mu}^E \> $ (satifying $ \> [ \gamma_{\mu}^E, \gamma_{\nu}^E ]_+ = 2 \delta_{\mu \nu} \> $ ) and the naive discretization of the derivative
\bea
\partial_{\mu} \, \cy{\psi}(n) \E \frac{1}{2 a} 
\lsp \cy{\psi}(n + \mu) - \cy{\psi}(n - \mu) \rsp \> . 
\eea

\edes
\edes    
\edes
\newpage

\pagestyle{myheadings}
\markboth{\textcolor{green}{Exercises: Solutions}}{\textcolor{green}{R. Rosenfelder : Path Integrals in  
Quantum Physics}}

\bce
{\large \purpur{\bf Solutions for the Problems in "Path Integrals in Quantum Physics"}}
\ece

\vspace{0.8cm}

\noindent
Note: These are only {\bf sketches} of the solutions -- straightforward calculational details are omitted.

\vspace{0.6cm}

\setcounter{ueb}{0}
\noindent
{\bf Problem \ref{nr frei Prop}:}  
\vspace{ 0.2cm}

\refstepcounter{ueb}    
\setcounter{equation}{0}
\renewcommand{\theequation}{\footnotesize\mbox{S }\arabic{ueb}.\arabic{equation}}    

\renewcommand{\baselinestretch}{0.9}
\scriptsize
\noindent
{\bf a)} Perform the differentiations in the explicit form \eqref{freier Prop 2}.
\vspace{0.2cm}

\noindent
{\bf b)} When doing the same for the free Green function there is now an additional term
\be  
\lsp \frac{\partial}{\partial t_b} \, \Theta  \lrp t_b - t_a \rrp \rsp \, U_0 \lrp x_b,t_b; x_a, t_a \rrp \E  \delta \lrp t_b - t_a \rrp \,  U_0 \lrp x_b,t_b; x_a, t_b \rrp \E \delta \lrp t_b - t_a \rrp \, 
\delta \lrp x_b - x_a \rrp \> ,
\ee
since $ \> U(x_b, t_b; x_a, t_b) \E <x_b | \hat U(t_b,t_b) | x_a > \E < x_b | x_a > \E \delta \lrp x_b - x_a \rrp \> $. Alternatively, one can use the fact that
\be 
\lim_{a \to \infty}  \sqrt{\frac{ a}{\pi}} \, e^{  -a \lrp x_b - x_a \rrp^2 } \E \delta \lrp x_b - x_a 
\rrp
\ee
with $ a = - im/(2 \hbar (t_b - t_a)) $
is a (well-known) representation of Dirac's $\delta$-function. 

\noindent
The inverse Fourier transform of $ G_0(x=x_b-x_a,T=t_b-t_a) $ is
\bea 
\tilde G_0(E,p) \EA \int_{-\infty}^{+\infty} dT dx \> \lrp -\frac{i}{\hbar}
\rrp \Theta(T) \sqrt{\frac{m}{2 \pi i \hbar T}} \, \exp \lrp \frac{im}{2 \hbar T} x^2 \rrp 
\, e^{-i p \cdot x/\hbar + i E T/\hbar} \E
 \int_0^{+\infty} dT \> \exp \lsp \frac{i}{\hbar} \lrp E + i \epsilon  
- \frac{p^2}{2m} \rrp T \rsp \non
\EA \frac{1}{E - p^2/(2m) + i \epsilon}
\eea
where the Gaussian $x$-integral has been performed first and the $T$-integral after adding a small positive imaginary part $ i \epsilon $ for convergence at the upper limit of the integral. The sign of this imaginary part determines whether one has forward or backward propagation in time: 
By closing the integration contour in the upper and lower $E$-plane, respectively, one finds from
the theorem of residues
\be 
\int_{-\infty}^{+\infty} dE \> \frac{1}{E- p^2/(2m) \pm i \epsilon} \, e^{-i E T/\hbar} \E 
\mp 2 \pi i \, \Theta(\pm T) \, e^{-i p^2 T/(2m \hbar)} \> .
\ee

\renewcommand{\baselinestretch}{1.2}
\normalsize
\vspace{0.4cm}
 
\noindent
{\bf Problem \ref{klass Wirk}:}  
\vspace{0.2cm}
\refstepcounter{ueb}    
\setcounter{equation}{0}
\renewcommand{\baselinestretch}{0.9}
\scriptsize

\noindent
{\bf a)} The classical trajectory obeying the boundary conditions is 
a straight-line path $ \> x(t) = x_a + (x_b - x_a) (t - t_a)/(t_b - t_a) \> $. Since the velocity 
is constant one obtains
\be 
S_{\rm cl}^{\rm free} \E \int_{t_a}^{t_b} dt \> \frac{m}{2} \dot x^2 \E \frac{m}{2} \, \frac{(x_b - x_a)^2}{t_b - t_a} \> .
\ee
\vspace{0.2cm}

\noindent
{\bf b)} The solution of the classical equation of motion $ \ddot x + \omega^2 x = 0 $ may be written
as any combination of $ \sin \omega t $ and $ \cos \omega t $ but taking
\be 
x_{\rm cl} (t) \E A \, \sin \lsp \omega (t_b - t) \rsp  + B \sin \lsp \omega (t-t_a) \rsp
\ee
has the advantage that the boundary conditions determine the coefficients easily:
$ \> A = x_a/\sin \omega T \> , \> \> B = x_b/\sin \omega T \> , \> \> (T \equiv t_b - t_a ) \> $.
Inserting that into the expression for the action, using trigonometric identities (
$ \> \sin 2x = 2 \sin x \, \cos x \> , \> \> \cos(x+y) = \cos x \cos y - \sin x \sin y\> $) gives
\be 
S_{\rm cl}^{\rm h.o.} \E \frac{m}{2} \omega^2 \, \int_{t_a}^{t_b} dt \, \Big \{ A^2 \cos \lsp 2 \omega 
(t_b - t) \rsp + B^2 \cos \lsp 2 \omega (t-t_a) \rsp - 2 A B \cos \lsp \omega (t_a + t_b - 2t) \rsp 
\Big \} \> .
\ee
Performing the integrations and inserting the values for the coefficients $ \> A , B \> $ one obtains
with little algebra
\be 
S_{\rm cl}^{\rm h.o.} \E \frac{m \omega}{2 \sin \omega T} \, \Big [ \lrp x_a^2 + x_b^2 \rrp 
\cos \omega T - 2 x_a x_b \Big ] \> ,
\label{S cl ho}
\ee
i.e.  Eq. \eqref{S kl fuer HO}.
\vspace{0.2cm}

\noindent
{\bf c)} The general solution of the equation of motion 
\be 
\ddot x_{\rm cl}(t) + \omega^2 x_{\rm cl}(t) = -\frac{e(t)}{m}
\label{cl eq} 
\ee 
is a particular solution + the general solution of the homogenous equation. A particular solution
is obtained by Fourier transformation 
$ x(t) = \int_{-\infty}^{+\infty} dE \, \tilde x(E) \, \exp (i E t ) \> $ etc. as
\be 
\tilde x(E)_{\rm part} \E - {\cal P} \, \frac{\tilde e(E)/m}{\omega^2 - E^2} \> \Longrightarrow \> 
x_{\rm part}(t) \E - \int_{-\infty}^{+\infty} d\tau \> \frac{e(\tau)}{m} \, \frac{\sin \omega |t - \tau|}{2 \omega} \> .
\label{part sol forced ho}
\ee
Here $ {\cal P} $ denotes the principal value which may be chosen to get a real particular solution  (one also could take the $ \pm i 0^+ $-prescription in the denominator but this would render the particular solution complex which is then compensated by a complex homogeneous solution).
Since one is only considering the time evolution in the interval
$ \> \lsp t_a, t_b \rsp \> $ one may cut off $ e(\tau) $ for $ \tau < t_a $ and $ \tau > t_b $ .
Indeed, $ \> g(t-\tau) := \sin [\omega |t-\tau|]/(2 \omega) \> $ is a Green function of the harmonic oscillator obeying $ \> \ddot g + \omega^2 g = \delta (t-\tau) \> $. The calculational effort is
reduced by writing $ \> g(t-\tau) =  \tilde g(t-\tau) - 
\sin [\omega(t-\tau)]/(2 \omega) \> $ and realizing that the last term just leads to  a modification of the homogenous solution. In other words: One can use the Green function $ \> \tilde g(t-\tau) = 
\Theta(t-t_a) \sin [\omega (t-\tau)]/\omega \> $ and take
\be 
x_{\rm cl} (t) \E   A \, \sin \omega (t_b - t) + B \, \cos \omega (t-t_a)- f(t) \> , \quad
f(t) \Def \int_{t_a}^t d\tau \> \frac{e(\tau)}{m} \, \frac{\sin \omega (t - \tau)}{ \omega} 
\label{cl sol forced ho}
\ee
as {\it ansatz} for the classical solution of the forced harmonic oscillator (it can be verified that it also fulfills the equation of motion \eqref{cl eq}).
Here  the homogenous solution has been written again in such a way that the boundary conditions are easily implemented:
$ \> A = x_a/\sin \omega T \> , \> \>  B = (x_b + f(t_b)/\sin \omega T \> $ . Note that use of the (less symmetric) Green function $ \tilde g $ implies $ f(t_a) = \dot f(t_a) = 0 $, i.e. generates less terms for the coefficients $A, B $.
It is now a straight-forward, albeit tedious 
calculational task to evaluate the classical action by 
inserting Eq. \eqref{cl sol forced ho} into the corrsponding expression and performing the
time integrations when possibe. Some simplification arises if the equation of motion
\eqref{cl eq} is used after an integration by parts
\bea 
S_{\rm cl}\! \! \!&=& \!\! \! \frac{m}{2} x_{\rm cl}(t) \dot x_{\rm cl}(t)  \Bigg |_{t_a}^{t_b} \!\! - \int\limits_{t_a}^{t_b} dt \, \Big[ \frac{m}{2} x_{\rm cl}(t)  \ddot x_{\rm cl}(t) + \frac{m}{2} \omega^2 x_{\rm cl}^2(t)+ e(t)x_{\rm cl}(t) \Big ] =
  \frac{m}{2} \Big [ x_{\rm cl}(t_b)  \dot x_{\rm cl}(t_b) - x_{\rm cl}(t_a)  
 \dot x_{\rm cl}(t_a) \Big ] - \int\limits_{t_a}^{t_b} dt  \frac{e(t)}{2} \, x_{\rm cl}(t) \non
 \EA \frac{m}{2 \sin \omega T} \, x_b \lsp - x_a \omega  + \lrp x_b + f(t_b) \rrp \omega \cos \omega T
 - \dot f(t_b) \sin \omega T \rsp - \frac{m}{2 \sin \omega T} \, x_a \Big [ - x_a \omega  \cos \omega T + \lrp x_b + f(t_b) \rrp \omega \Big ] \non
 && - \frac{1}{2 \sin \omega T} \int_{t_a}^{t_b} dt \> e(t) \, \Big \{ x_a \sin \lsp \omega (t_b-t)\rsp
 + \lrp x_b + f(t_b) \rrp \sin \lsp \omega (t-t_a) \rsp - f(t) \sin \omega T \Big \} \> .
\eea
It is seen that the terms which do not contain the external force $ e(t) $ (or $ f(t) $ ) add up to
the classical harmonic oscillator action \eqref{S cl ho}. The terms linear in $e(t) $ are
\be 
\frac{m \omega}{2 \sin \omega T} x_b f(t_b) \omega \cos \omega T - \frac{m}{2} x_b \dot f(t_b)
- \frac{m \omega}{2 \sin \omega T} f(t_b) - \frac{1}{2 \sin \omega T} \, 
\int_{t_a}^{t_b} dt \> e(t) \, \Big [ x_a \sin \lsp \omega (t_b-t) \rsp + x_b \sin \lsp \omega (t-t_a)
\rsp \Big ] \> .
\ee
Using the definition of $ f(t) $ in Eq. \eqref{cl sol forced ho} this 
can be brought into the form
$ \> - \int_{t_a}^{t_b} dt \, e(t) \, \Big \{  x_a \sin \lsp \omega (t_b - t) 
\rsp + x_b \sin \lsp \omega (t-t_a) \rsp \Big \}/\sin \omega T , $ 
in agreement with the corresponding term in Eq. \eqref{erzwung HO}. Finally the term quadratic in $ e(t) $ is
\be 
 -\frac{1}{2 \sin \omega T}  \, \int\limits_{t_a}^{t_b} dt \, e(t) \, \Big \{ f(t_b) \sin \lsp \omega (t-t_a) \rsp - f(t) \sin \omega T \Big \} 
\deF 
-\frac{1}{2 m \omega \sin \omega T}  \, \int_{t_a}^{t_b} dt \, e(t) \, \int_0^t d\tau \, e(\tau) 
\, G(t,\tau,T)
\ee
with 
\be  
G(t,\tau,T) \E \sin \lsp \omega (t_b-t) \rsp \,\sin \lsp \omega (\tau-t_a) \rsp + \sin \lsp \omega (t_b-\tau) \rsp \,\sin \lsp \omega (t-t_a) \rsp - \sin \omega T \, \sin\lsp \omega (t-\tau) \rsp \> .
\ee
Here the relation $ \> \int_{t_a}^{t_b} dt d\tau \, F(t,\tau) = \int_{t_a}^{t_b} dt \, \int_{t_a}^t d\tau \, [ F(t,\tau) + F(\tau,t) ] \> $ and the definition \eqref{cl sol forced ho} have been used. 
Rewrite this in terms of the variables $ \> x = \omega(t-t_a) \> , \> y = \omega (\tau-t_a) \> , \> 
z = \omega T \> $ as $ \> 
G(x,y,z) \E \sin(z-x) \sin y + \sin (z-y) \sin x - \sin z \sin(x-y) \> $
and use the addition theorems for the trigonometric functions. This leads to
\bea
G(x,y,z) \EA 2 \sin z \sin y \cos x - 2 \cos z \sin x \sin y \E 2 \sin y \, \lrp \sin z \cos x - \cos z \sin x \rrp = 2 \sin y \sin (z-x) \non
&\EQ& 2 \sin [ \omega (t_b -t) ] \, \sin [ \omega (\tau-t_a) ] \> , \\
\mbox{i.e. to the last term of Eq. \eqref{erzwung HO}.} \no
\eea
\renewcommand{\baselinestretch}{1.2}
\normalsize
\vspace{0.5cm}

\noindent
{\bf Problem \ref{U fuer zeitabh. Pot}:}  
\vspace{0.2cm}

\refstepcounter{ueb}    
\setcounter{equation}{0}
\renewcommand{\baselinestretch}{0.9}
\scriptsize

\noindent
{\bf a)} From Eq. \eqref{eq for U} one finds for $ \hat H^{\dagger}(t) = \hat H(t) $
\be 
i \hbar \frac{\partial}{\partial t} \lrp \hat U(t,t_0) \hat U^{\dagger}(t,t_0) \rrp
\E \lsp \hat H(t), \hat U(t,t_0) \hat U^{\dagger}(t,t_0) \rsp \> ,
\ee
for which $ \hat U(t,t_0) \hat U^{\dagger}(t,t_0) = {\rm const.} $ is a solution. From the initial condition $ \hat U(t_0,t_0) = \hat 1 $ one determines that $ \hat U^{\dagger} \hat U = 1 $.
Similarly for $ \hat U \hat U^{\dagger} = 1 $.
Conversely, if $ \hat U $ is unitary then $ \hat U^{-1} = \hat U^{\dagger} $ and
$ \hat H^{\dagger}\! = ( i\hbar (\partial_t \hat U) \hat U^{\dagger} )^{\dagger}\!=\! - i \hbar 
\hat U \partial_t \hat U^{\dagger}\! = i \hbar (\partial_t U) \hat U^{\dagger}\! - i \hbar \partial_t 
(\hat U \hat U^{\dagger} ) = i \hbar (\partial_t \hat U) \hat U^{\dagger}\! =\! \hat H $,
 i.e. $ \hat H $ is hermitean.
\vspace{0.2cm}

\noindent
{\bf b)} $ \hat U(t,t_1) \hat U(t_1,t_0) $ is a solution of the Schr\"odinger equation 
\eqref{eq for U}. For it to be equal to $ \hat U(t,t_0) $ for any value of $ t $, it is sufficient
that it be equal for a particular value of $ t $, say $ t = t_1 $ which is obvious because then the
first factor $ \hat U(t_1,t_1) = \hat 1 $.
\vspace{0.2cm}

\noindent
{\bf c)} Use of the composition law allows to write the time-sliced time-evolution 
operator as in Eq. \eqref{Trotter}:  
$ \> \> 
\hat U(t_b,t_a) \E \lim_{N \to \infty} \prod_{k=1}^N \, \hat U(t_k,t_{k-1}) , $ 
($ \>  t_k = t_a + k \epsilon \> , \quad \epsilon \E (t_b - t_a)/N $ ).
Instead of Eq. \eqref{kurz Zeit 1} we now have for
small time-steps  $ \> \hat U(t_k,t_{k-1}) \simeq \exp \lsp - i \epsilon \hat T/\hbar \rsp \, 
\exp \lsp - i \epsilon V(\hat x, t_{k-1})/\hbar \rsp \> $. Proceeding as before we end up with the (slightly generalized) Lagrange form \eqref{Lagrange diskret}
of the path integral
\bea
\hspace{-1cm} U(x_b, t_b; x_a, t_a) \EA \lim_{N \to \infty} \>
\left (\frac{m}{2 \pi i \epsilon \hbar} \right )^{N/2}
\int_{-\infty}^{+\infty} dx_1 \> dx_2 \> \ldots \> dx_{N-1} \> 
 \cdot  \> \exp \left \{ \frac{i \epsilon}{\hbar} \sum_{j=0}^{N-1} \left [
\frac{m}{2} \left ( \frac{x_{j+1} - x_j}{\epsilon} \right )^2 - 
V(x_j,t_j) \right ] \> \right \}  \non
&\EQ & \int_{x(t_a)=x_a}^{x(t_b)=x_b} {\cal D}x(t) \> \exp \lcp \frac{i}{\hbar} \int_{t_a}^{t_b} dt
\> \lsp \frac{m}{2} \dot x^2 - V(x,t) \rsp \rcp \> .
\eea
\renewcommand{\baselinestretch}{1.2}
\normalsize
\vspace{0.5cm}

\noindent
{\bf Problem \ref{functionalableit}:}  
\vspace{0.2cm}

\refstepcounter{ueb}    
\setcounter{equation}{0}
\renewcommand{\baselinestretch}{0.9}
\scriptsize
\noindent
{\bf a)} Applying the chain rule gives
\be 
\frac{\delta S}{\delta x(\sigma)} \E \int_{t_a}^{t_b} dt \lsp m \dot x(t) \, \frac{\delta \dot x(t)}{\delta x(\sigma)} - V'(x(t)) \, \frac{\delta x(t)}{\delta x(\sigma)} \rsp \> .
\ee
The last functional derivative is  $ \> \delta(t-\sigma)\>  $, while the first one may be written as
$ \> 
\frac{d}{dt} \frac{\delta x(t)}{\delta x(\sigma)} \E \frac{d}{dt} \, \delta (t-\sigma ) \E 
- \frac{d}{d\sigma} \, \delta (t-\sigma ) \>. $
This gives 
\be 
\frac{\delta S}{\delta x(\sigma)} \E - \frac{d}{d \sigma} \, \int_{t_a}^{t_b} dt \>  m \dot x(t)  \, \delta (t - \sigma) - \int_{t_a}^{t_b} dt \> V'(x(t)) \, \delta (t-\sigma) = - m \ddot x(\sigma) - V'(x(\sigma )
\label{delta S}
\ee
and Newton's equation when demanding $ \delta S = 0 $ .
\vspace{0.1cm}

\noindent
{\bf b)} Differentiate Eq. \eqref{delta S} again functionally and use the basic rule 
\eqref{func x x delta} to obtain
\be 
\frac{\delta^2 S}{\delta x(\sigma) \delta x(\sigma')} \E \lsp - m \, \frac{d^2}{d \sigma^2} 
- V''(x(\sigma)) \rsp \, \delta (\sigma - \sigma') \> .
\ee
The functional Taylor expansion \eqref{functional Taylor} around the classical path then becomes
\be
S[x_{\rm cl}(t) + y(t)] \E  S[x_{\rm cl}(t)] + \int d\sigma \> \lsp \frac{m}{2} \dot y^2 - \frac{1}{2} V''(x_{\rm cl}(\sigma)) \,   y^2(\sigma) \rsp + \ldots
\ee
\vspace{0.1cm}
 
\noindent
{\bf c)} Apply the chain rule and the basic rule \eqref{func x x delta} of functional differentiation to get the generalized result
\be 
\frac{\delta^n Z[J]}{\delta J(\sigma_1) \ldots \delta J(\sigma_n)} \E \lrp \frac{i}{\hbar} \rrp ^n \, 
\int {\cal D}x \> x(\sigma_1) \ldots x(\sigma_n) \> \exp \lrp \frac{i}{\hbar} S[x(t)] + 
\frac{i}{\hbar} \int dt \, x(t) J(t) \rrp \> .
\ee
If after differentiation the source is set to zero, the result allows to calculate  
integrals of arbitrary monomials of $ x $ weighted with the exponential of the action (the Green functions).
\renewcommand{\baselinestretch}{1.2}
\normalsize
\vspace{0.2cm}

\noindent
{\bf Problem \ref{Wigner Trans}:}  
\vspace{0.2cm}

\refstepcounter{ueb}    
\setcounter{equation}{0}
\renewcommand{\baselinestretch}{0.9}
\scriptsize
\noindent
{\bf a)} In the definition \eqref{def Wigner} of the Wigner transform, the matrix element is
$ \> \> \la x - \frac{y}{2} | V(\hat x) | x + \frac{y}{2} \ra = V(x + y/2)  \delta \lrp x - \frac{y}{2} - (x+ \frac{y}{2}\rrp = 
V \lrp x + \frac{y}{2} \rrp) \, \delta(-y) = V(x) \, \delta(y)\>  $ and therefore $ V_W(x,p) = V(x) $. For momentum-dependent operators transform \eqref{def Wigner} into
\be 
A_W(x,p) \E \int dy \, dk \, dk' \> \la x- \frac{y}{2} \Bigg | \, k \ra \, \la k \,  | \hat A | k' \ra \, 
\la k' \Bigg | x + \frac{y}{2} \ra \, e^{i p y/\hbar} = \int dq \, \la p - \frac{q}{2} \Bigg | \, \hat A  \, \Bigg | \, p + \frac{q}{2} \ra \, \exp \lrp - i q x/\hbar \rrp  
 \label{Wigner mom}
\ee
using the transformation bracket $ \la x | p \ra = \exp \lrp i p x \rrp /(2 \pi \hbar)^{1/2} \> $
between position and momentum eigenstates. From Eq. \eqref{Wigner mom} it is then clear that
$ T_W (x,p) = T(p) $, in particular for the non-relativistic kinetic energy $ T(p) = p^2/(2 m) $.
\vspace{0.1cm}

{\bf b)} Eq.  \eqref{def Wigner} is a Fourier transformation w.r.t. $ y $ and therefore can be inverted
in the usual way: Multiply by  $ \exp(-ip z /\hbar)$ and integrate over $p$. Shifting the arguments
in the matrix element gives Eq. \eqref{Wigner reverse}. Insert $ \> A_W(\frac{x+x'}{2},p) = \lrp \frac{x+x'}{2} \rrp^m \, p^n\> $ and use the binomial theorem. This gives
\be 
\la x \Big | \, \hat A \, \Big | \, y \ra \E \frac{1}{2^m} \sum_{l=0}^m {m \choose l} \, \int_{-\infty}^{+\infty} dp \> \frac{\exp(i p x/\hbar)}{\sqrt{2 \pi \hbar}} \, x^l \, p^n \, y^{m-l}
\, \frac{\exp(-i p y/\hbar)}{\sqrt{2 \pi \hbar}} \EQ  \frac{1}{2^m} \sum_{l=0}^m {m \choose l} \,
\la x \, \Big | \, \hat x^l \, \hat p^n \, \hat x^{m-l} \, \Big | \, y \ra \quad  q.e.d.
\ee
\renewcommand{\baselinestretch}{1.2}
\normalsize
\vspace{0.4cm}

\noindent
{\bf Problem \ref{klass Anfang/End}:}  
\vspace{0.2cm}

\refstepcounter{ueb}    
\setcounter{equation}{0}
\renewcommand{\baselinestretch}{0.9}
\scriptsize
\noindent
The path $ x(t) $ depends parametrically on $ x_a, t_a, x_b, t_b $. Using the rules for differentiation of integrals one therefore has for the (partial, not functional!) derivative of the classical
action w.r.t. the initial time
\be 
\frac{\partial S_{\rm cl}}{\partial t_a} \E -L \lrp x_a,\dot x_a \rrp + \int_{t_a}^{tb} dt
\lsp \frac{\partial L}{\partial x} \frac{\partial x(t)}{\partial t_a} + 
\frac{\partial L}{\partial \dot x} \frac{\partial \dot x(t)}{\partial t_a} \rsp
\ee
where $ \dot x_a = \dot x|_{t=t_a} $. Write the last factor in the last term as $ \frac{d}{dt} \, \frac{\partial x(t)}{\partial t_a} $, perform an integration by parts
and use the Euler-Lagrange equations to obtain
\be 
\frac{\partial S_{\rm cl}}{\partial t_a} \E -L \lrp x_a,\dot x|_{t=t_a} \rrp + \frac{\partial L}{\partial \dot x} \Big |_{t_b}  \, \lim_{t \to t_b} \frac{\partial}{\partial t_a} x(t) -  
\frac{\partial L}{\partial \dot x} \Big |_{t_a} \, \lim_{t \to t_a} \frac{\partial}{\partial t_a}  
x(t) \E - L \lrp x_a,\dot x_a \rrp +  \frac{\partial L}{\partial \dot x} \Big |_{t_a}
\EQ H \Big |_{t_a} = E_a  \> .
\ee
Here the Taylor expansion of the paths near initial and final time 
$ x(t) = x_{a/b} + \dot x_{a/b} (t -t_{a/b}) + \ldots $ has been used to determine
$  \frac{\partial}{\partial t_a} x(t) \big |_{t_b} = 0 $ and 
$  \frac{\partial}{\partial t_a} x(t)\big |_{t_a} = -\dot x_a $.
Similarly for 
\be 
\frac{\partial S}{\partial x_a} \E \frac{\partial L}{\partial \dot x} \frac{\partial x(t)}{\partial x_a} \, \Bigg |_{t_a}^{t_b} \E - \frac{\partial L}{\partial \dot x} \Big |_{t_a} \, \lim_{t \to t_a} \frac{\partial }{\partial x_a} \lsp x_a + \dot x_a (t - t_a) + \ldots \rsp \E - \frac{\partial L}{\partial \dot x} \Big |_{t_a} \EQ - p_a  \> .
\ee
With the explicit expressions for the classical action determined in {\bf Problem \ref{klass Wirk} a)} and {\bf b)},
it is easy to verify these relations 
for the free particle and the particle in a harmonic potential.
\renewcommand{\baselinestretch}{1.2}
\normalsize
\vspace{0.3cm}

\noindent
{\bf Problem \ref{Magnetfeld Eich}:}  
\vspace{0.2cm}

\refstepcounter{ueb}    
\setcounter{equation}{0}
\renewcommand{\baselinestretch}{0.9}
\scriptsize
\noindent
{\bf a)} Minimal substitution for a particle in an electromagnetic field $ {\bf A} $
means replacing the momentum of the particle by the canonical momentum:
$ \fp \longrightarrow \mbox{\boldmath $\Pi$} = \fp - e {\bf A}/c $. To derive the path-integral representation 
complete the square in the exponent of the phase space path integral
\be 
\fp \cdot \fx - H(\mbox{\boldmath $\Pi$}) \E - \frac{\fp^2}{2 m} + \fp \cdot \lrp \dot \fx + \frac{e}{mc} {\bf A} \rrp - \frac{e^2}{2m c^2} \, {\bf A}^2 = - \frac{1}{2 m} \lrp \fp - \lrp m \dot \fx + 
\frac{e}{c} {\bf A}\rrp \rrp^2 + \frac{m}{2} \dot \fx^2 + \frac{e}{c} {\bf A} \cdot \dot \fx \> 
\ee
and perform the Gaussian integal over {\boldmath $ \> \Pi$} - $ m \dot \fx \> $. This gives a constant while the remaining path integal reads $ \int {\cal D}^3 x(t) \exp \lrp i S[\fx]/\hbar \rrp $ with the action
\be 
S[\fx(t)] \E \int_{t_a}^{t_b} dt \> \lsp \frac{m}{2} \dot \fx^2 + \frac{e}{c} {\bf A} \cdot \dot \fx \rsp \> .
\label{S mit A}
\ee

\noindent
{\bf b)} Under the gauge transformation $ {\bf A} \to {\bf A} + \nabla \Lambda $ the action
\eqref{S mit A} changes into
\be 
S[\fx(t)] \> \longrightarrow \> S[\fx(t)] + \int_{t_a}^{t_b} dt \> \frac{e}{c} \nabla \Lambda(\fx) \cdot \dot \fx \E 
S[\fx(t)] + \int_{t_a}^{t_b} dt \> \frac{e}{c} \frac{d}{dt} \Lambda(\fx) \E 
S[\fx(t)] + \frac{e}{c} \, \Bigl [ \>  \Lambda(\fx_b] - \Lambda(\fx_a) \> \Bigr ] \> .
\ee
This means that
\be 
U(\fx_b,t_b;\fx_a,t_a) \E \sum_n \psi_n^{\star}(\fx_b) e^{-i E_n(t_b - t_a)/\hbar} \psi_n(\fx_a) \> \longrightarrow \> U(\fx_b,t_b;\fx_a,t_a) \cdot \exp \lcp \frac{i e}{\hbar c} \Bigl [\> \Lambda(\fx_b)
- \Lambda(\fx_a) \> \Bigr ] \rcp \>  ,
\ee
i.e. that the wave functions acquire a (position-dependent) phase factor
\be 
\psi_n(\fx) \> \longrightarrow \> \psi_n(\fx) \> \cdot  e^{ \frac{i e}{\hbar c} \Lambda(\fx)} \> .
\ee

\noindent
{\bf c)} Following the derivation of Schr\"odinger's equation for a particle in a potential 
in chapter \ref{sec1: Lagr,Ham} one may use 
 \be 
S \E \epsilon \sum_{j=0}^N \lsp \frac{m}{2} \lrp \frac{\fx_{j+1}-\fx_j}{\epsilon} \rrp^2 + \frac{e}{c} 
{\bf A} \lrp \lambda \fx_j + (1-\lambda) \fx_{j+1} \rrp \cdot  \frac{\fx_{j+1}-\fx_j}{\epsilon} \rsp
\ee
as discretized version of the action \eqref{S mit A}. Here
the parameter $ \lambda \in [0,1] $ determines the ordering prescription: $ \lambda = 1/2 $ 
gives the midpoint rule. 
 In 3 dimensions Eq. \eqref{psi etwas spaeter} now reads
\be
\psi(\fx,t+\epsilon) \E \left ( \frac{m}{2 \pi i \epsilon \hbar} 
\right )^{3/2} \, \int_{-\infty}^{+\infty}  d^3\xi \> \exp \left ( 
\frac{i m \fxi^2}{2 \epsilon \hbar} \right ) \, \exp \lsp - \frac{i e}{\hbar c} \fxi \cdot 
{\bf A} (x+ \lambda \fxi) \rsp \, 
\psi(\fx+\xi,t) \> .
\ee
and again one has to expand in powers of $ \epsilon $ and $ \fxi $.
 Use the Gaussian integrals
$
\int d^3 \xi \> \lrp 1, \xi_k, \xi_k \xi_l \rrp \, \exp (-a \fxi^2 ) \E \lrp 1, 0, \frac{\delta_{kl}}{2 a} 
\rrp \, \lrp \frac{\pi}{a} \rrp^{3/2} \> $, where  $ a \E -\frac{im}{2 \hbar \epsilon} $
and $ k, l $ are cartesian components, to derive
\vspace*{-0.2cm}

\be
\psi(\fx,t+\epsilon) \E \psi(\fx,t) + \epsilon \frac{i \hbar}{2 m} \, \Delta \psi(\fx,t) + \epsilon 
\frac{\lambda e}{m c} \, \lrp \nabla \cdot {\bf A}(\fx)\rrp  \, \psi(\fx,t) 
- \epsilon \frac{i e^2}{2 \hbar m c^2} {\bf A}^2(\fx) \, 
\psi(\fx,t) + \epsilon \frac{e}{mc} {\bf A}(\fx) \cdot \nabla \psi(\fx,t) + {\cal O}(\epsilon^2) \> .
\ee
In the limit $ \epsilon \to 0 $ one thus obtains
\be
i \hbar \, \frac{\partial \psi(\fx,t)}{\partial t} \E \lsp - \frac{\hbar^2}{2 m} \Delta + \frac{i \hbar e}{m c}
{\bf A}(\fx) \cdot \nabla +\lambda  \frac{i \hbar e}{m c} \, \lrp \nabla \cdot {\bf A}(\fx)\rrp + 
\frac{e^2}{2 m c^2} {\bf A}^2(\fx) \rsp \, \psi(\fx,t)
\ee
and one sees that only for $ \lambda = 1/2 $ the r.h.s. can be written as 
\be 
\frac{1}{2m} \lrp \frac{\hbar}{i} \nabla 
- \frac{e}{c} {\bf A}(\fx) \rrp^2 \psi(\fx,t) \EQ \frac{1}{2m} \hat {\mbox{\boldmath$\Pi$}}^2 \psi(\fx,t)  
\ee
(be careful when squaring the differential operator: $ (d/dx + f)^2 g = g'' + 2 f g' + g'f + f^2 g $).
Note that  "outer averaging'' $ \> \lrp {\bf A}(\fx_j) + 
{\bf A}(\fx_{j+1}) \rrp/2 \> $  would lead to the same result.

\renewcommand{\baselinestretch}{1.2}
\normalsize
\vspace{0.4cm}

\noindent
{\bf Problem \ref{Jacobi Eq.}:}  
\vspace{0.2cm}

\refstepcounter{ueb}    
\setcounter{equation}{0}
\renewcommand{\baselinestretch}{0.9}
\scriptsize

\noindent
{\bf a)} Differentiate the Euler-Lagrange eq. $ \> \frac{d}{dt} \lrp \frac{\partial L}{\partial \dot x} \rrp - 
\frac{\partial L}{\partial x} = 0 \> $ w.r.t. to the momentum $ p $. 
This gives
\be
\frac{d}{dt} \lrp \frac{\partial^2 L}{\partial \dot x^2} \, \dot J \rrp + \lsp \frac{d}{dt} \lrp 
\frac{\partial^2 L}{\partial x \, \partial \dot x} \rrp - \frac{\partial L}{\partial x^2} \rsp \, J \E 0 \> ,
\label{Jacobi Gl}
\ee
i.e. the Jacobi equation. The initial conditions are $ x(p,t) \> \longrightarrow \> x_a + \frac{p}{m} t 
+ \ldots \> \Longrightarrow \> J(p,t) \to \frac{1}{m} t + \ldots $ for $ t \to 0 $ and therefore 
$ J(p,0) = 0 $ and $ \partial J/\partial t |_{t=0} = 1/m $ .
\vspace{0.1cm}

\noindent
{\bf b)} For $ L = m \dot x^2/2 - c(t) x^2/2 - e(t) x $ we find from Eq. \eqref{Jacobi Gl}
$ \> \> m \ddot J + c(t) J \E 0 \> $, i.e. the Gel'fand-Yaglom equation for $  m J(t) $. 
Since $ m J(p,0) = 0 $ and $ m \dot J(p,t)|_{t=0} = 1 $ are precisely the initial conditions to determine the 
functional determinant in the prefactor as solution of the GY equation,
one obtains the result 
$ m J(p,t) \EQ f_{\rm GY}(t) \> .$
\vspace{0.1cm}

\noindent
{\bf c)} From {\bf Problem \eqref{klass Anfang/End}} we know that the initial momentum is given by 
$ p \E p(0) = - \frac{\partial S}{\partial x_a} \> $ .
Then 
\be
\frac{\partial p}{\partial x_b} \E \frac{1}{\partial x_b/\partial p} \E \frac{1}{J(p,t_b)} \E 
- \frac{\partial^2 S}{\partial x_a \partial x_b} \qquad q.e.d. 
\ee
Combined with the previous result this verifies Eq. \eqref{Vor durch S cl}.

\renewcommand{\baselinestretch}{1.2}
\normalsize
\vspace{0.4cm}

\noindent
{\bf Problem \ref{Maslov}:}  
\vspace{0.2cm}

\refstepcounter{ueb}    
\setcounter{equation}{0}
\renewcommand{\baselinestretch}{0.9}
\scriptsize

\noindent
Plugging the harmonic oscillator result \eqref{S kl fuer HO} for the time-evolution operator into the r.h.s., one has
\be 
\frac{m \omega}{2 \pi i \hbar} \, \frac{1}{\sqrt{s_1 s_2}} \, \int_{-\infty}^{+\infty} dx_c \> 
\exp \lcp \frac{i m \omega}{2 s_2 \hbar} \lsp \lrp x_b^2 + x_c^2 \rrp c_2 - 2 x_b x_c \rsp \rcp \, 
\exp \lcp \frac{i m \omega}{2 s_1 \hbar} \lsp \lrp x_c^2 + x_a^2 \rrp c_1 - 2 x_c x_a \rsp \rcp \
\ee
where $ s_k\Def \sin \omega T_k \> , \> c_k \Def \cos \omega T_k \> , \quad k = 1,2 $. 
By assumption both $s_1$ and $s_2$ are positive. The $x_c$-integral
is of the form of Eq. \eqref{erw Fresnel} with 
\be 
a \E \frac{ m \omega}{2 s_2 \hbar} \, \lrp \frac{c_2}{s_2} + \frac{c_1}{s_1} \rrp \E 
\frac{ m \omega}{2 s_2 \hbar} \, \frac{s_{1+2}}{s_1 s_2} \> , \quad b \E - \frac{ m \omega}{ s_2 \hbar}
\, \lrp \frac{x_b}{s_2} + \frac{x_a}{s_1} \rrp \> .
\ee
It now depends on the sign of $ s_{1+2} \EQ  \sin \omega (T_1 + T_2) $
whether the phase of the prefactor is $ \pi/4 $ (for $ s_{1+2} > 0 $, i.e. the focal point has not
been reached) or $\>  -\pi/4 = \pi/4 - \pi/2 \> $ (for  $ s_{1+2} < 0 $, i.e. the particle has passed the first focal point) as Eq. \eqref{Fresnel} clearly shows. Careful algebra then also demonstrates that all trigonometric 
functions on the r.h.s. have the argument $ \omega (T_1 + T_2 ) = \omega (t_b - t_a) $ -- no dependence on the intermediate time $t_c$ is left -- and that indeed
\be 
U^{\rm h.o.}\lrp x_b,t_b; x_a,t_c \rrp \E \sqrt{\frac{m \omega}{2 \pi i \hbar |s_{1+2}|}} \, 
e^{-i n \pi/2} \, e^{i S_{\rm cl}^{\rm h.o.}}
\ee
acquires a Maslov phase ($ n = 1 $ ) if a focal point is passed 
(cf. Ref.\cite{Pech}
\vspace{0.1cm}

\renewcommand{\baselinestretch}{1.2}
\normalsize 
\vspace{0.4cm}

\noindent
{\bf Problem \ref{Null Mode}$^{\star}$:}  
\vspace{0.2cm}

\refstepcounter{ueb}    
\setcounter{equation}{0}
\renewcommand{\baselinestretch}{0.9}
\scriptsize
\noindent
{\bf a)}  Inserting $ V(x) = (m \omega (x^2 - a^2)/a)^2/8 $ into Eq. \eqref{class sep} one obtains
for $ |x| < a $
\be
\tau - \tau_0 \E \frac{2 a}{\omega} \, \int_{x_1(\tau_0)}^{x_1(\tau)} dx \> \frac{1}{a^2 - x^2}
\E \frac{1}{\omega} \, \ln \frac{a+x}{a-x} \Big |_{x_1(\tau_0}^{x_1(\tau)}
\ee
from which $ \> 
x(\tau) = a (A \exp[\omega(\tau-\tau_0)] - 1)/(A \exp[\omega(\tau-\tau_0)] + 1) \> $
follows. Here $ A = (a + x_1(\tau_0))/(a - x_1(\tau_0) $.
Thus
\be 
x_1(\tau) \E a \tanh \lsp \frac{\omega}{2} (\tau - \tau_0) + \gamma \rsp \> , \quad 
\gamma \E \frac{1}{2} \ln \lrp \frac{a + x_1(\tau_0)}{a - x_1(\tau_0)} \rrp \> .
\label{pre kink}
\ee
If one chooses $ x_1(\tau_0) = 0 $, i.e. the particle should traverse the axis at time $ \tau_0$, then 
$ \gamma = 0 $ and Eq. \eqref{pre kink} turns into Eq. \eqref{kink}.
\vspace{0.5cm}

\noindent
{\bf b)} Use $ V''(x) = m \omega^2 (3 x^2/a^2 - 1 )/2 \> $ and evaluate 
$ m \partial^2 x_0(\tau-\tau_0)/\partial \tau^2  $ to 
verify that  $ {\cal O}_{V''} y_0 = 0 $ . Observe that this also follows if the classical equation of 
motion \eqref{klass Bewegungsgl}
for the instanton is differentiated again w.r.t. $ \tau_0 $: 
\be
m \frac{d^2}{d\tau^2} \, \frac{\partial x_1}{\partial \tau_0} - V''(x_1) \, 
 \frac{\partial x_1}{\partial\tau_0} \E 0
\> .
\ee
Normalizing the zero mode  
\be 
y_0(\tau) \E  A_0 \partial x_1/\partial \tau_0 = - A_0 \dot x_1  \E
\frac{A_0 a \omega}{ 2 \cosh^2 [\omega (\tau-\tau_0) ]}
\ee 
to unity one finds
\be
A_0^2 \int_{-\infty}^{+\infty} d\tau \lrp \frac{d x_1(\tau)}{d \tau} \rrp^2 \E A_0^2 \, \frac{1}{m} \, S_1
\E 1 \> \Longrightarrow \> A_0 \E \sqrt{\frac{m}{S_1}}
\ee
by means of Eq. \eqref{S1} in which the action $ S_1 $ of the one-instanton solution has been 
calculated.
Note that (in the $\beta \to \infty$-limit) the zero mode is orthogonal to the instanton solution
\be
\int_{-\infty}^{+\infty} d\tau \> y_0(\tau) \, x_1(\tau) \E \int_{-\infty}^{+\infty} d\tau \> ( - A_0 )
\frac{dx_1(\tau)}{d \tau} \, x_1(\tau) \E - A_0 \, \int_{-\infty}^{+\infty} d\tau \> \frac{d}{d \tau} 
\frac{1}{2} \lrp x_1(\tau) \rrp^2 \E  - \frac{1}{2} A_0 \, \lrp x_1(\tau) \rrp^2 
\Big |_{\tau \to -\infty}^{\tau \to +\infty} \E 0 \> .
\ee
\vspace{0.5cm}

\noindent
{\bf c)} Write down the linear differential equations $ \> -m \ddot f^{(i)}(\tau) + V''(x_{\rm cl}) \, 
f^{(i)}(\tau) = 0 \> $ 
for $ i = 1,2 $, multiply the first equation by $ \> f^{(2)}(\tau) \> $, the second one by $ \> f^{(1)}(\tau) \> $ and 
subtract. This gives
\be
0 \E -m \ddot f^{(1)}(\tau) f^{(2)}(\tau) + m \ddot f^{(2)}(\tau) f^{(1)}(\tau) \E m \frac{d}{d \tau} \lsp f^{(1)}(\tau) \dot f^{(2)}(\tau) - \dot f^{(1)}(\tau) f^{(2)}(\tau) \rsp \EQ m \, \frac{d}{d \tau}
W \lrp f^{(1)}, f^{(2)} \rrp \>.
\ee
Integration then shows that the Wronskian is a constant whose value $ \> 8 \omega^3 a^2 A_0^2 \> $
is determined from the asymptotic behaviour of the functions $ \> f^{(1)}, f^{(2)} \> $.
\vspace{0.5cm}

\noindent
{\bf d)} Differentiating $ \> g(\tau,\tau) \deF \Theta(\tau-\tau') \, \tilde g(\tau,\tau') \> $ twice w.r.t. $ \tau $, using $ \> \delta(\tau-\tau') \, \tilde g(\tau,\tau') = \delta(\tau-\tau') \, \tilde g(\tau,\tau) = 0 \> $ and the definition of the Wronskian shows that $ \> {\cal O}_{\rm V''}
\, g(\tau,\tau') = \delta(\tau-\tau') \> $ (similar as in {\bf Problem \ref{klass Wirk} c)} for the forced harmonic oscillator).
Thus Eq. \eqref{eigen e0} is an integral equation for the 
would-be-zero-mode eigen function $ \> Y_0(\tau) \> $ which by construction fulfills the boundary condition at $ \> \tau = - \beta/2 \> $ but not at $ \> \tau = \beta/2 \> $. Requiring that, one obtains for the eigenvalue
\be 
\frac{f_{\rm GY}(+\beta/2)}{e_0}  \E -\int_{-\beta/2}^{+\beta/2} d\tau' \, g(\beta/2,\tau') \, Y_0(\tau') \>\simeq \> - \int_{-\beta/2}^{+\beta/2} d\tau' \, g(\beta/2,\tau') \, f_{\rm GY}(\tau')
\ee
in a first-order expansion in powers of the exponentially small eigenvalue $ e_0 $ (similar as the Born
approximation in the Lippmann-Schwinger equation).
Inserting the expressions for the Green function and the Gel'fand-Yaglom function, the r.h.s.
of the equation becomes
\bea 
&& - \frac{1}{m W} \int_{-\beta/2}^{+\beta/2} d\tau' \, \lsp f^{(1)}(\beta/2) f^{(2)}(\tau') - 
f^{(2)}(\beta/2) f^{(1)}(\tau') \rsp \, \frac{1}{W}\lsp f^{(1)}(-\beta/2) f^{(2)}(\tau') - 
f^{(2)}(-\beta/2) f^{(1)}(\tau') \rsp \non
\EA - \frac{1}{m W^2} \, \lsp f^{(1)}(-\beta/2) f^{(1)}(\beta/2) \la f^{(2)} \big | f^{(2)} \ra - 
f^{(2)}(-\beta/2) f^{(2)}(\beta/2) \la f^{(1)} \big | f^{(1)} \ra \rsp \non
& \stackrel{\beta \to \infty}{\longrightarrow} & - \frac{1}{m} \lrp \frac{2 \omega a A_0}{W} \rrp^2 \, \lsp e^{-\omega \beta}
\, \la f^{(2)} \big | f^{(2)} \ra - e^{\omega \beta} \, \la f^{(1)} \big | f^{(1)} \ra \rsp 
\eea
since the mixed terms $ \> \la f^{(1)} \big |  f^{(2)} \ra  \>  $ vanish because of the different parity of the solutions. In the last line their asymptotic behaviour has been inserted which also allows to estimate the normalization $ \> \la f^{(2)} \big | f^{(2)} \ra \sim \int^{\beta/2} d\tau' \exp( 2 \omega 
\tau') \sim \exp(\omega \beta) \> $ while in the second term $ \> \la f^{(1)} \big | f^{(1)} \ra = \la y_0 \big | y_0 \ra = 1 \> $ has already been normalized to unity as zero-mode solution. Thus at large
 $ \beta $ the second term is dominant and with $ \> W = 8 \omega^3 a^2 A_0^2 \> , \> A_0^2 = m/S_1 
 = 3/(2 \omega a^2) \> $ one obtains
\be 
\fdet' {\cal O}_{\rm V''} \E 
\frac{f_{\rm GY}(+\beta/2)}{e_0}  \> \longrightarrow \> \frac{1}{m} \lrp \frac{2 \omega a A_0}{W} \rrp^2
\, e^{\omega \beta} \E \frac{1}{16 m \omega^4 a^2 A_0^2} \, e^{\omega \beta} 
\E \frac{1}{24 m \omega^3} \, e^{\omega \beta} \> .
\ee

\renewcommand{\baselinestretch}{1.2}
\normalsize
\vspace{0.4cm}

\noindent
{\bf Problem \ref{Feyn Polaron}$^{\star}$:}  
\vspace{0.2cm}

\refstepcounter{ueb}    
\setcounter{equation}{0}
\renewcommand{\baselinestretch}{0.9}
\scriptsize
\noindent
{\bf a)} For $ v = w(1+\epsilon) $ one obtains by expanding in $ \epsilon $, 
$ \> 
\mu^2(u) \E u - 2 \epsilon u + 2 \epsilon \lrp 1 - e^{-w u} \rrp/w  + {\cal O} \lrp \epsilon^2 \rrp
\> $ 
and therefore 
\be
E_F(w,\epsilon) \E \frac{3}{4} w \epsilon^2 - \frac{2 \alpha}{\sqrt{\pi}} \int_0^{\infty} dt \> e^{-t^2} \, 
\lsp 1 + \epsilon - \frac{\epsilon}{w t^2} \lrp 1 - e^{-w t^2} \rrp + \ldots \rsp \E - \alpha 
+ \frac{3}{4} w \epsilon^2 -
\frac{\alpha \epsilon}{w} \, 
\lrp 1  - \sqrt{1+w} \rrp^2 + {\cal O} \lrp \epsilon^3, \alpha \epsilon^2 \rrp \> .
\label{Feyn klein}
\ee
Here the substitution $ u = t^2 $ and an integration by part have been made to reduce the occuring 
integrals to the (ubiquitous) Gaussian one. Variation w.r.t. $\epsilon $ is straightforward since 
Eq. \eqref{Feyn klein} is a quadratic form in $\epsilon$ and gives $ \epsilon_0 = 2 \alpha ( 1 - 
\sqrt{1+w} )^2/(3 w^2) $. Plugging that back into the expression \eqref{Feyn klein} for the Feynman energy
gives
\be
E_F(w) \E - \alpha - \frac{\alpha^2}{3 w^3} \, \lrp 1 - \sqrt{1+w} \rrp^4 + \ldots
\ee
which can be varied w.r.t. $w$. This leads to the equation $ w_0^2 = 3 w_0 $ with two solutions: 
$w_0 = 0 $ (no  retardation) leads to the first-order result $ E_F = - \alpha $ whereas the second solution $ w_0 = 3 $
gives the lower ground-state energy $ E_F = -\alpha - \alpha^2/81 - \ldots $ .
\vspace{0.1cm}

\noindent
{\bf b)} Assuming $ v \gg w $ for large coupling immediately leads to $
 \mu^2(u) = 1/\sqrt(v) + {\cal O}(1/v^{3/2}) $ and 
thus 
\be 
E_F(v) \E \frac{3}{4} v - \frac{\alpha}{\sqrt{\pi}} \, v^{1/2} + \ldots
\ee
is only a function of $v$  in this limit. Variation is ``{\it kinderleicht}'' (easy as pie) and gives 
$ v_0 = 4 \alpha^2/(9 \pi)$
and therefore
\be
E_F \> \stackrel{\alpha \to \infty}{\longrightarrow} \> - \frac{\alpha^2}{3 \pi} \> .
\ee
\renewcommand{\baselinestretch}{1.2}
\normalsize
                                                                                     
\noindent
{\bf Problem \ref{kohaer Zust}:}  
\vspace{0.2cm}

\refstepcounter{ueb}    
\setcounter{equation}{0}
\renewcommand{\baselinestretch}{0.9}
\scriptsize
\noindent
Using
\be
\hat x \E \sqrt{\frac{\hbar}{2 m \omega}} \, \lrp \hat a + \hat a^{\dagger} \rrp \> , \quad
\hat p \E i \sqrt{\frac{m \hbar \omega}{2}} \, \lrp \hat a^{\dagger} - \hat a \rrp \>
\ee
gives, for example 
\be 
\la z | \, \hat x \, | z \ra \E \sqrt{\frac{\hbar}{2 m \omega}} \, \lrp z + z^{\star} \rrp \, \la z |  z \ra
\> , \quad \la z | \, \hat p \, | z \ra \E i \sqrt{\frac{m \hbar \omega}{2}} \, \lrp  z^{\star} - z \rrp
\, \la z |  z \ra
\ee
and similar for the squared operators:
\be
\lrp \Delta x \rrp^2 = \frac{\hbar}{2 m \omega} \, \Big [ z^2 + z^{\star \, 2} + 2 z z^{\star} + 1 
-  ( z + z^{\star} )^2 \Big ] = \frac{\hbar}{2 m \omega} \> , \>
\lrp \Delta p \rrp^2 = - \frac{m \hbar \omega}{2} \, \Big [ z^2 + z^{\star \, 2} - 2 z z^{\star} - 1 
-  ( z - z^{\star} )^2 \Big ] \E \frac{m \hbar \omega}{2} \> .
\ee
Therefore the equality sign holds in the uncertainty relation
\be
\Delta x \cdot \Delta p \E \sqrt{\frac{\hbar}{2 m \omega}} \cdot \sqrt{\frac{m \hbar \omega}{2}} \E 
\frac{\hbar}{2} \> .
\ee

\renewcommand{\baselinestretch}{1.2}
\normalsize
\vspace{0.4cm}

\noindent
{\bf Problem \ref{Det Spur}:}  
\vspace{0.2cm}

\refstepcounter{ueb}    
\setcounter{equation}{0}
\renewcommand{\baselinestretch}{0.9}
\scriptsize
\noindent
{\bf a)} Assume that the matrix $ \A$ can be diagonalized: 
$ \A \deF diag \lrp a_1, a_2 \ldots \rrp $. Then 
\be
\det \A \E \prod_k \, a_k \E \exp \lsp \sum_k \ln \lrp a_k \rrp \rsp \E 
\exp \lsp {\rm tr} \ln \A \rsp \> .
\label{Det ln}
\ee
\vspace{0.1cm}

\noindent
{\bf b)} Differentiate Eq. \eqref{Det ln} w. r. t. the parameter $\lambda$ to obtain
\be
\frac{\partial}{\partial \lambda} \, \det \A(\lambda) \E \lrp \sum_k \frac{1}{a_k(\lambda)} \, 
\frac{\partial a_k(\lambda)}{\partial \lambda} \rrp \cdot \det \A(\lambda) \E {\rm tr} \lrp
\A^{-1}(\lambda) \, \frac{\partial \A(\lambda)}{\partial \lambda} \rrp \cdot \det \A(\lambda) \> .
\label{diff Det}
\ee
\vspace{0.1cm}

\noindent
{\bf c)} Use the previous result \eqref{diff Det} to evaluate
\be
\frac{\partial}{\partial g'} \, \ln \det \lrp \A_0 + g' \A_1 \rrp \E {\rm tr} \lsp \lrp \A_0 + 
g' \A_1 \rrp^{-1} \, \A_1 \rsp \> .
\ee
Integrate both sides of this equation w.r.t. $ g' $ from $ 0 $ to $ g $ with the initial condition 
$ \> \det \lrp \A_0 + g' \A_1 \rrp \Big |_{g'=0} = \det \A_0 \> $; this proves the assertion. If $ g $ 
is a small (coupling) constant one may expand in powers of it and obtains
\be 
\det \lrp \A_0 + g \A_1 \rrp \E \det \lrp \A_0 \rrp \, \cdot \, \exp \lcp g \, {\rm tr} \lrp \A_1 \A_0^{-1} \rrp - \frac{g^2}{2} {\rm tr} \lrp \A_1 \A_0^{-1} \A_1 \A_0^{-1} \rrp + \ldots \rcp \> .
\ee

\renewcommand{\baselinestretch}{1.2}
\normalsize
\vspace{0.4cm}

\noindent
{\bf Problem \ref{Grassmann}:}  
\vspace{0.2cm}

\refstepcounter{ueb}    
\setcounter{equation}{0}
\renewcommand{\baselinestretch}{0.9}
\scriptsize
\noindent
Calculate
\be 
\int d\cy{\eta} \> e^{- \cy{\eta} ( \cy{\xi} - \cy{\xi'}) }\E  \int d\cy{\eta} \> \lsp 1 - \cy{\eta} ( \cy{\xi} - \cy{\xi'} ) \rsp \E -( \cy{\xi} - \cy{\xi'}) 
\ee
using the property that the square of a Grassmann variable vanishes and the Berezin integration rules.
That this is indeed a Grassmann $\delta$-function can be seen by evaluating its integral with a general
function $ \> f(\cy{\xi'}) = f_0 + f_a \cy{\xi'} $
\be 
\int d\cy{\xi'} \> ( - ) \, (\cy{\xi} - \cy{\xi'} )  \> \lsp f_0 + f_1 \cy{\xi'} \rsp \E 
- \int d\cy{\xi'} \> \cy{\xi} f_1 \cy{\xi'} + \int d\cy{\xi'} \> \cy{\xi'} f_0 \E f_1 \cy{\xi} + f_0 
\EQ f(\cy{\xi}) \> .
\ee

\renewcommand{\baselinestretch}{1.2}
\normalsize
\vspace{0.4cm}

\noindent
{\bf Problem \ref{fermion ho}:}  
\vspace{0.2cm}

\refstepcounter{ueb}    
\setcounter{equation}{0}
\renewcommand{\baselinestretch}{0.9}
\scriptsize
\noindent
{\bf a)} Eq. \eqref{Z boson/ferm} with $ \> H = \hbar \omega \bar \xi \, \xi \> $ 
leads to the expression \eqref{Z omega} for 
 the partition function of the harmonic oscillator. Berezin integration gives
$ \> Z_{\omega} \E {\rm const.} \, \fdet \, {\cal O}_{\omega} \> $ where $ \> {\cal O}_{\omega} \Def
\frac{\partial}{\partial \tau} + \omega \> $. The determinant is calculated as product of the eigenvalues  $ \lambda_n $ of the operator $ \> {\cal O}_{\omega} \> $ in the space of functions with anti-periodic boundary conditions at $\tau = 0 $ and $\tau = \hbar \beta $. The eigenfunctions are $\>$ const. $ \exp [(\lambda-\omega) \tau ] $ and therefore the 
eigenvalues $ \> \lambda_n \E \omega + \frac{2n + 1}{\hbar \beta} \pi i \> , n = 0, \pm 1 \ldots $
This leads to
\be
\fdet \, {\cal O}_{\omega} \E \fdet \, {\cal O}_0 \> \prod_{n=0,\pm 1, \pm 2 \ldots} \lrp 1 - \frac{\omega \hbar \beta}{(2 n +1) \pi} \> i \rrp \E \fdet \, {\cal O}_0 \> \prod_{n=0}^{\infty} \lrp 1 + \frac{\omega^2 \hbar^2 \beta^2}{(2 n +1)^2 \pi^2} \rrp \E \fdet \,  {\cal O}_0 \> \cosh \lrp \frac{1}{2} \hbar \omega \beta \rrp \> .
\ee
\vspace{0.1cm}

\noindent
{\bf b)} In the discretized version one has to evaluate
\be
Z_0(\beta) \E \lim_{N \to \infty} \> \prod_{n=0}^N \lrp \int d \bar \xi_n \, d \xi_n \rrp \> \exp \lsp - \epsilon \sum_{n=1}^N \bar \xi_n \frac{\xi_n - \xi_{n-1}}{\epsilon} \rsp \E \lim_{N \to \infty} \>
{\rm det}_{N+1} \He
\ee
where the determinant can be calculated by an expansion along the first row
\be
{\rm det}_{N+1} \He \E
\begin{vmatrix} 1 & 0 & \ldots & 0  & 1 \\
                -1 & 1 & \ldots & 0  & 0 \\
           \vdots &   &        &    &   \\
                0 & 0 & \ldots &  -1 & 1
\end{vmatrix} \E 1 + 1^{N+1} \E 2 \> .
\ee
Thus $ Z_0 \E 2 $ and
\be 
Z_{\omega} \E 2 \, \cosh \lrp \frac{1}{2} \hbar \omega \beta \rrp \> .
\label{Z ho ferm}
\ee
\vspace{0.1cm}

\noindent
{\bf c)} From Eq. \eqref{Z ho ferm} one finds
\be
Z_{\omega}(\beta \to \infty ) \> \to \> \exp \lrp \frac{1}{2} \hbar \omega\beta \rrp \> \Longrightarrow \> E_0 \E - \frac{1}{2} \hbar \omega \> .
\ee
That the fermionic "partners" cancel the (infinite) ground-state energy of the (infinite) harmonic oscillators (representing the scalar fields) in the Universe is a nice idea.
However, \textsf{"the great tragedy of Science is the slaying of a beautiful hypothesis by an ugly fact"} (T. H. Huxley). Here the "ugly facts" are
(i) The fermionic (or more generally, supersymmetric) partners have not been found experimentally.
(ii) The cosmological constant is not zero  but has a small, non-zero value determined from the observed acceleration of far-away galaxies (Nobel prize in Physics 2011).

\renewcommand{\baselinestretch}{1.2}
\normalsize
\vspace{0.4cm}

\noindent
{\bf Problem \ref{freie Energie}:}  
\vspace{0.2cm}

\refstepcounter{ueb}    
\setcounter{equation}{0}
\renewcommand{\baselinestretch}{0.9}
\scriptsize
\noindent
Write the free energy as
\be 
F(\beta) \E - \frac{1}{\beta} \, \ln Z(\beta) \E  - \frac{1}{\beta} \, \ln \lsp e^{-\beta E_0} + 
\sum_{n=1} e^{-\beta E_n} \rsp \E E_0  - \frac{1}{\beta} \, \ln  \lsp 1 + \sum_{n=1} e^{-\beta (E_n - E_0)} \rsp
\label{free en}
\ee
and observe that $ \> \ln (1 + x ) \ge 0 \> $ for $ \> x \ge 0 \> $ as can be seen from a graph of the function. Differentiating Eq. \eqref{free en} w.r.t $ \beta $ one obtains
\be 
F'(\beta) \E \frac{1}{\beta^2} \, \ln  \lsp 1 + \sum_{n=1} e^{-\beta (E_n - E_0)} \rsp + \frac{1}{\beta}
\, \sum_{n=1} \lrp (E_n - E_0 \rrp \, e^{-\beta (E_n - E_0)} \, \lsp 1 +  \sum_{n=1} e^{-\beta (E_n - E_0)} \rsp^{-1}
\ee
and sees that all terms are positive since $ E_n > E_0 $. Thus the free energy is monotonicllay growing
from $ F(0) = -\infty $ to $ F(\infty) = E_0 $ .
\renewcommand{\baselinestretch}{1.2}
\normalsize
\vspace{0.4cm}

\noindent
{\bf Problem \ref{G_0 Vielteil}:}  
\vspace{0.2cm}

\refstepcounter{ueb}    
\setcounter{equation}{0}
\renewcommand{\baselinestretch}{0.9}
\scriptsize
\noindent
Because of time-translation invariance and the absence of interactions one can write
 $ G_0(\alpha,\tau;\alpha',\tau') = \delta_{\alpha \alpha'} \, g_{\alpha}(\tau-\tau') $ where $ g_{\alpha}(t \Def \tau - \tau') $ fulfills
\be
\lrp \partial_t + \epsilon_{\alpha} - \mu \rrp \, g_{\alpha} \E \delta (t) \> .
\ee
One can solve this differential equation by the method of "variation of the constant", i.e. by the ansatz
\be
g_{\alpha}(t) \E C(t) \cdot \exp \lsp - \lrp \epsilon_{\alpha} - \mu \rrp t \rsp \> .
\ee
This gives $ \> \dot C(t) = \delta(t) \cdot \exp [ (\epsilon_{\alpha} - \mu) t ] = \delta(t) \> $, i.e.
$ \> C(t) = a + \Theta(t) \> $. The boundary condition $ \> g_{\alpha}(\beta) = \zeta g_{\alpha}(0) \> $ 
determine the constant as $ a = n_{\alpha} \, \lcp \zeta - \Theta(0) \, \exp [(\epsilon_{\alpha} - \mu )\beta ] \rcp \> $ where $ \> n_{\alpha} \> $ is the occupation probability \eqref{Besetz}. It remains to give a meaning to the step function at zero argument, i.e. to consider the time-ordered product at equal time. This is defined to be equal to a normal-ordered product at equal time so that in the time-sliced expression for 
$ \> \la \hat a_{\alpha'}^{\dagger}(\tau') \, \hat a_{\alpha}(\tau) \ra \> $ one has terms like
\be 
e^{-\epsilon \hat H} \, \lvl z_k \ra \la z_k \rvl \hat a_{\alpha}^{\dagger} \hat a_{\alpha'} \, 
e^{- \epsilon \hat H} \, \lvl z_{k-1} \ra \la z_{k-1} \rvl \ldots \E e^{-\epsilon \hat H} \, \lvl z_k \ra \lrp  z_{\alpha,k}^{\star} z_{\alpha',k-1} \, 
e^{- \epsilon H(z_k^{\star},z_{k-1}) } + {\cal O}(\epsilon) \rrp  \la z_{k-1} \rvl \ldots 
\ee
In other words: The creation operator is evaluated one time step later than the annihilation operator
($ \tau' = \tau + \epsilon $)
and therefore $ \Theta(0) \longrightarrow  \lim_{\epsilon \to 0} \Theta(\tau - (\tau + \epsilon) = 0 $.
The discretized version of the path integral thus gives the unambigous result
\be 
g_{\alpha}(t) \E e^{-( \epsilon_{\alpha} - \mu) t } \, \lcp \lrp 1 + \zeta n_{\alpha} \rrp \, 
\Theta(t-\epsilon) + \zeta n_{\alpha} \, \Theta (\epsilon - t) \rcp \> .
\ee
\renewcommand{\baselinestretch}{1.2}
\normalsize
\vspace{0.4cm}

\noindent
{\bf Problem \ref{Pekar}$^{\star}$:}  
\vspace{0.2cm}

\refstepcounter{ueb}    
\setcounter{equation}{0}
\renewcommand{\baselinestretch}{0.9}
\scriptsize
\noindent
{\bf a)} 
From Eq. \eqref{Euler Gamma} one finds $ \int_0^{\infty} dy \, 
y^n \exp(-\lambda y) = n!/\lambda^{n+1} $. Some algebra then gives 
the normalization as $ C_1 = \sqrt{2/a} $ and 
determines the kinetic and potential  terms as $ \> <T> = 1/(2 a^2) \> , \> <V> = - 5 \sqrt{2}/(16 \kappa a) \> $, respectively. Therefore the lowest value of $ \> <T> + <V> \> $ occurs at 
$ \> a_0 \E 8 \sqrt 2  \kappa / 5 \> $ and the optimal value of Pekar's constant is independent of the scaling parameter $ \kappa $
\be 
\gamma_P(a_0) \E - \frac{25}{256} \> .
\ee
\vspace{0.1cm}

\noindent
{\bf b)} Substitute in Eq. \eqref{Euler Gamma} $ t = \lambda^2 y^2 $ to derive 
\be 
\int_0^{\infty} dy \, y^{2n} \exp (- \lambda^2 y^2) \E \frac{ 1\cdot 3 \ldots (2n-1)}{2^{n+1} \lambda^{2n+1}} \sqrt \pi 
\ee 
and find thereby $ \> C_2^2 = 4/(\sqrt \pi b^3) \> $, 
$ \> <T> = 3/(4 b^2) $ (compare with the virial theorem for the 3-dim. harmonic oscillator!) and 
%
$ \> \la V \ra = - 1/(\sqrt \pi \kappa b) $. This gives the optimal value
\be 
b_0 \E \frac{3}{2} \sqrt \pi \kappa \quad \Longrightarrow \quad \gamma_P(b_0) \E - \frac{1}{3 \pi} \> .
\ee
\vspace{0.1cm}

\noindent
{\bf c)} Declare all real-valued variables in the beginning of the program as "DOUBLE PRECISION" or insert\\ "IMPLICIT REAL*8 (A-H,O-Z)" which declares all variables as double-precision unless they begin with "I,J,K,L,M,N" in which case they are integer.
To avoid loss of accuracy replace numerical constants by their double-precision value, e.g. "$ \, 1. \to 1.{\rm d}0 \, $" etc.

\renewcommand{\baselinestretch}{1.2}
\normalsize
\vspace{0.4cm}

\noindent
{\bf Problem \ref{Selbstenergie phi4}:}  
\vspace{0.2cm}

\refstepcounter{ueb}    
\setcounter{equation}{0}
\renewcommand{\baselinestretch}{0.9}
\scriptsize
\noindent
Use the Schwinger representation \eqref{Fock-Schwinger}
to perform the Gaussian $k$-integration in Minkowski space
\bea 
\int \frac{d^d k}{(2 \pi)^d} \> e^{i a k^2 + i b \cdot k} \EA e^{-i b^2/(4a)} \int \frac{d^d k'}{(2 \pi)^d} \> e^{i a k'^2} \E
\E e^{-i b^2/(4a)} \int \frac{dk_0'}{2 \pi} \> e^{ia k_0'^2} \prod_{i=1}^{d-1}
\lrp \int \frac{d^i k'}{(2 \pi)^i} \> e^{-ia k'^2_i} \rrp \non
\EA e^{-i b^2/(4a)} \frac{1}{2 \pi} \sqrt{\frac{\pi}{-ia}} \,
\frac{1}{(2 \pi)^{d-1}} \sqrt{\frac{\pi}{ia}}^{d-1} \E 
 \frac{i}{(4 \pi i a)^{d/2}} \, e^{-i b^2/(4a)}
\label{Gauss Minkowski}
\eea
with $ a = T $ and $ b_{\mu} = 0 $.
The $T$-integral is of the form $ \> \int_0^{\infty} dT \, T^{-d/2} \, \exp (-im^2 T) \> $ and can be 
performed by means of Eq. \eqref{Euler Gamma}. This gives the result \eqref{Sigma1 phi4}.
As the Gamma function has poles at argument $= 0, -1, -2 \ldots$  the integral diverges not only 
for the relevant case $ d = 4 $ but also for $ d = 2  $. This signals a quadratic divergence when the
momentum integral is  cut off at $ k = \Lambda $:
\be 
\int^{\Lambda} d^4 k \> \frac{1}{k^2 - m^2 + i0^+} \> \sim \> \int^{\Lambda} dk \, k^3 \, \frac{1}{k^2}
\> \sim \Lambda^2 \> .
\ee
Near the physical dimension one sets $ \> d = 4 - 2 \epsilon \> $ and finds with 
$ \> \Gamma(1-d/2) = \Gamma(-1 + \epsilon) = \Gamma(\epsilon)/(-1 + \epsilon) \> $ and the expansions
$ 
\Gamma(\epsilon) \E 1/\epsilon - \gamma_E + {\cal O}(\epsilon) \> , \quad x^{-\epsilon} \E 
\exp(-\epsilon \ln x) \E 1 - \epsilon \ln x + {\cal O}(\epsilon^2) 
$
($\gamma_E = 0.57721566...$ is Euler's constant) that the first-order self energy in $\Phi^4$  theory
is given in dimensional regularization by
\be 
\Sigma^{(1)}(p) \E - \frac{\lambda}{32 \pi^2} \,  \frac{m^2}{\epsilon} \lsp 1  - \epsilon \lrp \gamma_E - 1 + \ln \frac{m^2}{4 \pi \mu_0^2} \rrp + {\cal O}(\epsilon^2) \rsp \> ,
\ee
consisting of a divergent and a finite part when $ \epsilon \to 0 $.
\renewcommand{\baselinestretch}{1.2}
\normalsize
\vspace{0.4cm}

\noindent
{\bf Problem \ref{N skalare Teil}:}  
\vspace{0.2cm}

\refstepcounter{ueb}    
\setcounter{equation}{0}
\renewcommand{\baselinestretch}{0.9}
\scriptsize
\noindent
For $ N = 2 $ define 
\be 
\Phi \Def \frac{1}{\sqrt 2} \, \lrp \Phi_1 + i \Phi_2 \rrp \> , \quad \Phi^{\star} \Def 
\frac{1}{\sqrt 2} \, \lrp \Phi_1 -i \Phi_2 \rrp
\label{def komplex phi}
\ee 
so that
 $ \> \Phi_1 = (\Phi + \Phi^{\star})/\sqrt 2 \> , \Phi_2 =  (\Phi - \Phi^{\star})/(\sqrt 2 i) \> $. Then 
\be 
{\cal L}_0^{(2)} \E \frac{1}{2} \lsp (\partial_{\mu} \Phi + \partial \Phi^{\star} )^2
- (\partial_{\mu} \Phi -\partial \Phi^{\star} )^2 \rsp - \frac{m^2}{2} \lsp ( \Phi + \Phi^{\star} )^2
- ( \Phi - \Phi^{\star} )^2 \rsp \E \lvl \partial_{\mu} \Phi \rvl^2 - m^2 \lvl \Phi \rvl^2 \> ,
\ee
i.e. the factor $ 1/2 $ of the neutral case has been removed. The Jacobian of the transformation
\eqref{def komplex phi} is 
\be 
\begin{vmatrix}
\frac{\partial \Phi_1}{\partial \Phi} & \frac{\partial \Phi_2}{\partial \Phi} \\
& \\
\frac{\partial \Phi_1}{\partial \Phi^{\star}} & \frac{\partial \Phi_2}{\partial \Phi^{\star}}
\end{vmatrix} \E 
\begin{vmatrix}
\frac{1}{\sqrt 2} & \frac{1}{\sqrt 2 i} \\
& \\
\frac{1}{\sqrt 2} & - \frac{1}{\sqrt 2 i}
\end{vmatrix}
\E i
\ee
which just modifies the irrelevant normalization of the path integral for the generating functional
\bea
Z_0 \lsp J^{\star}, J \rsp \EA \int {\cal D} \Phi^{\star} \, {\cal D} \Phi \, \exp \lsp i \lrp \Phi^{\star}, K
K \Phi \rrp + \lrp J^{\star}\!, \Phi \rrp + \lrp \Phi^{\star}\!, J \rrp \rsp
\E {\rm const.} \int {\cal D} \chi^{\star} {\cal D} \chi \> \exp \lsp i \lrp \chi^{\star}\!,K^{-1} \chi \rrp - i \lrp J^{\star}\!, K^{-1} J \rrp \rsp \non
\EA {\rm const.'} \, \exp \lsp - i \lrp J^{\star}\!, K^{-1} J \rrp \rsp \> ,
\eea
where
the square in the exponent has been completed with
$ \> \chi \Def K \Phi + J \> , \> \chi^{\star} \Def \Phi^{\star} K + J^{\star}  \> $ .
$\> K \Def \partial^2 - m^2 \> $ is the same kernel as in the neutral case and its inversion just 
gives the usual scalar propagator $ \Delta_F $. Thus  by functional differentiation w.r.t. the sources
$ J $ and $ J^{\star} $ one obtains
\be 
\la \Phi^{\star}(x_1) \, \Phi(x_2) \ra \E (-i)^2 \frac{\delta^2}{\delta J(x_1) \delta J^{\star} (x_2)} \, Z_0 \lsp J^{\star}, J \rsp \, \Bigg |_{J^{\star}=J=0} \, \frac{1}{Z_0[0,0]} \E i \, \Delta(x_1,x_2) \> .
\ee
In the same way one finds $ \> \la \Phi(x_1) \, \Phi(x_2) \ra \E \la \Phi^{\star}(x_1) \, \Phi^{\star}(x_2) \ra \E 0 \> $.
\renewcommand{\baselinestretch}{1.2}
\normalsize
\vspace{0.4cm}

\noindent
{\bf Problem \ref{Kumulante}:}  
\vspace{0.2cm}

\refstepcounter{ueb}    
\setcounter{equation}{0}
\renewcommand{\baselinestretch}{0.9}
\scriptsize
\noindent
{\bf a)}
Differentiate Eqs. \eqref{mom entwick} and \eqref{cum entwick} w.r.t. the variable $t$ and equate
\be 
\sum_{n=1} m_n \, i \, n \, \frac{(it)^{n-1}}{n!} \E \sum_{k=0} m_k  \frac{(it)^k}{k!} \, \cdot \sum_{l=1} \lambda_l \, i \, l \, \frac{(it)^{l-1}}{l!} \> .
\ee
Writing the product of the two sums on the r.h.s. as 
\be 
\sum_{n=0} (it)^n \sum_{m=0}^n \lambda_{n+1} \, m_{n-m} \,  \frac{i (m+1)}{(n-m)! \, (m+1)!} 
\ee
and comparing powers of $ t $ on both sides establishes 
Eq. \eqref{cum from mom}.

Writing $ \> \la x^n \ra \Def m_n/m_0 \> $ one finds
\bea 
\lambda_1 \EA \la x \ra \> , \quad \lambda_2 \E \la x^2 \ra - \la x \ra^2 \E \la \lrp x - \la x \ra \rrp^2 \ra \> , \quad 
\lambda_3 \E \la x^3 \ra - 3 \la x^2 \ra \, \la x \ra + 2 \la x \ra^3 \E \la \lrp x - \la x \ra \rrp^3 \ra \non
\lambda_4 \EA \la x^4 \ra - 4 \la x^3 \ra \, \la x \ra + 12 \la x^2\ra \, \la x \ra^2 - 3 \la x^2 \ra^2 - 6 \la x \ra^4 \E \la \lrp x - \la x \ra \rrp^4 \ra - 3 \la \lrp x - \la x \ra \rrp^2 \ra^2 \> .
\eea
The simple structure of the first cumulants suggests a simpler relation in terms of the central moments
(see below).
\vspace{0.1cm}

\noindent
{\bf b)} Multiply Eqs. \eqref{mom entwick} and \eqref{cum entwick} both by $ \exp ( - i \la x \ra t) $
and equate: The l.h.s. then becomes the expansion in central moments whereas on the r.h.s only the first
cumulant is modified. Thus one can take over the final moment result \eqref{cum from mom} with the modifications
$ \> m_n \longrightarrow c_n \> , \> \lambda_n \longrightarrow 
\lambda_n - \la x \ra \, \delta_{n,1} \> $. Since by construction $ \> c_0 \EQ m_0 $ and $ c_1 = 0 \> $, one obtains $ \lambda_1 = \la x \ra $ and
\be 
\lambda_{n+1} \E \frac{c_{n+1}}{c_0} - \sum_{k=1}^{n-2} \,  {n \choose k} \>  \frac{c_{n-k}}{c_0} 
\> \lambda_{k+1} \> ,  \quad n = 1 , 2 \, \mbox{(the sum is then empty)}, 3  \ldots
\> .
\ee
Thus $ \> \lambda_2 = c_2/c_0 \> , \quad \lambda_3 = c_3/c_0 \> , \quad \lambda_4 = c_4/c_0 - 3 (c_2/c_0)^2 \> , \quad \lambda_5 = c_5/c_0 - 10 \, c_2 c_3/c_0^2 \> $ etc.

\vspace{0.1cm}

\noindent
{\bf c)} The moment and cumulant expansion of the generating functional \eqref{def erzeug functional} for the Green functions in a scalar theory read
\be
Z[J] \E \sum_{n=0} \frac{i^n}{n!} \, \int d^4x_1 \ldots d^4x_n \, m_n \lrp x_1 \ldots x_n \rrp \, J(x_1) \ldots J(x_n)
\E m_0 \, \exp \lsp \sum_{n=1} \frac{i^n}{n!}  \int d^4x_1 \ldots d^4x_n G_c^{(n)}\lrp x_1 \ldots x_n \rrp \, J(x_1) \ldots J(x_n) \rsp
\ee
where the moments 
\be 
m_n \lrp x_1 \ldots x_n \rrp \E \int {\cal D} \Phi \, \Phi(x_1) \ldots \Phi(x_n) \, e^{i S[\Phi]} \> , \quad
m_0 \E \int {\cal D} \Phi \, e^{i S[\Phi]}
\ee
are the Green functions up to a normalization: $ \> G_n \lrp x_1 \ldots x_n \rrp = m_n \lrp x_1 \ldots x_n\rrp /m_0 \deF \la 1 \ldots n \ra \> $.  The connected Green functions
are then just the corresponding cumulants (see Eq. \eqref{verbundene GF}) and for an even action where all odd moments/Green functions vanish, the first ones are found as
\bea 
G_c^{(1)}\lrp x_1\rrp \EA  \la 1 \ra \E 0 \> , \quad G_c^{(2)}\lrp x_1,x_2 \rrp \E \la 1 2  \ra \> , \quad
G_c^{(3)}\lrp x_1, x_2, x_3 \rrp \E 0 \non
\quad G_c^{(4)}\lrp x_1, x_2, x_3, x_4\rrp \EA 
\la 1 2 3 4  \ra - \la 1 2  \ra \,  \la 3 4  \ra 
- \la 1 3  \ra \,  \la 2 4 \ra  - \la 1 4  \ra \,  \la 2 3  \ra \> .
\eea

\renewcommand{\baselinestretch}{1.2}
\normalsize
\vspace{0.4cm} 

\noindent
{\bf Problem \ref{eff Wirk}$^{\star}$:}  
\vspace{0.2cm}

\refstepcounter{ueb}    
\setcounter{equation}{0}
\renewcommand{\baselinestretch}{0.9}
\scriptsize
\noindent
Differentiate Eq. \eqref{def Gamma} w.r.t. $ \> \Phi_{\rm cl}(x) \> $ and use the chain rule to obtain
\be 
\frac{\delta \Gamma}{\delta \Phi_{\rm cl}(x)} \E \frac{\delta W}{\delta \Phi_{\rm cl}(x)} - 
\int d^4y \, \frac{\delta J(y)}{\delta \Phi_{\rm cl}(x)} \, \Phi_{\rm cl}(y) - J(x) \E
\int d^4y \, \frac{\delta W[J]}{\delta J(y))} \, \frac{\delta J(y)}{\delta \Phi_{\rm cl}(x)}
- \int d^4y \, \frac{\delta J(y)}{\delta \Phi_{\rm cl}(x)} \, \Phi_{\rm cl}(y) - J(x) \E - J(x)
\ee
due to the definition in Eq. \eqref{def Phi_kl}.

Next differentiate the defining Eq. \eqref{def Phi_kl} for the classical field w.r.t. 
$ \> \Phi_{\rm cl}(y) \> $ . This gives 
\be 
\delta^{(4)}(x-y) \E \frac{\delta^2 W}{\delta J(x) \delta \Phi_{\rm cl}(y)} \E \int d^4z \, 
\frac{\delta J(z)}{\delta \Phi_{\rm cl}(y)} \, \frac{\delta^2 W[J]}{\delta J(z) \delta J(x)} \> ,
\ee
where again the chain rule has been applied. One can now use
the result $ \delta \Gamma/\delta \Phi_{\rm cl} = - J $ (just derived) and find
\be 
\delta^{(4)}(x-y) \E - \int d^4z \, 
\frac{\delta^2 \Gamma[\Phi_{\rm cl}]}{\delta \Phi_{\rm cl}(y) \delta \Phi_{\rm cl}(z) } \, \frac{\delta^2 W[J]}{\delta J(z) \delta J(x)} \> .
\label{Gamma W}
\ee
Recall that the connected Green functions are obtained from Eq. \eqref{verbundene GF} as $ G_c^{(n)} = (1/i)^n \delta^n \, i W \> $. Hence, 
if one sets $ \> \Phi_{\rm cl} = 0 \> $ one finds that the connected 2-point function $ \> G_c^{(2)}(x_1,x_2) = -i [ \delta^2 W/\delta J(x_1) \delta J(x_2) ]_{J=0} $ is the inverse (in operator sense) of
$ - \Gamma^{(2)}(x_1,x_2) = - [ \delta^2 \Gamma/\delta \Phi_{\rm cl}(x_1) \delta \Phi_{\rm cl}(x_2) ]_{\phi_{\rm cl} = 0} \> $. Due to translation invariance these functiona only depend on the difference
$ \> x_1 - x_2 \> $ so that in momentum space the relation reads
\be 
\tilde \Gamma^{(2)}(p) \E i \lsp \tilde G^{(2)}(p) \rsp^{-1} \> \Longrightarrow \> \tilde \Gamma_2(p) \E p^2 - m^2 - \Sigma(p)
\ee
where $ \> \Sigma(p) \> $ is the self-energy defined in Eq. \eqref{prop 1a}. From the 
expansion of the 2-point function
\be 
\tilde G^{(2)}(p) \E \frac{i}{p^2-m^2} + \frac{i}{p^2-m^2} \frac{1}{i} \Sigma(p) \frac{i}{p^2-m^2} + 
\frac{i}{p^2-m^2}  \frac{1}{i} \Sigma(p) \frac{i}{p^2-m^2}  \frac{1}{i} \Sigma(p) \frac{i}{p^2-m^2} 
+ \ldots
\ee
one sees that $ - i \Sigma $ is the 1-particle irreducible perturbative contribution to the the 2-point function. The interpretation of higher functions $ \> \Gamma_n \> $ as 1-particle irreducible (proper) functions can be derived by successive differentiations of Eq. \eqref{Gamma W} at
 $ \> \Phi_{\rm cl} = 0 \> $.

\renewcommand{\baselinestretch}{1.2}
\normalsize
\vspace{0.4cm}

\noindent
{\bf Problem \ref{Weltlin Pot}$^{\star}$:}  
\vspace{0.2cm}

\refstepcounter{ueb}    
\setcounter{equation}{0}
\renewcommand{\baselinestretch}{0.9}
\scriptsize
\noindent
{\bf a)}
With the {\it ansatz} \eqref{ansatz nr} the action becomes
\be
S_a \E \frac{1}{2m} \int d^4x \, \lsp \lvl -i m \varphi + \dot \varphi \rvl^2 - \lrp m^2 + 2m V(x) \rrp \lvl \varphi \rvl^2
\rsp \E \frac{1}{2m} \int d^4x \, \lsp im \varphi^{\star} \varphi - im \dot \varphi^{\star} \varphi - \lvl \nabla \varphi
\rvl^2 - 2m V(x) \lvl \varphi \rvl^2 + \lvl \dot \varphi \rvl^2 \rsp \, .
\ee
Neglecting the last term in the square bracket for large mass and performing appropriate integration by parts gives 
the non-relativistic action 
\be
S_a\lsp\varphi^{\star}, \varphi \rsp \E \int dt \int d^3x \> \varphi^{\star}(\fx,t) \lsp i \partial_t + \frac{\Delta}{2 m} - V(\fx,t) \rsp \, \varphi(\fx,t)
\ee
for a particle in a potential.
\vspace{0.1cm}

\noindent
{\bf b)} The generating functional can be calculated exactly by completing the square and performing the Gaussian integral since the full action is quadratic in the fields
\be
Z[J^{\star},J] \E \int {\cal D} \Phi^{\star} \, {\cal D} \Phi \> \exp \lcp i \lsp S_a[\Phi^{\star},\Phi] + \int d^4x \lrp 
J^{\star} \Phi + \Phi^{\star} J \rrp \rsp \rcp \E {\rm const.} \, \exp \lsp - i \lrp J^{\star}, \frac{1}{-\Box - m^2 - 2m V(x) } \, J \rrp \rsp \> .
\ee
Here the usual short-hand notation 
is used. 
This gives for the 2-point function (the only non-vanishing $n$-point function)
\be
G_2\lrp x_2,x_1 \rrp \E (-i)^2 \, \frac{\delta^2}{\delta J^{\star}(x_2) \delta J(x_1)} \, \frac{Z[J^{\star},J]}{Z[0,0]}
\Bigg |_{J^{\star}=J=0} \E i \la x_2 \lvl \frac{1}{-\Box - m^2 - 2m V(x)} \rvl x_1 \ra \> .
\ee
\vspace{0.1cm}

\noindent
{\bf c)} The Fock-Schwinger representation for the 2-point function reads
\be
G_2 \lrp x_2,x_1 \rrp \E \kappa \int_0^{\infty} dT \> e^{i \kappa (-m^2 + i 0^+)} \, \la x_2 \lvl e^{-i T \, H}
 \rvl x_1 \ra \> , \quad H \Def \kappa \Box + 2 m \kappa V(x) \> .
\ee
Choose $ \> \kappa = 1/(2m) \> $, then $ \> H = - \hat p^2/(2m) + V(x) \> $ with 
$ \> \hat p_{\mu} = \partial/(i\partial x^{\mu}) \> $ 
is formally the Hamiltonian of a non-relativistic
particle moving in 4 dimensions with a mass $ - m $. Therefore one can use the standard path-integral representation for the matrix element of the time-evolution operator and immediately obtain Eqs. \eqref{G2 eigen}, \eqref{S eigen}

\renewcommand{\baselinestretch}{1.2}
\normalsize
\vspace{0.4cm}

\noindent
{\bf Problem \ref{Ableit erzeug Funk}$^{\star}$:}  
\vspace{0.2cm}

\refstepcounter{ueb}    
\setcounter{equation}{0}
\renewcommand{\baselinestretch}{0.9}
\scriptsize
\noindent
Fourier transformation gives
\be
F\lrp \frac{1}{i} \frac{\partial}{\partial x} \rrp \, e^{-i a x^2/2} \E \int_{-\infty}^{+\infty} 
\frac{dt}{2 \pi} \> \tilde F(t) \, 
\underbrace{e^{t \frac{\partial}{\partial x} } \, e^{-ia x^2/2}}_{=\exp[-ia (x+t)^2/2]} 
\ee 
using the shift property of $ \> \exp (t \partial/\partial x) \> $ . Writing 
\be
\exp \lsp - \frac{i}{2} (x + t )^2 \rsp \E e^{-ia x^2/2} \, \exp\lsp - i \lrp  at^2/2 + a x t \rrp \rsp \E
 e^{-ia x^2/2} \, \exp\lsp - \frac{i}{2} a \lrp \frac{1}{a} \frac{\partial}{\partial x} \rrp^2 \rsp \, e^{-i a x t }
\ee
one finds Eq. \eqref{Ableit Bezieh}. Generalizing this result to the infinite-dimensional case and applying it to the generating functional in Eq. \eqref{Z aus Z0} one just has to set $ \> F[y] = \exp [ - i \int d^4x V(y(x)) ] \> $ and $ a \to \Delta_F $ to obtain Eq. \eqref{Kopitz result}.
\vspace{0.1cm}

\noindent
In $\Phi^4$-theory  one has $ \> S_{\rm int}[\Phi] \Def - \lambda \int d^4x \,\Phi^4(x)/4! \> $  and up to first order
\be 
Z[J] \E Z_0[J] \, \lsp 1 + A + \frac{1}{2} A^2 + \ldots \rsp \, \lsp 1 + i \, S_{\rm int}[\Phi] + {\cal O}(\lambda^2) \rsp \E Z_0[J] \, \Bigl [ \, 1 + \lambda \omega_1[J] + \ldots \, \Bigr ] \> ,
\ee
where the operator $ \> A \Def \lrp i \frac{\delta}{\delta \Phi}, \Delta_F \frac{\delta}{\delta \Phi} \rrp/2 \> $ contains 2 functional derivatives -- in first order one therefore only needs terms up 
to $ A^2 $.
Performing these derivatives one obtains $ \> A \, S_{\rm int} = -i \lambda \Delta_{xx} \Phi_x^2/4 \> $ and $ \> A^2 S_{\rm int} = \lambda \Delta_{xx}^2/4 \>$  in the abreviated notation used when calculating the second order $ \> \omega_2[J] \> $. When inserting $ \Phi_x = \Delta_{xy} J_y \> $  this gives Eq. \eqref{omega1}.
\renewcommand{\baselinestretch}{1.2}
\normalsize
\vspace{0.3cm}

\noindent
{\bf Problem \ref{Bi-Quelle}:}  
\vspace{0.2cm}

\refstepcounter{ueb}    
\setcounter{equation}{0}
\renewcommand{\baselinestretch}{0.9}
\scriptsize
\noindent
{\bf a)} As
\be 
\frac{\delta}{\delta K(x)} \, \exp \lsp - \frac{i}{2} \int d^4 y \, K(y) \sum_i^N \Phi_i^2(y) \rsp \E
 - \frac{i}{2} \, \sum_i^N \Phi_i^2(x) \, \cdot \, \exp \lsp - \frac{i}{2} \int d^4 y \, K(y) \sum_i^N \Phi_i^2(y) \rsp 
\ee 
Eq. \eqref{Z von Z0 bi} is (nearly) self-evident. 
\vspace{0.1cm}

\noindent
{\bf b)} With vanishing interaction the generating functional is just the $N$-fold product of a 
Gaussian integral where the prefactor (the determinant, see {\bf Problem \ref{Det Spur} a)}) cannot be subsumed in an irrelevant constant since it depends on the source $ K $
\be 
Z_0[J_i,K] \E {\rm const.} \, \prod_i^N \lrp \exp \lsp - \frac{i}{2} \la J_i \left | {\cal O}_K^{-1} \right | J_i \ra - \frac{1}{2} \, {\rm tr} \ln {\cal O}_K \rsp \rrp \> .
\ee
\vspace{0.1cm}

\noindent
{\bf c)} Expand Eq. \eqref{Z von Z0 bi} to first order in $ \lambda $ and use the analogue of the
relation $ \> (e^{if})'' = (if'' - f'^2) e^{if} \> $ for the case of functional derivatives in order  
to verify Eq. \eqref{omega1 von W0}.
Normalizing the determinant to the no-source case , i.e. $ \>  \ln {\cal O}_K \longrightarrow 
\ln \lsp {\cal O}_0^{-1} {\cal O}_K \rsp = \ln \lsp 1 - {\cal O}_0^{-1} K \rsp \> $,
and expanding the different terms up to $ {\cal O} (K^2) $ one obtains explicitly
\be 
W_0[J_i,K] \E - \frac{1}{2} \la J_i \lvl \Delta_F + \Delta_F K \Delta_F +  \Delta_F K \Delta_F  \Delta_F K \Delta_F + \ldots \rvl J_i \ra + \frac{i N}{2} {\rm tr} \lsp -  \Delta_F K  - \frac{1}{2}  \Delta_F K  \Delta_F K  - \ldots \rsp \> .
\ee
Here $ {\cal O}_0^{-1} \EQ \Delta_F $. 
Functional differentiation w.r.t. $ K(x) $ is just a matter of book-keeping and one obtains
\bea 
\frac{\delta W_0}{\delta K(x)} \Bigg |_{K=0} \EA - \frac{1}{2} \sum_i^N \la J_i \lvl \Delta_F \rvl x \ra \, \la x \lvl \Delta_F \rvl J_i \ra - \frac{iN}{2} \Delta_F(0)\\
\frac{\delta^2 W_0}{\delta K(x)}^2 \Bigg |_{K=0} \EA - \sum_i^N \la J_i \lvl \Delta_F \rvl x \ra 
\la x \lvl \Delta_F \rvl J_i \ra \, \Delta_F(0) - \frac{iN}{2} \Delta_F^2(0) \> .
\eea
This gives the following symmetry factors in first-order perturbation theory: For the vacuum energy density $ N^2 + 2 N $ (in agreement with Eq. \eqref{omega1} for $ N = 1 $), for the 2-point function 
$ 2N + 4 $ ($ = 6 $ for $ N = 1 $) and $ 1 $ for the 4-point function (unchanged).
\renewcommand{\baselinestretch}{1.2}
\normalsize
\vspace{0.4cm}

\noindent
{\bf Problem \ref{Vektor-Teilchen}:}  
\vspace{0.2cm}

\refstepcounter{ueb}    
\setcounter{equation}{0}
\renewcommand{\baselinestretch}{0.9}
\scriptsize
\noindent
{\bf a)} Multiplying out the field-strength term in the Lagrangian one has
\be 
S_0[V] \E -2 \frac{1}{4} \int d^4x \> \lsp \partial_{\mu} V_{\nu} \, \partial^{\mu} V^{\nu} - 
\partial_{\mu} V_{\nu} \, \partial^{\nu} V^{\mu} - m^2_V V_{\mu} V^{\mu} \rsp
\E \frac{1}{2} \int d^4x \> V_{\mu} \, \lsp \lrp \Box + m_V^2 \rrp g^{\mu \nu} - \partial^{\mu}
\partial^{\nu} \rsp \, V_{\nu}
\ee
with the help of an integration by parts. Thus the Gaussian integration gives for the free 
generating functional 
\bea 
Z_0[J_{\mu}] \EA \int \lrp \prod_{\rho} {\cal D} V_{\rho} \rrp \, \exp \lcp i S_0[V] + i \int d^4x \, 
J_{\mu}(x) V^{\mu} (x) \rcp \non
\EA {\rm const.} \exp \lcp - \frac{i}{2} \int d^4x \, d^4y \, J^{\mu}(y) \la 
y \lvl
\lsp \lrp   \Box + m_V^2 \rrp g^{\mu \nu} - \partial^{\mu}
\partial^{\nu} \rsp^{-1} \rvl x \ra \, J^{\nu}(x) \rcp \> .
\eea
In momentum space the free propagator therefore is
\be 
\tilde D_{\mu \nu}(k) \E \lsp \lrp -k^2 + m_V^2 \rrp g^{\mu \nu} +k^{\mu}k^{\nu} \rsp^{-1} \> .
\ee
For the inversion in Lorentz space one can make the {\it ansatz} 
$ \> \tilde D_{\mu \nu}(k) = A(k) \, g_{\mu \nu} + B(k)\, k_{\mu} k_{\nu} \> $ (these 
are the only available Lorentz structures).
Requiring that 
$ \> \lsp  \lrp -k^2 + m_V^2 \rrp g_{\mu \rho} +k_{\mu}k_{\rho }\rsp \lrp  A(k) \, g^{\rho \nu} 
+ B(k) \, k^{\rho} k^{\nu} \rrp = \delta_{\mu}^{\nu} \> $ one finds $ A(k) = - 1/(k^2 - m_V^2) \> , 
\> B(k) = - A(k)/m_V^2 \> $ leading to Eq. \eqref{vec prop}.
\vspace{0.1cm}

\noindent
{\bf b)} For the photon propagator in covariant gauge one has to evaluate
\be
\tilde \Delta_{\mu \nu}(k) \E \lsp - k^2 g^{\mu \nu} + \lrp 1 - \frac{1}{\lambda} \rrp k^{\mu} k^{\nu} 
\rsp^{-1} \, \stackrel{!}{=} \, A(k) \, g_{\mu \nu} + B(k) \, k_{\mu} k_{\nu}
\ee
and finds by a similar procedure $ \> A(k) = -1/k^2 \> , \> B(k) = (\lambda - 1 ) A(k)/k^2 \> $.
Note the different behaviour of the massive and massless vector propagators for large $k$, i.e. the 
different terms $ B(k) $.

\renewcommand{\baselinestretch}{1.2}
\normalsize
\vspace{0.4cm}

\noindent
{\bf Problem \ref{Noether Feld}:}  
\vspace{0.2cm}

\refstepcounter{ueb}    
\setcounter{equation}{0}
\renewcommand{\baselinestretch}{0.9}
\scriptsize
\noindent
If the transformation parameter depends on the spacetime then the Lagrangian changes as follows
\be 
{\cal L}\lrp \Phi,\partial_{\mu} \Phi \rrp \> \longrightarrow \> {\cal L}\lrp \Phi,\partial_{\mu} \Phi \rrp + \lrp \partial_{\mu} \alpha(x) \rrp \, \Delta \Phi(x) \, \frac{\partial {\cal L}}{\partial \lrp \partial_{\mu} \Phi \rrp } + \alpha(x) \partial_{\mu} \Lambda^{\mu} \> ,
\ee
since terms which do not contain derivatives of $ \alpha(x) $ vanish due to the assumed invariance under constant transformations or combine into the total derivative. Assume that the symmetry transformation 
is not "anomalous", i.e. that the Jacobian of the transformation is unity, then one has 
\be 
\int {\cal D} \Phi \, i \int d^4x \, \Delta {\cal L} \, e^{i S[\Phi]} \E 
\int {\cal D} \Phi \, i \int d^4x \lsp \lrp \partial_{\mu} \alpha(x) \rrp \, \Delta \Phi(x)
\, \frac{\partial {\cal L}}{\partial \lrp \partial_{\mu} \Phi \rrp } + \alpha(x) \partial_{\mu} \Lambda^{\mu} \rsp \, e^{i S[\Phi]} \E 0 \> .
\label{Pfad Feld Noether}
\ee
Perform an integration by parts in the  term containing $ \> \partial_{\mu} \alpha(x) \> $ and conclude from the arbitrariness of $ \alpha(x) $ that 
\be 
\partial_{\mu} \, \la J^{\mu}(x) \ra \E 0 \> , \quad {\rm with} \quad J^{\mu}(x) \Def C \lsp 
\frac{\partial {\cal L}}{\partial \lrp \partial_{\mu} \Phi \rrp } \, \Delta \Phi - \partial_{\mu} \Lambda^{\mu}  \rsp \> .
\label{Noether Strom}
\ee
The normalization $ C $ of this conserved current is arbitrary.
\vspace{0.1cm}

\noindent
{\bf a1)} For $ \> {\cal L}_0 = | \partial  \Phi |^2 - m^2 | \Phi |^2 \> $ one has  $ \> \Delta \Phi = i \Phi \, , \> \Delta \Phi^{\star} = - i \Phi^{\star} \>, \> \Lambda_{\mu} = 0 $ and therefore (summing over the two independent fields) 
\be 
J_{\mu}(x) \E \lrp \partial_{\mu} \Phi^{\star}(x) \rrp \, i \Phi(x) - i \Phi^{\star}(x) \lrp \partial_{\mu} \Phi(x) \rrp \> .
\ee
If $ C = 1/(2m) $ is taken, the result is in agreement with {\bf \{Itzykson-Zuber\}} eq. (2-5).
This current is obviously also conserved if there is an interaction of the form $ \> V \lrp \Phi^{\star} \, \Phi \rrp \> $.
\vspace{0.1cm}

\noindent
{\bf a2)} For $ \> {\cal L}_0 = i \bar \psi \gamma \cdot \partial \psi/2 - i \lrp \partial_{\mu} \bar \psi \rrp \gamma^{\mu} \psi/2  - m \bar \psi \psi \> $ one has
$ \Delta \Psi = i \psi \> , \> \Delta \bar \psi = - i \bar \Psi \>, \Lambda_{\mu} = 0 \> $ and thus
\be 
J_{\mu} \E i \frac{i}{2} \bar \psi \gamma_{\mu} \psi + (-i) \frac{-i}{2} \bar \psi \gamma_{\mu} \psi \E 
- \bar \psi \gamma_{\mu} \psi \> .
\ee
The same result is obtained if the equivalent Lagrangian $ \> {\cal L}_0 = \bar \psi \lrp i \gamma \cdot \partial - m \rrp \psi \> $ is used. The normalization $ C = - 1 $ gives the standard result (see, e.g.
 {\bf \{Itzykson-Zuber\}} eq. (2-13)).
\vspace{0.1cm}

\noindent
{\bf a3)$^{\star}$} For the Walecka Lagrangian the nucleonic parts $ \> \bar \Psi \lrp i \dslash - M \rrp \Psi - g_{\sigma} \bar \Psi \Psi \sigma - g_{\omega} \bar \Psi \gamma_{\mu} \Psi \omega^{\mu} \> $ do not
contain additional derivatives of the nucleon field. Hence the same  Dirac current 
$ \> \bar \Psi \gamma_{\mu} \Psi \> $ remains conserved.
\vspace{0.1cm}

\noindent
{\bf a4)} For the non-relativistic Lagrangian for particles interacting via a local 2-body potential one
can write \\
$ \> {\cal L} = \Phi^{\star}(\fx,t) i \hbar \partial_t \Phi(\fx,t) - {\cal T} - {\cal V} \> $ .
After an integration by parts in the Lagrange function the kinetic energy density reads 
 $ \> {\cal T} = \hbar^2 \lrp \nabla \Phi^{\star}(\fx,t)\rrp  \cdot  \lrp \nabla \Phi(\fx,t) \rrp/(2m) \> $, and the potential energy density is $ \> {\cal V} = - (1/2) \int d^3x' \, \Phi^{\star}(\fx,t) \Phi^{\star}(\fx',t) \,  V(\fx - \fx') \, \Phi(\fx',t) \Phi(\fx,t) \> $.
 Therefore the Lagrangian is invariant under the global phase transformation \eqref{phase transf} although it 
is non-local, i.e. does not depend only on the field $ \Phi(\fx,t) $ and its derivative as was assumed when deriving the Noether current \eqref{Noether Strom}. However, for a local potential the $x$-dependent transformation does not give a contribution and one can apply the same procedures:
 \be 
 \frac{\partial {\cal L}}{\partial (\partial_t \Phi)} \E i \hbar \Phi^{\star} \> , \quad 
 \frac{\partial {\cal L}}{\partial (\nabla\Phi)} \E - \frac{\hbar^2}{2 m} \nabla \Phi^{\star} \>,
 \quad \frac{\partial {\cal L}}{\partial (\nabla \Phi^{\star})} \E - \frac{\hbar^2}{2 m} \Phi \> .
 \ee
It then follows that 
\be 
J_0 \E -\hbar \Phi^{\star} \Phi \> , \quad J_k \E i \frac{\hbar^2}{2m} \lsp \Phi^{\star} \nabla_k \Phi 
- \lrp \nabla_k \Phi^{\star} \rrp \Phi \rsp \> , \> k = 1, 2, 3
\ee
is conserved: $ \partial J_0 + \nabla_k J_k = 0 $. If the normalization is taken as $ C = -1/\hbar $ 
one has agreement with standard quantum-mechanical results (see, e.g. {\bf \{Messiah 1\}}, 
eqs. (IV.9) and (IV.11)). 

\noindent
In the most general case of a (non-local) interaction 
$ \> V = \int d^3 x_1 d^3 x_2 d^3x_3 d^3x_4 \, \Phi^{\star}(\fx_1,t) \Phi^{\star}(\fx_2,t) \, 
V(\fx_1,\fx_2,\fx_3,\fx_4) \, \Phi(\fx_4) \Phi(\fx_3) \> $
there is a new source of contributions for a $x$-dependent transformation -- the current
is modified by a non-local interaction. This is well-known in Nuclear Physics where special non-local
(so-called Yamaguchi) potentials have been frequently used to describe the two-nucleon bound state,
the deuteron.
\vspace{0.1cm}

\noindent
{\bf b)} The shift operator is $ \> \exp(a_{\mu} \, \partial^{\mu}) \> $. Therefore the change 
of the Lagrangian for infinitesimal, constant  $ a_{\mu} $ is a total derivative
$ \> \Delta {\cal L} \Big |_a  \E a_{\mu} \, \partial^{\mu} {\cal L} \> $.
Thus expanding 
$ \> {\cal L}  \lrp \Phi + \Delta \Phi, \partial \Phi + \partial \Delta \Phi \rrp \> $ one obtains
\be
\Delta {\cal L}\Big |_{a(x)} \E 
\frac{\partial {\cal L}}{\partial (\partial_{\mu} \Phi)} \, \lrp \partial_{\mu} a_{\nu} \rrp \, 
\partial^{\nu} \Phi + a_{\mu} \, \partial^{\mu} {\cal L}\> .
\ee
The only difference to Eq. \eqref{Pfad Feld Noether} is that the infinitesimal parameter now has a Lorentz index so that after an integration by parts one obtains
\be 
0 \E \int {\cal D} \Phi \, e^{i S[\Phi]}  \int d^4x (- \Delta {\cal L} )\Big |_{a(x)}   \E \int {\cal D} \Phi \, e^{i S[\Phi]} \int d^4x \, a_{\nu}(x) \, \partial_{\mu} \Big [ \> 
\underbrace{\frac{\partial {\cal L}}{\partial \lrp \partial_{\mu} \Phi \rrp } \partial^{\nu} \Phi 
- g^{\mu \nu} {\cal L} }_{\deF T^{\mu \nu}} \> \Big ]   \> .
\ee
For a Lagrangian of the form \eqref{skalare Lagrangedichte} the energy-momentum tensor is symmetric
$ \> T^{\mu \nu} \E \partial^{\mu} \Phi \partial^{\nu} \Phi - g^{\mu \nu} {\cal L} \>$ .
In particular, one finds
$ \> T^{00} =  [ (\partial_0 \Phi)^2 + (\nabla \Phi)^2 + m^2 \Phi^2]/2 + V(\Phi) \EQ {\cal H} \> $,
i.e. the Hamilton density is a special component of the energy-momentum tensor. In contrast, the Lagrange density (Lagrangian) is a Lorentz scalar.
 
\renewcommand{\baselinestretch}{1.2}
\normalsize
\vspace{0.4cm}

\noindent
{\bf Problem \ref{Dirac nonrel}$^{\star}$:}  
\vspace{0.2cm}

\refstepcounter{ueb}    
\setcounter{equation}{0}
\renewcommand{\baselinestretch}{0.9}
\scriptsize
\noindent
{\bf a)} Since $ \> \vvslash \, \vvslash = v^2 = 1 \> $ it is clear that $ \> \Phi, \chi \> $ are eigenfunctions of $ \> \vvslash \> $ with eigenvalue $ \> +1, \, -1 \> $, respectively. 
Similarly, one also has $ \> \bar \phi/\bar \chi \, \vvslash = \pm \bar \phi/\chi \> $ from
the corrsponding definitions of  $ \bar \Phi, \bar \chi $
(note that this cannot be ``derived'' as $ \bar \Phi, \bar \chi $ are independent Grassmann-valued fields over which one has to integrate and {\bf not} 
$ \> \Phi^{\dagger} \gamma_0, \chi^{\dagger} \gamma_0 \> $!). 
Using Eq. \eqref{gamma gamma} the anticommutator becomes
\be 
[ \dddslash^{\perp}, \vvslash ]_+ \E \lrp \dddslash - v \cdot D \vvslash \rrp \vvslash + \vvslash \lrp \dddslash - v \cdot D \vvslash \rrp \E \dddslash \, \vvslash + \vvslash \, \dddslash - 2 v \cdot D \E - i \sigma_{\mu \nu} D^{\mu} v^{\nu} - i \sigma_{\mu \nu} v^{\mu} D^{\nu} \E 0
\ee
due to the antisymmetry of $ \> \sigma_{\mu \nu} \> $ 
(cf. Ref. \cite{MaRoRy}).

\vspace{0.1cm}

\noindent
{\bf b)} Substitute $ \> \Psi = \exp(-i M v \cdot x) \, ( \Phi + \chi )\> $ into Eq. \eqref{Dirac scalar+vector} and use
$ \> i \dddslash \, \exp(-i M v \cdot x) \phi/\chi = \exp(-i M v \cdot x) \, ( i \dddslash + M \vslash ) \, \phi/\chi = (i \dddslash \pm M ) \phi/\chi \> $. 
This gives $ \> S = S_{\bar \phi,\phi} + S_{\bar \chi,\chi} + S_{\bar \phi, \chi} + S_{\bar \chi,\phi} \> $. With $ \> \dddslash = \vvslash \, v \cdot D + \dddslash^{\perp} \> $ the first term is
\be 
S_{\bar \phi,\phi}  \E \lrp \bar \phi , \lsp i ( \vvslash v \cdot D + \dddslash^{\perp} ) + M - M^{\star} 
\rsp \, \phi \rrp \E \lrp \bar \phi, \lsp i v \cdot D + M - M^{\star} \rsp \, \phi \rrp
\ee
as $ \> \bar \phi \dddslash^{\perp} \phi =  \bar \phi \dddslash^{\perp} \vvslash \phi = 
- \bar \phi \vvslash \dddslash^{\perp} \phi = - \bar \phi \dddslash^{\perp} \phi \> \Longrightarrow 
\bar \phi \dddslash^{\perp} \phi = 0 \> $ . Similar one finds 
\be 
S_{\bar \chi,\chi}  \E \lrp \bar \chi , \lsp i ( \vvslash v \cdot D + \dddslash^{\perp} ) - M - M^{\star} 
\rsp \, \chi \rrp \E - \lrp \bar \chi, \lsp i v \cdot D + M + M^{\star} \rsp \, \chi \rrp \> .
\ee
The mixed terms simplify to
$ \> 
S_{\bar \chi,\phi}  \E \lrp \bar \chi , i \dddslash^{\perp} \phi \rrp \> , \> \>   
S_{\bar \phi,\chi}  \E \lrp \bar \phi , i \dddslash^{\perp} \chi \rrp
\> $
since, e.g. $ \> \bar \chi \, \lrp M \, {\rm or} M^{\star} \rrp \, \phi =  \bar \chi \, \lrp M \,  {\rm or} \, M^{\star} \rrp \, \vvslash \phi =  \bar \chi \, \vvslash \lrp M \, {\rm or} \, M^{\star} \rrp \,  \phi  = -  \bar \chi \, \lrp M \, {\rm or} \, M^{\star} \rrp \, \phi 
\> \Longrightarrow \quad  \bar \chi \, \lrp M \, {\rm or} \, M^{\star} \rrp \, \phi = 0 \> $. Alltogether this gives Eq. \eqref{S phi chi}.
One may integrate out the "small" component $ \> \chi \> $ by using the extended Gaussian Grassmann integral \eqref{Gauss ferm/boson}. This leads to
\be 
S_{\rm eff}[\bar \Phi, \phi] \E \int d^4x \> \bar \Phi \, \lsp i v \cdot D + M - M^{\star} - \dddslash^{\perp} \, \frac{1}{i v \cdot D + M + M^{\star}} \, \dddslash^{\perp} \rsp \, \Phi - i
{\rm tr} \ln \lsp i v \cdot D + M + M^{\star} \rsp \> ,
\ee 
where the last term from the determinant is a constant for external potentials which drops out 
when calculating Green functions as ratios of path integrals.

Now assume that the mass is large compared to the potentials and the derivatives and that one can exand
\be 
S_{\rm eff}[\bar \Phi, \phi] \E \int d^4x \> \bar \Phi \, \lsp i v \cdot \partial - v \cdot A(x) - U_S(x) - \frac{\dddslash_{\perp}^2}{2 M} + \frac{1}{4 M^2} \dddslash^{\perp}  \lrp i v \cdot D + U_S(x) - v \cdot A(x) \rrp  \, \dddslash^{\perp} + {\cal O} \lrp 1/M^3 \rrp \rsp \, \Phi \> .
\label{S eff expanded}
\ee
\vspace{0.1cm}

\noindent
{\bf c)} For $ \> v_{\mu}  = (1,{\bf 0}) \> $ and $ \> A_{\mu}(x) = (U_V(\fx),{\bf 0}) \> $ 
one has $ \, \dddslash_{\perp} = - \vec{\gamma} \cdot \nabla \quad  \Longrightarrow \quad 
\dddslash_{\perp}^2 = - \Delta \, $ and Eq. \eqref{S eff expanded} simplifies to
\be 
S_{\rm eff}[\bar \Phi, \Phi] \E \int d^4x \> \bar \Phi \, \hat L \, \Phi \quad {\rm with} \quad
\hat L \E i\partial_t - U_+
+ \frac{\Delta}{2 M}  + \frac{1}{4 M^2} \lrp  i \partial_t \, \Delta + \vec{\sigma} \cdot \nabla
 U_- \, \vec{\sigma} \cdot \nabla \rrp
\ee
where the explicit form \eqref{gamma explizit} of the Dirac matrices and the abbreviation 
$ U_{\pm} \Def U_S \pm U_V \, $ has been used. In the $ {\cal O}(1/M^2) $-
term there is an additional time-derivative which can be removed by a rescaling of the fields:
$ \Phi =: (1 - \Delta/(4 M^2 )^{-1/2} \varphi \> , \quad \bar \Phi =: \bar \varphi (1 - \Delta/(4 M^2 )^{-1/2} \> $ (the Jacobian is an irrelevant constant) so that one can read off the 
equivalent Hamiltonian $ \hat L =: i \partial_t - \hat H \> $ as
\be
\hat H \E - \frac{\Delta}{2 M} + U_+ + \frac{1}{8 M^2} \lrp \Delta U_+ + U_+ \Delta \rrp +
\frac{1}{4 M^2} \vec{\sigma} \cdot \nabla U_- \, \vec{\sigma} \cdot \nabla + {\cal O} (1/M^3) 
\> .
\ee
acting between $ \bar \varphi $ and $ \varphi $.
The 3rd term arises from the transformation $ \Phi \longrightarrow \varphi $ and explicitly reads
$ \> [ \, (\Delta U_+) + 2 (\nabla U_+) \cdot \nabla + 2 U_+ \Delta \, ]/(8 M^2) \> $
when taking into account that the derivatives also act on the field. Similar, the last term
has the form \\
$ \> [ \, (\nabla U_-) \, \cdot \nabla + i \vec{\sigma} \cdot ((\nabla U_-) \times \nabla )\, ]/(4 M^2) \> $
and contains the spin-orbit interaction. Alltogether one finds
\be
\hat H \E - \frac{\Delta}{2M} + \frac{1}{2 M^2} \lsp  
 U_S \Delta + (\nabla U_S) \cdot \nabla \rsp + U_+ + \frac{1}{8 M^2} (\Delta U_+) + V_{LS}
\label{H 1/M^2}
\ee
where for  spherically symmetric potentials  
\be 
V_{LS} \E  \frac{1}{4 M^2} \, \frac{1}{r} \frac{\partial}{\partial r} \lsp U_V(r) - U_S(r) \rsp
\, \vec{\sigma} \cdot \lrp \fr \times \frac{\nabla}{i} \rrp  \> .
\ee
The first three terms in Eq. \eqref{H 1/M^2} can be concisely written 
 as $ - \nabla (M + U_S)^{-1} \cdot \nabla /2  \> $ (up to $ {\cal O}(1/M^3) $ ),
i.e. as kinetic energy with an effective, position-dependent mass so that one finally obtains
\be
\hat H \E - \frac{1}{2} \nabla \frac{1}{M^{\star}(r)} \cdot \nabla + 
\underbrace{U_S(r) + U_V(r)}_{=: V_c(r)}
+ \frac{1}{8 M^2} \Delta \lsp U_S(r) + U_V(r) \rsp + V_{LS} + {\cal O} \lrp 1/M^3 \rrp
\ee
in agreement with {\bf \{Bjorken-Drell 1\}}, eqs. (4.5),(4.6) (no scalar potential there, 
no vector potential ${\bf A}$ here!).
The term $ \Delta U_+/(8 M^2) $ is the so-called Darwin term. 
Since $ U_V $ arises from the exchange of vector particles it is repulsive (i.e. positive) 
whereas the scalar potential is attractive (i.e. negative). Thus these potentials mostly cancel 
in the central potential but add in magnitude in the spin-orbit potential.

\renewcommand{\baselinestretch}{1.2}
\normalsize
\newpage

\noindent
{\bf Problem \ref{x Prop}:}  
\vspace{0.2cm}

\refstepcounter{ueb}    
\setcounter{equation}{0}
\renewcommand{\baselinestretch}{0.9}
\scriptsize
\noindent
{\bf a)} By Fourier transformation and Schwinger's trick one has
\bea
\Delta_F(x-y) \EA \int \frac{d^dk}{(2 \pi)^d} \, e^{i k \cdot (x-y)} \, \frac{1}{k^2-m^2 + i 0^+}
\E -i \int_0^{\infty}dT \, e^{-i m^2 T} \, \int \frac{d^dk}{(2 \pi)^d} \, \exp \lsp i k^2 T + i k \cdot (x-y)\rsp \non
\EA \frac{1}{(4 \pi i)^{d/2}} \,  \int_0^{\infty}dT \, \frac{1}{T^{d/2}}
\exp \lrp -i m^2 T - i \frac{(x-y)^2}{4 T}\rrp \> .
\eea
Here the Gaussian integral in Minkowski space \eqref{Gauss Minkowski} has been used with $ a = T $ and
 $ b_{\mu} = (x-y)_{\mu} $.
The remaining $T$-integral can be expressed by  a modified Bessel function of second kind 
(see {\bf \{Gradsteyn-Ryzhik\}}, eq. 8.432.6) so that
\be 
\Delta_F(x-y) \E \frac{-i}{(2 \pi)^{d/2}} \, \lrp \frac{m^2}{z}\rrp^{d/2-1} \, K_{d/2-1}(z) 
\> , \quad z \Def \sqrt{-m^2 (x-y)^2} \> .
\ee
Note that the argument of this Bessel function is real for space-like separation and purely imaginary for time-like $ r = x - y $ . For small arguments one has (see {\bf \{Handbook\}}, eq. 9.6.9 and 9.6.8)
 $ \> K_{\nu}(z) \to \frac{1}{2} \Gamma(\nu) \lrp \frac{1}{2} z \rrp^{-\nu} \> $ valid for Re $ \nu > 0 $ and $ \> K_0(z) \to - \ln z \> $. Thus 
 \be 
 \Delta_F(r) \> \stackrel{r \to 0}{\longrightarrow} \>  \left \{ \begin{array}{r@{\quad{\rm for}\quad}l} 
  \frac{-i}{4 \pi^{d/2} }\, \Gamma(d/2-1) \, (- r^2 )^{1-d/2} \quad & d \neq 2 \\
  \frac{i}{4 \pi} \, \ln \lrp -m^2 r^2 \rrp & d = 2 \> \> .        
                 \end{array} \right. 
\label{prop r to 0}
\ee
Notice that the small-distance behaviour is independent of the mass of the particle (for $ d = 2 $ 
the mass term is just a constant) which is to be expected as this corresponds to large momenta $ |k^2| \gg m^2 $.
The dependence on (spacelike) $r$ illustrates the frequent statement that the Coulomb potential -- as solution of the Poisson equation with a point-like source, i.e. as inverse Laplacian -- is linear in one dimension, logarithmic in two, of $1/r$-type in three dimensions etc.
\vspace{0.1cm}

\noindent
For large distances one can use (see {\bf \{Handbook\}}, eq. 9.7.2) the asymptotic expansion $ \> K_{\nu}(z) \to \sqrt{\pi/(2z)} \exp(-z) \> $ to obtain an oscillatory (time-like $r$ ) or exponentially
decaying (spacelike $ r $ ) behaviour of the free Feynman propagator.
\vspace{0.1cm}

\noindent
{\bf b)} For the calculation of the space-time behaviour of the photon propagator one proceeds in a similar way: Zero mass simplifies the algebra but for the gauge-dependent term in the propagator one has to use an extended Schwinger representation
$ \> 1/(a+i0^+)^2 = - \int_0^{\infty} dT \, T \, \exp(i a T) \> $. Then
\be 
\Delta_{\mu \nu}(r) \E  \int_0^{\infty} dT \, \int \frac{d^dk}{(2 \pi)^d} \, \lsp i g_{\mu \nu} + (1 - \lambda) T \, \frac{\partial^2}{\partial r^{\mu} \partial r^{\nu}} \rsp \, e^{i k^2 T +  i k \cdot r}
\> ,
\ee
where the terms $ \> k_{\mu} k_{\nu} \> $ have been written as derivatives w.r.t. $ r^{\mu} $ and $ r^{\nu} $ . The $k$-integration can then be performed as before and the final $T$-integration is even
simpler: Substituting $ \tau = 1/T $ one only encounters integrals of the type
 $ \> \int_0^{\infty} d\tau \, \tau^{\nu -1} e^{-a \tau} = \Gamma(\nu)/a^{\nu} \> $ in the massless case. After some algebra one obtains
\bea 
\Delta_{\mu \nu}(r) \EA \frac{i}{(4 \pi i)^{d/2}} \, \lsp i g_{\mu \nu} \, \Gamma \lrp \frac{d}{2}-1 \rrp \, 
\lrp \frac{4}{i r^2} \rrp^{d/2-1} + (1-\lambda) \, \Gamma\lrp \frac{d}{2} - 2 \rrp
\frac{\partial^2}{\partial r^{\mu} \partial r^{\nu}} \lrp \frac{4}{i r^2} \rrp^{d/2-2} \rsp \non
\EA \frac{i^{3-d}}{4 \pi^{d/2}} \, \Gamma \lrp \frac{d}{2} - 1 \rrp \, \lrp \frac{1}{r^2} \rrp^{d/2-1} \, \lsp \frac{1+\lambda}{2} g_{\mu \nu} + (1-\lambda)  \lrp \frac{d}{2} - 1 \rrp \, 
\frac{r_{\mu} r_{\nu}}{r^2} \rsp \> .
\label{photon prop x}
\eea
\renewcommand{\baselinestretch}{1.2}
\normalsize
\vspace{0.4cm}

\noindent
{\bf Problem \ref{Bloch-Nordsieck}$^{\star}$:}  
\vspace{0.2cm}

\refstepcounter{ueb}    
\setcounter{equation}{0}
\renewcommand{\baselinestretch}{0.9}
\scriptsize
\noindent
Integrating out the electron fields in the full generating functional of QED with no external photons
($ J = 0 $) one gets
\be
Z[\cy{\bar \eta}, \cy{\eta}] \E \int {\cal D} A \, {\cal D} \cy{\bar \psi} \, 
{\cal D}\cy{\psi} \> \exp \lcp i S_0[A] + i \lrp \cy{\bar \psi}, {\cal O}_A \cy{\psi} \rrp + 
i (\cy{\bar \psi}, \cy{\eta}) + i (\cy{\bar \eta}, \cy{\psi}) \rcp \E  \int {\cal D} A \, 
\fdet {\cal O}_A \, \exp \lcp i S_0[A] - i \lrp \cy{\bar \eta}, {\cal O}_A^{-1} \cy{\eta} \rrp \rcp \> 
\ee
with $ \> S_0[A] = \lrp A^{\mu}, \Delta_{\mu \nu}^{-1} A^{\nu} \rrp \> $ and 
$ \> {\cal O}_A =  \pslash - e \Aslash - m \>  $.
The 2-point function (the electron propagator) therefore is
\be 
\bar G_2(p) \E
\int d^4(x-y)  \, 
e^{-i p \cdot (x-y) } \int {\cal D} A \, e^{i S_0[A] + tr \ln {\cal O}_A} \, \int_0^{\infty} dT \, 
 \la y \lvl e^{i {\cal O}_A T} \rvl x \ra \> ,
\label{G2 BN1}
\ee
when the Fock-Schwinger representation is being used.

The Bloch-Nordsieck approximation consists in neglecting the trace terms (vacuum polarization) and to
replace the Dirac matrices $ \> \gamma_{\mu} \> \longrightarrow \> v_{\mu} = p_{\mu}/ m_{\rm phys} \> $, i.e. by a constant velocity made up by the external momentum $ \> p_{\mu} \> $. Then
\be 
 e^{i {\cal O}_A T} \> \simeq \> \exp \lsp i \lrp i v \cdot \partial - e v \cdot A(x) - m \rrp T \rsp 
\deF e^{-i m T} \exp ( - T v \cdot \partial ) \, F(T)
\ee
and by solving the differential equation $ \> \partial F/\partial T = - ie v \cdot A(x+vT) \, F \> $
one finds an eikonal-like expression
\be
 \la y \lvl e^{i {\cal O}_A^{\rm BN} T} \rvl x \ra \E e^{-i m T} \, \delta(y-x - v T) \, 
\exp \lsp -i e \int_0^T d\tau \, v \cdot A(x + v \tau) \rsp\> .
\ee
Inserting that into Eq. \eqref{G2 BN1}  one sees that the functional integral over the photon field 
is now a Gaussian  and can be performed. In addition, due to the $\delta$-function 
the Fourier transformation can be done as well.
Some algebra then gives
\be 
\bar G_2^{\rm BN}(p) \E \int_0^{\infty} \! dT \> e^{i ( p \cdot v - m + i0^+) T} \, e^{X(T)}\> , \quad
X(T) \E -\frac{i e^2}{2} \int_0^T d\tau_1 \, d\tau_2  \, v^{\mu} \, v^{\nu} \, \Delta_{\mu \nu}
\lrp v (\tau_1 - \tau_2) \rrp  \> ,
\label{G2 BN}
\ee
where $ \> \Delta_{\mu \nu}(r) \> $ is the free photon propagator in $x$-space studied in {\bf Problem 
\ref{x Prop}}. As the integrand only depends on $ \tau^2 = (\tau_1 - \tau_2)^2 \> $ one can replace
$ \int_0^T d\tau_1 \, d\tau_2 \, \longrightarrow \, 2 \int_0^T d\tau\,  (T- \tau) \> $.
Using Eq. \eqref{photon prop x}, $ \> v^2 = 1 \> $ and replacing in $d$ dimensions $ \> e^2 \longrightarrow e^2 \mu_0^{d-4} \> $  one obtains
\be 
X(T) \E -\mu_0^{4-d} e^2 \, \frac{i^{4-d}}{4 \pi^{d/2}} \Gamma \lrp \frac{d}{2} -1 \rrp \, 
\lsp \frac{1+\lambda}{2} + (1-\lambda)  \lrp \frac{d}{2} -1 \rrp \rsp \, \int_0^T d\tau \, 
( T - \tau ) \tau^{2-d} \> .
\label{X(T) massless}
\ee
In dimensional regularization it is assumed that $d$ is such that the integrals converge followed by an analytic continuation to $  \> d = 4 - 2 \epsilon \> $. Thus in the remaining $\tau$-integral there is no contribution (or divergence) at the lower limit and one finds
\be
X(T) \E -\frac{e^2}{16  \pi^2}\lrp i \sqrt{\pi} \mu_0 T \rrp^{2 \epsilon} \, 
\frac{\Gamma(-\epsilon)}{1 - 2 \epsilon} \, \Bigg [ 3 - \lambda - 2 (1-\lambda) \epsilon \Bigg ]
\E  \frac{\kappa}{2 \epsilon} + \kappa \, \ln (\mu_0 T) + C + {\cal O}(\epsilon) \>, 
\ee
where
\be 
\kappa \Def \frac{e^2}{8 \pi^2} (3 - \lambda) \> , \quad 
C \Def  \frac{e^2}{16 \pi^2} \lsp 4 + (3-\lambda) \lrp \gamma_E + \ln (\pi) + i \pi \rrp \rsp \> .
\ee
The proper-time integration can now be performed by means of Euler's Gamma-function integral \eqref{Euler Gamma} so that one has
\be 
\bar G_2^{\rm BN}(p) \E \frac{\Gamma(1+\kappa)}{\mu_0} \, \lrp i \mu_0 m_{\rm phys} \rrp^{1+\kappa} \, 
\frac{\exp \lsp \kappa/(2 \epsilon)+ C \rsp}{(p^2 - m \, m_{\rm phys} + i 0^+)^{1+ \kappa}} \> .
\ee
One sees that the pole of the free propagator has turned into a branch cut unless the gauge fixing parameter
is chosen as $ \, \lambda = 3  $ (Yennie gauge). There is no mass renormalization: Requiring the singularity to happen at $ p^2 = m^2_{\rm phys} $ implies $ m = m_{\rm phys} $. For the wave function renormalization constant
one finds 
\be 
Z_2 \E \lrp \frac{\mu_0}{m} \rrp^{\kappa} \, \Gamma(1+\kappa) \, \exp \lsp 
\frac{\kappa}{2} \lrp \frac{1}{\epsilon} + \gamma_E + \ln \pi + 2 i \pi \rrp 
+ \frac{e^2}{4 \pi^2} \rsp \> ,
\ee 
an expression with contains all divergences for $ \epsilon \to 0 $ and becomes unity with no interaction 
($ e^2 = 0) $.\\
\noindent
The singularity of the electron propagator at $ p^2 = m^2_{\rm phys} $ (pole or branch point) is determined by the large-$T$ behaviour
of the proper-time integrand in Eq. \eqref{G2 BN}, i.e. of the quantity $ \, X(T) \, $. Disregarding irrelevant Lorentz structures this behaviour
can be traced back to the short- or long-distance limits of the scalar Feynman propagator, viz. the corresponding limits of the modified Bessel functions:
For massless  ($ \mu = 0 $) photons the {\it logarithmic} behaviour of $ X(T) $ is caused by the $ \mu \tau \to 0 $-limit for the Feynman propagator 
as given in the first line of Eq. \eqref{prop r to 0}. This then leads to the $ \> (T-\tau) \tau^{-2 + 2 \epsilon} \> $
behaviour of the integrand in Eq. \eqref{X(T) massless}, i.e. after integration and the limit $ \, \epsilon \to 0 \, $ to the logarithmic 
growth of $ X(T) $ and thereby to the branch-type singularity of the electron propagator. In contrast, for $ \mu \neq 0 $ the $ \mu \tau \to \infty$-limit of the Feynman propagator matters at large $ T $ and $ \tau $; the integrand of $ X(T) $ then behaves like 
$  \> \sqrt{\mu} \,  (T - \tau) \tau^{-3/2} \exp(-i \mu \tau) \> $ making the integral convergent 
for $ T \to \infty $. This leads to $ \> X(T) \stackrel{T \to \infty}{\longrightarrow} \, C_1 T + C_0 $. Thus for massive photons the electron propagator 
would still display a pole as
the linear term in $ T $ just gives a mass renormalization whereas the constant $ C_0 $ contributes to the wave function
renormalization constant.

\renewcommand{\baselinestretch}{1.2}
\normalsize
\vspace{0.4cm}

\noindent
{\bf Problem \ref{BRST}$^{\star}$:}  
\vspace{0.2cm}

\refstepcounter{ueb}    
\setcounter{equation}{0}
\renewcommand{\baselinestretch}{0.9}
\scriptsize
\noindent
Compare the infinitesimal gauge transformation for the fermion field in Eq. \eqref{Eich nichtabelsch lokal}
$ \> \delta \psi(x)  = - i g \Theta^a(x) T^a \, \psi(x) \, $ with the BRST transformation 
$ \> \delta \psi(x) = -i g \omega \chi^a(x) T^a \, \psi(x) \> $: This allows to identify the
parameter of the gauge transformation as $ \> \Theta^a(x) \EQ - \omega \chi^a(x) \> $. That value 
inserted into the gauge transformation \eqref{Eich A nichtabelsch} for the gauge field also reproduces the BRST transformation of the field:
\be 
\delta A_{\mu}^a \E \partial_{\mu} \lsp \omega \chi^a(x) \rsp - g f^{abc} \omega  \chi^b \, A_{\mu}^c
\E \omega \big [ \, \underbrace{\partial_{\mu} \delta^{ab} + g f^{acb} A_{\mu}^c}_{\EQ D_{\mu}^{ab}} \,
\big ] \, \chi^b \> .
\ee
Here $ \> D_{\mu}^{ab} (x) \> $ exactly is the covariant derivative \eqref{kov ableit adjung} in the adjoint representation under which the gauge fields transform. As by construction the fermionic  and gauge-field
part of the Lagrangian are invariant under a local gauge transformation, one therefore also 
has invariance of $ \> {\cal L}_f + {\cal L}_g \> $ under BRST transformations.
\vspace{0.1cm}

Consequently one only has to prove the invariance of the gauge-fixing Lagrangian 
$ {\cal L}_{\rm gauge} $ and of the Faddeev-Popov part $ {\cal L}_{FP} $:
\be
\delta  \lrp {\cal L}_{\rm gauge} +  {\cal L}_{FP} \rrp = \delta \lsp -B^a \partial^{\mu} A_{\mu}^a
- \lrp \partial^{\mu} \bar \chi^a \rrp \, \lrp D_{\mu} \chi \rrp^a \rsp =
 -B^a \omega \partial^{\mu} \lrp D_{\mu} \chi \rrp^a - \omega \lrp \partial^{\mu} B^a \rrp 
\lrp D_{\mu} \chi \rrp^a + \lrp \partial^{\mu} \bar \chi^a \rrp \, \delta \lrp (D_{\mu} \chi)^a \rrp 
\> .
\ee
The first two terms cancel (after an intgration by parts in the action) and thus one has to evaluate
\bea 
\delta \lrp (D_{\mu} \chi)^a \rrp \EA D_{\mu}^{ab} \delta \chi^b + D_{\mu}^{ab} \delta \chi^b + g f^{acb} \delta A_{\mu}^c \chi^b \E \lsp \partial_{\mu} \delta^{ab} 
+ g f^{acb} A_{\mu}^c \rsp 
\, \lsp - \frac{g}{2} \omega f^{bed} \chi^e \chi^d \rsp \non
&& + g f^{abc} \omega \lsp \lrp \partial_{\mu} \delta^{cd} + g f^{ced} \A_{\mu}^e \rrp \chi^d \rsp \chi^b  \deF \frac{g}{2} \omega  X_1^a + \frac{g^2}{2} \omega X_2^a \> .
\eea
The terms $ \, {\cal O} (g) \, $
\be 
X_1^a \E - f^{acd} \partial_{\mu} \lrp \chi^c \chi^d \rrp + 2 f^{acb} \lrp \partial_{\mu} \chi^c \rrp \chi^b \E 0 
\ee
cancel if product rule and anticommutativity of the ghost fields $ \chi $ are used. 
The terms  $ \, {\cal O} (g^2) \, $ read
\be 
X_2^a \E -f^{acb} f^{bed} \, \chi^e \chi^d A_{\mu}^c + 2 f^{abc} f^{ced} A_{mu}^e \chi^d \chi^b \> .
\ee
By carefully renaming indices one finds
\be 
 X_2^a \E \lsp 2 f^{acd} f^{deb} + f^{aed} f^{dbc} \rsp \,  \chi^b \chi^c A_{\mu}^e \E 
 \lsp - f^{ade} f^{bcd} - 2 f^{bde} f^{cad} \rsp \, \chi^b \chi^c A_{\mu}^e
\ee
where the antisymmetry properties of the structure constants again has been used. Finally in the last term one can exchange $ b \leftrightarrow c $ and use  $ \chi^c \chi^b = -\chi^b \chi^c $. Adding it to the unchanged term and dividing by two gives
\be 
 X_2^a\E \lsp - f^{ade} f^{bcd} -  f^{bde} f^{cad} - f^{cde} f^{abd} \rsp \, \chi^b \chi^c A_{\mu}^e \EQ 0 
 \ee
due to the Jacobi identity \eqref{Jacobi ident}.
\renewcommand{\baselinestretch}{1.2}
\normalsize
\vspace{0.4cm}

\noindent
{\bf Problem \ref{Gitter 2-Punkt}$^{\star}$:}  
\vspace{0.2cm}

\refstepcounter{ueb}    
\setcounter{equation}{0}
\renewcommand{\baselinestretch}{0.9}
\scriptsize
\noindent
{\bf a)} The free lattice action may be written as (the factor $a^2$ is needed to make the action dimensionless)
\be 
S_E^{[0)} \E  
 \deF  \frac{a^2}{2} \sum_{l,l'} \Phi_l \, K_{l l'} \, \Phi_{l'} \> , \qquad {\rm with} \quad
K_{l l'} \E - \sum_{\mu} \lrp \delta_{l',l+\mu} + \delta_{l',l-\mu} \rrp + \delta_{l l'} \lrp 8 + m^2 a^2 \rrp \>. 
\label{K ll'}
\ee
(Note $ \> l = (l_1,l_2,l_3,l_4) \> $ is an
euclidean 4-vector made of integer numbers and $ \mu $ stands for a similar unit vector 
in direction $ \mu $).
Thus the free partition function is given by
\be 
Z_0[J] \E \prod_k d\Phi_k \, \exp \lsp - S_E^{(0)} + \sum_l J_l \Phi_l \rsp  \E {\rm const.} \, 
\exp \lsp \frac{1}{2 a^2} \sum_{l, l'} J_l K^{-1}_{l l'} J_{l'} \rsp
\ee
and the free correlation function by
\be 
\la \Phi_n \Phi_{n'} \ra_0 \E \frac{\partial^2 \, \ln Z_0(J)}{\partial J_n \partial J_{n'}} \E \frac{1}{a^2} \, K^{-1}_{n n'} \> .
\ee
As in the continuum case the inverse is calculated by Fourier transform methods which have to be slightly modified as $x$-space is now discrete:
The inversion of the Fourier transform of the propagator  is achieved by multiplying 
both sides of  Eq. \eqref{lattice FT} by $ \>  \exp( - i p \cdot j a ) \> $ and summing over 
$ j = l - l' $. Use of Poisson's summation formula ( see, e.g. {\bf \{Lighthill\}}, ch. 5.4)
for the 4-dimensional case
\be 
\sum_{j=-\infty}^{+\infty} e^{\pm i (k-p) \cdot j a} \E \lrp \frac{2 \pi}{a} \rrp^4 \, \sum_{j=-\infty}^{+\infty} \delta^{(4)} \lrp k - p \pm \frac{2 \pi j}{a} \rrp
\ee
gives 
\be 
\sum_j K_j \, e^{-i p \cdot j a} \E \int_{-\pi/a}^{+\pi/a} d^4k \, \tilde K(k) \sum_j \delta^{(4)} \lrp k - p - \frac{2 \pi j}{a} \rrp \E \tilde K(p) \> ,
\ee
since only the term $ j = 0 $ contributes in the sum on the r.h.s.  if $ \> |p| < \pi/a \> $
which is the case one is interested in Eq. \eqref{lattice FT}. Inserting Eq. 
\eqref{K ll'} on the l.h.s gives $ \> 
\tilde K(p) \E 
8 + m^2 a^2 - \sum_{\mu} \lsp \exp \lrp i p_{\mu} a  \rrp  + \exp \lrp -i p_{\mu} a \rrp \rsp 
\> $
and thus one obtains Eq. \eqref{Gitter 2-Punkt-Funktion}.
\vspace{0.1cm}

\noindent
{\bf b)} Here one writes (the Dirac fields $\psi$ and $\bar \psi$ have dimension length$^{-3/2} $ )
\be 
S_E^{(0)} [\bar \psi, \psi] \E a^3 \sum_{l,l',\alpha,\beta} \bar \psi_{\alpha}(l) \, K_{\alpha \beta}(l,l') \, \psi_{\beta}(l') 
\ee
with 
\be
K_{\alpha \beta} (l,l') \E \frac{1}{2} \sum_{\mu} \lrp \gamma_{\mu}^E \rrp_{\alpha \beta} \, 
\lsp \delta_{l,l'+\mu} - \delta_{l,l'-\mu} \rsp + m a \, \delta_{l l'} \,\delta_{\alpha \beta} \> .
\ee
The free generating functional is easily calculated by performing the Grassmann Gaussian integral and 
the free 2-point function is obtained as
\be 
\la \psi_{\alpha}(n) \bar \psi_{\beta}(n') \ra_0 \E \frac{1}{a^3} \, \lrp K^{-1} \rrp_{\alpha \beta}(n,n')
\> .
\ee
As in the scalar case the inverse of the kernel $ K $  can be calculated in Fourier space and one gets
$ \> \tilde K^{-1}(p) = 1/\tilde K(p) \> $ with
\be 
\tilde K_{\alpha \beta}(p) \E \sum_j K_{\alpha \beta}(j) \, e^{-i p \cdot j a} \E 
   m  a \, \delta_{\alpha \beta} + \frac{1}{2} \sum_{\mu}
\lrp \gamma_{\mu}^E \rrp_{\alpha \beta} \, \lsp \exp \lrp i p_{\mu} a  \rrp  - \exp \lrp -i p_{\mu} a \rrp \rsp \> .
\ee
Similar as in the continuum case,
the inverse in Dirac matrix space can be found by multiplying both the denominator and numerator by 
$ \> \tilde K^{\dagger}(p) = (m a - i \sum_{\mu} \gamma_{\mu}^E \sin (p_{\mu} a ) \> $. The denominator then becomes diagonal
\be 
\tilde K^{\dagger}(p) \, \tilde K(p)  \E 
m^2 a^2 + \sum_{\mu, \nu} \gamma_{\nu}^E \gamma_{\nu}^E\, \sin (p_{\mu} a) \sin (p_{\nu} a) \E 
m^2 a^2 + \sum_{\mu, \nu} \frac{1}{2} \lsp \gamma_{\nu}^E \gamma_{\nu}^E + \gamma_{\nu}^E \gamma_{\mu}^E \rsp \, \sin (p_{\mu} a) \sin (p_{\nu} a) \E m^2 a^2 + \sum_{\mu}  \, \sin^2 (p_{\mu} a) 
\ee
leading to 
\be 
\lrp \tilde K^{-1} \rrp_{\alpha \beta}(p)  \E \frac{\tilde K^{\dagger}_{\alpha \beta}(p) } {
[\tilde K^{\dagger}(p) \tilde K(p)]}
\ee
i.e. Eq. \eqref{Gitter 2-Punkt fermion}.

\edes


\begin{thebibliography}{99}

\bibitem{Styer} D. F. Styer, M. S. Balkin, K. M. Becker, M. R. Burns, C. E. Dudley, 
   S. T. Forth, J. S. Gaumer, M. A. Kramer, D. C. Oertel, L. H. Park, M. T. Rinkoski, 
   C. T. Smith, and T. D. Wotherspoon:
   ``Nine formulations of Quantum Mechanics'', Am. J. Phys. {\bf 70} (2002), 288. 

\bibitem{Feyn1} R. P. Feynman: ``Space-time approach to
non-relativistic Quantum Mechanics'', Rev. Mod. Phys. {\bf 20} (1948), 26.

\bibitem{Dirac} P. A. M. Dirac: ``The Lagrangian in 
Quantum Mechanics'', Phys. Zeit. d. Sowjetunion, Band 3, Heft 1 (1933). 

\bibitem{Schwing} J. Schwinger (ed.): ``Selected papers on Quantum Electrodynamics'', 
Dover (1958). 

\bibitem{GiZa} T. L. Gill and W. W. Zachary: ``Constructive representation theory for the Feynman 
   operator calculus'', arXiv: math-ph/0701039.

\bibitem{Klau} I. Daubechies and J. R. Klauder: ``Quantum mechanical path integrals
   with Wiener measures for all polynomial Hamiltonians'', J. Math. Phys. {\bf 26} (1985), 2239.

\bibitem{Horv} P.~A.~Horvathy: ``The Maslov correction in the semiclassical 
     Feynman integral,'' Central Eur. J. Phys. {\bf 9} (2011), 1 [arXiv:quant-ph/0702236].

\bibitem{YLUGP} K. Yeon, K. K. Lee, C. I. Um, T. F. George and L. N. Pandey: ``Exact quantum 
theory of a 
    time-dependent bound quadratic Hamiltonian systen'', Phys. Rev. A {\bf 48} (1993), 2716.

\bibitem{GelYag} I. M. Gel'fand and A. M. Yaglom: ``Integration in functional spaces 
      and its applications in Quantum Physics'', J. Math. Phys. {\bf 1} (1960), 48 .

\bibitem{DuKi} G. V. Dunne and K. Kirsten: ``Functional determinants 
     for radial operators'', J. Phys. A {\bf 39} (2006), 11915 [arXiv: hep-th/0607066]. 
 
\bibitem{DuKl} I. H. Duru and H. Kleinert: ``Solution of path integral for H atom'', 
      Phys. Lett. B {\bf 84} (1979), 30.

\bibitem{Ros1} R.~Rosenfelder: ``Path integrals for potential scattering'',
    Phys. Rev. A {\bf 79} (2009), 012701 
    [arXiv:0806.3217 [nucl-th]].

\bibitem{CaRo} J.~Carron and R.~Rosenfelder: ``A new path-integral representation of the 
    $T$-matrix in potential scattering'', Phys. Lett. A {\bf 375} (2011), 3781
    [arXiv:1107.3034 [nucl-th]].

\bibitem{Rosen} R.~Rosenfelder: "Scattering Theory with Path Integrals", J. Math. Phys. 
     {\bf 55} (2014), 032106 [arXiv:1302.3419[nucl-th]].

\bibitem{AI} H. D. I. Abarbanel and C. Itzykson: ``Relativistic eikonal expansion'',  
      Phys. Rev. Lett. {\bf 23} (1969), 53.

\bibitem{Wal} S. J. Wallace: ``	Eikonal expansion'', Ann. Phys. {\bf 78} (1973), 190. 

\bibitem{Sar} S. Sarkar: ``Higher-order terms in the eikonal expansion of the 
    T matrix for potential scattering'', Phys. Rev. D {\bf 21} (1980), 3437. 

\bibitem{GGSW} W. Gl\"ockle, J. Golak, R. Skibi\'{n}ski and H. Wita{\l}la: ``Exact three-dimensional
   wave function and the on-shell t matrix for the sharply cut-off Coulomb potential: Failure of the
   standard renormalization factor'', Phys. Rev. C {\bf 79} (2009), 044003 
   [arXiv:0903.0343 [nucl-th]].

\bibitem{MacK} R. MacKenzie: "Path integral methods and applications", arXiv: quant-ph/0004090.

\bibitem{Ros1a} R.~Rosenfelder: "Quasielastic electron scattering from nuclei", Ann. Phys.
    (N.Y.) {\bf 128} (1980), 188;
    erratum: {\it ibid.} {\bf 140} (1982), 203.

\bibitem{SSBa} G. Scher, M. Smith and M. Baranger: ``Numerical calculations in elementary 
      quantum mechanics using Feynman path integrals'', Ann. Phys. {\bf 130} (1980), 290.

\bibitem{HsCh} Ch.-Sh. Hsue and J. L. Chern: ``Two-step approach to one-dimensional 
      anharmonic oscillators'', Phys. Rev. D {\bf 29} (1984), 643.
   
\bibitem{Turb} M. A. Escobar-Ruiz, E. Shuryak and A. V. Turbiner: "Three-loop correction to the
       instanton density. I. The quartic double well potential", Phys. Rev. D {\bf 92} (2015), 025046 
       [arXiv:1501.03993]

\bibitem{Feyn2} R. P. Feynman: ``Slow electrons in a polar crystal'', 
     Phys. Rev. {\bf 97} (1955), 660.

\bibitem{HoMu} G. H\"ohler and A. M\"ullensiefen: ``St\"orungstheoretische Berechnung der 
      Selbstenergie and der Masse des Polarons'', Z. Phys. {\bf 157} (1959), 159.

\bibitem{Smon} M.~A. Smondyrev: ``Diagrams in the polaron model'', Theor. Math. Phys. 
      {\bf 68} (1986), 653.

\bibitem{Ros3} R.~Rosenfelder: ``Perturbation theory without diagrams: The 
    polaron case'', Phys. Rev. E {\bf 79} (2009), 016705
    [arXiv:0805.4525 [hep-th]].

\bibitem{RoSch} R.~Rosenfelder and A. W. Schreiber: ``Polaron variational 
    methods in the particle representation of field theory: I. General 
    formalism'', Phys. Rev. D {\bf 53} (1996), 3337 
    [arXiv:nucl-th/9504002].

\bibitem{AlRo} C. Alexandrou and R.~Rosenfelder : ``Stochastic solution to highly nonlocal 
    actions: The polaron problem'', Phys. Rep. {\bf 215} (1992), 1.

\bibitem{TPC} J. T. Titantah, C. Pierleoni and S. Ciuchi:
    ``Free energy of the Fr\"ohlich polaron in two and three dimensions'', 
     Phys. Rev. Lett. {\bf 87} (2001), 206406 [arXiv:cond-math/0010386].

\bibitem{Ing} G.-L. Ingold: ``Path integrals and their application to dissipative systems'',  
   in: {\it Coherent Evolution in Noisy Environments}, 
   Lecture Notes in Physics, vol. 611, pp. 1-53, Springer (2002) [arXiv:quant-ph/0208026].

\bibitem{CaLeg} A. O. Caldeira and A. J. Leggett: ``Quantum tunnelling in a 
       dissipative system'', Ann. Phys. {\bf 149} (1983), 374.

\bibitem{HaZw} A. Hanke and W. Zwerger: ``Density of states of a damped 
     quantum oscillator'', Phys. Rev. E {\bf 52} (1995), 6875. 

\bibitem{Ros2} R.~Rosenfelder: ``Structure function of a damped harmonic 
    oscillator'', Phys. Rev. C {\bf 68} (2003), 034602  
    [arXiv:nucl-th/0303003].

\bibitem{SuZh} Jun-Chen Su and Fu-Hou Zheng: ``Correct path-integral formulation of the quantum 
   thermal field theory in the coherent state representation'', Commun. Theor. Phys. {\bf 43} (2005), 
   641 [arXiv: hep-th/0510131].

\bibitem{Lutt} J. M. Luttinger: ``The asymptotic evaluation of a class of path integrals'',
    J. Math. Phys. {\bf 24} (1983), 2070; {\bf 23} (1982), 1011.

\bibitem{AdGeLe} J. Adamowski, B. Gerlach and H. Leschke: ``Strong-coupling limit of 
    polaron energy, revisited'', Phys. Lett. {\bf 79 A} (1980), 249.

\bibitem{Pekar} S. I. Pekar: ``Untersuchungen \"uber die Elektronentheorie der Kristalle'',
    Akademie-Verlag, Berlin (1954).

\bibitem{LiebThom} E. H. Lieb and L. E. Thomas: ``Exact ground state energy of the 
    strong-coupling polaron'', Comm. Math. Phys. {\bf 183} (1997), 511. 

\bibitem{ToSev} T. Tomoda and A. Sevgen: ``Path integral approach to nuclear pairing field'', 
Z. Phys. A {\bf 304} (1982) 221.

\bibitem{KPKB} H. Kleinert, A. Pelster, B. Kastening and M. Bachmann: ``Recursive graphical 
   construction of Feynman diagrams and their multiplicities in $\phi^4$- and in 
   $\phi^2 A$-theory'', 
   Phys. Rev. E {\bf 62} (2000), 1537 [arXiv:hep-th/9907168].

\bibitem{HHL} L. T. Hue, H. T. Hung and H. N. Long: ``General formula for symmetry factors of 
   Feynman diagrams'', arXiv:1011.4142 [hep-th].


\bibitem{Kop} P. Kopietz: ``Two-loop $\beta$-function from the exact renormalization group'',
    arXiv: hep-th/0007128; (in the published version 
    Nucl. Phys. B {\bf 595} (2001), 493 the appendix has been removed!)

\bibitem{Wale} J. D. Walecka: " A theory of highly condensed matter", Ann. Phys. {\bf 83} (1974), 491.

\bibitem{SeWa} B. D. Serot and J. D. Walecka: ``The relativistic nuclear many-body problem'', 
    Adv. Nucl. Phys. {\bf 16} (1986), 1.
    
\bibitem{HoSe} C. J. Horowitz and B. D. Serot: "Selfconsistent Hartree description of finite nuclei in 
    a relativistic quantum field theory", Nucl. Phys. A {\bf 368} (1981), 503.

\bibitem{CoWei} S. Coleman and E. Weinberg: ``Radiative corrections as the origin of 
    spontaneous symmetry breaking'', Phys. Rev. D {\bf 7} (1973), 1888. 

\bibitem{Schub} Ch. Schubert: ``Perturbative quantum field theory in the string inspired 
         formalism'', Phys. Rep. {\bf 355} (2001), 73.

\bibitem{RSS} M. Reuter, M. G. Schmidt and Ch. Schubert: ``Constant external fields 
    in gauge theory and the spin 0, 1/2, 1 path integrals'', Ann. Phys. {\bf 259} 
    (1997), 313.
 
\bibitem{FrGi} E. S. Fradkin and D. M. Gitman: ``Path-integral representation for the 
     relativistic particle propagators and BFV quantization'', 
     Phys. Rev. D {\bf 44} (1991), 3230.

\bibitem{ARS} C. Alexandrou, R. Rosenfelder and A. W. Schreiber: ``Worldline path integral 
      for the massive Dirac propagator: A four-dimensional approach'',
      Phys. Rev. A {\bf 59} (1999), 1762.

\bibitem{Wil} R. M. Wilcox: ``Exponential operators and parameter differentiation in 
     quantum physics'', J. Math. Phys. {\bf 8} (1967), 962.

\bibitem{BDZVH} L. Brink, S. Deser, B. Zumino, P. Di Vecchia and P. Howe: 
   ``Local supersymmetry for spinning particles'', Phys. Lett. B {\bf 64} (1976), 435.

\bibitem{DHGa} E. D'Hoker and D. G. Gagn\'{e}: "Worldline path integrals for fermions with general  
       couplings", Nucl. Phys. B {\bf 467} (1996), 297 [arXiv: hep-th/9512080v2].

\bibitem{Lun} F. A. Lunev: "Pure bosonic worldline path integral representation for the fermionic     
      determinant, non-abelian Stokes theorem and quasiclassical approximation in QCD", 
      Nucl. Phys. B {\bf  494} (1997), 433 [arXiv: hep-th/9609166].

\bibitem{Fuji} K. Fujikawa: ``Path-integral measure for gauge-invariant fermion theories'',
      Phys. Rev. Lett. {\bf 42} (1979), 1195. 

\bibitem{Reu} M. Reuter: ``Chiral anomalies and zeta-function regularization'',  
     Phys. Rev. D {\bf 31} (1985), 1374. 

\bibitem{Stein} J. Steinberger:  ``On the use of subtraction fields and the lifetimes of some 
     types of meson decay'', Phys. Rev. {\bf 76} (1949), 1180.

\bibitem{BaeWie} O. B\"ar and U.-J. Wiese: ``Can one see the number of colors?'',
      Nucl. Phys. B {\bf 609} (2001), 225.

\bibitem{Lep} G. P. Lepage: ``Redesigning Lattice QCD'', in: {\it Perturbative and
         nonperturbative aspects of quantum field theory},
         Schladming 1996, Lecture Notes in Physics, vol. 479,  p. 1 - 48, Springer.

\bibitem{ABLS} G. Arnold, B. Bunk, T. Lippert and K. Schilling: ``Compact QED under 
     scrutiny: it's first order'', Nucl. Phys. Proc. Suppl. {\bf 119} (2003), 864 
     [arXiv: hep-lat/0210010].


\bibitem{Pech} Ph. Pechukas: ``Time-dependent semiclassical scattering theory. I. Potential scattering'',
      Phys. Rev. {\bf 181} (1969) 166
      
\bibitem{MaRoRy} Th. Mannel, W. Roberts and Z. Ryzak: "A derivation of the heavy quark effective        
           lagrangian from QCD", Nucl. Phys. B {\bf 368} (1992) 204.


\end{thebibliography}
\end{document}